\RequirePackage[T1]{fontenc}
\RequirePackage[utf8]{inputenc} 
\RequirePackage[dvipsnames]{xcolor}
\documentclass[12pt,a4paper]{article}
\pdfoutput=1
\usepackage[hyphens]{url} 
\definecolor{linkc}{rgb}{.8,.15,1}
\usepackage[pagebackref=true,colorlinks=true,linkcolor=linkc,citecolor=linkc,urlcolor=linkc,linktoc=all,pdfusetitle=true]{hyperref}
\renewcommand*{\backref}[1]{}
\renewcommand*{\backrefalt}[4]{{%
		\ifcase #1 
		\or [Cited: pg.~#2.]%
		\else [Cited: pgs. #2.]%
		\fi%
	}}

\usepackage[top=80pt,bottom=85pt,left=54pt,right=54pt]{geometry} 
\usepackage[utf8]{inputenc} 
\usepackage[T1]{fontenc}

\input{epsf}
\usepackage{epsfig}
\usepackage{amssymb}
\usepackage{amsfonts}
\usepackage{amsbsy}
\usepackage[all]{xy}
\usepackage{amsmath}
\usepackage{float}
\usepackage{accents}
\usepackage{empheq}
\usepackage{bm}
\usepackage{tikz-cd}
\usetikzlibrary{shapes.geometric}
\usepackage{soul}

\usepackage{amssymb,amscd}
\usepackage{mathrsfs}
\usepackage{amsmath,amsthm}
\usepackage{stmaryrd}
\usepackage{longtable}

\makeatletter
\newbox\LT@firstfoot
\def\endfirstfoot{\LT@end@hd@ft\LT@firstfoot}
\newdimen\LT@footdiff
\def\LT@start{%
  \let\LT@start\endgraf
  \endgraf\penalty\z@
  \vskip\LTpre\endgraf
  \LT@footdiff-\ht\LT@foot
  \advance\LT@footdiff\ht\LT@firstfoot
  \dimen@\pagetotal
  \advance\dimen@ \ht\ifvoid\LT@firsthead\LT@head\else\LT@firsthead\fi
  \advance\dimen@ \dp\ifvoid\LT@firsthead\LT@head\else\LT@firsthead\fi
  \advance\dimen@ \ht\ifvoid\LT@firstfoot\LT@foot\else\LT@firstfoot\fi
  \dimen@ii\vfuzz
  \vfuzz\maxdimen
  \setbox\tw@\copy\z@
  \setbox\tw@\vsplit\tw@ to \ht\@arstrutbox
  \setbox\tw@\vbox{\unvbox\tw@}%
  \vfuzz\dimen@ii
  \advance\dimen@ \ht
      \ifdim\ht\@arstrutbox>\ht\tw@\@arstrutbox\else\tw@\fi
  \advance\dimen@\dp
      \ifdim\dp\@arstrutbox>\dp\tw@\@arstrutbox\else\tw@\fi
  \advance\dimen@ -\pagegoal
  \ifdim \dimen@>\z@\vfil\break\fi
  \global\@colroom\@colht
  \ifvoid\LT@firstfoot
    \ifvoid\LT@foot
    \else
      \advance\vsize-\ht\LT@foot
      \global\advance\@colroom-\ht\LT@foot
      \dimen@\pagegoal\advance\dimen@-\ht\LT@foot\pagegoal\dimen@
      \maxdepth\z@
    \fi
  \else
    \advance\vsize-\ht\LT@firstfoot
    \global\advance\@colroom-\ht\LT@firstfoot
    \dimen@\pagegoal\advance\dimen@-\ht\LT@firstfoot\pagegoal\dimen@
    \maxdepth\z@
  \fi
  \ifvoid\LT@firsthead\copy\LT@head\else\box\LT@firsthead\fi\nobreak
  \output{\LT@output}%
}
\def\LT@output{%
  \ifnum\outputpenalty <-\@Mi
    \ifnum\outputpenalty > -\LT@end@pen
      \LT@err{floats and marginpars not allowed in a longtable}\@ehc
    \else
      \setbox\z@\vbox{\unvbox\@cclv}%
      \ifdim \ht\LT@lastfoot>\ht\LT@foot
        \dimen@\pagegoal
        \advance\dimen@-\ht\LT@lastfoot
        \ifdim\dimen@<\ht\z@
          \setbox\@cclv\vbox{\unvbox\z@\copy\LT@foot\vss}%
          \@makecol
          \@outputpage
          \setbox\z@\vbox{\box\LT@head}%
        \fi
      \fi  
      \global\@colroom\@colht
      \global\vsize\@colht   
      \vbox
        {\unvbox\z@\box\ifvoid\LT@lastfoot\LT@foot\else\LT@lastfoot\fi}%
    \fi
  \else
    \ifvoid\LT@firstfoot
      \setbox\@cclv\vbox{\unvbox\@cclv\copy\LT@foot\vss}%
      \@makecol
      \@outputpage
      \global\vsize\@colroom
    \else
      \setbox\@cclv\vbox{\unvbox\@cclv\box\LT@firstfoot\vss}%
      \@makecol
      \@outputpage
      \global\advance\@colroom\LT@footdiff
      \global\vsize\@colroom
    \fi
    \copy\LT@head\nobreak
  \fi
}

\def\be{ \begin{equation} }
\def\ee{ \end{equation}}

\def\Aut{{\rm Aut}}

\def\det{{\rm det}}
\def\dim{{\rm dim\,}}

\def\vol{{\mathsf{vol}}}

\def\one{{\hbox{ 1\kern-.8mm l}}}

\def\CA{{\cal A}}
\def\CB{{\cal B}}
\def\CC {{\cal C}}

\def\CF {{\cal F}}
\def\CG {{\cal G}}

\def\CJ {{\cal J}}

\def\CL {{\cal L}}
\def\CM {{\cal M}}
\def\CN {{\cal N}}
\def\CO {{\cal O}}

\def\CX {{\cal X}}
\def\CO {{\cal O}}

\def\CG {{\cal G}}

\def\CB {{\cal B}}
\def\CQ {{\cal Q}}

\def\CT {{\cal T}}

\def\CX {{\cal X}}

\def\IA{\mathbb{A}}
\def\IB{\mathbb{B}}
\def\IC{\mathbb{C}}

\def\IE{\mathbb{E}}

\def\IG{\mathbb{G}}
\def\IH{\mathbb{H}}

\def\IM{\mathbb{M}}

\def\IP{\mathbb{P}}
\def\IQ{\mathbb{Q}}
\def\IR{{\mathbb{R}}}
\def\IS{{\mathbb{S}}}

\def\IV{{\mathbb{V}}}
\def\IX{{\mathbb{X}}}
\def\IZ{{\mathbb{Z}}}

\def\fg{\mathfrak{g}}

\def\rmk#1{\bigskip\noindent{\bf Remarks} }

\usepackage[most]{tcolorbox}

\newcommand{\eqa}[1]{\begin{alignat}{4}#1\end{alignat}}
\newcommand{\eqafour}[1]{\begin{alignat}{6}#1\end{alignat}}
\newcommand{\beqa}[1]{\begin{empheq}[box=\fbox]{alignat=4}#1\end{empheq}} 
\newcommand{\eqas}[1]{\begin{equation}\begin{alignedat}{3}#1\end{alignedat}\end{equation}}
\newcommand{\eqasfour}[1]{\begin{equation}\begin{alignedat}{4}#1\end{alignedat}\end{equation}}
\newcommand{\beqas}[1]{\begin{equation}\boxed{\begin{alignedat}{3}#1\end{alignedat}}\end{equation}} 

\newcommand{\sfd}{\mathsf{d}}
\newcommand{\adsf}{\mathsf{ad}\,}
\newcommand{\MET}{\mathsf{Met}}
\newcommand{\DIFF}{\mathsf{Diff}^{+}}
\newcommand{\diff}{\mathsf{diff}}
\newcommand{\Diff}{{\ensuremath \mathsf{Diff}^+}}
\newcommand{\LIE}{\mathsf{Lie\,}}
\newcommand{\sfZ}{\mathsf{Z}}

\newcommand*{\dt}[1]{%
  \accentset{\mbox{\large\bfseries .}}{#1}}

\DeclareUnicodeCharacter{0393}{\Gamma}
\DeclareUnicodeCharacter{0394}{\Delta}
\DeclareUnicodeCharacter{0398}{\Theta}
\DeclareUnicodeCharacter{039B}{\Lambda}
\DeclareUnicodeCharacter{039E}{\Xi}
\DeclareUnicodeCharacter{03A0}{\Pi}
\DeclareUnicodeCharacter{03A3}{\Sigma}
\DeclareUnicodeCharacter{03A5}{\Upsilon}
\DeclareUnicodeCharacter{03A6}{\Phi}
\DeclareUnicodeCharacter{03A8}{\Psi}
\DeclareUnicodeCharacter{03A9}{\Omega}
\DeclareUnicodeCharacter{03B1}{\alpha}
\DeclareUnicodeCharacter{03B2}{\beta}
\DeclareUnicodeCharacter{03B3}{\gamma}
\DeclareUnicodeCharacter{03B4}{\delta}
\DeclareUnicodeCharacter{03B5}{\epsilon}
\DeclareUnicodeCharacter{03B6}{\zeta}
\DeclareUnicodeCharacter{03B7}{\eta}
\DeclareUnicodeCharacter{03B8}{\theta}
\DeclareUnicodeCharacter{03D1}{\vartheta}
\DeclareUnicodeCharacter{03B9}{\iota}
\DeclareUnicodeCharacter{03BA}{\kappa}
\DeclareUnicodeCharacter{03BB}{\lambda}
\DeclareUnicodeCharacter{03BC}{\mu}
\DeclareUnicodeCharacter{03BD}{\nu}
\DeclareUnicodeCharacter{03BE}{\xi}
\DeclareUnicodeCharacter{03C0}{\pi}
\DeclareUnicodeCharacter{03C1}{\rho}
\DeclareUnicodeCharacter{03C3}{\sigma} 
\DeclareUnicodeCharacter{03C4}{\tau}
\DeclareUnicodeCharacter{03C5}{\upsilon}
\DeclareUnicodeCharacter{03C6}{\phi}
\DeclareUnicodeCharacter{03D5}{\varphi}
\DeclareUnicodeCharacter{03C7}{\chi}
\DeclareUnicodeCharacter{03C8}{\psi}
\DeclareUnicodeCharacter{03C9}{\omega}
\DeclareUnicodeCharacter{21D0}{\Leftarrow}
\DeclareUnicodeCharacter{0212B}{\AA}
\DeclareUnicodeCharacter{00B7}{\cdot}
\DeclareUnicodeCharacter{00B0}{^{\circ}}
\DeclareUnicodeCharacter{266A}{\eighthnote}
\DeclareUnicodeCharacter{266B}{\twonotes}
\DeclareUnicodeCharacter{2202}{\partial}

\newcommand{\dA}{\dt{A}}
\newcommand{\dB}{\dt{B}}
\newcommand{\dC}{\dt{C}}
\newcommand{\dD}{\dt{D}}
\newcommand{\dE}{\dt{E}}
\newcommand{\dF}{\dt{F}}
\newcommand{\dG}{\dt{G}}

\newcommand{\sq}{\sqrt{2}\,}
\newcommand{\bn}{\bm{η}}

\newcommand{\nn}{\nonumber\\}
\newcommand{\veps}{\varepsilon}
\newcommand{\n}{\bm{\nabla}}
\newcommand{\sym}{\text{sym}}
\newcommand{\eps}{\bm{\epsilon}}

\newcommand{\bu}[1]{\overrightarrow{\bm{#1}}}

\newcommand{\ha}{a}
\newcommand{\hb}{b}
\newcommand{\hg}{c}
\newcommand{\hd}{d}

\newcommand{\sfA}{\mathsf{A}}
\newcommand{\sfB}{\mathsf{B}}
\newcommand{\sfC}{\mathsf{C}}
\newcommand{\sfF}{\mathsf{F}}
\newcommand{\sfG}{\mathsf{G}}
\newcommand{\sfH}{\mathsf{H}}
\newcommand{\sfP}{\mathsf{P}}
\newcommand{\sfQ}{\mathsf{Q}}
\newcommand{\sfR}{\mathsf{R}}
\newcommand{\sfS}{\mathsf{S}}
\newcommand{\sfX}{\mathsf{X}}
\newcommand{\sfY}{\mathsf{Y}}
\newcommand{\sfL}{\mathsf{\Lambda}}

\newcommand{\wt}{\widetilde}
\newcommand{\wh}{\widehat }

\newcommand{\bcd}{\bm{\mathcal{D}}}
\newcommand{\scd}{\bm{\mathsf{D}}}

\usepackage[normalem]{ulem}
\usepackage{cancel}

\newcommand{\highlightYellow}[1]{%
  \colorbox{yellow!50}{$\displaystyle#1$}}

  \newcommand{\highlightOliveGreen}[1]{%
  \colorbox{OliveGreen!20}{$\displaystyle#1$}}

        \newcommand{\highlightOrange}[1]{%
  \colorbox{orange!40}{$\displaystyle#1$}}

\usepackage{stackrel}

\newcommand{\ov}[1]{\overline{#1}}

\newcommand{\ThetaD}{\bm{\Theta}^{\rm{(}\mathsf{D}\rm{)}}}
\newcommand{\ThetaR}{\bm{\Theta}^{\rm{(}\mathsf{R}\rm{)}}}
\newcommand{\ARsym}{A^{\rm{(}\mathsf{R}\rm{)}}}

\newcommand{\masd}{{\ensuremath {\mathcal{M}}_{\mathsf{ASD}}}}

\usepackage{framed}
\definecolor{shadecolor}{gray}{0.9}

\usepackage{multirow}
\usepackage[bottom,hang,flushmargin]{footmisc} 

\newcommand{\SCB}{\mathscr{B}}

\numberwithin{equation}{section}
\numberwithin{table}{section}
\usepackage{cite}
\usepackage{mcite}

\begin{document}

\begin{titlepage}  

    \vskip .5in
	\noindent
 
    \begin{flushright}
        \begin{tabular}{@{}l@{}}
           YITP-SB-2023-32
    \end{tabular}
    \end{flushright}
       
    \vskip 1.5in

    \begin{center}
        {\Large \bf{SUPERCONFORMAL GRAVITY AND  
        \\ \vspace{3mm} THE TOPOLOGY OF DIFFEOMORPHISM GROUPS}}

        \bigskip\medskip

        {Jay Cushing$^1$, Gregory W. Moore$^1$, Martin Ro\v{c}ek$^2$, Vivek Saxena$^1$}

        \bigskip\medskip

        {\small $^1$ New High Energy Theory Center and Department of Physics and Astronomy,\\Rutgers University, 126 Frelinghuysen Rd., Piscataway, NJ 08855-0849, USA 
        \\
        \vspace{3mm}
        $^2$ C.N. Yang Institute for Theoretical Physics and Department of Physics and Astronomy,\\Stony Brook University, Stony Brook, NY 11794-3840, USA
        }

        \vskip 5mm
        {\small \tt \href{mailto:jay.b.cushing@gmail.com}{jay.b.cushing@gmail.com}, \href{mailto:gmoore@physics.rutgers.edu}{gmoore@physics.rutgers.edu}, \href{mailto:martin.rocek@stonybrook.edu}{martin.rocek@stonybrook.edu}, \href{mailto:vivek.hepth@gmail.com}{vivek.hepth@gmail.com}}

        \vskip 9mm
        {\bf Abstract}
        \vskip .1in
    \end{center}

    \noindent Twisted four-dimensional supersymmetric Yang-Mills theory
    famously gives a useful point of view on the Donaldson and Seiberg-Witten invariants of
    four-manifolds. In this paper we generalize the construction to include a
    path integral formulation of generalizations of Donaldson invariants
    for smooth \underline{families} of four-manifolds. Mathematically these are equivariant
    cohomology classes for the action of the oriented diffeomorphism group on
    the space of metrics on the manifold. In principle these cohomology classes should
    contain nontrivial information about the topology of the diffeomorphism
    group of the four-manifold.  We show that the invariants may be
    interpreted as the standard topologically twisted path integral of four-dimensional
    $\CN=2$ supersymmetric Yang-Mills coupled to topologically twisted background
    fields of conformal supergravity. 
    \\\\
    November 14, 2023.

    \vfill
    \eject
    
\end{titlepage}

\tableofcontents

\section{Introduction And Conclusion}\label{sec:IntroductionConclusion}

This paper is a contribution to the
ongoing story of the mathematics-physics dialogue concerning
the relation between mathematical invariants of smooth four-dimensional
manifolds and the physics of supersymmetric Yang-Mills theory. 

In \cite{DonaldsonDurham} Donaldson proposed an interesting generalization
of his eponymous invariants of smooth manifolds. The main idea is to generalize from a single four-manifold to \underline{smooth families} of four-manifolds.
In the subsequent years there has been some mathematical work on this topic \cite{Hatcher:2012,Morava:2000yk,li2001family,Nakamura2003,*Nakamura2003corr,Ebert2011,Galatius2012stable,Galatius2013detecting,Ebert2014,Galatius2017homologicalI,Galatius2017homologicalII,Galatius2017TautRings,Galatius2019moduli,Baraglia2023}. Nevertheless, 
in his wonderful review of Seiberg-Witten theory \cite{DonaldsonReview} Donaldson
concludes by reminding the reader that the family invariants have been relatively
unexplored. Similarly, while the concluding paragraphs of \cite{Moore:1997pc} indicate that a path integral viewpoint on the family Donaldson invariants should follow
by coupling twisted super Yang-Mills to supergravity, no further investigation of
this idea has -- to our knowledge -- yet appeared. In this paper we fill this
25-year old gap.

To set the stage let us review the field-theoretic formulation of Donaldson invariants initiated by Witten in \cite{Witten:1988ze}. 
Consider a supersymmetric four-dimensional Yang-Mills theory; like any Yang-Mills theory coupled to fermionic and scalar fields, it can be formulated on an arbitrary smooth (pseudo-) Riemannian four-manifold $\IX$ equipped (for the moment) with a spin structure. In general, the background fields break supersymmetry. To preserve the supersymmetry one should couple to supergravity.
One of the background supergravity fields is, of course, the metric on the four-manifold. Theories with $\CN=2$ supersymmetry have a global $\mathsf{R}$-symmetry, and in general, one should also introduce background fields that couple to this symmetry. 
  Mathematically, one introduces a principal $\mathsf{SU(2)}$ bundle $\mathsf{P}_{\mathsf{R}} \to \IX$ known as the ``$\mathsf{R}$-symmetry bundle,'' and equips it with a connection 
$\n_{\mathsf{R}}  =d + V^{\mathsf{R}} $ known as the ``$\mathsf{R}$-symmetry connection.''  Indeed, this connection is part of the $\CN = 2$ superconformal gravity multiplet, so in effect, we are ``turning on'' another field in the background supergravity multiplet. Having phrased things this way it is natural to introduce other background fields in the $\CN=2$ superconformal gravity multiplet and indeed that will be a central theme of the present paper. For the moment we will only ``turn on'' the background metric and $\mathsf{R}$-symmetry connection.
In this way, the partition function 
of an $\CN = 2$ theory becomes a functional 
\be\label{eq:UntwistPF}
\mathsf{Z}\big[g_{\mu\nu}, V^{\mathsf{R}}_{\mu} \big] = \int [d\Phi_{\rm vm}] e^{S[\Phi_{\rm vm}; g_{\mu\nu}, V^{\mathsf{R}}_{\mu}]} ~.
\ee

When the metric $g_{\mu\nu}$ and $\mathsf{R}$-symmetry connection $V^{\mathsf{R}}_{\mu}$ are related to each other in a very special way (``topological twisting'') the path integral \eqref{eq:UntwistPF} becomes a topological invariant. 
We now review the idea of topological twisting.\footnote{What follows is a slightly novel way of phrasing topological twisting. It has independently been noted by Dan Freed, who furthermore put it in the framework of the Erlangen program.}
The above remarks apply to theories in both Lorentzian and Euclidean signatures. In this paper, we will focus on the Euclidean signature theory. 
Thus, on a spin four-manifold, the theory can be coupled to a bundle with connection, with structure group $\mathsf{SU(2)_+} \times \mathsf{SU(2)_-}  \times \mathsf{SU(2)}_{\mathsf{R}}$. 
We note that the spinor 
fields of the model transform in the representation 
\be\label{eq:FermionRep}
(\textbf{2}, \textbf{1}, \textbf{2}) \oplus  (\textbf{1}, \textbf{2}, \textbf{2}) ~,
\ee
and therefore factor through a representation of
\be 
B := (\mathsf{SU(2)_+} \times \mathsf{SU(2)_-}  \times \mathsf{SU(2)}_{\mathsf{R}})/\langle (-1,-1,-1) \rangle ~,
\ee
where $\langle (-1,-1,-1) \rangle \cong \IZ_2$ 
is a central subgroup.   Thus there is an injective homomorphism $H\hookrightarrow B$, 
where 
\be 
H  := (\mathsf{SU(2)_+} \times \mathsf{SU(2)_-})/\langle (-1,-1) \rangle ~,
\ee
defined by 
\be\label{eq:TwistingHomomorphism}
[(u_1, u_2)] \mapsto [(u_1, u_2, u_1)] ~ . 
\ee
The square brackets denote the equivalence class from the quotient by the respective central subgroups. 
Note that $H$ is the structure group of the oriented tangent bundle $T\IX$. Therefore, given a Riemannian metric, 
we can specialize the $\mathsf{R}$-symmetry bundle and connection so that the $B $-bundle with connection reduces to an
$H$-bundle with its Levi-Civita connection under the homomorphism \eqref{eq:TwistingHomomorphism}. Locally we have chosen 
an isomorphism of the principal $\mathsf{SU(2)_+}$ spinor bundle with the $\mathsf{SU(2)}_{\mathsf{R}}$  $\mathsf{R}$-symmetry bundle and an isomorphism of the Levi-Civita spin connection for $\mathsf{SU(2)_+}$ with the $\mathsf{R}$-symmetry connection. Reference  \cite{Witten:1988ze}  makes two remarkable observations about such backgrounds. First of all, the fields in the twisted vectormultiplet are all differential forms and therefore can be defined on any smooth manifold, even if it is not spin. Thus the twisted theory can be extended to nonspin manifolds. Henceforth in this paper, we no longer require the four manifold $\IX$ to be spinnable.  Moreover,  the dependence of the action in equation \eqref{eq:UntwistPF} on the background fields is $\CQ$-exact, where $\CQ$ is an odd vector field on the supermanifold of fields such that $\CQ^2=0$ on gauge invariant fields.  Indeed the action for a vectormultiplet in such a background is of the form 
\be\label{eq:TwistLag1}
\mathsf{S}_{\rm twisted}[\Phi_{\rm vm}; g_{\mu\nu} ]  := \mathsf{S}_{\mathsf{SYM}}[\Phi_{\rm vm}; g_{\mu\nu}, V^{\mathsf{R}}_{\mu} \cong \omega_{\mu}^+] =   \int_{\IX} \theta\, \mathsf{Tr\,} F \wedge F + \CQ\IV ~,
\ee
for a suitable $\IV$. Here  $\Phi_{\rm vm}$ stands for the vectormultiplet fields, 
$\omega^{+}$ is the self-dual part of the spin connection,
$F$ is the field strength of the dynamical vectormultiplet, and $\mathsf{Tr\,}$ is a suitably normalized invariant form on the Lie algebra, and in this paper, 
$\theta$ will be a coupling constant taken to be independent of spacetime. 
Since $\int_{\IX} \mathsf{Tr\,} F \wedge F$ only depends on the topology of the gauge bundle all the metric dependence is in the $\CQ$-exact term $\CQ\IV$.   Therefore, if we consider the path integral with 
the background fields restricted according to the topological twisting procedure   
\be\label{eq:TopTwistPF}
\mathsf{Z}_{\mathsf{DW}}[g_{\mu\nu}] := \int [d\Phi_{\rm vm}] e^{-S_{\rm twisted}[\Phi_{\rm vm}; g_{\mu\nu}] } ~,
\ee
the resulting functional is -- at least formally -- independent of the metric: 
\be\label{eq:MetricIndep} 
\sfd_{\mathsf{Met}} \sfZ_{\mathsf{DW}} = 0 ~,
\ee
where $\sfd_{\mathsf{Met}}$ is the exterior derivative along $\MET(\IX)$, the space of 
smooth metrics on $\IX$. The reason for the qualifier ``formally'' (aside from the 
fact that the path integral itself is not mathematically rigorously defined) is the 
following. The proof of \eqref{eq:MetricIndep} is that
\be\label{eq:MetricIndep2}
\sfd_{\mathsf{Met}} \sfZ_{\mathsf{DW}} = \int [d\Phi_{\rm vm}] \CQ(e^{-S'} ) = 0 ~,
\ee
where $e^{-S'}$ is a local gauge invariant expression in the fields. In the 
second equality above we have used an integration by parts in fieldspace.  
In fact, further analysis has revealed that for manifolds $\IX$ with $b_2^+ \leq 1$ the integration 
by parts can fail. For $b_2^+=1$  the path integral is only piecewise constant 
as a function of the metric \cite{Moore:1997pc}. That is, it undergoes wall-crossing. The case with $b_2^+=0$ is
interesting but somewhat unexplored. It is likely that there will be continuous
metric dependence in this case, but that the continuous metric dependence is 
exactly calculable. 

The topologically twisted theory is closely related to the mathematics of
equivariant cohomology \cite{Atiyah:1990tm,Baulieu:1988xs,Cordes:1994fc,Labastida:2005zz}, 
and that viewpoint will be important in the present paper. Recall that, 
given a principal bundle  $P \to \IX$ for the gauge fields there is a group 
$\CG=\Aut(P)$ of automorphisms of $P$ (a.k.a.~the group of gauge transformations) 
which acts on the affine space  $\CA(P)$ of connections on $P$. The Cartan model of 
$\CG$-equivariant cohomology of $\CA(P)$ introduces a degree two generator $\phi$ of 
the Lie algebra of $\CG$ and a complex\footnote{In this paper all the relevant Lie algebras are equipped with a nondegenerate invariant metric. We use this metric to identify the Lie algebra and its dual throughout the paper.}
\be\label{eq:VM-Cartan-Complex}
\big( \Omega^\bullet(\CA(P)) \otimes S^\bullet[{\mathsf{Lie\,}}\CG]\big)^{\CG} ~,
\ee
with differential (written in physics notation) 
%
\eqas{
\label{eq:VM-EquivCoho} 
&d_{\CC} A_\mu &&= \psi_\mu ~, \\ 
&d_{\CC} \psi_\mu &&= - D_\mu \phi ~, \\ 
&d_{\CC} \phi && = 0 ~,\\ 
}
The field $\psi_\mu dx^\mu$ can be interpreted as a cotangent vector on $\CA(P)$. 

The relation of the twisted super-Yang-Mills theory to equivariant cohomology begins to emerge when we recognize that not only the spinor fields but also the supersymmetry operators transform under the 
 representation \eqref{eq:FermionRep} of $B$. The representation \eqref{eq:FermionRep} pulls back under the homomorphism $H\hookrightarrow B$ of 
equation \eqref{eq:TwistingHomomorphism} to the representation
\be\label{eq:TwistedFermionRep}
(\textbf{2}, \textbf{2}) \oplus  \Big((\textbf{1},\textbf{1}) \oplus (\textbf{3}, \textbf{1})\Big) ~,
\ee
of $H$. 
When applied to supersymmetry transformations the middle term corresponds to the scalar nilpotent symmetry,
$\CQ$. When applied to the fermionic fields the first term defines the odd one-form field $\psi_\mu(x) dx^\mu$ which appears in \eqref{eq:VM-EquivCoho}. In this way, we recover the differential and the fields used to define the complex for the Cartan model of equivariant cohomology \eqref{eq:VM-Cartan-Complex}. (The field $\phi$ is the usual adjoint-valued scalar in the vectormultiplet.) Moreover the supersymmetry transformation generated by the action of $\CQ$ on these fields is precisely the set of equations 
\eqref{eq:VM-EquivCoho} that defines the Cartan differential $d_{\CC}$. Thus, we may identify $d_{\CC}$  with $ \CQ$. 

As explained in detail in \cite{Atiyah:1990tm,Cordes:1994fc,Labastida:2005zz,Losev:1997tp}, localization 
of the path integral using the $\CQ$-symmetry shows that \eqref{eq:TopTwistPF} reduces to an integral of $1$ over the moduli space of instantons.
It thus counts, with signs the contributions of $0$-dimensional moduli spaces of instantons.
These are the simplest of the Donaldson invariants: They simply count -- with signs -- the instantons. 

By including point and surface observables in the action one can generalize this to a metric-independent
function on the homology of $\IX$ which turns out to be a generating function for the Donaldson polynomials.
That is, for $x\in \sfH_{\bullet}(\IX)$ one can define dimensionless $\CQ$-closed observables $\CO(x)$ so that, using the above 
path integrand as a measure the expectation value may be expressed as
\be\label{eq:IncludeObs-NoFamily}
\langle e^{\CO(x)} \rangle = \int_{\CM} e^{\mu_D(x)} ~.
\ee
Here $\CM$ is the moduli space of instantons with nonnegative instanton number 
\be\label{eq:Comp-MASD}
\CM = \coprod_{k\geq 0} \CM_k ~,
\ee
where $\CM_k$ is the moduli space of solutions to the anti-self-dual instanton 
equations $F^+=0$ with instanton number $k$. As explained in \cite{DK}, and elsewhere, for $b_2^+>0$ and generic metric 
it is a smooth manifold of dimension $\dim \CM_{k} = 4h^\vee k - \dim G_{\rm gauge} (b_2^+ - b_1 +1)$. Finally, 
$\mu_{D}: \sfH_\bullet(\IX) \rightarrow \sfH^{4-\bullet}(\CM)$ is the famous Donaldson $\mu$-map. 

Following the work of Seiberg and Witten \cite{Seiberg:1994rs,Seiberg:1994aj}, 
when evaluating physical quantities that only involve the low-energy sector of the theory, 
the path integral of the strongly coupled non-Abelian theory can, in fact,
be equated to the path integral of an effective low energy theory based on an \underline{Abelian}
gauge symmetry. This insight led to the formulation of the renowned Seiberg-Witten invariants 
\cite{Witten:1994cg}, a development that had a profound impact on the development of four-manifold differential topology.
Using the IR formulation of the UV theory one can give path integral derivations of the wall-crossing formula, the blow-up formula, and the ``Witten conjecture'' that expresses the Donaldson invariants 
in terms of the Seiberg-Witten invariants \cite{Moore:1997pc}.  For reviews see 
\cite{Labastida:2005zz,Moore:2017SCGP}.\footnote{The Witten conjecture has since been put on a more rigorous mathematical footing in \cite{Gottsche2010donaldson,Feehan:2014gea,Feehan:2014gsa}.}  

The main aim of this paper is the generalization of the above story to the 
case of smooth families of four-manifolds. From a physical viewpoint, 
we will do this by coupling the theory to a suitably twisted and 
truncated version of the \underline{conformal} supergravity BRST complex.
(The twisted and truncated version is however not a conformal theory.) From the mathematical viewpoint, 
we will be studying the equivariant cohomology of the space of Riemannian metrics $\MET(\IX)$
under the action of the oriented diffeomorphism group $\DIFF(\IX)$.

In more detail, we formulate the $\DIFF(\IX)$-equivariant cohomology of $\MET(\IX)$
by introducing   an odd, degree one, symmetric tensor $\Psi_{\mu\nu}$, which plays the 
role of a one-form on  $\MET(\IX)$ together with an even, degree two, 
vector field $\Phi^\mu$ associated with the Lie algebra of $\DIFF(\IX)$.
The Cartan complex is then -- according to the usual rules -- 
\be\label{eq:CartanGravityComplex}
\left( \Omega^\bullet(\MET(\IX)) \otimes S^\bullet({\LIE}(\DIFF(\IX) ))\right)^{\DIFF(\IX)} ~,
\ee
with differential
%
\eqas{
\label{eq:Metric-EquivCoho}
&d_{\CC} g_{\mu\nu}  && = \Psi_{\mu\nu} ~, \\
&d_{\CC} \Psi_{\mu\nu} && = \n_\mu \Phi_\nu + \n_\nu \Phi_\mu ~,\\
&d_{\CC} \Phi^\mu && = 0 ~.
}
Once again the fields $(g_{\mu\nu}, \Psi_{\mu\nu}, \Phi^\mu)$ have an 
interpretation in terms of a twisted supermultiplet, but now we must look 
to twisted $\CN=2$ superconformal gravity. The field $\Psi_{\mu\nu}$  
arises from components of the gravitini and $\Phi^\mu$ is a bosonic 
BRST ghost for a vector-supersymmetry in superconformal gravity.  
(See section \ref{sec:TwistedScfmlGrav} below for more details.)

Our approach to writing  $\DIFF(\IX)$-equivariantly closed forms on  $\MET(\IX)$
will be to consider the path integral of the theory of a dynamical twisted $\CN=2$ vectormultiplet coupled to a background twisted and 
truncated $\CN=2$ superconformal gravity multiplet. Thus, in the path integral when we integrate out the vectormultiplet, we will obtain an element of the gravity complex \eqref{eq:CartanGravityComplex}.

From the mathematical viewpoint, the procedure can be described as follows. 
The vectormultiplet theory has fields that generate a complex that computes the equivariant cohomology of the group of gauge transformations $\CG = {\rm Aut}(P)$ acting on the space of connections $\CA(P)$.
This complex is described in detail in equations 
\eqref{eq:complex-GaugeCartan} and \eqref{eq:BigVM-Complex} and, before projection to the gauge-invariant subcomplex, has the form:
\eqa{ 
& \widetilde{\IE}^{\rm gauge} &&= \Omega^\bullet(\CA(P)) \otimes S^\bullet[\mathsf{Lie\,}\CG] \otimes \cdots 
\label{eq:CartanAlgebraEllipsis}}
where the $\otimes \cdots$ refers to the other fields from the twisted vectormultiplet.\footnote{There is a 
subtlety here since among the twisted vectormultiplet fields there are self-dual fields, including the 
odd, degree one,  self-dual tensors $\chi_{\mu\nu}$  (arising from the third term in \eqref{eq:TwistedFermionRep}) 
and self-dual auxiliary fields $H_{\mu\nu}$  (arising from the $\CN=2$ D-terms). Thus our fieldspace is in fact fibered over the space of metrics. These subtleties are addressed in  Appendix \ref{app:variation-of-self-dual-fields}.}
All the factors in \eqref{eq:BigVM-Complex} have a nice geometrical meaning. 
We now take a tensor product with a complex $\widetilde{\mathbb{E}}^{\rm diffeo}$ that computes
(after projection to the invariant subcomplex) the $\DIFF(\IX)$-equivariant cohomology of $\MET(\IX)$. 
See equation \eqref{eq:Gravity-DiffCartan} below for more details. The complex $\widetilde{\mathbb{E}}^{\rm diffeo}$ is generated by the fields of the twisted and truncated superconformal gravity multiplet. The product complex $\widetilde{\IE}^{\rm gauge}\otimes \widetilde{\mathbb{E}}^{\rm diffeo}$ computes the 
equivariant cohomology for the natural action of the group 
\be
 \IG := \CG \rtimes \DIFF(\IX) ~, \label{eq:SemiDirect-Intro}
\ee
on $\CA(P) \times \MET(\IX)$.  The full differential on the
$\IG$-invariant subspace of the product complex, denoted
$\IQ$,  is given in 
equations \eqref{eq:CartanAlgebra-1}--\eqref{eq:CartanAlgebra-7}, 
and takes the form 
\eqa{
\label{eq:CartanAlgebra-Intro1}&\IQ g_{\mu\nu} &&= \Psi_{\mu\nu} ~, &&\IQ A_{\mu} &&=\psi_\mu ~,\\
\label{eq:CartanAlgebra-Intro2}&\IQ \Psi_{\mu\nu} &&= \n_{\mu} \Phi_{\nu} + \n_{\nu} \Phi_{\mu} ~, \qquad && \IQ \psi_\mu &&= -D_\mu \phi + \Phi^\sigma F_{\sigma\mu} ~,\\
\label{eq:CartanAlgebra-Intro3}&\IQ \Phi^\sigma &&= 0 ~, \quad && \IQ \phi &&= -\Phi^\sigma \psi_\sigma ~, 
} 
together with transformations for the fields omitted in \eqref{eq:CartanAlgebraEllipsis}. 
Note carefully the terms in the right-hand column of equations 
\eqref{eq:CartanAlgebra-Intro1}-\eqref{eq:CartanAlgebra-Intro3} involving $\Phi^\mu$.
These arise from the important fact that $\IG$ is a semi-direct product. These models of equivariant cohomology are independently derived from the viewpoint of $\CN=2$ supergravity in sections \ref{subsec:Diff-Cartan-from-TwistedSugra} and \ref{subsec:Full-Cartan-from-TwistedSugra}.

Now that we have an algebraic framework for describing the twisted gauge theory vectormultiplet coupled to a twisted and truncated background superconformal gravity multiplet we can view the path integral over the vectormultiplet fields as a pushforward on cohomology. To describe this pushforward we formulate a generalization of Witten's UV action in equations \eqref{eq:GeneralAction}--\eqref{eq:GeneralActionTotDer} below. 
The path integral of a twisted vectormultiplet coupled to a background twisted and truncated superconformal gravity multiplet is:
\be\label{eq:FamilyTopTwistPF}
\mathsf{Z}_{\mathsf{family}}[g_{\mu\nu}, \Psi_{\mu\nu}, \Phi^\mu] = \int [d\Phi_{\rm vm}] e^{-S[\Phi_{\rm vm}; g_{\mu\nu}, \Psi_{\mu\nu}, \Phi^\mu]} ~.
\ee
On formal grounds $\mathsf{Z}_{\mathsf{family}}[g_{\mu\nu}, \Psi_{\mu\nu}, \Phi^\mu] $ is  expected to be an equivariantly closed form for the Cartan model of
$\DIFF(\IX)$-equivariant cohomology of $\MET(\IX)$. The degree zero 
part is just $Z_{\mathsf{DW}}$ described above, and the equivariant closure generalizes the equation 
\eqref{eq:MetricIndep}.  The $\Psi_{\mu\nu}$
represent cotangent vectors on  $\MET(\IX)$. Expanding the partition function as a formal
series in $\Psi_{\mu\nu}$ and $\Phi^\mu$ gives a series of forms on $\MET(\IX)$ of increasing 
degree in the Cartan complex: 
\eqa{
& \sfZ_{\mathsf{family}} &&= \mathsf{Z}^{(0)} + \sum_{\substack{p,q\,=\,0 \\ p+q\,\geq\,1}}^{\infty}\int_{\IX^{p+q}}\mathsf{Z}^{(p,q)}_{\mu_1\nu_1\cdots\mu_p\nu_p;\rho_1\cdots\rho_q}(x_1,\ldots,x_{p+q})\prod_{j=1}^{p}\Psi^{\mu_j \nu_j}(x_j)\prod_{k=1}^{q} \Phi^{\rho_k}(x_{p+k})\prod_{\ell=1}^{p+q}\vol(g(x_\ell))\nn
& &&= \sfZ^{(0)} + \int_{\IX}\sfZ_{\mu\nu}^{(1,0)} \Psi^{\mu\nu}\vol(g) \nn
& &&\quad + \left(\int_{\IX\times\IX}\sfZ_{\mu_1\nu_1\mu_2\nu_2}^{(2,0)}(x_1, x_2) \Psi^{\mu_1\nu_1}(x_1)\Psi^{\mu_2\nu_2}(x_2)\vol(g(x_1))\vol(g(x_2)) + \int_{\IX} \sfZ_{\mu}^{(0,1)}(x) \Phi^{\mu}(x) \vol(g(x)) \right) \nn
& &&\quad + \cdots \label{eq:Psi-Phi-Expansion}
}
For example, we can view $\int_{\IX} \mathsf{Z}^{(1,0)}_{\mu\nu} \Psi^{\mu\nu} {\rm vol}(g)$ in the first line of 
\eqref{eq:Psi-Phi-Expansion} as an equivariantly closed $1$-form on $\MET(\IX)$ which descends to a closed $1$-form on 
$\MET(\IX)/\DIFF(\IX)$. This closed $1$-form could in principle be non-exact, thus defining a nontrivial cohomology class of degree one. 
Similarly, the expression in line 2 is expected to be equivariantly closed on general grounds, and could, in principle define a nontrivial cohomology class of 
degree two, etc. 

Neglecting subtleties related to manifolds with isometries, the quotient
$\MET(\IX)/\DIFF(\IX)$ is a model for $\mathsf{BDiff}^+(\IX)$ since $\MET(\IX)$ is contractible.\footnote{For some basic facts about the differential geometry of $\MET(\IX)/\DIFF(\IX)$  see \cite{FreedGroisser:1989}.}
In this way, the path integral is expected to define elements of the cohomology of $\mathsf{BDiff}^+(\IX)$. 
    In particular, the integration of the degree $n$
component of \eqref{eq:FamilyTopTwistPF} over $n$-dimensional families of metrics that
project to nontrivial cycles in   $\mathsf{BDiff}^+(\IX)$ are expected to depend only
on the homology class of the cycle in $\mathsf{BDiff}^+(\IX)$. This is the generalization of the metric independence of Witten's path integral \eqref{eq:TopTwistPF}. Moreover, since\footnote{We are ignoring fixed points of the $\DIFF$ action on $\MET$, which we will return to in section \ref{sec:Observables}.} 
$\pi_{k}(\mathsf{BDiff}(\IX)) = \pi_{k-1}(\mathsf{Diff}(\IX))$, 
%
these invariants encode information about the topology of the orientation-preserving diffeomorphism group $\DIFF(\IX)$, hence the title of this paper.

As in the case of a single 4-manifold, we expect there to be a UV and IR formulation of the path integrals that compute these equivariant cohomology classes, and the equality of these formulations 
is expected to give rather nontrivial mathematical predictions, generalizing the ``Witten conjecture'' mentioned above. We comment more about this in our brief remarks on wall-crossing below. 
To perform computations and test ideas we will need explicit actions. One of the main results of this 
paper is a complete formulation of suitable actions, both in the IR and the UV.

 We construct the actions in several 
different ways. One way  begins with the standard actions which are degree zero in the 
gravity complex and then successively adds terms of higher order in an  expansion in $\Psi_{\mu\nu}$ 
and $\Phi^\mu$ using $\IQ$-closure as a guide.  The expansion is initiated by noting that, at degree zero the 
energy-momentum tensor of $\Phi_{\rm vm}$ is $\CQ$-exact, $T_{\mu\nu} = \{ \CQ, \Lambda_{\mu\nu} \}$. 
Since the variation of the twisted vectormultiplet action with respect to the metric is 
\eqa{
& \delta \IS_{\rm twisted}[\Phi_{\rm vm} ; g_{\mu\nu} ] &&= \frac{1}{2} \int_{\IX} \vol(g) \delta g^{\mu\nu} T_{\mu\nu} ~,
}
it follows that the action coupled to the twisted-truncated background supergravity up to first order must 
be of the form  
\eqa{
& \IS[\Phi_{\rm vm}; g_{\mu\nu}, \Psi_{\mu\nu}, \Phi^\mu] &&= \IS_{\rm twisted}[\Phi_{\rm vm} ; g_{\mu\nu} ] + 
\int_{\IX} \vol(g) \Psi^{\mu\nu} \Lambda_{\mu\nu}  + \cdots ~. \label{eq:S gravity expansion}
}
%
%
A minimal action with this leading behavior can be obtained by choosing a suitable primitive $\IV$ so that the total action is of the form of a nonexact topological action plus $\IQ(\IV )$. This action terminates at degree two in the degree as measured by the gravitational Cartan complex. The actions derived in this way are presented in equation \eqref{eq:GeneralAction} et. seq.


%
%
Motivated by methods for writing actions for general $\CN=2$ matter coupled to $\CN=2$ supergravity, we propose an action in section \ref{sec:TwistedSugraAction} that reproduces the actions in \cite{Witten:1988ze,Moore:1997pc,Marino:1998bm} at degree zero and is $\IQ$-closed (up to surface terms). The action derived in this way terminates at gravitational degree four and is presented in \eqref{eq:TwistedSugraAction-General-Proposal} et. seq. After some field redefinitions 
the two sets of actions are shown to be the same -- up to $\IQ$-exact terms in 
section \ref{sec:CompareActions}. 

Given the explicit expressions for the action we can now give quite explicit expressions for the 
expansion in \eqref{eq:Psi-Phi-Expansion} as correlation functions for the theory of a topologically twisted vectormultiplet. For example letting $\sfZ^{(p,q)}$ denote the expression appearing with $p$ factors of $\Psi^{\mu\nu}$ and $q$ factors of $\Phi^\mu$ in \eqref{eq:Psi-Phi-Expansion}, we can say that 
\be \label{eq:Zn0}
\sfZ^{(n,0)}_{\mu_1\nu_1\cdots \mu_n \nu_n}(x_1,\dots, x_n) = 
\biggl\langle \Lambda_{\mu_1\nu_1}(x_1) \cdots \Lambda_{\mu_n\nu_n}(x_n) \biggr\rangle + \cdots 
\ee 
where the correlation function is computed in the twisted vectormultiplet theory coupled to a Riemannian metric with $\Phi=\Psi=0$ and the $+\cdots$ arises from terms in the action of gravity degree two. 

The expressions \eqref{eq:Zn0} present a striking contrast to the usual expressions for Donaldson invariants (the zeroth order term in the expansion of $\sfZ_{\mathsf{family}}$). Here we are working with the correlation functions of operators, such as $\Lambda_{\mu\nu}(x)$, which are \underline{not} $\CQ$-closed. Therefore we have no reason  to believe that these correlation functions will localize to integrals over the   subspace of 
$\CA/\CG$ defined by the moduli space of instantons. One might have guessed that family Donaldson invariants would be based on families of instanton moduli spaces parametrized by $\MET(\IX)/\DIFF(\IX)$, and indeed some papers in the literature approach the topic of family invariants from this viewpoint \cite{Morava:2000yk,Baraglia2023}.
However, we are not aware of any formulation of our correlators in terms of families of instanton moduli spaces.    This provides both a challenge and an opportunity. The challenge is that new methods must be brought to bear to compute these terms in any explicit form.\footnote{One possible approach is to develop some of the ideas in \cite{Frenkel:2008vz}.}
The opportunity is that the invariants provided herein might not just depend on the topology of instanton moduli space and hence might detect information invisible to the Donaldson and Seiberg-Witten invariants.

Let us amplify on the previous sentence. The Donaldson invariants for a simply-connected manifold $\IX$  are intersection numbers of even-degree classes on $\mathsf{SO(3)}$ instanton moduli space, which has (virtual) dimension 
$8k - 3(b_2^+ +  1)$ where $k$ is an integer. Evidently, such intersection numbers must vanish if $b_2^+ + 1$ is odd.\footnote{For simply-connected $\IX$ the condition that $b_2^+$ is odd is equivalent to the condition that $\IX$ admits an almost complex structure. See \cite{Scorpan:2005}, p. 377, section 10.1 for an explanation.}
In this sense, Donaldson theory is blind to ``half'' the world of four-manifolds. We are not aware of any such parity restriction on $b_2^+$ for our cohomology classes. 

As we have mentioned, in our topological field theory it is expected that the computations of the UV theory can be reproduced in the IR theory. Based on our knowledge of the zeroth order term we might ask whether there is a relation of the form 
\eqa{\label{eq:UVfamily-to-IRfamily}
 & \sfZ_{\mathsf{family}}^{\mathsf{UV}} &&=  \sfZ_{\mathsf{family}}^{\mathsf{u-plane}} +  \sfZ_{\mathsf{family}}^{\mathsf{SW}} ~,
}
with the LHS computed from correlation functions in the nonabelian UV gauge theory. The expression 
$\sfZ_{\mathsf{family}}^{\mathsf{u-plane}}$ would be the family invariants computed in the abelian IR theory, 
using the actions derived below, 
and $\sfZ_{\mathsf{family}}^{\mathsf{SW}}$ would be a sum over $\sfH^{2}(\IX,\IZ)$ of expressions weighted by 
family SW invariants. In \cite{Moore:1997pc} the zeroth order term of $\sfZ_{\mathsf{family}}^{\mathsf{u-plane}}$ is computed, but \underline{not} using the method of localization. 
Rather, it is argued in section 2.3 of \cite{Moore:1997pc}  that, for $b_2^+(\IX)> 1$ the expression vanishes, for $b_2^+(\IX) = 1$  the tree-level approximation  is exact, and for $b_2^+(\IX)=0$ the one-loop approximation is exact. Because tree-level evaluation of a path integral is relatively straightforward, for $b_2^+(\IX)=1$ the path integral can accordingly be computed quite explicitly in a useful form. In reference \cite{ScalingWCF-ToAppear} 
we have made progress on extending the statements of section 2.3 of \cite{Moore:1997pc}  to the full family invariants.

One of the novel aspects of family invariants is that for $n$-dimensional families there is an analog of the wall-crossing phenomenon associated to families of four-manifolds with $b_2^+ = n+1$. The general framework is explained in section \ref{sec:WallCrossing} and is based on the existence of an Abelian instanton in the family.  Using the generalization of section 2.3 of \cite{Moore:1997pc} to the family invariants in the IR path integral we find a fairly explicit formula for the wall crossing in \eqref{eq:FinalFamilWCF}. Our expression for the overall normalization of the wall-crossing is somewhat cumbersome and it would be very nice to have a more effective formula. 
If an expression of the form \eqref{eq:UVfamily-to-IRfamily} is valid one could then use our wall-crossing formula to generalize the derivation of the Witten conjecture given in  \cite{Moore:1997pc}  to the family case.

In the case of a single four-manifold, Donaldson's invariants are considerably enriched by including observables, as in equation \eqref{eq:IncludeObs-NoFamily} above. The physical interpretation of the observables was explained in Witten's original paper \cite{Witten:1988ze} and is based on a sequence of local $j$-form gauge invariant observables $\CO^{(j)}$ that obey the descent equation 
\eqa{
& \CQ \CO^{(j)} &&= d \CO^{(j+1)} ~ . 
}
It follows from this simple equation that the integral over a closed $j$-cycle $\Sigma$ is both $\CQ$-closed and only depends on the homology class of $\Sigma$. 
In the family case there is an important variation on this identity: 
\eqa{
& \IQ \CO^{(j)} &&= d\CO^{(j-1)} + \iota_{\Phi}\CO^{(j+1)} ~, \label{eq:asc-des-Intro}
} 
and consequently, the naive generalization of Witten's observables will not work. 
In particular $\CO^{(0)}(p) \sim \mathsf{Tr\,} \phi^2(p)$  (where $p$ is a point) is not 
$\IQ$-closed! A simple way to get around this is to consider equivariant cohomology 
for subgroups of the diffeomorphism group that are restricted to be the identity on subspaces of the four-manifold. For example, by stipulating that the diffeomorphism 
fix the point $p$ and that its differential act as the identity on the tangent space at $p$ we enforce $\Phi^\mu(p)=0$ and now $\CO^{(0)}(p)$ is a $\IQ$-closed observable. This idea can easily be generalized to include any insertion of collections of Witten's local operators $\int_{\Sigma_j} \CO^{(j)}$ for closed cycles $\Sigma_j$.
These ideas are sketched in section \ref{sec:toobsfd} below. In section 
\ref{subsec:ReviewPreviousLit} we review some literature that is potentially related to the construction of observables in our setting. (We also discuss briefly  the observables in the  topological gravity setting, since it is not unrelated.) In particular, we briefly recall the definition of the famous generalized MMM classes and comment on some of the known results about them. An interesting question for the future is whether the observables of section \ref{sec:toobsfd} have any relation to the generalized MMM classes.

Regrettably, in this paper, we will not perform explicit computations on explicit families of four-manifolds, much less will we give concrete applications of our invariants to problems in four-manifold theory. We hope that such applications will be the topic of future research.
%
%
Indeed, the only reason we can adduce that our cohomology classes are not all zero is the wall-crossing formula. 

A brief outline of the remaining sections is as follows: In section 
\ref{sec:CartanModels} we review some aspects of equivariant cohomology and apply it to formulate the equivariant cohomology complexes of interest here. In section 
\ref{sec:ProposalForTheAction} we give our first derivation of the UV and IR actions. In sections \ref{sec:TwistedScfmlGrav} and \ref{sec:SuperconformalGrav-twistedSYM} we approach the problem from the more physical viewpoint of conformal supergravity. In section \ref{sec:TwistedScfmlGrav} we begin with a brief introduction to conformal supergravity and then describe the twisted and truncated multiplet of interest here. Then in section \ref{sec:SuperconformalGrav-twistedSYM} we explain the coupling to the twisted vectormultiplet and the relation of the resulting formulae to the expressions arising from equivariant cohomology. In section \ref{sec:TwistedSugraAction} we use the results of seections \ref{sec:TwistedScfmlGrav} and \ref{sec:SuperconformalGrav-twistedSYM} to make a proposal for the UV and IR actions of the twisted VM coupled to the twisted-truncated supergravity background. Somewhat alarmingly, the result looks a bit different from that derived in section 
\ref{sec:ProposalForTheAction}. For example, it is quartic in gravity-Cartan degree. In section \ref{sec:CompareActions} we demonstrate that the differences between the actions are $\IQ$-exact and therefore (barring anomalies) harmless. 
In section \ref{sec:WallCrossing} we discuss the wall-crossing formula. In section \ref{sec:Observables} we discuss how some observables can be defined to enrich the invariants discussed herein. We also discuss related observables that have appeared in the previous literature. A number of Appendices contain reference material and technical details. Appendix \ref{app:SymbolLists} contains reference tables for the notation used throughout the paper. Appendix \ref{app:conventions} summarizes some conventions for differential forms, Hodge duals, self-dual projections, gamma matrices, and so forth. Appendix \ref{app:spinormethods} briefly reviews the very well-known conversion to local frame coordinates and spinor indices of various differential-geometric objects in the twisted theory. Appendix \ref{app:useful-identities} summarizes various useful identities. Appendix \ref{app:variation-of-self-dual-fields} discusses one of the more subtle aspects of variational computations with respect to the metric in the twisted theory: Some of the fields in the twisted multiplet are self-dual. The space of fields is therefore fibered over the space of metrics. 
Appendix \ref{app:csugra-misc} summarizes some technical aspects of conformal supergravity that we need for defining the twisted and truncated conformal supergravity.  
Appendix \ref{app:DetailedComputations} gives many technical details in the computations used to verify the closure of algebras, the consistency of the truncation and twisting, and the comparison of the different versions of the action for the twisted vectormultiplet in a twisted truncated conformal supergravity background.

\section*{Acknowledgements} 
We thank Daniel Butter, Clay C\'{o}rdova, Kevin Costello, Thomas Dumitrescu, Dan Freed, Michael Freedman, Ahsan Khan, Zohar Komargodski, Jan Manschot, Sameer Murthy, Peter van Nieuwenhuizen, Warren Siegel, Shehryar Sikander, Dennis Sullivan, Alessandro Tomasiello, Loring Tu, and Edward Witten for various discussions.

The work of J.C., G.M., and V.S. was supported by the US Department of Energy under grant DE-SC0010008. The work of M.R. was supported in part by NSF grants PHY-1620628 and PHY-2210533.

\section{Equivariant Cohomology And The Cartan Models}\label{sec:CartanModels}

The physics of cohomological field theories is closely related to the mathematics of equivariant cohomology. We briefly revisit some elementary aspects of this subject and refer the reader to the references \cite{Cordes:1994fc,GuilleminSternberg:1999,Tu:2020} for additional details.\footnote{Perhaps the first application of these methods to cohomological quantum field theory is due to \cite{Atiyah:1990tm}. See also \cite{Witten:1988xj,Witten:1990bs,vanBaal:1989aw,Stora:1993zw,Stora:1995eq,Stora:1996ip,Stora:1996yc} for detailed expositions.} 
%

 Simply stated, equivariant cohomology accords a proper definition to the cohomology ring of a quotient space realized as the orbit space of a manifold or topological space admitting a group action. Since the group action need not be free, the resulting quotient space is not well-defined as a topological manifold (even if the parent space prior to quotienting is), and the naive quotient cannot be used to define cohomology classes as it tends to be a singular space.  
 
Let $\IM$ be a topological space admitting a continuous action of a compact Lie group $\IG$. If $\IG$ acts freely on $\IM$, the orbit space $\IM/\IG$ is a well-defined topological space with the quotient topology, and one defines the $\IG$-equivariant cohomology of $\IM$, denoted by $\sfH^{\bullet}_{\IG}(\IM)$ simply as the cohomology of the quotient space, i.e., $\sfH^{\bullet}_{\IG}(\IM) = \sfH^{\bullet}(\IM/\IG)$. However, if the $\IG$ action is not free, the quotient space $\IM/\IG$ is singular, and the putatively defined equivariant cohomology $H_{\IG}^{\bullet}(\IM)$ is the correct substitute for $\sfH^{\bullet}(\IM/\IG)$. One way to proceed is to replace $\IM$ by the space $\mathbb{E}\IG \times \IM$, where $\mathbb{EG}$ is the total space of the universal $\IG$-bundle, a contractible space on which $\IG$ acts freely, and consider instead of $\IM/\IG$ the quotient space
\eqa{
& \mathbb{EG}\times_{\IG} \IM &&:= \big( \mathbb{EG}\times \IM \big)/\IG ~, \label{eq:homotopy-quotient}
}
where the quotient by the $\IG$ action identifies $(e, m) \in \mathbb{EG}\times \IM$ with $g\cdot(e,m) = (e g^{-1}, g\cdot m)$, for every $g \in \IG$. This quotient space is also called the \textit{homotopy quotient} or \textit{Borel quotient} and denoted by $\IM_{\IG}$. The cohomology groups of two homotopy equivalent spaces are isomorphic, and since $\mathbb{EG}$ is contractible, the space $\mathbb{EG}\times \IM$ is homotopy equivalent to $\IM$. Therefore, the equivariant cohomology of $\IM$ with respect to $\IG$ (or succinctly, the $\IG$-equivariant cohomology of $\IM$) can be defined as\footnote{The coefficient ring is typically suppressed in notation but for us, it will mostly be $\IR$.}
\eqa{
& \sfH_{\IG}^{\bullet}(\IM) &&:= \sfH^{\bullet}\big( \big( \mathbb{EG}\times \IM \big)/\IG \big) ~. \label{eq:borel construction}
}
This homotopy quotient construction, also called the Borel construction \cite{Borel:1960}, in fact, appeared about ten years after Cartan's description in \cite{Cartan:1950} of a complex of equivariant differential forms, which can be viewed as a model of forms on the homotopy quotient $\IM_{\IG}$. In what follows, we will make contact with the Borel construction via Cartan's more practical \textit{algebraic} approach based on equivariant differential forms, for this is the one more directly related to the fields--equations--symmetries paradigm of cohomological quantum field theories. This will lead us to a discussion of the Cartan model for equivariant cohomology, which is the correct framework for discussing aspects of topologically twisted $\CN=2$ theories in four dimensions, and in particular, the Donaldson invariants.

Let us motivate the algebraic model slightly. As mentioned above, we are interested in the cohomology rings of $\IM_{\IG}$. As a naive guess, one might consider de Rham cohomology of the complex $\Omega^{\bullet}(\IE\IG) \otimes \Omega^{\bullet}(\IM)$. 
%
%
It turns out that a good substitute for $\Omega^{\bullet}(\IE\IG)$ is a graded commutative superalgebra $\mathscr{A}$ that is equipped with a representation of a Lie superalgebra (this superalgebra is generated\footnote{A smooth $\IG$ action on $\IM$ implies that to every element of $\LIE\IG$, one can canonically associate a vector field on $\IM$ -- this vector field acts on the de Rham complex $\Omega^{\bullet}(\IM)$ by an interior derivative and a Lie derivative.} by an interior derivative $\iota_{\IA}$, a Lie derivative $L_{\IA}$, where $\IA$ is an index labeling the generators of $\mathsf{Lie}(\IG)$, and an exterior differential on $\Omega^{\bullet}(\IM)$) and satisfies \underline{two} conditions: (1) it is acyclic with respect to the exterior differential (i.e., the cohomology of the complex is trivial -- this reflects contractibility of $\IE\IG$), and (2) there are elements $\theta^{\IA}$ of degree one in the superalgebra $\mathscr{A}$ that satisfy $\iota_{\IA}\theta^{\IB} = \delta_{\IA}^{\IB}$ (this property reflects the free action of $\IG$ on $\IE\IG$). Assuming such a superalgebra $\mathscr{A}$ exists, it is a model for $\Omega^{\bullet}(\IE\IG)$, meaning that we can replace $\Omega^{\bullet}(\IE\IG) \otimes \Omega^{\bullet}(\IM)$ by $\mathscr{A} \otimes \Omega^{\bullet}(\IM)$. Then the algebraic analog of the quotient by $\IG$ in \eqref{eq:borel construction} is via a restriction to a subcomplex $(\mathscr{A} \otimes \Omega^{\bullet}(\IM))_{\text{basic}}$ of basic forms: these are forms that are $\IG$-invariant and are annihilated by the interior derivatives (i.e., the $\iota_{\IA}$'s). This is the essence of the algebraic approach to equivariant cohomology.

There are several models of equivariant cohomology for topological field theories -- these involve different commutative graded superalgebras $\mathscr{A}$ of the previous paragraph, and hence different differential graded complexes. To physicists familiar with the BRST formalism, these correspond to different choices of ``antighosts,'' and ``auxiliary fields,'' (more properly, contractible pairs) resulting in different parametrizations of complexes with different closure relations of the BRST differential. Differing only by contractible pairs, these complexes are homotopically equivalent (i.e., they have isomorphic equivariant cohomology groups). The standard procedure is to construct the Weil model or BRST model and use the Mathai-Quillen procedure \cite{Cordes:1994fc,Mathai:1986tc,GuilleminSternberg:1999} to arrive at the (simpler) Cartan model. 

To apply the above formalism to the physical situations of interest to us, it is necessary to allow $\IG$ to be an infinite-dimensional Lie group, such as the group of gauge transformations or the group of diffeomorphisms of a four-manifold, or, as in our case, a semidirect product of these groups. Likewise, we will allow $\IM$ to be an infinite-dimensional topological space, such as the space of connections on a principal bundle or the space of Riemannian metrics (in our case, a direct product of these spaces). So we emphasize that we are quotienting an infinite-dimensional topological space by an infinite-dimensional group. The homotopy quotients that arise are still amenable to the Borel construction; however, we will adopt a bottom-up algebraic approach based on \cite{Cordes:1994fc}.

The fields of a topologically twisted $\CN=2$ vectormultiplet can be viewed as the elements of a differential graded complex for the Cartan Model of $\CG$-equivariant cohomology of $\CA$ augmented by so-called ``antighost multiplets,'' \cite{Witten:1990bs} (also called ``contractible pairs,'' for they have contractible equivariant cohomology). The Cartan model is adequate for studying observables in Donaldson theory, but to write down an off-shell action that is equivariantly closed under the Cartan differential, one needs to augment the Cartan model with a choice of antighost multiplets. The collection of Cartan $+$ antighost fields assembles into the fields of the topologically twisted $\CN=2$ vectormultiplet.

As remarked in the introduction, the ultimate goal of this section is to describe the Cartan model for the $\IG$-equivariant cohomology of $\IM$, where now
\eqa{
& \IM &&= \CA(P) \times \MET(X) ~, \label{eq:IM family}
}
and
\eqa{
& \IG &&= \CG \rtimes \Diff(\IX) ~. \label{eq:IG family}
}
We describe the Cartan models of $\CG$-equivariant cohomology of $\CA$ in section \ref{subsec:GaugeCartan}, the $\Diff(\IX)$-equivariant cohomology of $\MET(\IX)$ in section \ref{subsec:CartanModelDiff}, and finally, combining them to yield the $\IG$-equivariant cohomology of $\IM$ in section \ref{subsec:HGIM}.  In the last step, it is important that $\IG$ is a semidirect product. In section \ref{subsec:Full-Cartan-from-TwistedSugra}, we give a first principles derivation of using $\CN=2$ supergravity.

\subsection{The Cartan Model For $\sfH_{\CG}^{\bullet}(\CA(P))$\label{subsec:GaugeCartan}}
As mentioned above, the fields of the four-dimensional $\CN=2$ twisted vectormultiplet furnish a presentation of the differential graded complex of $\CG$-equivariant cohomology of $\CA(P)$, where the scalar supercharge $\CQ$ (realized in Witten's twisted $\CN=2$ SYM \cite{Witten:1988ze} as the BRST differential) plays the role of the equivariant differential. Additionally, two pairs of fields in the twisted vectormultiplet form modules for $\sfH_{\CG}^{\bullet}(\CA(P)$ and are referred to as ``antighost multiplets,'' called the localization and projection multiplets. Both these multiplets are equivariantly contractible so their inclusion does not affect the equivariant cohomology. The total complex of fields is
\eqa{
&\IE_{g}^{\text{gauge}} = \big(\widetilde{\IE}_{g}^{\text{gauge}}\big)^{\CG} ~, \label{eq:complex-GaugeCartan}
}
where $g \in \MET(\IX)$ denotes a (fixed) Riemannian metric on $\IX$, and the superscript $\CG$ indicates the restriction to the $\CG$-invariant subcomplex of

\eqa{
&\widetilde{\IE}_{g}^{\text{gauge}} &&:= \underbrace{\underbrace{\Omega^{\bullet}(\CA(P))}_{A} \otimes \underbrace{\Pi\Omega^{\bullet}(\CA(P))}_{\psi} \otimes \underbrace{S^{\bullet}(\fg^{\vee})}_{\phi}}_{\text{Cartan Model}}  \bigotimes \underbrace{\underbrace{\Omega_{g}^{2,+}(\IX, \adsf P)}_{H} \otimes \underbrace{\Pi\Omega_{g}^{2,+}(\IX, \adsf P)}_{\chi}}_{\text{Localization Multiplet}} \nonumber \\ & && \quad \bigotimes \underbrace{\underbrace{\Omega^{0}(\IX, \adsf P)}_{\lambda} \otimes \underbrace{\Pi\Omega^{0}(\IX, \adsf P)}_{\eta}}_{\text{Projection Multiplet}} 
\label{eq:BigVM-Complex}
}
and where we have indicated by underbraces the fields that serve as generators of various tensor factors. Here $\fg^{\vee}$ denotes the dual of the Lie algebra $\fg$, and for a vector bundle $V \to \IX$, $\Pi V$ denotes the superspace in which the coordinates of the fiber of $V$ are considered odd,\footnote{There is an isomorphism $C^{\infty}(\Pi T\IX) \cong \Omega^{\bullet}(\IX)$.} $\Omega_{g}^{2,+}$ is the space of self-dual 2-forms (with respect to the fixed metric $g$), and $S^{\bullet}(\fg^\vee)$ is the symmetric algebra of $\fg^\vee$.
The complex is graded, with the grading referred to as the \textit{homological degree} (or ghost number), summarized in Table \ref{tbl:HomDegree} below. 
\begin{small}
\begin{table}[H]
\centering
\begin{tabular}{|c|c|}\hline
Field & Homological Degree \\ \hline
$A_\mu$ & $0$ \\ \hline
$\psi_\mu$ & $1$ \\ \hline
$\phi$ & $2$ \\ \hline
$\lambda$ & $-2$ \\ \hline
$\eta$ & $-1$ \\ \hline
$H_{\mu\nu}$ & $0$ \\ \hline
$\chi_{\mu\nu}$ & $-1$ \\ \hline
\end{tabular}
\caption{\label{tbl:HomDegree}The homological degree of the fields of $\sfH_{\CG}^{\bullet}(\CA(P))$ and its modules.}
\end{table}
\end{small}
The action of the Cartan differential $\CQ$ is given by
\eqafour{
\label{eq:GaugeCartan-1}&\CQ A_\mu &&= \psi_\mu ~, \qquad && \CQ \psi_\mu &&= -D_\mu \phi ~, \qquad&& \CQ \phi &&= 0~, \\
\label{eq:GaugeCartan-2}&\CQ \lambda &&= \eta ~, \qquad && \CQ \eta &&= [\phi, \lambda] ~,\\
\label{eq:GaugeCartan-3}&\CQ \chi_{\mu\nu} &&= H_{\mu\nu} ~, \qquad && \CQ H_{\mu\nu} &&=[\phi,\chi_{\mu\nu}] ~.
}
where $D_{\mu}$ is the (metric + gauge)-covariant derivative. The Cartan differential squares to a gauge transformation by $\phi$, i.e.,
\eqa{\label{eq:GaugeCartanSquared}
&\CQ^2 &&= \delta_\phi ~.
}
Therefore on the $\CG$-invariant subcomplex $\CQ$ squares to zero.   The set of fields $\{A_\mu$, $\psi_\mu$, $\phi$, $\lambda$, $\eta$, $\chi_{\mu\nu}$, $H_{\mu\nu}\}$ is the twisted $\CN=2$ vectormultiplet with a shifted auxiliary field compared to the standard parametrization in Wess-Zumino gauge.\footnote{\label{foot:auxfield-param}After a simple field redefinition to define another adjoint-valued 2-form $D_{\mu\nu} := F_{\mu\nu}^{+} - H_{\mu\nu}$,
where $F_{\mu\nu}^{+}$ is the self-dual part of the Yang-Mills field strength ($F_{\mu\nu} = \partial_{\mu}A_{\nu} - \partial_{\nu}A_{\mu} + [A_\mu, A_\nu]$) we recognize $(A_\mu, \psi_\mu, \phi, \lambda, \eta, D_{\mu\nu}, \chi_{\mu\nu})$ as fields of the twisted $\CN=2$ vectormultiplet in Wess-Zumino gauge. Here $D_{\mu\nu}$ is the twisted version of the auxiliary field. This induces minor changes in the transformation laws but of course, does not change \eqref{eq:GaugeCartanSquared}.}
%

%
%

\subsection{The Cartan Model For $\sfH_{\DIFF(\IX)}^{\bullet}(\MET(\IX))$\label{subsec:CartanModelDiff}}
The second well-known model of equivariant cohomology is the Cartan Model of $\DIFF(\IX)$-equivariant cohomology of $\MET(\IX)$. The model consists of a Riemannian metric $g \in \MET(\IX)$, a symmetric gravitino $\Psi \in \Gamma(\Pi \mathsf{Sym}^2 T^{*}\IX)$ and a vector field $\Phi \in \Gamma(T\IX)$. The complex can be represented as
\eqa{
& \IE^{\text{diffeo}} &&= \big(\widetilde{\IE}^{\text{diffeo}}\big)^{\DIFF(\IX)} ~, \label{eq:complex-DiffCartan}
}
where the superscript $\DIFF(\IX)$ indicates the restriction to the diffeomorphism invariant subcomplex of 
\eqa{
& \widetilde{\IE}^{\text{diffeo}} &&:= \underbrace{\Omega^{\bullet}(\MET(\IX))}_{g} \otimes \underbrace{\Pi\Omega^{\bullet}(\MET(\IX))}_{\Psi} \otimes \underbrace{S^{\bullet}( \diff(\IX)^\vee)}_{\Phi} ~.
\label{eq:Gravity-DiffCartan}
}

 These fields are endowed with a grading, which we call \textit{gravity degree}, and this simply reflects their degree as differential forms on $\MET(\IX)$. This is summarized in Table \ref{tbl:GravDegree} below.
 \begin{small}
\begin{table}[H]
\centering
\begin{tabular}{|c|c|}\hline
Field & Gravity Degree \\ \hline
$g_{\mu\nu}$ & $0$ \\ \hline
$\Psi_{\mu\nu}$ & $1$ \\ \hline
$\Phi^{\mu}$ & $2$ \\ \hline
\end{tabular}
\caption{\label{tbl:GravDegree}The gravity degree of the fields of $\sfH_{\DIFF(\IX)}(\MET(\IX))$.}
\end{table}
\end{small}
We denote the differential of this model by $\sfd$ and its action is given by\footnote{Note that the equivariant differential $\sfd$ is should be distinguished from the exterior differential $\sfd_{\MET}$ on $\MET(\IX)$.}
\eqa{
\label{eq:GravityCartan-1}&\sfd g_{\mu\nu} &&= \Psi_{\mu\nu} ~,\quad  && \sfd\Psi_{\mu\nu} &&= \n_{\mu}\Phi_{\nu} + \n_{\nu}\Phi_{\mu} ~,\\
\label{eq:GravityCartan-2}&\sfd\Phi^\mu &&= 0 ~.
}
The equivariant differential $\sfd$ squares to a diffeomorphism on $\IX$ by the vector field $\Phi$, i.e.,
\eqa{
\sfd^2 = \CL_{\Phi} ~, \label{eq:sfd squared}
}
where $\CL_{\Phi}$ is a Lie derivative on $\IX$ along $\Phi$. Hence, $\mathsf{d}^2=0$ on our invariant subcomplex \eqref{eq:complex-DiffCartan}. 

The presence of a gravitino suggests the following natural question: is there is a model of $\CN=2$ supergravity that can be used to arrive at the diffeomorphism Cartan Model? We answer this in the affirmative in section \ref{subsec:Diff-Cartan-from-TwistedSugra}. For now, let us note that this model has previously appeared in the physics literature in the context of topological gravity and two-dimensional gravity \cite{Witten:1988xi,Witten:1989ig,Myers:1989dn,Myers:1990sa,Myers:1990hc,Wu:1993ef}. Notably, the model in \cite{Witten:1988xi} consisted of a traceless gravitino, but the version in \cite{Witten:1989ig} is, in fact, identical to \eqref{eq:GravityCartan-1}--\eqref{eq:GravityCartan-2}.

\subsection{The $\CG \rtimes \DIFF(\IX)$-equivariant cohomology of $\CA \times \MET(\IX)$\label{subsec:HGIM}}

Combining with the Cartan model for $\DIFF(\IX)$-equivariant cohomology of $\MET(\IX)$ and taking into account that $\IG$ is a semi-direct product 
one can immediately write down the full Cartan complex.\footnote{In \cite{CushingThesis} the relation of the Cartan model to other models of equivariant cohomology known as the Weil and BRST models is discussed in the present context.} We denote the action of the Cartan differential by $\IQ$ (rather than $d_{\CC}$), we are finally led to the following algebra for the Cartan Model:
\eqa{
& \label{eq:CartanSum1}\IQ g_{\mu\nu} &&= \Psi_{\mu\nu} ~, \qquad && \IQ A_{\mu} &&= \psi_{\mu} ~,\\
& \label{eq:CartanSum2}\IQ \Psi_{\mu\nu} &&= \n_{\mu}\Phi_{\nu} + \n_{\nu}\Phi_{\mu} ~, \qquad && \IQ \psi_{\mu} &&= -D_{\mu}\phi + \Phi^{\sigma}F_{\sigma\mu} ~,\\
& \label{eq:CartanSum3}\IQ \Phi^{\sigma} &&= 0 ~, \qquad && \IQ\phi &&= -\Phi^{\sigma}\psi_{\sigma} ~.
}
The Cartan differential $\IQ$ is no longer nilpotent. Instead,
\eqa{
\label{eq:CartanSquared}
&\IQ^2 &&= \delta_{\phi +\Phi^\sigma A_\sigma} + \CL_{\Phi} = \delta_{\phi} + \CL_{\Phi}^{(A)} ~,
}
where $\CL_{\Phi}^{(A)}$ a gauge-covariantized Lie derivative by the vector field $\Phi$, and $\delta_{\phi}$ is the usual gauge transformation by $\phi$. As usual, on the invariant subcomplex this differential \textit{is} nilpotent.

 %
%

\subsubsection{Antighost Multiplets}

We now need to extend \eqref{eq:CartanSum1}-\eqref{eq:CartanSum3} to include the antighost multiplets of section \ref{subsec:GaugeCartan}, namely the projection multiplet $(\lambda, \eta)$ and the localization multiplet $(\chi, H)$. These will be understood as two modules for the equivariant cohomology $H_{\IG}(\IM)$. And as mentioned before, these multiplets are equivariantly contractible.

Note that on the gauge Cartan base $(A,\psi,\phi)$,
\eqa{
& \left.\IQ\right|_{\Psi, \Phi = 0} &&= \CQ ~,
}
where $\CQ$ is the Cartan differential for $\sfH_{\CG}^{\bullet}(\CA(P))$, introduced in section \ref{subsec:GaugeCartan}. The antighost multiplets are contractible pairs for $\CQ$. For the projection multiplet in $\sfH_{\CG}^{\bullet}(\CA(P))$, 
\eqa{
& \IQ \lambda &&= \eta ~, \qquad Q\eta &&= [\phi, \lambda] ~,
}
where $\lambda$ and $\phi$ are respectively commuting and anticommuting fields that are both adjoint-valued scalars. There is a simple extension to $\IQ$, namely
\eqa{
&\IQ \lambda &&= \eta~ , \qquad && \IQ\eta &&= [\phi, \lambda] +\Phi^\sigma D_\sigma \lambda ~,
}
which happens to satisfy \eqref{eq:CartanSquared}. 

The localization multiplet is more complicated. In $\sfH_{\CG}^{\bullet}(\CA(P))$, 
\eqa{
&\CQ\chi_{\mu\nu} &&= H_{\mu\nu} ~,\qquad \CQ H_{\mu\nu} &&= [\phi, \chi_{\mu\nu}] ~,
}
where $\chi$ and $H$ are respectively anticommuting and commuting $2$-form fields that are both self-dual and adjoint-valued. The subtlety in extending this to $\IQ$ is that $\IQ$ acts nontrivially on the metric (unlike $\CQ$), see for example \eqref{eq:CartanSum1}, and thus does not commute with the Hodge star. This issue is discussed in detail in Appendix \ref{app:variation-of-self-dual-fields}, and there we learn that a minimal extension is
\eqa{\label{eq:CartanChi}
\IQ \chi_{\mu\nu} = H_{\mu\nu} -(\Psi_{[\mu}{}^{\sigma}\chi_{\nu]\sigma})^- ~,
}
where the second term on the right-hand side is needed for compatibility with the self-duality condition. It is worth noting that the second term in \eqref{eq:CartanChi} does not arise if one uses frame indices on $\chi$ and $H$. This is natural in the supergravity-inspired approach of the later sections but we will not do so here.\footnote{The version of \eqref{eq:CartanChi} with frame indices (cf.  Appendices \ref{app:spinormethods} and \ref{app:variation-of-self-dual-fields}) has no $\Psi H$ term. But now the differential acts not on $\MET(\IX)$, but rather on the space of orthonormal frames. It is straightforward to write down two-component versions of all the transformations in this section. But in that case, the square of the differential will involve local Lorentz transformations.} Now to determine $\IQ H_{\mu\nu}$, we compute $\IQ^2\chi_{\mu\nu}$ and require closure of the algebra via \eqref{eq:CartanSquared}. Defining the 2-form,
\eqa{
& \sfX_{\mu\nu} := \Psi_{[\mu}{}^{\sigma}\chi_{\nu]\sigma} ~,
}
we find, after using the identities \eqref{eq:identB} and \eqref{eq:identA refined}, that
\eqa{
& \IQ^2\chi_{\mu\nu} &&= \IQ H_{\mu\nu} +((\n_\mu \Phi^\sigma)\chi_{\sigma\nu} + (\n_\nu \Phi^\sigma) \chi_{\mu\sigma})^- +(\Psi^\sigma{}_{[\mu} H_{\nu]\sigma})^- + (\Psi^\sigma{}_{[\mu} \sfX_{\nu]\sigma}^-)^+ ~.
}
Requiring that \eqref{eq:CartanSquared} hold on $\chi_{\mu\nu}$, we find the following transformation law for $H_{\mu\nu}$:
\eqa{
\label{eq:CartanH}
& \IQ H_{\mu\nu} &&= [\phi, \chi_{\mu\nu}] + \Phi^\sigma D_\sigma \chi_{\mu\nu} +\left((\n_\mu \Phi^\sigma)\chi_{\sigma\nu} - (\n_\nu\Phi^\sigma)\chi_{\sigma\mu}\right)^+ \nonumber\\
& &&\quad -(\Psi_{[\mu}{}^{\sigma}H_{\nu]\sigma})^- +(\Psi_{[\mu}{}^{\sigma}(\Psi_{[\sigma}{}^{\rho}\chi_{\nu]]\rho})^-)^+ ~.
}    

\subsubsection{Summary Of The Cartan Model For $\sfH_{\CG\rtimes\DIFF(\IX)}(\CA(P)\times \MET(\IX))$}\label{subsec:FullEqCohCartanModelSummary}
The complete Cartan Model with antighost multiplets is given by 
\eqa{
\label{eq:CartanAlgebra-1}&\IQ g_{\mu\nu} &&= \Psi_{\mu\nu} ~, &&\IQ A_{\mu} &&=\psi_\mu ~,\\
\label{eq:CartanAlgebra-2}&\IQ \Psi_{\mu\nu} &&= \n_{\mu} \Phi_{\nu} + \n_{\nu} \Phi_{\mu} ~, \qquad && \IQ \psi_\mu &&= -D_\mu \phi + \Phi^\sigma F_{\sigma\mu} ~,\\
\label{eq:CartanAlgebra-3}&\IQ \Phi^\sigma &&= 0 ~, \quad && \IQ \phi &&= -\Phi^\sigma \psi_\sigma ~,\\
\label{eq:CartanAlgebra-4}&\IQ \lambda &&= \eta ~,\\
\label{eq:CartanAlgebra-5}&\IQ \eta &&= [\phi, \lambda] + \Phi^\sigma D_\sigma \lambda ~,\\
\label{eq:CartanAlgebra-6}&\IQ \chi_{\mu\nu} &&= H_{\mu\nu} -(\Psi_{[\mu}{}^{\sigma}\chi_{\nu]\sigma})^-,\\
\label{eq:CartanAlgebra-7}&\IQ  H_{\mu\nu} &&= [\phi, \chi_{\mu\nu}] + \Phi^\sigma D_\sigma \chi_{\mu\nu} - 2\left((\n_{[\mu} \Phi^\sigma)\chi_{\nu]\sigma}\right)^+ \nn
& &&\quad -(\Psi_{[\mu}{}^{\sigma}H_{\nu]\sigma})^- +(\Psi_{[\mu}{}^{\sigma}(\Psi_{[\sigma}{}^{\rho}\chi_{\nu]]\rho})^-)^+ ~.
}
Let us make a few observations about this model. First of all, there is an alternate parametrization of the auxiliary field -- along the lines of footnote \ref{foot:auxfield-param} -- which makes the connection to the twisted $\CN=2$ vectormultiplet (in Wess-Zumino gauge) more explicit, at the expense of modifying the transformation laws of $\chi_{\mu\nu}$ and the (new) auxiliary field. Defining
\eqa{
& H_{\mu\nu} &&= F_{\mu\nu}^+ - D_{\mu\nu} ~,  \label{eq:CartanModel-AlternateAuxField}
}
where $F_{\mu\nu} := 2\,\partial_{[\mu}A_{\nu]} + [A_\mu, A_\nu]$ is the Yang-Mills field strength, we find
\eqa{
&\IQ\chi_{\mu\nu} = F^+_{\mu\nu}-D_{\mu\nu} -(\Psi_{[\mu}{}^{\sigma}\chi_{\nu]\sigma})^- ~, \label{eq:CartanAlgebra-6prime}\\
&\IQ D_{\mu\nu} =2(D_{[\mu}\psi_{\nu]})^+ -[\phi, \chi_{\mu\nu}] - \Phi^\sigma D_\sigma \chi_{\mu\nu} -\left((\n_\mu \Phi^\sigma)\chi_{\sigma\nu} - (\n_\nu\Phi^\sigma)\chi_{\sigma\mu}\right)^+\nonumber\\
&\quad\quad\quad\quad+(\Psi_{[\mu}{}^{\sigma}F_{\nu]\sigma}^-)^+-(\Psi_{[\mu}{}^{\sigma}D_{\nu]\sigma})^- +\frac{1}{2}\Psi^{\rho\sigma}\Psi_{\rho[\mu}\chi_{\nu]\sigma} ~. \label{eq:CartanAlgebra-7prime}
}
We will prefer use the original parametrization for explorations of the algebra (which we shall carry out momentarily) and the alternate version involving $D_{\mu\nu}$ when we work with the action. 

In section \ref{sec:TwistedScfmlGrav} we will arrive at this model quite independently from the viewpoint of $\CN=2$ supergravity. Specifically, in section \ref{subsec:Full-Cartan-from-TwistedSugra}, we show that these methods land us on the alternative parametrization \eqref{eq:CartanAlgebra-1}--\eqref{eq:CartanAlgebra-5} and \eqref{eq:CartanAlgebra-6prime}--\eqref{eq:CartanAlgebra-7prime} of the Cartan Model for $H_{\CG\rtimes\DIFF(\IX)}(\CA(P)\times \MET(\IX))$ involving the auxiliary field $D_{\mu\nu}$.

\subsubsection{Closure Of The Cartan Model Algebra}
Next, we wish to discuss the closure of the algebra. Let us begin by noting that 
\eqa{
&\CQ^2 && = \delta_{\phi} ~,\\
&\mathsf{d}^2 &&= \CL_{\Phi} ~,\\
\label{eq:IQ2}&\IQ^2 &&= \CL^{(A)}_{\Phi} + \delta_{\phi} ~.
}
In this section, we will justify $\eqref{eq:IQ2}$ as a check that we have written down the correct 
differentials.\footnote{Our \eqref{eq:IQ2} should be compared with (1.1) of \cite{Imbimbo:2018duh} $\left.\IQ\right|_{\text{here}} = \left.S\right|_{\text{there}}$. The authors of \cite{Imbimbo:2018duh} considered the space $\IM = \CA \times \mathsf{Met}(\IX)$ and remarked that their equivariant differential $S$ is equivariant with respect to the action of \textit{both} gauge transformations and diffeomorphisms. Our direct supergravity approach of section \ref{sec:TwistedScfmlGrav} appears to be new.}

Turning off all supergravity fields (except the metric), we see that 
\eqa{
&\IQ\big|_{\Psi=0,\Phi=0} &&= \CQ ~, \label{eq:IQ-restriction-to-metric-only}
}    
where $\CQ$ is the Cartan differential for  $\sfH_{\CG}^{\bullet}(\CA(P))$. Similarly, by dropping all vectormultiplet fields, we have
\eqa{
& \IQ\big|_{g, \Psi, \Phi} &&= \sfd ~,\label{eq:IQ-restriction-to-sugra-fields}
}
where $\sfd$ is the Cartan differential for $H_{\DIFF(\IX)}^{\bullet}(\MET(\IX))$. This can be aptly summarized by splitting $\IQ$ into bidegrees $(p,q)$ where $p$ is the homological degree inherited from the $\sfH_{\CG}^{\bullet}(\CA(P))$ model and $q$ is the gravity degree from the $H_{\DIFF(\IX)}^{\bullet}(\MET(\IX))$ model. We can then write
\eqa{
&\IQ &&= \IQ^{(1,0)} + \IQ^{(0,1)} + \IQ^{(-1,2)} ~, \label{eq:bidegree-Q-cartan-variables}
}
where the bidegree differentials are given by
\eqa{
&\IQ^{(1,0)} &&=\CQ ~,\\
\label{eq:tilded}&\IQ^{(0,1)} &&=\widetilde{\sfd}~,\\
&\IQ^{(-1,2)} && =\mathsf{K} + \bm{\Delta}_{H} ~.
}
Here, $\CQ$ acts as in \eqref{eq:GaugeCartan-1}--\eqref{eq:GaugeCartan-3}. Next, $\widetilde{\sfd}$ is the lift of the differential $\mathsf{d}$ to the total space of the projected bundle of adjoint-valued self-dual 2-forms $\Omega_g^{2,+}(\IX, \adsf{P})$ over $\MET(\IX)$. It acts as $\mathsf{d}$ on $g$ and $\Psi$ as in \eqref{eq:GravityCartan-1} and \eqref{eq:GravityCartan-2} and on our self-dual fields $\chi$ and $H$ via the induced projected connection.\footnote{See Appendix \ref{app:variation-of-self-dual-fields} for a discussion of self-duality and the projected connection.} Specifically,
\eqa{
&\widetilde{\sfd}\chi_{\mu\nu} &&= -(\Psi_{[\mu}{}^{\sigma} \chi_{\nu]\sigma})^- ~, \qquad && \widetilde{\sfd}H_{\mu\nu} &&= -(\Psi_{[\mu}{}^{\sigma}H_{\nu]\sigma})^- ~, \label{eq:dtilde-action}
}
and $\widetilde{\sfd}$ acts trivially on all other fields. Finally, $\mathsf{K}$ acts as
\eqa{
&\mathsf{K}\psi_{\mu} &&= \Phi^\sigma F_{\sigma\mu} ~, \label{eq:Kaction-1}\\
&\mathsf{K}\phi &&= -\Phi^\sigma \psi_\sigma ~, \label{eq:Kaction-2}\\
&\mathsf{K} \eta &&= \Phi^\sigma D_\sigma \lambda ~,\label{eq:Kaction-3}\\
&\mathsf{K} H_{\mu\nu} &&= \Phi^\sigma D_\sigma \chi_{\mu\nu} +\left((\n_\mu \Phi^\sigma)\chi_{\sigma\nu} - (\n_\nu\Phi^\sigma)\chi_{\sigma\mu}\right)^+ ~, \label{eq:Kaction-4}
}
whereas $\bm{\Delta}_{H}$ acts nontrivially only on $H$, its action given by
\eqa{
&\bm{\Delta}_{H}H_{\mu\nu} &&=(\Psi_{[\mu}{}^{\sigma}(\Psi_{[\sigma}{}^{\rho}\chi_{\nu]]\rho})^-)^+ ~. \label{eq:DeltaH-action}
}
On all fields -- \underline{except} $(\chi, H)$ --  we can summarize the algebra via the relations
\eqa{
&\CQ^2 &&=\delta_{\phi} ~, \qquad &&\widetilde{\sfd}^2\big|_{g, \Psi, \Phi} = \CL_{\Phi} ~,\\
&\mathsf{K}^2 &&=\{\CQ, \widetilde{\sfd}\}=\{\widetilde{\sfd}, \mathsf{K}\} = 0 ~, \qquad &&
\{\CQ, \mathsf{K}\}\big|_{F_A,\psi, \phi, \eta, \lambda} =\CL_{\Phi}^{(A)} ~,
}
where the final relation is only true on fields that transform homogeneously in the adjoint representation (e.g., not on the YM connection $A$, but rather on the curvature $F_A$).

As previously mentioned, the self-dual fields $\chi$ and $H$ present subtleties due to the metric variation of the self-duality condition. We still have 
\eqa{
&\CQ^2 &&= \delta_\phi ~, \qquad
&& \mathsf{K}^2 &&=\{\CQ, \widetilde{\sfd}\}=0 ~,
}
but now the other relations change. First of all, due to $\bm{\Delta}_{H}$, we have
\eqa{
\label{eq:QDeltaH-1}&\{\CQ, \bm{\Delta}_{H}\}\chi_{\mu\nu}&&=(\Psi_{[\mu}{}^{\sigma}(\Psi_{[\sigma}{}^{\rho}\chi_{\nu]]\rho})^-)^+ ~, \qquad  && \bm{\Delta}_{H}^2\chi_{\mu\nu} &&=\{\bm{\Delta}_{H}, \mathsf{K}\}\chi_{\mu\nu}=0 ~,\\
\label{eq:QDeltaH-2}&\{\CQ, \bm{\Delta}_{H}\}H_{\mu\nu}&&=(\Psi_{[\mu}{}^{\sigma}(\Psi_{[\sigma}{}^{\rho}H_{\nu]]\rho})^-)^+ ~, \qquad &&\bm{\Delta}_{H}^2H_{\mu\nu} &&=\{\bm{\Delta}_{H}, \mathsf{K}\}H_{\mu\nu}=0 ~.
}
Moreover,
\eqa{
&\{\CQ, \mathsf{K}\}\chi_{\mu\nu} = (\CL_{\Phi}^{(A)}\chi_{\mu\nu})^+, \qquad \{\CQ, \mathsf{K}\}H_{\mu\nu} = (\CL_{\Phi}^{(A)}H_{\mu\nu})^+.
}
As explained in Appendix \ref{app:d2Omega-d2chi}, we have
\eqa{
\label{eq:dSquaredChi}&\widetilde{\sfd}^2\chi_{\mu\nu} &&= \frac{1}{2}\sqrt{g}\epsilon_{\mu\nu\alpha\beta}\left(\frac{1}{2}(\n_\sigma \Phi^\sigma)g^{\alpha\alpha'}g^{\beta\beta'} - (\n^\alpha \Phi^{\alpha'} + \n^{\alpha'}\Phi^{\alpha})g^{\beta\beta'}\right)\chi_{\alpha'\beta'} +\frac{1}{2}\Psi^{\rho\sigma}\Psi_{\rho[\mu}\chi_{\nu]\sigma}\nonumber\\
& &&=(\CL_{\Phi}^{(A)}\chi)_{\mu\nu}-(\CL_{\Phi}^{(A)}\chi)_{\mu\nu}^+ +\frac{1}{2}\Psi^{\rho\sigma}\Psi_{\rho[\mu}\chi_{\nu]\sigma} ~,\\
\label{eq:dSquaredH}&\widetilde{\sfd}^2 H_{\mu\nu} &&= \frac{1}{2}\sqrt{g}\epsilon_{\mu\nu\alpha\beta}\left(\frac{1}{2}(\n_\sigma \Phi^\sigma)g^{\alpha\alpha'}g^{\beta\beta'} - (\n^\alpha \Phi^{\alpha'} + \n^{\alpha'}\Phi^{\alpha})g^{\beta\beta'}\right)H_{\alpha'\beta'} +\frac{1}{2}\Psi^{\rho\sigma}\Psi_{\rho[\mu}H_{\nu]\sigma}\nonumber\\
& &&=(\CL_{\Phi}^{(A)}H)_{\mu\nu}-(\CL_{\Phi}^{(A)}H)_{\mu\nu}^+ +\frac{1}{2}\Psi^{\rho\sigma}\Psi_{\rho[\mu}H_{\nu]\sigma} ~.
}
The first term in each expression is precisely what is needed to change the self-dual part of a Lie derivative to the Lie derivative of a self-dual field. It turns out that the additional terms are precisely the opposite of those in \eqref{eq:QDeltaH-1} and \eqref{eq:QDeltaH-2} respectively -- see \eqref{eq:Psi Psi Chi Diff SD}.  This is desirable, but we must verify that the remaining anticommutators do not contribute any additional terms. On $\chi$, we trivially find
\eqa{
&\{\widetilde{\sfd}, \mathsf{K}\} \chi_{\mu\nu} &&= \{\widetilde{\sfd}, \bm{\Delta}_{H}\}\chi_{\mu\nu}=0 ~.
}
On the other hand, on $H$,
\eqa{
\{\widetilde{\sfd}, \mathsf{K}\}H_{\mu\nu} &=\frac{1}{2}\Psi_{\sigma[\mu}(\n^\rho \Phi^\sigma + \n^\sigma \Phi^\rho)\chi_{\nu]\rho} - \frac{1}{2}\Psi^{\sigma \rho}(\n_{[\mu}\Phi_{\rho} + \n_\rho \Phi_{[\mu})\chi_{\nu]\sigma} ~,\\
\{\widetilde{\sfd}, \bm{\Delta}_{H}\}H_{\mu\nu} &=-\frac{1}{2}\Psi_{\sigma[\mu}(\n^\rho \Phi^\sigma + \n^\sigma \Phi^\rho)\chi_{\nu]\rho} + \frac{1}{2}\Psi^{\sigma \rho}(\n_{[\mu}\Phi_{\rho} + \n_\rho \Phi_{[\mu})\chi_{\nu]\sigma} ~,
}
so that 
\eqa{
&\{\widetilde{\sfd}, \mathsf{K}\} &&=-\{\widetilde{\sfd}, \bm{\Delta}_{H}\} ~.
}
This nontrivial relation is proved in detail in Appendix \ref{app:DKH}. It ensures that the localization multiplet can be consistently included in the equivariant cohomology model of diffeomorphisms and gauge transformations. 
The additional terms in \eqref{eq:dSquaredChi} and \eqref{eq:dSquaredH} are a consequence of the self-duality constraint, and in fact, these terms arise from the curvature of the projected connection (see Appendices \ref{app:d2Omega-d2chi} and \ref{app:variation-of-self-dual-fields}).

All together, these relations ensure that algebra \eqref{eq:CartanAlgebra-1}--\eqref{eq:CartanAlgebra-7} indeed closes correctly, in accordance with \eqref{eq:IQ2}.

\section{A Proposal For The Action}\label{sec:ProposalForTheAction}
Our goal is to construct an invariant
\eqa{ \label{eq:FamilyInvariant}
&\mathsf{Z}_{\mathsf{family}}[g, \Psi, \Phi] &&= \int[d\Phi_{{\rm vm}}] e^{-\IS[g, \Psi, \Phi]}~,
}
where $[d\Phi_{{\rm vm}}]$ is the formal path integral measure for the vectormultiplet, which can be schematically written as 
\eqa{
&\left[d\Phi_{{\rm vm}}\right] &&= \left[dA\,d\psi\,d\phi\,d\eta\,d\lambda\,d\chi\,dH\right]~.
}
Here $\IS$ is a $\IQ$-closed local functional of the fields of our Cartan model with antighosts and is invariant under gauge transformations and diffeomorphisms (provided the background fields transform as well). In this section, we will propose an expression for $\IS$, which will meet these requirements. This will be interpreted as the general action of an $\CN=2$ vectormultiplet coupled to the supergravity fields $(g_{\mu\nu}, \Psi_{\mu\nu}, \Phi^{\mu})$. We will follow \cite{Moore:1997pc,Marino:1998bm} and include an arbitrary prepotential and also allow for a generic gauge group $G$, usually taken to be a compact Lie group with an invariant form $\mathsf{Tr}$. For \eqref{eq:FamilyInvariant} to be an appropriate generalization of both Donaldson-Witten (UV) and Seiberg-Witten invariants (IR), we require that the gravity degree zero part of $\IS$ coincide with an off-shell formulation of Witten's original action (2.13) of \cite{Witten:1988ze} in the UV and the action of Moore and Witten (2.15) \cite{Moore:1997pc} in the IR. Denoting these actions by $S_{\mathsf{UV}}$ and $S_{\mathsf{IR}}$ respectively and our prepotentials by $\CF$ (a function of $\phi$ or $a$) and $\overline{\CF}$ (a function of $\lambda$ or $\overline{a}$), $\IS$ must satisfy
\eqa{\label{eq:ISConstraints}
&\IQ\IS && = 0 ~, \quad \IS\big|_{\text{quad}, \Psi=0,\Phi=0} = S_{\mathsf{UV}}~, \quad \IS\big|_{G=\mathsf{U(1)},\Psi=0,\Phi=0} = S_{\mathsf{IR}}~,
}
where the subscript `quad' is short for quadratic prepotential in the UV:
\eqa{
 \text{quad }(\text{UV}) &: \qquad \CF &&=\frac{1}{2}\tau_0\mathsf{Tr}(\phi^2) ~, \qquad && \ov{\CF} &&=\frac{1}{2}\ov{\tau}_0\mathsf{Tr}(\lambda^2) ~, \label{eq:quadratic prepotential}
}
where $\tau_0$ is a constant complex coupling $\ov{\tau}_0$ is its complex conjugate.
We remark that $\CF$ and $\ov{\CF}$ are \underline{not} complex conjugates in the topologically twisted theory (as $\phi$ and $\lambda$ are not complex conjugates), but we will stick to this notation as it is standard in the literature. Reality conditions on the fields, although important for various purposes, will play no role in our discussion.

To identify a strategy to construct $\IS$, we will briefly review the form and properties of both $S_{\mathsf{UV}}$ and $S_{\mathsf{IR}}$, before proposing a form of $\IS$ that is consistent with \eqref{eq:ISConstraints}. The reader familiar with the actions of \cite{Witten:1988ze,Moore:1997pc,Marino:1998bm} may skip to section \ref{sec:minimalgeneralaction}. 

\subsection{Degree Zero UV Action\label{subsec:degree-0-UV-action}}
The off-shell action of topologically twisted $\CN=2$ super Yang Mills theory for gauge group $G$ in the UV \cite{Witten:1988ze,Labastida:2005zz} is given by\footnote{The translation to the conventions of \cite{Witten:1988ze} is: $\psi \mapsto i\psi^{\mathsf{W}}  $, $\phi\mapsto i\phi^{\mathsf{W}}$, $\lambda\mapsto -\frac{i}{2}\lambda^{\mathsf{W}}$, and $\chi_{\mu\nu}\mapsto \frac{1}{2}\chi_{\mu\nu}^{\mathsf{W}} $. We also need to impose the equation of motion for the auxiliary field.}
\eqa{
S_{\mathsf{UV}}&=\frac{1}{g_0^2}\int_{\IX}d^4x\sqrt{g}\,\mathsf{Tr\,}\bigg[\frac{1}{4}F_{\mu\nu}F^{\mu\nu} - \frac{1}{4}D_{\mu\nu}D^{\mu\nu} +2(D_\mu \psi_\nu)\chi^{\mu\nu} + \frac{1}{2}\chi_{\mu\nu}[\phi, \chi^{\mu\nu}]-\eta D_{\mu}\psi^\mu\nonumber\\
&\qquad\qquad\qquad\qquad- \lambda[\psi_\mu, \psi^\mu] + \lambda D_\mu D^\mu\phi-\frac{1}{2}\phi[\eta,\eta] -\frac{1}{2}[\phi,\lambda]^2\bigg]\nonumber\\
&\qquad + \frac{i\theta_0}{64\pi^2}\int_{\IX}d^4x\textup{Tr}[\epsilon^{\mu\nu\rho\sigma} F_{\mu\nu}F_{\rho\sigma}]~, \label{eq:UVActionDeg0}
}
where $\mathsf{Tr\,}$ denotes a suitably normalized invariant form on $\mathsf{Lie}(G)$. The equation of motion for the self-dual auxiliary field is $D_{\mu\nu}=0$. Together with the $\CQ$-fixed point equation for $\chi$, this leads to the instanton equation $F_A^+ = 0$, and thus the theory localizes to the moduli space of anti-self-dual connections $\masd$. Introducing a constant complex coupling
\eqa{
&\tau_0 &&= \frac{4\pi i}{g_0^2}+\frac{\theta_0}{2\pi} ~, \label{eq:constant coupling}
}
the action \eqref{eq:UVActionDeg0} can be expressed as
\be
S_{\mathsf{UV}}=\CQ \IV_{\mathsf{UV}} + \frac{i\tau_0}{16\pi}\int_{\IX}\mathsf{Tr\,}F\wedge F~,
\ee
where $\CQ$ is the differential of \eqref{eq:GaugeCartan-1}--\eqref{eq:GaugeCartan-3} and 
\be\label{eq:UVV}
\IV_{\mathsf{UV}}=\frac{1}{g_0^2}\int_{\IX}d^4x\sqrt{g}\,\mathsf{Tr}\left[ \frac{1}{2}(F_{\mu\nu}^+ + D_{\mu\nu})\chi^{\mu\nu} - \lambda D_\mu \psi^\mu -\frac{1}{2} \eta[\phi, \lambda]\right]~.
\ee
Therefore, on a closed $\IX$, we have 
\eqa{
&\CQ S_{\mathsf{UV}} &&=0~.
}
In addition, the energy-momentum tensor of this action is $\CQ$ exact:
\be
T_{\mu\nu}^{\mathsf{UV}} =\CQ \Lambda_{\mu\nu}^{\mathsf{UV}}~, \label{eq:TeeUV}
\ee
where $T_{\mu\nu}^{\mathsf{UV}}$ is defined by
\be
\delta S_{\mathsf{UV}} = \frac{1}{2}\int_{\IX}d^4x \sqrt{g}\delta g^{\mu\nu} T_{\mu\nu}^{\mathsf{UV}}~.
\ee

\subsection{Degree Zero IR Action\label{subsec:degree-0-IR-action}}
For the IR theory, we consider $G=\mathsf{U(1)}$ and an arbitrary holomorphic prepotential.\footnote{This means $\CF(a)$ and $\ov{\CF}(\ov{a})$ are formally holomorphic functions of $a$ and $\ov{a}$ respectively.} Here, following tradition, we denote $\phi$ by $a$ and $\lambda$ by $\overline{a}$ so that \eqref{eq:GaugeCartan-1}--\eqref{eq:GaugeCartan-3} now take the form
\eqafour{
\label{eq:GaugeCartanAbelian-1}&\CQ A_\mu &&= \psi_\mu ~, \qquad && \CQ \psi_\mu &&= -\n_\mu a ~, \qquad \CQ a &&= 0 ~, \\
\label{eq:GaugeCartanAbelian-2}&\CQ \overline{a} &&= \eta ~, \qquad && \CQ \eta &&= 0 ~,\\
\label{eq:GaugeCartanAbelian-3}&\CQ \chi_{\mu\nu} &&= F_{\mu\nu}^+ - D_{\mu\nu} ~, \qquad && \CQ D_{\mu\nu} &&=2(\n_{[\mu}\psi_{\nu]})^+~.
}
We also define $\tau =\frac{\partial^2\CF(a)}{\partial a^2}$ and $\overline{\tau}=\frac{\partial^2\overline{\CF}(\overline{a})}{\partial \bar{a}^2}$. The IR action of \cite{Moore:1997pc} reads\footnote{\label{foot:MWContact}To obtain the IR action (2.15) of ref. \cite{Moore:1997pc} (with fields denoted by subscript $\mathsf{MW}$) (or (3.4) of ref. \cite{Marino:1998bm}), we need the following redefinitions: $\left.a_{\rm}\right|_{\rm here} = -4\sqrt{2}\left.a\right|_{\mathsf{MW}}$, $\left.\ov{a}\right|_{\rm here} = \frac{1}{2\sqrt{2}}\left.\ov{a}\right|_{\mathsf{MW}}$, $\left.\eta\right|_{\rm here} = \frac{i}{2}\left.\eta\right|_{\mathsf{MW}}$, $\left.\psi_{\mu}\right|_{\rm{here}} = \left.\psi_{\mu}\right|_{\mathsf{MW}}$, $\left.A_{\mu}\right|_{\rm{here}} = \left.A_{\mu}\right|_{\mathsf{MW}}$, $\left.\chi_{\mu\nu}\right|_{\rm{here}} = -i\left.\chi_{\mu\nu}\right|_{\mathsf{MW}}$, and $\left.D_{\mu\nu}\right|_{\rm{here}} = \left.D_{\mu\nu}\right|_{\mathsf{MW}}$. For more details, see Appendix \ref{app:conventions}.}
\eqa{
S_{\mathsf{IR}} &= \frac{i}{16\pi}\int_{\IX}d^4x\sqrt{g}\left[ \frac{1}{2}\overline{\tau}F_{\mu\nu}^+F^{\mu\nu}_+ - \frac{1}{2}\tau F_{\mu\nu}^-F^{\mu\nu}_- + 4i\mathsf{Im\,} \tau (\n_\sigma a)(\n^\sigma \overline{a}) +i\mathsf{Im\,}\tau D_{\mu\nu}D^{\mu\nu}\right.\nonumber\\
&\qquad\qquad\qquad\qquad\left. + 2\tau \psi_\sigma \n^\sigma \eta - 2\overline{\tau}\eta \n_\sigma \psi^\sigma - 2\tau \psi_{\mu}\n_{\nu}\chi^{\mu\nu} + 2\overline{\tau}(\n_{\mu}\psi_\nu)^+\chi^{\mu\nu}\right.\nonumber\\
&\qquad\qquad\qquad\qquad\left. +\frac{1}{2}\frac{\partial \overline{\tau}}{\partial \overline{a}} \eta (F^+_{\mu\nu} +D_{\mu\nu})\chi^{\mu\nu} + \frac{\partial\tau}{\partial a}\psi_{\mu}\psi_{\nu}(F_-^{\mu\nu} - D^{\mu\nu}) \right.\nonumber\\
&\qquad\qquad\qquad\qquad\left. +\frac{1}{12}\sqrt{g}^{-1}\frac{\partial^2\tau}{\partial a^2}\epsilon^{\mu\nu\rho\sigma}\psi_{\mu}\psi_{\nu}\psi_{\rho}\psi_{\sigma}+\CQ \left(\frac{i}{12}\frac{\partial \overline{\tau}}{\partial \overline{a}} \chi_{\mu}{}^\rho \chi^{\mu\sigma}\chi_{\rho\sigma}\right) \right] ~,\label{eq:IRActionDeg0}
}
where we have used the conventions of Appendix \ref{app:conventions} and \eqref{eq:GaugeCartan-1}--\eqref{eq:GaugeCartan-3}. As with the UV theory, we can write the action as a $\CQ$ exact part and a topological term:
\eqa{
& S_{\mathsf{IR}} &&= \CQ \IV_{\mathsf{IR}} + \mathsf{C}_{\mathsf{IR}} ~,
}
where\footnote{Our conventions for differential forms are spelled out in Appendix \ref{app:conventions}.}
\eqa{
& \IV_{\mathsf{IR}} &&= \frac{i}{16\pi}\int_{\IX}d^{4}x\sqrt{g}\left[-\frac{1}{2}(\tau-\ov{\tau})(F_{\mu\nu}^{+} + D_{\mu\nu})\chi^{\mu\nu} - \tau \psi_{\sigma}\n^{\sigma}\ov{a} - \frac{\partial\ov{\CF}}{\partial{\ov{a}}}\n_{\sigma}\psi^{\sigma} \right.\nonumber\\
& &&\qquad\qquad\qquad\qquad \left. + \frac{\partial\tau}{\partial a}\psi_{\mu}\psi_{\nu}\chi^{\mu\nu} + \frac{i}{12}\frac{\partial\ov{\tau}}{\partial\ov{a}}\chi_{\mu}{}^{\rho}\chi^{\mu\sigma}\chi_{\rho\sigma} \right] ~,\label{eq:IRV} \\
%
&\mathsf{C}_{\mathsf{IR}} &&= \frac{i}{16\pi}\int_{\IX}\left[\tau F \wedge F - \frac{\partial\tau}{\partial a} \psi \wedge \psi \wedge F +\frac{1}{12}\frac{\partial^2\tau}{\partial a^2}\psi \wedge \psi \wedge \psi \wedge \psi\right]~. \label{eq:IRC}
}
To prove that $\CQ S_{\mathsf{IR}} = 0$ on a closed $\IX$, we need to verify that $\CQ\mathsf{C}_{\mathsf{IR}}$ vanishes. We will actually prove this explicitly and in more generality in section \ref{sec:minimalgeneralaction} below.

\subsection{Construction Of $\IS$\label{sec:minimalgeneralaction}}
Guided by the preceding discussion, we propose the following action for Donaldson-Witten theory coupled to the fields of the diffeomorphism Cartan Model:
\eqa{
&\IS_{\text{Cartan}} &&= \IQ\IV + \sfC ~, \label{eq:minimalgeneralactioneqn}
}
where
\eqa{
&\IV &&= \frac{i}{16\pi}\int_{\IX}d^{4}x\sqrt{g}\big(-i\big(\mathsf{Im\,}\tau_{IJ}\big)\big(F_{\mu\nu}^{+}{}^{I} + D_{\mu\nu}{}^{I}\big)\chi^{\mu\nu}{}^{J} - 2\,\CF_{IJ}\psi_{\sigma}{}^{I}D^{\sigma}\lambda^{J} - 2\,\ov{\CF}_{I}D_{\sigma}\psi^{\sigma}{}^{I} \nn
   & &&\qquad\qquad + \CF_{IJK}\psi_{\mu}{}^{I}\psi_{\nu}{}^{J}\chi^{\mu\nu}{}^{K} + \tfrac{i}{12}\ov{\CF}_{IJK}\chi_{\mu}{}^{\rho}{}^{I}\chi^{\mu\sigma}{}^{J}\chi_{\rho\sigma}{}^{K}  + 2\,\CF_{I}[\lambda,\eta]^{I} - 2\,\ov{\CF}_{I}[\phi, \eta]^{I}\big) ~, \label{eq:IV}\\
&\sfC &&= \frac{i}{16\pi}\int_{\IX}d^{4}x\left(\CF_{IJ} F^{I} \wedge F^{J} - \CF_{IJK}\psi^{I}\wedge\psi^{J}\wedge F^{K} + \frac{1}{12}\CF_{IJKL}\psi^{I}\wedge\psi^{J}\wedge\psi^{K}\wedge\psi^{L}\right) ~. \label{eq:IC}
}
Here $D_{\mu}$ denotes the (metric+gauge)-covariant derivative, and $I, J, K, \ldots$ serve as adjoint-valued indices for the gauge Lie algebra in the UV, and simply index the different Abelian vectormultiplets in the IR. We have defined\footnote{Note that $\tau_{IJ} := \CF_{IJ}$ and $\ov{\tau}_{IJ} := \ov{\CF}_{IJ}$ are not complex conjugates in the topologically twisted Euclidean theory.} $\mathsf{Im\,}\tau_{IJ} := \frac{\CF_{IJ}-\ov{\CF}_{IJ}}{2i}$. Note that \eqref{eq:IV} and \eqref{eq:IC} bear a similarity to the gauge fermion and non-exact part used in \cite{Marino:1998bm,Moore:1997pc}, except that the gauge fermion $\IV$ contains commutator terms chosen so as to reproduce the degree $=0$ Donaldson-Witten non-Abelian (UV) action of \cite{Witten:1988ze}. Both $\IV$ and $\sfC$ have gravity degree zero. By design, $\IV$ is a gauge- and diffeomorphism- invariant scalar density constructed out of fields of the vectormultiplet and the metric, and since the algebra closes on these fields (cf. \eqref{eq:IQ2}), it is clear that $\IQ^2 \IV$, being the Lie derivative of a \textit{density}, is a total derivative. The $\IQ$-closure of $\sfC$ is less obvious.\footnote{From the construction in \cite{Marino:1998bm,Moore:1997pc}, it is clear that $\sfC$ is closed under the action of $\CQ $, the Cartan differential for connections modulo gauge transformations.} 
To illustrate it, we begin by denoting the integrand of \eqref{eq:IC} by $\sfC_{\mathsf{form}}$ and acting on it with $\IQ$, yielding
\eqa{
&\IQ\sfC_{\mathsf{form}} &&= \iota_{\Phi}\psi^{I}\left(-\CF_{IJK}F^{J}\wedge F^{K} + \CF_{IJKL}\psi^{J}\wedge \psi^{K}\wedge F^{L} - \tfrac{1}{12}\CF_{IJKLM}\psi^{J}\wedge\psi^{K}\wedge\psi^{L}\wedge\psi^{M}\right)\nn
 & &&\quad  + 2\,\CF_{IJ}\big(D\psi^{I}\big) \wedge F^{J}  + 2\,\CF_{IJK}\big(D\phi^{I} - \iota_{\Phi}F^{I}\big)\wedge\psi^{J}\wedge F^{K} - \CF_{IJK}\psi^{I}\wedge\psi^{J}\wedge \big(D\psi^{K}\big) \nn
 & &&\quad + \tfrac{1}{3}\CF_{IJKL}\big(-D\phi^{I} + \iota_{\Phi}F^{I}\big)  \wedge \psi^{J} \wedge \psi^{K} \wedge \psi^{L}  \nn
 & &&= \textcolor{blue}{2\,\CF_{IJ}\big(D\psi^{I}\big)\wedge F^{J}} + \textcolor{blue}{2\,\CF_{IJK}(D\phi^{I}\big)\wedge\psi^{J}\wedge F^{K}} - \textcolor{red}{\CF_{IJK}\big(D\psi^{I}\big)\wedge \psi^{J}\wedge \psi^{K}} \nn
  & &&\quad - \textcolor{red}{\tfrac{1}{3}\CF_{IJKL}\big(D\phi^{I}\big)\wedge \psi^{J} \wedge \psi^{K} \wedge \psi^{L}} \nn
  & &&= d\left(\textcolor{blue}{2\,\CF_{IJ}\psi^{I} \wedge F^{J}} - \textcolor{red}{\tfrac{1}{3}\CF_{IJK}\psi^{I}\wedge \psi^{J} \wedge \psi^{K}}\right) ~,
}
where $d$ is the (de Rham) exterior derivative on $\Omega^{\bullet}(\IX)$. In the second equality, we have used some identities that arise from the elementary fact that a $p$-form on a 4-manifold $\IX$ vanishes if $p > 4$; the \textcolor{red}{red} and \textcolor{blue}{blue} terms separately combine.\footnote{The required identities are: (1) $\CF_{IJK}\big(\iota_{\Phi}F^{I}\big)\wedge \psi^{J} \wedge F^{K} = -\frac{1}{2}\CF_{IJK}\big(\iota_{\Phi}\psi^{I}\big)F^{J}\wedge F^{K}$, (2) $\CF_{IJKL}\big(\iota_{\Phi}F^{I}\big)\wedge \psi^{J} \wedge \psi^{K} \wedge \psi^{L} = -3\,\CF_{IJKL}\big(\iota_{\Phi}\psi^{I}\big)F^{J}\wedge \psi^{K} \wedge \psi^{L}$, and (3) $ \CF_{IJKLM}\big(\iota_{\Phi}\psi^{I}\big)\psi^{J}\wedge\psi^{K}\wedge\psi^{L}\wedge\psi^{M} = 0$.} We have thus shown that $\IQ\sfC$ is a total derivative (and hence integrates to zero over a closed 4-manifold $\IX$).

We can expand the action \eqref{eq:minimalgeneralactioneqn} in gravity degree, as the sum of the standard (degree zero) Donaldson-Witten action and contributions up to gravity degree $=2$:\footnote{For a differential form version of $\mathscr{L}_0$, see \eqref{eq:L0-form-our-conv}.}
\begin{empheq}[box=\fbox]{align}
\IS_{\text{Cartan}} &= \frac{i}{16\pi}\int_{\IX}d^{4}x\left[\mathscr{L}_{0} + \sqrt{g}\,\Psi^{\mu\nu}\Lambda_{\mu\nu} + \sqrt{g}\,\Phi^{\sigma}Z_{\sigma} + \sqrt{g}\,\Psi^{\mu\sigma}\Psi^{\nu}{}_{\sigma}\Upsilon_{\mu\nu} \right] + \int_{\IX}d^{4}x \sqrt{g}\,\n_{\mu}\mathscr{K}^{\mu} ~, \label{eq:GeneralAction}
\end{empheq}
where 
\eqa{
&\mathscr{L}_{0} &&= \frac{1}{2}\sqrt{g}\big(\ov{\tau}_{IJ}F_{\mu\nu}^{+}{}^{I}F_{+}^{\mu\nu}{}^{J} - \tau_{IJ}F_{\mu\nu}^{-}{}^{I}F^{\mu\nu}_{-}{}^{J}\big)  + 4i\sqrt{g}\big(\mathsf{Im\,}\tau_{IJ}\big)\big(D_{\mu}\phi^{I}\big)\big(D^{\mu}\lambda^{J}\big)  + i\sqrt{g}\big(\mathsf{Im\,}\tau_{IJ}\big)D_{\mu\nu}{}^{I}D^{\mu\nu}{}^{J} \nn
& &&\quad + 2\sqrt{g}\,\tau_{IJ}\psi_{\sigma}{}^{I} D^{\sigma}\eta^{J} - 2\sqrt{g}\,\ov{\tau}_{IJ}\eta^{I}D_{\sigma}\psi^{\sigma}{}^{J}  - 2\sqrt{g}\,\tau_{IJ}\psi_{\mu}{}^{I}D_{\nu}\chi^{\mu\nu}{}^{J} + 2\sqrt{g}\,\ov{\tau}_{IJ} \big(D_{[\mu}\psi_{\nu]}{}^{I}\big)^{+}\chi^{\mu\nu}{}^{J} \nn
& &&\quad + \frac{1}{2}\sqrt{g}\,\ov{\CF}_{IJK}\eta^{I}\big(F_{\mu\nu}^{+}{}^{J} + D_{\mu\nu}{}^{J}\big)\chi^{\mu\nu}{}^{K} + \sqrt{g}\,\CF_{IJK}\psi_{\mu}{}^{I}\psi_{\nu}{}^{J}\big(F_{-}^{\mu\nu}{}^{K} -D^{\mu\nu}{}^{K}\big) \nn
& &&\quad + \frac{1}{12}\CF_{IJKL}\veps^{\mu\nu\rho\sigma}\psi_{\mu}{}^{I}\psi_{\nu}{}^{J}\psi_{\rho}{}^{K}\psi_{\sigma}{}^{L} +   \frac{i}{12}\sqrt{g}\,\ov{\CF}_{IJKL}\eta^{I}\chi_{\mu}{}^{\rho}{}^{J}\chi^{\mu\sigma}{}^{K}\chi_{\rho\sigma}{}^{L} \nn
& &&\quad  + \frac{i}{4}\sqrt{g}\,\ov{\CF}_{IJK}\big(F_{\mu\rho}^{+}{}^{I} - D_{\mu\rho}{}^{I}\big)\chi^{\mu\sigma}{}^{J}\chi^{\rho}{}_{\sigma}{}^{K} + i\sqrt{g}\big(\mathsf{Im\,}\tau_{IJ}\big)[\phi,\chi_{\mu\nu}]^{I}\chi^{\mu\nu}{}^{J} \nn
& &&\quad + 2\sqrt{g}\,\tau_{IJ}\psi_{\sigma}{}^{I}[\psi^{\sigma}, \lambda]^{J} - 2\sqrt{g}\,\ov{\CF}_{I}[\psi_\sigma, \psi^\sigma]^{I} + 2\sqrt{g}\,\CF_{I}[\eta, \eta]^{I} \nn
& &&\quad - 2\sqrt{g}\,\ov{\tau}_{IJ}\eta^{I}[\phi, \eta]^{J} + 2\sqrt{g}\,\CF_{I}\big[\lambda, [\phi, \lambda]\big]^{I} - 2\sqrt{g}\,\ov{\CF}_{I}\big[\phi, [\phi, \lambda]\big]^{I} ~, \label{eq:GeneralActionDegree0}\\
&\Lambda_{\mu\nu} &&= -2i\big(\mathsf{Im\,}\tau_{IJ}\big)F^{-}{}_{\!\!\!\!\!\!\rho(\mu}{}^{I}\chi_{\nu)}{}^{\rho}{}^{J} - 2i g_{\mu\nu}\big(\mathsf{Im\,}\tau_{IJ}\big)\psi_{\rho}{}^{I}D^{\rho}\lambda^{J} + 2i\big(\mathsf{Im\,}\tau_{IJ}\big)\psi_{(\mu}{}^{I}D_{\nu)}\lambda^{J} \nonumber\\
& &&\quad + \frac{1}{2}\CF_{IJK}\big(\psi_{\mu}{}^{I}\psi_{\rho}{}^{J}\big)^{-}\chi^{\rho}{}_{\nu}{}^{K} + \frac{1}{2}\CF_{IJK}\big(\psi_{\nu}{}^{I}\psi_{\rho}{}^{J}\big)^{-}\chi^{\rho}{}_{\mu}{}^{K} - \frac{i}{48}g_{\mu\nu}\ov{\CF}_{IJK}\chi_{\kappa}{}^{\rho}{}^{I}\chi^{\kappa\sigma}{}^{J}\chi_{\rho\sigma}{}^{K}\nonumber\\
& &&\quad + g_{\mu\nu}\CF_{I}[\lambda,\eta]^{I} - g_{\mu\nu}\ov{\CF}_{I}[\phi, \eta]^{I} ~,\label{eq:GeneralActionDegree1}\\
 &Z_{\sigma} &&= \frac{1}{2}\CF_{IJK}\psi_{\sigma}{}^{I}(F^{+}_{\mu\nu}{}^{J} + D_{\mu\nu}{}^{J}\big)\chi^{\mu\nu}{}^{K} + i\big(\mathsf{Im\,}\tau_{IJ}\big)\big(D_{\sigma}\chi_{\mu\nu}{}^{I}\big)\chi^{\mu\nu}{}^{J} \nn
 & &&\quad + \big(\CF_{IJK}D_{\mu}\phi^{I} - \ov{\CF}_{IJK}D_{\mu}\lambda^{I}\big)\chi_{\nu\sigma}{}^{J}\chi^{\mu\nu}{}^{K} + 2i\big(\mathsf{Im\,}\tau_{IJ}\big)\big(D_{\mu}\chi_{\nu\sigma}{}^{I}\big)\chi^{\mu\nu}{}^{J} \nn
 & &&\quad + 2i\big(\mathsf{Im\,}\tau_{IJ}\big)\chi_{\nu\sigma}{}^{I}D_{\mu}\chi^{\mu\nu}{}^{J} + 2\,\CF_{IJK}\psi_{\sigma}{}^{I}\psi_{\rho}{}^{J}D^{\rho}\lambda^{K}  \nn
 & &&\quad - 4i\big(\mathsf{Im\,}\tau_{IJ}\big)F_{\sigma\rho}{}^{I}D^{\rho}\lambda^{J} - \CF_{IJKL}\psi_{\sigma}{}^{I}\psi_{\mu}{}^{J}\psi_{\nu}{}^{K}\chi^{\mu\nu}{}^{L}\nn
 & &&\quad + 2\,\CF_{IJK}F_{\sigma\mu}{}^{I}\psi_{\nu}{}^{J}\chi^{\mu\nu}{}^{K} - 2\,\tau_{IJ}\psi_{\sigma}{}^{I}[\lambda, \eta]^{J} + 2\,\CF_{I}[\lambda, D_{\sigma}\lambda]^{I} + 2\,\ov{\CF}_{I}[\psi_\sigma, \eta]^{I} \nn
 & &&\quad  - 2\,\ov{\CF}_{I}[\phi, D_\sigma\lambda]^{I} ~,\label{eq:GeneralActionDegree2-1}\\
 &\Upsilon_{\mu\nu} &&= \frac{i}{2}\big(\mathsf{Im\,}\tau_{IJ}\big)\chi_{\mu\rho}{}^{I}\chi_{\nu}{}^{\rho}{}^{J} ~,\label{eq:GeneralActionDegree2-2}\\
& \mathscr{K}^{\mu} &&= \frac{i}{16\pi}\left[2\,\ov{\CF}_{I}D^{\mu}\phi^{I}-2\,\tau_{IJ}\psi_{\nu}{}^{I}\chi^{\mu\nu}{}^{J} + 2\,\ov{\CF}_{I}\Psi^{\mu\nu}\psi_{\nu}{}^{I} - \ov{\CF}_{I}\Psi_{\sigma}{}^{\sigma}\psi^{\mu}{}^{I} + 2\,\ov{\CF}_{I}\Phi_{\nu}F^{\mu\nu}{}^{I}\right.\nonumber\\
& &&\qquad\qquad  \left. - 2\,i\,\big(\mathsf{Im\,}\tau_{IJ}\big)\Phi^{\sigma}\chi_{\nu\sigma}{}^{I}\chi^{\mu\nu}{}^{J}\right] ~.\label{eq:GeneralActionTotDer}
}

In section \ref{sec:CompareActions}, we will compare \eqref{eq:GeneralAction} with an action obtained using methods of twisted superconformal gravity, which we review next.

\section{Twisted Superconformal Gravity}\label{sec:TwistedScfmlGrav}

\subsection{Review Of Superconformal Gravity}\label{subsec:SuperconformalGravRev}

In this section, we give an expository introduction to superconformal gravity, also known as conformal supergravity -- the gauge theory of the superconformal group.\footnote{The gauge theory of the four-dimensional $\CN=1$ Lorentzian superconformal group was developed in a series of papers by Kaku, Townsend, and van Nieuwenhuizen \cite{Kaku:1977pa,Kaku:1977rk,Kaku:1978nz} (and, with Ferrara \cite{Ferrara:1977ij}). This was extended to the $\CN=2$ (Lorentzian) case in a series of papers by de Wit et al. \cite{deWit:1983xhu,deWit:1983xe,deWit:1984rvr,deWit:1979dzm,deWit:1980lyi,deWit:1984wbb}. See \cite{VanNieuwenhuizen:1981ae,Fradkin:1985am,Freedman:2012zz,Lauria:2020rhc} for reviews.} This is a field theory model of constrained principal bundles (with connections) for the generators of the superconformal algebra. The presentation of the theory in terms of physical fields is replete with indices, and in fact, quantum supergravity poses various technical challenges concerning proper definitions of supermoduli spaces.
%
%
 In this work, we will restrict our attention to classical non-dynamical field theory models of supergravity. We remind the reader of some interesting work by Lott \cite{Lott:1990zh,Lott:2001st} that uses the language of $G$-structures to define supergravity theories in diverse dimensions.\footnote{For more basic aspects of supersymmetry we refer our more mathematically-inclined readers to \cite{Deligne:1999qp,Deligne:1999ur,Freed:1999mn,Varadarajan:2004yz}.}

As explained in the introduction (cf. the discussion around \eqref{eq:FermionRep}), topologically twisting a four-dimensional $\CN=2$ theory involves specifying a local isomorphism of the $\mathsf{SU(2)}_+$ bundle and (self-dual) spin connection with the $\mathsf{SU(2)}_{\mathsf{R}}$ symmetry bundle and connection. A natural setting for this is $\CN=2$ conformal supergravity where $\mathsf{SU(2)}_{\mathsf{R}}$ is a local symmetry.\footnote{We distinguish between a `local symmetry,' and a `gauged symmetry.' Gauging involves summing over isomorphism classes of bundles with connection.   The $\mathsf{R}$-symmetry here is a nondynamical background gauge symmetry (i.e., local symmetry) of the classical background supergravity theory.}

We will restrict our attention to four-dimensional spacetimes with Euclidean signature since our focus is topological twisting on smooth, compact Riemannian 4-manifolds. The Euclidean conformal group is $\mathsf{SO(5,1)}$ and it includes the connected component of the Poincar\'{e} group, namely $\mathsf{ISO_0(4)} \cong \mathsf{SO_0(4)} \ltimes \IR^4$ (a semidirect product of the connected or proper orthochronous Lorentz group and the translation group), and additionally includes the dilatation group $\mathsf{SO(1,1)} \cong \IR$ and special conformal transformations.\footnote{A special conformal transformation is a composition of an inversion, a translation, and a second inversion.} The generators of the conformal algebra are translations $P_{a}$, Lorentz transformations $M_{[ab]}$, dilatations $\mathcal{D}$, and special conformal transformations $K_{a}$. They satisfy certain Lie bracket relations listed in Appendix \ref{app:csugra-misc}, see \eqref{eq:ConfAlgebra}.

The Euclidean conformal algebra\footnote{Here $\mathfrak{su}^{*}(4)$ is a particular real form of the algebra $\mathfrak{a}_{3}$ with maximal compact subalgebra $\mathfrak{usp(4)}$. See \cite{Gilmore:2012,Kugo:1982bn,Dahm:1996gz} for more details, including the isomorphism with $\mathfrak{sl}(2,\IH)$.} $\mathfrak{so(5,1)} \cong \mathfrak{su^*(4)}$ admits an $\CN=2$ supersymmetric extension to a superalgebra $\mathfrak{su^*(4|2)} \supset \mathfrak{so(5,1)} \oplus \mathfrak{su(2)}_{\mathsf{R}} \oplus \mathfrak{so(1,1)}_{\mathsf{R}}$ \cite{Frappat:1996pb}.\footnote{\label{foot:realforms}By contrast, the Lorentzian $\CN=2$ superconformal algebra is $\mathfrak{su}(2,2|2) \supset \mathfrak{su}(2,2) \oplus \mathfrak{su(2)}_{\mathsf{R}} \oplus \mathfrak{u(1)}_{\mathsf{R}}$ \cite{Kac:1977em}. Note that $\mathfrak{u(1)}_{\mathsf{R}} \cong i\IR$ and $\mathfrak{so(1,1)}_{\mathsf{R}} \cong \IR$, but in this paper we allow for the complexification of supergravity fields before restricting to an appropriate real section. Therefore, this distinction in real forms will not be important. The appearance of a noncompact factor in the Euclidean $\mathsf{R}$-symmetry group is well known, cf. \cite{Zumino:1977yh,Pestun:2007rz}.} The generators of the $\CN=2$ superconformal algebra include the bosonic generators of the conformal algebra, in addition to fermionic generators $\bm{Q}^{i}$ for supersymmetry and $\bm{S}^{i}$ for conformal supersymmetry (both $\mathfrak{su(2)}_{\mathsf{R}}$-valued Dirac spinors), bosonic generators $U$ for the $\mathfrak{su(2)}_{\mathsf{R}}$ symmetry and  $\mathcal{T}$ for the $\mathfrak{so(1,1)}_{\mathsf{R}}$ symmetry. See \eqref{eq:SupConfAlgebra} for the (super)Lie bracket relations of $\mathfrak{su^*(4|2)}$. 

Euclidean superconformal gravity  is formulated as follows: 
One begins by introducing a principal bundle with connection for the superconformal group. We will not enter into the definition of a principal bundle for a supergroup since our computations will be local, so we can replace the supergroup by its super Lie algebra. The connection is thus a $\IZ_2$-graded 1-form valued in the superconformal algebra that transforms inhomogeneously under gauge transformations. (Since we are working with gravity such gauge transformations include the action of diffeomorphisms and frame rotations.)  To form a proper supermultiplet we must introduce other fields in various associated bundles. Altogether the initial set of fields we begin with contain redundant degrees of freedom.  
We next impose constraints that express some components of the connection in terms of others. We call these components  ``composite fields.''  As we have stressed, for us, superconformal gravity is a nondynamical background theory. 
There exists a systematic procedure for writing general actions of matter fields coupled to a background off-shell conformal supergravity multiplet (known as the ``superconformal tensor calculus'' \cite{VanNieuwenhuizen:1981ae,VanProeyen:1983wk}).\footnote{Poincar\'{e} supergravity can be obtained from conformal supergravity coupled to special vector and hypermultiplets called compensators by a process called conformal gauge fixing and compensation (see, e.g., \cite{Freedman:2012zz}). In this sense, conformal supergravity is a big model from which other supergravity theories (with fewer symmetries or gauge redundancies) can be derived in different regimes.} We will exploit this to write actions generalizing Witten's twisted $\CN=2$ vectormultiplet action.

In \cite{Karlhede:1988ax}, Witten's twisted $\CN=2$ theory \cite{Witten:1988ze} was rederived by coupling the $\CN=2$ twisted vectormultiplet to a (twisted, bosonic) conformal supergravity background defined by a constant Killing spinor that realized the nilpotent BRST scalar supercharge corresponding to the middle term in \eqref{eq:TwistedFermionRep}. We extend this idea, by using $\CN=2$ superconformal gravity to arrive at a suitable twisted and truncated version that, in fact, reproduces the Cartan Model of $\DIFF$-equivariant cohomology of $\MET$, and upon coupling with vectormultiplet matter, the Cartan Model for the $\IG$-equivariant cohomology of $\IM$ with antighost multiplets. Therefore, we give a first-principles supergravity construction of the Cartan model for $\IG$-equivariant cohomology of $\IM$ of section \ref{sec:CartanModels}.

In Table \ref{tbl:ConformalSugraConnections} we list the various generators, parameters, and connections of $\CN=2$ superconformal gravity in local coordinates. Our index conventions for frames, coordinates, spinors, and $\mathsf{R}$-symmetry, are listed in Table \ref{tbl:index conventions}. Let $\widehat{\IX}$ denote a smooth, oriented Riemannian 4-manifold equipped with a spin structure: this is the spacetime for $\CN=2$ superconformal gravity.\footnote{Superconformal gravity consists of various spinorial fields, and thus a spin structure is needed. The spin 4-manifold with spin structure $\widehat{\IX}$ should be distinguished from the 4-manifold $\IX$ (which need not be spin) that we will switch to after topological twisting (for Donaldson theory).} The tangent space $T_{x}\widehat{\IX}$ at a point $x \in \IX$ is spanned by a set of independent vector fields $\{e_a\}_{a=1}^{4}$, each of which can be expanded in local coordinates as $e_{a} = e_{a}{}^{\mu}\partial_{\mu}$. The dual vector space is generated by the set of vielbein one-forms $\{e^a\}_{a=1}^{4}$ where $e^{a} = e_{\mu}{}^{a}dx^{\mu}$, and these satisfy $e^{a}e_{b} = e_{\mu}{}^{a}e^{\mu}{}_{b} = \delta_{b}^{a}$.\footnote{In the literature, $\{e_a\}$ are called vielbeins (tetrads) and $\{e^a\}$ are called co-vielbeins (co-tetrads). We adopt the more prevalent physics terminology and simply call them both vielbeins.} The spaces of connections of $\CN=2$ superconformal gravity are torsors for the sections of various associated bundles on $\IX$. We denote a torsor of the space of sections of a bundle $P \to \IX$ by $\mathsf{Tors\,}\Gamma(P)$.
The various connections are: the spin connection $\omega \in \mathsf{Tors\,}\Gamma\big(T^*\widehat{\IX}\otimes \Lambda^2 T\widehat{\IX}\big)$, the dilatation connection $b \in \mathsf{Tors\,}\Gamma\big(T^*\widehat{\IX}\big)$, the special conformal connection $f \in \mathsf{Tors\,}\Gamma\big(T^*\widehat{\IX} \otimes T\widehat{\IX}\big)$, the $\mathfrak{su(2)}_{\mathsf{R}}$ connection $V \in \mathsf{Tors\,}\Gamma\big(T^*\widehat{\IX}\otimes \mathsf{Sym}^2 \mathsf{K}_{\mathsf{R}}\big)$ (here $\mathsf{K}_{\mathsf{R}} \to \IX$ is the vector bundle associated (via the fundamental pseudoreal representation) to the principal $\mathsf{SU(2)}_{\mathsf{R}}$ bundle $\mathsf{P}_{\mathsf{R}} \to \IX$), 
the $\mathfrak{so(1,1)}_{\mathsf{R}}$ connection $A^{(\mathsf{R})} \in \mathsf{Tors\,}\Gamma\big(T^*\widehat{\IX}\big)$, the gravitino $\bm{\Psi} \in \mathsf{Tors\,}\Pi\Gamma\big(T^*\widehat{\IX}\otimes S\widehat{\IX} \otimes \mathsf{K}_{\mathsf{R}}\big)$ (where $S\widehat{\IX}$ is the spin bundle on $\widehat{\IX}$), and S-gravitino $\bm{S} \in \mathsf{Tors\,}\Pi\Gamma\big(T^*\widehat{\IX}\otimes S\widehat{\IX}\otimes\mathsf{K}_{\mathsf{R}}\big)$.

\begin{small}
\begin{table}[h]
\centering
\begin{tabular}{|c|c|c|c|c|c|}\hline
Local Symmetry & \begin{tabular}{@{}c@{}}Generator\end{tabular} & Parameter & \multicolumn{2}{c|}{Connection} & Supercurvature \\ \hline
Translations & $P_\ha$ & $\bm{\xi}^\ha$ & $e_{\mu}{}^{\ha}$ & vielbein (frame field) & $R(P)_{\mu\nu}{}^{\ha}$ \\ \hline
\begin{tabular}{@{}c@{}}Lorentz\\ Transformations\end{tabular} & $M_{\ha\hb}$ & $\bm{\lambda}^{\ha\hb}$ & $\omega_{\mu}{}^{\ha\hb}$ & spin connection & $R(M)_{\mu\nu}{}^{\ha\hb}$\\ \hline
Dilatations & $\mathcal{D}$ & $\ThetaD$ & $b_{\mu}$ & dilatation connection & $R(D)_{\mu\nu}$\\ \hline
 \begin{tabular}{@{}c@{}}Special Conformal\\Transformations\end{tabular} & $K_{\ha}$ & $\bm{\Lambda}_{K}{}^{\ha}$ & $f_{\mu}{}^{\ha}$ & \begin{tabular}{@{}c@{}}special conformal\\connection\end{tabular} & $R(K)_{\mu\nu}{}^{a}$\\ \hline
 \begin{tabular}{@{}c@{}} $\mathfrak{su(2)}_{\mathsf{R}}$ $\mathsf{R}$-symmetry \\ Transformations \end{tabular} & $U^{j}{}_{i}$ & $\bm{\Theta}^{i}{}_{j}$ & $V_{\mu}{}^{i}{}_{j}$ & $\mathfrak{su(2)}_{\mathsf{R}}$ connection & $R(V)_{\mu\nu}{}^{i}{}_{j}$\\ \hline
 \begin{tabular}{@{}c@{}} $\mathfrak{so(1,1)}_{\mathsf{R}}$ $\mathsf{R}$-symmetry \\ Transformations \end{tabular} & $\mathcal{T}$ & $\ThetaR$ & $\ARsym_{\mu}$ & $\mathfrak{so(1,1)}_{\mathsf{R}}$ connection & $R(\ARsym)_{\mu\nu}$\\ \hline
\begin{tabular}{@{}c@{}} Supersymmetry\end{tabular} & $\bm{Q}^i$ & $\eps_i$ & $\bm{\Psi}_{\mu i}$ & \begin{tabular}{@{}c@{}}gravitino \\ ($\mathfrak{su(2)}_{\mathsf{R}}$ Dirac spinor)\end{tabular} & $\bm{R(Q)}_{\mu\nu}{}^{i}$\\ \hline
\begin{tabular}{@{}c@{}}Conformal (S)\\ Supersymmetry\end{tabular} & $\bm{S}^i$ & $\bn_i$ & $\bm{S}_{\mu i}$ & \begin{tabular}{@{}c@{}}S-gravitino \\ ($\mathfrak{su(2)}_{\mathsf{R}}$ Dirac spinor)\end{tabular} & $\bm{R(S)}_{\mu\nu}{}^{i}$\\ \hline
\end{tabular}
\caption{\label{tbl:ConformalSugraConnections}Elements of four-dimensional $\CN=2$ superconformal gravity.}
\end{table}
\end{small}

As noted above, an important distinguishing feature of superconformal gravity is the imposition of constraints on certain curvatures that make some connections (namely, the spin connection $\omega_{\mu}{}^{\ha\hb}$, the connection for special conformal transformations $f_{\mu}{}^{\ha}$, and the (spinorial) connection for conformal supersymmetry $\bm{S}_{\mu}{}^{i}$) dependent on other gauge fields. These are the composite fields referred to above. In a superspace formulation, these supercurvature constraints arise from constraints on the supertorsion tensor, and define a superconformal structure \cite{Lott:1990zh,Lott:2001st}.

Closure of the algebra on fieldspace demands an off-shell equality of the number of bosonic and fermionic degrees of freedom. This requires a completion of the multiplet by adding non-dynamical ``auxiliary fields''. In general this process is non-unique, leading to different formulations of the conformal supergravity multiplet. We choose to work with the so-called standard Weyl multiplet, which, apart from the connections outlined in Table \ref{tbl:ConformalSugraConnections}, consists of an auxiliary scalar $\mathscr{D} \in \Omega^{0}(\widehat{\IX})$, an auxiliary 2-form $T \in \Omega^{2}(\widehat{\IX})$,\footnote{In Euclidean signature, the Hodge star squares to $\star^2 = 1$ and so the self-dual and anti-self-dual projections, $T_{\mu\nu}^{+}$ and $T_{\mu\nu}^{-}$, are independent and not related by complex conjugation.} and an auxiliary spinor $\bm{\Xi} \in \Pi\Gamma\big(S\widehat{\IX}\otimes \mathsf{K}_{\mathsf{R}}\big)$; it consists of $24$ bosonic and $24$ fermionic degrees of freedom, denoted as $24_{B}\oplus 24_{F}$. In  Table \ref{tbl:ConformalSugraConnections} we also list various supercurvatures -- these transform ``supercovariantly,'' meaning their supersymmetry transformations exclude derivatives of the parameters $\eps_i$ and $\bn_i$. They differ from ordinary (non-supercovariant) curvatures by the inclusion of terms involving fermionic fields.

The defining data of the off-shell Weyl $\CN=2$ superconformal gravity multiplet is, therefore:
\vspace{-0.9cm}
\begin{itemize}\itemsep -5pt
\item a set of fields:
\eqa{
  & &&\big( \underbrace{e_{\mu}{}^{\ha}, b_{\mu}, {V_\mu{}^i}{}_j, \ARsym_\mu, \bm{\Psi}_{\mu}{}^{i}}_{\text{independent connections}} ~;~ \underbrace{\omega_\mu{}^{\ha\hb}, f_\mu{}^\ha, \bm{S}_{\mu}{}^{i}}_{\text{composite connections}} ~;~ \underbrace{\mathscr{D}, T_{\ha\hb}, \bm{\Xi}^{i}}_{\text{auxiliary fields}}  \big) ~,
}
\item supersymmetry transformations of the independent connections and auxiliary fields,
\item supercurvature constraints (consistent with these transformation laws), the solutions to which yield expressions for the supercurvatures. 
\end{itemize}
We will encounter detailed expressions below.\footnote{See also \cite{deWit:2017cle,Fradkin:1985am,Freedman:2012zz,Lauria:2020rhc,deWit:1983xhu,deWit:1983xe,deWit:1984rvr,deWit:1979dzm,deWit:1980lyi,deWit:1984wbb} for further details, and Tables \ref{tbl:untwisted sugra notation},  \ref{tbl:untwisted VM notation}, \ref{tbl:untwisted chiral notation}, and \ref{tbl:untwisted antichiral notation} for notational differences.} The two Weyl spinor representations in four dimensions are unrelated by complex conjugation in Euclidean signature. This slightly changes the details of Euclidean supergravity relative to its Lorenztian version. However, regardless of signature, one can always work with Weyl spinor representatives of all spinorial objects (instead of reducible Dirac spinors), and then if one never invokes complex conjugation, the signature-based differences are benign for our purposes.\footnote{However, see \cite{vanNieuwenhuizen:1996ip,vanNieuwenhuizen:1996tv,Blau:1997pp,Belitsky:2000ii,Cortes:2003zd,Mohaupt:2007md,Mohaupt:2010du,Cassani:2012ri,Hama:2012bg,Klare:2012gn,Klare:2013dka,Witten:2015aba,Butter:2015tra,Freed:2016rqq,Freed:2019lec,Cortes:2019mfa,LopesCardoso:2019mlj} for some discussions of subtleties concerning analytic continuations of theories with spinors.} We extensively use these ``two-component'' or ``spinor'' methods for twisting as they are very useful for handling self-dual two-forms. We provide a self-contained introduction in Appendix \ref{app:spinormethods}, which may be reviewed at this point if needed.

\subsection{Truncating And Twisting}\label{subsec:Superconformal-TwistTruncate}
In this section, we show that there exists a conserved scalar supercharge (corresponding to a constant Killing spinor) on \underline{any} smooth, oriented Riemannian 4-manifold $\IX$ in $\CN=2$ conformal supergravity, when the $\mathfrak{su(2)}_+$ spin connection is identified with the $\mathfrak{su(2)}_{\mathsf{R}}$ gauge field (realizing \eqref{eq:FermionRep}--\eqref{eq:TwistingHomomorphism}) while retaining a chiral half of the gravitino that becomes a vector upon twisting. We use spinor methods for twisting, but present four-component transformation laws at the end to make contact with the diffeomorphism Cartan model (cf. \eqref{eq:GravityCartan-1}--\eqref{eq:GravityCartan-2}).

Traditionally, a supersymmetric bosonic background is determined by solving the generalized Killing spinor equations that result from setting the fermionic fields to zero and demanding that their supersymmetry variations also vanish. This leads generically to a system of PDEs which typically impose nontrivial geometric constraints. In a supergravity theory, this amounts to setting the gravitino (and any other fermionic fields of the supergravity multiplet, if present) to zero. The authors of \cite{Karlhede:1988ax} adopted this approach to make contact with \cite{Witten:1988xi}. This method has since been heavily expanded upon and generalized in many papers\footnote{A non-exhaustive list of references is \cite{Festuccia:2011ws,Klare:2012gn,Dumitrescu:2012ha,Cassani:2012ri,Dumitrescu:2012at,Klare:2013dka,Closset:2013vra,Closset:2014uda}.} to classify various supersymmetric backgrounds, and is crucial for the supersymmetric localization program \cite{Pestun:2016zxk,Teschner:2016yzf}. Our analysis differs from these works, since our interest is \underline{not} in a bosonic vacuum solution, but rather a specific truncation of twisted background conformal supergravity that leads to the Cartan model. As such, our background has a fermion (a chiral gravitino) which is part of its defining data as much as a metric is.\footnote{In fact, the key motivation for this pursuit, as spelled out in the conclusions of \cite{Moore:1997pc}, is to view the gravitino as a BRST partner of the metric, thus making them coexist democratically.}  We do not impose geometric constraints on the underlying compact, smooth manifold.\footnote{The constant Killing spinor found in \cite{Karlhede:1988ax} exists as a solution of the twisted conformal supergravity equations on \underline{any} smooth 4-manifold, and does not impose any restrictions on the tangential structure.}

Our starting point is the set of spinorial transformation laws (under susy, conformal susy, and special conformal transformations) of Euclidean $\CN=2$ superconformal gravity \cite{deWit:2017cle,Freedman:2012zz,Lauria:2020rhc}: 
\eqa{
&\hspace{-0.2cm}δ e_{μ}{}^{A\dA} &&= -i\sq\eps_{i}{}^{A}\Psi_{\mu}{}^{i\dA} + i\sq \eps_{i}{}^{\dA}\Psi_{\mu}{}^{iA} ~,\label{eq:csgura-susy-vielbein-2comp} \\
&\hspace{-0.2cm}δ Ψ_{μ}{}^{iA} &&= 2\,\bcd_{μ}\eps^{iA} - i\sq T^{A}{}_{B}e_{μ}{}^{B\dB}\eps^{i}{}_{\dB} - \sq e_{\mu}{}^{A\dA}\bn^{i}{}_{\dA} ~, \label{eq:csugra-susy-grav-left-2comp}\\
&\hspace{-0.2cm}δ Ψ_{μ}{}^{i\dA} &&= 2\,\bcd_{μ}\eps^{i\dA} - i\sq T^{\dA}{}_{\dB}e_{\mu}{}^{B\dB}\eps^{i}{}_{B} - \sq e_{\mu}{}^{A\dA}\bn^{i}{}_{A} ~, \label{eq:csugra-susy-grav-right-2comp}\\
&\hspace{-0.2cm}δ b_{μ} &&= \tfrac{i}{2}\big(\eps_{iA}S_{\mu}{}^{iA} + \eps_{i}{}^{\dA}S_{\mu}{}^{i}{}_{\dA}\big) + \tfrac{3i}{2\sq}\big(\eps_{iA}\Xi^{i}{}_{\dA} - \eps_{i\dA}\Xi^{i}{}_{A}\big)e_{\mu}{}^{A\dA} + \tfrac{i}{2}\big(\bn_{iA}\Psi_{\mu}{}^{iA} + \bn_{i}{}^{\dA}\Psi_{\mu}{}^{i}{}_{\dA}\big) \nn
& &&\quad + \bm{\Lambda}_{K,A\dA}e_{\mu}{}^{A\dA} ~, \label{eq:csugra-susy-dilatation-2comp}\\
&\hspace{-0.2cm}δ \ARsym_{\mu} &&= -\tfrac{i}{2}\big(\eps_{iA}S_{\mu}{}^{iA} - \eps_{i}{}^{\dA}S_{\mu}{}^{i}{}_{\dA}\big) + \tfrac{3i}{2\sq}\big(\eps_{iA}\Xi^{i}{}_{\dA} + \eps_{i\dA}\Xi^{i}{}_{A}\big)e_{\mu}{}^{A\dA} -\tfrac{i}{2}\big(\bn_{iA}\Psi_{\mu}{}^{iA} - \bn_{i}{}^{\dA}\Psi_{\mu}{}^{i}{}_{\dA}\big) ~, \label{eq:csgura-susy-rsym-u1-2comp} \\
&\hspace{-0.2cm}δ V_{μ}{}^{i}{}_{j} &&= 2i\big(\eps_{jA}S_{\mu}{}^{iA} + \eps_{j}{}^{\dA}S_{\mu}{}^{i}{}_{\dA}\big) + 3i\sq \big(\eps_{jA}\Xi^{i}{}_{\dA} - \eps_{j\dA}\Xi^{i}{}_{A}\big)e_{\mu}{}^{A\dA} - 2i\big(\bn_{jA}\Psi_{\mu}{}^{iA} + \bn_{j}{}^{\dA}\Psi_{\mu}{}^{i}{}_{\dA}\big)\nn
&\hspace{-0.2cm} &&\quad - \frac{1}{2}\delta^{i}_{j}(\text{trace}) ~, \label{eq:csugra-susy-rsym-su2-2comp} \\
&\hspace{-0.2cm}δ T_{EF} &&= -\big( \eps_{iA}R(Q)_{EF}{}^{iA} + \eps_{i}{}^{\dA}R(Q)_{EF}{}^{i}{}_{\dA} \big) ~,\label{eq:csugra-susy-TSD-2comp} \\
&\hspace{-0.2cm}δ T_{\dE\dF} &&= -\big( \eps_{iA}R(Q)_{\dE\dF}{}^{iA} + \eps_{i}{}^{\dA}R(Q)_{\dE\dF}{}^{i}{}_{\dA} \big) ~,\label{eq:csugra-susy-TASD-2comp} \\
&\hspace{-0.2cm}δ \Xi^{iA} &&= \mathscr{D}\eps^{iA} -\frac{4i}{3\sq}\big(\scd_{B\dA}T^{AB}\big)\eps^{i\dA} - \frac{1}{3}R(V)^{A}{}_{B}{}^{i}{}_{j}\eps^{jB}   + \frac{2}{3}R(\ARsym)^{A}{}_{B}\eps^{iB} - \frac{2i}{3}T^{A}{}_{B}\bn^{iB} ~, \label{eq:csugra-susy-C-left-2comp} \\
&\hspace{-0.2cm}δ \Xi^{i\dA} &&= \mathscr{D}\eps^{i\dA}-\frac{4i}{3\sq}\big(\scd_{A\dB}T^{\dA\dB}\big)\eps^{iA} - \frac{1}{3}{R(V)^{\dA}{}_{\dB}{}^{i}}_{j}\eps^{j\dB}  - \frac{2}{3}R(\ARsym)^{\dA}{}_{\dB}\eps^{i\dB} - \frac{2i}{3}T^{\dA}{}_{\dB}\bn^{i\dB} ~, \label{eq:csugra-susy-C-right-2comp}\\
&\hspace{-0.2cm}δ \mathscr{D} &&= -i\sq \eps_{i}{}^{A}\scd_{A\dA}\Xi^{i\dA} + i\sq \eps_{i}{}^{\dA}\scd_{A\dA}\Xi^{iA} ~.\label{eq:csugra-susy-auxiliary-scalar-2comp} 
}
Note that the one-form index on every connection is written as a curved index, but every flat index is written as a bispinor index. Here, $\scd_{\mu}$ denotes the fully supercovariant derivative,
whereas $\bcd_{\mu}$ is covariant with respect to the metric, dilatations, and $\mathsf{R}$-symmetry in the \emph{untwisted} theory. (Note that $\scd_{A\dA} = e^{\mu}{}_{A\dA}\scd_{\mu}$.) 
The covariant derivatives appearing in \eqref{eq:csugra-susy-grav-left-2comp} and \eqref{eq:csugra-susy-grav-right-2comp} are:
\eqa{
&\bcd_{μ}\eps^{iA} &&= \partial_{μ}\epsilon^{iA} + \frac{1}{2}{\omega_{μ}{}^{A}}_{B}\eps^{iB}	+ \frac{1}{2}(b_μ + A_μ^{\rm{(}\mathsf{R}\rm{)}})\eps^{iA} + \frac{1}{2}{V_{μ}{}^{i}}_{j}\eps^{jA}\label{eq:covder 2 comp eps} ~,\\
&\bcd_{μ}\eps^{i\dA} &&= \partial_{μ}\epsilon^{i\dA} + \frac{1}{2}\omega_{μ}{}^{\dA}{}_{\dB}\epsilon^{i\dB} + \frac{1}{2}(b_μ - A_μ^{\rm{(}\mathsf{R}\rm{)}})\eps^{i\dA} + \frac{1}{2}{V_{μ}{}^{i}}_{j}\eps^{j\dA}\label{eq:covder 2 comp epsbar} ~.
}
The relevant expressions for various (super)covariant derivatives, composite fields, and (super)covariant curvatures can be found in Appendix \ref{app:csugra-misc}, cf. \eqref{eq:covder 2 comp eps repeated}--\eqref{eq:covder 2 comp Xibar}, \eqref{eq:conformal gravitino 2-comp left}--\eqref{eq:DeltaCont2}, and \eqref{eq:omega curv 4comp}--\eqref{eq:RM supercurvature dotted} respectively.

The theory is defined by the following supercurvature constraints.\footnote{In \cite{Freedman:2012zz,deWit:2017cle} these are referred to as ``conventional constraints,'' as they are algebraic in the composite fields, which are determined as solutions to these constraints. They are to be distinguished from differential constraints.} The first of these,
\eqa{
& R(P)_{\mu\nu}{}^{C\dC} &&= 0 ~,
}
where $R(P)_{\mu\nu}{}^{C\dC}$ is given by \eqref{eq:RP-supercurvature}, determines the components of the spin connection in terms of the vielbein, the gravitino, and the dilatation gauge field. The second set of constraints,
\eqa{
& R(Q)_{AB}{}^{iA} &&= -3\,\Xi^{i}{}_{B} ~, \qquad & R(Q)_{AB}{}^{i}{}_{\dA}  &= 0 ~,\\
& R(Q)_{\dA\dB}{}^{i\dA} &&= -3\,\Xi^{i}{}_{\dB} ~, \qquad & R(Q)_{\dA\dB}{}^{i\dA} &= 0 ~,
}
determines the components of the S-gravitino in terms of the vielbein, the gravitino, the 2-form $T$, and the auxiliary spinor $\bm{\Xi}$. (The components of $R(Q)$ are given by \eqref{eq:RQ undotted}--\eqref{eq:RQ dotted antiselfdual}. As written above, $R(Q)$ is symmetric in the first two lower spinor indices.) Finally, the third constraint,
\eqa{
& e^{\nu}{}_{B\dB}R(M)_{\mu\nu}{}^{A\dA,B\dB} - \widetilde{R}(\ARsym)_{\mu}{}^{A\dA} - \frac{1}{64}e_{\mu}{}^{C\dC}T^{A}{}_{C}T_{\dC}{}^{\dA} - \frac{3}{2}\mathscr{D}e_{\mu}{}^{A\dA} &&= 0 ~,\label{eq:curvature-constraint-3}
}
determines the special conformal connection.\footnote{$\widetilde{R}(\ARsym)_{\mu\nu}$ denotes the Hodge dual of the supercurvature $R(\ARsym)_{\mu\nu}$ with respect to a Riemannian metric, and $\widetilde{R}(\ARsym)_{\mu}{}^{A\dA} = \widetilde{R}(\ARsym)_{\mu}{}^{\nu}e_{\nu}{}^{A\dA}$. The curvature $R(M)_{\mu\nu}{}^{A\dA,B\dB}$ depends algebraically on $f_{\mu}{}^{a}$ (cf. \eqref{eq:RM supercurvature}).} Thus these constraints -- the imposition of which ensures that the other fields transform under covariant general coordinate transformations \cite{VanProeyen:1983wk} -- express the spin connection $\omega_{\mu}{}^{\ha\hb}$, the S-gravitino $\bm{S}_{\mu}{}^{i}$, and the special conformal connection $f_{\mu}{}^{\ha}$ in terms of the remaining fields. 
(Thus they are referred to as ``composite fields.'')   
One can interpret these constraints in superspace supergeometry as defining an $\CN=2$ superconformal structure \cite{Lott:1990zh,Lott:2001st}, but in the present component-based approach we will not pursue this direction.

Following \eqref{eq:TwistingHomomorphism} and \cite{Karlhede:1988ax} we twist by identifying $\mathfrak{su(2)}_+$ (undotted spinor) indices with $\mathfrak{su(2)}_{\mathsf{R}}$ indices, and set
\eqa{
	\omega_{\mu}{}^{AB} &= V_{\mu}{}^{AB} ~.\label{eq:toptwist sugra}
}
(This equation is symmetric under $A\leftrightarrow B$.)
We seek a minimal solution of the supersymmetry conditions which preserves a constant scalar supercharge, and a local vector supersymmetry, the connection for which is a vector gravitino $\Psi_{\mu}{}^{A\dA}$, which we recognize as the chiral half of a gravitino post twist: $\Psi_{\mu}{}^{i\dA} \rightarrow \Psi_{\mu}{}^{A\dA}$. (Note that $e_{\nu,A\dA}\Psi_{\mu}{}^{A\dA} = \Psi_{\mu\nu}$. The twisted gravitino so obtained is \emph{not} symmetric. In fact, this is an important point that we will return to below.)

First, we gauge away the dilatation connection $b_{\mu}$ using a special conformal transformation (SCT) -- this is a standard first step in supergravity ($b_{\mu}$ is non-dynamical, and the complete supergravity action is independent of $b_{\mu}$ \cite{Kaku:1977pa}). We use a compensating SCT to preserve $\delta b_{\mu} = 0$. Setting the RHS of \eqref{eq:csugra-susy-dilatation-2comp} to zero determines the appropriate $Λ_{K,\ha}$.\footnote{To obtain the complete transformation law of $b_{\mu}$, we add to \eqref{eq:csugra-susy-dilatation-2comp} the gauge transformation of $b_\mu$ itself (i.e., a local dilatation: $\delta_{\text{Weyl}}b_{\mu} = \partial_\mu\ThetaD$).} As no elementary field of interest in the sugra or matter multiplets transforms under SCTs, this does not introduce any compensating transformations. Therefore, in the equations below, we simply drop $b_{\mu}$.

We set $\Psi_{\mu}{}^{AB} = 0$ (which is neither symmetric nor antisymmetric under $A\leftrightarrow B$), and demand $\delta \Psi_{\mu}{}^{AB} = 0$ for consistency.\footnote{All four-dimensional torsion terms in supergravity vanish once a chiral half of the gravitino is set to zero.} From \eqref{eq:csugra-susy-grav-left-2comp}, this condition translates to
\eqa{
&2\,\bcd_{\mu}\eps^{AB} - i\sq T^{B}{}_{C}\eps^{A}{}_{\dC}e_{\mu}{}^{C\dC} - \sq e_{\mu}{}^{B\dC}\bn^{A}{}_{\dC} &&= 0 ~, \label{eq:susy twisted psi set to zero}
}
where \eqref{eq:covder 2 comp eps} is modified (due to \eqref{eq:toptwist sugra}) to
\eqa{
&\bcd_{μ}\eps^{AB} &&= ∂_{μ}\eps^{AB} + \frac{1}{2}{\omega_{μ}{}^{A}}_{C}\eps^{CB} + \frac{1}{2}{\omega_{μ}{}^{B}}_{C}\eps^{AC} + \frac{1}{2}A_{μ}^{\rm{(}\mathsf{R}\rm{)}}\eps^{AB} ~.\label{eq:covder eps twist}
}
Contracting \eqref{eq:susy twisted psi set to zero} with $e^{\mu}{}_{B\dA}$ yields
\eqa{
   &\bn_{A\dA} &&= \frac{1}{\sq}\bcd_{B\dA}\eps_{A}{}^{B} ~. \label{eq:vec conformal}
}
Substituting \eqref{eq:vec conformal} into \eqref{eq:susy twisted psi set to zero} yields
\eqa{
	&2\,\bcd_{C\dC}\eps^{AB} - i\sq {T^{B}}_{C}\eps^{A}{}_{\dC} - δ_{C}{}^{B}\bcd_{D\dC}\eps^{AD} &&= 0 ~. \label{eq:delta psi zero 2}
}
Since we want a field-independent local vector superysmmetry to be preserved, we set \underline{$T_{AB} = 0$} (or equivalently, \underline{$T_{\ha\hb}^{+} = 0$}) and hence \eqref{eq:delta psi zero 2} reduces to
\eqa{
	&2\,	\bcd_{C\dC}\eps^{AB} - δ_{C}{}^{B}\bcd_{D\dC}\eps^{AD} &&= 0 ~. \label{eq:delta psi zero 3}
}
We now \emph{require} that a scalar supercharge and a vector supercharge be preserved. That is,
\eqas{
   &\eps^{AB} &&= \tfrac{i}{2\sq}\veps^{AB}\eps ~, \quad \text{where } \eps \neq 0 ~, \quad \text{ and } \quad  
		\eps_{A\dA} &&\neq 0 ~.
}
So \eqref{eq:delta psi zero 3} reduces to $\bcd_{μ}\eps = 0$, i.e., there exists a \emph{covariantly constant fermionic scalar} (which is an irreducible component of the twisted spinor). Using \eqref{eq:covder eps twist} this yields $∂_{μ}\eps = -\frac{1}{2}A_{μ}^{\rm{(}\mathsf{R}\rm{)}} \eps$. Upon setting \underline{$A_{\mu}^{\rm{(}\mathsf{R}\rm{)}} = 0$}, the covariantly constant scalar becomes a \emph{constant scalar}. Then \eqref{eq:vec conformal} tells us that the vector conformal supersymmetry parameter vanishes: \underline{$\bn_{A\dA} = 0$}. 

This computation is similar to the one carried out in \cite{Karlhede:1988ax} except that we have \underline{not} set as many background fields (esp.~fermionic fields such as all components of the gravitino) to zero. Therefore this entails a careful consistency check of the supersymmetry (susy) transformations of all fields set equal to zero since their susy variations must necessarily vanish too. We begin with the (gauge + susy) variation of the \emph{Abelian} $\mathsf{R}$-symmetry connection $A_{\mu}^{\rm{(}\mathsf{R}\rm{)}}$. From \eqref{eq:csgura-susy-rsym-u1-2comp},
\eqa{
 &δ\ARsym_{\mu} &&= ∂_{μ}\bm{\Theta}^{\rm{(}\mathsf{R}\rm{)}} - \tfrac{i}{2}\left(\tfrac{i}{2\sq}\eps\,S_{μ,A}{}^{A} + \eps^{A\dA}S_{μ,A\dA}\right)+ \tfrac{3i\sq}{4}\left(\tfrac{i}{2\sq}\eps\,\Xi_{C\dC} + \eps_{A\dC}{\Xi^{A}}_{C}\right)e_{μ}{}^{C\dC} = 0 ~,
}
which upon using \eqref{eq:SgravTrunc-1} and \eqref{eq:SgravTrunc-2}, reduces to\footnote{Here $\CJ_{μν,ρ} = \n_{[μ}Ψ_{ν]ρ}$ is the curl of the twisted gravitino. The truncation and twist reduce $\bcd_{μ}Ψ_{νρ}$ to $\n_{μ}Ψ_{νρ}$. As the curl shows up frequently in twisted supergravity, it deserves its own symbol and Appendix \ref{subsec:curl-twisted-gravitino}.}
\eqa{
 &∂_{μ}\ThetaR -\frac{1}{2}\eps\left[\Xi_{C\dC} - \frac{1}{2}\left(\CJ^{A}{}_{C,\dA\dC} + \frac{1}{3}{\CJ_{\dC}{}^{\dA}}_{CA}\right)\right]e_{μ}{}^{C\dC} - i\sq\eps^{A}{}_{\dC}\Xi_{AC}e_{μ}{}^{C\dC} &&= 0 ~.
}
As we elected to make the Abelian $\mathsf{R}$-symmetry \emph{global}, we posit that each of the three terms in this equation vanishes separately. Without loss of generality, we set \underline{$\ThetaR = 0$}. The components of the (twisted) auxiliary field $\Xi$ thus have the following profiles:
\eqa{
	&\Xi_{C\dC} &&= \frac{1}{2}{\CJ^{A}}_{C,A\dC} + \frac{1}{6}\CJ_{\dC}{}^{\dA}{}_{C\dA} ~,\label{eq:chi vec} \quad &\text{or equivalently,} &\quad \Xi_{μ} &&= \CJ_{μρ}^{+}{}^{ρ} + \frac{1}{3}\CJ_{μρ}^{-}{}^{ρ} ~,\\
	&\Xi_{AB} &&= 0 ~,\quad &\text{or equivalently,} &\quad \Xi_{μν} &&= 0 \label{eq:chi nonvec} ~,
}
where the coordinate versions of these expressions are obtained using \eqref{eq:contJbig1} and \eqref{eq:contJbig2}. Next, we consider the constraint $\delta T_{AB} = 0$, which, from \eqref{eq:csugra-susy-TSD-2comp}, reads
\eqa{
  &δT_{AB} &&= -\tfrac{i}{2\sq}\eps R(Q)_{AB,C}{}^{C} + \eps^{C\dC}R(Q)_{AB,C\dC} = 0 ~,
}
which is an identity, because after truncation and twisting, the supercurvatures $R(Q)_{AB,CD}$ (and hence, the ``trace'' $R(Q)_{AB,C}{}^{C} : = \veps^{CD}R(Q)_{AB,CD}$) and $R(Q)_{AB,C\dC}$ vanish identically -- see Appendix \ref{app:misc-expr-twisted} for a careful derivation. Therefore, no constraints are imposed by $δT_{AB} = 0$. Finally, we impose $δ\Xi^{AB} = 0$ for consistency with \eqref{eq:chi nonvec}. Using \eqref{eq:csugra-susy-C-left-2comp}, this condition reads
\eqa{
 &\left(\mathscr{D}\veps^{AB} + \frac{1}{3}R(\omega)^{B}{}_{C}{}^{AC}\right)\eps = 0 ~, \label{eq:delta CAB zero}
}
and this must hold for any constant fermionic scalar $\eps$. (See \eqref{eq:riemann spinor} and \eqref{eq:RABCD} for a definition of $R(\omega)_{ABCD}$.) This implies that the expression in the parenthesis must identically vanish.\footnote{The symmetrization of \eqref{eq:delta CAB zero} under $A\leftrightarrow B$ is an identity due to \eqref{eq:contracted R}.} 
Multiplying \eqref{eq:delta CAB zero} throughout by $\veps_{AB}$ and using \eqref{eq:contracted R}, we get
\eqa{
&\mathscr{D} &&= \frac{1}{6}R(\omega)_{AB}{}^{AB} = \frac{1}{6} \mathscr{R}\label{eq:Dscr} ~,
}
where $\mathscr{R}$ is the Ricci scalar on $\IX$.\footnote{Here we used the fact that \eqref{eq:toptwist sugra} and \eqref{eq:contracted R} together imply that $R(V)_{AB}{}^{AB} = R(ω)_{AB}{}^{AB} = \mathscr{R}$.} Indeed, \eqref{eq:Dscr} is identical to the result obtained in \cite{Karlhede:1988ax} relating the auxiliary scalar in the Weyl multiplet to the Ricci scalar of the 4-manifold. 

Decomposing the 2-form S-susy parameter $\bn_{AB}$ as $\bn_{AB} = \frac{1}{2\sq}\bn^{\text{sym}}_{(AB)} + \frac{1}{\sq}\eps_{AB} \bn_{0}$, the transformation law \eqref{eq:csugra-susy-grav-right-2comp} of the twisted gravitino takes the form
\eqa{
   &\delta \Psi_{\mu\nu} &&= -\frac{1}{2}\eps\,\Psi_{\mu}{}^{\rho}\Psi_{\nu\rho} - \eps\,T_{\mu\nu}^{-} + 2\, \n_{\mu}\eps_{\nu} - \bn_{\mu\nu}^{+} + \bn_{0} g_{\mu\nu} ~, \label{eq:twisted-grav-transflaw}
}
where $\bn_{\mu\nu}^{+} := \frac{1}{2}\bn^{\text{sym}}_{(AB)} e_{\mu}{}^{A}{}_{\dA}e_{\nu}{}^{B\dA}$, $T_{\mu\nu}^{-} := \frac{1}{2}T_{\dA\dB} e_{\mu,A}{}^{\dA}e_{\nu}{}^{A\dB}$, etc. The reader may be perturbed by the fact that the gravitino here is \emph{not} symmetric, in contrast to section \ref{subsec:CartanModelDiff} where it was symmetric \emph{by fiat}. 
We will address this in section \ref{subsec:Diff-Cartan-from-TwistedSugra}.

As $\bn_{A\dA} = 0$, the field $\Xi_{\mu}$ does not transform under S-supersymmetry -- see \eqref{eq:csugra-susy-C-right-2comp}. It may seem concerning that \eqref{eq:chi vec} involves $\Psi_{\mu\nu}$, which does transform under S-supersymmetry. However, by decomposing the gravitino\footnote{This elementary decomposition, explained in Appendix \ref{subsec:spinorial-decomposition-of-twisted-gravitino}, proves to be exceedingly useful.} into a symmetric traceless part ($\widehat{\Psi}_{\mu\nu}$ or $\widehat{\Psi}_{(AB),(\dA\dB)}$), a trace ($\Psi_{\rho}{}^{\rho}$ or $\Psi$), a SD part ($\Psi_{[\mu\nu]}^{+}$ or $\Psi_{(AB)}$) and an ASD part ($\Psi_{[\mu\nu]}^{-}$ or $\Psi_{(\dA\dB)}$), we find that \eqref{eq:chi vec} becomes
\eqas{
    & \Xi_{\mu} &&= -\frac{1}{3}\n_{\rho}\wh{\Psi}_{\mu}{}^{\rho} + \frac{1}{4}\n_{\mu}\Psi_{\rho}{}^{\rho} - \frac{2}{3}\n^{\rho}\Psi_{[\mu\rho]}^{-} ~, \\
    & \Xi_{A\dA} &&= -\frac{1}{3}\n_{E\dE}\wh{\Psi}_{A}{}^{E,\dE}{}_{\dA} + \frac{1}{4}\n_{A\dA}\Psi - \frac{1}{3}\n_{A\dE}\Psi_{\dA}{}^{\dE} ~,
    \label{eq:chi vec rewrite}
}
which is independent of $\Psi_{[\mu\nu]}^{+}$ or $\Psi_{AB}$ -- the \underline{only} component of the gravitino that transforms under 2-form S-susy -- see, e.g., \eqref{eq:tsugra-gravitino-irred-2comp} and \eqref{eq:tsugra-gravitino-irred-4comp}. Therefore the RHS of \eqref{eq:chi vec} (or equivalently, of \eqref{eq:chi vec rewrite}) is indeed invariant under 2-form S-susy. Moreover, the only irreducible component transforming under 0-form S-susy is the trace $\Psi$. In particular, this implies that
\eqa{
   \delta_{\substack{\rm 0-form \\ \rm S-susy}}(\bn_0)\Xi_{\mu} = 0 = \n_{\mu}\bn_{0}  \implies \bn_{0} = \text{constant.} 
}
Therefore, $\bn_{0}$ is a constant fermionic scalar, the S-susy partner of the constant scalar $\eps$.

This concludes the analysis of the constraints from the supersymmetry conditions. However, we must establish the consistency of the truncation and topological twist by proving that the constraints are robust under a supersymmetry variation. We do so in Appendix \ref{sec:ConsistencyTwist}.

To summarize, we have obtained a truncation of the Weyl multiplet of superconformal gravity to three independent fields: the vielbein $e_{\mu}{}^{a}$ (or metric $g_{\mu\nu}$), the gravitino $\Psi_{\mu\nu}$ and the bosonic anti-self-dual auxiliary field $T_{\mu\nu}^{-}$. In addition there are various constrained composite fields such as the $\mathfrak{su(2)}_{\mathsf{R}}$ connection given by \eqref{eq:toptwist sugra}, the auxiliary field $\Xi_{\mu}$ given by \eqref{eq:chi vec} (and the vanishing $0$-form and $2$-form versions, cf. \eqref{eq:chi nonvec}), and the S-gravitinos, given by \eqref{eq:SgravTrunc-5} and \eqref{eq:SgravTrunc-7}. 

The supersymmetry transformation laws of this model in 4-component notation are:
\eqas{
   &\delta g_{\mu\nu} &&= \eps\,\Psi_{(\mu\nu)} ~, \qquad \delta \Psi_{\mu\nu} = -\frac{1}{2}\eps\,\Psi_{\mu}{}^{\rho}\Psi_{\nu\rho} - \eps\,T_{\mu\nu}^{-} + 2\, \n_{\mu}\eps_{\nu} - \bn_{\mu\nu}^{+} + \bn_{0} g_{\mu\nu} ~, \\
   &\delta T_{\mu\nu}^{-} &&= -\eps\,\Psi_{[\mu}{}^{\rho}T_{\nu]\rho}^{-} + \eps^{\rho}R(Q)_{\mu\nu,\rho}^{-} ~, \label{eq:delta-twisted-conformal-sugra}
}
where the twisted (anti-self-dual) supercurvature $R(Q)_{\mu\nu,\rho}^{-}$ expanded upon in \eqref{eq:RQ twisted 4-component}--\eqref{eq:RQ twisted 4-component irred}, is: 
\eqas{
  &R(Q)_{\mu\nu,\rho}^{-} &&= 2\,\big(\n_{[\mu}\Psi_{\nu]\rho}\big)^{-} - \tfrac{1}{2}\left[g_{\rho\mu}\left(\n^{\sigma}\Psi_{[\nu\sigma]}^{+} -  \n^{\sigma}\Psi_{[\nu\sigma]}^{-}\right) - \left(\mu \leftrightarrow \nu\right) \right]^{-}  \label{eq:RQ-twisted-general-mainbody} \\
 & &&= 2\big(\n_{[\mu}\wh{\Psi}_{\nu]\rho}\big)^{-} - \tfrac{1}{2}\big(g_{\rho[\mu}\n_{\nu]}\Psi_{\sigma}{}^{\sigma}\big)^{-} + 2\big(\n_{[\mu}\Psi_{\nu]\rho}^{-}\big)^{-} + 2\big(g_{\rho[\mu}\n^{\sigma}\Psi^{-}_{\nu]\sigma}\big)^{-} ~,
}
where, in the second equality, we used \eqref{eq:4comp}.\footnote{Note that the supercurvature $R(Q)^{-}_{\mu\nu,\rho}$ is independent of the self-dual component $\Psi_{[\mu\nu]}^{+}$ of the antisymmetric gravitino. This is less clear from the first line of \eqref{eq:RQ-twisted-general-mainbody}, but is obvious from the second line. This is another useful application of \eqref{eq:4comp}.} In Table \ref{tbl:twisted-weyl-multiplet} we list the supersymmetry parameters, basic fields of the twisted Weyl multiplet, and their Weyl weights.
\begin{small}
\begin{table}[H]
\centering
\def\arraystretch{1.1}
\begin{tabular}{|c|c|c|c|c|}\hline
 \begin{tabular}{@{}c@{}}Component in\\ a local chart\end{tabular} & Description & \begin{tabular}{@{}c@{}} Element/ \\ section of \end{tabular} & \begin{tabular}{@{}c@{}}Grassmann\\ Parity\end{tabular} & \begin{tabular}{@{}c@{}} Weyl \\ Weight \end{tabular}\\ \hline\hline
   $e_{\mu}{}^{\ha}$ & vielbein ($e^a = e_{\mu} dx^{\mu}$) & $T_{x}^*\IX$ & even & $-1$ \\ \hline
   $g_{\mu\nu}$ & metric & $\MET(\IX)$ & even & $-2$ \\ \hline
   $\Psi_{\mu\nu}$ & gravitino & $\Pi\mathsf{Sym}^2(T\IX)$ & odd & $-\frac{3}{2}$ \\ \hline
   $T_{\mu\nu}^{-}$ & ASD auxiliary field & $\Omega^{2,-}(\IX)$ & even & $-1$ \\ \hline \hline
   $\eps$ & (const.) scalar susy parameter & $\Pi\Omega^{0}(\IX)$ & odd & $-\frac{1}{2}$ \\ \hline
   $\eps^{\mu}$ & vector susy parameter & $\Pi\Omega^{0}(\IX)$ & odd & $+\frac{1}{2}$ \\ \hline
   $\bn_{0}$ & (const.) 0-form S-susy parameter & $\Pi\Omega^{0}(\IX)$ & odd & $+\frac{1}{2}$ \\ \hline
   $\bn_{\mu\nu}^{+}$ & 2-form S-susy parameter & $\Pi\Omega^{2,+}(\IX)$ & odd & $-\frac{3}{2}$ \\ \hline
\end{tabular}
\caption{\label{tbl:twisted-weyl-multiplet}Basic fields and supersymmetry parameters of the twisted $\CN=2$ Weyl multiplet.}
\end{table}
\end{small}

The twist \eqref{eq:toptwist sugra} breaks local Weyl covariance. Although we began with a bigger model with local conformal invariance, the truncated model does not exhibit this property. But this poses no problem for our work. Local Weyl transformations do not appear on the right-hand side of the algebra of the truncated model, and moreover, they play no role in the Cartan model of equivariant cohomology. In untwisted $\CN=2$ conformal supergravity, the spin connection is a composite field but it is Weyl-covariant by design, and all the $b_{\mu}$-dependence drops out of the action. Nevertheless, the untwisted theory is still Weyl-invariant because it is invariant under diffeomorphisms and special conformal transformations (the commutator of these two is a dilatation plus a local Lorentz transformation). This is compatible with the observation by Witten in \cite{Witten:1988xi}  that the energy-momentum tensor of twisted $\CN=2$ SYM coupled to ``ordinary'' gravity (i.e., just a metric) is not traceless. Since it is not traceless there is a breakdown of conformal invariance. Since the trace is a total divergence the theory has a global scale invariance. This is consistent with the absence of a mass gap in Donaldson theory \cite{Witten1995}.\footnote{In the case of K\"{a}hler manifolds, it is possible to reduce to a situation with a mass gap and compute Donaldson invariants in terms of an $\CN=1$ theory. See \cite{Witten:1994ev,Witten1995} for details.} 

\subsection{The \textsf{Diffeomorphism} Cartan Model From Twisted Supergravity}\label{subsec:Diff-Cartan-from-TwistedSugra}

In this section, we rederive the Cartan Model for diffeomorphisms on the space of metrics (which was discussed in section \ref{subsec:CartanModelDiff}) from twisted supergravity. We have already met the gravitino $\Psi_{\mu\nu}$ (which, in the sugra formulation, has no definite symmetry under $\mu\leftrightarrow\nu$ yet),
but the degree $2$ bosonic vector field $\Phi^{\mu}$ of the Cartan Model is notably absent in section \ref{subsec:Superconformal-TwistTruncate}. In twisted supergravity, the field $\Phi^{\mu}$ is the (bosonic) BRST ghost for the local vector supersymmetry. Since the background supergravity fields transform under background BRST transformations, it makes sense to introduce ghosts for the residual local supersymmetries, namely the 1-form (or vector) supersymmetry and the self-dual 2-form conformal supersymmetry, and as we shall see, these ghosts play a crucial role. To this end, we construct a BRST complex for background supergravity, which incorporates $\Phi^{\mu}$ while also rendering the gravitino symmetric.

Let us begin by examining the obstruction to the naive truncation $\Psi_{[\mu\nu]} = 0$ in the supergravity transformations \eqref{eq:delta-twisted-conformal-sugra}. Here it is helpful to split the gravitino into irreducible components using \eqref{eq:tsugra-gravitino-irred-2comp}--\eqref{eq:tsugra-gravitino-irred-4comp}.
The SD and ASD components of $\Psi_{[\mu\nu]}$ transform respectively as\footnote{The bilinear actually decomposes as $\Psi_{\mu}{}^{\rho}\Psi_{\nu\rho} = \wh{\Psi}_{\mu}{}^{\rho}\wh{\Psi}_{\nu\rho} + 2\,\underbrace{\wh{\Psi}_{[\mu}{}^{\rho}{\Psi}^{+}_{\nu]\rho}}_{\text{purely ASD}} + 2\,\underbrace{\wh{\Psi}_{[\mu}{}^{\rho}\Psi^{-}_{\nu]\rho}}_{\text{purely SD}} + \underbrace{\Psi^{+}_{\mu}{}^{\rho}\Psi_{\nu\rho}^{+}}_{\text{purely SD}} + \underbrace{\Psi^{-}_{\mu}{}^{\rho}\Psi_{\nu\rho}^{-}}_{\text{purely ASD}}$.}
\eqa{
  &\delta\Psi_{[\mu\nu]}^{+} &&= -\eps\big(\wh{\Psi}_{[\mu}{}^{\rho}\Psi^{+}_{\nu]\rho}\big) - \frac{\eps}{2}\big(\wh{\Psi}_{[\mu}{}^{\rho}\wh{\Psi}_{\nu]\rho}\big)^{+}  - \frac{\eps}{2}\big(\Psi^{+}_{\mu}{}^{\rho}\Psi^{+}_{\nu\rho}\big) + 2\big(\n_{[\mu}\eps_{\nu]}\big)^{+} - \bn_{\mu\nu}^{+}~,\\
   &\delta\Psi_{[\mu\nu]}^{-} &&= -\eps\big(\wh{\Psi}_{[\mu}{}^{\rho}\Psi^{-}_{\nu]\rho}\big) - \frac{\eps}{2}\big(\wh{\Psi}_{[\mu}{}^{\rho}\wh{\Psi}_{\nu]\rho}\big)^{-}  - \frac{\eps}{2}\big(\Psi^{-}_{\mu}{}^{\rho}\Psi^{-}_{\nu\rho}\big) + 2\big(\n_{[\mu}\eps_{\nu]}\big)^{-} - \eps\,T_{\mu\nu}^{-}~,
}
where $\wh{\Psi}_{\mu\nu}$ is the symmetric traceless part of the gravitino. We could gauge away the self-dual component $\Psi_{[\mu\nu]}^{+}$, at the expense of introducing compensating 2-form S-susy transformations on matter fields with a parameter
\eqa{
   &\bn_{\mu\nu}^{+} &&= -\frac{1}{2}\eps\,\big(\Psi_{\mu}{}^{\rho}\Psi_{\nu\rho}\big)^{+} + 2 \big(\n_{[\mu}\eps_{\nu]}\big)^{+} ~. \label{eq:eta-2-form-if}
}
However, the vanishing condition on the variation of the anti-self-dual part $\Psi_{[\mu\nu]}^{-}$,
\eqa{
  & -\frac{1}{2}\eps\,\big(\Psi_{\mu}{}^{\rho}\Psi_{\nu\rho}\big)^{-} + 2 \big(\n_{[\mu}\eps_{\nu]}\big)^{-} - \eps\,T_{\mu\nu}^{-} &&\stackrel{?}{=} 0 ~, \label{eq:obstruct sym grav}
}
would then constrain the anti-self-dual part of the curl of the 1-form susy\footnote{We will frequently refer to this as vector supersymmetry, exploiting the duality between 1-forms and vector fields on a Riemannian manifold.} parameter $\eps_{\mu}$. Thus the classical Weyl multiplet of $\CN=2$ superconformal gravity (i.e., the multiplet \emph{without} BRST ghosts) does not admit a truncation to a purely symmetric gravitino without gauge fixing the vector supersymmetry. Instead of this, we eliminate $\Psi_{[\mu\nu]}$ using a BRST procedure.

To construct the BRST complex, we proceed as follows: the Grassmann-odd scalar supersymmetry parameter $\eps$ is formally replaced by the product of a Grassmann-odd constant $\bm{\Lambda}$ and a commuting (constant) scalar supersymmetry ghost $\mathsf{c}_{\eps}$, i.e., $\eps \to \bm{\Lambda} \mathsf{c}_{\eps}$.  Similarly, for vector supersymmetry, we introduce a commuting ghost -- a vector field $\Phi \in \Gamma(T\IX)$ which, in a local chart, has components $\Phi^{\mu}$ -- and formally replace $\eps^\mu \rightarrow \bm{\Lambda} \Phi^{\mu}$. We also replace the self-dual 2-form S-susy parameter with its self-dual (bosonic) 2-form ghost $\mathsf{c}_{\mu\nu}^{+}$, i.e., $\bn_{\mu\nu}^{+} \to \bm{\Lambda} \mathsf{c}_{\mu\nu}^{+}$. We restrict to the subspace of transformations for which $\bn_0 = 0$, and no ghost is needed for 0-form S-susy. (This is consistent with the inference above that $\bn_{0}$ is a constant. The Cartan Model does not admit a transformation that rotates the symmetric gravitino into the metric, though it is quite natural in supergravity, as simply the conformal susy counterpart of the susy transformation.) We also do not add ghosts for local diffeomorphisms.\footnote{Doing so would lead to the Weil model or the BRST model.} We do \underline{not} view the vector supersymmetry as a quantum gauge symmetry that needs gauge fixing.\footnote{Quantum gauge fixing in supergravity is quite a delicate matter, see \cite{Freed:2002qp,Freed:2004yc} for mathematical expositions.} The fields in the gravitational BRST complex are background fields that are \underline{not} quantized. 

We \emph{choose} $\mathsf{c}_{\eps} = 1$. This is an explicit choice of scale for the constant scalar susy ghost field, made to match the transformation laws of the Cartan model. We define the `BRST variation' of a field by $\delta_{B}(\text{field}) := \bm{\Lambda} \, \IQ(\text{field})$ where $\IQ$ is the `BRST differential.'\footnote{The BRST differential is an odd derivation on the algebra of local functionals of the fields which is nilpotent. The quotes around `BRST' are meant to emphasize that $\IQ$ is nilpotent on the relevant invariant subcomplexes.}

%
Closure of the algebra dictates that $\IQ \Phi^{\mu} = 0$. The transformation laws \eqref{eq:delta-twisted-conformal-sugra} thus yield the following `BRST' transformation laws:
\eqa{
   &\IQ g_{\mu\nu} &&= \Psi_{(\mu\nu)} ~, \label{eq:BRST metric general gravitino}\\
   &\IQ \Psi_{(\mu\nu)} &&= \n_{\mu}\Phi_{\nu} + \n_{\nu}\Phi_{\mu} ~,  \label{eq:BRST symmetric part of general gravitino}\\
   &\IQ \Phi^{\mu} &&= 0 ~, \label{eq:BRST Phi}\\
   &\IQ \Psi_{[\mu\nu]} &&= \n_{\mu}\Phi_{\nu} - \n_{\nu}\Phi_{\mu} - \tfrac{1}{2}\Psi_{\mu}{}^{\rho}\Psi_{\nu\rho} - T_{\mu\nu}^{-} - \mathsf{c}_{\mu\nu}^{+} ~, \label{eq:BRST antisymmetric grav}\\
   &\IQ T_{\mu\nu}^{-} &&= -\Psi_{[\mu}{}^{\rho}T_{\nu]\rho}^{-} + \Phi^{\rho} R(Q)_{\mu\nu,\rho}^{-} \label{eq:BRST T transformation}~,
}
where $R(Q)_{\mu\nu,\rho}^{-}$ is given by \eqref{eq:RQ-twisted-general-mainbody}. This construction endows $\IQ$ as well as the fields a natural ``gravity degree,'' namely $\IQ$ has degree $=1$, $g_{\mu\nu}$ has degree $=0$, $\Psi_{\mu\nu}$ has degree $=1$, and $\Phi^{\mu}$, $T_{\mu\nu}^{-}$ and $\mathsf{c}_{\mu\nu}^{+}$ each have degree $=2$. 

Having constructed the BRST complex, we now set $\Psi_{[\mu\nu]} = 0$ and impose the constraint $\IQ \Psi_{[\mu\nu]} = 0$, which using \eqref{eq:BRST antisymmetric grav}, leads to the following two conditions:
\eqa{
    &T_{\mu\nu}^{-} &&= \big(\n_{\mu}\Phi_{\nu} - \n_{\nu}\Phi_{\mu}\big)^{-} - \tfrac{1}{2}\big(\Psi_{\mu}{}^{\rho}\Psi_{\nu\rho}\big)^{-} ~, \label{eq:sym-grav-T}\\
    &\mathsf{c}_{\mu\nu}^{+} &&= \big(\n_{\mu}\Phi_{\nu} - \n_{\nu}\Phi_{\mu}\big)^{+} - \tfrac{1}{2}\big(\Psi_{\mu}{}^{\rho}\Psi_{\nu\rho}\big)^{+} ~. \label{eq:sym-grav-ghost}
}
The $\mathsf{c}_{\mu\nu}^{+}$ ghost profile is therefore constrained and we can simply \emph{define} $\IQ \mathsf{c}_{\mu\nu}^{+}$ to be equal to the $\IQ$-variation of the RHS of \eqref{eq:sym-grav-ghost}.\footnote{This is also consistent with the self-dual S-susy parameter $\bn_{\mu\nu}^{+}$ being otherwise unconstrained.} However, we must verify that the prescribed profile \eqref{eq:sym-grav-T} is consistent with the $\IQ$-transformation of $T_{\mu\nu}^{-}$ given by \eqref{eq:BRST T transformation}. To do so, we define $\sfP_{\mu\nu} :=2\n_{[\mu}\Phi_{\nu]}$ and $\sfQ_{\mu\nu} := \Psi_{\mu}{}^{\rho}\Psi_{\nu\rho}$, so that $T_{\mu\nu}^{-} = \sfP_{\mu\nu}^{-} - \frac{1}{2}\sfQ_{\mu\nu}^{-}$. Now,
\eqa{
   &\IQ \sfP_{\mu\nu} 
   &&= -2\big(\Psi_{[\mu}{}^{\rho}\n_{\nu]}\Phi_{\rho}\big) +  2 \Phi^{\rho}\big(\n_{[\mu}\Psi_{\nu]\rho}\big) ~,\\
   &\IQ \sfQ_{\mu\nu} 
   &&= -\Psi^{\rho\sigma}\Psi_{\mu\rho}\Psi_{\nu\sigma} - 2 \big(\Psi_{[\mu}{}^{\rho}\n_{\nu]}\Phi_{\rho}\big) - 2\big(\Psi_{\rho[\mu}\n^{\rho}\Phi_{\nu]}\big) ~,
}
and so
\eqa{
 &\IQ\left(\sfP_{\mu\nu} - \tfrac{1}{2}\sfQ_{\mu\nu}\right) &&= -\big(\Psi_{[\mu}{}^{\rho}\sfP_{\nu]\rho}\big) + 2\Phi^{\rho}\big(\n_{[\mu}\Psi_{\nu]\rho}\big) + \tfrac{1}{2}\Psi^{\rho\sigma}\Psi_{\mu\rho}\Psi_{\nu\sigma} ~.
}
Recalling the identity \eqref{eq:identA refined}, we have
\eqa{
   &\IQ \sfA_{\mu\nu}^{-} 
  & &&= \big(\IQ \sfA_{\mu\nu}\big)^{-}  - \big(\Psi_{[\mu}{}^{\rho}\sfA_{\nu]\rho}^{-}\big)^{+} + \big(\Psi_{[\mu}{}^{\rho}\sfA_{\nu]\rho}^{+}\big)^{-} ~,\label{eq:IQ on ASD}
}
for a generic 2-form $\sfA_{\mu\nu}$. Applying \eqref{eq:IQ on ASD} to the 2-form $\sfA_{\mu\nu} := \sfP_{\mu\nu}^{-} - \frac{1}{2}\sfQ_{\mu\nu}^{-}$ and using \eqref{eq:identB} successively, we get
\eqa{
&\IQ\left(\sfP_{\mu\nu}^{-} - \tfrac{1}{2}\sfQ_{\mu\nu}^{-}\right) &&= -\textcolor{blue}{\big(\Psi_{[\mu}{}^{\rho}\sfP_{\nu]\rho}\big)^{-}} + 2\Phi^{\rho}\big(\n_{[\mu}\Psi_{\nu]\rho}\big)^{-} +  \tfrac{1}{2}\big(\Psi^{\rho\sigma}\Psi_{\mu\rho}\Psi_{\nu\sigma}\big)^{-}\nn
& &&\quad - \big(\Psi_{[\mu}{}^{\rho}T_{\nu]\rho}^{-}\big)^{+} + \textcolor{blue}{\big(\Psi_{[\mu}{}^{\rho}\sfP_{\nu]\rho}^{+}\big)^{-}} - \textcolor{red}{\tfrac{1}{2}\big(\Psi_{[\mu}{}^{\rho}Q_{\nu]\rho}^{+}\big)^{-}} \nn
& && =  - \big(\Psi_{[\mu}{}^{\rho}T_{\nu]\rho}^{-}\big)^{+} + 2\Phi^{\rho}\big(\n_{[\mu}\Psi_{\nu]\rho}\big)^{-} - \underline{\textcolor{blue}{\big(\Psi_{[\mu}{}^{\rho}\sfP_{\nu]\rho}^{-}\big)^{-}}} \nn
& &&\quad -\textcolor{red}{\tfrac{1}{2}\big(\Psi_{[\mu}{}^{\rho}Q_{\nu]\rho}\big)^{-}} + \underline{\textcolor{red}{\tfrac{1}{2}\big(\Psi_{[\mu}{}^{\rho}Q_{\nu]\rho}^{-}\big)^{-}}} +  \tfrac{1}{2}\big(\Psi^{\rho\sigma}\Psi_{\mu\rho}\Psi_{\nu\sigma}\big)^{-} \nn
& &&\stackrel{\eqref{eq:identB}}{=\joinrel=\joinrel=} -\textcolor{OliveGreen}{\big(\Psi_{[\mu}{}^{\rho}T_{\nu]\rho}^{-}\big)^{+}} + 2\Phi^{\rho}\big(\n_{[\mu}\Psi_{\nu]\rho}\big)^{-} + \underline{\textcolor{OliveGreen}{\tfrac{1}{4}\Psi_{\rho}{}^{\rho}T_{\mu\nu}^{-}}}\nn
& &&\qquad - \dashuline{\tfrac{1}{4}\big(\Psi_{\mu}{}^{\rho}\Psi_{\nu}{}^{\sigma}\Psi_{\rho\sigma} - \Psi_{\nu}{}^{\rho}\Psi_{\mu}{}^{\sigma}\Psi_{\rho\sigma}\big)^{-}} + \dashuline{\tfrac{1}{2}\big(\Psi^{\rho\sigma}\Psi_{\mu\rho}\Psi_{\nu\sigma}\big)^{-}} \nn
& &&\stackrel{\eqref{eq:identB}}{=\joinrel=\joinrel=} -\Psi_{[\mu}{}^{\rho}T_{\nu]\rho}^{-} + \Phi^{\rho}R(Q)_{\mu\nu,\rho}^{-} ~,
} 
which is identical to the RHS of \eqref{eq:BRST T transformation}, as was desired to be shown.\footnote{Terms that combine with each other in a single step or are related to the preceding step are underlined or colored. In the penultimate step, the dashed terms add up to zero because $\Psi_{\mu}{}^{\rho}\Psi_{\nu}{}^{\sigma}\Psi_{\rho\sigma}$ is antisymmetric in $\mu, \nu$ as the gravitino is Grassmann-odd and symmetric. In the final step, we used \eqref{eq:RQ-twisted-general-mainbody} or equivalently, its vastly simplified version \eqref{eq:sym-grav-RQ-twisted-4-component} for the symmetric gravitino.} 

We have thus proved the consistency of truncating the antisymmetric part of the gravitino. To summarize, the $\IQ$-transformations of our twisted and truncated supergravity model with a symmetric gravitino are
\eqas{
      &\IQ g_{\mu\nu} &&= \Psi_{\mu\nu} ~, \\
      &\IQ \Psi_{\mu\nu} &&= \n_{\mu}\Phi_{\nu} + \n_{\nu}\Phi_{\mu} ~, \\
      &\IQ \Phi^{\mu} &&= 0 ~, \label{eq:TwistedSUGRA-Cartan-Diffeo}
}
which are identical to \eqref{eq:GravityCartan-1}--\eqref{eq:GravityCartan-2} with the understanding that $\sfd = \left.\IQ\right|_{g,\Psi,\Phi}$ (cf. \eqref{eq:IQ-restriction-to-sugra-fields}). The profiles for $T_{\mu\nu}^{-}$ and $\mathsf{c}_{\mu\nu}^{+}$ are constrained by \eqref{eq:sym-grav-T} and \eqref{eq:sym-grav-ghost}.\footnote{In gauge-fixing and compensating untwisted $\CN=2$ superconformal gravity to arrive at $\CN=2$ Poincar\'{e} supergravity, a field equation of $T_{\ha\hb}$ is imposed, which identifies $T_{\ha\hb}$ with the graviphoton field strength (the graviphoton connection is the connection in an Abelian vectormultiplet compensator). By contrast, in our model, $T_{\ha\hb}^{-}$ is constrained off-shell to make contact with a model of equivariant cohomology.} The expressions for various composite fields and curvatures simplify tremendously once we truncate to a symmetric gravitino -- see Appendix \ref{app:csugra-misc} for these simplifications. We remark that for \eqref{eq:TwistedSUGRA-Cartan-Diffeo}, $\IQ$ squares to a Lie derivative along the vector field $\Phi$, i.e., $\left.\IQ\right|_{g,\Psi,\Phi}^2 = \CL_{\Phi}$, consistent with \eqref{eq:sfd squared}. (The result for $\IQ^2$ is modified by matter fields as we saw in section \ref{subsec:CartanModelDiff}, and will see again in section \ref{sec:SuperconformalGrav-twistedSYM}.)

We have obtained the Cartan Model for orientation-preserving diffeomorphisms $\DIFF(\IX)$ on the space of metrics $\MET(\IX)$, by a truncation and twisting of $\CN=2$ superconformal gravity. 

It is interesting to compare our findings with Witten's model of four-dimensional topological gravity \cite{Witten:1988xi}. Witten began with a model consisting of a metric, a symmetric traceless gravitino, and a bosonic vector field $C$ (our $\Phi$). This minimal multiplet can be obtained by linearizing around a metric with an anti-self-dual curvature. Indeed, this is the diffeomorphism Cartan model (albeit with a traceless gravitino; the traceless condition was relaxed in section 2.2 of \cite{Witten:1989ig}). We have derived this model using conformal supergravity, enabling us to use the superconformal tensor calculus to write general matter couplings.

The discord with local Weyl invariance that was mentioned above can be understood as follows. Witten's model does not have an anti-self-dual 2-form field $T$ (or for that matter, many of the other composite objects that appear in conformal supergravity) and differs from ours in the Weyl weight assignments. But the Weyl weight of $\Phi_{\mu}$ ($C_\mu$ for Witten) in \textit{either} formulation can never be zero (for Witten it is $-2$, and for us, it is $-1$) and so the right-hand sides of \eqref{eq:BRST symmetric part of general gravitino}, \eqref{eq:BRST antisymmetric grav}, and \eqref{eq:sym-grav-T} transform inhomogenously under local Weyl transformations (i.e., pick up a derivative of the Weyl transformation parameter) whereas the left-hand sides do not. This is again consistent with a breakdown of local Weyl invariance in the twisted theory.

The term ``twisted supergravity'' has a different meaning in the work of Costello and Li \cite{Costello:2016mgj} and related follow-up works such as \cite{Costello:2018zrm,Costello:2020jbh}. The bosonic background BRST ghost $\Phi^{\mu}$ in our model is reminiscent of the ghost field for a fermionic background symmetry in \cite{Costello:2016mgj,Distler:1991mf,Govindarajan:1991pn,Fujitsu:1991gx}. However, as frequently noted above, our gravitational theory is purely classical: We make no attempt to integrate over the fields $g_{\mu\nu}, \Psi_{\mu\nu}, \Phi^\mu$.   Another approach to twisting in supergravity -- which also aims to sum over quantum fluctuations -- is developed in \cite{deWit:2018dix,Jeon:2018kec}. These papers develop a BRST formalism for quantizing gauge theories with a soft algebra (i.e., closure involves field-dependent structure functions rather than structure constants, as is standard in supergravity) and use them for supersymmetric localization. Equivariant cohomology appears in these papers -- as is natural whenever a BRST complex is involved \cite{Kanno1989}, and consequently, there are some structural similarities to our transformations.\footnote{Refs. \cite{Frenkel:2020djn,Frenkel:2020dic,Frenkel:2020ixs} describe a BRST complex that bears similarities to ours, though they do not use equivariant cohomology or make contact with supergravity. If $\xi^{\mu}$ denotes the diffeomorphism ghost (absent in the Cartan Model, but present in the BRST or Weil models), then with the identifications $\left.g_{\mu\nu}\right|_{\text{here}} \mapsto \left.g_{ij}\right|_{\text{there}}$, $\left.\Psi_{\mu\nu}\right|_{\text{here}} \mapsto \left.\Psi_{ij}\right|_{\text{there}}$, $\left.\Phi^{\mu}\right|_{\text{here}} \mapsto \left.\xi^{\mu}\right|_{\text{there}}$, and $\left.\xi^\mu\right|_{\text{here}} \mapsto \left.c^i\right|_{\text{there}}$, and setting $\left.n^i\right|_{\text{there}} = 0$, $\left.\Psi^{i}\right|_{\text{there}} = 0$, $\left.\dt{c}^{i}\right|_{\text{there}} = 0$ and $\left.\phi^{i}\right|_{\text{there}} = 0$, we recover the BRST model for $\DIFF(\IX)$-equivariant cohomology of $\MET(\IX)$ as a particular truncation in a static limit (no dependence on the Hamilton-Perelman Ricci flow parameter $t$).} However, these papers do not make contact with the Cartan model for $\DIFF(\IX)$ equivariant cohomology of $\MET(\IX)$, and moreover, their analysis applies only to spaces that have isometries (i.e., admit Killing vectors). Finally, their notion of an invertible twist (a field redefinition) differs from the Donaldson-Witten twist of \cite{Witten:1988xi}. Our approach also makes contact with some earlier work of Imbimbo and Rosa \cite{Imbimbo:2018duh} where the BRST algebra underlying a supergravity theory was presented without a first-principles derivation. 

\section{Coupling Twisted SYM To A Twisted Supergravity Background}\label{sec:SuperconformalGrav-twistedSYM}

In this section, we study the coupling of twisted $\CN=2$ super Yang-Mills (SYM) to the twisted supergravity background of section \ref{sec:TwistedScfmlGrav}.  The untwisted $\CN=2$ vectormultiplet consists (in Wess-Zumino gauge) of scalar fields $\lambda, \phi \in \Omega^{0}(\widehat{\IX}, \adsf P\big)$, the YM connection $A \in \CA(P)$,
%
%
an $\mathfrak{su(2)}_{\mathsf{R}}$-doublet spinor $\bm{\Omega} \in \Gamma\big(S\widehat{\IX}\otimes \adsf P \otimes \mathsf{K}_{\mathsf{R}}\big)$ (with Weyl spinor components $\Omega^{iA}$ and $\Omega^{i\dA}$), and an $\mathfrak{su(2)}_{\mathsf{R}}$-triplet auxiliary field $D \in \Gamma(\adsf P \otimes \mathsf{Sym}^2\mathsf{K}_{\mathsf{R}}\big)$.\footnote{See Table \ref{tbl:untwisted VM notation} for a comparison of our vectormultiplet notation with other supergravity literature.}
\eqa{
&\hspace{-0.4cm} \delta \lambda &&= \tfrac{i}{\sq}\eps_{iA}\Omega^{iA} ~, \label{eq:untwisted vm first}\\
&\hspace{-0.4cm}  \delta \phi &&= -\eps^{i\dA}\Omega_{i\dA} ~,\\
&\hspace{-0.4cm}  \delta A_{\mu} &&= \big(\eps_{iA}\Omega^{i}{}_{\dA} + i\sq \eps_{i\dA}\Omega^{i}{}_{A}\big)e_{\mu}{}^{A\dA} + i\sq\,\phi\,\eps_{iA}\Psi_{\mu}{}^{iA} + \lambda\,\eps_{i}{}^{\dA}\Psi_{\mu}{}^{i}{}_{\dA} ~,\\
&\hspace{-0.4cm}  \delta \Omega^{iA} &&= -i\sq\big(\scd^{A\dA}\lambda\big)\eps^{i}{}_{\dA} + \tfrac{i}{2\sq}\big(\wh{F}^{A}{}_{B} + 4\,\phi\,T^{A}{}_{B}\big)\eps^{iB} + \tfrac{i}{2\sq}D^{i}{}_{j}\eps^{jA} + \tfrac{i}{\sq}[\lambda,\phi]\eps^{iA} + i\,\lambda\,\bn^{iA} ~,\\
&\hspace{-0.4cm}  \delta \Omega^{i}{}_{\dA} &&= -\big(\scd_{A\dA}\phi\big) \eps^{iA} - \tfrac{1}{2}\big(\wh{F}_{\dA}{}^{\dB} + \lambda\,T_{\dA}{}^{\dB}\big)\eps^{i}{}_{\dB} + \tfrac{1}{2}D^{i}{}_{j}\eps^{j}{}_{\dA} - [\lambda,\phi]\eps^{i}{}_{\dA} + 2\,i\,\phi\,\bn^{i}{}_{\dA} ~,\\
&\hspace{-0.4cm}  \delta D^{ij} &&= \eps^{(i A}\scd_{A\dA}\Omega^{j)\dA} - i\sq\,\eps^{(i\dA}\scd_{A\dA}\Omega^{j)A} + 8\,i\,\eps^{(i}{}_{A}[\phi, \Omega^{j)A}] - 4\,i\,\eps^{(i\dA}[\lambda, \Omega^{j)}{}_{\dA}] ~.
}
Here, the final equation is symmetrized over $i,j$ indices, $\wh{F}$ denotes the supercovariant YM field strength, and $\scd_{\mu}$ is the supercovariant derivative, which is now also gauge-covariant. Explicitly,
\eqa{
& \scd_{\mu}\lambda &&= D_{\mu}\lambda - \tfrac{i}{2\sq}\Psi_{\mu,iA}\Omega^{iA} ~,\\
& \scd_{\mu}\phi &&= D_{\mu}\phi + \tfrac{1}{2}\Psi_{\mu}{}^{i\dA}\Omega_{i\dA} ~,\\
& \scd_{\mu}\Omega^{iA} &&= D_{\mu}\Omega^{iA} + \tfrac{i}{\sq}\big(\scd^{A\dA}\lambda\big)\Psi_{\mu}{}^{i}{}_{\dA} - \tfrac{i}{4\sq}\big(\wh{F}^{A}{}_{B} + 4\,\phi\,T^{A}{}_{B}\big)\Psi_{\mu}{}^{iB} - \tfrac{i}{4\sq}D^{i}{}_{j}\Psi_{\mu}{}^{jA} \nonumber\\
& &&\quad - \tfrac{i}{2\sq}[\lambda,\phi]\Psi_{\mu}{}^{iA} - \tfrac{i}{2}\lambda\,S_{\mu}{}^{iA} ~,\\
& \scd_{\mu}\Omega^{i}{}_{\dA} &&= D_{\mu}\Omega^{i}{}_{\dA} + \tfrac{1}{2}(\scd_{A\dA}\phi)\Psi_{\mu}{}^{iA} + \tfrac{1}{4}\big(\wh{F}_{\dA}{}^{\dB} + \lambda\,T_{\dA}{}^{\dB}\big)\Psi_{\mu}{}^{i}{}_{\dB} - \tfrac{1}{4}D^{i}{}_{j}\Psi_{\mu}{}^{j}{}_{\dA} \nonumber\\
& &&\quad + \tfrac{1}{2}[\lambda, \phi]\Psi_{\mu}{}^{i}{}_{\dA} - i\,\phi\,S_{\mu}{}^{i}{}_{\dA} ~,\\
&\wh{F}_{\mu\nu} &&= \underbrace{2\,\partial_{[\mu}A_{\nu]} + [A_{\mu}, A_{\nu}]}_{F(A)_{\mu\nu}} - i\sq\,\Psi_{[\mu,i}{}^{\dA}\Omega^{iA}e_{\nu]\,A\dA} + i\sq\,\Psi_{[\mu,iA}\Omega^{i}{}_{\dA}e_{\nu]}{}^{A\dA} \nonumber\\
& &&\quad - 2\,\Psi_{\mu,iA}\Psi_{\nu}{}^{iA}\,\phi + \tfrac{1}{2}\Psi_{\mu,i\dA}\Psi_{\nu}{}^{i\dA}\,\lambda ~.
}
To twist, we write $\Omega^{iB} \mapsto \Omega^{AB} :=  \frac{i}{2\sq}\chi^{AB} -\frac{i}{\sq}\veps^{AB} \eta$, where $\chi^{AB}$ is symmetric under $A \leftrightarrow B$, and $\Omega_{i\dA} \mapsto \Omega_{A\dA} := \psi_{A\dA}$. The fields of the twisted vectormultiplet obtained this way are listed in the Table \ref{tbl:twisted-vector-multiplet} (which differs from Table \ref{tbl:HomDegree} by a different auxiliary field choice).
\begin{small}
\begin{table}[H]
\centering
\def\arraystretch{1.1}
\begin{tabular}{|c|c|c|c|c|c|}\hline
 \begin{tabular}{@{}c@{}}Component in\\ a local chart\end{tabular} & Description & Element of &   \begin{tabular}{@{}c@{}} Homological \\ Degree \end{tabular} &  \begin{tabular}{@{}c@{}} Grassmann \\ Parity \end{tabular} &  \begin{tabular}{@{}c@{}}Weyl \\Weight \end{tabular}\\ \hline\hline
   $\lambda$ & scalar & $\Omega^{0}(\IX,\adsf P)$ & $-2$ & even & $+1$\\ \hline
   $\phi$ & scalar & $\Omega^{0}(\IX,\adsf P)$ & $2$ & even & $+1$ \\ \hline
   $A_\mu$ & Yang-Mills connection & $\CA(P)$ & $0$ & even & $0$\\ \hline
   $\eta$ & $0$-form gaugino & $\Pi\Omega^{0}(\IX,\adsf P)$ & $-1$ & odd & $+\frac{3}{2}$ \\ \hline
   $\psi_\mu$ & $1$-form gaugino & $\Pi\Omega^{1}(\IX,\adsf P)$ &  $1$ & odd & $+\frac{1}{2}$ \\ \hline
   $\chi_{\mu\nu}$ & SD $2$-form gaugino & $\Pi\Omega^{2}(\IX,\adsf P)$ & $-1$ & odd & $-\frac{1}{2}$ \\ \hline
   $D_{\mu\nu}$ & SD $2$-form auxiliary field & $\Omega^{2}(\IX,\adsf P)$ & $0$ & even & $0$\\ \hline
\end{tabular}
\caption{\label{tbl:twisted-vector-multiplet}Fields of the twisted $\CN=2$ vectormultiplet, homological degrees, and Weyl weights.}
\end{table}
\end{small}
The transformation laws of the twisted $\CN=2$ vectormultiplet coupled to the twisted and truncated supergravity multiplet of Sec. \ref{sec:TwistedScfmlGrav} are:
\eqa{
   &\delta \lambda &&= \eps \, \eta ~, \label{eq:delta-twisted-vec lambda}\\
   &\delta \phi  &&= -\eps^{\mu} \, \psi_{\mu} ~, \label{eq:delta-twisted-vec phi}\\
   &\delta A_{\mu} &&= \eps\,\psi_{\mu} + \eps^{\rho}\,\chi_{\rho\mu} - \eps_{\mu}\,\eta - \eps^{\rho}\Psi_{\mu\rho}\lambda ~, \label{eq:delta-twisted-vec ymconnection}\\
   &\delta \eta &&= \eps^{\rho}\,\scd_{\rho}\lambda - \eps\,[\lambda,\phi] - \bn_{0}\,\lambda ~, \label{eq:delta-twisted-vec eta}\\
   &\delta \psi_{\mu} &&= \frac{1}{2}\eps\,\Psi_{\mu}{}^{\rho}\psi_{\rho} - \eps\,\scd_{\mu}\phi + \eps^{\rho}\big( \wh{F}_{\rho\mu}^{-} + \lambda\,T_{\rho\mu}^{-} + D_{\rho\mu} \big) - \eps_{\mu}[\lambda,\phi] ~, \label{eq:delta-twisted-vec psi}\\
   &\delta \chi_{\mu\nu} &&= -\eps\big(\Psi_{[\mu}{}^{\rho}\chi_{\nu]\rho}\big) - 4\big(\eps_{[\mu}\scd_{\nu]}\lambda\big)^{+} + \eps \big(\wh{F}_{\mu\nu}^{+} - D_{\mu\nu}\big) + \bn_{\mu\nu}^{+}\,\lambda ~, \label{eq:delta-twisted-vec chi}\\ 
   &\delta D_{\mu\nu} &&= -\eps\big(\Psi_{[\mu}{}^{\rho}D_{\nu]\rho}\big) + 2\, \eps \big(\scd_{[\mu}\psi_{\nu]}\big)^{+} + 2 \big( \eps_{[\mu}\scd^{\rho}\chi_{\nu]\rho} \big)^{+}  - 2\big(\eps_{[\mu}\scd_{\nu]}\eta\big)^{+} \nn
   & && \quad - \eps [\phi, \chi_{\mu\nu}] + 4 \big( \eps_{[\mu}[\lambda, \psi_{\nu]}] \big)^{+} ~. \label{eq:delta-twisted-vec aux-field}
}
The supercovariant + gauge-covariant derivatives appearing above are:
\eqa{
  &\scd_{\mu}\lambda &&:= D_{\mu}\lambda  ~, \label{eq:scd lambda}\\
  &\scd_{\mu}\phi &&:= D_{\mu}\phi + \frac{1}{2}\Psi_{\mu}{}^{\rho}\psi_{\rho} ~, \label{eq:scd phi}\\
  &\scd_{\mu}\eta &&:= D_{\mu}\eta - \frac{1}{2}\Psi_{\mu}{}^{\rho}D_{\rho}\lambda + \frac{1}{2\sq}S_{\mu} \lambda ~, \label{eq:scd eta}\\
  &\scd_{\mu}\psi_{\nu} &&:= D_{\mu}\psi_{\nu} - \frac{1}{2}\Psi_{\mu}{}^{\rho}\big(\wh{F}_{\rho\nu}^{-} + \lambda\,T_{\rho\nu}^{-} + D_{\rho\nu}\big) + \frac{1}{2} \Psi_{\mu\nu}\,[\lambda,\phi] ~, \label{eq:scd psi}\\
  &\scd_{\mu}\chi_{\nu\rho} &&:= D_{\mu}\chi_{\nu\rho} + 2 \big(\Psi_{\mu[\nu}D_{\rho]}\lambda\big)^{+} - \sq S_{\mu,[\nu\rho]}^{+}\lambda ~. \label{eq:scd chi}
}
where $D_{\mu}$ denotes the (metric + gauge)-covariant derivative, $S_{\mu}$ denotes the composite 1-form S-gravitino \eqref{eq:SgravTrunc-5}, and $S_{\mu,[\nu\rho]}$ denotes the composite self-dual 2-form S-gravitino \eqref{eq:SgravTrunc-7}. The twisted supercovariant Yang-Mills field strength is given by
\eqa{
   &\wh{F}(A)_{\mu\nu} &&:= \underbrace{2\,\partial_{[\mu}A_{\nu]} + [A_{\mu}, A_{\nu}]}_{F(A)_{\mu\nu}} + \Psi_{[\mu}{}^{\rho}\chi_{\nu]\rho} + \Psi_{[\mu\nu]}\eta + \frac{1}{2}\Psi_{\mu}{}^{\rho}\Psi_{\nu\rho}\lambda  ~.\label{eq:tsugra-supercov-ym-curvature}
}
To make contact with the Cartan model of equivariant cohomology, we follow the procedure outlined in section \ref{subsec:Diff-Cartan-from-TwistedSugra} to construct the BRST version of these transformations with the gravitino truncated to be symmetric by using the constraints \eqref{eq:sym-grav-T} and \eqref{eq:sym-grav-ghost}. This enables us to set $\Psi_{[\mu\nu]} = 0$ in all expressions that follow. Using \eqref{eq:sym-grav-T}, \eqref{eq:sym-grav-ghost} and \eqref{eq:tsugra-supercov-ym-curvature}, we can write
\eqa{
  &\wh{F}_{\mu\nu}^{-} + \lambda\,T_{\mu\nu}^{-} &&= F_{\mu\nu}^{-} + \big(\Psi_{[\mu}{}^{\rho}\chi_{\nu]\rho}\big)^{-} + 2\,\lambda\big(\n_{[\mu}\Phi_{\nu]}\big)^{-} ~, \label{eq:Fhat minus lambda-Tminus}\\
  &\wh{F}_{\mu\nu}^{+} + \lambda\,\mathsf{c}_{\mu\nu}^{+} &&= F_{\mu\nu}^{+} + \big(\Psi_{[\mu}{}^{\rho}\chi_{\nu]\rho}\big)^{+} + 2\,\lambda\big(\n_{[\mu}\Phi_{\nu]}\big)^{+} ~.  \label{eq:Fhat plus lambda-cplus}
}
These structures appear in the BRST versions of \eqref{eq:delta-twisted-vec psi} and \eqref{eq:delta-twisted-vec chi} respectively. It is useful to expand
all supercovariant derivatives ($\scd_{\mu}$) appearing in the transformation laws, and replace them with (metric + gauge)-covariant derivatives ($D_\mu$) by using \eqref{eq:scd lambda}--\eqref{eq:scd chi}. In particular, we claim that (with a symmetric gravitino), the S-gravitino terms do not contribute to (the BRST version of) the auxiliary field transformation law \eqref{eq:delta-twisted-vec aux-field}. Such terms could arise only from the terms $-\sq g^{\mu\rho}S_{\mu,[\nu\rho]}^{+} \subset \scd^{\rho}\chi_{\nu\rho}$ (see \eqref{eq:scd chi}) and $\frac{1}{2\sq}S_{\nu} \subset \scd_{\nu}\eta$ (see \eqref{eq:scd psi}).  However, there is a remarkably useful identity that holds when the gravitino is symmetric, namely
\eqa{
   &g^{\mu\rho}S_{\mu,[\nu\rho]}^{+} &&= -\frac{1}{4}S_{\nu} ~, \label{eq:contraction-identity-S-susy-connections}
}
due to which the S-gravitino contribution cancels between the third and fourth terms of the BRST version of \eqref{eq:delta-twisted-vec aux-field} (cf. Appendix \ref{app:misc-expr-twisted} for a proof of \eqref{eq:contraction-identity-S-susy-connections}). The 2-form S-susy ghost profile \eqref{eq:sym-grav-ghost} is crucial for simplifying various supercovariant fermionic terms in \eqref{eq:delta-twisted-vec chi}, due to \eqref{eq:Fhat plus lambda-cplus}. We present below the maximally simplified versions of the BRST transformations.\footnote{Note the compensating transformations for self-duality, specifically the first term on the RHS in \eqref{eq:CartanSugra-6} and \eqref{eq:CartanSugra-7}. They arise from the conversion of frame indices to coordinate indices, see Appendix \ref{app:variation-of-self-dual-fields}.}
\eqa{
	&\IQ \lambda &&= \eta ~, \label{eq:CartanSugra-1}\\
	&\IQ \phi &&= -\Phi^{\rho}\psi_{\rho} ~, \label{eq:CartanSugra-2}\\
	&\IQ A_{\mu} &&= \psi_{\mu} + \Phi^{\rho}\chi_{\rho\mu} - \Phi_{\mu}\eta - \Phi^{\rho}\Psi_{\mu\rho}\lambda ~, \label{eq:CartanSugra-3}\\
	&\IQ \eta &&= \Phi^{\rho}D_{\rho}\lambda - [\lambda, \phi] ~, \label{eq:CartanSugra-4}\\
	&\IQ \psi_{\mu} &&= -D_{\mu}\phi + \Phi^{\rho}\big(F_{\rho\mu}^{-} + D_{\rho\mu} + \big(\Psi_{[\rho}{}^{\sigma}\chi_{\mu]\sigma}\big)^{-} + 2 (\n_{[\rho}\Phi_{\mu]}\big)^{-}\lambda \big) - \Phi_{\mu}[\lambda,\phi] ~, \label{eq:CartanSugra-5}\\
	&\IQ \chi_{\mu\nu} &&= -\big(\Psi_{[\mu}{}^{\rho}\chi_{\nu]\rho}\big)^{-} - 4\big(\Phi_{[\mu}D_{\nu]}\lambda\big)^{+} + (F_{\mu\nu}^{+} - D_{\mu\nu}) + 2\big(\n_{[\mu}\Phi_{\nu]}\big)^{+}\lambda ~, \label{eq:CartanSugra-6}\\
	&\IQ D_{\mu\nu} &&= - \big(\Psi_{[\mu}{}^{\rho}D_{\nu]\rho}\big)^{-} +  2\big(D_{[\mu}\psi_{\nu]}\big)^{+} + \big(\Psi_{[\mu}{}^{\rho}F_{\nu]\rho}^{-}\big)^{+} + 2\big(\Phi_{[\mu}D^{\rho}\chi_{\nu]\rho}\big)^{+} - 2\big(\Phi_{[\mu}D_{\nu]}\eta\big)^{+} \nonumber\\
     &	&&\quad - \tfrac{1}{2}\big[\Psi_{\mu}{}^{\rho}\big(\Psi_{[\rho}{}^{\sigma}\chi_{\nu]\sigma}\big)^{-} - \Psi_{\nu}{}^{\rho}\big(\Psi_{[\rho}{}^{\sigma}\chi_{\mu]\sigma}\big)^{-}\big]^{+} -\big[\Psi_{\mu}{}^{\rho}\big(\n_{[\rho}\Phi_{\nu]}\big)^{-}\lambda - \Psi_{\nu}{}^{\rho}\big(\n_{[\rho}\Phi_{\mu]}\big)^{-}\lambda\big]^{+} \nonumber\\
	& &&\quad+ 2\big(\Phi_{[\mu}\Psi_{\nu]}{}^{\rho}\big)^{+}D_{\rho}\lambda - \Psi_{\rho}{}^{\rho}\big(\Phi_{[\mu}D_{\nu]}\lambda\big)^{+} -  [\phi,\chi_{\mu\nu}] + 4\big(\Phi_{[\mu}[\lambda,\psi_{\nu]}]\big)^{+} ~.\label{eq:CartanSugra-7}
}
Note the absence of the supercovariant curvatures and derivatives (all of which have been expanded out), as well as the antisymmetric part of the gravitino. For subsequent computations with the action, we use \eqref{eq:scd lambda}--\eqref{eq:scd chi} along with \eqref{eq:Fhat minus lambda-Tminus}, and take the gravitino to be symmetric.

In this field parametrization, the differential $\IQ$ admits the following $(p,q)$ bidegree decomposition (where $p = $ homological degree, and $q = $ gravity degree),
\eqa{
& \IQ &&= \IQ^{(1,0)} + \IQ^{(0,1)} + \IQ^{(-1,2)} + \IQ^{(-2,3)} + \IQ^{(-3,4)} ~, \label{eq:bidegree-Q-sugra-variables}
}
which should be contrasted with \eqref{eq:bidegree-Q-cartan-variables}. For each summand, $p+q = 1$, as it must be.

In section \ref{subsec:Closure-of-TwistedSugra-Algebra}, we demonstrate the closure of the algebra in the present parametrization of fields, and in section \ref{subsec:Full-Cartan-from-TwistedSugra}, we make contact with the full Cartan Model of section \ref{subsec:FullEqCohCartanModelSummary}.

\subsection{Closure Of The Twisted Supergravity Algebra}\label{subsec:Closure-of-TwistedSugra-Algebra}
Consider the twisted transformations \eqref{eq:CartanSugra-1}--\eqref{eq:CartanSugra-7}. The $\IQ$-closure on $\lambda$, $\phi$ and $\eta$ is simple:
\eqa{
	&\IQ^2\lambda &&= \IQ\eta = \Phi^{\rho}D_{\rho}\lambda + [\phi, \lambda] ~.\\
	&\IQ^2\phi &&= -\Phi^{\mu}\IQ\psi_{\mu} \nonumber\\
	&              &&= \Phi^{\mu}D_{\mu}\phi - \Phi^{\mu}\Phi^{\rho}\underbrace{\big(F_{\rho\mu}^{-} + D_{\rho\mu} - \big(\Psi_{[\rho}{}^{\sigma}\chi_{\mu]\sigma}\big)^{-} + 2(\n_{[\rho}\Phi_{\mu]}\big)^{-}\lambda \big)}_{\text{antisymmetric in $\rho,\mu$}} + \Phi^{\mu}\Phi_{\mu}[\lambda,\phi] \nonumber\\
	&               &&= \Phi^{\mu}D_{\mu}\phi + [\Phi^{\mu}\Phi_{\mu}\lambda,\phi] ~.\\
        &\IQ^2\eta &&= \Phi^{\rho}\IQ D_{\rho}\lambda - [\eta,\phi] + \Phi^{\rho}[\lambda, \psi_{\rho}] \nonumber\\
	&              &&= \Phi^{\rho}\big(D_{\rho}\eta + [\psi_{\rho} + \Phi^{\sigma}\chi_{\sigma\rho} - \Phi_{\rho}\eta - \Phi^{\sigma}\Psi_{\rho\sigma}\lambda,\lambda]\big) + [\eta,\phi] +  \Phi^{\rho}[\lambda, \psi_{\rho}] \nonumber\\
	&              &&= \Phi^{\rho}D_{\rho}\eta + [\Phi^{\mu}\Phi_{\mu}\lambda + \phi,\eta] ~.
}
These formulae are special cases of equation \eqref{eq:IQ squared twisted-sugra} below and show that 
$\IQ^2$ is a sum of symmetry transformations (with field-dependent structure functions). On $\psi_{\mu}$ and $\chi_{\mu\nu}$ the closure is more intricate. We outline a few steps in each case, highlighting various identities used along the way, which are proved in the appendices. Let us compute $\IQ^2\psi_{\mu}$ first.  We require certain $\IQ$-variations from Appendix \ref{app:supporting-twisted-sugra-closure}, specifically \eqref{eq:H2-37}, \eqref{eq:dt curl phi ASD}, and \eqref{eq:dt Fminus plus D}. We note that
\begin{equation*}
\begin{aligned}
&\Phi^{\sigma}\IQ \left[F_{\sigma\mu}^{-} + D_{\sigma\mu} + \big(\Psi_{[\sigma}{}^{\rho}\chi_{\mu]\rho}\big)^{-} + 2 (\n_{[\sigma}\Phi_{\mu]}\big)^{-}\lambda\right]\nonumber\\
&= 2\Phi^{\sigma}\big(D_{[\sigma}\psi_{\mu]}\big) + \underbrace{\Phi^{\sigma}\big[\big(\n_{[\sigma}\Phi^{\rho} - \n^{\rho}\Phi_{[\sigma}\big)\chi_{\mu]\rho}\big]^{-}}_{=\,0 \text{ (due to \eqref{eq:identB antisym general})}} \underbrace{-2\Phi^{\sigma} \Phi^{\rho}\big(D_{[\sigma}\chi_{\mu]\rho}\big)^{-} + 2\Phi^{\sigma}\big(\Phi_{[\sigma}D^{\rho}\chi_{\mu]\rho}\big)^{+}}_{=\,0 \text{ (see Appendix \ref{sec:proof of vanishing of del chi terms in dt sq psimu} for a proof)}} \nonumber\\
&\quad + \underbrace{2\Phi^{\sigma}\big(\Phi_{[\sigma}D_{\mu]}\eta\big)^{-} - 2\Phi^{\sigma}\big(\Phi_{[\sigma}D_{\mu]}\eta\big)^{+}}_{=\,0 \text{ (due to \eqref{eq:contracted SD equals ASD}) } } - \Phi^{\sigma}[\phi,\chi_{\sigma\mu}] + \underbrace{4 \Phi^{\sigma}\big(\Phi_{[\sigma}[\lambda,\psi_{\mu]}]\big)^{+}}_{\text{use \eqref{eq:contracted SD equals ASD}}} \nonumber\\
\end{aligned}
\end{equation*}
\begin{equation*}
\hspace{-3.0cm}\underbrace{\boxed{
\begin{aligned}
&+ 2\Phi^{\sigma}\Phi^{\rho}\big[\Psi_{\rho[\sigma}D_{\mu]}\lambda\big]^{-} + 2\Phi^{\sigma}\big(\Phi_{[\sigma}\Psi_{\mu]}{}^{\rho}\big)^{+}D_{\rho}\lambda - \Phi^{\sigma}\Psi_{\rho}{}^{\rho}\big(\Phi_{[\sigma}D_{\mu]}\lambda\big)^{+} \nonumber\\
& + 2\Phi^{\sigma}\big[\Psi_{\sigma}{}^{\rho}\big(\Phi_{[\mu}D_{\rho]}\lambda\big)^{+} - \Psi_{\mu}{}^{\rho}\big(\Phi_{[\sigma}D_{\rho]}\lambda\big)^{+}\big]^{-}
\end{aligned}
}}_{=\,0 \text{ (see Appendix \ref{sec:proof of vanishing of del lambda terms in dt sq psimu} for a proof)} }
\end{equation*}
Therefore, after a cancellation of the highlighted terms, one is left with
\eqa{
&\Phi^{\sigma}\IQ \left[F_{\sigma\mu}^{-} + D_{\sigma\mu} + \big(\Psi_{[\lambda}{}^{\rho}\chi_{\mu]\rho}\big)^{-} + 2 (\n_{[\lambda}\Phi_{\mu]}\big)^{-}\lambda\right]\nonumber\\
 &= \Phi^{\sigma}D_{\sigma}\psi_{\mu} - \Phi^{\sigma}D_{\mu}\psi_{\sigma} - \Phi^{\sigma}[\phi, \chi_{\sigma\mu}] + \Phi^{\sigma}\Phi_{\sigma}[\lambda,\psi_{\mu}] - \Phi^{\sigma}\Phi_{\mu}[\lambda,\psi_{\sigma}] ~. \label{eq:dtsquared psi mu big piece 1}
}
Additionally,
\eqa{
	&\IQ \big(-D_{\mu}\phi - \Phi_{\mu}[\lambda,\phi]\big) &&= \Phi^{\sigma}D_{\mu}\psi_{\sigma} + \big(\n_{\mu}\Phi^{\sigma}\big)\psi_{\sigma} + [\phi,\psi_{\mu}] + \Phi^{\sigma}[\phi, \chi_{\sigma\mu}] + \Phi^{\sigma}\Phi_{\mu}[\lambda,\psi_{\sigma}] ~. \label{eq:dtsquared psi mu big piece 2}
}
Adding \eqref{eq:dtsquared psi mu big piece 1} and \eqref{eq:dtsquared psi mu big piece 2}, we get
\eqa{
	&\IQ ^2\psi_{\mu} &&= \Phi^{\sigma}D_{\sigma}\psi_{\mu} + \big(\n_{\mu}\Phi^{\sigma}\big)\psi_{\sigma} + \Phi^{\sigma}\Phi_{\sigma}[\lambda,\psi_{\mu}] + [\phi,\psi_{\mu}] ~.
}
Once again, this is of the form \eqref{eq:IQ squared twisted-sugra} below.

Next, we compute $\IQ^2\chi_{\mu\nu}$. For this computation, the $\IQ$-variations we require are \eqref{eq:H2-37}, \eqref{eq:dt curl phi SD}, \eqref{eq:dt Phi del lambda SD}, and \eqref{eq:dt Fplus minus D}. We find
\eqa{
	&\IQ ^2\chi_{\mu\nu} &&= \textcolor{blue}{-2\Phi^{\rho}\big(D_{[\mu}\chi_{\nu]\rho}\big)^{+} - 2 \big(\Phi_{[\mu}D^{\rho}\chi_{\nu]\rho}\big)^{+}} \nonumber\\
	& &&\quad\quad \textcolor{red}{-\big[\big(\n_{[\mu}\Phi^{\rho}\big)\chi_{\nu]\rho}\big]^{-} -\big[\big(\n^{\rho}\Phi_{[\mu}\big)\chi_{\nu]\rho}\big]^{-} - 2\big[\big(\n_{[\mu}\Phi^{\rho}\big)\chi_{\nu]\rho}\big]^{+}} \nonumber\\
	& &&\quad\quad \textcolor{OliveGreen}{- 2 \big[\Psi_{\mu}{}^{\rho}\big(\Phi_{[\nu}D_{\rho]}\lambda\big)^{-} - \Psi_{\nu}{}^{\rho}\big(\Phi_{[\mu}D_{\rho]}\lambda\big)^{-}\big]^{+}} \nonumber\\
	& &&\quad\quad \textcolor{OliveGreen}{- 2\Phi^{\rho}\big[\Psi_{\rho[\mu}\n_{\nu]}\lambda\big]^{+} - 2\big(\Phi_{[\mu}\Psi_{\nu]}{}^{\rho}\big)^{+}D_{\rho}\lambda + \Psi_{\rho}{}^{\rho}\big(\Phi_{[\mu}\n_{\nu]}\lambda\big)^{+} } \nonumber\\
	& &&\quad\quad + [\phi, \chi_{\mu\nu}] + \Phi^{\rho}\Phi_{\rho}[\lambda,\chi_{\mu\nu}] ~. \label{eq:dtsquared chi intermediate}
}
The last line is a gauge transformation by $\phi + \Phi^2 \lambda$. The two \textcolor{blue}{\text{blue}} terms combine to give $\Phi^{\rho}D_{\rho}\chi_{\mu\nu}$, the transport term in the (gauge-covariantized) Lie derivative $\CL^{(A)}_{\Phi}\chi_{\mu\nu}$. This is proved in Appendix  \ref{sec:blue terms in dtsquared chi}. The three \textcolor{red}{\text{red}} terms combine to give $\big(\n_{\mu}\Phi^{\rho}\big)\chi_{\rho\nu} + \big(\n_{\nu}\Phi^{\rho}\big)\chi_{\mu\rho}$, as proved in Appendix  \ref{sec:red terms in dtsquared chi}. Finally, the \textcolor{OliveGreen}{\text{green}} terms add up to zero, as proved in Appendix  \ref{sec:green terms in dtsquared chi}. Therefore, we have
\eqa{
	&\IQ ^2\chi_{\mu\nu} &&= \Phi^{\rho}D_{\rho}\chi_{\mu\nu} + \big(\n_{\mu}\Phi^{\rho}\big)\chi_{\rho\nu} + \big(\n_{\nu}\Phi^{\rho}\big)\chi_{\mu\rho} + [\phi, \chi_{\mu\nu}] + \Phi^{\rho}\Phi_{\rho}[\lambda,\chi_{\mu\nu}] ~.
}
Once again, this is of the form \eqref{eq:IQ squared twisted-sugra} below.  

The details of the computation of $\IQ^2$ on the remaining fields are left to the reader.

This concludes the proof of closure of the twisted supergravity algebra. To summarize, we find that in the supergravity parametrization of fields,
\beqa{
&\IQ^2 &&=  \CL_{\Phi}^{(A)} + \delta_{\text{gauge}}(\phi + \Phi^{\rho}\Phi_{\rho}\lambda) ~, \label{eq:IQ squared twisted-sugra}
}
whereas in the parametrization of section \ref{subsec:FullEqCohCartanModelSummary}, we recall from \eqref{eq:IQ2} that
\beqa{
&\IQ^2 &&=  \CL_{\Phi}^{(A)} + \delta_{\text{gauge}}(\phi) ~. \label{eq:IQ squared EqCoh}
}
In the next section, we comment on and reconcile the differences between the two parametrizations, and between \eqref{eq:IQ squared twisted-sugra} and \eqref{eq:IQ squared EqCoh}.

\subsection{The \textsf{Gauge} $\rtimes$ \textsf{Diffeomorphism} Cartan Model From Twisted Supergravity}\label{subsec:Full-Cartan-from-TwistedSugra}
A comparison of \eqref{eq:CartanSugra-1}--\eqref{eq:CartanSugra-7} with \eqref{eq:CartanAlgebra-1}--\eqref{eq:CartanAlgebra-5} and \eqref{eq:CartanAlgebra-6prime}--\eqref{eq:CartanAlgebra-7prime} reveals that the transformations of topologically twisted $\CN=2$ SYM  coupled to background $\CN=2$ supergravity are at variance with those of the Cartan Model with antighosts. Of course, all differences vanish when all supergravity fields except the metric are turned off, and also when only $\Phi^{\mu} = 0$. But when all background fields are present, there are important differences between \eqref{eq:CartanSugra-3} and \eqref{eq:CartanAlgebra-1} (the transformation of the YM connection $A_\mu$), and between \eqref{eq:CartanSugra-5} and \eqref{eq:CartanAlgebra-2} (the transformation of the gaugino $\psi_\mu$).
Clearly, the vectormultiplet fields couple differently to the background supergravity fields in the two models. The transformation laws of section \ref{subsec:FullEqCohCartanModelSummary} are simpler and amenable to an equivariant interpretation. 

As we have seen above, the closure of the $\IQ$ algebra in the supergravity parametrization differs from the equivariant version only by an additional Yang-Mills gauge transformation by the field-dependent parameter $\Phi^{\mu}\Phi_{\mu}\lambda$. The model obtained via the methods of twisted supergravity is, in fact, simply another realization of the Cartan Model for the semidirect product $\IG$ \eqref{eq:IG family}, that uses a different parametrization of fields. In this section, we will derive an invertible map between the two parametrizations, explaining how one can go back and forth between \eqref{eq:CartanSugra-1}--\eqref{eq:CartanSugra-7} and \eqref{eq:CartanAlgebra-1}--\eqref{eq:CartanAlgebra-5} + \eqref{eq:CartanAlgebra-6prime}--\eqref{eq:CartanAlgebra-7prime}. 

Let us begin by trying to simplify the transformation law \eqref{eq:CartanSugra-3} and remove the extra field-dependent gauge transformation by $\Phi^{\mu}\Phi_{\mu}\lambda$ from the algebra. To do so, we must redefine (or ``shift'') the Yang-Mills connection to remove the $\eta$- and $\lambda$- dependent terms in its transformation -- here we exploit the fact that $\IQ \lambda = \eta$ and $\IQ \Phi_{\mu} = \Phi^{\rho}\Psi_{\mu\rho}$. This induces nontrivial modifications both to the gauge-covariant derivative as well as the YM field strength. This shift, however, does not eliminate the $\chi$-dependent term in the transformation of $A_\mu$, and so we must shift $\psi_{\mu}$ to absorb this term. Naturally, this modifies the $\IQ$ transformation of $\psi_{\mu}$. But this does not quite get us to \eqref{eq:CartanAlgebra-6prime} as there is a compensating term in the (new) transformation law of $\psi_{\mu}$ involving $(\n_{[\mu}\Phi_{\nu]})^{+}\lambda$ due to the shift of $\psi_{\mu}$. However, an inspection of the transformation law \eqref{eq:CartanSugra-6} of $\chi_{\mu\nu}$ reveals that this term can be absorbed in the auxiliary field.

This chain of arguments is more aptly summarized by the following map:
\begin{empheq}[box=\fbox]{align}
 	  A_{\mu} &\longmapsto A_{\mu} - \Phi_{\mu} \lambda ~,  &\lambda &\longmapsto \lambda ~, & D_{\mu\nu} &\longmapsto D_{\mu\nu} - 2\big(\Phi_{[\mu}D_{\nu]}\lambda\big)^{+} ~, \label{eq:field-redef-1}\\
 	  \psi_{\mu} &\longmapsto \psi_{\mu} - \Phi^{\rho}\chi_{\rho\mu} ~,  & \eta &\longmapsto \eta ~, &\chi_{\mu\nu}  &\longmapsto \chi_{\mu\nu} ~, \label{eq:field-redef-2}\\
	   \phi  &\longmapsto \phi ~. \label{eq:field-redef-3}
\end{empheq}
where on the left are the fields of the twisted supergravity parametrization, and on the right, the fields of the (alternate) parametrization of the Cartan Model + antighosts of section \ref{subsec:FullEqCohCartanModelSummary}. 

Some remarks are in order. First, two sides and their respective transformations align when $\Phi^{\mu} = 0$. Second, the map is clearly invertible. Third, the parametrization favored by twisted supergravity actually mixes -- in the terminology of section \ref{subsec:FullEqCohCartanModelSummary} -- the subset ($A_{\mu}, \psi_{\mu}$) of the gauge Cartan multiplet with the field $\lambda$ of the projection multiplet and the field $\chi_{\mu\nu}$ of the localization multiplet. Recalling \eqref{eq:CartanModel-AlternateAuxField}, twisting also identifies the auxiliary field $D_{\mu\nu}$ with a combination of the localization multiplet field $H_{\mu\nu}$, the YM connection and field strength (constructed solely out of the Cartan multiplet field $A_{\mu}$), and the projection multiplet field $\lambda$. So we can interpret the field redefinitions \eqref{eq:field-redef-1}--\eqref{eq:field-redef-3} as changing the $\IG = \textsf{Gauge} \rtimes \textsf{Diff}$ action on the total complex of fields.

We now demonstrate \eqref{eq:field-redef-1}--\eqref{eq:field-redef-3} explicitly.  Under the shift of the YM connection, the covariant derivative on an adjoint-valued object $*$ is modified as:
\eqa{
 	&D_{\mu} *= \partial_{\mu} * +  [A_{\mu}, *] &&\longmapsto D_{\mu}* -  \Phi_{\mu}[\lambda, *] ~, \label{eq:shift-1}
}
 whereas the YM field strength is modified as:
\eqa{
 	&F_{\mu\nu} = 2\,\partial_{[\mu}A_{\nu]} + [A_{\mu}, A_{\nu}] &&\longmapsto F_{\mu\nu} - 2\,D_{[\mu}\big(\Phi_{\nu]}\lambda\big)\label{eq:shift-2}\\ 
 	& &&\qquad = F_{\mu\nu} - 2\,\big(\n_{[\mu}\Phi_{\nu]}\big)\lambda + 2\,\Phi_{[\mu}D_{\nu]}\lambda ~.\label{eq:shift-3}
}
In particular, 
\eqa{
	&F_{\mu\nu}^{+} - D_{\mu\nu} &&\longmapsto F_{\mu\nu}^{+} - D_{\mu\nu} - 2\big(\n_{[\mu}\Phi_{\nu]}\big)^{+}\lambda + 4\big(\Phi_{[\mu}D_{\nu]}\lambda\big)^{+} ~, \label{eq:shift-4}\\
	&F_{\mu\nu}^{-} + D_{\mu\nu} &&\longmapsto F_{\mu\nu}^{-} + D_{\mu\nu} - 2\big(\n_{[\mu}\Phi_{\nu]}\big)^{-}\lambda + 2\big(\Phi_{[\mu}D_{\nu]}\lambda\big)^{-} - 2\big(\Phi_{[\mu}D_{\nu]}\lambda\big)^{+} ~. \label{eq:shift-5}
}
By using \eqref{eq:shift-1}--\eqref{eq:shift-5} consistently on  \eqref{eq:CartanSugra-1}--\eqref{eq:CartanSugra-7}, we can arrive at the Cartan Model transformation laws  \eqref{eq:CartanAlgebra-1}--\eqref{eq:CartanAlgebra-5} + \eqref{eq:CartanAlgebra-6prime}--\eqref{eq:CartanAlgebra-7prime}. The reader willing to accept this assertion should feel free to skip to the paragraph below \eqref{eq:dt Dmunu shifted final} at this point.

We emphasize that the notion of `gravity degree' is not invariant under \eqref{eq:field-redef-1}--\eqref{eq:field-redef-3}: for example, a degree $=0$ term will be shifted to a degree $=0$ term and a degree $=2$ term.

The transformation laws of $\lambda$, $\phi$ and $\eta$ do not change under the field redefinitions.\footnote{For $\eta$, the gauge-covariant derivative on $\lambda$ is invariant under \eqref{eq:shift-1}.} Consider $\IQ\chi_{\mu\nu}$ from \eqref{eq:CartanSugra-6}. Upon using \eqref{eq:shift-3} and \eqref{eq:shift-4}, this is modified to
\eqa{
  &\left.\IQ \chi_{\mu\nu}\right|_{\text{shifted}} &&= -\big(\Psi_{[\mu}{}^{\rho}\chi_{\nu]\rho}\big)^{-} + \big(F_{\mu\nu}^{+} - D_{\mu\nu}\big) ~, \label{eq:dt chi shifted}
}
which is identical to \eqref{eq:CartanAlgebra-6prime}. Next, consider $\IQ\psi_{\mu}$, cf. \eqref{eq:CartanSugra-5}. Using \eqref{eq:shift-1}--\eqref{eq:shift-5} and \eqref{eq:dt chi shifted}, 
\eqa{
	&\left.\IQ\psi_{\mu}\right|_{\text{shifted}} + \Phi^{\rho}\big(\Psi_{[\rho}{}^{\sigma}\chi_{\mu]\sigma}\big)^{-} - \Phi^{\rho}\big(F_{\rho\mu}^{+} - D_{\rho\mu}\big) &&= -D_{\mu}\phi + \Phi_{\mu}[\lambda,\phi] + \Phi^{\rho}\big(F^{-}_{\rho\mu} + D_{\rho\mu}\big) \nonumber\\
	& &&\quad + \Phi^{\rho}\big(\Psi_{[\rho}{}^{\sigma}\chi_{\mu]\sigma}\big)^{-} - \Phi_{\mu}[\lambda,\phi]\nonumber\\
	& &&\quad \underbrace{- 2\Phi^{\rho}\big(\Phi_{[\rho}D_{\mu]}\lambda\big)^{+} + 2\Phi^{\rho}\big(\Phi_{[\rho}D_{\mu]}\lambda\big)^{-}}_{=\,0 \text{ (using \eqref{eq:contracted SD equals ASD})}}  ~,
}
from which it follows that
\eqa{
 & \left.\IQ\psi_{\mu}\right|_{\text{shifted}} &&= -D_{\mu}\phi + \Phi^{\rho}F_{\rho\mu} ~, \label{eq:dt psi shifted}
}
which is identical to \eqref{eq:CartanAlgebra-2}. For the variation \eqref{eq:CartanSugra-3} of the YM connection $A_{\mu}$, we find
\eqa{
	&\left.\IQ A_{\mu}\right|_{\text{shifted}} - \Phi^{\rho}\Psi_{\rho\mu}\lambda - \Phi_{\mu}\eta &&= \psi_{\mu} - \Phi^{\rho}\chi_{\rho\mu} + \Phi^{\rho}\chi_{\rho\mu} - \Phi_{\mu}\eta - \Phi^{\rho}\Psi_{\mu\rho}\lambda ~.
}
This simplifies to
\eqa{
	\left.\IQ A_{\mu}\right|_{\text{shifted}} &&= \psi_{\mu} ~, \label{eq:dt Amu shifted}
}
which is identical to \eqref{eq:CartanAlgebra-1}. This is a tremendous simplification of the sugra transformation law \eqref{eq:CartanSugra-3}, and leads to the desirable interpretation of $\psi_{\mu}dx^{\mu}$ being a cotangent vector on the space $\mathcal{A}$ of Yang-Mills connections. (Here $\psi_{\mu}$ on the RHS is, of course, in the \emph{shifted} parametrization.)

Finally, since the auxiliary field $D_{\mu\nu}$ has the most complicated transformation law \eqref{eq:CartanSugra-7}, it is worth explaining its shifted transformation in some detail. We begin by focusing on the following underlined terms in \eqref{eq:dt Phi del lambda SD}:
\eqa{
	&\left.\IQ \big(\Phi_{[\mu}D_{\nu]}\lambda\big)^{+}\right|_{\text{unshifted}} &&= \underline{\big(\Phi_{[\mu}[\psi_{\nu]},\lambda]\big)^{+}} + \big(\Phi_{[\mu}D_{\nu]}\eta\big)^{+} \nonumber\\
	& &&\quad +\tfrac{1}{2}\big(\Psi_{\mu}{}^{\rho}\big(\Phi_{[\nu}D_{\rho]}\lambda\big)^{-} - \Psi_{\nu}{}^{\rho}\big(\Phi_{[\mu}D_{\rho]}\lambda\big)^{-}\big)^{+} \nonumber\\
	& &&\quad -\tfrac{1}{2}\big(\Psi_{\mu}{}^{\rho}\big(\Phi_{[\nu}D_{\rho]}\lambda\big)^{+} - \Psi_{\nu}{}^{\rho}\big(\Phi_{[\mu}D_{\rho]}\lambda\big)^{+}\big)^{-} \nonumber\\
	& &&\quad + \Phi^{\rho}\big(\Psi_{\rho[\mu}D_{\nu]}\lambda\big)^{+} - \dashuline{\Phi^{\rho}\big(\Phi_{[\mu}[\chi_{\nu]\rho},\lambda]\big)^{+}} ~. \label{eq:dt Phi del lambda SD repeated}
}
The field redefinition \eqref{eq:field-redef-2} of $\psi_{\nu}$ in the first term in \eqref{eq:dt Phi del lambda SD repeated} spawns an equal and opposite version of the last term, which consequently drops out, leading to
\eqa{
   &\left.\IQ \big(\Phi_{[\mu}D_{\nu]}\lambda\big)^{+}\right|_{\text{shifted}} &&= +\big(\Phi_{[\mu}[\psi_{\nu]},\lambda]\big)^{+} + \big(\Phi_{[\mu}D_{\nu]}\eta\big)^{+} + \Phi^{\rho}\big(\Psi_{\rho[\mu}D_{\nu]}\lambda\big)^{+} \nonumber\\
	& &&\quad +\tfrac{1}{2}\big(\Psi_{\mu}{}^{\rho}\big(\Phi_{[\nu}D_{\rho]}\lambda\big)^{-} - \Psi_{\nu}{}^{\rho}\big(\Phi_{[\mu}D_{\rho]}\lambda\big)^{-}\big)^{+} \nonumber\\
	& &&\quad -\tfrac{1}{2}\big(\Psi_{\mu}{}^{\rho}\big(\Phi_{[\nu}D_{\rho]}\lambda\big)^{+} - \Psi_{\nu}{}^{\rho}\big(\Phi_{[\mu}D_{\rho]}\lambda\big)^{+}\big)^{-} ~.  \label{eq:Q Phi D Lambda}
}
Therefore, upon shifting, the LHS of \eqref{eq:CartanSugra-7} becomes
\eqa{
	&\left.\text{LHS of \eqref{eq:CartanSugra-7}}\right|_{\text{shifted}} 
         &&\stackrel{\eqref{eq:Q Phi D Lambda}}{=\joinrel=\joinrel=} \left.\IQ D_{\mu\nu}\right|_{\text{shifted}} - 2 \big(\Phi_{[\mu}[\psi_{\nu]},\lambda]\big)^{+} -2 \big(\Phi_{[\mu}D_{\nu]}\eta\big)^{+} \nonumber\\
	& &&\qquad\,\, -\big(\Psi_{\mu}{}^{\rho}\big(\Phi_{[\nu}D_{\rho]}\lambda\big)^{-} - \Psi_{\nu}{}^{\rho}\big(\Phi_{[\mu}D_{\rho]}\lambda\big)^{-}\big)^{+} \nonumber\\
	& &&\qquad\,\, +\big(\Psi_{\mu}{}^{\rho}\big(\Phi_{[\nu}D_{\rho]}\lambda\big)^{+} - \Psi_{\nu}{}^{\rho}\big(\Phi_{[\mu}D_{\rho]}\lambda\big)^{+}\big)^{-} \nonumber\\
	& &&\qquad\,\, -2 \Phi^{\rho}\big(\Psi_{\rho[\mu}D_{\nu]}\lambda\big)^{+} ~. \label{eq:LHS of dt Dmunu shifted intermediate}
}
As for the RHS of  \eqref{eq:CartanSugra-7}, the following terms are modified by the field redefinitions:
\eqas{
     2\big(D_{[\mu}\psi_{\nu]}\big)^{+} &\longmapsto  2\big(D_{[\mu}\psi_{\nu]}\big)^{+} + 2\big[\big(\n_{[\mu}\Phi^{\rho}\big)\chi_{\nu]\rho}\big]^{+} + 2\Phi^{\rho}\big(D_{[\mu}\chi_{\nu]\rho}\big)^{+} \\
	& \qquad - 2\big(\Phi_{[\mu}[\lambda, \psi_{\nu]}]\big)^{+} - 2\Phi^{\rho}\big(\Phi_{[\mu}[\lambda, \chi_{\nu]\rho}]\big)^{+} ~,\\
      -\big(\Psi_{[\mu}{}^{\rho}D_{\nu]\rho}\big)^{-} &\longmapsto -\big(\Psi_{[\mu}{}^{\rho}D_{\nu]\rho}\big)^{-} + \big(\Psi_{\mu}{}^{\rho}\big(\Phi_{[\nu}D_{\rho]}\lambda\big)^{+} - \Psi_{\nu}{}^{\rho}\big(\Phi_{[\mu}D_{\rho]}\lambda\big)^{+}\big)^{-} ~,\\
 \big(\psi_{[\mu}{}^{\rho}F_{\nu]\rho}^{-}\big)^{+} &\longmapsto \big(\psi_{[\mu}{}^{\rho}F_{\nu]\rho}^{-}\big)^{+} - \big(\Psi_{\mu}{}^{\rho}\big(\n_{[\nu}\Phi_{\rho]}\big)^{-}\lambda - \Psi_{\nu}{}^{\rho}\big(\n_{[\mu}\Phi_{\rho]}\big)^{-}\lambda\big)^{+}\\
& \qquad + \big(\Psi_{\mu}{}^{\rho}\big(\Phi_{[\nu}D_{\rho]}\lambda\big)^{-} - \Psi_{\nu}{}^{\rho}\big(\Phi_{[\mu}D_{\rho]}\lambda\big)^{-}\big)^{+} ~,\\
  2\big(\Phi_{[\mu}D^{\rho}\chi_{\nu]\rho}\big)^{+} &\longmapsto 2\big(\Phi_{[\mu}D^{\rho}\chi_{\nu]\rho}\big)^{+} - 2 \Phi^{\rho}\big(\Phi_{[\mu}[\lambda,\chi_{\nu]\rho}]\big)^{+} ~,\\
	4\big(\Phi_{[\mu}[\lambda,\psi_{\nu]}]\big)^{+} &\longmapsto 4\big(\Phi_{[\mu}[\lambda,\psi_{\nu]}]\big)^{+} + 4 \Phi^{\rho}\big(\Phi_{[\mu}[\lambda, \chi_{\nu]\rho}]\big)^{+} ~. \label{eq:shifted RHS terms of D}
}
Therefore, using \eqref{eq:shifted RHS terms of D}, the RHS of \eqref{eq:CartanSugra-7} reads
\eqa{
	&\left.\text{RHS of \eqref{eq:CartanSugra-7}}\right|_{\text{shifted}} &&= 2\big(D_{[\mu}\psi_{\nu]}\big)^{+} + 2\big[\big(\n_{[\mu}\Phi^{\rho}\big)\chi_{\nu]\rho}\big]^{+} + 2\Phi^{\rho}\big(D_{[\mu}\chi_{\nu]\rho}\big)^{+} \nonumber\\
	& &&\quad -\big(\Psi_{[\mu}{}^{\rho}D_{\nu]\rho}\big)^{-} + \big(\Psi_{\mu}{}^{\rho}\big(\Phi_{[\nu}D_{\rho]}\lambda\big)^{+} - \Psi_{\nu}{}^{\rho}\big(\Phi_{[\mu}D_{\rho]}\lambda\big)^{+}\big)^{-} \nonumber\\
	& &&\quad + \big(\psi_{[\mu}{}^{\rho}F_{\nu]\rho}^{-}\big)^{+} + \big(\Psi_{\mu}{}^{\rho}\big(\Phi_{[\nu}D_{\rho]}\lambda\big)^{-} - \Psi_{\nu}{}^{\rho}\big(\Phi_{[\mu}D_{\rho]}\lambda\big)^{-}\big)^{+}\nonumber\\
	& &&\quad + 2\big(\Phi_{[\mu}D^{\rho}\chi_{\nu]\rho}\big)^{+} -2\big(\Phi_{[\mu}D_{\nu]}\eta\big)^{+}  - \tfrac{1}{2}\big[\Psi_{\mu}{}^{\rho}\big(\Psi_{[\rho}{}^{\sigma}\chi_{\nu]\sigma}\big)^{-} - \Psi_{\nu}{}^{\rho}\big(\Psi_{[\rho}{}^{\sigma}\chi_{\mu]\sigma}\big)^{-}\big]^{+} \nonumber\\
	& &&\quad + 2\big(\Phi_{[\mu}\Psi_{\nu]}{}^{\rho}\big)^{+}D_{\rho}\lambda - \Psi_{\rho}{}^{\rho}\big(\Phi_{[\mu}D_{\nu]}\lambda\big)^{+}\nonumber\\
	& &&\quad -  [\phi,\chi_{\mu\nu}] + 2\big(\Phi_{[\mu}[\lambda,\psi_{\nu]}]\big)^{+} ~. \label{eq:RHS of dt Dmunu shifted intermediate}
}
Equating the LHS \eqref{eq:LHS of dt Dmunu shifted intermediate} and the RHS \eqref{eq:RHS of dt Dmunu shifted intermediate}, we find
\eqa{
	&\left.\IQ D_{\mu\nu}\right|_{\text{shifted}} &&= 2\big(D_{[\mu}\psi_{\nu]}\big)^{+} + 2\big[\big(\n_{[\mu}\Phi^{\rho}\big)\chi_{\nu]\rho}\big]^{+} \textcolor{blue}{+ 2\Phi^{\rho}\big(D_{[\mu}\chi_{\nu]\rho}\big)^{+} + 2\big(\Phi_{[\mu}D^{\rho}\chi_{\nu]\rho}\big)^{+}} -  [\phi,\chi_{\mu\nu}] \nonumber\\
	& &&\quad -\big(\Psi_{[\mu}{}^{\rho}D_{\nu]\rho}\big)^{-} + \big(\psi_{[\mu}{}^{\rho}F_{\nu]\rho}^{-}\big)^{+} - \tfrac{1}{2}\big[\Psi_{\mu}{}^{\rho}\big(\Psi_{[\rho}{}^{\sigma}\chi_{\nu]\sigma}\big)^{-} - \Psi_{\nu}{}^{\rho}\big(\Psi_{[\rho}{}^{\sigma}\chi_{\mu]\sigma}\big)^{-}\big]^{+} \nonumber\\
	& &&\quad \textcolor{red}{+ 2\big[\Psi_{\mu}{}^{\rho}\big(\Phi_{[\nu}D_{\rho]}\lambda\big)^{-} - \Psi_{\nu}{}^{\rho}\big(\Phi_{[\mu}D_{\rho]}\lambda\big)^{-}\big] + 2\big(\Phi_{[\mu}\psi_{\nu]}\big)^{+}D_{\rho}\lambda}\nonumber\\
	& &&\quad \textcolor{red}{+ 2\Phi^{\rho}\big(\Psi_{\rho[\mu}D_{\nu]}\lambda\big)^{+} - \Psi_{\rho}{}^{\rho}\big(\Phi_{[\mu}D_{\nu]}\lambda\big)^{+}} ~.
}
The \textcolor{red}{\text{red}} terms add up to zero: this is precisely \eqref{eq:eks plus wye identity}, an identity used in the computation of $\IQ ^2\chi_{\mu\nu}$ in the previous subsection (see Appendix \ref{sec:green terms in dtsquared chi} for a proof). The \textcolor{blue}{\text{blue}} terms add up to $-\Phi^{\rho}D_{\rho}\chi_{\mu\nu}$, a result we used earlier (see Appendix \ref{sec:blue terms in dtsquared chi} for a proof). Therefore, we find
\eqa{
	& \left.\IQ D_{\mu\nu}\right|_{\text{shifted}} &&= 2\big(D_{[\mu}\psi_{\nu]}\big)^{+} + 2\big[\big(\n_{[\mu}\Phi^{\rho}\big)\chi_{\nu]\rho}\big]^{+} - \Phi^{\rho}D_{\rho}\chi_{\mu\nu} -  [\phi,\chi_{\mu\nu}] \nonumber\\
	& &&\quad -\big(\Psi_{[\mu}{}^{\rho}D_{\nu]\rho}\big)^{-} + \big(\psi_{[\mu}{}^{\rho}F_{\nu]\rho}^{-}\big)^{+} - \tfrac{1}{2}\dashuline{\big[\Psi_{\mu}{}^{\rho}\big(\Psi_{[\rho}{}^{\sigma}\chi_{\nu]\sigma}\big)^{-} - \Psi_{\nu}{}^{\rho}\big(\Psi_{[\rho}{}^{\sigma}\chi_{\mu]\sigma}\big)^{-}\big]^{+}} \nonumber \\
 & &&\stackbin[\eqref{eq:Psi Psi Chi Diff SD}]{\eqref{eq:Psi Psi Chi ASD}}{=\joinrel=\joinrel=}  2\big(D_{[\mu}\psi_{\nu]}\big)^{+} + 2\big[\big(\n_{[\mu}\Phi^{\rho}\big)\chi_{\nu]\rho}\big]^{+} - \Phi^{\rho}D_{\rho}\chi_{\mu\nu}\nonumber\\
	& &&\qquad  -  [\phi,\chi_{\mu\nu}] -\big(\Psi_{[\mu}{}^{\rho}D_{\nu]\rho}\big)^{-} + \big(\psi_{[\mu}{}^{\rho}F_{\nu]\rho}^{-}\big)^{+} + \tfrac{1}{2}\dashuline{\Psi^{\rho\sigma}\Psi_{\sigma[\mu}\chi_{\nu]\rho}} ~, \label{eq:dt Dmunu shifted final}
}
which is identical to \eqref{eq:CartanAlgebra-7prime}. To summarize, we have derived the Cartan Model for $\CG \rtimes \DIFF(\IX)$-equivariant cohomology of $\CA(P)\times \MET(\IX)$  starting from the transformations of twisted supergravity. 

One may wonder whether some field redefinitions on the untwisted vectormultiplet might have led us \emph{directly} to the results of section \ref{subsec:FullEqCohCartanModelSummary}, without requiring \eqref{eq:field-redef-1}--\eqref{eq:field-redef-3}. However these field redefinitions only make sense in the topologically twisted theory: first, they involve the field $\Phi^{\mu}$ (which is a vector supersymmetry ghost obtained \emph{after} twisting), and second, they mix form fields originating from spinors of \emph{different} chiralities before the twist (e.g, see \eqref{eq:field-redef-2}).

\section{Action Of Twisted SYM Coupled To Twisted Supergravity} \label{sec:TwistedSugraAction}
In this section, we propose an action for the twisted vectormultiplet of section \ref{sec:SuperconformalGrav-twistedSYM} coupled to the twisted and truncated supergravity multiplet of section \ref{subsec:Diff-Cartan-from-TwistedSugra} (which has a symmetric gravitino).

In superconformal gravity, the action for a non-Abelian vectormultiplet (or for a collection of Abelian vectormultiplets) coupled to supergravity is traditionally developed via an intermediate step involving a superconformal chiral and antichiral multiplet \cite{deRoo:1980mm,deWit:1984rvr,deWit:1979dzm,deWit:1980lyi}. The most general action of a chiral and an antichiral multiplet coupled to superconformal gravity can be written using the \emph{chiral density formula}. This can be generalized to write the action of a collection of chiral and antichiral multiplets by introducing a homogeneous degree $=2$ function (which we recognize as the \emph{prepotential} in vectormultiplet language). An $\CN=2$ vectormultiplet can be regarded as a constrained combination of a chiral and an antichiral multiplet. This fact allows us to express the fields of the chiral and antichiral multiplets in terms of gauge-invariant local functionals of vectormultiplet fields. This method is referred to as the ``superconformal tensor calculus,'' and enables us to write very general actions for matter multiplets \cite{VanProeyen:1983wk}. In the present context, the distinction between the UV (Donaldson-Witten) and the IR (Seiberg-Witten) regimes is simply that in the UV, the prepotential is quadratic and the gauge group is (generically) non-Abelian, whereas in the IR, the prepotential is generic and the gauge group is Abelian. In both cases, the prepotential must be holomorphic and of homogeneity degree two. 

We will use these principles to derive an action for topologically twisted $\CN=2$ super Yang-Mills coupled to twisted supergravity. The action must be gauge and diffeomorphism invariant, and must reduce to the twisted UV and IR actions of \cite{Witten:1988ze} and \cite{Marino:1998bm,Moore:1997pc} respectively at gravity degree $=0$ (when all sugra fields except the metric are turned off). 
In section \ref{subsec:supconf-chiral-antichiral}, we introduce the chiral and antichiral multiplets, the chiral density formula, and exhibit the vectormultiplet as a constrained chiral $+$ antichiral multiplet. In section \ref{subsec:action-in-vm-language}, we present our proposal for the action. It has the form $\IS = \IQ\IV_{\text{sugra}} + \sfC_{\text{sugra}}$, 
where the Grassmann odd scalar $\IV_{\text{sugra}}$ and the Grassmann even scalar $\sfC_{\text{sugra}}$ (which is \underline{not} $\IQ$-exact) are respectively functionals of the chiral and antichiral multiplet (now viewed as gauge-invariant functionals of vectormultiplet). Since $\IQ(\sfC_{\text{sugra}}) = \text{a surface term}$ and the algebra closes, it follows that $\IQ\IS = 0$ up to surface terms. Readers interested only in the final results may skip to section \ref{subsec:action-in-vm-language} at this point.

\subsection{The Twisted Chiral And Antichiral Multiplets}\label{subsec:supconf-chiral-antichiral}
 A chiral multiplet is defined by the property that its bottom
 component, a scalar field, transforms into a chiral spinor under supersymmetry. The scalar field $\sfA_+$ (resp. $\sfA_-$) of the chiral (resp. antichiral) multiplet has a definite Weyl weight $w$, which is defined to be the Weyl weight of the multiplet. The remaining fields and transformations are inferred by supersymmetric completion, requiring consistency with the local superconformal algebra \cite{deRoo:1980mm}.

The untwisted $\CN=2$ chiral multiplet consists of a real scalar $\sfA_+ \in \Omega^{0}\big(\widehat{\IX}\big)$, a left-handed chiral spinor $\sfG{}_{+} \in \Pi\Gamma\big(S_{+}\widehat{\IX}\otimes \mathsf{K}_{\mathsf{R}}\big)$, a symmetric $\mathfrak{su(2)}_{\mathsf{R}}$-valued field $\sfB_{+} \in \Gamma\big(\mathsf{Sym}^2\mathsf{K}_{\mathsf{R}}\big)$, a self-dual 2-form $\sfF^{+} \in \Omega^{2,+}\big(\widehat{\IX}\big)$, a left-handed chiral spinor $\sfL_{+}\in \Pi\Gamma\big(S_{+}\widehat{\IX}\otimes \mathsf{K}_{\mathsf{R}}\big)$, and a real scalar $\sfC_{+} \in \Omega^{0}\big(\widehat{\IX}\big)$. The untwisted $\CN=2$ antichiral multiplet consists of a real scalar $\sfA_- \in \Omega^{0}\big(\widehat{\IX}\big)$, a right-handed chiral spinor $\sfG_{-} \in \Pi\Gamma\big(S_{-}\widehat{\IX}\otimes \mathsf{K}_{\mathsf{R}}\big)$, a symmetric $\mathfrak{su(2)}_\mathsf{R}$-valued 
field $\sfB_{-} \in \Gamma\big(\mathsf{Sym}^2\mathsf{K}_{\mathsf{R}}\big)$, an anti-self-dual 2-form $\sfF^{-} \in \Omega^{2,-}\big(\widehat{\IX}\big)$, a right-handed chiral spinor $\sfL_{-} \in \Pi\Gamma\big(S_{-}\widehat{\IX}\otimes \mathsf{K}_{\mathsf{R}}\big)$, and a real scalar $\sfC_{-} \in \Omega^{0}\big(\widehat{\IX}\big)$.\footnote{See Tables \ref{tbl:untwisted chiral notation} and \ref{tbl:untwisted antichiral notation} for a comparison of our (anti)chiral multiplet notation with other supergravity literature.} The chiral and antichiral multiplets are $8_B \oplus 8_F$ multiplets. 
%
%

In Euclidean signature, the chiral and antichiral multiplet are \underline{not} related by complex conjugation, so one must consider their twisted versions separately. The field content of the twisted multiplets is listed in Tables \ref{tbl:supconf-chiral} and \ref{tbl:supconf-antichiral}. Note that if the fields are written with spinor indices rather than coordinate (curved) indices, then in each table they are listed from top to bottom in order of nondecreasing Weyl weight. For example, $\sfG_{\mu\nu} = \frac{1}{2}\sfG^{\text{sym}}_{AB}e_{\mu}{}^{A}{}_{\dA}e_{\nu}{}^{B\dA}$ has Weyl weight $w-\frac{3}{2}$, but $\sfG^{\text{sym}}_{AB}$ -- an object obtained from twisting $\sfG^{i}{}_{+}$ -- has Weyl weight $w + \frac{1}{2}$. 

\begin{small}
\begin{table}[H]
\centering
\begin{tabular}{|c|c|c|c|c|}\hline
  Symbol & Field & Element of & Grassmann Parity & Weyl Weight \\ \hline
   $\sfA_+$ & bosonic 0-form & $\Omega^{0}(\IX)$ &  even & $w$ \\ \hline
   $\sfG$ & fermionic 0-form & $\Pi\Omega^{0}(\IX)$ & odd & $w + \frac{1}{2}$ \\ \hline
   $\sfG_{\mu\nu}$ & fermionic SD 2-form & $\Pi\Omega^{2,+}(\IX)$ & odd & $w- \frac{3}{2}$ \\ \hline
   $\sfB_{+\mu\nu}$ & bosonic SD 2-form & $\Omega^{2,+}(\IX)$ & even & $w-1$ \\ \hline 
   $\sfF^{+}_{\mu\nu}$ & bosonic SD 2-form & $\Omega^{2,+}(\IX)$ & even & $w-1$ \\ \hline
   $\sfL$ & fermionic 0-form & $\Pi\Omega^{0}(\IX)$ & odd & $w+\frac{3}{2}$ \\ \hline
   $\sfL_{\mu\nu}$ & fermionic SD 2-form & $\Pi\Omega^{2,+}(\IX)$ & odd & $w-\frac{1}{2}$ \\ \hline
   $\sfC_{+}$ & bosonic 0-form & $\Omega^{0}(\IX)$ & even & $w+2$ \\ \hline
\end{tabular}
\caption{\label{tbl:supconf-chiral}The fields of a twisted $\CN=2$ superconformal chiral multiplet of Weyl weight $w$.}
\end{table}

\begin{table}[H]
\centering
\begin{tabular}{|c|c|c|c|c|}\hline
  Symbol & Field & Element of & Grassmann Parity & Weyl Weight \\ \hline
  $\sfA_-$ & bosonic 0-form & $\Omega^{0}(\IX)$ & even & $w$ \\ \hline
  $\sfG_{\mu}$ & fermionic 1-form & $\Pi\Omega^{1}(\IX)$ & odd & $w-\frac{1}{2}$ \\ \hline 
  $\sfB_{-\mu\nu}$ & bosonic SD 2-form & $\Omega^{2,+}(\IX)$ & even & $w-1$ \\ \hline
  $\sfF_{\mu\nu}^{-}$ & bosonic ASD 2-form & $\Omega^{2,-}(\IX)$ & even & $w-1$ \\ \hline
  $\sfL_{\mu}$ & fermionic 1-form & $\Pi\Omega^{1}(\IX)$ & odd & $w+\frac{1}{2}$ \\ \hline
  $\sfC_{-}$ & bosonic 0-form & $\Omega^{0}(\IX)$ & even & $w+2$ \\ \hline
\end{tabular}
\caption{\label{tbl:supconf-antichiral}The fields of a twisted $\CN=2$ superconformal antichiral multiplet of Weyl weight $w$.}
\end{table}
\end{small}

Here $\sfF^{\pm}_{\mu\nu}$ denote SD/ASD 2-form fields which should not be confused with the SD/ASD YM field strengths $F_{\mu\nu}^{\pm}$. Later, as a result of constraints we will impose, the $\sfF^{\pm}_{\mu\nu}$
and $F_{\mu\nu}^{\pm}$ will in fact be closely related. See \eqref{eq:constrained-chiral-soln-5}--\eqref{eq:constrained-chiral-soln-6} and \eqref{eq:chiral-to-vm Fplus}   below. 
Also note that $\sfB_{+\mu\nu}$ and $\sfB_{-\mu\nu}$ are both self-dual. In what follows, we will be interested in (anti)chiral multiplets of Weyl weight $w = 1$ and $w = 2$. 
The transformation laws of the twisted chiral multiplet of Weyl weight $w$ are 
\eqa{
 &\delta \sfA_+ &&= -\frac{\eps}{2\sq}\sfG ~,\\
 &\delta \sfG   &&= -2\sq \eps^{\rho} \scd_{\rho}\sfA_+ + 2w\sq\bn_{0}\sfA_+ ~,\\
 &\delta \sfG_{\mu\nu} &&= -\eps\big(\Psi_{[\mu}{}^{\rho}\sfG_{\nu]\rho}\big) - 2\sq\big(\eps_{[\mu}\scd_{\nu]}\sfA_+\big)^+ + \frac{i\eps}{2\sq}\big(\sfF_{\mu\nu}^{+} - \sfB_{+\mu\nu}\big) + \frac{w}{\sq}\bn_{\mu\nu}^{+}\sfA_+ ~,\\
&\delta \sfB_{+\mu\nu} &&= -\eps\big(\Psi_{[\mu}{}^{\rho}\sfB_{+\nu]\rho}\big) - 4i\sq\big(\eps_{[\mu}\scd^{\rho}\sfG_{\nu]\rho}\big)^{+} - i\sq\big(\eps_{[\mu}\scd_{\nu]}\sfG\big)^{+} - \frac{\eps}{\sq}\sfL_{\mu\nu} \nn
& &&\quad -i(1-w)\sq \big(\bn_{[\mu}{}^{\rho}\sfG_{\nu]\rho}\big) - \frac{i(1-w)}{2\sq}\bn_{\mu\nu}\sfG + i(1-w)\sq \bn_{0}\sfG_{\mu\nu} ~, \label{eq:delta chiral Bplus}\\
&\delta \sfF^{+}_{\mu\nu} &&= -\eps\big(\Psi_{[\mu}{}^{\rho}\sfF^{+}_{\nu]\rho}\big) + 4i\sq\big(\eps^{\rho}\scd_{[\mu}\sfG_{\nu]\rho}\big)^{+} + i\sq\big(\eps_{[\mu}\scd_{\nu]}\sfG\big)^{+} - \frac{\eps}{\sq}\sfL_{\mu\nu} \nn
& &&\quad +i(1+w)\sq \big(\bn_{[\mu}{}^{\rho}\sfG_{\nu]\rho}\big) - \frac{i(1+w)}{2\sq}\bn_{\mu\nu}\sfG + i(1+w)\sq \bn_{0}\sfG_{\mu\nu} ~,\\
 &\delta \sfL &&= 2\sq \eps^{\rho}\scd^{\sigma}\big(\sfF_{\rho\sigma}^{+} + \sfB_{+\rho\sigma}\big) + \frac{i\eps}{\sq}\sfC_{+} - 4i\sq\eps^{\rho}\big[ \big(\scd^{\sigma}\sfA_+\big)T_{\rho\sigma}^{-} + w \sfA_+ \big(\scd^{\sigma}T_{\rho\sigma}^{-}\big) \big] \nn
 & &&\quad - 6i\eps^{\rho}\Xi^{\sigma}\sfG_{\rho\sigma} + \frac{3i}{2}\eps^{\rho}\Xi_{\rho}\sfG - \frac{(1+w)}{\sq}\sfB_{+\mu\nu}\bn^{\mu\nu} + \frac{(1-w)}{\sq}\sfF_{\mu\nu}^{+}\bn^{\mu\nu} ~,\\
 &\delta \sfL_{\mu\nu} &&= -\eps\big(\Psi_{[\mu}{}^{\rho}\sfL_{\nu]\rho}\big) + 2\sq\big(\eps_{[\mu}\scd^{\rho}\sfF_{\nu]\rho}^{+}\big)^{+} + 2\sq\big(\eps^{\rho}\scd_{[\mu}\sfB_{+\nu]\rho}\big)^{+} \nn
 & &&\quad -4i\sq\big[\big(\eps_{[\mu}\big(\scd^{\rho}\sfA_+\big)T_{\nu]\rho}^{-}\big)^{+} + w\sfA_+\big(\eps_{[\mu}\scd^{\rho}T_{\nu]\rho}^{-}\big)^{+} \big] - 6i\big(\eps_{[\mu}\Xi^{\rho}\sfG_{\nu]\rho}\big)^{+} + \frac{3i}{2}\big(\eps_{[\mu}\Xi_{\nu]}\big)^{+}\sfG \nn
 & &&\quad + \frac{(1+w)}{\sq}\big(\sfB_{+\rho[\mu}\bn_{\nu]}{}^{\rho}\big) + \frac{(1+w)}{\sq}\sfB_{+\mu\nu}\bn_0 + \frac{(1-w)}{\sq}\big(\sfF^{+}_{\rho[\mu}\bn_{\nu]}{}^{\rho}\big) - \frac{(1-w)}{\sq}\sfF_{\mu\nu}^{+}\bn_0 ~,\\
&\delta \sfC_{+} &&= -4i\sq\eps^{\rho}\scd_{\sigma}\sfL_{\rho}{}^{\sigma} - i\sq\eps^{\rho}\scd_{\rho}\sfL + 12\,i\,\eps^{\rho}\,\Xi^{\sigma}\sfB_{+\rho\sigma} \nn
& &&\quad -2\sq\big[ 4(w-1)\eps^{\rho}\big(\scd_{[\mu}T_{\nu]\rho}^{-}\big)\sfG^{\mu\nu} + (w-1)\eps^{\rho}\big(\scd^{\sigma}T_{\rho\sigma}^{-}\big)\sfG + 4\eps^{\rho}\big(\scd_{[\mu}\sfG_{\nu]\rho}\big)T^{\mu\nu-} + \eps^{\rho}T_{\rho\sigma}^{-}\scd^{\sigma}\sfG\big]\nn
& &&\quad + iw\sq \bn_{\mu\nu}\sfL^{\mu\nu} + iw\sq \bn_{0}\sfL ~,
}
and the transformations of the twisted antichiral multiplet of Weyl weight $w$ are
\eqa{
&\delta \sfA_- &&= -i\,\eps^{\rho}\,\sfG_{\rho} ~,\\
&\delta \sfG_{\mu} &&= \frac{1}{2}\eps\,\Psi_{\mu}{}^{\rho}\sfG_{\rho} + i\eps \scd_{\mu}\sfA_{-} + 2\eps^{\rho}\big(\sfF_{\rho\mu}^{-} + \sfB_{-\rho\mu}\big) ~,\\
&\delta \sfL_{\mu} &&= \frac{1}{2}\eps\,\Psi_{\mu}{}^{\rho}\sfL_{\rho} + i\eps\,\scd^{\rho}\big(\sfF^{-}_{\rho\mu} - \sfB_{-\rho\mu}\big) + \eps_{\mu}\sfC_{-} ~,\\
&\delta \sfB_{-\mu\nu} &&= -\eps\big(\Psi_{[\mu}{}^{\rho}\sfB_{-\nu]\rho}\big) + \eps\big(\scd_{[\mu}\sfG_{\nu]}\big)^{+} - 2i\big(\eps_{[\mu}\sfL_{\nu]}\big)^{+} ~, \label{eq:delta antichiral Bminus}\\
&\delta \sfF^{-}_{\mu\nu} &&= -\eps\big(\Psi_{[\mu}{}^{\rho}\sfF_{\nu]\rho}^{-}\big) + \eps\big(\scd_{[\mu}\sfG_{\nu]}\big)^{-} - 2i\big(\eps_{[\mu}\sfL_{\nu]}\big)^{-} ~,\\
&\delta \sfC_{-} &&= \eps\,\scd_{\mu}\sfL^{\mu} = \eps\left(\n_{\mu}\sfL^{\mu} - \frac{1}{2}\Psi_{\rho}{}^{\rho}\sfC_{-}\right) ~. \label{eq:delta antichiral Cminus}
}
Here $\Xi_{\mu}$ is the composite sugra field \eqref{eq:chi vec} (or \eqref{eq:chi vec rewrite}), and $T_{\mu\nu}^{-}$ is the independent (auxiliary) sugra field, $S_{\mu}$ is the composite S-susy connection for $0$-form S-susy, and $S_{\mu,[\nu\rho]} = S_{\mu,[\nu\rho]}^{+}$ is the composite S-susy connection for SD $2$-form S-susy. 
Below we list various supercovariant derivatives appearing in the transformation laws above.
\eqa{
&\scd_{\mu}\sfA_+ &&= \partial_{\mu}\sfA_+ ~,\\
&\scd_{\mu}\sfG &&= \partial_{\mu}\sfG + \sq \Psi_{\mu}{}^{\rho}\scd_{\rho}\sfA_{+} - w S_{\mu}\sf A_+ ~,\\
&\scd_{\mu}\sfG_{\nu\rho} &&= \n_{\mu}\sfG_{\nu\rho} + \sq\big(\Psi_{\mu[\nu}\partial_{\rho]}\sfA_+\big)^{+} - w S_{\mu,[\nu\rho]}\sfA_+ ~,\\
&\scd_{\mu}\sfB_{+\nu\rho} &&= \n_{\mu}\sfB_{+\nu\rho} + 2i\sq\big(\Psi_{\mu[\nu}\scd^{\sigma}\sfG_{\rho]\sigma}\big)^{+} + \frac{i}{\sq}\big(\Psi_{\mu[\nu}\scd_{\rho]}\sfG\big)^{+}\nn
& &&\quad + 2i(1-w)S_{\mu,[\nu}{}^{\sigma}\sfG_{\rho]\sigma} + \frac{i(1-w)}{2}S_{\mu,[\nu\rho]}\sfG - \frac{i(1-w)}{2}S_{\mu}\sfG_{\nu\rho} ~,\\
&\scd_{\mu}\sfF^{+}_{\nu\rho} &&= \n_{\mu}\sfF^{+}_{\nu\rho} - 2i\sq\big(\Psi_{\mu}{}^{\sigma}\scd_{[\nu}\sfG_{\rho]\sigma}\big)^{+} - \frac{i}{\sq}\big(\Psi_{\mu[\nu}\scd_{\rho]}\sfG\big)^{+} \nn
& &&\quad -2i(1+w)S_{\mu,[\nu}{}^{\sigma}\sfG_{\rho]\sigma} + \frac{i(1+w)}{2}S_{\mu,[\nu\rho]}\sfG  - \frac{i(1+w)}{2}S_{\mu}\sfG_{\nu\rho} ~,\\
&\scd_{\mu}\sfL_{\nu\rho} &&= \n_{\mu}\sfL_{\nu\rho} - \sq\big(\Psi_{\mu[\nu}\scd^{\sigma}\sfF^{+}_{\rho]\sigma}\big)^{+} - \sq\big(\Psi_{\mu}{}^{\sigma}\scd_{[\nu}\sfB_{+\rho]\sigma}\big)^{+} \nn
& &&\quad + 2i\sq\big[\big(\Psi_{\mu[\nu}\big(\partial^{\sigma}\sfA_{+}\big)T^{-}_{\rho]\sigma}\big)^{+} + w\sfA_+\big(\Psi_{\mu[\nu}\scd^{\sigma}T^{-}_{\rho]\sigma}\big)^{+}\big]\nn
& &&\quad + 3i\big(\Psi_{\mu[\nu}\Xi^{\sigma}\sfG_{\rho]\sigma}\big)^{+} - \frac{3i}{4}\big(\Psi_{\mu[\nu}\Xi_{\rho]}\big)^{+}\sfG - (1+w)\sfB_{+\sigma[\nu}S_{|\mu|\rho]}{}^{\sigma} \nn
& &&\quad - \frac{(1+w)}{4}\sfB_{+\nu\rho}S_{\mu} - (1-w)\sfF^{+}_{\sigma[\nu}S_{|\mu|,\rho]}{}^{\sigma} + \frac{(1-w)}{4}\sfF_{\nu\rho}^{+}S_{\mu} ~,\\
&\scd_{\mu}\sfA_{-} &&= \partial_{\mu}\sfA_{-} + \frac{i}{2}\Psi_{\mu}{}^{\rho}\sfG_{\rho} ~,\\
&\scd_{\mu}\sfG_{\nu} &&= \n_{\mu}\sfG_{\nu} - \Psi_{\mu}{}^{\rho}\big(\sfF^{-}_{\rho\nu} + \sfB_{-\rho\nu}\big) ~,\\
&\scd_{\mu}\sfL_{\nu} &&= \n_{\mu}\sfL_{\nu} - \frac{1}{2}\Psi_{\mu\nu}\sfC_{-} ~,\\
&\scd_{\mu}\sfB_{-\nu\rho} &&= \n_{\mu}\sfB_{-\nu\rho} + i\big(\Psi_{\mu[\nu}\sfL_{\rho]}\big)^{+} ~,\\
&\scd_{\mu}\sfF^{-}_{\nu\rho} &&= \n_{\mu}\sfF^{-}_{\nu\rho} + i\big(\Psi_{\mu[\nu}\sfL_{\rho]}\big)^{-} ~.
} 

\noindent The $\IQ$-transformations of the twisted chiral multiplet are
\eqa{
 &\IQ \sfA_+ &&= -\frac{1}{2\sq}\sfG ~,\label{eq:IQ chiral Aplus}\\
 &\IQ \sfG   &&= -2\sq \Phi^{\rho} \scd_{\rho}\sfA_+ ~,\\
 &\IQ \sfG_{\mu\nu} &&= -\big(\Psi_{[\mu}{}^{\rho}\sfG_{\nu]\rho}\big) - 2\sq\big(\Phi_{[\mu}\scd_{\nu]}\sfA_+\big)^+ + \frac{i}{2\sq}\big(\sfF_{\mu\nu}^{+} - \sfB_{+\mu\nu}\big) + \frac{w}{\sq}\mathsf{c}_{\mu\nu}\sfA_+ ~,\\
&\IQ \sfB_{+\mu\nu} &&= -\big(\Psi_{[\mu}{}^{\rho}\sfB_{+\nu]\rho}\big) - 4i\sq\big(\Phi_{[\mu}\scd^{\rho}\sfG_{\nu]\rho}\big)^{+} - i\sq\big(\Phi_{[\mu}\scd_{\nu]}\sfG\big)^{+} - \frac{1}{\sq}\sfL_{\mu\nu} \nn
& &&\quad -i(1-w)\sq \big(\mathsf{c}_{[\mu}{}^{\rho}\sfG_{\nu]\rho}\big) - \frac{i(1-w)}{2\sq}\mathsf{c}_{\mu\nu}\sfG  ~, \label{eq:Q chiral Bplus}\\
&\IQ \sfF^{+}_{\mu\nu} &&= -\eps\big(\Psi_{[\mu}{}^{\rho}\sfF^{+}_{\nu]\rho}\big) + 4i\sq\big(\Phi^{\rho}\scd_{[\mu}\sfG_{\nu]\rho}\big)^{+} + i\sq\big(\Phi_{[\mu}\scd_{\nu]}\sfG\big)^{+} - \frac{1}{\sq}\sfL_{\mu\nu} \nn
& &&\quad +i(1+w)\sq \big(\mathsf{c}_{[\mu}{}^{\rho}\sfG_{\nu]\rho}\big) - \frac{i(1+w)}{2\sq}\mathsf{c}_{\mu\nu}\sfG ~,\\
 &\IQ \sfL &&= 2\sq \Phi^{\rho}\scd^{\sigma}\big(\sfF_{\rho\sigma}^{+} + \sfB_{+\rho\sigma}\big) + \frac{i}{\sq}\sfC_{+} - 4i\sq\Phi^{\rho}\big[ \big(\scd^{\sigma}\sfA_+\big)T_{\rho\sigma}^{-} + w \sfA_+ \big(\scd^{\sigma}T_{\rho\sigma}^{-}\big) \big] \nn
 & &&\quad - 6i\Phi^{\rho}\Xi^{\sigma}\sfG_{\rho\sigma} + \frac{3i}{2}\Phi^{\rho}\Xi_{\rho}\sfG - \frac{(1+w)}{\sq}\sfB_{+\mu\nu}\mathsf{c}^{\mu\nu} + \frac{(1-w)}{\sq}\sfF_{\mu\nu}^{+}\mathsf{c}^{\mu\nu} ~, \label{eq:IQ chiral Lambda}\\
 &\IQ \sfL_{\mu\nu} &&= -\eps\big(\Psi_{[\mu}{}^{\rho}\sfL_{\nu]\rho}\big) + 2\sq\big(\Phi_{[\mu}\scd^{\rho}\sfF_{\nu]\rho}^{+}\big)^{+} + 2\sq\big(\Phi^{\rho}\scd_{[\mu}\sfB_{+\nu]\rho}\big)^{+} \nn
 & &&\quad -4i\sq\big[\big(\Phi_{[\mu}\big(\scd^{\rho}\sfA_+\big)T_{\nu]\rho}^{-}\big)^{+} + w\sfA_+\big(\Phi_{[\mu}\scd^{\rho}T_{\nu]\rho}^{-}\big)^{+} \big] - 6i\big(\Phi_{[\mu}\Xi^{\rho}\sfG_{\nu]\rho}\big)^{+} \nn
 & &&\quad + \frac{3i}{2}\big(\Phi_{[\mu}\Xi_{\nu]}\big)^{+}\sfG  + \frac{(1+w)}{\sq}\big(\sfB_{+\rho[\mu}\mathsf{c}_{\nu]}{}^{\rho}\big) + \frac{(1-w)}{\sq}\big(\sfF^{+}_{\rho[\mu}\mathsf{c}_{\nu]}{}^{\rho}\big)  ~,\\
&\IQ \sfC_{+} &&= -4i\sq\Phi^{\rho}\scd_{\sigma}\sfL_{\rho}{}^{\sigma} - i\sq\Phi^{\rho}\scd_{\rho}\sfL + 12\,i\,\Phi^{\rho}\,\Xi^{\sigma}\sfB_{+\rho\sigma} \nn
& &&\quad -2\sq\big[ 4(w-1)\Phi^{\rho}\big(\scd_{[\mu}T_{\nu]\rho}^{-}\big)\sfG^{\mu\nu} + (w-1)\Phi^{\rho}\big(\scd^{\sigma}T_{\rho\sigma}^{-}\big)\sfG\big]\nn
& && \quad -2\sq\big[4\Phi^{\rho}\big(\scd_{[\mu}\sfG_{\nu]\rho}\big)T^{\mu\nu-} + \Phi^{\rho}T_{\rho\sigma}^{-}\scd^{\sigma}\sfG\big] + iw\sq \mathsf{c}_{\mu\nu}\sfL^{\mu\nu} ~,
}
and the $\IQ$-transformations of the twisted antichiral multiplet are
\eqa{
&\IQ \sfA_- &&= -i\,\Phi^{\rho}\,\sfG_{\rho} ~,\\
&\IQ \sfG_{\mu} &&= \frac{1}{2}\Psi_{\mu}{}^{\rho}\sfG_{\rho} + i\scd_{\mu}\sfA_{-} + 2\Phi^{\rho}\big(\sfF_{\rho\mu}^{-} + \sfB_{-\rho\mu}\big) ~,\\
&\IQ \sfL_{\mu} &&= \frac{1}{2}\Psi_{\mu}{}^{\rho}\sfL_{\rho} + i\scd^{\rho}\big(\sfF^{-}_{\rho\mu} - \sfB_{-\rho\mu}\big) + \Phi_{\mu}\sfC_{-} ~,\\
&\IQ \sfB_{-\mu\nu} &&= -\big(\Psi_{[\mu}{}^{\rho}\sfB_{-\nu]\rho}\big) + \big(\scd_{[\mu}\sfG_{\nu]}\big)^{+} - 2i\big(\Phi_{[\mu}\sfL_{\nu]}\big)^{+} ~, \label{eq:Q antichiral Bminus}\\
&\IQ \sfF^{-}_{\mu\nu} &&= -\big(\Psi_{[\mu}{}^{\rho}\sfF_{\nu]\rho}^{-}\big) + \big(\scd_{[\mu}\sfG_{\nu]}\big)^{-} - 2i\big(\Phi_{[\mu}\sfL_{\nu]}\big)^{-} ~,\\
&\IQ \sfC_{-} &&= \scd_{\mu}\sfL^{\mu} = \n_{\mu}\sfL^{\mu} - \frac{1}{2}\Psi_{\rho}{}^{\rho}\sfC_{-} ~. \label{eq:IQ antichiral Cminus}
}
 

Note that \eqref{eq:IQ chiral Aplus}--\eqref{eq:IQ antichiral Cminus} should be understood and used with $T_{\mu\nu}^{-}$ substituted from \eqref{eq:sym-grav-T}, and $\mathsf{c}_{\mu\nu}^{+}$ substituted from \eqref{eq:sym-grav-ghost}. We have retained the $T_{\mu\nu}^{-}$ and $\mathsf{c}_{\mu\nu}^{+}$ dependence simply for brevity. As previously mentioned, there are two special values of the Weyl weight $w$ that are of interest to us. When $w = 1$, one can impose a set of constraints on a combination of a chiral and antichiral multiplet, which halve the degrees of freedom from $16_{B}\oplus 16_{F}$ to $8_{B} \oplus  8_{F}$, and allows us to recast the reduced set of fields as the fields of an Abelian vectormultiplet. This is referred to as a \emph{constrained chiral multiplet} \cite{deRoo:1980mm,deWit:1980lyi}. For the topologically twisted chiral + antichiral multiplet, a consistent set of constraints is\footnote{These constraints are obtained by decomposing the constraints on the combination of an untwisted chiral and antichiral multiplet in terms of chiral spinors, and then topologically twisting them and imposing the defining relations of the topologically twisted supergravity background of section \ref{subsec:Superconformal-TwistTruncate}.}
  \eqa{
    & \qquad\qquad\qquad\qquad\qquad\qquad\qquad\qquad \sfB_{+\mu\nu} - \sfB_{-\mu\nu} = 0 ~,\label{eq:constrained-chiral-1}\\
    &\sfL - \sq \scd^{\rho}\sfG_{\rho} = 0 ~, \qquad \sfL_{\mu\nu} + \sq\big(\scd_{[\mu}\sfG_{\nu]}\big)^{+} = 0 ~, \quad \sfL_{\mu} + 2\sq\scd_{\rho}\sfG_{\mu}{}^{\rho} + \frac{1}{\sq}\scd_{\mu}\sfG = 0 ~, \label{eq:constrained-chiral-4}\\
   &\qquad \qquad \qquad \qquad 2\sq\scd^{\rho}\big(\sfF^{+}_{\rho\mu} - \sfF^{-}_{\rho\mu} - 2i A_+ T^{-}_{\rho\mu}\big) + 6i\,\Xi^{\rho}\sfG_{\mu\rho} - \frac{3i}{2}\Xi_{\mu}\sfG = 0 ~, \label{eq:constrained-chiral-5}\\
    &\qquad\qquad 2\,\bm{\Box}_{C}\sfA_{-} + 2i\,\sfF^{-}_{\mu\nu}T^{\mu\nu-} - 3i\,\Xi^{\rho}\sfG_{\rho} - \sfC_{+} = 0 ~, \qquad 2\,\bm{\Box}_{C}\sfA_{+} - \sfC_{-} = 0 ~, \label{eq:constrained-chiral-7}
 }
 where $\bm{\Box}_{C} := \scd_{\ha}\scd^{\ha}$ denotes the superconformal d'Alembertian. These constraints are consistent with the BRST transformations only for $w = 1$. For example, comparing \eqref{eq:Q chiral Bplus} with \eqref{eq:Q antichiral Bminus}, it is clear that \eqref{eq:constrained-chiral-1} is only consistent for $w = 1$. Then the Weyl weight of $\sfF^{\pm}_{\mu\nu}$ equals zero (see Tables \ref{tbl:supconf-chiral} and \ref{tbl:supconf-antichiral}), and \eqref{eq:constrained-chiral-5} can be viewed as a Bianchi identity for a supercovariant field strength (specifically, we take as an ansatz, $\sfF_{\mu\nu}^{+} := \wh{F}_{\mu\nu}^{+}$ and $\sfF_{\mu\nu}^{-} := \wh{F}_{\mu\nu}^{-} - 2i A_+ T_{\mu\nu}^{-}$, where $\wh{F}_{\mu\nu}$ is supercovariant).\footnote{There is a slight conceptual leap here. On a contractible set, a closed 2-form can be identified with the field strength of a $\mathsf{U(1)}$ (or $\IR$)  connection. When generalized to supergravity, equation 
 \eqref{eq:constrained-chiral-5} implies that the 2-form $\sfF := \sfF^+ + \sfF^-$ 
 is the supercovariant curvature of a connection; working globally one must also take into account issues of normalization. On a general manifold, it is not true that any closed 2-form $\omega$ is the field strength of a $\mathsf{U(1)}$ connection. There is a quantization condition on the periods of $\omega$. In this sense, our identification of $\sfF := \sfF^+ + \sfF^-$ with the supercovariant curvature of a connection is an ansatz. It does not follow from local constraint equations.}
Denoting the fields of a constrained chiral multiplet with a subscript $\left.\right|_{V}$, the solution to these constraints consistent with the transformation laws is given by the following:
 \eqa{
    &\label{eq:constrained-chiral-soln-1}\left.\sfA_{+}{}^{I}\right|_{V} &&= \frac{i}{2}\lambda^{I} ~, \quad &\left.\sfA_{-}{}^{I}\right|_{V} &= 2i\,\phi^{I} ~,\\
    &\label{eq:constrained-chiral-soln-2}\left.\sfG_{\mu\nu}{}^{I}\right|_{V} &&= \frac{i}{2\sq}\chi_{\mu\nu}{}^{I} ~, \quad & \left.\sfG_{\mu}{}^{I}\right|_{V} &= 2\,\psi_{\mu}{}^{I} ~,\\
    &\label{eq:constrained-chiral-soln-3}\left.\sfG^{I}\right|_{V} &&= -i\sq\,\eta^{I} ~, \quad & \left.\sfB_{-\mu\nu}{}^{I}\right|_{V} &= D_{\mu\nu}{}^{I} ~,\\
    &\label{eq:constrained-chiral-soln-4}\left.\sfB_{+\mu\nu}{}^{I}\right|_{V} &&= D_{\mu\nu}{}^{I} ~, \quad &\left.\sfF^{-}_{\mu\nu}{}^{I}\right|_{V} &= \wh{F}(A)_{\mu\nu}^{-I} + \lambda^{I}\,T_{\mu\nu}^{-} ~,\\
   &\label{eq:constrained-chiral-soln-5}\left.\sfF^{+}_{\mu\nu}{}^{I}\right|_{V} &&= \wh{F}(A)_{\mu\nu}^{+}{}^{I} ~, \quad &\left.\sfL_{\mu}{}^{I}\right|_{V} &= -i\,\scd_{\rho}\chi_{\mu}{}^{\rho,I} + i\,\scd_{\mu}\eta^{I} ~,\\
   &\label{eq:constrained-chiral-soln-6}\left.\sfL_{\mu\nu}{}^{I}\right|_{V} &&= -2\sq\big(\scd_{[\mu}\psi_{\nu]}{}^{I}\big)^{+} ~, \quad &\left.\sfC_{-}{}^{I}\right|_{V} &= i\,\bm{\Box}_{C}\lambda^{I} ~,\\
   &\label{eq:constrained-chiral-soln7}\left.\sfL^{I}\right|_{V} &&= 2\sq\scd_{\mu}\psi^{\mu,I} ~,\\
   &\label{eq:constrained-chiral-soln-8}\left.\sfC_{+}{}^{I}\right|_{V} &&= 4i\,\bm{\Box}_{C}\phi^{I} + 2i\big(\wh{F}(A)_{\mu\nu}^{-}{}^{I} + \lambda^{I}\,T_{\mu\nu}^{-}\big)T_{\mu\nu}^{-} & \,-\, 6i\,\Xi^{\rho}\psi_{\rho}{}^{I} ~.
 }
 On the RHS of \eqref{eq:constrained-chiral-soln-1}--\eqref{eq:constrained-chiral-soln-8}, the notation of a twisted Abelian vectormultiplet is used, where the index $I$ labels the multiple Abelian multiplets. Indeed, upon substituting these relations into the transformation laws \eqref{eq:IQ chiral Aplus}--\eqref{eq:IQ antichiral Cminus} of the (anti)chiral multiplet with $w=1$, one recovers the Abelian version of the twisted vectormultiplet transformation laws \eqref{eq:CartanSugra-1}--\eqref{eq:CartanSugra-7}. The top components, $\sfC_+$ and $\sfC_-$, each have Weyl weight $3$.\footnote{The superconformal d'Alembertian $\bm{\Box}_{C} := \scd_{a}\scd^{a}$ has Weyl weight $+2$.}
 
 Next, let us consider the situation when $w = 2$. Now the top components $\sfC_+$ and $\sfC_-$ each have Weyl weight $+4$. Since $e = \sqrt{\det\,g}$ has Weyl weight $-4$, it follows that the combinations $e\sfC_+$ and $e\sfC_-$ each have Weyl weight $0$, so they are good candidates for Weyl-invariant terms in an action formula. In fact, for rigid supersymmetry, the most general action is a linear combination of $e\sfC_+$ and $e\sfC_-$ because each term transforms under (rigid) supersymmetry into a total derivative. However, for local supersymmetry, this is not the case and one does require contributions from other fields in the chiral and antichiral multiplets. The untwisted Lagrangians for chiral and antichiral multiplets are given by the so-called \emph{chiral density formula} \cite{deRoo:1980mm}. Upon twisting, the chiral density formula for the Lagrangian can be written as a linear combination 
 \eqa{
   &\mathscr{L}_{\text{cdf ($w=2$)}} &&= \alpha \mathscr{L}_{\text{chiral}} + \beta \mathscr{L}_{\text{antichiral}} ~, \label{eq:cdf}
 }
 where $\alpha$ and $\beta$ are arbitrary coefficients, and
%
 \eqa{
     \hspace{-0.50cm}\mathscr{L}_{\text{chiral}} 
&= e\sfC_+ - \tfrac{ie}{\sq}\Psi_{\rho}{}^{\rho}\sfL + 8e\sq\big(\n_{[\mu}\Phi_{\nu]}\big)^{-}\Psi^{\mu}{}_{\rho}G^{\rho\nu} - 16e\sfA_{+}\big(\n_{[\mu}\Phi_{\nu]}\big)^{-}\big(\n^{[\mu}\Phi^{\nu]}\big)^{-} \nn
    \hspace{-0.50cm} &\quad + \tfrac{3e}{\sq}\sfG^{\mu\nu}\big(\Psi_{\mu}{}^{\alpha}\Psi_{\nu\alpha}\big)^{+}\Psi_{\rho}{}^{\rho} - i e \Psi_{\mu}{}^{\rho}\Psi_{\nu\rho}\big(\sfF^{\mu\nu,+} + \sfB_{+}^{\mu\nu}\big)  + e \sfA_+\big(\Psi_{\mu}{}^{\alpha}\Psi_{\nu\alpha}\big)^{+}\big(\Psi^{\mu\beta}\Psi^{\nu}{}_{\beta}\big)^{+} ~, \label{eq:Lchiral}\\
  \hspace{-0.50cm} \mathscr{L}_{\substack{\text{anti-}\\\text{chiral}}} &= e\sfC_- ~. \label{eq:Lantichiral}
 }
Due to the truncation of section \ref{subsec:Superconformal-TwistTruncate}, only the top component of the antichiral multiplet $\sfC_-$ contributes to the action. Under a supersymmetry variation, $e\sfC_-$ is a total derivative (see \eqref{eq:delta antichiral Cminus} and \eqref{eq:IQ antichiral Cminus}) and so, $\mathscr{L}_{\text{antichiral}}$ is invariant. 
Further, note that \eqref{eq:IQ chiral Lambda} implies that $\IQ(-ie\sq\Lambda) = -\frac{ie}{\sq}\Psi_{\rho}{}^{\rho}\Lambda + e \sfC_+ + \cdots$, which produces two terms of $\mathscr{L}_{\text{chiral}}$.\footnote{\label{foot:motivateaction}If we work with $\CQ$ rather than $\IQ$ and retain $\Psi_{[\mu\nu]}$ and $T_{\mu\nu}^{-}$, then one can show that $\mathscr{L}_{\text{chiral}}$ is $\CQ$-exact. We conjecture that $\mathscr{L}_{\text{chiral}}$ is $\IQ$-exact, i.e., $\mathscr{L}_{\text{chiral}} = \IQ(-ie\sq\Lambda - e \sqrt{2}\Psi_{\mu\rho}\Psi_{\nu}{}^{\rho}\sfG^{\mu\nu})$. Then our choice, namely $\IQ(-ie\sq\Lambda)$, is simply a modification of the chiral density formula by $\IQ$-exact terms. It is a minimal choice that discards some $\IQ$-exact contributions of higher gravitational degree. This is supported by the intuition that there are no nontrivial $\IQ$-closed susy invariants with several factors of the gravitino.} 
We use this to motivate our action for the topologically twisted vectormultiplet coupled to topologically twisted supergravity. The Abelian vectormultiplet (viewed as a constrained chiral multiplet) cannot be used to write a Weyl-invariant action: for this, we at least need a top component with Weyl weight $+4$.
 

This is precisely where the prepotential enters. In rigid supersymmetry, the prepotential is an arbitrary holomorphic function of the vectormultiplet scalar. But in supergravity, the prepotential is further restricted to be homogenous of degree $2$. (The Weyl weight of each scalar in the vectormultiplet is $+1$.) Given a (twisted) vectormultiplet, $\CF(\phi)$ and $\ov{\CF}(\lambda)$ each have Weyl weight $+2$, and are themselves the bottom components of (twisted) chiral and antichiral multiplets, each of Weyl weight $+2$. We can readily write down actions for such multiplets. 
 
 \subsection{The Vectormultiplet Action From Twisted Supergravity}\label{subsec:action-in-vm-language}
 In this section, we build on the ideas above to propose an action in terms of the vectormultiplet fields. We begin with a translation from chiral + antichiral language to vectormultiplet language. Let $\ov{\CF} \equiv \overline{\CF}(\lambda)$ and $\CF \equiv \CF(\phi)$ denote homogeneous degree $=2$ functions of $\lambda$ and $\phi$ respectively (a.k.a. the prepotentials). By setting $\sfA_+ \sim \ov{\CF}(\lambda)$ and $\sfA_- \sim \CF(\phi)$, and demanding compatibility of the susy transformations of a chiral and antichiral multiplet (each with Weyl weight $w = 2$) with those of a vectormultiplet, and using the homogeneity relations,
 \eqa{
   &2\ov{\CF} &&= \ov{\CF}_{I}\lambda^{I} ~, \qquad &\ov{\CF}_{IJ}\lambda^{J} &= \ov{\CF}_{I} ~, \qquad &\ov{\CF}_{IJK}\lambda^{K} &= 0 ~,\\
   &2\CF &&= \CF_{I}\phi^{I} ~, \qquad &\CF_{IJ}\phi^{J} &= \CF_{I} ~, \qquad &\CF_{IJK}\phi^{K} &= 0 ~,
 }
 we can build a $w=2$ chiral and antichiral multiplet out of the prepotential like so
 \eqa{
 &\label{eq:chiral-to-vm Aplus}\sfA_+ &&= \frac{1}{4}\ov{\CF}(\lambda) ~,\\
 &\label{eq:chiral-to-vm Gmunu}\sfG_{\mu\nu} &&= \frac{1}{4\sq}\ov{\CF}_{I}\chi_{\mu\nu}{}^{I} ~,\\
 &\label{eq:chiral-to-vm G}\sfG &&= -\frac{1}{\sq}\ov{\CF}_I\eta^I ~,\\
 &\label{eq:chiral-to-vm Bplus}\sfB_{+\mu\nu} &&= -\frac{i}{2}\ov{\CF}_{I}D_{\mu\nu}{}^{I} + \frac{i}{8}\ov{\CF}_{IJ}\left(\chi_{\mu}{}^{\rho,I}\chi_{\nu\rho}{}^{J} - 2\,\chi_{\mu\nu}{}^{I}\eta^{J}\right) ~,\\
 &\label{eq:chiral-to-vm Fplus}\sfF^{+}_{\mu\nu} &&= -\frac{i}{2}\ov{\CF}_{I}\wh{F}^{+}_{\mu\nu}{}^{I} + \frac{i}{8}\ov{\CF}_{IJ}\left(\chi_{\mu}{}^{\rho,I}\chi_{\nu\rho}{}^{J} + 2\,\chi_{\mu\nu}{}^{I}\eta^{J}\right) ~,\\
 &\label{eq:chiral-to-vm Lmunu}\sfL_{\mu\nu} &&= i\sq \ov{\CF}_{I}\big(\scd_{[\mu}\psi_{\nu]}{}^{I}\big)^{+} + \frac{i}{2\sq}\ov{\CF}_{IJ}\left( \big(D_{[\mu}{}^{\rho,I} + \wh{F}_{[\mu}{}^{\rho,+I}\big)\chi_{\nu]\rho}{}^{I} - \big(\wh{F}_{\mu\nu}^{+}{}^{I} - D_{\mu\nu}{}^{I}\big)\eta^{J}\right) \nn
 & &&\quad - \frac{i}{\sq}\ov{\CF}_{I}[\phi, \chi_{\mu\nu}]^{I} - \frac{i}{4\sq}\ov{\CF}_{IJK}\chi_{\mu}{}^{\rho,I}\chi_{\nu\rho}{}^{J}\eta^{K} ~,\\
 &\label{eq:chiral-to-vm L}\sfL &&= -i\sq\,\ov{\CF}_{I}\scd_{\mu}\psi^{\mu,I} + \frac{i}{2\sq}\ov{\CF}_{IJ}\big(\wh{F}_{\mu\nu}^{+}{}^{I} + D_{\mu\nu}{}^{I}\big)\chi^{\mu\nu,J} - \frac{1}{12\sq}\ov{\CF}_{IJK}\chi_{\mu}{}^{\rho}{}^{I}\chi^{\mu\sigma}{}^{J}\chi_{\rho\sigma}{}^{K} \nn
 & &&\quad - i\sq\ov{\CF}_{I}[\phi, \eta]^{I} ~,\\
 &\label{eq:chiral-to-vm Cplus}\sfC_{+} &&= \ov{\CF}_{I}\left(2\,\bm{\Box}_{C}\phi^{I} + \big(\wh{F}^{-}_{\mu\nu}{}^{I}T^{\mu\nu-} + \lambda^{I}T^{-}_{\mu\nu}T^{\mu\nu-}\big) - 3\,\Xi^{\rho}\psi_{\rho}{}^{I} \right) \nn
 & &&\quad - 2\,\ov{\CF}_{I}\left[\phi,\left[\phi,\lambda\right]\right]^{I} - 2\,\ov{\CF}_{I}[\psi_{\mu},\psi^{\mu}]^{I} - 2\,\ov{\CF}_{IJ}\eta^{I}[\phi,\eta]^{J} - \frac{1}{2}\ov{\CF}_{IJ}[\phi,\chi_{\mu\nu}]^{I}\chi^{\mu\nu,J} \nn
 & &&\quad + \frac{1}{2}\ov{\CF}_{IJ}\left( \wh{F}_{\mu\nu}^{+}{}^{I}\wh{F}^{\mu\nu,+J} - D_{\mu\nu}{}^{I}D^{\mu\nu,J} + 4\big(\scd_{[\mu}\psi_{\nu]}{}^{I}\big)^{+}\chi^{\mu\nu,J} + 4\big(\scd_{\mu}\psi^{\mu,I}\big)\eta^{J} \right)\nn
 & &&\quad - \frac{i}{8}\ov{\CF}_{IJK}\left( \big(\wh{F}^{\mu\nu,+I} - D^{\mu\nu,I}\big)\chi_{\mu}{}^{\rho,I}\chi_{\nu\rho}{}^{K} + 2\big(\wh{F}^{\mu\nu,+I} + D^{\mu\nu,I}\big)\chi_{\mu\nu}{}^{J}\eta^{K}\right) \nn
 & &&\quad + \frac{1}{12}\ov{\CF}_{IJKL}\chi_{\mu}{}^{\rho}{}^{I}\chi^{\mu\sigma}{}^{J}\chi_{\rho\sigma}{}^{K}\eta^{L} ~.
 }
 
 \eqa{
  &\label{eq:antichiral-to-vm Aminus}\sfA_- &&= -4\,\CF(\phi) ~,\\
  &\label{eq:antichiral-to-vm Gmu}\sfG_{\mu} &&= 4i\,\CF_{I}\psi_{\mu}{}^{I} ~,\\
  &\label{eq:antichiral-to-vm Bminus}\sfB_{-\mu\nu} &&= 2i\,\CF_{I}D_{\mu\nu}{}^{I} - i\,\CF_{IJ}\big(\psi_{\mu}{}^{I}\psi_{\nu}{}^{J}\big)^{+} ~,\\
  &\label{eq:antichiral-to-vm Fminus}\sfF^{-}_{\mu\nu} &&= 2i\,\CF_{I}\big(\wh{F}_{\mu\nu}^{-}{}^{I} + \lambda^{I}T_{\mu\nu}^{-}\big) - i\, \CF_{IJ}\big(\psi_{\mu}{}^{I}\psi_{\nu}{}^{J}\big)^{-} ~,\\
  &\label{eq:antichiral-to-vm Lmu}\sfL_{\mu} &&= 2\,\CF_{I}\scd_{\rho}\chi_{\mu}{}^{\rho,I} - 2\,\CF_{I}\scd_{\mu}\eta^{I} - 2\,\CF_{IJ}\big(\wh{F}_{\mu\rho}^{-}{}^{I} - D_{\mu\rho}{}^{I}\big)\psi^{\rho,I} - \frac{4}{3}\CF_{IJK}\big(\psi_{\mu}{}^{I}\psi_{\rho}{}^{J}\big)^{+}\psi^{\rho,K} \nn
  & &&\quad -4\,\CF_{I}[\lambda, \psi_{\mu}]^{I} ~,\\
  &\label{eq:antichiral-to-vm Cminus}\sfC_{-} &&= -2\,\CF_{I}\bm{\Box}_{C}\lambda^{I} + 2\,\CF_{I}\left[\lambda,\left[\phi, \lambda\right]\right]^{I} + \frac{1}{2}\CF_{IJ}[\phi,\chi^{\mu\nu}]^{I}\chi^{\mu\nu}{}^{J} + 2\,\CF_{I}[\eta,\eta]^{I} + 2\,\CF_{IJ}\psi_{\mu}{}^{I}[\psi^{\mu},\lambda]^{J} \nn
  & &&\quad - \frac{1}{2}\CF_{IJ}\left( \big(\wh{F}^{-}_{\mu\nu}{}^{I} + \lambda^{I}T_{\mu\nu}^{-}\big)\big(\wh{F}^{\mu\nu,-J} + \lambda^{J}T^{\mu\nu-}\big) - D_{\mu\nu}{}^{I}D_{\mu\nu}{}^{J} - 4\big(\scd_{\rho}\chi_{\sigma}{}^{\rho,I}\big)\psi^{\sigma,J} + 4\big(\scd_{\rho}\eta^{I}\big)\psi^{\rho,J}\right) \nn
  & &&\quad + \CF_{IJK}\big(\wh{F}^{\mu\nu,-I} + \lambda^{I}T^{\mu\nu-} - D^{\mu\nu,I}\big)\psi_{\mu}{}^{J}\psi_{\nu}{}^{K} + \frac{1}{12}e^{-1}\CF_{IJKL}\veps^{\mu\nu\rho\sigma}\psi_{\mu}{}^{I}\psi_{\nu}{}^{J}\psi_{\rho}{}^{K}\psi_{\sigma}{}^{L} ~.
 }
Again, these expressions should be used with $T_{\mu\nu}^{-}$ substituted from \eqref{eq:sym-grav-T} (which leads to \eqref{eq:Fhat minus lambda-Tminus}). It is worth remarking that the 1-form $S$-susy connection $S_{\mu}$ contributes neither to $\sfC_{-}$ nor to $\sfL_{\mu}$\footnote{The $S_{\mu}$ dependence cancels between the supercovariant derivatives that enter these expressions. In particular, $\scd_{\rho}\chi_{\sigma}{}^{\rho} = D_{\rho}\chi_{\sigma}{}^{\rho} + \frac{1}{2}\Psi_{\sigma}{}^{\rho}D_{\rho}\lambda - \frac{1}{2}\Psi_{\rho}{}^{\rho}D_{\sigma}\lambda + \frac{1}{2\sq}S_{\sigma}\lambda$, and $\scd_{\sigma}\eta = D_{\sigma}\eta - \frac{1}{2}\Psi_{\sigma}{}^{\rho}D_{\rho}\lambda + \frac{1}{2\sq}S_{\sigma}\lambda$.}. This is nicely consistent with the fact that the 1-form $S$-susy parameter is constant (a fact that allowed us to set it to zero for making contact with the Cartan Model). The superconformal d'Alembertians in \eqref{eq:chiral-to-vm Cplus} and \eqref{eq:antichiral-to-vm Cminus} are derived in Appendix \ref{app:misc-expr-twisted} and have the following form:
\eqa{
 &\bm{\Box}_{C}\phi &&=D_{\mu}D^{\mu}\phi + \frac{1}{2}\big(\n_{\mu}\Psi^{\mu\nu}\big)\psi_{\nu} + \Psi^{\mu\nu}\big(D_{\mu}\psi_{\nu}) + \frac{1}{2}\Psi_{\rho}{}^{\rho}[\eta,\phi] + \frac{1}{4}\Psi_{\mu}{}^{\rho}\Psi_{\nu\rho}\big(F^{\mu\nu-} + D^{\mu\nu}\big) \nonumber\\
& &&\quad + \frac{1}{4}\Psi_{\mu}{}^{\rho}\Psi_{\nu\rho}\big(\Psi^{\sigma[\mu}\chi^{\nu]}{}_{\sigma}\big)^{-} + \frac{1}{2}\Psi_{\mu}{}^{\rho}\Psi_{\nu\rho}\big(\n^{[\mu}\Phi^{\nu]})^{-}\lambda ~, \label{eq:box phi}\\
& \bm{\Box}_{C}\lambda &&= D_{\mu}D^{\mu}\lambda + \frac{1}{2}\Psi_{\rho}{}^{\rho}[\eta, \lambda] ~. \label{eq:box lambda}
}
Note that the map \eqref{eq:chiral-to-vm Aplus}--\eqref{eq:antichiral-to-vm Cminus} endows the fields of the chiral and antichiral multiplets a homological degree grading, which is induced from the homological degree grading of the vectormultiplet. These homological degrees are listed in Table \ref{tbl:ghost number chiral antichiral}.
 \begin{small}
 \begin{table}[!htb]
	\centering
	\begin{tabular}[t]{|c|c|} \hline
		   field & homological degree \\ \hline\hline
		    $\sfA_+$ & $-4$ \\ \hline
		    $\sfG_{\mu\nu}$ & $-3$ \\ \hline
		    $\sfG$ & $-3$ \\ \hline
		    $\sfB_{+\mu\nu}$ & $-2$ \\ \hline
		    $\sfF_{\mu\nu}^{+}$ & $-2$ \\ \hline
		    $\sfL_{\mu\nu}$ & $-1$ \\ \hline
		    $\sfL$ & $-1$ \\ \hline
		    $\sfC_+$ & $0$ \\ \hline
	\end{tabular} \qquad 
   	\begin{tabular}[t]{|c|c|} \hline
   	field & homological degree \\ \hline\hline
   	$\sfA_-$ & $4$ \\ \hline
   	$\sfG_{\mu}$ & $3$ \\ \hline
   	$\sfB_{-\mu\nu}$ & $2$ \\ \hline
   	$\sfF_{\mu\nu}^{-}$ & $2$ \\ \hline
   	$\sfL_{\mu}$ & $1$ \\ \hline
   	$\sfC_-$ & $0$ \\ \hline
   \end{tabular}
  \caption{\label{tbl:ghost number chiral antichiral}Homological degrees of the $w=2$ twisted chiral (left) and antichiral (right) multiplets.}
\end{table}
\end{small}

We now have all the ingredients to write down the action of topologically twisted $\CN=2$ SYM coupled to twisted supergravity. Such an action should satisfy the following requirements:
\begin{enumerate}\itemsep -3pt  
\item It should be $\IQ$-closed, and should have zero homological degree.
\item It should be diffeomorphism invariant and gauge invariant. 
\item At gravity degree $0$ (or equivalently, when all sugra fields except the metric are turned off), it should reduce to the known actions of \cite{Witten:1988ze} in the UV, and of \cite{Moore:1997pc,Marino:1998bm} in the IR.
\end{enumerate}
An action for a cohomological QFT is determined only up to to $\IQ$-exact pieces. It is clear that the antichiral multiplet always contributes precisely $e\sfC_-$ (everything else is killed by the truncation and twist), whereas the third requirement also forces the minimal $\IQ$-exact part to be proportional to $\IQ(e\Lambda)$, as this is all that is left at degree $=0$ anyway. Therefore, motivated by the above considerations (and footnote \ref{foot:motivateaction}), we propose the following \underline{minimal} action:\footnote{The factor of $\frac{i}{16\pi}$ is useful for making contact with \cite{Moore:1997pc}. See Appendix \ref{app:conventions}.}
\beqa{
& \mathbb{S}_{\text{sugra}} &&= \IQ \IV_{\text{sugra}} + \sfC_{\text{sugra}} := \frac{i}{16\pi}\int_{\IX} d^{4}x\left(\IQ\Upsilon + e\sfC_-\right) = \frac{i}{16\pi}\int_{\IX} d^{4}x\,\mathscr{L} ~, \label{eq:TwistedSugraAction-General-Proposal}
}
where we defined $\Upsilon := -ie\sqrt{2}\,\sfL$, $\IV_{\text{sugra}} := \frac{i}{16\pi}\int_{\IX} d^{4}x\Upsilon$, and $\sfC_{\text{sugra}} := \frac{i}{16\pi}\int_{\IX}\sqrt{g}\,\sfC_{-}$, so that\footnote{In the following, we have used \eqref{eq:tsugra-supercov-ym-curvature}, the homogeneity relation $\ov{\CF}_{IJ}\lambda^{I} = 2\ov{\CF}_{J}$, and $\scd_{\mu}\psi^{\mu} = D_{\mu}\psi^{\mu} + \frac{1}{2}\Psi_{\mu}{}^{\mu}[\lambda,\phi]$.}
\eqa{
 & \IV_{\text{sugra}} 
  &&= \frac{i}{16\pi}\int_{\IX} d^{4}x\sqrt{g}\left( -2\,\ov{\CF}_{I}D_{\mu}\psi^{\mu}{}^{I} + \frac{1}{2}\ov{\CF}_{IJ}\big(F_{\mu\nu}^{+}{}^{I} + D_{\mu\nu}{}^{I}\big)\chi^{\mu\nu}{}^{J} + \frac{i}{12}\ov{\CF}_{IJK}\chi_{\mu}{}^{\rho}{}^{I}\chi^{\mu\sigma}{}^{J}\chi_{\rho\sigma}{}^{K}\right. \nn 
 & &&\qquad\qquad \left. + 2\,\ov{\CF}_{I}[\eta, \phi]^{I} - \Psi_{\rho}{}^{\rho}\ov{\CF}_{I}[\lambda,\phi]^{I} + \frac{1}{2}\Psi_{\mu}{}^{\rho}\ov{\CF}_{IJ}\chi_{\nu\rho}{}^{I}\chi^{\mu\nu}{}^{J} + \frac{1}{2}\Psi_{\mu}{}^{\rho}\Psi_{\nu\rho}\ov{\CF}_{I}\chi^{\mu\nu}{}^{I}\right)\label{eq:TwistedSugraUpsilon} ~,
}
and $\sfC_-$ is given by \eqref{eq:antichiral-to-vm Cminus}, which we can split by gravity degree as follows:
  \eqa{
  &\left.\sfC_{-}\right|_{\text{deg}\,=\,0} &&= -2\,\CF_{I}D_{\mu}D^{\mu}\lambda^{I} + 2\,\CF_{I}\left[\lambda,\left[\phi, \lambda\right]\right]^{I} + \frac{1}{2}\CF_{IJ}[\phi,\chi_{\mu\nu}]^{I}\chi^{\mu\nu}{}^{J} + 2\,\CF_{I}[\eta,\eta]^{I} + 2\,\CF_{IJ}\psi_{\mu}{}^{I}[\psi^{\mu},\lambda]^{J}\nn
  & &&\quad - \frac{1}{2}\CF_{IJ}\left( F^{-}_{\mu\nu}{}^{I}F^{\mu\nu,-J} - D_{\mu\nu}{}^{I}D_{\mu\nu}{}^{J} - 4\big(D_{\rho}\chi_{\sigma}{}^{\rho,I}\big)\psi^{\sigma,J} + 4\big(D_{\rho}\eta^{I}\big)\psi^{\rho,J}\right) \nn
  & &&\quad + \CF_{IJK}\big(F^{\mu\nu,-I} - D^{\mu\nu,I}\big)\psi_{\mu}{}^{J}\psi_{\nu}{}^{K} + \frac{1}{12}e^{-1}\CF_{IJKL}\veps^{\mu\nu\rho\sigma}\psi_{\mu}{}^{I}\psi_{\nu}{}^{J}\psi_{\rho}{}^{K}\psi_{\sigma}{}^{L} ~,\label{eq:Cminus deg 0}\\
  &\left.\sfC_{-}\right|_{\text{deg}\,=\,1} &&= -\CF_{I}\Psi_{\sigma}{}^{\sigma}[\eta,\lambda]^{I} - \CF_{IJ}F_{\mu\nu}^{-}{}^{I}\big(\Psi^{\rho[\mu}\chi^{\nu]}{}_{\rho}{}^{J}\big)^{-} + 2\,\CF_{IJ}\Psi_{\sigma}{}^{\rho}\big(D_{\rho}\lambda^{I}\big)\psi^{\sigma,J}\nn
  & &&\quad - \CF_{IJ}\Psi_{\rho}{}^{\rho}\big(D_{\sigma}\lambda^{I}\big)\psi^{\sigma,J} + \CF_{IJK}\big(\Psi^{\rho[\mu}\chi^{\nu]}{}_{\rho}{}^{I}\big)^{-}\psi_{\mu}{}^{J}\psi_{\nu}{}^{K} ~,\label{eq:Cminus deg 1} \\
  &\left.\sfC_{-}\right|_{\text{deg}\,=\,2} &&= -2\,\big(\n^{[\mu}\Phi^{\nu]}\big)^{-}\,\CF_{IJ}F_{\mu\nu}^{-}{}^{I}\lambda^{J} - \frac{1}{2}\CF_{IJ}\big(\Psi_{[\mu}{}^{\rho}\chi_{\nu]\rho}{}^{I}\big)^{-}\big(\Psi^{\sigma[\mu}\chi^{\nu]}{}_{\sigma}{}^{J}\big)^{-} \nn
  & &&\quad + 2\,\big(\n^{[\mu}\Phi^{\nu]}\big)^{-}\CF_{IJK}\lambda^{I}\psi_{\mu}{}^{J}\psi_{\nu}{}^{K} ~,\label{eq:Cminus deg 2}\\
  &\left.\sfC_{-}\right|_{\text{deg}\,=\,3} &&= -2\,\big(\n^{[\mu}\Phi^{\nu]}\big)^{-}\CF_{IJ}\lambda^{I}\big(\Psi_{[\mu}{}^{\rho}\chi_{\nu]\rho}{}^{J}\big)^{-} ~,\label{eq:Cminus deg 3}\\
  &\left.\sfC_{-}\right|_{\text{deg}\,=\,4} &&= -2\,\big[\big(\n_{[\mu}\Phi_{\nu]}\big)^{-}\big]^{2}\CF_{IJ}\lambda^{I}\lambda^{J} ~. \label{eq:Cminus deg 4}
}
Note that $\IQ$ increases the homological degree by $1$; $\Upsilon$ and $\sfC_-$ have homological degrees $-1$ and $0$ respectively.\footnote{The homological degrees of derivatives of the prepotentials are $\text{gh \#}(\ov{\CF}, \ov{\CF}_I, \ov{\CF}_{IJ},\ov{\CF}_{IJK}, \ov{\CF}_{IJKL}) = (-4, -2, 0, +2, +4)$ and $\text{gh \#}(\CF, \CF_I, \CF_{IJ}, \CF_{IJK}, \CF_{IJKL}) = (+4, +2, 0, -2, -4)$.} Furthermore, note that $\Upsilon$ is a scalar density on which $\IQ^2 = \CL_{\Phi}$ evaluates to a total derivative, whereas $\IQ(e\sfC_-)$ is a total derivative. So, on a  compact 4-manifold without boundary  $\IQ\mathbb{S}_{\text{sugra}} = 0$. Note that $\sfC_{\text{sugra}}$ is not $\IQ$-exact.
It is clear that $\IV_{\text{sugra}}$ in the IR has the same form as the primitive of \cite{Moore:1997pc} or \cite{Marino:1998bm}, as expected. In the UV, the action reproduces the Donaldson-Witten action \cite{Witten:1988xi,Labastida:2005zz}. 

The Lagrangian density $\mathscr{L}$ defined in \eqref{eq:TwistedSugraAction-General-Proposal} can be expanded in gravity degree, and we list below its components up to gravity degree $=4$, where the expansion terminates. As before, we define $\mathsf{Im\,}\tau_{IJ} := \frac{\CF_{IJ} - \ov{\CF}_{IJ}}{2i}$, where $\tau_{IJ} := \CF_{IJ}$ and $\ov{\tau}_{IJ} := \ov{\CF}_{IJ}$.

 \eqa{
 &\left.\mathscr{L}\right|_{\text{deg}\,=\,0} &&= \frac{e}{2}\big(\ov{\tau}_{IJ}F_{\mu\nu}^{+}{}^{I}F^{\mu\nu+}{}^{J} - \tau_{IJ}F_{\mu\nu}^{-}{}^{I} F^{\mu\nu-}{}^{J}\big) + i\,e\big(\mathsf{Im\,}\tau_{IJ}\big)D_{\mu\nu}{}^{I}D^{\mu\nu,J}  \nn
 & &&\quad +2\,e\,\ov{\CF}_{I}D_{\mu}D^{\mu}\phi^{I} - 2\,e\,\CF_{I}D_{\mu}D^{\mu}\lambda^{I}  - 2\,e\,\tau_{IJ}(D_{\sigma}\eta^{I}\big)\psi^{\sigma}{}^{J} - 2\,e\,\ov{\tau}_{IJ}\eta^{I}D_{\sigma}\psi^{\sigma,J} \nn
  & &&\quad + 2\,e\,\tau_{IJ}\big(D_{\rho}\chi_{\sigma}{}^{\rho}{}^{I}\big)\psi^{\sigma}{}^{J} + 2\,e\,\ov{\tau}_{IJ}\big(D_{[\mu}\psi_{\nu]}{}^{I}\big)\chi^{\mu\nu}{}^{J}   \nn
  & &&\quad + \frac{e}{2}\ov{\CF}_{IJK}\eta^{I}\big(F_{\mu\nu}^{+}{}^{J} + D_{\mu\nu}{}^{J}\big)\chi^{\mu\nu}{}^{K} + e\,\CF_{IJK}\,\big(F^{\mu\nu-}{}^{I} - D^{\mu\nu}{}^{I}\big)\psi_{\mu}{}^{J}\psi_{\nu}{}^{K} \nn 
  & &&\quad + \frac{1}{12}\CF_{IJKL}\veps^{\mu\nu\rho\sigma}\psi_{\mu}{}^{I}\psi_{\nu}{}^{J}\psi_{\rho}{}^{K}\psi_{\sigma}{}^{L} \nn
  & &&\quad +\frac{i}{12}\sqrt{g}\,\ov{\CF}_{IJKL}\eta^{I}\chi_{\mu}{}^{\rho}{}^{J}\chi^{\mu\sigma}{}^{K}\chi_{\rho\sigma}{}^{L}  + \frac{i}{4}\sqrt{g}\,\ov{\CF}_{IJK}\big(F_{\mu\rho}^{+}{}^{I} - D_{\mu\rho}{}^{I}\big)\chi^{\mu\sigma}{}^{J}\chi^{\rho}{}_{\sigma}{}^{K} \nn
& &&\quad  - \frac{e}{2}\ov{\tau}_{IJ}[\phi, \chi_{\mu\nu}]^{I}\chi^{\mu\nu,J} + \frac{e}{2}\tau_{IJ}[\phi,\chi_{\mu\nu}]^{I}\chi^{\mu\nu}{}^{J} - 2\,e\,\ov{\CF}_{I}[\psi_{\mu}, \psi^{\mu}]^{I} + 2\,\tau_{IJ}\psi_{\mu}{}^{I}[\psi^{\mu},\lambda]^{J} \nn
 & &&\quad + 2\,e\,\CF_{I}[\eta,\eta]^{I} + 2\,e\,\ov{\tau}_{IJ}\eta^{I}[\eta, \phi]^{J}  + 2\,e\,\CF_{I}\big[\lambda, [\phi, \lambda]\big]^{I}  - 2\,e\,\ov{\CF}_{I}\big[\phi, [\phi, \lambda]\big]^{I} ~. \label{eq:Lsugra-deg0-rescaled}
 }

 \eqa{
  &\left.\mathscr{L}\right|_{\text{deg}\,=\,1} &&= -e\,\Psi_{\sigma}{}^{\sigma}\ov{\CF}_{I}D_{\mu}\psi^{\mu,I} + 2\,e\,\ov{\CF}_{I}\Psi^{\mu\nu}D_{\mu}\psi_{\nu}{}^{I} + 2\,e\,\big(\n_{\mu}\Psi^{\mu\nu}\big)\ov{\CF}_{I}\psi_{\nu}{}^{I} - e\,\big(\n^{\rho}\Psi_{\sigma}{}^{\sigma}\big)\ov{\CF}_{I}\psi_{\rho}{}^{I} \nn
& &&\quad +\frac{e}{2}\Psi_{\sigma}{}^{\sigma}\ov{\tau}_{IJ}\big(F_{\mu\nu}^{+}{}^{I} + D_{\mu\nu}{}^{I}\big)\chi^{\mu\nu,J} + e\,\Psi_{\mu}{}^{\rho}\ov{\tau}_{IJ}F^{-}_{\nu\rho}{}^{I}\chi^{\mu\nu,J} \nn
& &&\quad  + \left.\IQ\left(\frac{ie}{12}\ov{\CF}_{IJK}\chi_{\mu}{}^{\rho,I}\chi^{\mu\sigma,J}\chi_{\rho\sigma}{}^{K}\right)\right|_{\text{deg}\,=\,1} \nn
& &&\quad + 2\,e\,\Psi_{\sigma}{}^{\sigma}\ov{\CF}_{I}[\eta, \phi]^{I} + e\,\Psi_{\sigma}{}^{\sigma}\ov{\tau}_{IJ}\eta^{I}[\lambda, \phi]^{J} - \frac{e}{2}\Psi_{\mu}{}^{\rho}\ov{\CF}_{IJK}\eta^{I}\chi_{\nu\rho}{}^{J}\chi^{\mu\nu,K} \nn
& && \quad -e\,\Psi_{\sigma}{}^{\sigma}\CF_{I}[\eta,\lambda]^{I} - e \tau_{IJ}F^{-}_{\mu\nu}{}^{I}\big(\Psi^{\rho[\mu}\chi^{\nu]}{}_{\rho}{}^{J}\big)^{-} + 2\,e\,\tau_{IJ}\Psi_{\sigma}{}^{\rho}\big(D_{\rho}\lambda^{I}\big)\psi^{\sigma,J} - e\,\tau_{IJ}\Psi_{\rho}{}^{\rho}\big(D_{\sigma}\lambda^{I}\big)\psi^{\sigma,J} \nn
& &&\quad + \CF_{IJK}\big(\Psi^{\rho[\mu}\chi^{\nu]}{}_{\rho}{}^{I}\big)^{-}\psi_{\mu}{}^{J}\psi_{\nu}{}^{K} ~. \label{eq:Lsugra-deg1-rescaled}
 }

   \eqa{
  &\left.\mathscr{L}\right|_{\text{deg}\,=\,2} &&= + 2\,e\,\big(\n^{[\mu}\Phi^{\nu]}\big)\ov{\CF}_{I}\left(F^{-}_{\mu\nu}{}^{I} + D_{\mu\nu}{}^{I}\right) + 2\,e\,\Phi^{\nu}\ov{\CF}_{I} D^{\mu}\left(F^{-}_{\mu\nu}{}^{I} + D_{\mu\nu}{}^{I}\right) \nn
 & &&\quad  + 4\,e\,\Phi^{\mu}\ov{\CF}_{I}[D_{\mu}\lambda,\phi]^{I} + 2\,e\,\Phi^{\mu}\ov{\CF}_{I}[\lambda, D_{\mu}\phi]^{I}  - 2\,e\,\Phi^{\rho}\ov{\CF}_{I}[\chi_{\rho\mu}, \psi^{\mu}]^{I} + 4\,e\,\Phi_{\mu}\ov{\CF}_{I}[\eta, \psi^{\mu}]^{I} \nn
 & &&\quad - e\,\ov{\tau}_{IJ}\big[\big(\n_{[\mu}\Phi^{\rho}\big)\chi_{\nu]\rho}{}^{I}\big]^{+}\chi^{\mu\nu,J} -  e\,\ov{\tau}_{IJ}\Phi^{\rho}\big(D_{[\mu}\chi_{\nu]\rho}{}^{I}\big)^{+}\chi^{\mu\nu,J} - e\,\ov{\tau}_{IJ} \big(\n_{[\mu}\Phi_{\nu]}\big)^{+}\eta{}^{I}\chi^{\mu\nu,J} \nn
& &&\quad + e\,\ov{\tau}_{IJ}\big(\Phi_{[\mu}D^{\rho}\chi_{\nu]\rho}{}^{I}\big)^{+}\chi^{\mu\nu,J} - \frac{e}{4}\ov{\tau}_{IJ}\big[\Psi_{\mu}{}^{\rho}\big(\Psi_{[\rho}{}^{\sigma}\chi_{\nu]\sigma}{}^{I}\big)^{-} - \Psi_{\nu}{}^{\rho}\big(\Psi_{[\rho}{}^{\sigma}\chi_{\mu]\sigma}{}^{I}\big)^{-}\big]^{+}\chi^{\mu\nu,J} \nn
& &&\quad + 2\,e\,\ov{\tau}_{IJ}\big(\Phi_{[\mu}[\lambda,\psi_{\nu]}]^{I}\big)^{+}\chi^{\mu\nu,J} - 2\,e\,\Phi^{\mu}\ov{\tau}_{IJ}\big(F_{\mu\nu}^{+}{}^{I} + D_{\mu\nu}{}^{I}\big)D^{\nu}\lambda^{J}\nn
& &&\quad  + e\,\big(\n^{\mu}\Phi^{\nu}\big)\,\ov{\tau}_{IJ}(F_{\mu\nu}^{+}{}^{I} + D_{\mu\nu}{}^{I}\big)\lambda^{J} +  \left.\IQ\left(\frac{ie}{12}\ov{\CF}_{IJK}\chi_{\mu}{}^{\rho,I}\chi^{\mu\sigma,J}\chi_{\rho\sigma}{}^{K}\right)\right|_{\text{deg}\,=\,2}  \nn
& &&\quad  + \frac{e}{2}\big(\Psi_{\mu\nu}\Psi^{\nu\rho} + \n_{\mu}\Phi^{\rho} + \n^{\rho}\Phi_{\mu}\big)\ov{\tau}_{IJ}\chi_{\nu\rho}{}^{I}\chi^{\mu\nu,J} \nn
   & &&\quad + e\,\Psi_{\mu}{}^{\rho}\ov{\tau}_{IJ}\big(\Psi_{[\nu}{}^{\sigma}\chi_{\rho]\sigma}{}^{I}\big)^{-}\chi^{\mu\nu,J}  \nn
   & &&\quad + \frac{e}{2}\Psi_{\mu}{}^{\rho}\Psi_{\nu\rho}\ov{\tau}_{IJ}\eta^{I}\chi^{\mu\nu,J} + \frac{e}{2}\Psi_{\mu}{}^{\rho}\Psi_{\nu\rho}\ov{\CF}_{I}\big(F^{\mu\nu,+I} - D^{\mu\nu,I}\big) \nn
   & &&\quad - 2\,e\,\big(\n^{[\mu}\Phi^{\nu]}\big)^{-}\tau_{IJ}F_{\mu\nu}^{-}{}^{I}\lambda^{J} - \frac{e}{2}\tau_{IJ}\big(\Psi_{[\mu}{}^{\rho}\chi_{\nu]\rho}{}^{I}\big)^{-}\big(\Psi^{\sigma[\mu}\chi^{\nu]}{}_{\sigma}{}^{J}\big)^{-} \nn
   & &&\quad + 2\,e\,\big(\n^{[\mu}\Phi^{\nu]}\big)^{-}\CF_{IJK}\lambda^{I}\psi_{\mu}{}^{J}\psi_{\nu}{}^{K} ~. \label{eq:Lsugra-deg2-rescaled}
  }

   \eqa{
  &\left.\mathscr{L}\right|_{\text{deg}\,=\,3} &&=  + 2\,e\,\big(\n^{[\mu}\Phi^{\nu]}\big)\ov{\CF}_{I}\big(\Psi_{[\mu}{}^{\sigma}\chi_{\nu]\sigma}{}^{I}\big)^{-} + 2\,e\,\Phi^{\nu}\ov{\CF}_{I}D^{\mu}\big(\Psi_{[\mu}{}^{\sigma}\chi_{\nu]\sigma}{}^{I}\big)^{-} + 2\,e\,\Phi^{\rho}\Psi_{\mu\rho}\ov{\CF}_{I}[\lambda, \psi^{\mu}]^{I} \nn
  & &&\quad - e\,\ov{\tau}_{IJ}\big[\big(\n_{[\mu}\Phi^{\rho}\big)\Psi_{\nu]\rho}\big]^{+}\lambda^{I}\chi^{\mu\nu,J}  - e\,\ov{\tau}_{IJ}\Phi^{\rho}\big(\n_{[\mu}\Psi_{\nu]\rho}\big)^{+}\lambda^{I}\chi^{\mu\nu,J}   \nn
& &&\quad - e\,\ov{\tau}_{IJ}\Phi^{\rho}\big[\Psi_{\rho[\mu}D_{\nu]}\lambda^{I}\big]^{+}\chi^{\mu\nu,J} - e\,\ov{\tau}_{IJ}\big[\Psi_{\mu}{}^{\rho}\big(\n_{[\rho}\Phi_{\nu]}\big)^{-}\lambda^{I} - \Psi_{\nu}{}^{\rho}\big(\n_{[\rho}\Phi_{\mu]}\big)^{-}\lambda^{I}\big]^{+}\chi^{\mu\nu,J} \nn
& &&\quad + e\,\ov{\tau}_{IJ}\big(\Phi_{[\mu}\Psi_{\nu]}{}^{\rho}\big)^{+}\big(D_{\rho}\lambda^{I}\big)\chi^{\mu\nu,J} - \frac{e}{2} \ov{\tau}_{IJ}\Psi_{\rho}{}^{\rho}\big(\Phi_{[\mu}D_{\nu]}\lambda^{I}\big)^{+} \chi^{\mu\nu,J} \nn
& &&\quad + \left.\IQ\left(\frac{ie}{12}\ov{\CF}_{IJK}\chi_{\mu}{}^{\rho,I}\chi^{\mu\sigma,J}\chi_{\rho\sigma}{}^{K}\right)\right|_{\text{deg}\,=\,3} - e\,\Psi_{\rho}{}^{\rho}\Phi^{\sigma}\ov{\CF}_{I}[\lambda,\psi_{\sigma}]^{I} \nn
   & &&\quad - \frac{e}{2}\Psi_{\mu\sigma}\Psi_{\nu\rho}\Psi^{\sigma\rho}\ov{\CF}_{I}\chi^{\mu\nu,I}  + e\big(\n_{[\mu}\Phi^{\rho} + \n^{\rho}\Phi_{[\mu}\big)\Psi_{\nu]\rho}\ov{\CF}_{I}\chi^{\mu\nu,I} \nn 
   & &&\quad + \frac{e}{2}\Psi_{\mu}{}^{\rho}\Psi_{\nu\rho}\ov{\CF}_{I}\big(\Psi^{\sigma[\mu}\chi^{\nu]}{}_{\sigma}{}^{I}\big)^{-} - 2\,e\,\big(\n^{[\mu}\Phi^{\nu]}\big)^{-}\tau_{IJ}\big(\Psi_{[\mu}{}^{\rho}\chi_{\nu]\rho}{}^{I}\big)^{-}\lambda^{J} ~. \label{eq:Lsugra-deg3-rescaled}
  }

\eqa{
  &\left.\mathscr{L}\right|_{\text{deg}\,=\,4} &&= + 8\,e\,\left[\big(\n_{[\mu}\Phi_{\nu]}\big)^{-}\right]^2\,\ov{\CF} + 8\,e\,\Phi^{\nu}\big(\n^{\mu}\big(\n_{[\mu}\Phi_{\nu]}\big)^{-}\big)\,\ov{\CF} + 4\,e\,\Phi^{\nu}\big(\n_{[\mu}\Phi_{\nu]}\big)^{-}\ov{\CF}_{I}D^{\mu}\lambda^{I} \nn
  & &&\quad + \left.\IQ\left(\frac{ie}{12}\ov{\CF}_{IJK}\chi_{\mu}{}^{\rho,I}\chi^{\mu\sigma,J}\chi_{\rho\sigma}{}^{K}\right)\right|_{\text{deg}\,=\,4} \nn
  & &&\quad - 2\,e\,\Psi_{\mu}{}^{\rho}\Psi_{\nu\rho}\ov{\CF}_{I}\big(\Phi^{[\mu}D^{\nu]}\lambda^{I}\big)^{+} + 2\,e\,\Psi_{\mu}{}^{\rho}\Psi_{\nu\rho}\big(\n^{[\mu}\Phi^{\nu]}\big)^{+}\,\ov{\CF} \nn
  & &&\quad - 2\,e\,\left[\big(\n_{[\mu}\Phi_{\nu]}\big)^{-}\right]^2\tau_{IJ}\lambda^{I}\lambda^{J} ~. \label{eq:Lsugra-deg4-rescaled}
 }

\section{Comparison Of The Actions\label{sec:CompareActions}}

In this section, we compare the minimal action of section \ref{sec:ProposalForTheAction} with the supergravity action of section \ref{sec:TwistedSugraAction}, and show that their difference is the sum of a surface term (the integral of a total derivative) and a $\IQ$-exact term. To compare the actions, they must first be written in the same field parametrization. 
The transformations of section \ref{sec:CartanModels} are simpler than those of section \ref{sec:SuperconformalGrav-twistedSYM}, so we elect to shift the twisted supergravity action to the Cartan Model parametrization. This is carried out using the field redefinitions of section \ref{subsec:Full-Cartan-from-TwistedSugra}, equations 
\eqref{eq:field-redef-1}--\eqref{eq:field-redef-3}.

The minimal action of section \ref{sec:ProposalForTheAction} is
\eqa{
& \IS_{\text{Cartan}} &&= \IQ \IV + \sfC ~,
}
where $\IV$ (cf. \eqref{eq:IV}) depends on both unbarred ($\CF \equiv \CF(\phi)$) and barred ($\ov{\CF} \equiv \ov{\CF}(\lambda)$) prepotentials, and $\sfC$ (cf. \eqref{eq:IC}) depends \textit{only} on the unbarred prepotential.

The supergravity action of section \ref{sec:TwistedSugraAction} is
\eqa{
& \IS_{\text{sugra}} &&= \IQ \IV_{\text{sugra}} + \sfC_{\text{sugra}} ~, 
}
where $\IV_{\text{sugra}}$ (cf. \eqref{eq:TwistedSugraUpsilon}) depends \textit{only} on the barred prepotential, and $\sfC_{\text{sugra}} := \frac{i}{16\pi}\int_{\IX}\sqrt{g}\sfC_{-}$ depends only on the unbarred prepotential (cf. \eqref{eq:Cminus deg 0}--\eqref{eq:Cminus deg 4}). Here it is important to remember is that $\sfC_{\text{sugra}}$ is not $\IQ$-exact. 
%

We shift \eqref{eq:TwistedSugraUpsilon} and \eqref{eq:Cminus deg 0}--\eqref{eq:Cminus deg 4} using \eqref{eq:field-redef-1}--\eqref{eq:field-redef-3}. It is natural to ask if the difference between the (shifted) sugra action and the minimal action is $\IQ$-exact. This difference is
\eqa{
& \IS_{\text{sugra}}^{\text{shifted}} - \IS_{\text{Cartan}}  &&= \underbrace{\IQ\big(\IV_{\text{sugra}}^{\text{shifted}} - \IV^{\text{barred}}\big)}_{\text{barred}} + \underbrace{\sfC_{\text{sugra}}^{\text{shifted}} - \IQ\IV^{\text{unbarred}} - \sfC}_{\text{unbarred}} ~.
}
Clearly, the barred terms contribute only to a $\IQ$-exact difference. Our approach is to focus on the unbarred difference $\big(\sfC_{\text{sugra}}^{\text{shifted}} - \IQ\IV^{\text{unbarred}} - \sfC\big)$, 
and show that this equals a $\IQ$-exact term plus a total derivative. The details of this rather involved computation are relegated to Appendix \ref{app:ActionComparisonDetails}, and the result is
\beqas{
& \IS_{\text{sugra}}^{\text{shifted}} - \IS_{\text{Cartan}}  &&= \IQ\big(\IV_{\text{sugra}}^{\text{shifted}} - \IV^{\text{barred}} + \Delta_{\mathsf{diff}} \big)  + \int_{\IX}d^{4}x\sqrt{g}\,\n_{\mu}\mathscr{P}^{\mu} ~,
}
where
\begin{small}
\beqas{\label{eq:DiffActionFinal}
\mathscr{P}^{\mu} &:= -2\,\CF_{I}D^{\mu}\lambda^{I} + 2\,\CF_{IJ}\psi_{\nu}{}^{I}\chi^{\mu\nu}{}^{J} + \CF_{IJ}\Phi^{\sigma}\chi_{\nu\sigma}{}^{I}\chi^{\mu\nu}{}^{J} ~,\\
  \Delta_{\mathsf{diff}} &:= \int_{\IX}d^{4}x\sqrt{g}\left(2\,\CF_{IJ}\Phi_{\mu}\chi^{\mu\nu}{}^{I}D_{\nu}\lambda{}^{J} + \CF_{IJK}\Phi^{\alpha}\chi_{\alpha\mu}{}^{I}\chi^{\mu\sigma}{}^{J}\psi_{\sigma}{}^{K} + \tfrac{1}{3}\CF_{IJKL}\Phi^{\alpha}\Phi^{\beta}\chi^{\rho\sigma}{}^{I}\chi_{\alpha\rho}{}^{J}\chi_{\beta\sigma}{}^{K}\right) ~.
}
\end{small}

The nontrivial field redefinitions \eqref{eq:field-redef-1}--\eqref{eq:field-redef-3} make it manifest that the higher gravity degree terms in the sugra action become cohomologically uninteresting (being $\IQ$-exact). On a closed 4-manifold, the difference is entirely $\IQ$-exact. 
Naturally, due to the relative simplicity of the transformation laws and the action, we will prefer to use the Cartan parametrization in actual applications of this model.

\section{A First Look At Wall-Crossing}\label{sec:WallCrossing}



Donaldson invariants famously undergo the phenomenon of wall crossing \cite{KotschickMorgan,Gottsche,Moore:1997pc}, a phenomenon closely related to the existence of \textit{reducible connections} \cite{Scorpan:2005,DK,Moore:2012SCGP,Moore:2017SCGP}. We begin by reviewing the basic idea here, before describing the same phenomenon in the setup with smooth families.
Let $\IX$ be a fixed compact, smooth, oriented Riemannian 4-manifold. In  Donaldson theory, one is   interested in gauge theory with structure group $G = \mathsf{SU(2)}$ or $\mathsf{SO(3)}$
%
%
and an associated moduli space $\masd_{g}$ of anti-self-dual (ASD) connections on principal $G$-bundles. 
The invariants are obtained as intersection numbers in $\masd_{g}$ of classes in the image of the   Donaldson map,
\eqa{
& \mu_D : \sfH_{\bullet}(\IX) &\rightarrow \sfH^{4-\bullet}(\masd_{g}) ~. \label{eq:DonaldsonMap}
}
Defining these intersection numbers is nontrivial since the moduli space is singular and noncompact. 
See, for instance, \cite{DK} for details. Since the moduli space is defined using a metric $g$ the intersection numbers will, a priori, depend on the metric. Surprisingly, they are metric-independent for $b_2^+>1$. 
 When  $b_2^+=1$ there is piecewise constant metric dependence. This dependence is related 
to the existence of   line bundles $L \to \IX$ with ASD connections, known as reducible solutions or \textit{Abelian instantons} \cite{DK}. Recall that the space of all self-dual projections of $\mathsf{U(1)}$ curvatures on $L$ forms an \textit{affine} subspace of the space $\Omega^{2,+}(\IX)$ of all self-dual $2$-forms, and a reducible solution exists if and only if this space becomes a \textit{linear} subspace of $\Omega^{2,+}(\IX)$. When such a solution exists, the associated rank two bundle $E$ splits\footnote{This is equivalent to the $\mathsf{SU(2)}$ connection $A$ having the form $A = \mathsf{a} \oplus -\mathsf{a}$ (as a matrix, $A = \mathsf{diag}(\mathsf{a},-\mathsf{a})$) 
where $\mathsf{a}$ denotes a $\mathsf{U(1)}$ connection on $L$.} as $E = L \oplus L^{-1}$. Investigating reducible solutions on $\masd_{g}$ is thus equivalent to studying the existence of such line bundles admitting ASD connections, over conformal classes of metrics.\footnote{The ASD condition is conformally invariant, so we must work with a conformal class of metrics.} If, while varying the metric from $g_1$ to $g_2$, no reducible connections are encountered along the way, then the intersection numbers in the moduli spaces $\masd_{g_1}$ and $\masd_{g_2}$ for the classes defined by the Donaldson map \eqref{eq:DonaldsonMap} will be the same.

Whether or not a generic path of metrics contains a point admitting a line bundle with ASD connection depends on the value of $b_2^{+} = \dim \sfH^{2,+}(\IX)$.  When $b_2^+ \geq 1$, for a generic Riemannian metric, $L$ admits no ASD connections. There are three cases to consider:\footnote{For a detailed proof of these statements about $b_2^+$ and line bundles, see \cite{Taubes82,DK}.}
\begin{itemize} 
\item When $b_2^+ \geq 2$, any two Riemannian metrics $g_0$ and $g_1$ on $M$ (each not admitting ASD connections on $L$) can be connected by a \underline{one-parameter path} in a conformal class of metrics $\{g(t)\}$ (parametrized by $t$) such that no representative $g(t)$ along this path admits ASD connections on $L$. In this case, Donaldson invariants become metric-independent for $b_2^+ > 1$. 
\item When $b_2^+ = 0$, then given \textit{any} Riemannian metric, $L$ necessarily admits ASD connections. In this case, path integral representations of the generating function for  Donaldson invariants exhibit continuous metric dependence.
\item  The key case is when $b_2^+ = 1$, for this is when one can encounter reducible ASD connections over a generic \underline{one-dimensional family} of conformal classes of Riemannian metrics, and along such a family, the Donaldson invariants will jump.
\end{itemize} 
We will extend these remarks to $n$-parameter families of conformal classes of metrics below.

We discuss the piecewise-constant behavior for $b_2^+(\IX) = 1$ in a little more detail. 
In this case the intersection form on $\sfH^2(\IX,\IZ)$ has Lorentzian signature and one can speak 
of the ``light-cone'' in the cohomology space. A conformal class of a metric will define a period 
point $J$, which is self-dual and satisfies $(J,J)=1$. The definition of Donaldson invariants 
involves some choices of orientations, in particular, a ``time-orientation'' on $\sfH^2(\IX,\IZ)$
and we choose $J$ to be in the ``forward light cone.'' The metric dependence of the Donaldson 
invariants enters only through $J$. 
Being self-dual, $J$ depends on the conformal class of the metric. A path of metrics $g(t)$ defines a path of period points $J(t)$ inside the cone $V_{+} := \{ J \in \sfH^{2}(\IX; \IR) : (J, J) > 0 \}$. Along the path 
there will be points where there exists   $\zeta \in \sfH^{2}(\IX; \mathbb{Z})$ such that $(\zeta, J) = 0$, $(\zeta, \zeta) < 0$. 
%
%
%
Such points correspond to  
line bundles admitting an ASD connection for a metric with period point $J$. 
%
%
Turning this around, for each $\zeta$ we can introduce a real codimension one wall in   $V_+$:\footnote{To switch to the notation used in \cite{Moore:1997pc}, note that $\left.J\right|_{\text{here}} = \left.\omega\right|_{\text{there}}$, $\left.\zeta\right|_{\text{here}} = \left.\lambda\right|_{\text{there}}$.}
\eqa{
& W_{\zeta} &&= \{ J : (\zeta, J) = 0 \} ~, \label{eq:real-codim-one-wall}
}
and the connected components of $V_{+} \setminus W_{\zeta}$ are called \textit{chambers}. Donaldson invariants are metric-independent in each chamber and jump discontinuously when $J$ crosses a wall $W_{\zeta}$.  

One can derive the $J$-dependence of the generating function of Donaldson invariants, and in particular, 
the discontinuities across walls $W_{\zeta}$ using path integral techniques \cite{Moore:1997pc} and this 
is the computation we will generalize below.  The discontinuity may be expressed as  
\eqa{
& \mathsf{WC}_{\zeta}(p, \Sigma, \delta') &&:= \mathsf{Z}_{\mathsf{DW},+}^{\zeta}(p, \Sigma, \delta') - \mathsf{Z}_{\mathsf{DW},-}^{\zeta}(p, \Sigma, \delta') ~, \label{eq:wc-discontinuity}
}
where  $p$, $\Sigma$, $\delta'$ denote the representatives of homology classes $\sfH_{0}(\IX)$, $\sfH_{2}(\IX)$ and $\sfH_{1}(\IX)$ corresponding to points, surfaces, and curves respectively.\footnote{Note that $\left.\Sigma\right|_{\text{here}} = \left.S\right|_{\text{ref. \cite{Moore:1997pc}}}$.}) It was conjectured in \cite{KotschickMorgan} that $\mathsf{WC}_{\zeta}(p, S, \delta)$ depends only on the homology ring of $\IX$, and a universal formula for it was found for simply connected $\IX$ in \cite{Gottsche}. The path integral methods of  \cite{Moore:1997pc} confirm the conjecture of \cite{KotschickMorgan} and reproduce the formula obtained in \cite{Gottsche}.

\subsection{Wall-Crossing For Multiparameter Families Of Metrics \label{sec:WC-Multiparameter}}

We now generalize the discussion of the previous section to multiparameter families of metrics. 
The path integral $\mathsf{Z}_{\mathsf{DW}}[g_{\mu\nu}, \Psi_{\mu\nu}, \Phi^{\mu}]$ is equivariantly 
closed and hence -- at least morally -- descends to an inhomogeneous form 
$\overline{\mathsf{Z}_{\mathsf{DW}}[g_{\mu\nu}, \Psi_{\mu\nu}]}$ on $\MET(\IX)/\DIFF(\IX)$. The Donaldson 
``invariants'' are integrals over cycles $\gamma \subset \MET(\IX)/\DIFF(\IX)$
\be\label{eq:DonaldsonPeriod}
\int_{\gamma} \overline{\mathsf{Z}_{\mathsf{DW}}[g_{\mu\nu}, \Psi_{\mu\nu}]} ~,
\ee
but they can only be considered to be ``invariant'' if they only depend on the \underline{homology class} of 
$\gamma$ and not on other details of $\gamma$. Suppose $\gamma_+$ and $\gamma_-$ are two representative $(n-1)$ 
cycles of  $[\gamma]$. Then $\gamma_+ - \gamma_-= \partial \Upsilon$ where $\Upsilon$ is an $n$-chain. Stokes' 
theorem will guarantee that 
\be 
\int_{\gamma_+} \overline{\mathsf{Z}_{\mathsf{DW}}[g_{\mu\nu}, \Psi_{\mu\nu}]}=
\int_{\gamma_-} \overline{\mathsf{Z}_{\mathsf{DW}}[g_{\mu\nu}, \Psi_{\mu\nu}]} ~,
\ee
provided $d \overline{\mathsf{Z}_{\mathsf{DW}}[g_{\mu\nu}, \Psi_{\mu\nu}]}$ has no singularity on $\Upsilon$. 
What we will show, using the Coulomb branch integral of the IR theory is that it in fact \underline{does} have a
delta-function singularity for $n$-chains $\Upsilon$ that contain a metric $g^{(0)}$ admitting a $\mathsf{U(1)}$-line bundle with connection with anti-self-dual (ASD) curvature. 

A simple codimension argument demonstrates that generic $n$-chains $\Upsilon$ will in fact contain a metric $g^{(0)}$ admitting a $\mathsf{U(1)}$-line bundle with ASD connection  precisely when $n=b_2^+(\IX)$. 
%
%
%
%
Recall that $\sfH^{2}(\IX,\IZ)/\mathsf{Tors}\big(\sfH^{2}(\IX,\IZ)\big)$ is a lattice in $\sfH^{2}(\IX,\IR)$ of full dimension. 
For every lattice point $c \in \sfH^{2}(\IX,\IZ)/\mathsf{Tors}\big(\sfH^{2}(\IX,\IZ)\big)$, there exists a $\mathsf{U(1)}$ line bundle  $L \to \IX$ with connection such that $\left[\frac{F}{2\pi}\right] = c$. This connection will be ASD if $c \in \sfH^{2,-}(\IX,\IR)$. 
 Let $\{g(t)\}$ denote an $n$-parameter family of metrics parametrized by an $n$-dimensional space $P$. For each $g(t)$ we therefore have a linear subspace $\sfH^{2,-}(\IX,\IR)(t) \subset \sfH^{2}(\IX,\IR)$. We are seeking a condition on $n$-parameter families such that the  subset
\eqa{
  & \bigcup_{t \in P} \sfH^{2,-}(\IX, \IR)(t) \subset \sfH^{2}(\IX, \IR) ~.
}
will contain a lattice point $c \in \sfH^{2,-}(\IX,\IR)(t)$ for some $t\in P$. 
Now, $\sfH^{2,-}(\IX, \IR)$ is a linear subspace of $\sfH^{2}(\IX,\IR)$ of codimension $b_{2}^{+}$. 
An infinitesimal variation of the metric will vary $\sfH^{2,-}(\IX,\IR)$ in the Grassmannian of subspaces. An infinitesimal variation of $\sfH^{2,-}(\IX,\IR)$ in this Grassmannian corresponds to an  element of $\varphi \in \mathsf{Hom}\big( \sfH^{2,-}(\IX,\IR), \sfH^{2,+}(\IX,\IR)\big)$. The perturbed space will be the graph of $\varphi$. Generically a set of $n$ independent variations of the metric will produce $n$ linearly  independent 
elements of $\mathsf{Hom}\big( \sfH^{2,-}(\IX,\IR), \sfH^{2,+}(\IX,\IR)\big)$, at least for  $n\leq b_2^+ b_2^-$. 
If $n$ is the codimension of $\sfH^{2,-}(\IX,\IR)$ in $\sfH^{2}(\IX,\IR)$, i.e. if $n=b_2^+$ and $b_2^-\geq 1$  the union of the graphs will be an open set in $\sfH^{2}(\IX,\IR)$. A sufficiently large open set will contain a lattice point. 
The same argument shows that when $n < b_{2}^{+}$, the generic $n$-dimensional family of metrics will not include a lattice point. It is for this reason that we expect that \eqref{eq:DonaldsonPeriod} will only depend on the homology class of an $(n-1)$-cycle $\gamma$ for $n < b_{2}^{+}$.

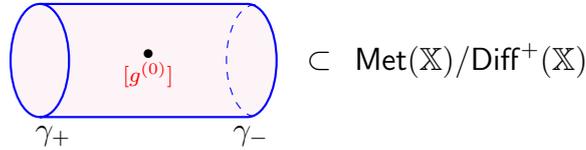
\begin{figure}[H]
\centering
\begin{tikzpicture}
\node[cylinder, cylinder uses custom fill,draw=blue,text = purple,cylinder body fill=magenta!05,cylinder end fill=magenta!05,minimum width = 1.5cm,
    minimum height = 3.5cm,shape border rotate=180,thick] (c) at (0,0) {$\vphantom{\substack{\textcolor{black}{\bullet}\\ [g^{(0)}]}}$};
    \node[] at (-.25,-0.1) {$\substack{\textcolor{black}{\bullet}\\ \textcolor{red}{[g^{(0)}]}}$};
    \node[] at (-1.5,-1) {$\gamma_+$};
    \node[] at (1.1,-1) {$\gamma_-$};
    \node[] at (2,0) {$\subset$};
    \node[] at (4.0,0) {$\MET(\IX)/\DIFF(\IX)$};
    \draw[dashed,blue] (1.16,0.74) arc (90:270:0.38 and 0.74);
\end{tikzpicture}
\caption{\label{fig:one-param}A one-parameter family of deformations of a closed loop in $\MET(\IX)/\DIFF(\IX)$.}
\end{figure}

As an example of the above discussion consider closed loops in $\MET(\IX)/\DIFF(\IX)$, and their associated invariants. As explained above, wall-crossing for invariants associated with $1$-cycles will only occur for $\IX$ with $b_2^+(\IX) = 2$. (For simplicity we will put $b_1(\IX)=0$ in our computations below.) 
%
%
Let $\gamma$ denote a nontrivial $1$-cycle in $\MET(\IX)/\DIFF(\IX)$, and let $\mathsf{Z}_{\mathsf{DW}}[g_{\mu\nu}, \Psi_{\mu\nu}, \Phi^{\mu}]$ denote the generating functional. Expanding it in gravity degree, we have
\eqa{
& \mathsf{Z}_{\mathsf{DW}}[g_{\mu\nu}, \Psi_{\mu\nu}, \Phi^{\mu}] &&= \mathsf{Z}^{(0)}_{\mathsf{DW}}[g_{\mu\nu}] + \mathsf{Z}^{(1)}_{\mathsf{DW}}[g_{\mu\nu}, \Psi_{\mu\nu}] + \cdots ~,
}
where $\mathsf{Z}^{(1)}_{\mathsf{DW}}[g_{\mu\nu}, \Psi_{\mu\nu}]$ is order $1$ in the gravitino $\Psi_{\mu\nu}$, and the ellipsis denotes higher-order contributions.\footnote{In this section, we use the parametrization of section \ref{sec:ProposalForTheAction} in which the action terminates in gravity degree $= 2$.} (Recall \eqref{eq:Psi-Phi-Expansion}.) Therefore, the relevant ``Donaldson invariant,'' is 
\eqa{
& \oint_{\gamma}\mathsf{Z}^{(1)}_{\mathsf{DW}}[g_{\mu\nu}, \Psi_{\mu\nu}] ~, \label{eq:loop-zdw}
}
and it deserves to be called an ``invariant'' only if the result depends on the homology class of $\gamma$
and no other details of $\gamma$. 
%
 %
 %
%
Let $\gamma_+$ and $\gamma_-$ be two nontrivial $1$-cycles in $\MET(\IX)/\DIFF(\IX)$ whose difference is a 2-chain. For simplicity, assume there is a homotopy of $\gamma_+$ to $\gamma_-$  as illustrated schematically in Fig. \ref{fig:one-param}.
As we have just explained, if $b_2^+(\IX)=2$ and we deform $\gamma_+$ to $\gamma_-$ through this one-parameter family of loops, generically there will exist an isolated metric that admits an ASD Abelian instanton, and thus \eqref{eq:loop-zdw} might be discontinuous along the path of deformation. More precisely, suppose we make the deformation of the loop small enough so that the cylinder defined by $\gamma_{\pm}$ encloses a single such metric $g^{(0)}$ (actually, a conformal class of metrics, $[g^{(0)}]$). In this case,
a special case of the computation below shows that  $\mathsf{d}\mathsf{Z}^{(1)}$ has a delta-function singularity at $g^{(0)}$. Equivalently, in a small neighborhood of $g^{(0)}$,  $\mathsf{Z}^{(1)}_{\mathsf{DW}}[g_{\mu\nu}, \Psi_{\mu\nu}]$  contains a nonzero multiple of the angular form.   The line integral of $\mathsf{Z}^{(1)}_{\mathsf{DW}}$ over the boundary $\Delta_{\epsilon}$ of a small disk $D_{\epsilon}$ of radius $\epsilon$ around $g^{(0)}$ in the $\epsilon \rightarrow 0$ limit is
\eqa{
& \lim_{\epsilon \rightarrow 0}\int_{\Delta_\epsilon}\mathsf{Z}^{(1)}_{\mathsf{DW}}[g_{\mu\nu}, \Psi_{\mu\nu}] &&= \oint_{\gamma_+}\mathsf{Z}^{(1)}_{\mathsf{DW}}[g_{\mu\nu}, \Psi_{\mu\nu}] - \oint_{\gamma_-}\mathsf{Z}^{(1)}_{\mathsf{DW}}[g_{\mu\nu}, \Psi_{\mu\nu}] ~, \label{eq:zdw-limit}
}
In particular, this means \eqref{eq:loop-zdw} is, in fact, \underline{not} independent of the choice of representative of $\sfH_{1}\big(\MET(\IX)/\DIFF(\IX)\big)$. This is precisely the phenomenon of wall crossing for $b_{2}^+(\IX) = 2$. 

The above picture generalizes readily to $n$-dimensional chains separating representatives of $(n-1)$-dimensional homology classes. We next turn to the computation that demonstrates that when $b_2^+(\IX)=n$ the pullback of $\overline{\mathsf{Z}^{(n-1)}_{\mathsf{DW}}[g_{\mu\nu}, \Psi_{\mu\nu}]}$ to a generic $n$-dimensional ball of metrics containing $g^{(0)}$ can be expressed as a nonzero multiple of the angular form plus a smooth form.

\subsection{A General Formula For The Leading Singularity\label{sec:WC-LeadingSingularity}}
In this section, we perform a preliminary computation of the wall-crossing formula arising 
from a given $\lambda \in \sfH^{2}(\IX, \IZ)$ for a manifold $\IX$ with general $b_2^+(\IX)=n$. 
Our computation should be viewed as exploratory since a full discussion, including the justification of the use of tree-level evaluation\footnote{This will be the subject of \cite{ScalingWCF-ToAppear}.} of the Coulomb branch path integral,  is beyond the scope of this paper.

As before, let $g^{(0)}$ 
be a metric such that $\lambda_{+} = 0$ (where by $\lambda$ we henceforth mean the two-form harmonic representative of the cohomology class). To simplify the notation below, we let $b := b_{2}^{+}(\IX)$. We construct an $n$-dimensional neighborhood of the metric $g^{(0)}$ by choosing a set of $n$ linearly independent symmetric tensors $p_\alpha$ so that we can form a family of metrics,
\eqa{
& g(t)_{\mu\nu} &&= g_{\mu\nu}^{(0)} + \sum_{\alpha=1}^{n}t^{\alpha}(p_{\alpha})_{\mu\nu} ~. \label{eq:n-param-family}
}
In a sufficiently small neighborhood of $t=0$, the tensor $g(t)$ will be positive definite and we will henceforth work in such a neighborhood. 
Indeed, we are ultimately only interested in the singularity of $\sfd \mathsf{Z}_{\mathsf{DW}}[g_{\mu\nu}, \Psi_{\mu\nu}, \Phi^{\mu}]$
at $t=0$ so the neighborhood should be viewed as a germ around $t=0$. We assume for simplicity that the $p_{\alpha}$'s are traceless with respect to $g^{(0)}$. Then, since $\sfd g_{\mu\nu} = \Psi_{\mu\nu}$, to lowest order in $t$, the gravitino pulled back to this family is just 
\eqa{
& (\Psi_{0})_{\mu\nu} &&= \sum_{\alpha=1}^{n} dt^{\alpha}(p_{\alpha})_{\mu\nu} ~.\label{eq:gravpullback}
}
We wish to examine the behavior of the gravity-degree $(n-1)$ component of $\mathsf{Z}_{\mathsf{DW}}[g_{\mu\nu}, \Psi_{\mu\nu}, \Phi^{\mu}]$.  We focus on the Coulomb branch integral (``$u$-plane'') contribution to the evaluation in the IR theory. In searching for a delta-function contribution to the pull-back of
$\sfd\mathsf{Z}_{\mathsf{DW}}[g_{\mu\nu}, \Psi_{\mu\nu}, \Phi^{\mu}] $ to the $n$-dimensional family we will 
assume that the contribution comes entirely from $\mathsf{Z}_{\mathsf{DW}}^{(n-1,0)}$, i.e., from the correlation function of an insertion of $(n-1)$ factors of  $\int_{\IX}\mathsf{vol(}g\mathsf{)} \Psi^{\mu\nu} \Lambda_{\mu\nu}$,
where $\Lambda_{\mu\nu}$ was derived in section \ref{sec:minimalgeneralaction} (cf. \eqref{eq:GeneralActionDegree1}) and takes the following simpler form in the IR for gauge group $G_{\mathsf{IR}} =\mathsf{U(1)}$ (and after using \eqref{eq:GoodRescaling} to switch to the notation of \cite{Moore:1997pc})
\eqa{
&\Lambda_{\mu\nu} &&= -2\big(\mathsf{Im\,}\tau \big)F^{-}{}_{\!\!\!\!\!\!\rho(\mu}\chi_{\nu)}{}^{\rho} - \frac{i}{\sq} g_{\mu\nu}\big(\mathsf{Im\,}\tau \big)\psi_{\rho}\n^{\rho}\ov{a} + \frac{i}{\sq}\big(\mathsf{Im\,}\tau\big)\psi_{(\mu}\n_{\nu)}\ov{a} \nonumber \\
& &&\quad + \frac{i}{8\sq}\CF'''\big(\psi_{\mu}\psi_{\rho}\big)^{-}\chi^{\rho}{}_{\nu} + \frac{i}{8\sq}\CF'''\big(\psi_{\nu}\psi_{\rho}\big)^{-}\chi^{\rho}{}_{\mu} + \frac{1}{12\sq}g_{\mu\nu}\ov{\CF}'''\chi_{\kappa}{}^{\rho}\chi^{\kappa\sigma}\chi_{\rho\sigma}  ~.\label{eq:GeneralActionDegree1-U1-repeated}
}
Of the many terms in $\Lambda_{\mu\nu}$ we only consider the ones that can soak up the relevant $\chi$-zeromodes and that can contribute a singularity to the $u$-plane integral. Therefore we make the replacement 
\be 
\Psi_0^{\mu\nu} \Lambda_{\mu\nu} \rightarrow -2\big(\mathsf{Im\,}\tau \big){\rm tr} (\Psi_0 F^- \chi)
\ee
where ${\rm tr}$ refers to a trace on the spacetime indices (with indices raised by $g(t)$). 
 
Following the general discussions on the $u$-plane integral \cite{Moore:1997pc,Korpas:2019cwg,Manschot:2019pog,Manschot:2021qqe,Moore:2012SCGP,Moore:2017SCGP}  
we are led to the integral over the Coulomb branch parametrized by $a,\bar a$:  
\eqa{
&\int y\,da\,d\ov{a}\int \frac{d\eta}{y^{1/2}}&&\prod_{i=1}^{b}\frac{d\chi_i}{y^{1/2}}\int\prod_{i=1}^{b}(y^{1/2}dD_{i})e^{-y D_i^2}\left(\frac{du}{da}\right)^{\chi/2}\Delta^{\sigma/8}y^{-1/2} \nonumber\\
&  && \qquad \sum_{\lambda' \in H^2(\IX,\IZ)}e^{-i\pi\ov{\tau}(\lambda'_+)^2 - i\pi\tau(\lambda'_-)^2}(-1)^{(\lambda'-\lambda_0)\cdot w_2(\IX)}
\nonumber\\
& &&\qquad \qquad\,\, \left(-\frac{i\sq}{16\pi}\right)  \frac{d\ov{\tau}}{d\ov{a}}\eta\int_{\IX}\chi(F^{+} + D)\left(-2 \int_{\IX}y\,\,\mathsf{tr}\left[\Psi_{0}F^-\chi \right]\right)^{b-1} ~,
}
Here the integral over $d a d\bar a$ is to be converted to an integral over the modular curve for $\Gamma^0(4)$ 
for $\tau(a) =  x + i y$ using $da = \frac{da}{d\tau} d\tau$. The integrals $d \eta$ and $\prod_i d \chi_i$ are over fermion zeromodes and the 
integrals over $\prod_i dD_i$ are over $\IR^b$. The sum over $\lambda' \in \sfH^2(\IX,\IZ)$ is a sum over the topological classes of the IR $\mathsf{U(1)}$ gauge bundle. (For simplicity we put the 't Hooft flux to zero.)  For the remaining notation see 
\cite{Moore:1997pc,Korpas:2019cwg,Manschot:2019pog,Manschot:2021qqe,Moore:2012SCGP,Moore:2017SCGP}.

The integral over the $\eta$ zeromode is straightforward. To integrate out the $\chi$ zeromodes we parametrize 
them as  
\eqa{
& \chi &&= \sum_{i=1}^{b}\chi_{i}\omega_{i}(t) ~, \label{eq:chi-zero-modes-gen}
}
where the $\omega_{i}(t)$'s are an orthonormal basis of self-dual $2$-forms with respect to $g(t)$, i.e.,
\eqa{
& \star_{g(t)} \omega_{i}(t) &&= \omega_{i}(t) ~,
}
chosen to satisfy
\eqa{
& \int_{\IX} \omega_{i}(t) \star_{g(t)} \omega_{j}(t) &&= \delta_{ij} ~, \quad i, j \in 1, \ldots, b ~.
}
We will have $\omega_i(t) = \omega_i + \CO(t)$. In computing the leading singularity we can replace $F^-\rightarrow F$
and then 
\eqa{
& \int \mathsf{tr}(\Psi_{0}\chi F) &&\rightarrow  dt^{\alpha}\chi_{i}\int_{\IX}(p_{\alpha})_{\mu}{}^{\rho}\lambda_{\rho}{}^{\nu}(\omega_i)_{\nu}{}^{\mu}\mathsf{vol}(g) ~,
}
where indices are raised and lowered with $g(t)$, and we have used \eqref{eq:gravpullback}, \eqref{eq:chi-zero-modes-gen} and $F = 4\pi \lambda$. Motivated by this we define a square matrix $Q^{\lambda}(t)$, with elements $Q_{i\alpha}^{\lambda}(t)$ given by
\eqa{
& Q_{i\alpha}^{\lambda}(t) &&:= \int_{\IX}(p_{\alpha})_{\mu}{}^{\rho}\lambda_{\rho}{}^{\nu}(\omega_i)_{\nu}{}^{\mu}\mathsf{vol}(g) ~. \label{eq:WCF-QMatrix}
}
For generic $p_{\alpha}$, the matrix $Q^{\lambda}(t=0) \neq 0$, but the contribution to $Q^{\lambda}(t=0)$ from the trace part of $p_{\alpha}$ indeed vanishes. 

To focus on the leading singular contribution we note that it must come from the noncompact region of the $u$-plane, hence $y= \mathsf{Im}(\tau) \to \infty$. We define the coefficients $c(n)$ of a modular form by 
\be 
\frac{da}{d\tau} \left( \frac{du}{da}\right)^{\chi/2} \Delta^{b/2} =: \sum_{n\in \mathbb{Z}/8} c(n)q^n 
\ee
and then perform the integration over $x=\mathsf{Re}(\tau)$. This sets $n= \lambda^2/2$ in the sum over $\lambda$. 
The coefficients $c(\lambda^2/2)$ are the Fourier coefficients of the modular form that appears in the standard wall crossing formula for the Donaldson polynomials \cite{Moore:1997pc}.
Next, one performs the $D_i$ integrations, which are simply Gaussian. The leading singularity comes from 
an integration over $y$ of the form:   
\eqa{
& \int^{\infty}   \, y^{\frac{1}{2}b - 1} e^{-2\pi y \lambda_{+}^2} dy && = 
\frac{\Gamma(b/2)}{\left(\sqrt{2\pi} \vert \lambda_+ \vert \right)^b } \left(1 + \cal{O}(\vert \lambda_+\vert) \right)    ~.
}
for $\vert \lambda_+ \vert \to 0$. 
This integral multiplies the integral over $\chi$. Our formula for the leading singularity becomes 
(up to some trivial numerical factors): 
\be 
\frac{\Gamma(b/2)}{\left( \sqrt{2\pi} \vert \lambda_+ \vert \right)^b }  
c(\lambda^2/2) (-1)^{(\lambda-\lambda_0)\cdot w_2(X)} 
\int \prod_{i=1}^b d\chi_i \left( \sum_i \chi_i \int_{\IX} \omega_i \lambda^+\right) 
\left( \sum_{i,\alpha} dt^\alpha \chi_i Q^\lambda_{i\alpha} \right)^{b-1} 
\ee
Now both $\vert \lambda_+ \vert$ and  $\int_{\IX} \omega_i \lambda_+$ will vanish linearly in the $t^\alpha$ and hence 
our wall-crossing formula is of the form 
\beqa{ \label{eq:FinalFamilWCF}
&\mathsf{Z}^{\mathsf{sing}}  &&= 
\frac{\Gamma(b/2)}{\left(\sqrt{2\pi}  \right)^b }  
c(\lambda^2/2) (-1)^{(\lambda-\lambda_0)\cdot w_2(X)} 
K_{\alpha_1\ldots\alpha_b}\frac{t^{\alpha_1}dt^{\alpha_2} \wedge \cdots \wedge dt^{\alpha_b}}{|t|^b} ~,
}
where $K_{\alpha_1\ldots\alpha_b}$ is a totally antisymmetric tensor that depends on the family 
and on $\lambda$.
For generic neighborhoods \eqref{eq:n-param-family} defined by the $p_{\alpha}$'s
we expect that 
$K_{\alpha_1\ldots\alpha_b}\frac{t^{\alpha_1}dt^{\alpha_2} \wedge \cdots \wedge dt^{\alpha_b}}{|t|^b}$
will be  a nonvanishing multiple of the angular form \cite{Bott1982} in $t$ around $t = 0$.
Thus we finally obtain a nontrivial wall-crossing contribution. 
In the case of Donaldson invariants for $b_2^+(\IX)=1$ the wall-crossing expression only depended on 
the wall. In our case, there is potential dependence on specific details of the family - such as the choice of 
$p_\alpha$. It would be very desirable to improve our computation to give a more tractable and natural expression for the multiple of the angular form,  as a function of the family and $\lambda$.

%

To summarize, the partition function $\mathsf{Z}$ is a $\mathsf{Diff}^{+}(\IX)$-equivariant form on $\mathsf{Met}(\IX)$, and the family invariants are integrals of $\mathsf{Z}$ around closed $(n-1)$-cycles in $\mathsf{Met}(\IX)/\mathsf{Diff}^{+}(\IX)$.  When $n<b_2^+(\IX)$ we expect the integrals around the cycles to depend only on the homology class of the cycles. When $n=b_2^+(\IX)$ the dependence on the cycle can in principle be computed using the ``wall-crossing'' formula \eqref{eq:FinalFamilWCF}. It would be interesting to know if one can say anything precise about the case $n > b_2^+(\IX)$. The first interesting case is standard Donaldson theory, i.e. the case $n=1$ and $b_2^+(\IX)=0$.

%
%
%
%
%
%
%

\section{Comments On Observables}\label{sec:Observables}


\subsection{Review Of Observables In Donaldson-Witten Theory\label{subsec:ReviewObservablesDonaldson}}
Before discussing observables in the context of family Donaldson theory, let us briefly recall the story of observables in standard Donaldson theory \cite{DonaldsonDurham,DonaldsonReview,DK,Witten:1988ze,Moore:2012SCGP,Moore:2017SCGP}. We have a principal $G$ bundle $P$ over $\IX$, and a principal $\CG$ bundle $\CA \times \adsf P \xrightarrow{\pi_1} (\CA \times \adsf P)/\CG$, where $\CG$ is the group of gauge transformations and $\CA$ is the space of $G$-connections on $P$. We denote the quotient space $\CA/\CG$ by $\SCB$. The base of $\pi_1$ is itself a principal $G$ bundle $(\CA \times \adsf P)/\CG \to \SCB \times \IX$, known as the \textit{universal bundle}. The Donaldson polynomials are obtained via characteristic classes of the universal bundle, in a way that we shall describe momentarily.\footnote{\label{foot:framedconnections}There is an important subtlety in this discussion because the $\CG$ action on $\CA$ is not free even when the connection is irreducible. So one should really work with \textit{framed} irreducible connections. The space $\SCB = \CA/\CG$ is replaced by the space of gauge equivalence classes of framed connections $\widetilde{\SCB} = (\CA\times\mathsf{Hom}(G,P_{x_0}))/\CG$ where $P_{x_0}$ is the fiber of the principle $G$-bundle $P \to \IX$ over $x_0 \in \IX$. (One can equivalently define $\widetilde{\SCB} = \CA/\CG_{0}$ where $\CG_0 = \{g \in \CG ~|~ g(x_0) = 1\} \subset \CG$ is the subgroup of gauge transformations that trivialize at $x_{0} \in \IX$.) We further restrict to the space of gauge equivalence classes of framed \textit{irreducible} connections $\widetilde{\SCB}^* \subset \widetilde{\SCB}$. We can replace $\SCB$ with $\widetilde{\SCB}^*$ in the main text to properly account for this, treating $\widetilde{\SCB}^* \times \IX$ as the base of the universal bundle instead of $\SCB \times \IX$. See \cite{Cordes:1994fc,DK} for details. The treatment of $\mathsf{BDiff}(\IX)$ in section \ref{sec:toobsfd} will have a close parallel.}

Given a degree $d$ invariant polynomial on $\LIE G$, we get a degree $2d$ class $\varpi \in \sfH^{2d}(BG)$ in the cohomology of the classifying space $BG$. Pulling this back by the classifying map $\mathscr{B} \times \IX \xrightarrow{f} BG$, we obtain a degree $2d$ cohomology class on $\SCB \times \IX$, i.e., $f^{*}(\varpi) \in \sfH^{2d}(\SCB \times \IX)$. Then one can take a slant product with homology classes in $\IX$ to get cohomology classes on $\SCB$ -- in this way, we have a map $\sfH_{j}(\IX) \to \sfH^{2d-j}(\SCB)$, and the restriction of the image of this map to the (compactified) moduli space of ASD connections $\wh{\CM}_{\mathsf{ASD},g} \in \SCB$ yields the famed Donaldson map
\eqas{
\mu_{D} :\quad  \sfH_{j}(\IX) &\rightarrow \sfH^{4-j}(\wh{\CM}_{\mathsf{ASD},g}) \\
 \Sigma_{j} &\mapsto \mu_{D}(\Sigma_j) := \int_{\Sigma_j} f^{*}(\varpi) ~. \label{eq:slant}
}
The cohomology class on $\wh{\CM}_{\mathsf{ASD},g}$ depends only on the homology class of $\Sigma_{j}$.

For $G = \mathsf{SU(2)}$, we take $\varpi$ to be a generator of $\sfH^{4}(\mathsf{BSU(2)}; \mathbb{Z}) \cong \mathbb{Z}$, and this leads to the point and surface observables,
\eqa{
& \wp &&\rightarrow \mu_{D}(\wp) \in \sfH^{4}(\wh{\CM}_{\mathsf{ASD},g}) ~,\\
& \Sigma &&\rightarrow \mu_{D}(\Sigma) \in \sfH^{2}(\wh{\CM}_{\mathsf{ASD},g}) ~,
}
assuming $\IX$ to be simply connected. The Donaldson polynomials,
\eqas{
 P_D : \sfH_{0}(\IX) \oplus \sfH_{2}(\IX) &\rightarrow \IQ  ~, \label{eq:DonaldsonPoly}
}
are then defined as
\eqa{
& \wp^{\ell}\Sigma^{r} &&\mapsto P_D(\wp^{l}\Sigma^{r}) := \int_{\wh{\CM}_{\mathsf{ASD},g}}\mu_D(\wp)^{\ell} \mu_{D}(\Sigma)^{r} ~,
}
for nonnegative integers $\ell$, $r$. Recall that the moduli space has infinitely many connected components as in \eqref{eq:Comp-MASD} and the integral picks out the component with   
\eqa{
& 4\ell + 2 r &&= \mathsf{vdim\,} \wh{\CM}_{\mathsf{ASD},g,k} ~. \label{eq:vdim}
}
%
As explained in \cite{Cordes:1994fc}, the invariants are certain integrals of equivariant cohomology classes in $\sfH_{\CG}(\CA)$ over the instanton moduli space. 

From the viewpoint of topologically twisted $\CN=2$ SYM \cite{Witten:1988ze,Moore:2012SCGP,Moore:2017SCGP}, we are interested in correlation functions of $\CQ $-closed and gauge invariant operators. Using any invariant polynomial $\mathcal{P}$ on $\mathsf{Lie}(G)$, we can form a $\CQ$-closed and gauge invariant observable $\mathcal{P}(\phi)$. This is a $0$-observable, because it is a local operator defined at a point. For $G = SU(N)$, a natural basis of point observables is
\eqa{
& \CO_{s}^{(0)}(\wp) &&:= \mathsf{Tr\,}\phi^{s}(\wp)  \qquad s = 2, \ldots, N ~. \label{eq:point-observable}
}
In general, there is a ring of $0$-observables, generated by the Casimirs of $\mathsf{Lie}(G)$. Since the twisted theory is equipped with a globally-defined one-form supersymmetry operator $K$ such that $\{\CQ , K\} = d$ \cite{Moore:1997pc}, one can begin with a zero-observable $\CO^{(0)}$ and define
\eqa{
& \CO^{(1)} &&:= K \CO^{(0)} ~. \label{eq:O1}
}
In particular, for any $1$-chain,
\eqa{
& \CQ \int_{\gamma}\CO^{(1)} &&= \int_{\gamma}\{\CQ, K\}\CO^{(0)} = \int_{\gamma} d\CO^{(0)} = \left.\CO^{(0)}\right|_{\partial\gamma} ~. \label{eq:1-chain-argument}
}
Note that if $\partial\gamma = \wp_1 - \wp_2$, it follows that $\CO^{(0)}(\wp_1) = \CO^{(0)}(\wp_2) + \CQ \int_{\gamma}\CO^{(1)}$, so changing $\wp$ changes the $0$-observable by a $\CQ $-exact piece. And of course, it immediately follows that if $\gamma$ is a cycle (i.e., it is closed) then $\int_{\gamma}\CO^{(1)}$ is $\CQ $-closed.

We can generalize \eqref{eq:O1} and define certain $j$-forms on $\IX$, the so-called ``observable densities,''
\eqa{
& \CO^{(j)} &&:= K^{j}\CO^{(0)} ~,
}
and integrate them over $j$-cycles $\Sigma_j$ where $[\Sigma_j] \in H_{j}(\IX)$, to get ``$j$-observables,''\footnote{When an argument in $\CO^{(j)}$ is suppressed, one is referring to the observable \textit{density}. When the argument is included, one is referring to the pairing of the density with a homology class. Note that $\CO^{(0)}$ is a $0$-observable density as well as a $0$-observable.}
\eqa{
 & \CO(\Sigma_j) &&:= \int_{\Sigma_j}\CO^{(j)} ~. \label{eq:OJ}
} 
Now \eqref{eq:OJ} is a $\CQ $-closed and gauge-invariant observable that depends only on the homology class of $\Sigma_j$. The $\CQ $-closure follows from the descent equation,
\eqa{
& \CQ \CO^{(j+1)} &&= d\CO^{(j)} ~, \label{eq:descenteq}
}
and Stokes' Theorem, as usual.


Starting from \eqref{eq:point-observable}, one can construct the higher observable densities by using the descent equation \eqref{eq:descenteq}. For $G = \mathsf{SU(2)}$, these are
\eqa{
& \label{eq:obs-den-0}\CO^{(0)} &&= \frac{1}{2}\mathsf{Tr\,}(\phi^2) ~,\\
& \label{eq:obs-den-1}\CO^{(1)} &&= -\mathsf{Tr\,}(\phi\,\psi_\mu) dx^{\mu} ~,\\
& \label{eq:obs-den-2}\CO^{(2)} &&= \frac{1}{2}\mathsf{Tr\,}(\phi\,F_{\mu\nu} + \psi_{\mu}\,\psi_{\nu})dx^{\mu}\wedge dx^{\nu} ~,\\
& \label{eq:obs-den-3}\CO^{(3)} &&= -\frac{1}{2}\mathsf{Tr\,}(\psi_{\mu}\,F_{\nu\rho}\big)dx^{\mu}\wedge dx^{\nu} \wedge dx^{\rho} ~,\\
& \label{eq:obs-den-4}\CO^{(4)} &&= \frac{1}{4}\mathsf{Tr\,}(F_{\mu\nu}\,F_{\rho\sigma})dx^{\mu}\wedge dx^{\nu} \wedge dx^{\rho}\wedge dx^{\sigma} ~.
}
The Cartan model for $\CG$-equivariant cohomology of $\CA$ (i.e., $\sfH_{\CG}(\CA)$) is described by a set of fields $(A_\mu, \psi_\mu, \phi)$ satisfying the algebra \eqref{eq:GaugeCartan-1}--\eqref{eq:GaugeCartan-2}. From the viewpoint of equivariant cohomology, observables are constructed out of basic classes \cite{Cordes:1994fc}. In particular, being a basic class means that there is no dependence on vertical fields of the $\CN=2$ vectormultiplet. In fact, these omitted fields contribute to contractible equivariant cohomology \cite{Deligne:1999qp,Witten:1990bs}, and hence the observables are constructed entirely out of the fields of the Cartan model.\footnote{In equivariant cohomology, these correspond to basic classes (horizontal and gauge invariant) \cite{Cordes:1994fc}.} 

The two notions presented above -- the description of the Donaldson map $\mu_D$ via the universal bundle, and the descent formalism leading to $\IQ$-cohomology classes, i.e., equivariant cohomology classes, on the base of the universal bundle, fit together nicely via the equivariant localization formula. We do not review this here but refer the reader to \cite{Baulieu:1988xs,Atiyah:1990tm,Cordes:1994fc,Moore:2012SCGP,Moore:2017SCGP}.

\subsection{Some Observables In Family Donaldson Theory\label{sec:toobsfd}}

The key moral from the equivariant viewpoint of observables in standard Donaldson theory is that they are constructed out of basic classes in equivariant cohomology. Therefore, to upgrade this idea to family Donaldson theory, one would seek basic classes for the full Cartan model of $\IG$-equivariant cohomology of $\IM$ that was developed in section \ref{subsec:HGIM}. In this section, we comment on some ideas and efforts in this direction.

Family Donaldson invariants are invariants of smooth families of smooth, closed, oriented four-manifolds. To describe them, we introduce a smooth family $\CX$ of closed, compact manifolds $\IX$ fibered over a base $\CB$, exhibited as
\eqa{
& \IX &&\rightarrow \CX \xrightarrow{\pi_f} \CB ~. \label{eq:family_fib}
}
Each fiber of $\pi_f$ is diffeomorphic to $\IX$.\footnote{In the literature on families Seiberg-Witten invariants, $\CX$ and $\CB$ are taken to be smooth, compact manifolds, $\pi_f$ is a proper submersion, and the fibers of $\CX \xrightarrow{\pi_f}\CB$ are four-manifolds with positive-definite intersection form. The present discussion does not require such specializations.} 
When $\CB$ is $\mathsf{BDiff}^{+}(\IX)$, the classifying space of orientation-preserving diffeomorphisms of $\IX$, the bundle \eqref{eq:family_fib} has the form
\eqa{
& \IX &&\rightarrow \CX \xrightarrow{\pi_f} \mathsf{BDiff}^{+}(\IX) ~, \label{eq:family_fib2}
}
and is called the \textit{universal bundle of $\IX$ manifolds}. 
It can also be presented as a  homotopy quotient
\eqa{
& U_{\IX} &&= \mathsf{EDiff}^{+}(\IX) \times_{\Diff(\IX)} \IX = \frac{\mathsf{EDiff}^{+}(\IX) \times \IX}{\Diff(\IX)} ~. \label{eq:universal bundle UX}
}
where  $\pi_f$ is just the projection to the first factor in the definition of $U_{\IX}$.

It is now time to take into account that, in fact, the action of $\Diff(\IX)$ on  $\MET(\IX)$  is not 
free, because some metrics have isometries. This technical difficulty is can be addressed as follows.  Fix a point $p \in \IX$, and consider the subgroup of diffeomorphisms that leave this point fixed, i.e.,
\eqa{
&\mathsf{Diff}^{+}_{p}(\IX) &&:= \{ \varphi \in \Diff(\IX) ~|~ \varphi(p) = p  \text{ and } d\varphi_{p}: T_{p}\IX \to T_{p}\IX \text{ is the identity map.} \} ~, \label{eq:diffp observer}
}
Then $\mathsf{Diff}^{+}_{p}(\IX)$ acts freely\footnote{Let $\IX$ be a compact, connected, smooth Riemannian manifold, and $g$ be a Riemannian metric on $\IX$, and $p \in \IX$ a marked point. To see why $\mathsf{Diff}^{+}_{p}(\IX)$ acts freely on $\MET(\IX)$, assume the contrary. If $\varphi \in \mathsf{Diff}_{p}(\IX)$ has a fixed point on $\MET(\IX)$, $\varphi$ is an isometry of $(\IX, g)$. By definition, $\varphi$ fixes $p$ and the tangent space at $p$. As $\varphi$ is an isometry, it should also fix any geodesic out of $p$. But since $\IX$ is compact, any generic point of $\IX$ lies on a geodesic out of $p$. But this means $\varphi$ fixes all of $\IX$, i.e., is the identity map on $\IX$.} on $\MET(\IX)$ and therefore $\MET(\IX)/\mathsf{Diff}_{p}^{+}(\IX)$ is homotopy equivalent to $\mathsf{BDiff}^{+}_{p}(\IX)$. This is reminiscent of how one treats irreducible framed connections in Yang-Mills theory and Donaldson theory (see footnote \ref{foot:framedconnections}).\footnote{\label{foot:observermodspace}In the mathematical literature, $\mathsf{BDiff}_{p}(\IX)$ is sometimes called the observer moduli space. See \cite{Tuschmann2015,Botvinnik2019} for recent discussions. However, $\MET(\IX)/\mathsf{Diff}_{p}(\IX)$ also appeared earlier, e.g., in \cite{Wu:1993ef}.} We can generalize this discussion and \eqref{eq:family_fib2}. Consider the universal family of 4-manifolds with marked points $p_1, \ldots, p_n$ where $n \geq 1$, 
\eqa{
& \IX &&\rightarrow \CX \xrightarrow{\pi_f} \mathsf{BDiff}^{+}_{p_1, \ldots, p_n}(\IX) \cong \MET(\IX)/\mathsf{Diff}^{+}_{p_1,\ldots,p_n}(\IX) ~. \label{eq:family_fib3}
}
As we shall see shortly, this is also a natural family to consider from a physical viewpoint.

As we reviewed above, in Donaldson theory the observable densities were obtained by descent $\CQ \CO^{(n)} = d\CO^{(n-1)}$, and were listed in \eqref{eq:obs-den-0}--\eqref{eq:obs-den-4}.  In our larger equivariant cohomology model for $\IM/\IG$ (where $\IM = \CA \times \MET(\IX)$ and $\IG = \CG \rtimes \mathsf{Diff}^+(\IX)$), the  densities \eqref{eq:obs-den-0}--\eqref{eq:obs-den-4} now satisfy an \textit{ascent-descent equation},
\beqa{
& \IQ \CO^{(n)} &&= d\CO^{(n-1)} + \iota_{\Phi}\CO^{(n+1)} ~. \label{eq:asc-des}
}
Although this can be explicitly checked on  \eqref{eq:obs-den-0}--\eqref{eq:obs-den-4} with $\CO^{(-1)} = \CO^{(5)} = 0$, this equation can be conceptually understood as follows. On the basic complex of fields, the Cartan differential $\IQ$ acts as $d + \iota_{\Phi}$ (see, for example, \cite{Cordes:1994fc,GuilleminSternberg:1999}). Therefore, one can write $\IQ \CO^{(\bullet)} = d\CO^{(\bullet)} + \iota_{\Phi}\CO^{(\bullet)}$, and specialize to form degree $n$, using the fact that $\IQ$ does not change differential form degree. This leads precisely to \eqref{eq:asc-des}.

Now, if we attempt to use $\CO^{(n)}$ to form observables in the family context by writing $\int_{\Sigma_n}\CO^{(n)}$ for an  $n$-cycle $\Sigma_{n}$ in $\IX$, the ascent-descent equation 
\eqref{eq:asc-des} leads to difficulties, because  the term involving $\Phi$ spoils the $\IQ$-closure. 
Moreover, if $\Sigma_n$ and $\Sigma_n'$ are homologous then
\be
\int_{\Sigma_n}\CO^{(n)} - \int_{\Sigma_n'}\CO^{(n)} 
\ee
is no longer $\IQ$-exact.  
The above two remarks imply that one cannot immediately use the standard descent formalism to construct observables.  Fortunately, there is a simple remedy. As an example of the remedy, restrict   attention to $\mathsf{Diff}^{+}_{\wp_0}(\IX)$ and consider the subcomplex of $\mathsf{Diff}^{+}_{\wp_0}(\IX)$-equivariant cohomology of $\MET(\IX)$. Equivalently, 
consider a restricted subcomplex of the Cartan model where $\Phi^{\mu}(\wp_0) = 0$. In such a subcomplex,
\eqa{
 & \CO^{(0)}(\wp_0) := \mathsf{tr\,}\phi^2(\wp_0) ~,
}
is $\IQ$-closed and thus defines a cocycle for the $\IG$-equivariant cohomology of $\IM$. The proof that the class is independent of the point $\wp_0$ does not generalize, even if we use the subcomplex. 
The idea is easily 
generalized: If we restrict attention to $\mathsf{Diff}^{+}_{\wp_1,\wp_2}(\IX)$, then the correlation function
\eqa{
& \left\langle\!\!\left\langle \mathsf{tr\,}\phi^2(\wp_1) \mathsf{tr\,}\phi^2(\wp_2) \right\rangle\!\!\right\rangle_{g,\Psi,\Phi} ~,
}
is a cocycle for $\MET(\IX)/\mathsf{Diff}^{+}_{\wp_1,\wp_2}(\IX)$. Here $\left\langle\!\left\langle \cdot,\cdot\right\rangle\!\right\rangle_{g,\Psi,\Phi}$ denotes the path integral computed in the background of the subcomplex, and we are using the fact that $\left\langle\!\left\langle \IQ \cdot,\cdot\right\rangle\!\right\rangle_{g,\Psi,\Phi} = \mathsf{d}\left\langle\!\left\langle \cdot,\cdot\right\rangle\!\right\rangle_{g,\Psi,\Phi}$, by an equivariant generalization of the chain map, where $\mathsf{d}$ is the $\DIFF(\IX)$-equivariant differential on $\MET(\IX)$. 

It is now entirely straightforward to generalize this to $\MET(\IX)/\mathsf{Diff}^{+}_{\wp_1,\ldots,\wp_n}(\IX)$, for which
\eqa{
& \left\langle\!\!\left\langle \mathsf{tr\,}\phi^2(\wp_1) \cdots \mathsf{tr\,}\phi^2(\wp_n) \right\rangle\!\!\right\rangle_{g,\Psi,\Phi} ~,
}
is a cocycle, or to $\MET(\IX)/\mathsf{Diff}^{+}_{\wp_1,\ldots,\wp_n,\Sigma^{(k_1)}_{1},\ldots,\Sigma^{(k_r)}_{r}}(\IX)$, for which
\eqa{
& \left\langle\!\!\left\langle \mathsf{tr\,}\phi^2(\wp_1) \cdots \mathsf{tr\,}\phi^2(\wp_n) \prod_{i=1}^{r} \int_{\Sigma_{i}^{(k_i)}}\CO^{(k_i)} \right\rangle\!\!\right\rangle_{g,\Psi,\Phi} ~,
}
is a cocycle. Here $\Sigma_{i}^{(k_i)}$ is a $k_{i}$-cycle ($0 \leq k_{i} \leq 4$) for each $i = 1, \ldots, r$, and further, $\Phi^{\mu}(\wp_j) = 0$ for each $j=1, \ldots, n$ and the pullback of $\Phi^{\mu}$ to $\Sigma_{i}^{(k_i)}$ vanishes.

We leave the investigation of these correlators for future work. 

\subsection{Review Of Some Related Literature}\label{subsec:ReviewPreviousLit}

There has been some work in the mathematical literature which is potentially relevant to the issue of defining observables in family Donaldson theory\footnote{In the mathematical literature, this is also referred to as parametric Donaldson theory, e.g., \cite{ruberman1999polynomial,Morava:2000yk}.} \cite{Kontsevich1994,januszkiewicz2002characteristic,Hatcher:2012,Galatius2012stable,Galatius2013detecting,Ebert2014,Galatius2017homologicalI,Galatius2017homologicalII,Galatius2017TautRings,Galatius2019moduli,Berglund2020characteristic,Konno2021,Baraglia2023}. We will now very briefly comment on this work.

To begin with, it is worthwhile considering the two-dimensional analog. Since the moduli space of Riemann surfaces can be viewed as the quotient of the space of metrics on closed compact oriented topological surfaces by the semidirect product of the group of Weyl transformations and diffeomorphisms, one can view the study of the cohomology of the moduli space of curves as a two-dimensional analog of the subject of this paper. In this context, the physical results related to two-dimensional topological gravity models have been extremely effective \cite{Labastida:1988zb,Witten:1989ig,Distler:1989ax,Witten:1990hr,Dijkgraaf:1990nc,Verlinde:1990ku,Dijkgraaf:1990rs,Witten:1991mn,Witten:1993mgm,Dijkgraaf:2018vnm}. In this case, it has been shown that the cohomology ring is generated by the so-called ``Miller-Morita-Mumford (MMM) classes,'' \cite{Mumford1983,Miller1986,Morita1987} and thanks to the results on intersection numbers, the ring relations are known. In particular, Madsen and Weiss \cite{Madsen2007} have shown that the rational polynomial ring generated by the MMM classes is isomorphic to the rational cohomology of the moduli space of curves. While the 2d case can serve as a guide to our present considerations, a fundamental difference is that the moduli space of curves is finite-dimensional, while $\mathsf{BDiff}^{+}(\IX)$ is not. 

The MMM classes have higher-dimensional generalizations known as the generalized Miller-Morita-Mumford classes or tautological classes \cite{Galatius2017homologicalII,Berglund2020characteristic,Baraglia2023}. These can be viewed as characteristic classes of manifold bundles. For a family \eqref{eq:family_fib3} of 4-manifolds that we momentarily denote by $\CX \xrightarrow{\pi_f} \CB$, these classes can be obtained as pushforwards of cohomology classes of the relative (co)tangent bundle of the family. More precisely, let  
\eqa{
& T(\CX/\CB) &&:= \mathsf{ker}\left( \pi_{f*} : T\CX \to T\CB \right) ~,
}
denote the relative or vertical tangent bundle. This is a real oriented rank $4$ vector bundle over the total space $\CX$ of the family, and one can consider, for example, powers of the rational Pontryagin class $p_{j}\big(T(\CX/\CB)\big) \in \sfH^{4j}(\CX; \IQ)$, and integrate them over the fibers of the relative tangent bundle (or equivalently, pushforward by $\pi_{f}$), yielding cohomology classes on the base $\CB$ of the family:
\eqa{
& \kappa_{p_j;k}(\CX) &&:= \pi_{f*}p_{j}\big(T(\CX/\CB)\big)^{k} = \int_{\CX/\CB} p_{j}\big(T(\CX/\CB)\big)^{k} \in \sfH^{4jk-4}(\CB; \IQ) ~.
}
This provides a way to obtain cohomology classes on $\CB = \mathsf{BDiff}^{+}(\IX)$ or one of its variants considered above, such as $\mathsf{BDiff}_{\wp_0}^{+}(\IX)$. We remark that this is reminiscent of the slant product \eqref{eq:slant} appearing in the Donaldson map \cite{Cordes:1994fc,DK}. 

There is evidence in the mathematical literature for the nontriviality of these cohomology classes \cite{Galatius2017TautRings,Konno2021,Baraglia2023,Berglund2020characteristic}. In particular, in a famous series of papers, Galatius,  Randal-Williams, and collaborators \cite{Galatius2019moduli} have shown that the generalized MMM classes generate rational cohomology of the moduli space of a distinguished class of even-dimensional manifolds of dimension $\geq 6$. Berglund \cite{Berglund2020characteristic} takes a slightly different approach, by considering the classifying space for the fibration of the family of vector bundles formed by the tangent bundles of the four-manifold fibers of the original family \eqref{eq:family_fib3}, and showing that the polynomial ring of the generalized MMM classes generates the cohomology ring of this classifying space.

To our knowledge, at the time of writing this paper, it is not known whether there is an isomorphism between the cohomology of the classifying space of the diffeomorphism group of 4-manifolds, i.e., $\sfH^{\bullet}(\mathsf{BDiff}(\IX);\IQ)$, and the rational polynomial ring generated by the generalized MMM classes (called the tautological ring), i.e., whether the four-dimensional analog of the Madsen-Weiss statement holds. However, it is certainly true that the tautological ring is a subring of $\sfH^{\bullet}(\mathsf{BDiff}(\IX);\IQ)$, and in the case of families of four-manifolds, some results are available. Baraglia \cite{Baraglia2023} has recently computed the tautological rings of $\IC\IP^{2}$ and $\IC\IP^{2}\#\IC\IP^{2}$ and found that they are generated by generalized MMM classes obtained in terms of powers of the first Pontryagin class $p_{1}$ of the relative tangent bundle. For 4-manifold families satisfying some mild technical restrictions and with fibers satisfying $b_{2}(\IX) = n \geq 1$, he has given an explicit construction of the tautological classes. Roughly speaking, Baraglia's gauge theory inspiration involves a fibration of instanton moduli spaces over a base, under the assumption that the moduli spaces are smooth (away from reducible connections) and that their topology does not vary with motion along the base. There are technical obstructions in extending this to more general families because one will inevitably encounter reducible solutions (i.e., Abelian instantons) and in general the family will define a nontrivial bordism of moduli spaces.  

It is worth remarking that connections to generalized MMM classes for four-manifolds in the context of $\mathsf{Diff}^{+}$-equivariant cohomology of $\mathsf{Met}$ have been hinted at in the past \cite{Myers:1989dn,Myers:1990sa,Birmingham:1991nq,Birmingham:1991tu,Birmingham:1992ur,Birmingham:1993ch,Birmingham:1993eh}, inspired by Witten's four-dimensional topological gravity \cite{Witten:1988xi,Perry:1992ta}, and by earlier work on two-dimensional topological gravity by Labastida-Pernici-Witten \cite{Labastida:1988zb}. In \cite{Myers:1990sa}, the authors described a physical model for equivariant cohomology of diffeomorphisms and local Lorentz transformations but needed to impose severe constraints on the four-manifolds, namely the vanishing of the self-dual part of the Weyl tensor and the constancy of the Ricci scalar. (See \cite{Birmingham:1992ur} for a similarly constrained approach.) They retained local Lorentz transformations anticipating a supergravity model that might be useful for spin manifolds. In our conception of family Donaldson invariants, we do not wish to impose any such restrictions on the underlying four-manifold fibers of our smooth families. Secondly, \cite{Myers:1989dn,Myers:1990sa,Wu:1993ef} suggested that diffeomorphism-invariant observables can only be constructed out of operators integrated over the entire manifold because generic diffeomorphisms would perturb generic points and cycles, with the possible exception of punctures or marked points that are mapped to themselves under such diffeomorphisms. Our proposal above suggests that \underline{potentially} nontrivial diffeomorphism-invariant point and surface observables \textit{can} nevertheless be obtained as long as operator insertions are along loci of vanishing $\Phi^{\mu}$ -- the fundamental vector field generating diffeomorphisms on the space of metrics.\footnote{This is consistent also with the observations of \cite{Birmingham:1993eh}.} We note that in the two-dimensional context, correlation functions of certain BRST-closed operators in (dynamical) topological gravity with insertions of puncture operators (to restrict to diffeomorphisms that fix marked points) in the Labastida-Pernici-Witten model lead to (nontrivial) 2d MMM classes \cite{Witten:1989ig,Labastida:1988zb}. 
Finally, we note that it might also be interesting to make contact with the work of \cite{Tan:2009qq}.

We emphasize that our approach is conceptually on a different footing than the works on two-dimensional gravity cited above: we use dynamical vectormultiplets coupled to a non-dynamical supergravity background to produce classes on $\mathsf{BDiff}(\IX)$, \underline{without} integrating over supergravity fields. While we do not provide a formal proof of the nontriviality of our proposed observables, our discussion of wall crossing on the space of metrics suggests that the partition function $\mathsf{Z}[g_{\mu}, \Psi_{\mu\nu},\Phi^{\mu}]$ is a nontrivial $\mathsf{Diff}^{+}(\IX)$-equivariantly closed form on $\mathsf{Met}(\IX)$. 


In view of our findings and the abovementioned results, an interesting open problem for the future is whether our cohomology classes have any relation to those that already appear in the mathematical literature. A related zeroth-order question is: for $n$-dimensional families of four-manifolds $\IX$ with $b_{2}^{+}(\IX) = n$, do the pairings of the classes of \cite{Baraglia2023} with homology cycles on $\mathsf{BDiff}(\IX)$ illustrate wall-crossing phenomena? We leave these questions for further work. 
%
%

In another direction, a natural set of questions arises when we consider four-manifolds $\IX$ with non-empty boundary $\partial \IX \neq \emptyset$. Traditionally, one has to define relative Donaldson invariants of $\IX$ that are valued in the Floer groups of $\partial \IX$. It would be interesting to have a suitable family version of relative Donaldson invariants. With an eye to the future, we have been careful to keep track of all the boundary terms in our supergravity analysis.

\appendix

\section{Symbol Lists}\label{app:SymbolLists}



\begin{small}
\begin{table}[H]
\centering 
\begin{tabular}{|c|l|}\hline
  $\mu, \nu, \rho, \ldots$ & \begin{tabular}{@{}l@{}} coordinate indices for a local chart of $\IX$ (a.k.a. curved spacetime indices) \\ range: $\{1, 2, 3, 4\}$ \end{tabular} \\ \hline
  $a, b, \ldots$ & \begin{tabular}{@{}l@{}} indices for a local frame of $\IX$ (a.k.a. flat indices or frame indices) \\ range: $\{1, 2, 3, 4\}$ \end{tabular} \\ \hline
  $A, B, \ldots$ & \begin{tabular}{@{}l@{}} doublet indices of $\mathfrak{su(2)}_+$ (a.k.a. undotted spinor indices) \\ range: $\{1,2\}$ \end{tabular} \\ \hline
  $\dt{A}, \dt{B}, \ldots$ & \begin{tabular}{@{}l@{}} doublet indices of $\mathfrak{su(2)}_-$ (a.k.a. dotted spinor indices) \\ range: $\{1,2\}$ \end{tabular} \\ \hline
  $\mathbb{A}, \mathbb{B}$ & indices labeling generators of $\mathsf{Lie}(\IG)$ \\ \hline
  $i, j, k, \ldots$ & \begin{tabular}{@{}l@{}} doublet indices of $\mathfrak{su(2)}_{\mathsf{R}}$ (a.k.a. $\mathsf{R}$-symmetry indices) \\ range: $\{1,2\}$ \end{tabular} \\ \hline
  $I, J, K, \ldots$ & \begin{tabular}{@{}ll@{}} UV &: adjoint-valued indices for $\fg := \LIE(G)$, the Yang-Mills gauge Lie algebra \\ & \,\, range: $1$ to $\dim \fg$ \\ \hline IR &: indices labeling the Abelian vectormultiplets in the IR \\ & \,\, range: 1 to $\rm{rk}(G)$\end{tabular} \\ \hline
\end{tabular}
\caption{\label{tbl:index conventions}Conventions for indices.}
\end{table}

\renewcommand*{\arraystretch}{1.12}
\begin{longtable}[c]{|c|l|c|}
\hline
Symbol & Description & \begin{tabular}{@{}l@{}} Reference \end{tabular} \\
\hline\hline
 \endhead
 \caption{\label{tbl:symbol list}List of symbols used in the paper.}
 \endfirstfoot
 \caption[]{(Continued) List of symbols used in the paper.}
 \endfoot
 $a$ & IR notation for  vectormultiplet scalar $\phi$ & Sec. \ref{subsec:degree-0-IR-action} \\ \hline
 $\ov{a}$ & IR notation for  vectormultiplet scalar $\lambda$ & Sec. \ref{subsec:degree-0-IR-action} \\ \hline
 $\CA(P)$ & the affine space of connections on $P$ & Sec. \ref{subsec:GaugeCartan} \\ \hline
 $\adsf P$ & the adjoint bundle $P \times_{G} \fg$ & Sec. \ref{subsec:GaugeCartan} \\ \hline
 $\mathscr{A}$ & a graded commutative superalgebra & Sec. \ref{sec:CartanModels} \\ \hline
 $A_{\mu}$ & \begin{tabular}{@{}l@{}} bosonic degree 0 field of the Cartan Model \\ connection on principal $G$-bundle (Yang-Mills gauge field) \end{tabular}  & \begin{tabular}{@{}c@{}} Sec. \ref{subsec:CartanModelDiff} \\ Sec. \ref{sec:SuperconformalGrav-twistedSYM} \end{tabular} \\ \hline
 $\sfA_+$ & bosonic real $0$-form in the untwisted and twisted chiral multiplet & \begin{tabular}{@{}c@{}} Sec. \ref{subsec:supconf-chiral-antichiral} \\ \eqref{eq:chiral-to-vm Aplus} \end{tabular} \\ \hline
 $\sfA_-$ & bosonic real $0$-form in the untwisted and twisted antichiral multiplet & \begin{tabular}{@{}c@{}} Sec. \ref{subsec:supconf-chiral-antichiral} \\ \eqref{eq:antichiral-to-vm Aminus} \end{tabular} \\ \hline
 $\ARsym_{\mu}$ & $\mathfrak{so(1,1)}_{\mathsf{R}}$ R-symmetry connection & Sec. \ref{subsec:SuperconformalGravRev} \\ \hline
 $b$ & shorthand for $b_2^+(\IX)$ Sec. \ref{sec:WC-LeadingSingularity} \\ \hline
 $b_1$ & dimension of $\sfH^{1}(\IX; \IR)$ & Sec. \ref{sec:IntroductionConclusion} \\ \hline
 $b_2^+$ & dimension of $\sfH^{2,+}(\IX; \IR)$ &  Sec. \ref{sec:WallCrossing} \\\hline
 $b_{\mu}$ & dilatation gauge field in untwisted conformal supergravity & Sec. \ref{subsec:SuperconformalGravRev}\\ \hline
 $\sfB_{+(ij)}$ & bosonic symmetric field in the untwisted chiral multiplet & Sec. \ref{subsec:supconf-chiral-antichiral} \\ \hline
 $\sfB_{-(ij)}$ & bosonic symmetric field in the untwisted antichiral multiplet & Sec. \ref{subsec:supconf-chiral-antichiral} \\ \hline
 $\sfB_{+[\mu\nu]}$ & bosonic SD 2-form in the twisted chiral multiplet & \begin{tabular}{@{}c@{}} Sec. \ref{subsec:supconf-chiral-antichiral} \\ \eqref{eq:chiral-to-vm Bplus} \end{tabular} \\ \hline
 $\sfB_{-[\mu\nu]}$ & bosonic SD 2-form in the twisted antichiral multiplet & \begin{tabular}{@{}c@{}} Sec. \ref{subsec:supconf-chiral-antichiral} \\ \eqref{eq:antichiral-to-vm Bminus} \end{tabular} \\ \hline
 $\mathsf{BDiff}^{+}(\IX)$ & classifying space of the orientation-preserving diffeomorphism group of $\IX$ & Sec. \ref{sec:IntroductionConclusion} \\ \hline
 $\mathsf{BDiff}_{p}^{+}$ & the observer moduli space of $\IX$ with a fixed $p \in \IX$ the observer & Sec. \ref{sec:toobsfd} \\ \hline
 $B\IG$ & the classifying space of $\IG$ & Sec. \ref{sec:CartanModels} \\\hline
 $\mathscr{B}$ & gauge equivalence classes of connections $\CA/\CG$ & Sec. \ref{subsec:ReviewObservablesDonaldson}\\ \hline
  $\widetilde{\mathscr{B}}$ & gauge equivalence classes of framed connections $\CA/\CG_{0}$ & Sec. \ref{subsec:ReviewObservablesDonaldson}\\ \hline
    $\widetilde{\mathscr{B}}^*$ & gauge equivalence classes of irreducible framed connections & Sec. \ref{subsec:ReviewObservablesDonaldson}\\ \hline
 $\CB$ & base of the universal bundle of the family & Sec. \ref{sec:toobsfd} \\ \hline
 $\sfC_{\mathsf{IR}}$ & $\CQ$-non-exact part of the degree zero IR action & \eqref{eq:IRC} \\ \hline
 $\sfC_+$ & bosonic $0$-form in the untwisted and twisted chiral multiplet & \begin{tabular}{@{}c@{}} Sec. \ref{subsec:supconf-chiral-antichiral} \\ \eqref{eq:chiral-to-vm Cplus} \end{tabular} \\ \hline
 $\sfC_-$ & bosonic $0$-form in the untwisted and twisted antichiral multiplet & \begin{tabular}{@{}c@{}} Sec. \ref{subsec:supconf-chiral-antichiral} \\ \eqref{eq:Cminus deg 0}--\eqref{eq:Cminus deg 4} \end{tabular} \\ \hline
 $\sfC_{\text{sugra}}$ & $\IQ$-non-exact part of twisted supergravity action, $\sfC_{\text{sugra}} := \frac{i}{16\pi}\int_{\IX}\sqrt{g}\sfC_{-}$ & Sec. \ref{sec:TwistedSugraAction} \\ \hline
 $\mathsf{c}_{\mu\nu}^{+}$ & bosonic SD $2$-form S-susy ghost & \eqref{eq:sym-grav-ghost} \\ \hline
 \begin{tabular}{@{}c@{}} $C_{ABCD}$ \\ $C_{\dA\dB\dC\dD}$  \end{tabular} & building blocks of $R_{A\dA B\dB C\dC D\dD}(\omega)$ & \eqref{eq:riemann spinor} \\ \hline
 $\chi_{\mu\nu}$ & \begin{tabular}{@{}l@{}} odd degree $-1$ field in the Localization Multiplet for $\sfH_{\CG}^{\bullet}(\CA(P))$ \\ adjoint-valued two-form twisted gaugino \end{tabular} & \begin{tabular}{@{}c@{}} Sec. \ref{subsec:GaugeCartan} \\ Sec. \ref{sec:SuperconformalGrav-twistedSYM} \end{tabular} \\ \hline
 $\chi_{i}$ & Grassman-odd linear expansion coefficients for zeromodes of $\chi$ & \eqref{eq:chi-zero-modes-gen}  \\ \hline
 $\mathcal{D}$ & generator of dilatations & Sec. \ref{sec:TwistedScfmlGrav} \\ \hline
 $d$ & de Rham exterior differential on $\IX$ & Sec. \ref{sec:IntroductionConclusion} \\ \hline
 $\sfd$ & the Cartan differential for $\sfH_{\DIFF(\IX)}\big(\MET(\IX)\big)$ & Sec. \ref{sec:CartanModels} \\\hline
 $\sfd_{\mathsf{Met}}$ & de Rham exterior differential on $\MET(\IX)$ & Sec. \ref{subsec:CartanModelDiff}\\ \hline
 $\widetilde{\sfd}$ & lift of $\sfd$ to projected bundle of $\Omega_{g}^{2,+}(\MET(\IX))$ over $\MET(\IX)$ & \eqref{eq:dtilde-action} \\ \hline
 $\bcd_{\mu}$ &\begin{tabular}{@{}l@{}} (metric + dilatation + $\mathsf{R}$-symmetry)-covariant \\ derivative in the untwisted theory\end{tabular} &  \begin{tabular}{@{}c@{}} Sec. \ref{subsec:Superconformal-TwistTruncate} \\ App. \ref{app:csugra-misc} \end{tabular} \\ \hline
 $D_{\mu}$ & (metric + gauge)-covariant derivative & Sec. \ref{subsec:GaugeCartan}\\ \hline
 $\scd_{\mu}$ & (supercovariant + gauge-covariant) derivative & \begin{tabular}{@{}c@{}} Sec. \ref{subsec:Superconformal-TwistTruncate} \\ App. \ref{app:csugra-misc} \end{tabular} \\ \hline
 $\bm{\Box}_{C}$ & superconformal d'Alembertian & \begin{tabular}{@{}c@{}}\eqref{eq:box phi}--\eqref{eq:box lambda} \\ App. \ref{app:misc-expr-twisted} \end{tabular} \\ \hline
 $\mathscr{D}$ & auxiliary scalar field in twisted and untwisted supergravity & \begin{tabular}{@{}c@{}} Sec. \ref{subsec:SuperconformalGravRev} \\ \eqref{eq:Dscr} \end{tabular} \\ \hline
 $D_{\mu\nu}$ & \begin{tabular}{@{}l@{}}adjoint-valued self-dual auxiliary field \\ in the twisted $\CN=2$ vectormultiplet \end{tabular} & \begin{tabular}{@{}c@{}} Sec. \ref{subsec:FullEqCohCartanModelSummary} \\ Sec. \ref{sec:SuperconformalGrav-twistedSYM} \end{tabular}\\ \hline
 $\diff(\IX)$ & the Lie algebra of $\DIFF(\IX)$ & Sec. \ref{subsec:CartanModelDiff} \\ \hline  
 $\DIFF(\IX)$ & the group of orientation-preserving diffeomorphisms of $\IX$ & Sec. \ref{sec:IntroductionConclusion} \\ \hline
  $\DIFF_{p}(\IX)$ & \begin{tabular}{@{}l@{}} subgroup of $\DIFF(\IX)$ that fixes $p \in \IX$ with \\ induced tangent map at $x_{0}$ equal to the identity map \end{tabular}  & \eqref{eq:diffp observer} \\ \hline
  $D_{\epsilon}$ & small disk of radius $\epsilon$ & Sec. \ref{sec:WC-Multiparameter} \\ \hline
 $\delta$ & variation & Sec. \ref{sec:TwistedScfmlGrav} \\ \hline
 $\delta'$ & representative of $\sfH_{1}(\IX)$ & Sec. \ref{sec:WallCrossing} \\ \hline
 $\delta_\phi$ & gauge transformation by $\phi$ & Sec. \ref{subsec:GaugeCartan} \\ \hline
 $\Delta_{\epsilon}$ & boundary of $D_{\epsilon}$ & Sec. \ref{sec:WC-Multiparameter} \\ \hline
 $\bm{\Delta}_{H}$ & a part of $\IQ^{(-1,2)}$ & \eqref{eq:DeltaH-action} \\ \hline 
 $\bm{\Delta}_{A\dA,B\dB}{}^{i}$ & building blocks of $\bm{S}_{\mu}{}^{i}$ in untwisted conformal supergravity & \eqref{eq:untwisted building block Delta left}--\eqref{eq:untwisted building block Delta right} \\ \hline
 $\Delta_{\mathsf{diff}}$ & integrated part of the gauge fermion for the comparison in Section \ref{sec:CompareActions} & \eqref{eq:DiffActionFinal} \\ \hline
 $\widetilde{\Delta}_{\mathsf{diff}}$ & part of the gauge fermion for the comparison in Appendix \ref{app:ActionComparisonDetails} & \eqref{eq:DiffActionAppendix} \\ \hline
 $e_{\mu}{}^{a}$ & vielbein (frame field) on $\IX$ & Sec. \ref{subsec:SuperconformalGravRev} \\ \hline
 $e$ & $e := \sqrt{\det\,g} = \sqrt{g}$ & Sec. \ref{sec:TwistedSugraAction} \\ \hline
 $\veps_{AB}$ & invariant tensor of $\mathfrak{su(2)}_{+}$ & App. \ref{app:conventions}\\ \hline
 $\veps_{\dA\dB}$ & invariant tensor of $\mathfrak{su(2)}_{-}$ & App. \ref{app:conventions} \\ \hline
 \scalebox{0.9}{$\veps_{A\dA,B\dB,C\dC,D\dD}$} & Levi-Civita tensor with spinor indices & \eqref{eq:vareps} \\ \hline
 $\veps_{\mu\nu\rho\sigma}$ & Levi-Civita symbol with curved indices (invariant density) & App. \ref{app:conventions} \\ \hline
 $\veps_{\ha\hb\hg\hd}$ & Levi-Civita symbol with flat indices (invariant tensor) & App. \ref{app:conventions} \\ \hline
 $\epsilon$ & an infinitesimal real number (non-boldfaced) & Sec. \ref{sec:WC-Multiparameter} \\ \hline
 $\eps$ & scalar supersymmetry parameter in twisted supergravity (boldfaced) & Sec. \ref{subsec:Superconformal-TwistTruncate} \\ \hline
 $\eps_{\mu}$ & 1-form supersymmetry parameter in twisted supergravity & Sec. \ref{subsec:Superconformal-TwistTruncate} \\ \hline
 $\eps^{i}$ & \begin{tabular}{@{}l@{}}supersymmetry parameter in untwisted \\ conformal supergravity (Dirac spinor) \end{tabular} & Sec. \ref{subsec:Superconformal-TwistTruncate} \\ \hline
  $\eta$ & \begin{tabular}{@{}l@{}} degree $-1$ field in the Projection Multiplet for $\sfH_{\CG}^{\bullet}(\CA(P))$ \\ adjoint-valued zero-form twisted gaugino \end{tabular} & 
 \begin{tabular}{@{}c@{}} Sec. \ref{subsec:GaugeCartan} \\ Sec. \ref{sec:SuperconformalGrav-twistedSYM} \end{tabular} \\ \hline
  $\bn^{i}$ & \begin{tabular}{@{}l@{}}conformal supersymmetry parameter in untwisted \\ conformal supergravity (Dirac spinor) \end{tabular} & Sec. \ref{subsec:Superconformal-TwistTruncate} \\ \hline
 $E$ & rank two vector bundle associated to $\mathsf{SU(2)}$ bundle over $\IX$ & Sec. \ref{sec:WallCrossing} \\ \hline
 $\IE_{g}^{\text{gauge}}$ & $\CG$-invariant subcomplex of $\widetilde{\IE}_{g}^{\text{gauge}}$ &   \eqref{eq:complex-GaugeCartan} \\ \hline
 $\widetilde{\IE}_{g}^{\text{gauge}}$ & complex for $\sfH_{\CG}^{\bullet}(\CA(P))$ with (anti)ghosts + auxiliary fields &   \eqref{eq:BigVM-Complex} \\ \hline
 $\IE^{\text{diffeo}}$ & $\DIFF(\IX)$-invariant subcomplex of $\widetilde{\IE}^{\text{diffeo}}$ &   \eqref{eq:complex-DiffCartan} \\ \hline
 $\widetilde{\IE}^{\text{diffeo}}$ & complex for $\sfH_{\DIFF(\IX)}^{\bullet}(\MET(\IX))$ with (anti)ghosts + auxiliary fields &   \eqref{eq:Gravity-DiffCartan} \\ \hline
 $E\IG$ & the total space of the universal bundle over $B\IG$ & Sec. \ref{sec:CartanModels} \\ \hline
 $E\IG \times_{\IG} \IM$ & \begin{tabular}{@{}l@{}}the quotient $(E\IG \times \IM)/\IG$, where the $\IG$ action is \\ $(h, x) \mapsto g\cdot(h,x) = (hg^{-1}, gx)$,\\ for $(h, x) \in E\IG \times \IM$ and $g \in \IG$ \end{tabular} & Sec. \ref{sec:CartanModels} \\ \hline
 $F$ & \begin{tabular}{@{}l@{}} adjoint-valued 2-form Yang-Mills field strength \\ curvature of a $G$ connection on $P$ \end{tabular} & \begin{tabular}{@{}c@{}} Sec. \ref{subsec:FullEqCohCartanModelSummary} \\ Sec. \ref{sec:SuperconformalGrav-twistedSYM} \end{tabular} \\ \hline
 $\widehat{F}$ & \begin{tabular}{@{}l@{}} supercovariant Yang-Mills field strength \end{tabular} &   \eqref{eq:tsugra-supercov-ym-curvature}\\ \hline
 $\sfF_{\mu\nu}^+$ & bosonic SD $2$-form in the untwisted and twisted chiral multiplet & \begin{tabular}{@{}c@{}} Sec. \ref{subsec:supconf-chiral-antichiral} \\ \eqref{eq:chiral-to-vm Fplus} \end{tabular} \\ \hline
 $\sfF_{\mu\nu}^-$ & bosonic ASD $2$-form in the untwisted and twisted antichiral multiplet & \begin{tabular}{@{}c@{}} Sec. \ref{subsec:supconf-chiral-antichiral} \\ \eqref{eq:antichiral-to-vm Fminus} \end{tabular} \\ \hline
 $\CF$ & homogeneous degree $2$ prepotential $\CF \equiv \CF(\phi)$ or $\CF(a)$ & Sec. \ref{subsec:action-in-vm-language} \\ \hline
 $\ov{\CF}$ & homogeneous degree $2$ prepotential $\ov{\CF} \equiv \CF(\lambda)$ or $\ov{\CF}(\ov{a})$ & Sec. \ref{subsec:action-in-vm-language} \\ \hline
 $1/g^2$ & spacetime-independent coupling constant & Sec. \ref{subsec:degree-0-UV-action} \\ \hline
 $g_{\mu\nu}$ & Riemannian metric on $\IX$ & Sec. \ref{sec:IntroductionConclusion} \\ \hline
 $g^{(0)}$ & Special metric with $\lambda_{+} = 0$ & Sec. \ref{sec:WC-LeadingSingularity} \\ \hline
 $\fg$ & the Lie algebra of $G$, i.e. $\fg := \LIE(G)$ & Sec. \ref{subsec:CartanModelDiff} \\ \hline
 $G$ & the Yang-Mills gauge group; a compact Lie group & Sec. \ref{subsec:ReviewObservablesDonaldson}\\ \hline
 $\CG$ & \begin{tabular}{@{}l@{}} group of gauge transformations, i.e. the set of fiber-preserving \\ automorphisms of $\adsf P$, that cover the identity map over $\IX$\end{tabular} & Sec. \ref{sec:IntroductionConclusion}\\\hline
 $\CG_0$ & subgroup of $\CG$ that trivialize at a point in $\IX$ & Sec. \ref{subsec:ReviewObservablesDonaldson}\\\hline
 $\IG$ & the semidirect product group $\CG \rtimes \DIFF(\IX)$ &   \eqref{eq:SemiDirect-Intro} \\ \hline
 $\sfG_+$ & left-handed chiral spinor in the untwisted chiral multiplet & Sec. \ref{subsec:supconf-chiral-antichiral} \\ \hline
 $\sfG_-$ & right-handed chiral spinor in the untwisted antichiral multiplet & Sec. \ref{subsec:supconf-chiral-antichiral} \\ \hline
 $\sfG$ & fermionic $0$-form in the twisted chiral multiplet & \begin{tabular}{@{}c@{}} Sec. \ref{subsec:supconf-chiral-antichiral} \\ \eqref{eq:chiral-to-vm G} \end{tabular} \\ \hline
 $\sfG_{\mu}$ & fermionic $1$-form in the twisted antichiral multiplet & \begin{tabular}{@{}c@{}} Sec. \ref{subsec:supconf-chiral-antichiral} \\ \eqref{eq:antichiral-to-vm Gmu} \end{tabular} \\ \hline
 $\sfG_{\mu\nu}$ & fermionic SD $2$-form in the twisted chiral multiplet & \begin{tabular}{@{}c@{}} Sec. \ref{subsec:supconf-chiral-antichiral} \\ \eqref{eq:chiral-to-vm Gmunu} \end{tabular} \\ \hline
 $\gamma^a$ & Euclidean Dirac matrices with flat indices & \eqref{eq:euclidean dirac matrices} \\ \hline
 $\gamma^5$ & Euclidean chiral gamma matrix & \eqref{eq:gamma5 in terms of euclidean dirac matrices} \\ \hline
 $H_{\mu\nu}$ & \begin{tabular}{@{}l@{}} even degree $0$ field in the Localization Multiplet for $\sfH_{\CG}^{\bullet}(\CA(P))$ \\ bosonic self-dual 2-form auxiliary field \end{tabular} & \begin{tabular}{@{}c@{}} \eqref{eq:CartanModel-AlternateAuxField} \end{tabular} \\ \hline
 $\sfH_{\bullet}(\circ)$ & homology groups of $\circ$ & Sec. \ref{sec:IntroductionConclusion} \\ \hline
 $h^\vee$ & dual Coxeter number of $G$ & Sec. \ref{sec:IntroductionConclusion} \\ \hline
 $\sfH^{\bullet}(\circ)$ & cohomology groups of $\circ$ & Sec. \ref{sec:IntroductionConclusion} \\ \hline
 $\sfH_{\IG}(\IM)$ & the $\IG$-equivariant cohomology of $\IM$ &   \eqref{eq:borel construction} \\ \hline
 $\iota$ & interior derivative on $\IX$ & Sec. \ref{sec:IntroductionConclusion} \\ \hline
 $\iota_{\mathbb{A}}$ & interior derivative on $\mathscr{A}$ & Sec. \ref{sec:CartanModels} \\ \hline
 $\CJ$ & the curl of the gravitino & \begin{tabular}{@{}c@{}} Sec. \ref{subsec:Superconformal-TwistTruncate} \\ App. \ref{subsec:curl-twisted-gravitino} \end{tabular}\\ \hline
 $J$ & period point & Sec. \ref{sec:WallCrossing} \\ \hline
 $\mathsf{K}$ & a part of $\IQ^{(-1,2)}$ & \eqref{eq:Kaction-1}--\eqref{eq:Kaction-4} \\ \hline
 $\mathsf{K}_{\mathsf{R}}$ & \begin{tabular}{@{}l@{}} total space of the vector bundle associated \\ (via the fundamental pseudoreal representation) to $\mathsf{P}_{\mathsf{R}} \to \IX$ \end{tabular} & Sec. \ref{sec:TwistedScfmlGrav} \\ \hline
 $K$ & one-form supersymmetry operator in standard Donaldson-Witten theory & Sec. \ref{subsec:ReviewObservablesDonaldson} \\ \hline
 $K_\ha$ & generator of special conformal transformations & Sec. \ref{sec:TwistedScfmlGrav} \\ \hline
  $\mathscr{K}^{\mu}$ & total derivative term in the Lagrangian & \eqref{eq:GeneralActionTotDer}  \\ \hline
  $L$ & line bundle on $\IX$ admitting an ASD connection & Sec. \ref{sec:WallCrossing} \\ \hline
 $\mathscr{L}_{\text{cdf}}$ & chiral density formula & \eqref{eq:cdf} \\ \hline
 $\mathscr{L}_{\text{chiral}}$ & chiral contribution to the chiral density formula & \eqref{eq:Lchiral} \\ \hline
 $\mathscr{L}_{\text{antichiral}}$ & antichiral contribution to the chiral density formula & \eqref{eq:Lantichiral} \\ \hline
 $\mathscr{L}_{k}$ & \begin{tabular}{@{}l@{}} gravitational degree $k$ part of the action of twisted SYM\\ coupled to twisted supergravity ($k \in \{ 0, 1, 2, 3, 4 \}$) \end{tabular}   & \eqref{eq:Lsugra-deg0-rescaled}--\eqref{eq:Lsugra-deg4-rescaled} \\ \hline
 $\lambda$ & \begin{tabular}{@{}l@{}} degree $-2$ field in the Projection Multiplet for $\sfH_{\CG}^{\bullet}(\CA(P))$ \\ adjoint-valued scalar in the twisted vectormultiplet \end{tabular} & \begin{tabular}{@{}c@{}} Sec. \ref{subsec:GaugeCartan} \\ Sec. \ref{sec:SuperconformalGrav-twistedSYM} \end{tabular} \\ \hline
 $\lambda'$ & class in $\sfH^{2}(\IX; \IZ)$ & Sec. \ref{sec:WC-LeadingSingularity} \\ \hline
 $\lambda_{0}$ & fixed element in $\sfH^{2}(\IX; \IZ) + \frac{1}{2}w_{2}(E)$ & Sec. \ref{sec:WC-LeadingSingularity} \\ \hline
 $\bm{\lambda}^{\ha\hb}$ & parameter for Lorentz transformations & Sec. \ref{sec:TwistedScfmlGrav} \\ \hline
 $\bm{\Lambda}_{K,A\dA}$ & \begin{tabular}{@{}l@{}} parameter for special conformal transformations in \\ untwisted conformal supergravity \end{tabular} & Sec. \ref{subsec:Superconformal-TwistTruncate} \\ \hline
 $\sfL_{+}$ & left-handed chiral spinor in the untwisted chiral multiplet & Sec. \ref{subsec:supconf-chiral-antichiral} \\ \hline
 $\sfL_{-}$ & right-handed chiral spinor in the untwisted chiral multiplet & Sec. \ref{subsec:supconf-chiral-antichiral} \\ \hline
 $\sfL$ & fermionic $0$-form in the twisted chiral multiplet & \begin{tabular}{@{}c@{}} Sec. \ref{subsec:supconf-chiral-antichiral} \\ \eqref{eq:chiral-to-vm L} \end{tabular} \\ \hline
 $\sfL_{\mu}$ & fermionic $1$-form in the twisted antichiral multiplet & \begin{tabular}{@{}c@{}} Sec. \ref{subsec:supconf-chiral-antichiral} \\ \eqref{eq:antichiral-to-vm Lmu} \end{tabular} \\ \hline
 $\sfL_{\mu\nu}$ & fermionic SD $2$-form in the twisted chiral multiplet & \begin{tabular}{@{}c@{}} Sec. \ref{subsec:supconf-chiral-antichiral} \\ \eqref{eq:chiral-to-vm Lmunu} \end{tabular} \\ \hline
 $\Lambda_{(\mu\nu)}$ & \begin{tabular}{@{}l@{}} coefficient of gravitational degree one in the Lagrangian \\ symmetric twisted supercurrent \end{tabular} & \begin{tabular}{@{}c@{}} \eqref{eq:S gravity expansion} \\ \eqref{eq:GeneralActionDegree1} \end{tabular} \\ \hline
 $L_{\mathbb{A}}$ & Lie derivative on $\mathscr{A}$ & Sec. \ref{sec:CartanModels} \\ \hline
 $\CL_\Phi$ & Lie derivative on $\IX$ along $\Phi$ &   \eqref{eq:sfd squared} \\ \hline
 $\CL_\Phi^{(A)}$ & gauge-covariantized Lie derivative on $\IX$ along $\Phi$ &   \eqref{eq:CartanSquared} \\ \hline
 $\mu_{D}$ & Donaldson map &   \eqref{eq:DonaldsonMap} \\ \hline
 $\IM$ & the product space $\CA(P) \times \MET(\IX)$ &   \eqref{eq:IM family} \\ \hline
 $\IM_{\IG}$ & the homotopy quotient & \eqref{eq:homotopy-quotient} \\ \hline
 \begin{tabular}{@{}c@{}}$\CM$ \\or $\CM_{\mathsf{ASD},g}$ \end{tabular} & \begin{tabular}{@{}l@{}}moduli space of anti-self-dual connections on $P$ \\ with respect to metric $g$ on $\IX$ \end{tabular} & \begin{tabular}{@{}c@{}} Sec. \ref{sec:IntroductionConclusion} \\ Sec. \ref{sec:WallCrossing} \end{tabular} \\ \hline
 \begin{tabular}{@{}c@{}}$\wh{\CM}_{\mathsf{ASD},g}$ \end{tabular} & \begin{tabular}{@{}l@{}} compactified moduli space of anti-self-dual connections on $P$ \\ with respect to metric $g$ on $\IX$ \end{tabular} & \begin{tabular}{@{}c@{}} Sec. \ref{sec:WallCrossing} \\ Sec. \ref{sec:Observables} \end{tabular} \\ \hline
 $M_{\ha\hb}$ & generator of Lorentz transformations & Sec. \ref{sec:TwistedScfmlGrav} \\ \hline
 $\CM_{k}$ & \begin{tabular}{@{}l@{}}component of the moduli space of instantons\\ with instanton number $k$ \end{tabular} & Sec. \ref{sec:IntroductionConclusion} \\ \hline
 $\MET(\IX)$ & the space of Riemannian metrics on $\IX$ & Sec. \ref{sec:IntroductionConclusion} \\  \hline
 $\CO^{(k)}$ & $k$-form observable density & \begin{tabular}{@{}c@{}} Sec. \ref{subsec:ReviewObservablesDonaldson} \\ \eqref{eq:obs-den-0}--\eqref{eq:obs-den-4} \end{tabular} \\ \hline
 $\omega_{\mu}{}^{[ab]}$ & spin connection on $\IX$ & Sec. \ref{sec:TwistedScfmlGrav} \\ \hline
 $\omega_{\mu}{}^{(AB)}$ & self-dual part of spin connection on $\IX$ & Sec. \ref{sec:TwistedScfmlGrav} \\ \hline
 $\omega_{\mu}{}^{(\dA\dB)}$ & anti-self-dual part of spin connection on $\IX$ & Sec. \ref{sec:TwistedScfmlGrav} \\ \hline
 $\omega_{i}(t)$ & basis of self-dual harmonic 2-forms ($i \in \{1, \ldots, b_2^+\}$)  & \eqref{eq:chi-zero-modes-gen}\\ \hline
 $\Omega^{\bullet}(\circ)$ & space of differential forms on $\circ$ & Sec. \ref{sec:IntroductionConclusion}\\ \hline
 $\n$ & affine connection (or metric covariant derivative $\n_\mu$) & Sec. \ref{sec:CartanModels} \\ \hline
 $\n_{\mathsf{R}}$ & $\mathsf{R}$-symmetry connection & Sec. \ref{sec:IntroductionConclusion} \\ \hline
 $\wp$ & point on $\IX$ & Sec. \ref{sec:Observables} \\ \hline
 $(p_{\alpha})_{\mu\nu}$ & \begin{tabular}{@{}l@{}}linearly independent symmetric tensors for $n$-dimensional \end{tabular} & \eqref{eq:n-param-family} \\ \hline
 $P_{\ha}$ & generator of translations & Sec. \ref{sec:TwistedScfmlGrav} \\ \hline
 $P_{D}$ & Donaldson polynomial & \eqref{eq:DonaldsonPoly} \\ \hline
 $P$ & the total space of a principal $G$-bundle $P \to \IX$ over $\IX$ & Sec. \ref{sec:IntroductionConclusion}  \\ \hline
 $\sfP_{\sfR}$ & the total space of the principal $\mathsf{SU(2)}_{\mathsf{R}}$ $\mathsf{R}$-symmetry bundle over $\IX$ & Sec. \ref{sec:IntroductionConclusion} \\ \hline
 $\varpi$ & class in $\sfH^{2d}(BG)$ & Sec. \ref{subsec:ReviewObservablesDonaldson} \\ \hline
 $\pi_f$ & projection defining the universal bundle & \eqref{eq:family_fib}--\eqref{eq:family_fib2} \\ \hline
 $\mathscr{P}^{\mu}$ & vector field entering the surface term in Section \ref{sec:CompareActions} & \eqref{eq:DiffActionFinal} \\ \hline
 $\widetilde{\mathscr{P}}^{\mu}$ & $\widetilde{\mathscr{P}}^{\mu} := e^{-1} \mathscr{P}^{\mu}$ where $e = \sqrt{g}$ & \eqref{eq:DiffActionAppendix} \\ \hline
 $\phi$ & \begin{tabular}{@{}l@{}}degree two bosonic generator in the Cartan model for $\sfH_{\CG}^{\bullet}(\CA(P))$ \\ adjoint-valued scalar in the $\CN=2$ twisted vectormultiplet  \end{tabular} & \begin{tabular}{@{}c@{}} Sec. \ref{subsec:GaugeCartan} \\ Sec. \ref{sec:SuperconformalGrav-twistedSYM} \end{tabular} \\ \hline
 $\Phi^{\mu}$ & \begin{tabular}{@{}l@{}} degree two bosonic generator of the Cartan model for $\sfH_{\DIFF(\IX)}^{\bullet}(\MET(\IX))$ \\ bosonic BRST ghost for vector supersymmetry  \end{tabular} & \begin{tabular}{@{}c@{}} Sec. \ref{subsec:CartanModelDiff} \\ Sec. \ref{subsec:Diff-Cartan-from-TwistedSugra} \end{tabular} \\ \hline
 $\Phi_{\rm vm}$ & vectormultiplet measure (or collection of vectormultiplet fields) & Sec. \ref{sec:IntroductionConclusion} \\ \hline
 $\bm{Q}^i$ & generator of supersymmetry transformations (Dirac spinor) & Sec. \ref{sec:TwistedScfmlGrav} \\ \hline
 $\CQ$ & \begin{tabular}{@{}l@{}} Cartan differential for $\sfH_{\CG}^{\bullet}(\CA(P))$  \end{tabular} &  \eqref{eq:GaugeCartan-1}--\eqref{eq:GaugeCartan-3} \\ \hline
 $\IQ$ & \begin{tabular}{@{}l@{}} Cartan differential for $\sfH_{\CG\rtimes\DIFF(\IX)}^{\bullet}(\CA(P)\times\MET(\IX))$ \\ BRST operator for twisted supergravity \end{tabular} & \begin{tabular}{@{}c@{}} \eqref{eq:CartanAlgebra-1}--\eqref{eq:CartanAlgebra-7prime} \\ \eqref{eq:CartanSugra-1}--\eqref{eq:CartanSugra-7} \end{tabular} \\ \hline
 $Q_{i\alpha}^{\lambda}(t)$ & square matrix entering the wall-crossing formula & \eqref{eq:WCF-QMatrix} \\ \hline
 \scalebox{1}{$R_{A\dA B\dB C\dC D\dD}$} & Riemann tensor with spinor indices & \eqref{eq:riemann spinor}\\ \hline
  $\mathscr{R}$ & Ricci scalar of $\IX$ & \eqref{eq:riemann spinor} \\ \hline
  \begin{tabular}{@{}c@{}} $R_{\dA\dB CD}$ \\ $R_{AB\dC\dD}$  \end{tabular} & building blocks of $R_{A\dA B\dB C\dC D\dD}(\omega)$ & \eqref{eq:riemann spinor} \\ \hline
  $R_{ABCD}(\omega)$ & all dotted contraction of $R_{A\dA B\dB C\dC D\dD}(\omega)$ & \eqref{eq:RABCD} \\ \hline
  $R(P)_{\mu\nu}{}^{a}$ & supercovariant curvature for translations & \begin{tabular}{@{}c@{}} Sec. \ref{sec:TwistedScfmlGrav} \\  \eqref{eq:RP-supercurvature} \end{tabular} \\ \hline
   $\bm{R(Q)}_{\mu\nu}{}^{i}$ & supercovariant curvature for $\bm{Q}^i$ in untwisted conformal supergravity & \begin{tabular}{@{}c@{}} Sec. \ref{sec:TwistedScfmlGrav} \\ \eqref{eq:RQ undotted}--\eqref{eq:RQ dotted} \end{tabular} \\ \hline
  $R(Q)_{\mu\nu}^{-}{}^{\rho}$ & supercovariant curvature for $\CQ$ in twisted supergravity & \eqref{eq:sym-grav-RQ-twisted-4-component} \\ \hline
   $\bm{R(S)}_{\mu\nu}{}^{i}$ & supercovariant curvature for $\bm{S}^i$ in untwisted conformal supergravity & \begin{tabular}{@{}c@{}} Sec. \ref{sec:TwistedScfmlGrav} \\ \eqref{eq:RQ undotted}--\eqref{eq:RQ dotted} \end{tabular} \\ \hline
 \begin{tabular}{@{}c@{}} $\sigma_a$ \\ $\widetilde{\sigma_a}$ \end{tabular} & sigma matrices  & \eqref{eq:sigma matrices} \\ \hline
 $\Sigma$ & $2$-cycle on $\IX$ & Sec. \ref{sec:Observables} \\ \hline
 $\Sigma_j$ & $j$-cycle on $\IX$ (where $j \in \{0, 1, 2, 3, 4\}$) & Sec. \ref{sec:Observables} \\ \hline
 $\mathsf{Sym}^{\bullet}(\circ)$ & $\bullet$--fold symmetric product of a vector space (or bundle) $\circ$ & Sec. \ref{subsec:CartanModelDiff} \\ \hline
 $S_{+}\wh{\IX}$ & left-handed or positive chiral spin bundle on $\wh{\IX}$ & Sec. \ref{subsec:supconf-chiral-antichiral}\\ \hline 
 $S_{-}\wh{\IX}$ & right-handed or negative chiral spin bundle on $\wh{\IX}$ & Sec. \ref{subsec:supconf-chiral-antichiral}\\ \hline
 $\bm{S}^i$ & generator of conformal supersymmetry transformations (Dirac spinor) & Sec. \ref{sec:TwistedScfmlGrav} \\ \hline
 $S^\bullet(\circ)$ & symmetric algebra of the Lie algebra $\circ$ & Sec. \ref{subsec:GaugeCartan} \\ \hline
 $\IS_{\text{Cartan}}$ & \begin{tabular}{@{}l@{}} the minimal action of twisted $\CN=2$ SYM \\ coupled to the diffeomorphism Cartan Model \end{tabular} & \eqref{eq:minimalgeneralactioneqn} \\ \hline
 $\IS_{\text{sugra}}$ & the action of twisted $\CN=2$ SYM coupled to twisted supergravity & \eqref{eq:TwistedSugraAction-General-Proposal} \\ \hline
 $\IS_{0}$ & the action of twisted $\CN=2$ SYM coupled to a metric & Sec. \ref{sec:IntroductionConclusion} \\ \hline
 $\mathsf{S}_{\mathsf{UV}}$ & degree $0$ UV Donaldson-Witten action & \eqref{eq:UVActionDeg0} \\ \hline
 $\mathsf{S}_{\mathsf{IR}}$ & degree $0$ IR Donaldson-Witten action & \eqref{eq:IRActionDeg0} \\ \hline
 $\bm{S}_{\mu}{}^{i}$ & \begin{tabular}{@{}l@{}}composite $S$-gravitino (Dirac spinor) \\ in untwisted conformal supergravity \end{tabular} & \eqref{eq:conformal gravitino 2-comp left}--\eqref{eq:DeltaCont2} \\ \hline
 $S_{\mu}$ & \begin{tabular}{@{}l@{}} twisted $1$-form S-gravitino in twisted supergravity \\ composite connection for 0-form S-susy \end{tabular} & \eqref{eq:sym-grav-0-form-S-susy-connection} \\ \hline
 $S_{\mu,[\rho\sigma]}^{+}$ & \begin{tabular}{@{}l@{}} twisted SD $2$-form S-gravitino in twisted supergravity \\ composite connection for 2-form SD 2-form S-susy \end{tabular} & \eqref{eq:sym-grav-2-form-S-susy-connection} \\ \hline
 $S_{\mu\nu}$ & \underline{vanishing} connection for $1$-form $S$-susy in twisted supergravity & \eqref{eq:SgravTrunc-4} \\ \hline
 $\psi_{\mu}$ & \begin{tabular}{@{}l@{}} degree one fermionic generator in the Cartan  model for $\sfH_{\CG}^{\bullet}(\CA(P))$ \\ adjoint-valued one-form gaugino in the $\CN=2$ twisted vectormultiplet \end{tabular} & \begin{tabular}{@{}c@{}} Sec. \ref{subsec:GaugeCartan} \\ Sec. \ref{sec:SuperconformalGrav-twistedSYM} \end{tabular} \\ \hline
 $\bm{\Psi}_{\mu}{}^{i}$ & untwisted gravitino (a Dirac spinor) & Sec. \ref{sec:TwistedScfmlGrav} \\ \hline
 $\Psi_{\mu}{}^{iA}$ & untwisted left chiral gravitino (a Weyl spinor) & Sec. \ref{sec:TwistedScfmlGrav}\\ \hline
 $\Psi_{\mu}{}^{i\dA}$ & untwisted right chiral gravitino (a Weyl spinor) & Sec. \ref{sec:TwistedScfmlGrav} \\ \hline
 $\Psi_{\mu\nu}$ & \begin{tabular}{@{}l@{}} degree $1$ fermionic generator in the Cartan Model for $H_{\DIFF(\IX)}^{\bullet}(\MET(\IX))$ \\ twisted gravitino in twisted supergravity  \end{tabular} & \begin{tabular}{@{}c@{}} Sec. \ref{subsec:CartanModelDiff}  \\ Sec. \ref{subsec:Diff-Cartan-from-TwistedSugra} \end{tabular} \\ \hline
 $\Psi_{A\dA,B\dB}$ & twisted gravitino with spinor indices & \eqref{eq:2comp} \\ \hline
 $\widehat{\Psi}_{(AB),(\dA\dB)}$ & symmetric traceless part of twisted gravitino & \eqref{eq:gravitino 2comp symtraceless} \\ \hline
 $\Psi$ & trace of the twisted gravitino & \eqref{eq:gravitino 2comp trace} \\ \hline
 \begin{tabular}{@{}c@{}} $\Psi_{(AB)}$\\ $\Psi_{[\mu\nu]}^{+}$ \end{tabular} & SD projection of the antisymmetric part of the twisted gravitino & \eqref{eq:gravitino 2comp SD} \\ \hline
 \begin{tabular}{@{}c@{}}$\Psi_{(\dA\dB)}$ \\ $\Psi_{[\mu\nu]}^{-}$ \end{tabular} & ASD projection of the antisymmetric part of the twisted gravitino & \eqref{eq:gravitino 2comp ASD} \\ \hline
 $\Psi_{0}$ & gravitino pulled back to the family & \eqref{eq:gravpullback} \\ \hline
 $t^\alpha$ & deformation parameters for an $n$-dimensional neighbhorhood of $g^{(0)}$ & \eqref{eq:n-param-family} \\ \hline
 $\bu{τ}$ & 3-vector of Pauli matrices $\bu{τ} = (τ_1, τ_2, τ_3)$ & \eqref{eq:pauli matrices} \\ \hline
 $\tau_{0}$ & constant complex coupling  &  \eqref{eq:constant coupling} \\ \hline
 $\ov{\tau}_{0}$ & complex conjugate of $\tau_{0}$ &  \eqref{eq:constant coupling} \\ \hline
 $\tau_{IJ}$ & Hessian of $\CF$, defined as $\tau_{IJ} := \frac{\partial^2\CF}{\partial \phi^I \partial \phi^J}$ & \begin{tabular}{@{}c@{}} Sec. \ref{sec:ProposalForTheAction} \\ Sec. \ref{sec:TwistedSugraAction} \end{tabular}\\ \hline
 $\ov{\tau}_{IJ}$ & Hessian of $\ov{\CF}$, defined as $\ov{\tau}_{IJ} := \frac{\partial^2\ov{\CF}}{\partial \lambda^I \partial \lambda^J}$ & \begin{tabular}{@{}c@{}} Sec. \ref{sec:ProposalForTheAction} \\ Sec. \ref{sec:TwistedSugraAction} \end{tabular}\\ \hline
 $\CT$ & generator of $\mathsf{so(1,1)}_{\mathsf{R}}$ $\mathsf{R}$-symmetry transformations & Sec. \ref{sec:TwistedScfmlGrav} \\ \hline
 $\mathsf{Tor\,}$ & torsion (in cohomology) & Sec. \ref{sec:WC-Multiparameter} \\ \hline
 $\mathsf{Tr}$ & suitably normalized invariant form on $\fg$ & Sec. \ref{sec:IntroductionConclusion} \\ \hline
 $T_{AB}$ & bosonic SD 2-form in untwisted conformal supergravity & Sec. \ref{subsec:SuperconformalGravRev} \\ \hline
 $T_{\dA\dB}$ & bosonic ASD 2-form in untwisted conformal supergravity & Sec. \ref{subsec:SuperconformalGravRev} \\ \hline
 $T_{\mu\nu}^{-}$ & bosonic ASD 2-form in twisted supergravity & \eqref{eq:sym-grav-T} \\ \hline
 $T_{\mu\nu}^{\mathsf{UV}}$ & energy-momentum tensor in the UV theory at degree zero & \eqref{eq:TeeUV} \\ \hline
 $\theta$ & spacetime-independent coupling constant & Sec. \ref{sec:IntroductionConclusion} \\ \hline
 $\bm{\Theta}^{(\mathsf{D})}$ & parameter for dilatations & Sec. \ref{sec:TwistedScfmlGrav} \\ \hline
 $\bm{\Theta}^{(\mathsf{R})}$ & parameter for $\mathfrak{so(1,1)}_{\mathsf{R}}$ $\mathsf{R}$-symmetry transformations & Sec. \ref{sec:TwistedScfmlGrav} \\ \hline
 $\bm{\Theta}^{i}{}_{j}$ & parameter for $\mathfrak{su(2)}_{\mathsf{R}}$ $\mathsf{R}$-symmetry transformations & Sec. \ref{sec:TwistedScfmlGrav} \\ \hline
 $U^{j}{}_{i}$ & generator of $\mathsf{su(2)}_{\mathsf{R}}$ $\mathsf{R}$-symmetry transformations & Sec. \ref{sec:TwistedScfmlGrav} \\ \hline
 $U_{\IX}$ & universal bundle of $\IX$ manifolds & \eqref{eq:universal bundle UX} \\ \hline
 $\Upsilon$ & functional entering the twisted supergravity action, $\Upsilon:= -ie\sqrt{2}\sfL$ & \eqref{eq:TwistedSugraAction-General-Proposal} \\ \hline
  $\Upsilon_{\mu\nu}$ & \begin{tabular}{@{}l@{}} coefficient of the gravitational degree two ($\Psi^{\mu\sigma}\Psi^{\nu}{}_{\sigma}$) term \\ in the Lagrangian \end{tabular} & \begin{tabular}{@{}c@{}} \eqref{eq:S gravity expansion} \\ \eqref{eq:GeneralActionDegree2-2} \end{tabular} \\ \hline
\begin{tabular}{@{}c@{}}  $V_\mu^{\mathsf{R}}$ \\ or $V_{\mu}{}^{i}{}_{j}$ \end{tabular} & $\mathsf{SU(2)}_{\mathsf{R}}$ R-symmetry connection & \begin{tabular}{@{}c@{}} Sec. \ref{sec:IntroductionConclusion} \\ Sec. \ref{sec:TwistedScfmlGrav} \end{tabular} \\ \hline
 $\IV$ & gauge-fermion for the minimal action of Section \ref{sec:minimalgeneralaction} & \eqref{eq:IV} \\ \hline
 $\IV_{\text{sugra}}$ & gauge-fermion for the supergravity action of Section \ref{subsec:action-in-vm-language} & \eqref{eq:TwistedSugraUpsilon} \\ \hline
  $\IV_{\mathsf{UV}}$ & gauge fermion for degree zero UV action & \eqref{eq:UVV} \\ \hline
  $\IV_{\mathsf{IR}}$ & gauge fermion for degree zero IR action & \eqref{eq:IRV} \\ \hline
  $V_{+}$ & cone inside $\sfH^{2}(\IX;\IR)$ & Sec. \ref{sec:WallCrossing} \\ \hline
  $W_{\zeta}$ & real codimension wall & \eqref{eq:real-codim-one-wall} \\ \hline
  $\mathsf{WC}_{\zeta}(p,\Sigma,\delta')$ & wall crossing discontinuity & \eqref{eq:wc-discontinuity} \\ \hline
 $\IX$ & a smooth, closed, oriented, Riemannian 4-manifold & Sec. \ref{sec:IntroductionConclusion} \\     \hline
 $\widehat{\IX}$ & a smooth, closed, Riemannian spin 4-manifold  & Sec. \ref{subsec:SuperconformalGravRev} \\     \hline
 $\CX$ & universal bundle of $\IX$ manifolds & \eqref{eq:family_fib}--\eqref{eq:family_fib2} \\ \hline
 $\bm{\xi}^{a}$ & parameter for translations & Sec. \ref{sec:TwistedScfmlGrav} \\ \hline
 $\bm{\Xi}^{i}$ & auxiliary Dirac spinor in untwisted conformal supergravity & Sec. \ref{subsec:SuperconformalGravRev} \\     \hline
 $\Xi_{\mu}$ & auxiliary fermionic 1-form in twisted supergravity & \eqref{eq:chi vec} \\     \hline
 $\Xi_{\mu\nu}$ & \underline{vanishing} auxiliary fermionic SD 2-form in twisted supergravity & \eqref{eq:chi nonvec} \\     \hline
 $\zeta$ & parameter labeling a wall & Sec. \ref{sec:WallCrossing} \\ \hline
  $Z_{\sigma}$ & \begin{tabular}{@{}l@{}} coefficient of the gravitational degree two ($\Phi^\mu$) term \\ in the Lagrangian \end{tabular} & \begin{tabular}{@{}c@{}} \eqref{eq:S gravity expansion} \\ \eqref{eq:GeneralActionDegree2-1} \end{tabular} \\ \hline
 $\mathsf{Z}_{\mathsf{DW}}$ & Donaldson-Witten path integral & Sec. \ref{sec:IntroductionConclusion} \\ \hline
 $\mathsf{Z}_{\mathsf{family}}$ & family Donaldson-Witten path integral & Sec. \ref{sec:IntroductionConclusion} \\ \hline
 $\mathsf{Z}^{(0)}$ & \begin{tabular}{@{}l@{}}family Donaldson-Witten path integral at gravitational degree $0$,\\ coincides with $\mathsf{Z}_{\mathsf{DW}}$\end{tabular} & Sec. \ref{sec:IntroductionConclusion} \\ \hline
 $\mathsf{Z}^{(p,q)}$ & \begin{tabular}{@{}l@{}}coefficient of term with $p$ $\Psi$'s and $q$ $\Phi$'s \\ in degree expansion of $\mathsf{Z}_{\mathsf{family}}$ \end{tabular} & Sec. \ref{sec:IntroductionConclusion} \\ \hline
 $\mathsf{Z}^{\zeta}_{\mathsf{DW},\pm}(p,\Sigma,\delta')$ & Donaldson-Witten generating functions on the two sides of a wall $W_{\zeta}$ & \eqref{eq:wc-discontinuity} \\ \hline
 $\ov{\mathsf{Z}_{\mathsf{DW}}[g_{\mu\nu},\Psi_{\mu\nu}]}$ & inhomogeneous form on $\MET(\IX)/\DIFF(\IX)$ & Sec. \ref{sec:WC-Multiparameter} \\ \hline
 \end{longtable}
\end{small}

\begin{small}
\begin{table}[H]\centering
	\begin{tabular}{|l|c|c|c|}\hline
		field & our notation & \begin{tabular}{@{}c@{}}other sugra literature \end{tabular} & \begin{tabular}{@{}c@{}} Grassmann parity \end{tabular}\\ \hline
       \begin{tabular}{@{}l@{}} vielbein \end{tabular} & $e_{\mu}{}^{\ha}$ & $e_{\mu}{}^{\ha}$ & even \\ \hline
       \begin{tabular}{@{}l@{}} spin connection  \end{tabular} & $\omega_{\mu}{}^{[\ha\hb]}$ & $\omega_{\mu}{}^{[\ha\hb]}$ & even  \\ \hline
       \begin{tabular}{@{}l@{}} dilatation connection \end{tabular} & $b_\mu$  & $b_\mu$ & even\\ \hline
       \begin{tabular}{@{}l@{}} special conformal connection  \end{tabular} & $f_{\mu}{}^{\ha}$ & $f_{\mu}{}^{\ha}$ & even\\ \hline
       \begin{tabular}{@{}l@{}} $\mathfrak{su(2)}_{\mathsf{R}}$ connection \end{tabular} & $V_{\mu}{}^{i}{}_{j}$ & $V_{\mu}{}^{i}{}_{j}$ & even\\ \hline
       \begin{tabular}{@{}l@{}} $\mathfrak{so}(1,1)_R$ connection \end{tabular} & $\ARsym_{\mu}$ & $A_{\mu}$ & even\\ \hline
       \begin{tabular}{@{}l@{}} gravitino \end{tabular} & $\bm{\Psi}_{\mu}{}^{i}$ & $\bm{\Psi}_{\mu}{}^{i}$ & odd\\ \hline
       \begin{tabular}{@{}l@{}} conformal gravitino \end{tabular} & $\bm{S}_{\mu}{}^{i}$ & $\bm{\phi}_{\mu}{}^{i}$ & odd \\ \hline\hline
       \begin{tabular}{@{}l@{}} auxiliary $2$-form \end{tabular} & $T_{\ha\hb}$ & $T_{\ha\hb}$ & even\\ \hline
       \begin{tabular}{@{}l@{}} auxiliary spinor \end{tabular} & $\bm{\Xi}^{i}$ & $\bm{\chi}^{i}$ & odd\\ \hline
       \begin{tabular}{@{}l@{}} auxiliary scalar \end{tabular} & $\mathscr{D}$ & $D$ & even \\ \hline
	\end{tabular}
	\caption{\label{tbl:untwisted sugra notation}Notation for the \underline{untwisted} $\CN=2$ conformal supergravity multiplet.}
\end{table}
\end{small}

\begin{small}
\begin{table}[H]\centering
	\begin{tabular}{|l|c|c|c|}\hline
		field & our notation & \begin{tabular}{@{}c@{}}other sugra literature \end{tabular} & \begin{tabular}{@{}c@{}} Grassmann parity \end{tabular}\\ \hline
       scalar & $\lambda$ & $X_+$ & even\\ \hline
       scalar & $\phi$ & $X_-$ & even\\ \hline
       connection & $A_\mu$ & $W_\mu$ & even\\ \hline
       gaugino   & $\bm{\Omega}^{i}$ & $\bm{\Omega}^{i}$ & odd \\ \hline
      self-dual auxiliary field & $D_{\mu\nu}$ & $Y_{\mu\nu}$ & even \\ \hline
	\end{tabular}\caption{\label{tbl:untwisted VM notation}Notation for the \underline{untwisted} $\CN=2$ vectormultiplet in Wess-Zumino gauge.}
\end{table}
\end{small}

\begin{small}
\begin{table}[H]\centering
	\begin{tabular}{|l|c|c|c|}\hline
		field & our notation & \begin{tabular}{@{}c@{}}other sugra literature \end{tabular} & \begin{tabular}{@{}c@{}} Grassmann parity \end{tabular}\\ \hline
       scalar & $\sfA_+$ & $A_+$ & even\\ \hline
       Weyl spinor & $\sfG_{+}^{i}$ & $\Psi^{i}_{+}$ & odd\\ \hline
       symmetric $\mathfrak{su(2)}_{\mathsf{R}}$ tensor & $\sfB_{+}^{ij}$ & $B_{+}^{ij}$ & even\\ \hline
       2-form   & $\sfF_{\mu\nu}^{+}$ (SD) & $F_{\mu\nu}^{-}$ (ASD) & even \\ \hline
      Weyl spinor & $\sfL_{+}^{i}$ & $\Lambda_{+}^{i}$ & odd \\ \hline
      scalar & $\sfC_{+}$ & $C_{+}$ & even \\ \hline
	\end{tabular}\caption{\label{tbl:untwisted chiral notation}Notation for the \underline{untwisted} $\CN=2$ superconformal chiral multiplet.}
\end{table}
\end{small}

\begin{small}
\begin{table}[H]\centering
	\begin{tabular}{|l|c|c|c|}\hline
		field & our notation & \begin{tabular}{@{}c@{}}other sugra literature \end{tabular} & \begin{tabular}{@{}c@{}} Grassmann  parity \end{tabular}\\ \hline
       scalar & $\sfA_-$ & $A_+$ & even\\ \hline
       Weyl spinor & $\sfG_{-}^{i}$ & $\Psi^{i}_{-}$ & odd\\ \hline
       symmetric $\mathfrak{su(2)}_{\mathsf{R}}$ tensor & $\sfB_{-}^{ij}$ & $B_{-}^{ij}$ & even\\ \hline
       2-form   & $\sfF_{\mu\nu}^{-}$ (ASD) & $F_{\mu\nu}^{+}$ (SD) & even \\ \hline
      Weyl spinor & $\sfL_{-}^{i}$ & $\Lambda_{-}^{i}$ & odd \\ \hline
      scalar & $\sfC_{-}$ & $C_{-}$ & even \\ \hline
	\end{tabular}\caption{\label{tbl:untwisted antichiral notation}Notation for the \underline{untwisted} $\CN=2$ superconformal antichiral multiplet.}
\end{table}
\end{small}

\section{Conventions\label{app:conventions}}
The reader is referred to Table \ref{tbl:index conventions} for our index conventions.
\paragraph{\underline{(Anti)symmetrization}:} We (anti)symmetrize with `strength 1'. For a rank-2 tensor $ω_{μν}$,
\eqa{
  &ω_{(μν)} && := \frac{ω_{μν} + ω_{νμ}}{2} ~,\quad &ω_{[μν]} && \,\,:= \frac{ω_{μν} - ω_{νμ}}{2} ~.
}

\paragraph{\underline{The completely antisymmetric (Levi-Civita) symbol (a.k.a. $\veps$ symbol)}:} In our conventions, $\varepsilon_{\ha\hb\hg\hd}$ (with frame indices) is an invariant tensor, whereas $\varepsilon_{\mu\nu\rho\sigma}$ (with curved indices) is an invariant density. The Levi-Civita symbol with raised curved indices given by
\eqa{
&\det(g)\,g^{\mu\mu'}g^{\nu\nu'}g^{\rho\rho'}g^{\sigma\sigma'} \varepsilon_{\mu'\nu'\rho'\sigma'} &&= \varepsilon^{\mu\nu\rho\sigma} ~. \label{eq:vareps-upper-and-lower}
}
The $\veps$ symbol satisfies the following useful contraction identity on a Riemannian $D$-manifold:
\eqa{
  &\veps_{μ_1 \ldots μ_n ν_1 \ldots ν_p}\veps^{μ_1 \ldots μ_n ρ_1 \ldots ρ_p} &&= p!n! \delta_{ν_1}^{[ρ_1}\delta_{ν_2}^{ρ_2}\cdots \delta_{ν_p}^{ρ_p]} ~,\label{eq:epsilon contraction identity}
}
where the antisymmetrization on the right is of `strength $1$' (i.e., it includes a factor of $1/p!$) and $n + p = D$. A frequently useful special case of this identity is for $D = 4$, $n = 1$ and $p=3$:
\eqa{
&\veps_{μνρσ}\veps^{μκλη} &&= 6 \,δ_{ν}^{[κ}δ_{ρ}^{λ}δ_{σ}^{η]} =  δ_{ν}^{κ}δ_{ρ}^{λ}δ_{σ}^{η} - δ_{ν}^{κ}δ_{ρ}^{η}δ_{σ}^{λ} + δ_{ν}^{λ}δ_{ρ}^{η}δ_{σ}^{κ} - δ_{ν}^{λ}δ_{ρ}^{κ}δ_{σ}^{η} + δ_{ν}^{η}δ_{ρ}^{κ}δ_{σ}^{λ} - δ_{ν}^{η}δ_{ρ}^{λ}δ_{σ}^{κ} ~.\label{eq:epsilon contraction identity spl case}
}
The four-component epsilon symbol and its two-component version are related by
\eqa{
	\varepsilon_{A\dA,B\dB,C\dC,D\dD} &= -(\varepsilon_{AC}\varepsilon_{BD}\varepsilon_{\dA\dD}\varepsilon_{\dB\dC} - \varepsilon_{AD}\varepsilon_{BC}\varepsilon_{\dA\dC}\varepsilon_{\dB\dD} ) ~, \label{eq:vareps}\\
	e_{\mu}{}^{A\dA}e_{\nu}{}^{B\dB}e_{\rho}{}^{C\dC}e_{\sigma}{}^{D\dD}\varepsilon_{A\dA,B\dB,C\dC,D\dD} &= \sqrt{g}  \,\varepsilon_{\mu\nu\rho\sigma} ~.
}

\paragraph{\uline{Self- and anti-self- dual projections}:} 
An antisymmetric tensor $T_{[μν]}$ (or $T_{[\ha\hb]}$) transforms reducibly under $\mathfrak{so}(4)$. Its irreducible self-dual part $T_{[μν]}^{+}$ (or $T_{[\ha\hb]}^{+}$) and anti-self- dual part $T_{[μν]}^{-}$ (or $T_{[\ha\hb]}^{-}$) are given by
\eqa{
	&T_{[\mu\nu]}^{\pm} &&= \tfrac{1}{2}\big(T_{[\mu\nu]} \pm \tfrac{1}{2}\sqrt{|g|}\varepsilon_{\mu\nu\rho\sigma}T^{[\rho\sigma]}\big)	 ~, \label{eq:Tmunu in terms of curved indices} \\
	&T_{[\ha\hb]}^{\pm} &&= \tfrac{1}{2}\big(T_{[\ha\hb]} \pm \tfrac{1}{2}\varepsilon_{\ha\hb\hg\hd}T^{[\hg\hd]}\big)	 ~. \label{eq:Tab in terms of frame indices}
}
so that $T_{μν} = T_{μν}^{+} + T_{μν}^{-}$ (or equivalently, $T_{\ha\hb} = T_{\ha\hb}^{+} + T_{\ha\hb}^{-}$). Here $g \equiv \det\,g_{\mu\nu}$. Moreover 
\eqa{
& {T^{[\mu\nu]}}^{\pm} &&= \tfrac{1}{2}\big(T^{[\mu\nu]} \pm \tfrac{1}{2}\tfrac{1}{\sqrt{|g|}}\varepsilon^{\mu\nu\rho\sigma}T_{[\rho\sigma]}\big) ~. \label{eq:SD ASD upper indices}
}
In two-component notation, the decomposition is
\eqa{
	& T_{A\dA,B\dB} &&= \tfrac{1}{2}T_{AB}\veps_{\dA\dB} + \tfrac{1}{2}T_{\dA\dB}\veps_{AB} ~. \label{eq:decomposition of antisym tensor in terms of self and anti self bispinor}
}
Here, the symmetric undotted component $T_{AB} = T_{BA}$ is the self-dual part (transforming in the adjoint of $\mathfrak{su(2)}_+$), whereas the symmetric dotted component $T_{\dA\dB} = T_{\dB\dA}$ is the anti-self-dual part (transforming in the adjoint of $\mathfrak{su(2)}_-$).\footnote{This convention is consistent with \cite{Gates:1983nr}, however, it is at variance with some supergravity literature where the naturally adopted convention associates symmetric undotted objects with \textit{anti-self-dual} components. 
} We can write $T_{\mu\nu}$ in various equivalent ways:
\eqa{
	& T_{\mu\nu} &&= e_{\mu}{}^{\ha}e_{\nu}{}^{\hb}T_{\ha\hb} = 
	\tfrac{1}{2}e_{\mu}{}^{\ha}e_{\nu}{}^{\hb}(σ_{\ha})^{A\dA}(σ_{\hb})^{B\dB}T_{A\dA,B\dB} 
	=  e_{\mu}{}^{A\dA}e_{\nu}{}^{B\dB}T_{A\dA,B\dB} ~.\label{eq:c43}
}
Also, in terms of \eqref{eq:sigma matrices} and \eqref{eq:sigma ha hb}--\eqref{eq:sigma tilde ha hb},
\eqas{ \label{eq:4-comp self and antiself dual in terms of 2-comp}
		&T_{[\mu\nu]}^{+} && =-\tfrac{1}{4}e_{\mu}{}^{A}{}_{\dA}e_{\nu}{}^{B\dA}({σ}^{\ha\hb})_{AB}T_{[\ha\hb]}^{+}  =  \tfrac{1}{2}e_{\mu}{}^{A}{}_{\dA}e_{\nu}{}^{B\dA}T_{AB}~, \\
		&T_{[\mu\nu]}^{-} &&= +\tfrac{1}{4}e_{\mu,A}{}^{\dA}e_{\nu}{}^{A\dB}(\wt{σ}^{\ha\hb})_{\dA\dB}T_{[\ha\hb]}^{-}   = \tfrac{1}{2}e_{\mu,A}{}^{\dA}e_{\nu}{}^{A\dB}T_{\dA\dB}~.
} 
\eqas{
		&T_{AB} &&= e^{\mu}{}_{A\dA}e^{\nu}{}_{B}{}^{\dA} T^{+}_{[\mu\nu]} ~, \quad 
         T_{\dA\dB} &&= e^{\mu}{}_{A\dA}e^{\nu,A}{}_{\dB} T^{-}_{[\mu\nu]} ~. \label{eq:2-comp self and antiself dual in terms of 4-comp}
}
The important minus sign in \eqref{eq:vareps} ensures consistency with \eqref{eq:4-comp self and antiself dual in terms of 2-comp} and \eqref{eq:2-comp self and antiself dual in terms of 4-comp}.\footnote{With a plus sign, the dotted $T_{\dA\dB}$ (resp. undotted $T_{AB}$) would have been the self-dual (resp. anti-self-dual) component instead, in agreement with \cite{Labastida:2005zz}. That convention is more suited to the twist that identifies dotted indices with $\mathsf{R}$-symmetry indices, and still results in \emph{self-dual} form fields in the twisted vectormultiplet.}  

A consequence of \eqref{eq:decomposition of antisym tensor in terms of self and anti self bispinor} is that for two antisymmetric tensors $U$ and $V$, 
\eqa{
	&U_{\mu\nu}V^{\mu\nu} &&= U_{\ha\hb}V^{\ha\hb} = U_{A\dA,B\dB}V^{A\dA,B\dB} = \tfrac{1}{2}U_{AB}V^{AB} + \tfrac{1}{2}U_{\dA\dB}V^{\dA\dB} ~.
}

\paragraph{\uline{(Bi)Spinors and Dirac matrices}:} See also Appendix \ref{app:spinormethods}.
$\,$\\$\,$\\
\noindent \underline{\textbf{$\bm{\sigma}$ matrices}}. 
We denote the triplet of standard Pauli matrices by $\bu{τ} = (τ_1, τ_2, τ_3)$, where
\eqa{
	\tau^1 &= \begin{pmatrix} 0 & 1 \\ 1 & 0 \end{pmatrix} ~, \quad 
	\tau^2 = \begin{pmatrix} 0 & -i \\ i & 0 \end{pmatrix} ~, \quad 
	\tau^3 = \begin{pmatrix} 1 & 0 \\ 0& -1 \end{pmatrix} ~, \label{eq:pauli matrices}
}
and define
\eqa{
	(σ_{\ha})^{A\dB} &= (\bu{τ},-i\mathbb{I}_2)^{A\dB} ~, \quad 
	(\wt{σ}_{\ha})_{\dA B} = (\bu{τ},i\mathbb{I}_2)_{\dA B} ~.\label{eq:sigma matrices}
}
These matrices satisfy the following relations
\eqa{
	σ_{\ha}\wt{σ}_{\hb} + σ_{\hb}\wt{σ}_{\ha} &= 2δ_{\ha\hb}\mathbb{I}_2 ~, \quad 
	\wt{σ}_{\ha}{σ}_{\hb} + \wt{σ}_{\hb}{σ}_{\ha} = 2δ_{\ha\hb}\mathbb{I}_2 ~, \label{eq:cliff sigma}\\
	(σ_{\ha})^{\dagger} &= +\wt{σ}_{\ha} ~, \quad (\wt{σ}^{\ha})^{\dA A} =  (σ^{\ha})^{A \dA} ~,\\
	(σ_{\ha})^{A\dB}(\wt{σ}_{\ha})_{\dC D} &= + 2\, δ_{D}{}^{A}δ_{\dC}{}^{\dB} \label{eq:completeness relation 1} ~.
}

\noindent \underline{\textbf{Bispinor notation}}. 
An object with a frame index is decomposed in terms of \eqref{eq:sigma matrices} as\footnote{In these conventions, there is a simple orthonormality condition, namely $e^{\mu}{}_{C\dC} e_{\mu}{}^{B\dB} = \delta_{C}^{B}\delta_{\dC}^{\dB}$.}
\eqa{
	V_{\ha} &= \tfrac{1}{\sqrt{2}}	V_{C\dC}(σ_{\ha})^{C\dC} ~, \quad 
	V_{C\dC} = \tfrac{1}{\sqrt{2}} (σ^{\ha})_{C\dC}V_{\ha}	~,\label{eq:bispinor 1}
}
where the second relation follows from \eqref{eq:completeness relation 1}. For the vielbein and its inverse,
\eqa{ 
	\begin{split}
		e^{\mu}{}_{\ha} &= \tfrac{1}{\sqrt{2}}e^{\mu}{}_{C\dC}(σ_{\ha})^{C\dC}~, \qquad e_{\mu}{}^{\ha} = \tfrac{1}{\sqrt{2}}e_{\mu}{}^{C\dC} (σ^{\ha})_{C\dC} ~,\\
		e^{\mu}{}_{C\dC} &= \tfrac{1}{\sqrt{2}}e^{\mu}{}_{\ha}(σ^{\ha})_{C\dC}~, \qquad e_{\mu}{}^{C\dC} = \tfrac{1}{\sqrt{2}} e_{\mu}{}^{\ha}(σ_{\ha})^{C\dC} ~.
	\end{split}\label{eq:bispinor 2}
}

\noindent \underline{\textbf{Raising and lowering spinor indices}}. 
Dotted and undotted spinor indices are raised and lowered using the invariant tensors $\veps^{AB}$ (resp. $\veps^{\dA\dB}$) and $\veps_{AB}$ (resp. $\veps_{\dA\dB}$) of $\mathfrak{su(2)}_{+}$ (resp. $\mathfrak{su(2)}_{-}$), following the north-west south-east (NWSE) convention of \cite{Gates:1983nr,Belitsky:2000ii,Belitsky:2000ws,Vandoren:2008xg}:
\eqa{
	\lambda^{A} &= \veps^{AB}\lambda_{B}~, \qquad \lambda_{B} = \lambda^{A}\veps_{AB}~, \quad \text{(idem for dotted spinors).} \label{eq:raising and lowering spinor indices}
}
where $\veps_{AB} = \veps^{AB} = -\veps_{\dA\dB} = -\veps^{\dA\dB}$. We choose $\veps_{12} = +1$ (\emph{without} factors of $i$, unlike \cite{Gates:1983nr}). In terms of the Pauli matrices, $\bu{τ} = (τ^1, τ^2, τ^3)$, these invariant tensors, as matrices, are
\eqa{
	[\veps_{AB}]	&= iτ^2 = \begin{pmatrix}
		0 & 1 \\ 
		-1 & 0
	\end{pmatrix} = [\veps^{AB}] ~,\label{eq:veps undotted} \qquad 
	[\veps_{\dA\dB}]	= -iτ^2 = \begin{pmatrix}
		0 & -1 \\ 
		1 & 0
	\end{pmatrix} = [\veps^{\dA\dB}] ~.
}
These matrices are real and antisymmetric (and hence also antihermitian) and satisfy\footnote{There is \underline{no} difference between $\delta_{A}{}^{B}$ and $\delta^{B}{}_{A}$ in this convention, which both equal $\delta_{A}^{B}$. The form $\delta_{A}{}^{B}$ is suited to the NWSE convention, since $\veps^{AC}\delta_{C}{}^{B} = \veps^{AB}$ and $\delta_{A}{}^{C}\veps_{CB} = \veps_{AB}$, without any minus signs.}
\eqa{\begin{split}
		\veps^{AB}\veps_{CB} &= \delta_{C}{}^{A}~, \quad \veps^{AB}\veps_{AB} = +2	~, \quad 
		\veps^{AB}\veps_{CD} = \delta_{C}{}^{A}\delta_{D}{}^{B} - \delta_{D}{}^{A}\delta_{C}{}^{B} ~.
	\end{split}
}

\noindent \underline{\textbf{Dirac matrices}}.
We define the four-dimensional Euclidean Dirac (``gamma'') matrices in terms of the $\sigma$ matrices of \eqref{eq:sigma matrices} as
\eqa{
	γ^{\ha} &:= \begin{pmatrix}
		0        & -i σ^{\ha,A\dB} \\
		i  \wt{σ}^{\ha}_{\dA B} & 0
	\end{pmatrix}	~. \label{eq:euclidean dirac matrices}
}
Owing to \eqref{eq:cliff sigma}, these matrices satisfy the Euclidean Clifford algebra $\mathcal{C}\ell(0,4)$,
\eqa{
	\{\gamma^\ha, \gamma^\hb\} = \gamma^\ha\gamma^\hb + \gamma^\hb\gamma^\ha &= 2 \delta^{\ha\hb}\mathbb{I}_4 ~. \label{eq:hermitcity of dirac matrices in euclidean signature}
}
The Dirac matrices \eqref{eq:euclidean dirac matrices} are hermitian:
\eqa{
	(\gamma^\ha)^\dagger &= \gamma^\ha ~\quad \text{for } \ha = 1, \ldots, 4.
}
\noindent We define the $\gamma_5$ matrix in terms of \eqref{eq:euclidean dirac matrices} as
\eqa{
	γ_{5} := -γ^{1}γ^{2}γ^{3}γ^{4} = \begin{pmatrix}
		\mathbb{I}_2 & 0 \\ 0 & -\mathbb{I}_2
	\end{pmatrix} \label{eq:gamma5 in terms of euclidean dirac matrices} ~.
}

\noindent For the choice \eqref{eq:gamma5 in terms of euclidean dirac matrices}, we have the following identities,
\eqa{
	\gamma_{\ha}\gamma_5 &= -\tfrac{1}{3!}\varepsilon_{\ha\hb\hg\hd}\gamma^{\hb\hg\hd} ~, \quad \gamma_{\ha\hb}\gamma_5 = \tfrac{1}{2}\varepsilon_{\ha\hb\hg\hd}\gamma^{\hg\hd} ~, \quad \gamma_{\ha\hb\hg}\gamma_5 = \varepsilon_{\ha\hb\hg\hd}\gamma^{\hd} ~.
}
We also have
\eqa{
	γ^{\ha\hb} \equiv \tfrac{1}{2}(γ^{\ha}γ^{\hb} - γ^{\hb}γ^{\ha}) 
	&= \begin{pmatrix}
		{(σ^{\ha\hb})^{A}}_{B} & 0 \\
		0 & {(\wt{σ}^{\ha\hb})_{\dA}}^{\dB}
	\end{pmatrix} \label{eq:gamma-ab} ~,
}
where we have defined
\eqa{
	{(σ^{\ha\hb})^{A}}_{B} &\equiv \tfrac{1}{2}{(σ^{\ha}\wt{σ}^{\hb} - σ^{\hb}\wt{σ}^{\ha})^{A}}_{B} = \tfrac{1}{2}\big[(σ^{\ha})^{A\dB}(\wt{σ}^{\hb})_{\dB B} - (σ^{\hb})^{A\dB}(\wt{σ}^{\ha})_{\dB B}\big] ~, \label{eq:sigma ha hb}\\
	{(\wt{σ}^{\ha\hb})_{\dA}}^{\dB} &\equiv \tfrac{1}{2}(\wt{σ}^{\ha}σ^{\hb} - \wt{σ}^{\hb}σ^{\ha})_{\dA}{}^{\dB} = \tfrac{1}{2}\big[(\wt{σ}^{\ha})_{\dA C}(σ^{\hb})^{C \dB} - (\wt{σ}^{\hb})_{\dA C}(σ^{\ha})^{C \dB}\big] ~.\label{eq:sigma tilde ha hb}
}
By explicit evaluation, one can verify that
\eqa{
	σ^{\ha\hb} = \begin{pmatrix}
		0     & iτ^3   & -iτ^2  & iτ^1 \\
		-iτ^3 &   0    &  iτ^1  & iτ^2 \\
		iτ^2  & -iτ^1  &   0    & iτ^3 \\
		-iτ^1 &  -iτ^2 &  -iτ^3 &  0
	\end{pmatrix}  ~, \quad 
	\wt{σ}^{\ha\hb} &= \begin{pmatrix}
		0     & iτ^3   & -iτ^2  & -iτ^1 \\
		-iτ^3 &   0    &  iτ^1  & -iτ^2 \\
		iτ^2  & -iτ^1  &   0    & -iτ^3 \\
		iτ^1 &  iτ^2   &  iτ^3  &  0
	\end{pmatrix} ~.\label{eq:sig and osig matrix}
}
It is also straightforward to check that $σ^{\ha\hb}$ is self-dual whereas $\wt{σ}^{\ha\hb}$ is anti-self-dual:\footnote{This in contrast to the conventions in Appendix A of \cite{Labastida:2005zz}.}
\eqa{
	σ^{\ha\hb}     &= \tfrac{1}{2}\varepsilon^{\ha\hb\hg\hd}σ_{\hg\hd} ~, \quad  	
	\wt{σ}^{\ha\hb} = -\tfrac{1}{2}\varepsilon^{\ha\hb\hg\hd}\wt{σ}_{\hg\hd} ~. \label{eq:self and anti-self dual sigmas}
}
The completeness relations involving $σ^{\ha\hb}$ and $\wt{σ}^{\ha\hb}$ are
\eqa{
	{(σ_{\ha\hb})^{A}}_{B}{(σ^{\ha\hb})^{C}}_{D} &= 4(δ^{A}_{B}δ^{C}_{D} - 2 δ^{A}_{D}δ^{C}_{B}) \label{eq:completeness relation 2} ~,\\
	{(\wt{σ}_{\ha\hb})_{\dA}}^{\dB}{(\wt{σ}^{\ha\hb})_{\dC}}^{\dD} &= 4(δ_{\dA}^{\dB}δ_{\dC}^{\dD} - 2 δ_{\dA}^{\dD}δ_{\dC}^{\dB}) ~. \label{eq:completeness relation 3}
}

\noindent \underline{\textbf{Decomposition of a Dirac spinor}}.
A 4-component spinor $\bm{\lambda}^{i}$ is decomposed in terms of Weyl spinors $\lambda^{iA}$ (left or positive chirality) and $\wt{\lambda}^{i}{}_{\dA}$ (right or negative chirality) as:
\eqa{\label{eq:our 4-comp spinor su2}
	&\bm{\lambda}^{i} &&= \begin{pmatrix}
		\lambda^{iA} \\
		\wt{\lambda}^{i}{}_{\dA}
	\end{pmatrix} ~.
}
The tilde indicates that $\lambda^{iA}$ and $\wt{\lambda}_{i\dA}$ are \underline{not} related by complex conjugation in Euclidean signature; we drop it in the main text to improve readability. The location of the $\mathfrak{su(2)}_{\mathsf{R}}$ index $i$ is \underline{not} linked to spinor chirality, and the index can be raised and lowered using $\veps^{ij}$ and $\veps_{ij}$ respectively.
\paragraph{\uline{Differential Forms}}.
Assume $\IX$ is a smooth oriented Riemannian $d$-manifold. We expand a $p$-form $A_{(p)} \in \Omega^{p}(\IX)$ (where $0 \leq p \leq d$) in a local chart as:
\eqa{
   A_{(p)} &= \frac{1}{p!}A_{\mu_1 \cdots \mu_p} dx^{\mu_1} \wedge \cdots \wedge dx^{\mu_p} ~.
} 
Let $B_{(q)} \in \Omega^{q}(\IX)$, for some $0 \leq q \leq d$. The wedge product of $A_{(p)}$ and $B_{(q)}$ is a $(p+q)$-form:
\eqa{
   &A_{(p)} \wedge B_{(q)} &&= \frac{1}{p!q!} A_{[\mu_1\ldots \mu_p}B_{\nu_1\ldots \nu_q]}dx^{\mu_1} \wedge \cdots \wedge dx^{\mu_p} \wedge dx^{\nu_1} \wedge \cdots \wedge dx^{\nu_p}  \label{eq:pq-wedge-1}\\
   & &&= \frac{1}{(p+q)!}\big(A_{(p)} \wedge B_{(q)}\big)_{\mu_1 \cdots \mu_{p+q}} dx^{\mu_1}\wedge \cdots \wedge dx^{\mu_{p+q}} ~. \label{eq:pq-wedge-2}
}
Equating \eqref{eq:pq-wedge-1} and \eqref{eq:pq-wedge-2}, we get
\eqa{
  &\big(A_{(p)} \wedge B_{(q)}\big)_{\mu_1 \cdots \mu_{p} \nu_{1} \cdots \nu_{p+q}} &&= \frac{(p+q)!}{p!q!} A_{[\mu_1\ldots \mu_p}B_{\nu_1\ldots \nu_q]} ~.
}
The exterior derivative of a $p$-form $A_{(p)}$ is a $(p+1)$-form, given by
\eqa{
  & dA_{(p)} &&= \frac{1}{p!}\n_{[\mu_1}A_{\mu_2\cdots\mu_{p+1}]}dx^{\mu_1}\wedge dx^{\mu_2}\wedge \cdots \wedge dx^{\mu_{p+1}} \label{eq:ext-der-1} \\
   & &&= \frac{1}{(p+1)!}(dA_{(p)})_{\mu_1\cdots\mu_{p+1}}dx^{\mu_1}\wedge dx^{\mu_2}\wedge \cdots \wedge dx^{\mu_{p+1}} \label{eq:ext-der-2} ~.
}
Equating \eqref{eq:ext-der-1} and \eqref{eq:ext-der-2} we get
\eqa{
  (dA_{(p)})_{\mu_1\cdots\mu_{p+1}} &= (p+1)\n_{[\mu_1}A_{\mu_2\cdots\mu_{p+1}]} ~.
}
The Hodge dual of $A_{(p)}$ is a $(d-p)$-form, which can be expanded in local coordinates as\footnote{Some authors prefer to write $\varepsilon_{\nu_{1} \cdots \nu_{d-p}\mu_1\cdots \mu_{p}}$ instead. The two conventions obviously differ by $(-1)^{p(d-p)}$. Note that for $(d,p) = (4,2)$ it does not matter, but for $(d,p) = (4,1)$ it does.}
\eqa{
 &\star A_{(p)} &&= \frac{1}{p!(d-p)!}\sqrt{g}\varepsilon_{\mu_{1} \cdots \mu_{p} \nu_1\cdots \nu_{d-p} }A^{\mu_{1}\cdots \mu_{p}} dx^{\nu_{1}} \wedge \cdots \wedge dx^{\nu_{d-p}} ~.
}
With these definitions, 
\eqa{
   &A_{(p)} \wedge \star A_{(p)} 
   &&= \frac{1}{p!} A_{\mu_1 \cdots \mu_p}A^{\mu_1 \cdots \mu_p} \sqrt{g}\, d^{d}x ~.
}
Henceforth, we set $\bm{d = \dim\IX = 4}$.
For $0$-forms $a$, $\ov{a}$, and $\eta$,
\eqas{
  &da &&= (\partial_{\mu} a)dx^{\mu} ~, \quad d\ov{a} = (\partial_{\mu} \ov{a})dx^{\mu} ~, \\
  &\star d\ov{a} &&= \frac{1}{3!}\sqrt{g} \varepsilon_{\mu_1 \nu_1 \nu_2 \nu_3}\,(\partial^{\mu_1} \ov{a})\,dx^{\nu_1} \wedge dx^{\nu_2} \wedge dx^{\nu_3} ~,\\
}
For a $1$-form $\psi \in \Omega^{1}(\IX)$, 
\eqas{
   &\psi &&= \psi_{\mu} dx^{\mu} ~, \quad \psi \wedge \psi = \frac{1}{2!}\big(2\,\psi_{[\mu_1}\psi_{\mu_2]}\big) dx^{\mu_1} \wedge dx^{\mu_2} ~,\\
   &d\psi &&= \frac{1}{2!}\big(2\,\n_{[\mu_1}\psi_{\mu_2]} \big) dx^{\mu_1} \wedge dx^{\mu_2} ~,\\
   &\star \psi &&= \frac{1}{3!}\sqrt{g}\,\varepsilon_{\mu_1 \mu_2 \mu_3 \mu_4}\psi^{\mu_1}dx^{\mu_2} \wedge dx^{\mu_3} \wedge dx^{\mu_4} ~,\\
   &d\star\psi   &&= \frac{1}{4!}(\n_{\mu}\psi^{\mu})\sqrt{g}\,\varepsilon_{\mu_1\mu_2\mu_3\mu_4}dx^{\mu_1}\wedge \cdots \wedge dx^{\mu_4} ~.
}
For a $2$-form $F \in \Omega^{2}(\IX)$,
\eqas{
   &F \wedge \star F &&= \frac{1}{2}F_{\mu\nu}F^{\mu\nu} \sqrt{g}\, d^{4}x ~, \quad F \wedge F = \frac{1}{4}\varepsilon^{\mu\nu\rho\sigma}F_{\mu\nu}F_{\rho\sigma}  d^{4}x ~,\\
  &F_+ \wedge F_+ &&= F_+ \wedge \star F_+ = \frac{1}{2}F_{\mu\nu}^{+}F_{+}^{\mu\nu} \sqrt{g}\,d^{4}x~,\\
  &F_- \wedge F_- &&= -F_- \wedge \star F_- = -\frac{1}{2}F_{\mu\nu}^{-}F_{-}^{\mu\nu} \sqrt{g}\,d^{4}x ~.
}
For a $2$-form $\chi \in \Omega^{2}(\IX)$, $\chi = \frac{1}{2}\chi_{\mu_1\mu_2}dx^{\mu_1} \wedge dx^{\mu_2}$, and
\eqa{
   &d\chi &&= \frac{1}{2}\n_{[\mu_1}\chi_{\mu_2\mu_3]}dx^{\mu_1}\wedge dx^{\mu_2} \wedge dx^{\mu_3} = \frac{1}{3!}(3\,\n_{[\mu_1}\chi_{\mu_2\mu_3]})dx^{\mu_1}\wedge dx^{\mu_2} \wedge dx^{\mu_3}~.
}
Finally, for Lie-algebra valued forms, $A_{(p)} \in \Omega^{p}(\IX) \otimes \mathsf{Lie}(G)$ and $B_{(q)} \in \Omega^{q}(\IX) \otimes \mathsf{Lie}(G)$,
\eqa{
   &\mathsf{Tr}(A_{(p)}\wedge B_{(q)}) &&= \frac{1}{p!q!}\mathsf{Tr\,}(A_{\mu_1\cdots\mu_p}B_{\nu_1\cdots\nu_q})dx^{\mu_1}\wedge \cdots dx^{\mu_p}\wedge dx^{\nu_1}\wedge \cdots \wedge dx^{\nu_q} ~,\\
  &\big[A_{(p)}, B_{(q)}\big] &&= \frac{1}{p!q!}\big[A_{\mu_1\cdots\mu_p}, B_{\nu_1\cdots\nu_q}\big] dx^{\mu_1}\wedge \cdots \wedge dx^{\mu_p} \wedge dx^{\nu_1} \wedge \cdots \wedge dx^{\nu_q} ~,
}
where $\mathsf{Tr\,}$ is a suitably normalized invariant form on $\mathsf{Lie}(G)$. Therefore, there is a natural inner product on the space of $\mathsf{Lie}(G)$-valued differential forms, given by 
\eqa{
& \langle A_{(p)}, B_{(p)}\rangle &&:= \int_{\IX}\mathsf{Tr\,}\big( A_{(p)} \wedge \star B_{(p)} \big) ~,
}
which enters every term of the action.

\noindent \underline{\textbf{The $u$-plane actions of refs. \cite{Moore:1997pc} and \cite{Marino:1998bm}}.} Our general gravity degree $=0$ action \eqref{eq:GeneralActionDegree0} as adapted to the IR reads
\eqa{
&\mathscr{L}_{0}^{\mathsf{IR}} &&= \frac{1}{2}\sqrt{g}\big(\ov{\tau}_{IJ}F_{\mu\nu}^{+}{}^{I}F_{+}^{\mu\nu}{}^{J} - \tau_{IJ}F_{\mu\nu}^{-}{}^{I}F^{\mu\nu}_{-}{}^{J}\big)  + 4i\sqrt{g}\big(\mathsf{Im\,}\tau_{IJ}\big)\big(\n_{\mu}a^{I}\big)\big(\n^{\mu}\ov{a}^{J}\big)\nn
& &&\quad + i\sqrt{g}\big(\mathsf{Im\,}\tau_{IJ}\big)D_{\mu\nu}{}^{I}D^{\mu\nu}{}^{J} + 2\sqrt{g}\,\tau_{IJ}\psi_{\sigma}{}^{I} \n^{\sigma}\eta^{J} - 2\sqrt{g}\,\ov{\tau}_{IJ}\eta^{I}\n_{\sigma}\psi^{\sigma}{}^{J} \nn
& &&\quad - 2\sqrt{g}\,\tau_{IJ}\psi_{\mu}{}^{I}\n_{\nu}\chi^{\mu\nu}{}^{J} + 2\sqrt{g}\,\ov{\tau}_{IJ} \big(\n_{[\mu}\psi_{\nu]}{}^{I}\big)^{+}\chi^{\mu\nu}{}^{J} \nn
& &&\quad + \frac{1}{2}\sqrt{g}\,\ov{\CF}_{IJK}\eta^{I}\big(F_{\mu\nu}^{+}{}^{J} + D_{\mu\nu}{}^{J}\big)\chi^{\mu\nu}{}^{K} + \sqrt{g}\,\CF_{IJK}\psi_{\mu}{}^{I}\psi_{\nu}{}^{J}\big(F_{-}^{\mu\nu}{}^{K} -D^{\mu\nu}{}^{K}\big) \nn
& &&\quad + \frac{1}{12}\CF_{IJKL}\veps^{\mu\nu\rho\sigma}\psi_{\mu}{}^{I}\psi_{\nu}{}^{J}\psi_{\rho}{}^{K}\psi_{\sigma}{}^{L} +   \CQ\left(\frac{i}{12}\sqrt{g}\,\ov{\CF}_{IJK}\chi_{\mu}{}^{\rho}{}^{I}\chi^{\mu\sigma}{}^{J}\chi_{\rho\sigma}{}^{K}\right) ~. \label{eq:GeneralActionDegree0-IR}
}
Using the above conventions for differential forms, this can be written as a differential form
\eqa{
&\left.\mathscr{L}_{0}^{\mathsf{IR}}\right|_{\rm{form}} &&= \ov{\tau}_{IJ}\,F_+^I \wedge F_+^J + \tau_{IJ}\, F_-^I \wedge F_-^J + 4i\big(\mathsf{Im\,}\tau_{IJ}\big)da^I \wedge \star d\,\ov{a}^J \nonumber\\
  & &&\quad + 2i\big(\mathsf{Im\,}\tau_{IJ}\big)D^I \wedge D^J + 2\,\tau_{IJ}\,\psi^I\wedge\star d\eta^J - 2\,\ov{\tau}_{IJ}\eta^I\wedge d\star\psi^J \nonumber\\
  & &&\quad - 2\,\tau_{IJ} \psi^{I}\wedge d\chi^{J} - 2\,\ov{\tau}_{IJ} \chi^{I} \wedge d\psi^{J} \nonumber\\
  & &&\quad + \ov{\CF}_{IJK}\eta^{I} \chi^{J} \wedge (F_+^K + D^K) - \CF_{IJK}\psi^I\wedge\psi^J\wedge(F_-^K + D^K) \nonumber\\
  & &&\quad + \frac{1}{12}\CF_{IJKL} \psi^I \wedge \psi^J \wedge \psi^K \wedge \psi^L + \CQ\left(\frac{i}{12}\ov{\CF}_{IJK}\chi_{\mu}{}^{\rho}{}^{I}\chi^{\mu\sigma}{}^{J}\chi_{\rho\sigma}{}^{K}\right)\sqrt{g}\,d^{4}x  ~. \label{eq:L0-form-our-conv}
}
If we rescale
\beqas{
& a &&\mapsto -4\sqrt{2}\,a ~, \quad \ov{a} \mapsto \frac{1}{2\sqrt{2}}\,\ov{a} ~, \quad F \mapsto F ~, \quad D \mapsto D ~, \\
 &\eta &&\mapsto \frac{i}{2}\eta ~, \quad \psi \mapsto \psi ~, \quad \chi \mapsto - i\chi ~, \label{eq:GoodRescaling}
}
and define a rescaled action $\left.\mathscr{L}_{0}^{\mathsf{IR}}\right|_{\rm{form}}' := \frac{i}{16\pi}\left.\mathscr{L}_{0}^{\mathsf{IR}}\right|_{\rm{form}}$, the result is
\eqa{
&\left.\mathscr{L}_{0}^{\mathsf{IR}}\right|_{\rm{form}}' &&= \frac{i}{16\pi}\big(\ov{\tau}_{IJ}\,F_+^I \wedge F_+^J + \tau_{IJ}\, F_-^I \wedge F_-^J\big) + \frac{1}{2\pi}\big(\mathsf{Im\,}\tau_{IJ}\big)da^I \wedge \star d\ov{a}^J  - \frac{1}{8\pi}\big(\mathsf{Im\,}\tau_{IJ}\big)D^I \wedge D^J \nonumber\\
  & &&\quad - \frac{1}{16\pi}\tau_{IJ}\,\psi^I\wedge\star d\eta^J + \frac{1}{16\pi}\ov{\tau}_{IJ}\eta^I\wedge d\star\psi^J \highlightYellow{-} \frac{1}{8\pi}\tau_{IJ} \psi^{I}\wedge d\chi^{J} - \frac{1}{8\pi}\ov{\tau}_{IJ} \chi^{I} \wedge d\psi^{J} \nonumber\\
  & &&\quad +\frac{i\sqrt{2}}{16\pi} \ov{\CF}_{IJK}\eta^{I} \chi^{J} \wedge (F_+^K + D^K) \highlightYellow{+} \frac{i\sqrt{2}}{2^7\pi} \CF_{IJK}\psi^I\wedge\psi^J\wedge(F_-^K + D^K) \nonumber\\
  & &&\quad + \frac{i}{3\cdot 2^{11}\pi}\CF_{IJKL} \psi^I \wedge \psi^J \wedge \psi^K \wedge \psi^L - \frac{i\sqrt{2}}{3\cdot 2^5\pi} \CQ\left(\ov{\CF}_{IJK}\chi_{\mu}{}^{\rho}{}^{I}\chi^{\mu\sigma}{}^{J}\chi_{\rho\sigma}{}^{K}\right)\sqrt{g}\,d^{4}x ~, \label{eq:MW-MM-Action}
}
which is identical to (3.4) of \cite{Marino:1998bm} up to some corrected signs (highlighted in yellow above) for a general Abelian Lie group as the IR gauge group, and to (2.15) of \cite{Moore:1997pc} for gauge group $\mathsf{U(1)}$ up to the same corrected signs but also the corrected prefactor of the scalar kinetic term. (The foregoing rescalings also motivate the overall factor of $\frac{i}{16\pi}$ in the actions of sections \ref{sec:ProposalForTheAction} and \ref{sec:TwistedSugraAction}.)

\section{Spinor Methods\label{app:spinormethods}}
A four-component (Dirac) spinor of $\mathsf{Spin}(4)$ is a pair of two-component (Weyl) spinors that are unrelated by complex conjugation, consistent with the exceptional isomorphism $\mathsf{Spin}(4) \cong \mathsf{SU(2)}_+ \times \mathsf{SU(2)}_-$. The Weyl spinors are labeled by $\mathfrak{su}(2)_{\pm}$ Lie algebra-valued indices (called ``spinor indices'' or ``dotted'' and ``undotted'' indices). So a Dirac spinor $\bm{\lambda}$ has components $\lambda_{A}$ and $\lambda_{\dA}$, where $A$ (resp. $\dA$) is a doublet index of $\mathfrak{su}(2)_{+}$ (resp. $\mathfrak{su}(2)_{+}$). This splitting was important in early studies of twisted supersymmetric theories and topological gravity \cite{Witten:1988xi,Witten:1988ze}. Topologically twisted \underline{pure} $\CN=2$ super Yang Mills theory has no spinors, only Lie algebra-valued differential forms. So one can always write (as we do) the resulting fields with coordinate indices by contracting with frame fields or inverse frame fields, thus dispensing with spinor indices. Nevertheless, the intermediate analysis is greatly simplified by the use of these methods.\footnote{Using spinor indices eliminates the need for manipulations involving Dirac ($\gamma$) matrices and Fierz identities (which are replaced by ``exchange identities,'' familiar from basic tensor methods of $\mathfrak{su}(2)$). It also vastly streamlines manipulations involving (anti-)self-dual fields.} We have therefore sketched a few details here for the benefit of readers unfamiliar with these methods (the so-called Van der Waerden notation). It may be useful to consult Appendix \ref{app:conventions} while reading this section.

Let $e^{a} = e_{\mu}{}^{a}dx^{\mu}$ denote the vielbein 1-forms (i.e., frame fields) on $\IX$ as expressed in local chart. Here $a$ is a tangent space index (or ``flat'' index) and $\mu$ is a coordinate index (or ``curved'' index) -- see Table \ref{tbl:index conventions} for our index conventions. Given a vector $V^{\mu}$, we can write $V^{a} = V^{\mu}e_{\mu}{}^{a}$, and conversely, $V^{\mu} = e^{\mu}{}_{a}V^{a}$ where $e^{\mu}{}_{a}$ denote the components (in the same chart) of the inverse vielbein.\footnote{Depending on which of $V^\mu$ or $V^a$ is regarded as more fundamental, this relation endows (or takes away) some metric/frame dependence. The vielbein dependence must be kept in mind while computing a \emph{variation}.} Here, $a$ is an index labeling the defining reducible representation $\bm{\underline{4}}$ of the Lorentz $\mathfrak{so}(4)$, which can be decomposed in terms of irreducible $\mathfrak{su}(2)_+$ (undotted) indices $A$ and $\mathfrak{su}(2)_-$ (dotted) indices $\dA$. Thus $V^{a}$ is equivalent to the bispinor $V^{A\dA}$.\footnote{One uses a set of intertwiner matrices $(\sigma_\ha)^{A\dA}$ to define $V^{A\dA} \sim V^{\ha}(\sigma_\ha)^{A\dA}$. See \eqref{eq:bispinor 1},\eqref{eq:bispinor 2}.} So there is a bijection:\footnote{Recall that $\mathfrak{su}(2)_+$ (resp. $\mathfrak{su}(2)_-$) has a completely antisymmetric invariant tensor $\veps_{AB}$ (resp. $\veps_{\dA\dB}$) that is used to lower undotted (resp. dotted) spinor indices, and there is an analogous object $\veps^{AB}$ (resp. $\veps^{\dA\dB}$) that is used to raise them. Our conventions for these are spelled out in \eqref{eq:raising and lowering spinor indices}.}
\be\label{eq:vector-bispinor-equivalence}
\begin{tikzcd}
   V^{\mu} \arrow[bend left=60]{r}{\times e_{\mu}{}^{a}} & V^{a} \arrow[bend left=60]{l}{\times e^{\mu}{}_{a}} \arrow[r,leftrightarrow]& V^{A\dA} ~.
\end{tikzcd} 
\ee
On a Riemannian 4-manifold $\IX$, the natural decomposition of two-forms $\Lambda^{2}T^{*}\IX \cong \Lambda^{2}_{+}T^{*}\IX \oplus \Lambda^{2}_{-}T^{*}\IX$ into self-dual (SD) and anti-self-dual (ASD) projections also has an elegant spinorial version. In view of \eqref{eq:vector-bispinor-equivalence}, an antisymmetric rank-2 tensor $T_{[\mu\nu]}$ can be equivalently written in terms of bispinor indices as a \emph{reducible} object $T_{[A\dA,B\dB]}$, which can in turn be decomposed as 
\eqa{
T_{[A\dA,B\dB]} &= \frac{1}{2}T_{AB}\veps_{\dA\dB} + \frac{1}{2}T_{\dA\dB}\veps_{AB} ~, \label{eq:sd-asd-2-comp}
}
in terms of its symmetric traceless SD and ASD components, $T_{AB}$ and $T_{\dA\dB}$ respectively. (See \eqref{eq:4-comp self and antiself dual in terms of 2-comp} and \eqref{eq:2-comp self and antiself dual in terms of 4-comp}.) Therefore, analogous to \eqref{eq:vector-bispinor-equivalence}, there are bijections $T_{\mu\nu}^{+} \leftrightarrow T_{AB}$ and $T_{\mu\nu}^{-} \leftrightarrow T_{\dA\dB}$. Note that the symmetry of $T_{AB}$ implies its tracelessness,  since $T_{A}{}^{A} = T_{AB}\veps^{AB} = 0$.

As noted above, a four-component Dirac spinor $\bm{\lambda}^{i}$ (with $i$ being a $\mathfrak{su(2)}_{\mathsf{R}}$ doublet index -- see Table \ref{tbl:index conventions}) can be decomposed in terms of Weyl spinors (see \eqref{eq:our 4-comp spinor su2})  $\lambda^{iA}$ and $\lambda^{i}{}_{\dA}$. 
The twisting homomorphism \eqref{eq:TwistingHomomorphism} -- which identifies $\mathsf{SU(2)}_+$ with $\mathsf{SU(2)}_R$ -- is implemented by identifying the indices $\{i\}$ with the indices $\{A\}$.\footnote{The other choice of twist identifies $\mathfrak{su(2)}_{\mathsf{R}}$ with $\mathfrak{su}(2)_+$, i.e., the index set $\{i\}$ with the index set $\{\dA\}$.} Therefore, upon twisting, the spinor $\bm{\lambda}^{a}$ splits into $\lambda^{AB}$ (where $A$ and $B$ are both $\mathfrak{su}(2)_+$ doublet indices) and $\lambda^{A}{}_{\dA}$. Note that $\lambda_{A\dA} = \lambda^{B}{}_{\dA}\veps_{BA}$. 

It should be noted that unlike $T_{AB}$ (or $T_{\dA\dB}$) of \eqref{eq:sd-asd-2-comp} which is automatically symmetric and hence irreducible by design (regardless of any twisting), the object $\lambda^{AB}$ that is obtained by twisting ($\lambda^{iA} \to \lambda^{AB}$), \emph{is not}. In fact, it admits a natural decomposition,
\eqa{
\lambda^{AB} &= \lambda_{\sym}^{AB} + \frac{1}{2}\lambda\,\veps^{AB} ~, \label{eq:TwistedFermionDecompositionSpinorIndices}
}
into a symmetric part $\lambda_{\sym}^{AB}$ and a trace $\lambda$. The symmetric object $\lambda_{\sym}^{AB}$ is the spinorial version of a SD 2-form $\lambda_{\mu\nu}^{+}$ (see \eqref{eq:4-comp self and antiself dual in terms of 2-comp} and \eqref{eq:2-comp self and antiself dual in terms of 4-comp}). The three summands of \eqref{eq:TwistedFermionRep} correspond, respectively, to a $1$-form $\lambda_{\mu}$ (which is the same as $\lambda_{A\dA}$, due to \eqref{eq:vector-bispinor-equivalence}), a $0$-form $\lambda$ (i.e., the trace in \eqref{eq:TwistedFermionDecompositionSpinorIndices}), and a SD $2$-form $\lambda_{\mu\nu}^{+}$ (i.e., the first term, $\lambda^{AB}_{\sym}$, in \eqref{eq:TwistedFermionDecompositionSpinorIndices}).
This is merely a spinorial reinterpretation of \eqref{eq:TwistedFermionRep}. The spin connection $1$-form has components $\omega_{\mu}{}^{ab}$, but we often refer to its SD part as $\omega_{\mu}{}^{(AB)}$ (the $\mathfrak{su}(2)_+$ spin connection), and ASD part as $\omega_{\mu}{}^{(\dA\dB)}$ (the $\mathfrak{su}(2)_-$ spin connections). \\

\noindent \underline{\textbf{The Riemann tensor in spinor notation}}.
The Riemann tensor can be written as: 
\eqa{
	&R_{A\dA B\dB C\dC D\dD}(\omega) &&=  \veps_{AB}\veps_{CD}C_{\dA\dB\dC\dD} + \veps_{\dA\dB}\veps_{\dC\dD} C_{ABCD} 
	 + \veps_{AB}\veps_{\dC\dD} R_{\dA\dB C D} + \veps_{\dA\dB}\veps_{CD}R_{AB\dC\dD} \nn
	& && \quad  + \frac{\mathscr{R}}{24}\left[\veps_{AB}\veps_{CD}(\veps_{\dA\dC}\veps_{\dB\dD} + \veps_{\dB\dC}\veps_{\dA\dD}) + \veps_{\dA\dB}\veps_{\dC\dD}(\veps_{AC}\veps_{BD} + \veps_{BC}\veps_{AD})\right] ~,\label{eq:riemann spinor}
}
where $\mathscr{R}$ denotes the scalar curvature. As a consequence of this decomposition, 
\eqa{
& R_{ABCD}(\omega) &&:= \veps^{\dA\dB}\veps^{\dC\dD}R_{A\dA B\dB C\dC D\dD}(\omega) = 4\,C_{ABCD} + \frac{\mathscr{R}}{6}(\veps_{AC}\veps_{BD} + \veps_{BC}\veps_{AD}) ~, \label{eq:RABCD}\\
& R_{AE}{}^{EF}(\omega) &&= \frac{\mathscr{R}}{2}\,\delta_{A}{}^{F} ~, \quad R_{EF}{}^{EF}(\omega) = \mathscr{R} \label{eq:contracted R} ~,
}
where $\mathscr{R}$ denotes the Ricci scalar. Similar properties hold for the dotted components. 

The exchange property of the Riemann tensor, $R_{A\dA B\dB C\dC D\dD} = R_{C\dC D\dD A\dA B\dB}$ translates into the following conditions on the individual pieces of \eqref{eq:riemann spinor}:
\eqasfour{
   & C_{ABCD} &&= C_{CDAB} ~, \quad & C_{\dA\dB\dC\dD} &&= C_{\dC\dD\dA\dB} ~, \quad  & R_{AB\dA\dB} &&= R_{\dA\dB AB} ~.
}

\section{Useful Identites}\label{app:useful-identities}
 
\subsection{Self- And Anti-self- Dual Projections}
In this appendix, we list and prove some identities involving self-dual (SD) and anti-self-dual (ASD) projections of composite antisymmetric tensors. We denote a generic rank-2 symmetric tensor by $\sfS \in \mathsf{Sym}^2 (T\IX)$ and a generic 2-form by $\sfA \in \Omega^{2}(\IX)$. 
We also use $\sfX, \sfY \in \Omega^{1}(\IX)$ to denote generic 1-forms on $\IX$. In a local chart the components are $\sfS^{\mu\nu}$, $\sfA_{\mu\nu}$, $\sfX_\mu$ and $\sfY_\nu$. The SD and ASD projections are given by \eqref{eq:Tmunu in terms of curved indices}.
\begin{itemize}
\item For two one-forms $\sfX = dx^{\mu}\sfX_\mu$ and $\sfY = dx^{\nu}\sfY_\nu$,
\eqa{
  & \sfX^{\mu}\big(\sfX_{[\mu}\sfY_{\nu]}\big)^{\pm} &= \frac{1}{2}\,\sfX^{\mu}\big(\sfX_{[\mu}\sfY_{\nu]}\big) \label{eq:contracted SD equals ASD} ~.
}
 \item For the symmetric gravitino $\Psi_{μν}$ and a generic antisymmetric tensor $\sfA_{μν}$, 
\eqa{
	&\big(\Psi_{[\mu}{}^{\rho}\sfA_{\nu]\rho}^{\pm}\big)^{\pm} &&= -\frac{1}{4}\Psi_{\rho}{}^{\rho}\sfA_{\mu\nu}^{\pm} ~. \label{eq:identB}
	}
\item The identity \eqref{eq:identB} can be generalized to \emph{any} symmetric tensor $\sfS_{μν}$:
\eqa{
  &\big(\sfS_{[\mu}{}^{\rho}\sfA_{\nu]\rho}^{\pm}\big)^{\pm} &&= -\frac{1}{4}\sfS_{\sigma}{}^{\sigma}\sfA_{\mu\nu}^{\pm} ~. \label{eq:identB general}
}
\item For two antisymmetric tensors $\sfA$ and $\sfA'$,
\eqa{
	&\big(\sfA'_{[\mu}{}^{\rho}\sfA_{\nu]\rho}^{\pm}\big)^{\mp} &&= 0 \label{eq:identB antisym general}~.
}
It follows that $\sfA'_{[\mu}{}^{\rho}\sfA_{\nu]\rho}^{+}$ is always self-dual and $\sfA'_{[\mu}{}^{\rho}\sfA_{\nu]\rho}^{-}$ is always anti-self-dual.
\item For the symmetric gravitino $\Psi_{\mu\nu}$ and the antisymmetric self-dual field $\chi_{\mu\nu}$,
\eqa{
\Psi^{\rho\sigma}\big(\Psi_{\rho[\mu}\chi_{\nu]\sigma}\big)^{-} &= 0 ~, \label{eq:Psi Psi Chi ASD}\\
\left[ \Psi_{\mu}{}^{\sigma}\big(\Psi_{[\nu}{}^{\rho}\chi_{\sigma]\rho}\big)^{-} - \Psi_{\nu}{}^{\sigma}\big(\Psi_{[\mu}{}^{\rho}\chi_{\sigma]\rho}\big)^{-}\right]^{+} &= \Psi^{\rho\sigma}\Psi_{\sigma[\mu}\chi_{\nu]\rho} ~.    \label{eq:Psi Psi Chi Diff SD} 
}
\end{itemize}
\noindent \underline{\textbf{Proof of \eqref{eq:contracted SD equals ASD}}}.
\eqa{
  & \sfX^{\mu}\big(\sfX_{[\mu}\sfY_{\nu]}\big)^{\pm} &&= \sfX^{\mu}\left(\frac{1}{2}\sfX_{[\mu}\sfY_{\nu]} \pm \frac{e}{4}\veps_{\mu\nu\rho\sigma}\sfX^{\rho}\sfY^{\sigma}\right) =  \frac{1}{2}\sfX^{\mu}\big(\sfX_{[\mu}\sfY_{\nu]}\big) ~. \nonumber 
} 
\noindent \underline{\textbf{Proof of \eqref{eq:identB} and \eqref{eq:identB general}}}. For a self-dual 2-form $\sfA$,
\eqa{
 &\big(\sfS_{[μ}{}^{ρ}\sfA_{ν]ρ}\big)^{+} &&= \frac{1}{2}\sfS_{[μ}{}^{ρ}\sfA_{ν]ρ} + \frac{e}{4}\veps_{μνρσ}\sfS^{ρ}{}_{λ}\sfA^{σλ}  \nn
 & &&\stackrel{\eqref{eq:SD ASD upper indices}}{=\joinrel=\joinrel=} \frac{1}{2}\sfS_{[μ}{}^{ρ}\sfA_{ν]ρ} - \frac{1}{8}\sfS^{ρ}{}_{λ}ε_{σμνρ}ε^{σλκη}\sfA_{κη} \nn
 & && \stackrel{\eqref{eq:epsilon contraction identity spl case}}{=\joinrel=\joinrel=}  \frac{1}{2}\sfS_{[μ}{}^{ρ}\sfA_{ν]ρ} - \frac{1}{8}\big(2\,\sfS_{μ}{}^{ρ}\sfA_{νρ} + 2\,\sfS_{ρ}{}^{ρ}\sfA_{μν} - 2\,\sfS_{ν}{}^{ρ}\sfA_{μρ}\big) \nn
 & &&= -\frac{1}{4}\sfS_{ρ}{}^{ρ}\sfA_{μν} ~. \nonumber
}
In the second equality, we used the fact that $\sfA$ is self-dual and introduced an $\veps$-symbol using \eqref{eq:SD ASD upper indices}, allowing the use of the identity \eqref{eq:epsilon contraction identity spl case} in the third equality, leading to the desired result. The corresponding result for an anti-self-dual $\sfA$ can be worked out similarly. The identity \eqref{eq:identB} is a special case of this one for $\sfS_{\mu\nu} \to \Psi_{\mu\nu}$.

For completeness, we provide a two-component proof of \eqref{eq:identB}. Let $\alpha_{μν} := \Psi_{[μ}{}^{ρ}\sfA_{ν]ρ}$, i.e., the LHS of \eqref{eq:identB}. Since $\sfA$ is self-dual by assumption, \eqref{eq:decomposition of antisym tensor in terms of self and anti self bispinor} tells us that $\sfA_{A\dA,B\dB} = \frac{1}{2}\sfA_{AB}\veps_{\dA\dB}$. Therefore,
\eqas{\nonumber
  &α_{A\dA,B\dB} &&= \frac{1}{4}\big(-Ψ_{A\dA}{}^{C}{}_{\dB}\sfA_{BC} + Ψ_{B\dB}{}^{C}{}_{\dA}\sfA_{AC}\big) ~,
}
and hence the self-dual part is
\eqa{\nonumber
  &α_{AB} &&= \veps^{\dA\dB}α_{A\dA,B\dB} =  -\frac{1}{4}\big(Ψ_{A\dA}{}^{C\dA}\sfA_{BC} + Ψ_{B\dA}{}^{C\dA}\sfA_{AC}\big) &\stackrel{\text{\eqref{eq:gravitino contraction trace}}}{=\joinrel=\joinrel=} -\frac{1}{4}Ψ\,\sfA_{AB} ~,
}
which directly yields \eqref{eq:identB}. 
 \hfill $\blacksquare$
\\\\
The 4-component method relies on well-known identities obeyed by the Levi-Civita symbol. On the other hand, the 2-component method enables us to simply read off the (A)SD part, because here one exploits manifest irreducibility of the expressions with respect to $\mathfrak{su}(2)_\pm$.
\\\\
\noindent \underline{\textbf{Proof of \eqref{eq:identB antisym general}}}. Let $\alpha_{\mu\nu} := \sfA'_{[\mu}{}^{\rho}\sfA_{\nu]\rho}^{+}$. Using \eqref{eq:decomposition of antisym tensor in terms of self and anti self bispinor} to write $\sfA_{A\dA,B\dB}$ and $\sfA'_{A\dA,B\dB}$, we find 
\eqa{
  & \alpha_{A\dA,B\dB} &&= \tfrac{1}{8}\left[\big(\sfA'_{A}{}^{C}\delta_{\dA}{}^{\dC} + \sfA'_{\dA}{}^{\dC}\delta_{A}{}^{C}\big)\sfA_{BC}\veps_{\dB\dC} - \big(\sfA'_{B}{}^{C}\delta_{\dB}{}^{\dC} + \sfA'_{\dB}{}^{\dC}\delta_{B}{}^{C}\big)\sfA_{AC}\veps_{\dA\dC}\right] \nn
   & &&= \tfrac{1}{8}\left[\big(\sfA'_{A}{}^{C}\sfA_{BC}\veps_{\dB\dA} - \sfA'_{\dA\dB}\sfA_{BA}\big) - \big(\sfA'_{B}{}^{C}\sfA_{AC}\veps_{\dA\dB} - \sfA'_{\dB\dA}\sfA_{AB}\big) \right] \nn
   & &&= -\tfrac{1}{8}\left[\sfA'_{A}{}^{C}\sfA_{BC} + \sfA'_{B}{}^{C}\sfA_{AC} \right]\veps_{\dA\dB} ~,
}
which, being symmetric under $A\leftrightarrow B$ and antisymmetric under $\dA\leftrightarrow \dB$, is purely self-dual. $\hfill\blacksquare$
\\\\
\noindent \underline{\textbf{Proof of \eqref{eq:Psi Psi Chi ASD}}}. 
  Let $\alpha_{\mu\nu} := \Psi^{\rho\sigma}\big(\Psi_{\rho[\mu}\chi_{\nu]\sigma}\big)$. In two-component notation, this is
\eqa{
     &\alpha_{A\dA,B\dB} &&= \tfrac{1}{4}\Psi^{C\dC,D\dD}\big(\Psi_{C\dC,A\dA}\chi_{BD}\varepsilon_{\dB\dD} - \Psi_{C\dC,B\dB}\chi_{AD}\varepsilon_{\dA\dD}\big) \nonumber\\
    & &&= \tfrac{1}{4}\big(-\Psi^{C\dC,D}{}_{\dB}\Psi_{C\dC,A\dA}\chi_{BD} + \Psi^{C\dC,D}{}_{\dA}\Psi_{C\dC,B\dB}\chi_{AD}\big) ~.
}
Therefore, the anti-self-dual part of $\alpha$ is
\eqa{
     &\alpha_{\dA\dB} = \varepsilon^{AB}\alpha_{A\dA,B\dB} &&= \tfrac{1}{4}\big(-\Psi^{C\dC,D}{}_{\dB}\Psi_{C\dC,A\dA}\chi^{A}{}_{D} - \Psi^{C\dC,D}{}_{\dA}\Psi_{C\dC,A\dB}\chi^{A}{}_{D}\big) \nonumber\\
     & &&=  \tfrac{1}{4}\big(\Psi^{C\dC}{}_{D\dB}\Psi_{C\dC,A\dA} - \Psi_{C\dC,A\dB}\Psi^{C\dC}{}_{D\dA}\big)\chi^{AD} \nonumber\\
     &  &&=  \tfrac{1}{4}\big(\Psi^{C\dC}{}_{D\dB}\Psi_{C\dC,A\dA} - \Psi_{C\dC,D\dB}\Psi^{C\dC}{}_{A\dA}\big)\chi^{AD} = 0 ~, 
}
  where, in the penultimate step, we used that $\chi^{AD}$ is symmetric.  \hfill $\blacksquare$
\\\\
\noindent \underline{\textbf{Proof of \eqref{eq:Psi Psi Chi Diff SD}}}. 
 \begin{align}
    \left[ \Psi_{\mu}{}^{\sigma}\big(\Psi_{[\nu}{}^{\rho}\chi_{\sigma]\rho}\big)^{-} - \Psi_{\nu}{}^{\sigma}\big(\Psi_{[\mu}{}^{\rho}\chi_{\sigma]\rho}\big)^{-}\right]^{+} &\stackrel{\text{\eqref{eq:identB}}}{=\joinrel=\joinrel=} \left[ \Psi_{\mu}{}^{\sigma}\big(\Psi_{[\nu}{}^{\rho}\chi_{\sigma]\rho} + \tfrac{1}{4}\Psi_{\kappa}{}^{\kappa}\chi_{\nu\sigma}\big) - (\mu \leftrightarrow \nu) \right]^{+} \nonumber\\
    &= \big[ \Psi^{\sigma\rho}\Psi_{\sigma[\mu}\chi_{\nu]\rho} - \tfrac{1}{2}\Psi_{\kappa}{}^{\kappa}\underbrace{\big(\Psi_{[\mu}{}^{\sigma}\chi_{\nu]\sigma}\big)}_{=-\tfrac{1}{4}\Psi_{\rho}{}^{\rho}\chi_{\mu\nu}}\big]^{+} \nonumber\\
    &\stackrel{\text{\eqref{eq:identB}}}{=\joinrel=\joinrel=} \Psi^{\sigma\rho}\big(\Psi_{\sigma[\mu}\chi_{\nu]\rho}\big)^{+} \nonumber \\
    &\stackrel{\text{\eqref{eq:Psi Psi Chi ASD}}}{=\joinrel=\joinrel=} \Psi^{\sigma\rho}\Psi_{\sigma[\mu}\chi_{\nu]\rho} ~,
 \end{align}
 where, in the first step, we wrote the ASD projection as the difference of the unprojected term and its SD projection, and used \eqref{eq:identB} on the latter, and in the last equality, we used \eqref{eq:Psi Psi Chi ASD} to drop the redundant SD projection.\hfill $\blacksquare$

\subsection{Spinorial Decomposition Of The Twisted Gravitino}\label{subsec:spinorial-decomposition-of-twisted-gravitino}
 The gravitino can be decomposed as 
\eqa{
	&\Psi_{\mu\nu} &&= \underbrace{\left(\Psi_{(\mu\nu)} - \frac{1}{4}\Psi_{\rho}{}^{\rho}g_{\mu\nu}\right)}_{\text{symmetric, traceless}} +  \underbrace{\frac{1}{4}\Psi_{\rho}{}^{\rho}g_{\mu\nu}}_{\text{trace contribution}} + \underbrace{\Psi_{[\mu\nu]}^{+}}_{\text{SD part}} +  \underbrace{\Psi_{[\mu\nu]}^{-}}_{\text{ASD part}} ~\\
	& &&= \textcolor{blue}{\wh{\Psi}_{\mu\nu}} + \frac{1}{4}\textcolor{red}{\Psi_{\rho}{}^{\rho}}g_{\mu\nu} + \Psi_{[\mu\nu]}^{+} +  \Psi_{[\mu\nu]}^{-} ~. \label{eq:4comp}
}
In two-component form,
\eqa{
	&\Psi_{A\dA,B\dB} &&= \textcolor{blue}{\widehat{\Psi}_{(AB),(\dA\dB)}} + \frac{1}{4}\veps_{AB}\veps_{\dA\dB}\textcolor{red}{\Psi} + \frac{1}{2}\Psi_{(AB)}\veps_{\dA\dB} + \frac{1}{2}\Psi_{(\dA\dB)}\veps_{AB} ~. \label{eq:2comp}
}
The translation between \eqref{eq:4comp} and \eqref{eq:2comp} is:\footnote{Note that $g_{\mu\nu} = e_{\mu}{}^{A\dA}e_{\nu,A\dA}$.}
\eqa{
	&\textcolor{blue}{\wh{\Psi}_{\mu\nu}} &&:= \Psi_{(\mu\nu)} - \frac{1}{4}\Psi_{\rho}{}^{\rho}g_{\mu\nu} = \textcolor{blue}{\widehat{\Psi}_{(AB),(\dA\dB)}} e_{\mu}{}^{A\dA}e_{\nu}{}^{B\dB} ~ , \label{eq:gravitino 2comp symtraceless}\\
	&\textcolor{red}{\Psi_{\rho}{}^{\rho}} &&:= \textcolor{red}{\Psi}  \label{eq:gravitino 2comp trace}  ~,\\
	&\Psi_{[\mu\nu]}^{+} &&:= \frac{1}{2}\Psi_{(AB)}\veps_{\dA\dB} e_{\mu}{}^{A\dA}e_{\nu}{}^{B\dB} ~, \label{eq:gravitino 2comp SD}\\
	&\Psi_{[\mu\nu]}^{-} &&:= \frac{1}{2}\Psi_{(\dA\dB)}\veps_{AB} e_{\mu}{}^{A\dA}e_{\nu}{}^{B\dB} ~. \label{eq:gravitino 2comp ASD}
}
Using this decomposition, and the identity \eqref{eq:identA refined}, we can write the 2-component versions of the transformation laws of the irreducible parts of the gravitino (as defined in \eqref{eq:2comp}) as
\eqas{\label{eq:tsugra-gravitino-irred-2comp}
  &\delta\Psi &&= 2\,\n_{A\dA}\eps^{A\dA} + 4\,\bn_0 ~,\\
  &\delta \wh{\Psi}_{AB,\dA\dB} &&= -\tfrac{\eps}{2}\left[\wh{\Psi}_{(A}{}^{C}{}_{|\dA\dB|}\Psi_{B)C} + \wh{\Psi}_{AB,(\dA}{}^{\dC}\Psi_{\dB)\dC}\right] + \big(2\,\n_{(A\dA}\eps_{B\dB)} - \tfrac{1}{2}\veps_{AB}\veps_{\dA\dB}\n_{\mu}\eps^{\mu}\big) ~, \\
  &\delta\Psi_{AB} &&= -\tfrac{\eps}{2}\left(-\wh{\Psi}_{(A}{}^{C,\dA\dB}\wh{\Psi}_{B)C,\dA\dB} + \tfrac{1}{2}\Psi_{A}{}^{C}\Psi_{BC} \right) - 2\,\n_{(A}{}^{\dA}\eps_{B)\dA} - \bn_{(AB)}^{\text{sym}} ~, \\
  &\delta\Psi_{\dA\dB} &&= -\tfrac{\eps}{2}\left(-\wh{\Psi}_{AB,(\dA}{}^{\dC}\wh{\Psi}^{AB}{}_{\dB)\dC} + \tfrac{1}{2}\Psi_{\dA}{}^{\dC}\Psi_{\dB\dC}\right) + 2\,\n_{A(\dA}\eps^{A}{}_{\dB)} - \eps\,T_{\dA\dB} ~,
}
or in 4-component notation, in terms of \eqref{eq:4comp}, as
\eqas{\label{eq:tsugra-gravitino-irred-4comp}
  &\delta\Psi_{\mu}{}^{\mu} &&= 2\,\n_{\mu}\eps^{\mu} + 4\,\bn_0 ~,\\
  &\delta\wh{\Psi}_{\mu\nu} &&= \n_{\mu}\eps_{\nu} + \n_{\nu}\eps_{\mu} - \frac{1}{2}g_{\mu\nu}\n_{\rho}\eps^{\rho} - \frac{1}{4}\eps\,\wh{\Psi}_{\mu\nu}\Psi ~,\\
  &\delta\Psi_{[\mu\nu]}^{+} &&= -\eps\big(\wh{\Psi}_{[\mu}{}^{\rho}\Psi^{+}_{\nu]\rho}\big) - \frac{\eps}{2}\big(\wh{\Psi}_{[\mu}{}^{\rho}\wh{\Psi}_{\nu]\rho}\big)^{+}  - \frac{\eps}{2}\big(\Psi^{+}_{\mu}{}^{\rho}\Psi^{+}_{\nu\rho}\big) + 2\big(\n_{[\mu}\eps_{\nu]}\big)^{+} - \bn_{\mu\nu}^{+}~,\\
   &\delta\Psi_{[\mu\nu]}^{-} &&= -\eps\big(\wh{\Psi}_{[\mu}{}^{\rho}\Psi^{-}_{\nu]\rho}\big) - \frac{\eps}{2}\big(\wh{\Psi}_{[\mu}{}^{\rho}\wh{\Psi}_{\nu]\rho}\big)^{-}  - \frac{\eps}{2}\big(\Psi^{-}_{\mu}{}^{\rho}\Psi^{-}_{\nu\rho}\big) + 2\big(\n_{[\mu}\eps_{\nu]}\big)^{-} - \eps\,T_{\mu\nu}^{-}~.
}
Note that the SD/ASD projections on certain bilinear terms on the RHS have been dropped, since they are redundant due to the identity \eqref{eq:identB antisym general}.

\subsection{Curl Of The Twisted Gravitino}\label{subsec:curl-twisted-gravitino}
Substituting \eqref{eq:2comp} into $\CJ_{[A\dA,B\dB],C\dC} \equiv \n_{[A\dA}\Psi_{B\dB],C\dC}$ yields
\eqa{
	&\CJ_{[A\dA,B\dB],C\dC} &&= \frac{1}{2}\left[(\n_{A\dA}\widehat{\Psi}_{BC,\dB\dC} - \n_{B\dB}\widehat{\Psi}_{AC,\dA\dC}) + \frac{1}{4}(\veps_{BC}\veps_{\dB\dC}\n_{A\dA}\Psi - \veps_{AC}\veps_{\dA\dC}\n_{B\dB}\Psi) \right.\nonumber\\
	& &&\qquad + \left. \frac{1}{2}(\veps_{\dB\dC}\n_{A\dA}\Psi_{BC} - \veps_{\dA\dC}\n_{B\dB}\Psi_{AC}) + \frac{1}{2}(\veps_{BC}\n_{A\dA}\Psi_{\dB\dC} - \veps_{AC}\n_{B\dB}\Psi_{\dA\dC})\right] ~.
}
In particular, the SD and ASD parts are $\CJ_{AB,C\dC} := \veps^{\dA\dB}\CJ_{[A\dA,B\dB],C\dC}$ and $\CJ_{\dA\dB,C\dC} := \veps^{AB}\CJ_{[A\dA,B\dB],C\dC}$, which evaluate respectively to
\eqa{
	&\CJ_{AB,C\dC} &&= \frac{1}{2}\left[(\n_{A\dA}\widehat{\Psi}_{BC}{}^{\dA}{}_{\dC} + \n_{B\dA}\widehat{\Psi}_{AC}{}^{\dA}{}_{\dC}) - \frac{1}{4}(\veps_{BC}\n_{A\dC}\Psi + \veps_{AC}\n_{B\dC}\Psi)\right. \nonumber\\
	& && \qquad  \left. -\frac{1}{2}(\n_{A\dC}\Psi_{BC} + \n_{B\dC}\Psi_{AC}) + \frac{1}{2}(\veps_{BC}\n_{A\dA}\Psi^{\dA}{}_{\dC} + \veps_{AC}\n_{B\dA}\Psi^{\dA}{}_{\dC})\right] ~, \label{eq:self-dual-J-2comp}\\
	& \CJ_{\dA\dB,C\dC} &&= \frac{1}{2}\left[(\n_{A\dA}\widehat{\Psi}^{A}{}_{C,\dB\dC} + \n_{A\dB}\widehat{\Psi}^{A}{}_{C,\dA\dC}) - \frac{1}{4}(\veps_{\dB\dC}\n_{C\dA}\Psi + \veps_{\dA\dC}\n_{C\dB}\Psi) \right.\nonumber\\
	& &&\qquad  \left. + \frac{1}{2}(\veps_{\dB\dC}\n_{A\dA}\Psi^{A}{}_{C} + \veps_{\dA\dC}\n_{A\dB}\Psi^{A}{}_{C}) - \frac{1}{2}(\n_{C\dA}\Psi_{\dB\dC} + \n_{C\dB}\Psi_{\dA\dC})\right] ~. \label{eq:anti-self-dual-J-2comp}
}
Two contractions of $\CJ$ that frequently arise are:
\eqa{
	&\label{eq:curly j sd contracted}\CJ^{+}_{\mu\nu,ρ}{}^{\rho}  &&:= \CJ_{[\mu\nu],\rho}^{+}g^{\rho\nu} = \frac{1}{2}\CJ_{A}{}^{E}{}_{E\dA}e_{\mu}{}^{A\dA} ~,\\
	&\label{eq:curly j asd contracted} \CJ^{-}_{\mu\nu,ρ}{}^{\rho}  &&:= \CJ_{[\mu\nu],\rho}^{-}g^{\rho\nu} = \frac{1}{2}\CJ_{\dA}{}^{\dE}{}_{A\dE}e_{\mu}{}^{A\dA} ~.
}
In terms of the two-component decomposition  \eqref{eq:2comp}, 
\eqa{
	& 2e^{\mu}{}_{A\dA} g^{\rho\nu}\mathcal{J}_{[\mu\nu],\rho}^{+} &&= \mathcal{J}_{A}{}^{E}{}_{E\dA} = \frac{1}{2}\left[-\n_{E\dE}\widehat{\Psi}_{A}{}^{E,\dE}{}_{\dA} + \frac{3}{4}\n_{A\dA}\Psi + \frac{1}{2}\n_{E\dA}\Psi_{A}{}^{E} - \frac{3}{2}\n_{A\dE} \Psi^{\dE}{}_{\dA}\right] ~, \label{eq:contJbig1}\\
	& 2e^{\mu}{}_{A\dA} g^{\rho\nu}\mathcal{J}_{[\mu\nu],\rho}^{+} &&= \mathcal{J}_{\dA}{}^{\dE}{}_{A\dE} = \frac{1}{2}\left[-\n_{E\dE}\widehat{\Psi}_{A}{}^{E,\dE}{}_{\dA} + \frac{3}{4}\n_{A\dA}\Psi + \frac{1}{2}\n_{A\dE}\Psi_{\dA}{}^{\dE} - \frac{3}{2}\n_{E\dA}\Psi^{E}{}_{A}\right]  \label{eq:contJbig2}~.
}
Therefore,
\eqa{
	& g^{\rho\nu}\mathcal{J}_{[\mu\nu],\rho}^{+} -  g^{\rho\nu}\mathcal{J}_{[\mu\nu],\rho}^{-} &&= \frac{1}{2}e_{\mu}{}^{A\dA}\left(\mathcal{J}_{A}{}^{E}{}_{E\dA} - \mathcal{J}_{\dA}{}^{\dE}{}_{A\dE}\right) \nonumber\\
	& &&=\frac{1}{2}e_{\mu}{}^{A\dA}\left(\n_{E\dA}\Psi_{A}{}^{E} - \n_{A\dE}\Psi^{\dE}{}_{\dA}\right) \nonumber\\
	& &&= -\n^{\rho}\Psi_{[\rho\mu]}^{+} + \n^{\rho}\Psi_{[\rho\mu]}^{-} ~. \label{eq:diff contJbig}
}
If the gravitino is symmetric, $\Psi_{[\mu\nu]}^{+} = 0$ and $\Psi_{[\mu\nu]}^{-} = 0$. This means that $\Psi_{AB} = 0$ and $\Psi_{\dA\dB} = 0$ in \eqref{eq:2comp}, and therefore,
\eqa{
&\text{symmetric gravitino :} \quad Ψ_{A\dA,B\dB} &&=  \textcolor{blue}{\widehat{\Psi}_{(AB),(\dA\dB)}} + \frac{1}{4}\veps_{AB}\veps_{\dA\dB}\textcolor{red}{\Psi} ~. \label{eq:spinorial-decomp-sym-gravitino}
}
In particular, for a symmetric gravitino, 
\eqas{
  Ψ_{A\dA,B}{}^{\dA} &= \frac{1}{2}\veps_{AB}\textcolor{red}{Ψ} ~, \qquad &Ψ_{A\dA}{}^{B\dA} &= \frac{1}{2}δ_{A}{}^{B}\textcolor{red}{Ψ} ~. \label{eq:gravitino contraction trace}
}
Furthermore, if the gravitino is symmetric, \eqref{eq:contJbig1} and \eqref{eq:contJbig2} are equal
\eqa{
	& \mathcal{J}_{A}{}^{E}{}_{E\dA} &&=  -\frac{1}{2}\n_{E\dE}\widehat{\Psi}_{A}{}^{E,\dE}{}_{\dA} + \frac{3}{8}\n_{A\dA}\Psi  \label{eq:contJbig1-sym}~,\\
	& \mathcal{J}_{\dA}{}^{\dE}{}_{A\dE} &&= -\frac{1}{2}\n_{E\dE}\widehat{\Psi}^{E}{}_{A,\dA}{}^{\dE} + \frac{3}{8}\n_{A\dA}\Psi ~, \label{eq:contJbig2-sym}
}
To summarize,
\eqa{
		&\text{symmetric gravitino } \implies \left\{ \begin{array}{lll} \mathcal{J}_{A}{}^{E}{}_{E\dA} &= \mathcal{J}_{\dA}{}^{\dE}{}_{A\dE} ~, & \\ \mathcal{J}_{[\mu\nu],\rho}^{+}g^{\rho\nu} &= \mathcal{J}_{[\mu\nu],\rho}^{-}g^{\rho\nu} &= -\frac{1}{4}\n^{\nu}\Psi_{\mu\nu} + \frac{1}{4}\n_{\mu}\Psi_{\rho}{}^{\rho} ~.\end{array}\right. \label{eq:symmetric-gravitino-simp-Jcont}
}

\section{Self-Duality And Projected Connections}\label{app:variation-of-self-dual-fields}
The vectormultiplet fieldspace is nontrivially fibered over the space of Riemannian metrics, and consists of self-dual differential form fields (bosonic and fermionic). The self-duality condition, of course, relies on the conformal class of the metric. Under a variation of the metric, a self-dual (resp. anti-self-dual) 2-form must receive a compensating anti-self-dual (resp. self-dual) transformation so as to preserve the self-duality (resp. anti-self-duality) condition. In our setup, since we \emph{do} vary the metric -- see, for example the action of $\mathsf{d}$ on $g_{\mu\nu}$ in \eqref{eq:GravityCartan-1}, or the action of $\IQ$ on $g_{\mu\nu}$ in \eqref{eq:CartanAlgebra-1} --  this is an important concept that we review here. We also illustrate how the use of frames handles this effortlessly. 

Let $\chi \in \Omega^{2,+}(\IX)$ denote a self-dual 2-form, so that 
\eqa{
  & \chi &&= \star \chi ~, \label{eq:SD condition of chi}
}
where $\star$  is the Hodge star operator on $\IX$. 
 Under a metric variation, $g \mapsto g' = g + δg$, the 2-form field $χ$ varies as $χ \mapsto χ' = χ + δ_{g}χ$. We require that \eqref{eq:SD condition of chi} be preserved under this metric variation. Since $\delta$ is a linear operator satisfying the Leibniz rule, \eqref{eq:SD condition of chi} in particular implies that $δ_{g}χ = \big(δ_{g}\star\big)χ + \star\big(δ_{g}χ\big)$, from which it follows that
\eqa{
  & \big(δ_{g}χ\big)^{-} := \frac{1}{2}\big(1-\star\big)δ_{g}χ &&= \frac{1}{2}\big(δ_{g}\star\big)χ ~. \label{eq:variation of SD condition}
}
Therefore the \emph{variation} $δ_{g}χ$ of a self-dual 2-form $χ$ has a \emph{constrained} \underline{anti}-self-dual part. The self-dual part of the variation is not constrained, however. This was observed already in \cite{Witten:1988ze}. In a local coordinate chart for $\IX$, \eqref{eq:SD condition of chi} is equivalent to
\eqa{
   &χ_{μν} &&= \frac{1}{2}e \,\varepsilon_{μνρσ}χ^{ρσ} ~,\label{eq:SD condition of chi components}
}
where $e = \sqrt{|\det\, g|}$. Using standard relations $δ_{g} g^{μν} =  - g^{μρ}g^{νσ}\delta g_{ρσ}$ and $δ_{g} e = \frac{1}{2}e\,g^{μν}\delta g_{μν}$,
the condition \eqref{eq:variation of SD condition} can be written as
\eqa{
 &\big(δ_{g}χ_{μν}\big)^{-} 
 & &&= \frac{1}{2}e\,\veps_{μνρσ}\big(δ_{g}g^{ρρ'}\big)χ_{ρ'}{}^{σ} - \frac{1}{8}e\,\veps_{μνρσ}χ^{ρσ}g_{ρ'σ'}δ_{g}g^{ρ'σ'} ~. \label{eq:constrained ASD part of SD variation}
}
Therefore, the \emph{compensating transformation} \eqref{eq:constrained ASD part of SD variation} (a manifestly anti-self-dual contribution) must be added to the variation of $χ_{μν}$ with respect to the \underline{unperturbed} metric, for the perturbed $χ_{μν}'$ to remain self-dual with respect to the perturbed metric $g_{μν}'$.\footnote{Similar results can be obtained for the variation of anti-self-dual 2-forms, by a series of straightforward sign flips in the preceding equations.}

It is instructive to review how the ``two-component approach'' handles the (anti)-self-duality condition rather equivariantly, and in fact, automatically introduces the required compensating transformations. If we choose a basis of orthonormal frames, we can use \eqref{eq:c43} to write
\eqa{
& \chi_{\mu\nu} &&= \chi_{ab}^{+}e_{\mu}{}^{a}e_{\nu}{}^{b} ~, \label{eq:chimunu-ab}
}
where $\chi_{ab}^{+}$ is antisymmetric and the superscript is meant to remind us that it is self-dual. Suppose $\delta e_{\mu}{}^{a} = \frac{1}{2}\alpha_{\mu}{}^{a}$, so that $\delta g_{\mu\nu} = \alpha_{(\mu\nu)}$.\footnote{The tensor $\alpha_{μν}$, introduced here just to improve readability, is \underline{not} required to be symmetric.} Varying \eqref{eq:chimunu-ab}, we get
\eqa{
 & δ_gχ_{μν} &&= \big(\delta_g\chi_{ab}^{+}\big)e_{\mu}{}^{a}e_{\nu}{}^{b} + \frac{1}{2}\chi_{ab}^{+}\big(\alpha_{\mu}{}^{a}e_{\nu}{}^{b} + e_{\mu}{}^{a}\alpha_{\nu}{}^{b}\big) \nn
 & &&= \big(\delta\chi_{ab}^{+}\big)e_{\mu}{}^{a}e_{\nu}{}^{b} + \frac{1}{2}\big(\alpha_{\mu}{}^{\rho}\chi_{\rho\nu}-\alpha_{\nu}{}^{\rho}\chi_{\rho\mu}\big) \nn
  & &&\stackrel{ \eqref{eq:identB}}{=\joinrel=\joinrel=} \big(\delta\chi_{ab}^{+}\big)e_{\mu}{}^{a}e_{\nu}{}^{b} + \frac{1}{4}α_{ρ}{}^{ρ}χ_{μν} - \big(α_{[μ}{}^{ρ}χ_{ν]ρ}\big)^{-} ~. \label{eq:delta-chi-munu-using-frames}
}
In the first equality, we used the Leibniz rule, in the second we used \eqref{eq:chimunu-ab} twice, and the third equality is due to antisymmetrizing with `strength 1'. In the final step we decomposed $α_{[μ}{}^{ρ}χ_{ν]ρ}$ into self-dual and anti-self-dual parts and used \eqref{eq:identB} on the former. 

We claim that the last term in \eqref{eq:delta-chi-munu-using-frames} is \underline{precisely} the compensating transformation \eqref{eq:constrained ASD part of SD variation}. In particular, using $δ_{g}g^{μν} = -α^{(μν)}$, \eqref{eq:constrained ASD part of SD variation} can be rewritten as
\eqa{
&\big(δ_{g}χ_{μν}\big)^{-} &&= +\frac{1}{8}e\,\veps_{μνρσ}χ^{ρσ}α_{λ}{}^{λ} - \frac{1}{2}e\,\veps_{μνρσ}α^{(ρρ')}χ_{ρ'}{}^{σ} \nonumber\\
& &&= \frac{1}{4}α_{ρ}{}^{ρ}χ_{μν} - \frac{1}{4}\big(\veps_{σμνρ}\veps^{σρ'κη}\big) α^{ρ}{}_{ρ'} χ_{κn} \nonumber\\
& &&\stackrel{\eqref{eq:epsilon contraction identity spl case}}{=\joinrel=\joinrel=} \frac{1}{4}α_{ρ}{}^{ρ}χ_{μν} - \big(α_{[μ}{}^{ρ}χ_{ν]ρ}\big) -  \frac{1}{2}α_{ρ}{}^{ρ}χ_{μν} \label{eq:delta-chi-munu-compensating-penultimate-step}\\
& &&\stackrel{ \eqref{eq:identB}}{=\joinrel=\joinrel=} - \big(α_{[μ}{}^{ρ}χ_{ν]ρ}\big)^{-} ~,\label{eq:constrained ASD part of SD variation simplified}
}
as claimed. In the second equality, we used self-duality of $\chi_{μν}$ to absorb the $\veps$ symbol in the first term and reintroduce an $\veps$ symbol in the second term, so as to use \eqref{eq:epsilon contraction identity spl case} in the next step.\footnote{The computation is simplified by the assumption that $\alpha_{μν}$ is symmetric. But such an assumption is unnecessary, as one observes that any antisymmetric part in $\alpha_{μν}$ never contributes to this computation. More explicitly, instead of \eqref{eq:delta-chi-munu-compensating-penultimate-step} one will have $-\frac{1}{4}α_{ρ}{}^{ρ}χ_{μν} - \frac{1}{2}\alpha^{ρ}{}_{[μ}χ_{ν]ρ} - \frac{1}{2}α_{[μ}{}^{ρ}χ_{ν]ρ}$. But the antisymmetric part of $\alpha_{μν}$ is killed in the sum of the second and third terms.} In the final step, we decomposed $\big(α_{[μ}{}^{ρ}χ_{ν]ρ}\big)$ into self- and anti-self- dual parts and used \eqref{eq:identB} as before.

For completeness, since frame indices can be split into bispinor indices (cf. Appendix \ref{app:spinormethods}),
\eqa{
   &χ_{μν} &&= \frac{1}{2}χ_{AB}e_{μ}{}^{A}{}_{\dA}e_{ν}{}^{B\dA} ~, \label{eq:chimunu-AB}
}
and consequently 
\eqa{
 &δ_gχ_{μν} &&= \frac{1}{2}\big(δ_g χ_{AB}\big)e_{μ}{}^{A}{}_{\dA}e_{ν}{}^{B\dA} + \frac{1}{4}α_{ρ}{}^{ρ}χ_{μν} - \big(α_{[μ}{}^{ρ}χ_{ν]ρ}\big)^{-} ~. \label{eq:delta-chi-munu-using-spinors}
}

\begin{tcolorbox}
Therefore, to vary a self-dual field, we write it as the manifestly self-dual object \eqref{eq:chimunu-ab} or \eqref{eq:chimunu-AB}, and vary it using the Leibniz rule. This automatically includes the compensating transformation \eqref{eq:constrained ASD part of SD variation} that must be added manually in the metric-based approach (where it originates from ``varying the self-duality condition'').
\end{tcolorbox}
Note that \eqref{eq:delta-chi-munu-using-frames} includes a term $δ_{g}χ_{ab}$, which played no role in the preceding discussion. But this term does contribute to the details of the variation of $χ_{μν}$ in specific physical situations, for instance in the case of the twisted $\CN=2$ vectormultiplet coupled to (super)gravity, where $δ_{g}χ_{ab}$ encodes the (local) supersymmetry transformations of the twisted gaugino. 

The bundle of the space of \uline{all} 2-forms on $\IX$ is a canonically trivial bundle as a bundle over the space $\MET(\IX)$ of Riemannian metrics. However, the subbundle of self-dual 2-forms (the space of sections of which, above a point $g \in \MET$ is $\Omega_{g}^{2,+}(\IX)$) is really a \uline{projected bundle} over $\MET(\IX)$ as explained in \cite{Moore:2017byz} and is \uline{not} a trivial bundle, because the projection operator depends continuously on the metric (the \textit{base} of the fiber bundle introduced above). We recall that a vector bundle can equivalently be  viewed as a continuous family of projection operators on a Hilbert space. Therefore, the setup here is identical to the one in \cite{Moore:2017byz}, where a topologically nontrivial bundle can be obtained by projection from a trivial bundle.\footnote{Let us remark, though, that not every projected bundle comes from the projection of a trivial bundle.} 
The computation of $\sfd^2$ (or rather $\widetilde{\sfd}^2$ -- the square of -- the differential $\sfd$ lifted to the above bundle) on self-dual fields is sensitive to the curvature of the projected connection. We provide a short self-contained introduction to this concept, which explains the presence of complicated fermionic bilinear terms in our equivariant cohomology transformation laws -- as derived explicitly in Appendix \ref{app:d2Omega-d2chi}.

Let $\IM$ be a topological space, with $\pi: \mathcal{H} \to \IM$ denoting a Hilbert bundle on $\IM$. (If $\mathcal{H}$ is trivial, it will have the form $\mathcal{H} = \IM \times \mathcal{H}_{0}$ for a fixed Hilbert space $\mathcal{H}_{0}$.) Denote by $P$ a continuous family of projection operators defined over $X$, so that the projection operator at $x \in X$ is $P(x)$. The fiber $E_{x} = \pi^{-1}(x) = \{(x,v) ~|~ v \in \text{Im } P(x)\}$ has the structure of a vector space. 

Given a section of $\mathcal{H}$, namely $\Psi: \IM \to \mathcal{H}$ satisfying $\pi\big(\Psi(x)\big) = x$, we have a continuous map $\psi: \IM \to \mathcal{H}$, given by $x \mapsto \psi(x) \in \mathcal{H}_x$. Using this, we can define a section $s(x)$ of $E$, satisfying $\pi\big(s(x)\big) = x$, by $s(x) = (x, P(x)\psi(x))$. In particular, the space of sections of $E$ is
\eqa{
	&\Gamma(E) &&:= \{ \Psi \in \Gamma(\mathcal{H}) ~:~ P(x)\psi(x) = \psi(x) \quad \forall~ x \in \IM\} ~.
}

The projected connection on $E$ is then given in local coordinates by
\eqas{
	&\n^{\text{P}} : \Gamma(E) &&\longrightarrow \Omega^{1}(X,E) \\
	&   (x, s(x)) &&\longmapsto dx^{\mu}\big(x, P(x)\partial_{\mu}(P(x)s(x))\big)
}
or equivalently, $\n_{X}^{P}s = X^{\mu}P\partial_{\mu}\big(P s\big)$. In differential form notation, this is $\n^{P} = P\circ d \circ P$, where $d$ denotes the exterior derivative.\footnote{We are assuming for calculational simplicity that the ``Hilbert bundle'' is trivial and is equipped with the canonically trivial connection. That is, $\n^{\mathcal{H}} = d$. Indeed one can have a nontrivial connection on the Hilbert bundle, as explained in \cite{Moore:2017byz}. Then one replaces $\partial_{\mu}$ by $\delta^{\alpha}_{\beta}\partial_{\mu} + A_{\mu}{}^{\alpha}{}_{\beta}$, and $s$ by $s^{\beta}$, etc.} The projection operator satisfies the following properties:
\begin{enumerate}\itemsep 0pt
\item $P^2 = P$. (And of course, $P_{\perp} := 1-P$ satisfies $P_{\perp}P = P P_{\perp} = 0$ and $P_{\perp}^2 = P_{\perp}$.)
\item $P(dP) = P (dP) P_{\perp}$ and $(dP)P = P_{\perp}(dP)P$.
\item $dP = P_{\perp}(dP)P + P(dP)P_{\perp}$.
\item $(\partial_{\mu}P)(\partial_{\nu}P) = P_{\perp}(\partial_{\mu}P)P(\partial_{\nu}P)P_{\perp} + P(\partial_{\mu}P)P_{\perp}(\partial_{\nu}P)P$.
\item If $s = P s$, then $P_{\perp} ds = (dP)s$, or equivalently $P_{\perp} \partial_{\mu}s = (\partial_{\mu}P)Ps$.
\end{enumerate}
We claim that the curvature 2-form is simply
\beqa{
	& \bm{F}^{P} &&= P dP\wedge dP P ~. \label{eq:curv-proj-connection}
}
This can be justified as follows. The standard differential geometric curvature has the form
\eqa{
	& F^P(X,Y)s &&= \n^{P}_{X}\n^{P}_{Y}s - \n^{P}_{Y}\n^{P}_{X}s - \n^{P}_{[X,Y]}s ~,
}
where $[X,Y]$ denotes the Lie bracket of vector fields $X, Y \in \Gamma\big(T\IM\big)$. Note that
\eqa{
&	\n_{Y}^{P}s &&= Y^{\mu}P\partial_{\mu}\big(P s\big) = Y^{\mu}P(\partial_{\mu}P\big)P_{\perp}s + Y^{\mu}P\partial_{\mu}s \equiv \widetilde{s} ~.\\
&	\n_{X}^{P}\n_{Y}^{P}s &&= \n_{X}^{P}\widetilde{s} = X^{\mu}\partial_{\mu}Y^{\nu} P (\partial_{\nu}P)P_{\perp}s + X^{\mu}PY^{\nu}(\partial_{\mu}P)(\partial_{\nu}P)P_{\perp}s \nonumber\\
&       &&\quad + X^{\mu}PY^{\nu}P(\partial_{\mu}\partial_{\nu}P)P_{\perp}s - X^{\mu}PY^{\nu}P(\partial_{\nu}P)(\partial_{\mu}P)s \nonumber\\
&	      &&\quad + X^{\mu}PY^{\nu}P(\partial_{\nu}P)P_{\perp}\partial_{\mu}s + X^{\mu}\partial_{\mu}Y^{\nu} P \partial_{\nu}s \nonumber\\
&	      &&\quad + X^{\mu}Y^{\nu}P(\partial_{\mu}P)P_{\perp}\partial_{\nu}s + X^{\mu}PY^{\nu}\partial_{\mu}\partial_{\nu}s ~.
}
Therefore,
\eqa{
&	[\n_{X}^{P},\n_{Y}^{P}]s &&= [X,Y]^{\mu}P(\partial_{\mu}P)P_{\perp}s + (X^{\mu}Y^{\nu} - Y^{\mu}X^{\nu})P(\partial_{\nu}P)P_{\perp}\partial_{\mu}s + [X,Y]^{\mu}P\partial_{\mu}s  ~,\\
&	-\n_{[X,Y]}^{P}s &&= -[X,Y]^{\nu}P\partial_{\nu}(Ps) = -[X,Y]^{\nu}P(\partial_{\nu}P)P_{\perp}s - [X,Y]^{\nu}P\partial_{\nu}s ~.
}
Finally, 
\eqa{
	& F^{P}(X,Y)s &&= (X^{\mu}Y^{\nu} - Y^{\mu}X^{\nu})P(\partial_{\nu}P)P_{\perp}\partial_{\mu}s = (X^{\mu}Y^{\nu} - Y^{\mu}X^{\nu})P(\partial_{\nu}P)(\partial_{\mu}P)s \nonumber\\
	& &&= X^{\mu}Y^{\nu}\big(P \partial_{[\mu}P\big)\big((\partial_{\nu]}P) P\big)s ~,
}
which agrees with \eqref{eq:curv-proj-connection}. $\hfill \blacksquare$
\begin{tcolorbox}
    The key idea here is that since the restriction to the complex of self-dual (resp. anti-self-dual) forms involves a projection operator, the square of the equivariant differential acting on self-dual (resp. anti-self-dual) two-forms picks up contributions from the curvature of the projected connection \eqref{eq:curv-proj-connection} -- these are the gravitino bilinears in Appendix \ref{app:d2Omega-d2chi}.
\end{tcolorbox}
We end this appendix by recording some useful variants of $\IQ$ acting on self-dual and anti-self-dual projections of generic $2$-forms.
\begin{itemize}
\item Since $\IQ g_{μν} = Ψ_{(μν)}$ (cf. \eqref{eq:CartanAlgebra-1}), we have the following equivalent forms of $\IQ \sfA_{μν}^{\pm}$, for a general element $\sfA \in \Omega^{2}(\IX)$: 
\eqa{
	&\IQ\sfA_{\mu\nu}^{\pm} &&= \big(\IQ\sfA_{\mu\nu}\big)^{\pm} \pm \frac{e}{8}\Psi_{\alpha}{}^{\alpha}\veps_{\mu\nu\rho\sigma}\sfA^{\rho\sigma} \mp \frac{e}{4}\veps_{\mu\nu\rho\sigma}\big(\Psi^{(\rho\rho')}\sfA_{\rho'}{}^{\sigma} - \Psi^{(\sigma\rho')}\sfA_{\rho'}{}^{\rho}\big) \nonumber\\
	& &&= \big(\IQ\sfA_{\mu\nu}\big)^{\pm} \pm \frac{e}{8}\Psi_{\alpha}{}^{\alpha}\veps_{\mu\nu\rho\sigma}\sfA^{\rho\sigma} \pm \frac{e}{2}\veps_{\mu\nu\rho\sigma}\Psi^{(\alpha\rho)}\sfA^{\sigma}{}_{\alpha}\nonumber\\
	& &&= \big(\IQ\sfA_{\mu\nu}\big)^{\pm} \pm \frac{1}{4}\Psi_{\sigma}{}^{\sigma}(\sfA_{\mu\nu}^{+} - \sfA_{\mu\nu}^{-}) \pm \big(\Psi^{\sigma}{}_{[\mu}\sfA_{\nu]\sigma}\big)^{+} \mp \big(\Psi^{\sigma}{}_{[\mu}\sfA_{\nu]\sigma}\big)^{-} ~. \label{eq:identA}
}
\item Using \eqref{eq:identB}, one can rewrite \eqref{eq:identA} as
\eqas{
	&\IQ\sfA_{\mu\nu}^{+} &&= \big(\IQ\sfA_{\mu\nu}\big)^{+} + \big(\Psi^{\sigma}{}_{[\mu}\sfA_{\nu]\sigma}^{-}\big)^{+} - \big(\Psi^{\sigma}{}_{[\mu}\sfA_{\nu]\sigma}^{+}\big)^{-} ~,\\
	&\IQ\sfA_{\mu\nu}^{-} &&= \big(\IQ\sfA_{\mu\nu}\big)^{-} + \big(\Psi^{\sigma}{}_{[\mu}\sfA_{\nu]\sigma}^{+}\big)^{-} - \big(\Psi^{\sigma}{}_{[\mu}\sfA_{\nu]\sigma}^{-}\big)^{+} ~.
	\label{eq:identA refined}
}
Due to \eqref{eq:GravityCartan-1}, \eqref{eq:identA refined} also holds for the differential $\mathsf{d}$ (and in fact, for any differential operator $\mathsf{d}_{B}$ satisfying $\mathsf{d}_{B}g_{μν} = Ψ_{(μν)}$). 
It is worth pointing out that only the symmetric part of the gravitino contributes to the $\big(\Psi \sfA^{\pm}\big)^{\mp}$ terms in \eqref{eq:identA refined}. This is consistent with \eqref{eq:identB antisym general}. As a further caveat, note that \eqref{eq:identA} and \eqref{eq:identA refined} also hold for the supergravity \emph{variation} $\delta$ if we multiply from the left by $\eps$.
\end{itemize}

\section{Miscellaneous Expressions In 4d $\CN=2$ Superconformal Gravity}\label{app:csugra-misc}
\subsection{Untwisted Theory}
\paragraph{\uline{Conformal and Superconformal Algebras}:} The generators of the Euclidean conformal algebra $\mathfrak{so(5,1)}$, namely translations $P_\ha$, Lorentz transformations $M_{[\ha\hb]}$, dilatations $\mathcal{D}$, and special conformal transformations $K_\ha$ satisfy the following commutation relations:
\eqas{\label{eq:ConfAlgebra}
 &[M_{\ha\hb}, M_{\hg\hd}] &&= \delta_{\ha\hg}M_{\hd\hb} - \delta_{\hb\hg}M_{\hd\ha} - \delta_{\ha\hd}M_{\hg\hb} + \delta_{\hb\hd}M_{\hg\ha} ~,\\
 &[P_\ha, M_{\hb\hg}] &&= \delta_{\ha\hb}P_{\hg} - \delta_{\ha\hg}P_{\hb} ~,\\
 &[K_\ha, M_{\hb\hg}] &&= \delta_{\ha\hb}K_{\hg} - \delta_{\ha\hg}K_{\hb} ~,\\
 &[\mathcal{D}, P_\ha] &&= P_\ha ~,\\
 &[\mathcal{D}, K_\ha] &&= -K_\ha ~,\\
 &[P_\ha, K_\hb] &&= 2\big(\delta_{\ha\hb}\mathcal{D} - 2 M_{\ha\hb}\big) ~,
}
where $\delta_{\ha\hb}$ denotes the flat Euclidean $4$-metric.  Here, $\ha, \hb, \hg, \hd$ are frame indices (or ``flat indices''). Our index conventions are spelled out in detail in Table \ref{tbl:index conventions}.

The generators of the $\CN=2$ superconformal algebra $\mathfrak{su^*(4|2)}$ are all the bosonic generators of the conformal algebra $\mathfrak{so(5,1)}$ above, and in addition, supersymmetry generators $\bm{Q}^{i}$, and conformal supersymmetry generators $\bm{S}^{i}$, bosonic generators for the $\mathfrak{su(2)}_{\mathsf{R}}$ symmetry, conventionally denoted by a triplet $\bu{U} = (U_1, U_2, U_3)$ with $U_{\ell} := -\frac{1}{2}\big(\tau_{\ell}\big)_{i}{}^{j}U_{j}{}^{i}$ where $\{\tau_{\ell}\}_{\ell=1}^{3}$ are the Pauli matrices (see \eqref{eq:pauli matrices}), and a bosonic generator $\mathcal{T}$ for the $\mathfrak{so(1,1)}_{\mathsf{R}}$ symmetry. In terms of a basis of Dirac matrices $\{\gamma_\ha\}_{\ha=1}^{4}$ that satisfy the Euclidean Clifford algebra $\mathcal{C}\ell(0,4)$, i.e., $\{\gamma_\ha, \gamma_\hb\} = 2\,\delta_{\ha\hb}\mathbb{I}_{4}$ (see also \eqref{eq:euclidean dirac matrices}, \eqref{eq:gamma5 in terms of euclidean dirac matrices} and \eqref{eq:gamma-ab}), the non-vanishing (anti)commutators involving $\bm{Q}^{i}$ and $\bm{S}^{i}$ are\footnote{Two-component versions of these (anti)commutators can be writen down using Appendices \ref{app:conventions} and \ref{app:spinormethods}. A twisted version can be found in eqn. (10.3) of \cite{Marino:1998bm}.}
\eqasfour{\label{eq:SupConfAlgebra}
   &[M_{\ha\hb}, \bm{Q}^{i}] &&= \frac{1}{4}\gamma_{\ha\hb}\bm{Q}^{i} ~, \qquad && [M_{\ha\hb}, \bm{S}^{i}] &&= \frac{1}{4}\gamma_{\ha\hb}\bm{S}^{i} ~, \\
   &[U_\ell, \bm{Q}]^{i} &&= i\big(\tau_\ell\big)^{i}{}_{j}\bm{Q}^{j} ~,\qquad && [U_\ell, \bm{S}]^{i} &&= i\big(\tau_\ell)^{i}{}_{j}\bm{S}^{j} ~,\\
   &[\mathcal{D}, \bm{Q}^i] &&= \frac{1}{2}\bm{Q}^{i} ~, \qquad && [\mathcal{D}, \bm{S}^{i}] &&= -\frac{1}{2}\bm{S}^{i} ~,\\
   &[\mathcal{T}, \bm{Q}^{i}] &&= \frac{1}{2}\gamma_5 \bm{Q}^i ~, \qquad && [\mathcal{T}, \bm{S}^i] &&= -\frac{1}{2}\gamma_5\bm{S}^i ~,\\
   &[K_\ha, \bm{Q}^{i}] &&= \frac{1}{2}\gamma_{\ha}\gamma_{5}\bm{S}^{i} ~, \qquad && [P_\ha, \bm{S}^{i}] &&= -\frac{1}{2}\gamma_{\ha}\gamma_{5}\bm{S}^{i} ~,\\
  &\{\bm{Q}^{i}, \bm{Q}_{j}\} &&= \gamma^{\ha}P_{\ha}\delta_{j}{}^{i} ~, \qquad && \{\bm{S}^{i}, \bm{S}_{j}\} &&= \gamma^{\ha}K_{\ha}\delta_{j}{}^{i} ~,\\
  &\{\bm{Q}^{i}, \bm{S}_{j}\} &&= -\big(\gamma^{\ha\hb}M_{\ha\hb} + \mathcal{D} - \gamma_{5}\mathcal{T}\big)\delta_{j}{}^{i} + 2\,U_{j}{}^{i} ~.
}

\paragraph{\uline{Covariant and Supercovariant Derivatives}:} The covariant derivatives appearing in \eqref{eq:csugra-susy-grav-left-2comp} and \eqref{eq:csugra-susy-grav-right-2comp} are:
\eqa{
&\bcd_{μ}\eps^{iA} &&= \partial_{μ}\epsilon^{iA} + \frac{1}{2}{\omega_{μ}{}^{A}}_{B}\eps^{iB}	+ \frac{1}{2}(b_μ + A_μ^{\rm{(}\mathsf{R}\rm{)}})\eps^{iA} + \frac{1}{2}{V_{μ}{}^{i}}_{j}\eps^{jA}\label{eq:covder 2 comp eps repeated} ~,\\
&\bcd_{μ}\eps^{i\dA} &&= \partial_{μ}\epsilon^{i\dA} + \frac{1}{2}\omega_{μ}{}^{\dA}{}_{\dB}\epsilon^{i\dB} + \frac{1}{2}(b_μ - A_μ^{\rm{(}\mathsf{R}\rm{)}})\eps^{i\dA} + \frac{1}{2}{V_{μ}{}^{i}}_{j}\eps^{j\dA}\label{eq:covder 2 comp epsbar repeated} ~.
}
The corresponding expressions for the conformal susy parameters are:
\eqa{
&\bcd_{μ}\bn^{iA} &&= \partial_{μ}\bn^{iA} + \frac{1}{2}{\omega_{μ}{}^{A}}_{B}\bn^{iB}	- \frac{1}{2}(b_μ + A_μ^{\rm{(}\mathsf{R}\rm{)}})\bn^{iA} + \frac{1}{2}{V_{μ}{}^{i}}_{j}\bn^{jA}\label{eq:covder 2 comp eta} ~,\\
&\bcd_{μ}\bn^{i\dA} &&= \partial_{μ}\bn^{i\dA} + \frac{1}{2}\omega_{μ}{}^{\dA}{}_{\dB}\bn^{i\dB} - \frac{1}{2}(b_μ - A_μ^{\rm{(}\mathsf{R}\rm{)}})\bn^{i\dA} + \frac{1}{2}{V_{μ}{}^{i}}_{j}\bn^{j\dA}\label{eq:covder 2 comp etabar} ~.
}
On the vielbein,
\eqa{
& \bcd_{μ}e_{ν}{}^{A\dA} &&= \n_{μ}e_{ν}{}^{A\dA} + b_{μ}e_{ν}{}^{A\dA} \nn
& &&= \partial_{μ}e_{ν}{}^{A\dA} - \Gamma_{μν}{}^{ρ}e_{ρ}{}^{A\dA} + \frac{1}{2}{\omega_{μ}{}^{A}}_{B}e_{\nu}{}^{B\dA} + \frac{1}{2}{\omega_{μ}{}^{\dA}}_{\dB}e_{ν}{}^{A\dB} + b_{μ}e_{ν}{}^{A\dA} ~. \label{eq:covder 2 comp vielbein}
}
The fully supercovariant derivatives on $T_{AB}$ and $T_{\dA\dB}$, which appear in \eqref{eq:csugra-susy-TSD-2comp} and \eqref{eq:csugra-susy-TASD-2comp} are:
\eqa{
&\scd_{\mu}T_{AB} &&= \partial_{\mu}T_{AB} - \frac{1}{2}\omega_{\mu,A}{}^{C}T_{CB} - \frac{1}{2}\omega_{\mu,B}{}^{C}T_{A}{}_{C} + (b_{\mu} + A_μ^{\rm{(}\mathsf{R}\rm{)}})T_{AB} \nn 
& &&\quad + 4i\Psi_{\mu,iC}R(Q)_{AB}{}^{iC} + 4i \Psi_{\mu,i}{}^{\dC}R(Q)_{AB}{}^{i}{}_{\dC} ~, \label{eq:covder 2comp TAB} \\
&\scd_{\mu}T_{\dA\dB} &&= \partial_{\mu}T_{\dA\dB} - \frac{1}{2}\omega_{\mu,\dA}{}^{\dC}T_{\dC\dB} - \frac{1}{2}\omega_{\mu,\dB}{}^{\dC}T_{\dA\dC} + (b_{\mu} - A_μ^{\rm{(}\mathsf{R}\rm{)}})T_{\dA\dB} \nn
& &&\quad + 4i \Psi_{\mu,iC}R(Q)_{\dA\dB}{}^{iC} + 4i \Psi_{\mu,i}{}^{\dC}R(Q)_{\dA\dB}{}^{i}{}_{\dC} ~. \label{eq:covder 2comp TAdotBdot}
}
The fully supercovariant derivatives on $\Xi^{iA}$ and $\Xi^{i\dA}$, which appear in \eqref{eq:csugra-susy-auxiliary-scalar-2comp}, are:
\eqa{
 &\scd_{\mu}\Xi^{iA} &&= \partial_{μ}\Xi^{iA} + \frac{1}{2}{\omega_{μ}{}^{A}}_{B}\Xi^{iB} + \left(\frac{3}{2}b_{\mu}-\frac{1}{2}A_μ^{\rm{(}\mathsf{R}\rm{)}}\right)\Xi^{iA} + \frac{1}{2}{V_{μ}{}^{i}}_{j}\Xi^{jA} - \frac{1}{2}\mathscr{D}\Psi_{μ}{}^{iA}  \nn
 & &&\quad + \frac{2i}{3\sq}\big(\scd_{B\dA}T^{AB}\big)\Psi_{μ}{}^{i\dA} + \frac{1}{6}R(V)^{A}{}_{B}{}^{i}{}_{j}\Psi_{\mu}{}^{jB} - \frac{1}{3}R(\ARsym)^{A}{}_{B}\Psi_{μ}{}^{iB} + \frac{i}{3}T^{A}{}_{B}S_{μ}{}^{iB} \label{eq:covder 2 comp Xi} ~,\\
 & \scd_{\mu}\Xi^{i\dA} &&= \partial_{μ}\Xi^{i\dA} + \frac{1}{2}{\omega_{μ}{}^{\dA}}_{\dB}\Xi^{i\dB} + \left(\frac{3}{2}b_{\mu}+\frac{1}{2}A_μ^{\rm{(}\mathsf{R}\rm{)}}\right)\Xi^{i\dA} + \frac{1}{2}{V_{μ}{}^{i}}_{j}\Xi^{j\dA} - \frac{1}{2}\mathscr{D}\Psi_{\mu}{}^{i\dA} \nn
 & &&\quad  + \frac{2i}{3\sq}\big(\scd_{A\dB}T^{\dA\dB}\big)\Psi_{μ}{}^{iA} + \frac{1}{6}R(V)^{\dA}{}_{\dB}{}^{i}{}_{j}\Psi_{\mu}{}^{j\dB}  - \frac{1}{3}R(\ARsym)^{\dA}{}_{\dB}\Psi_{μ}{}^{i\dB} + \frac{i}{3}T^{\dA}{}_{\dB}S_{μ}{}^{i\dB} \label{eq:covder 2 comp Xibar} ~.
}
\paragraph{\uline{Conformal Gravitino}:} The components of the conformal gravitino (a composite field) are:
\eqa{
   &S_{μ}{}^{iB} &&:= \sq\left(Δ^{B}{}_{A}{}^{i}{}_{\dA} - \frac{1}{3}\delta_{A}{}^{B}Δ_{\dA\dC}{}^{i\dC}\right)e_{μ}{}^{A\dA} ~, \label{eq:conformal gravitino 2-comp left}\\
   &S_{μ}{}^{i}{}_{\dB} &&:= \sq\left({Δ_{\dA\dB}{}^{i}}_{A} - \frac{1}{3}\veps_{\dA\dB}Δ_{AC}{}^{iC}\right)e_{μ}{}^{A\dA} ~. \label{eq:conformal gravitino 2-comp right}
}
Here $\bm{Δ}_{A\dA,B\dB}{}^{i}$ is a 4-component spinor whose components are given by
\eqa{
&Δ_{A\dA,B\dB}{}^{iE} &&= e^{\mu}{}_{A\dA}e^{\nu}{}_{B\dB}\bcd_{[\mu}\Psi_{\nu]}{}^{iE} - \frac{1}{16\sq}\big(T^{E}{}_{A}\Psi_{B\dB}{}^{i}{}_{\dA} - T^{E}{}_{B}\Psi_{A\dA}{}^{i}{}_{\dB}\big) \nn
  & &&\quad + \frac{1}{4}\veps_{\dA\dB}\big(\delta_{A}{}^{E}\Xi^{i}{}_{B} + \delta_{B}{}^{E}\Xi^{i}{}_{A}\big) ~, \label{eq:untwisted building block Delta left}\\
  &Δ_{A\dA,B\dB}{}^{i}{}_{\dE} &&= e^{\mu}{}_{A\dA}e^{\nu}{}_{B\dB}\bcd_{[\mu}\Psi_{\nu]}{}^{i}{}_{\dE} - \frac{1}{16\sq}\big(T_{\dE\dA}{\Psi_{B\dB}{}^{i}}_{A} - T_{\dE\dB}{\Psi_{A\dA}{}^{i}}_{B}\big) \nn
  & &&\quad + \frac{1}{4}\veps_{AB}\big(\veps_{\dA\dE}\Xi^{i}{}_{\dB} + \veps_{\dB\dE}\Xi^{i}{}_{\dA}\big) ~, \label{eq:untwisted building block Delta right}
}
and the self- and anti-self- dual parts of $Δ_{A\dA,B\dB}{}^{iE}$ and  $Δ_{A\dA,B\dB}{}^{a}{}_{\dE}$ (which appear in \eqref{eq:conformal gravitino 2-comp left} and \eqref{eq:conformal gravitino 2-comp right}) are given by
\eqa{
\label{eq:Delta1}  &Δ_{AB}{}^{iE} &&= \CJ_{AB}{}^{iE} + \tfrac{1}{16\sq}\big(T^{E}{}_{A}Ψ_{B\dA}{}^{i\dA} + T^{E}{}_{B}Ψ_{A\dA}{}^{i\dA}\big) + \tfrac{1}{2}\big(δ_{A}{}^{E}\Xi^{i}{}_{B} + δ_{B}{}^{E}\Xi^{i}{}_{A}\big) ~,\\
\label{eq:Delta2}  &Δ_{\dA\dB}{}^{iE} &&= \CJ_{\dA\dB}{}^{iE} - \tfrac{1}{16\sq}T^{E}{}_{A}\big(Ψ^{A}{}_{\dB}{}^{i}{}_{\dA} + Ψ^{A}{}_{\dA}{}^{i}{}_{\dB}\big) ~, \\
\label{eq:Delta3}  &Δ_{AB}{}^{i}{}_{\dE} &&= \CJ_{AB}{}^{i}{}_{\dE} - \tfrac{1}{16\sq}T_{\dE\dA}\big(Ψ_{B}{}^{\dA i}{}_{A} + Ψ_{A}{}^{\dA i}{}_{B}\big) ~,\\
\label{eq:Delta4}  &Δ_{\dA\dB}{}^{i}{}_{\dE} &&= \CJ_{\dA\dB}{}^{i}{}_{\dE} + \tfrac{1}{16\sq}\big(T_{\dE\dA}Ψ_{A\dB}{}^{i A} + T_{\dE\dB}Ψ_{A\dA}{}^{i A}\big) + \tfrac{1}{2}(\veps_{\dA\dE}\Xi^{i}{}_{\dB} + \veps_{\dB\dE}\Xi^{i}{}_{\dA}\big) ~,
}
where $\bm{\CJ}_{μν}{}^{i} := \bcd_{[\mu}\bm{Ψ}_{ν]}{}^{i}$ denotes the curl of the 4-component gravitino, whose components 
appear in the preceding equations in terms of their self-dual and anti-self-dual parts. In particular,
\eqa{
   \label{eq:DeltaCont1}&Δ_{AB}{}^{iA} &= \CJ_{AB}{}^{iA} + \frac{1}{16\sq}T^{A}{}_{B}Ψ_{A\dA}{}^{i\dA} + \frac{3}{2}\Xi^{i}{}_{B} ~,\\
  \label{eq:DeltaCont2}  &Δ_{\dA\dB}{}^{i\dA} &= \CJ_{\dA\dB}{}^{i\dA} + \frac{1}{16\sq}T^{\dA}{}_{\dB}Ψ_{A\dA}{}^{iA} + \frac{3}{2}\Xi^{i}{}_{\dB} ~.
}
\paragraph{\uline{Curvature of the Spin Connection}:} The curvature for the spin connection $\omega$ is given by 
\eqa{
  &R(\omega)_{\mu\nu}{}^{\ha}{}_{\hb} &&= \partial_{\mu}\omega_{\nu}{}^{\ha}{}_{\hb} - \partial_{\nu}\omega_{\mu}{}^{\ha}{}_{\hb} - \big(\omega_{\mu}{}^{a}{}_{c}\,\omega_{\nu}{}^{c}{}_{b} - \omega_{\nu}{}^{a}{}_{c}\,\omega_{\mu}{}^{c}{}_{b}\big) ~. \label{eq:omega curv 4comp}
}
In two-component notation, the self-dual and anti-self-dual components are: 
\eqa{
   &R(\omega)_{\mu\nu}{}^{A}{}_{B} &&= \partial_{\mu}\omega_{\nu}{}^{A}{}_{B} - \partial_{\nu}\omega_{\mu}{}^{A}{}_{B} + \frac{1}{2}\big(\omega_{\mu}{}^{A}{}_{C}\,\omega_{\nu}{}^{C}{}_{B} - \omega_{\nu}{}^{A}{}_{C}\,\omega_{\mu}{}^{C}{}_{B}\big) ~, \label{eq:omega curv sd}\\
   &R(\omega)_{\mu\nu}{}^{\dA}{}_{\dB} &&= \partial_{\mu}\omega_{\nu}{}^{\dA}{}_{\dB} - \partial_{\nu}\omega_{\mu}{}^{\dA}{}_{\dB} + \frac{1}{2}\big(\omega_{\mu}{}^{\dA}{}_{\dC}\,\omega_{\nu}{}^{\dC}{}_{\dB} - \omega_{\nu}{}^{\dA}{}_{\dC}\,\omega_{\mu}{}^{\dC}{}_{\dB}\big) ~, \label{eq:omega curv asd}
}
%
\paragraph{\uline{Supercurvatures}:} The $P$-supercurvature $R(P)_{\mu\nu}{}^{a}$ in spinor indices reads
\eqas{
& R(P)_{\mu\nu}{}^{C\dC} &&= 2\,\bcd_{[\mu}e_{\nu]}{}^{C\dC} + i\sq\Psi_{[\mu|i}{}^{C}\Psi_{\nu]}{}^{i\dC}\\
& &&= 2\,\bcd_{[\mu}e_{\nu]}{}^{C\dC} + \frac{i}{\sq}\big( \Psi_{A\dA,i}{}^{C}\Psi_{B\dB}{}^{i\dC} - \Psi_{A\dA,i}{}^{\dC}\Psi_{B\dB}{}^{iC} \big)e_{\mu}{}^{A\dA}e_{\nu}{}^{B\dB} ~. \label{eq:RP-supercurvature}
}
Note that $R(P)_{\mu\nu}{}^{a}$ depends algebraically on the spin connection through the first term, cf. \eqref{eq:covder 2 comp vielbein}.
\\\\
The components of the $Q$-supercurvature are given by
\eqa{
  &R(Q)_{μν}{}^{iE} &&= 2\,\CJ_{μν}{}^{iE} - \frac{1}{\sq}\big(δ_{A}{}^{E}S_{B\dB}{}^{i}{}_{\dA} - δ_{B}{}^{E}S_{A\dA}{}^{i}{}_{\dB}\big)e_{μ}{}^{A\dA}e_{ν}{}^{B\dB} \label{eq:RQ undotted} \nn
  & &&\quad - \frac{1}{8\sq}\big(T^{E}{}_{A}Ψ_{B\dB}{}^{i}{}_{\dA} - T^{E}{}_{B}Ψ_{A\dA}{}^{i}{}_{\dB}\big)e_{μ}{}^{A\dA}e_{ν}{}^{B\dB} ~, \\
  &R(Q)_{μν}{}^{i}{}_{\dE} &&= 2\,\CJ_{μν}{}^{i}{}_{\dE} - \frac{1}{\sq}\big(\veps_{\dA\dE}S_{B\dB}{}^{i}{}_{A} - \veps_{\dB\dE}S_{A\dA}{}^{i}{}_{B}\big)e_{μ}{}^{A\dA}e_{ν}{}^{B\dB}  \nn
  & &&\quad - \frac{1}{8\sq}\big(T_{\dE\dA}{Ψ_{B\dB}{}^{i}}_{A} - T_{\dE\dB}{Ψ_{A\dA}{}^{i}}_{B}\big)e_{μ}{}^{A\dA}e_{ν}{}^{B\dB} ~. \label{eq:RQ dotted}
}
Their self-dual and anti-self-dual components are given by
\eqa{
  &R(Q)_{AB}{}^{iE} &&= 2\,\CJ_{AB}{}^{iE} + \frac{1}{\sq}\big(δ_{A}{}^{E}S_{B\dA}{}^{i\dA} + δ_{B}{}^{E}S_{A\dA}{}^{i\dA}\big) + \frac{1}{4\sq}T^{E}{}_{(A}Ψ_{B)\dA}{}^{i\dA} ~,\label{eq:RQ undotted selfdual} \\
  &R(Q)_{\dA\dB}{}^{iE} &&= 2\,\CJ_{\dA\dB}{}^{iE} - \frac{1}{\sq}\big(S^{E}{}_{\dB}{}^{i}{}_{\dA} + S^{E}{}_{\dA}{}^{i}{}_{\dB}\big)  - \frac{1}{8\sq}T^{E}{}_{A}\big(Ψ^{A}{}_{\dB}{}^{i}{}_{\dA} + Ψ^{A}{}_{\dA}{}^{i}{}_{\dB} \big) ~,\label{eq:RQ undotted antiselfdual} \\
  &R(Q)_{AB}{}^{i}{}_{\dE} &&= 2\,\CJ_{AB}{}^{i}{}_{\dE} - \frac{1}{\sq}\big({S_{A\dE}{}^{i}}_{B} + {S_{B\dE}{}^{i}}_{A}\big) - \frac{1}{8\sq}T_{\dE\dA}\big(Ψ_{B}{}^{\dA i}{}_{A} + Ψ_{A}{}^{\dA i}{}_{B}\big) ~, \label{eq:RQ dotted selfdual} \\
  &R(Q)_{\dA\dB}{}^{i}{}_{\dE} &&= 2\,\CJ_{\dA\dB}{}^{i}{}_{\dE} + \frac{1}{\sq}\big(\veps_{\dA\dE}S_{A\dB}{}^{iA} + \veps_{\dB\dE}S_{A\dA}{}^{iA}\big) + \frac{1}{4\sq}T_{\dE(\dA}{Ψ^{A}{}_{\dB)}{}^{i}}_{A} ~. \label{eq:RQ dotted antiselfdual}
}
Note that $R(Q)_{\mu\nu}{}^{i}$ depends algebraically on the conformal gravitinos $S_{\mu}{}^{iA}$, $S_{\mu}{}^{i\dA}$.\\\\
The $D$- and $\ARsym$-supercurvatures are
\eqa{
  & R(D)_{\mu\nu} &&= 2\,\partial_{[\mu}b_{\nu]} - 2\,f_{[\mu}{}^{a}e_{\nu]a} - \frac{i}{2}\big( \Psi_{[\mu\,iA}S_{\nu]}{}^{iA} + \Psi_{[\mu\,i}{}^{\dA}S_{\nu]}{}^{i}{}_{\dA} \big)\nn
  & &&\quad - \frac{3i}{2\sqrt{2}}\big( \Psi_{[\mu\,iA}e_{\nu]}{}^{A\dA}\Xi^{i}{}_{\dA} - \Psi_{[\mu\,i\dA}e_{\nu]}{}^{A\dA}\Xi^{i}{}_{A} \big) ~,\\
   & R(\ARsym)_{\mu\nu} &&= 2\,\partial_{[\mu}\ARsym_{\nu]} + \frac{i}{2}\big( \Psi_{[\mu\,iA}S_{\nu]}{}^{iA} - \Psi_{[\mu\,i}{}^{\dA}S_{\nu]}{}^{i}{}_{\dA} \big) \nn
   & &&\quad - \frac{3i}{2\sqrt{2}}\big( \Psi_{[\mu\,iA}e_{\nu]}{}^{A\dA}\Xi^{i}{}_{\dA} + \Psi_{[\mu\,i\dA}e_{\nu]}{}^{A\dA}\Xi^{i}{}_{A} \big) ~. \label{eq:RARsym}
}
The Lorentz or $M$-supercurvature is $R(M)_{\mu\nu}{}^{ab}$ or in spinor indices,
\eqa{
& R(M)_{\mu\nu}{}^{C\dC,D\dD} &&= R(\omega)_{\mu\nu}{}^{C\dC,D\dD} - 4 f_{[\mu}{}^{[C\dC}e_{\nu]}{}^{D\dD]} + \frac{i}{2}\veps^{\dC\dD}\big(\Psi_{[\mu\,i}{}^{C}S_{\nu]}{}^{iD} + \Psi_{[\mu\,i}{}^{D}S_{\nu]}{}^{iC}\big) \nn
& &&\quad - \frac{i}{2}\veps^{CD}\big(\Psi_{[\mu\,i}{}^{\dC}S_{\nu]}{}^{i\dD} + \Psi_{[\mu\,i}{}^{\dD}S_{\nu]}{}^{i\dC}\big) -\frac{i}{8}\big( \Psi_{[\mu\,iA}\Psi_{\nu]}{}^{iA} + \Psi_{[\mu\,i}{}^{\dA}\Psi_{\nu]}{}^{i}{}_{\dA} \big)T^{C\dC,D\dD} \nn
& &&\quad -\frac{3i}{2\sq}\left[ \Psi_{[\mu\,iB}e_{\nu]}{}^{B\dB}\veps^{CD}\big(\delta_{\dB}{}^{\dC}\Xi^{i\dD} + \delta_{\dB}{}^{\dD}\Xi^{i\dC}\big)\right.\nn
& &&\qquad \qquad \quad \quad \left. - \Psi_{[\mu\,i\dB}e_{\nu]}{}^{B\dB}\veps^{\dC\dD}\big(\delta_{B}{}^{C}\Xi^{iD} + \delta_{B}{}^{D}\Xi^{iC}\big) \right] \nn
& &&\quad + i\sq\big( \Psi_{[\mu\,iB}e_{\nu]}{}^{B\dB} R(Q)^{C\dC,D\dD}{}^{,i}{}_{\dB} - \Psi_{[\mu\,i\dB}e_{\nu]}{}^{B\dB}R(Q)^{C\dC,D\dD}{}^{,i}{}_{B} \big) ~. \label{eq:RM supercurvature}
}
Its self-dual and anti-self-dual components (with respect to the upper indices) are
\eqa{
 & R(M)_{\mu\nu}{}^{CD} &&= R(\omega)_{\mu\nu}{}^{CD} - 4 f_{[\mu}{}^{(C}{}_{|\dC|}e_{\nu]}{}^{D)\dC} + i\big(\Psi_{[\mu\,i}{}^{C}S_{\nu]}{}^{iD} + \Psi_{[\mu\,i}{}^{D}S_{\nu]}{}^{iC}\big) \nn
 & &&\quad -\frac{i}{8}\big( \Psi_{[\mu\,iA}\Psi_{\nu]}{}^{iA} + \Psi_{[\mu\,i}{}^{\dA}\Psi_{\nu]}{}^{i}{}_{\dA} \big)T^{CD}  + \frac{3i}{\sq}\Psi_{[\mu\,i\dB}e_{\nu]}{}^{B\dB}\big(\delta_{B}{}^{C}\Xi^{iD} + \delta_{B}{}^{D}\Xi^{iC}\big)\nn
& &&\quad + i\sq\big( \Psi_{[\mu\,iB}e_{\nu]}{}^{B\dB} R(Q)^{CD}{}^{,i}{}_{\dB} - \Psi_{[\mu\,i\dB}e_{\nu]}{}^{B\dB}R(Q)^{CD}{}^{,i}{}_{B} \big) ~, \label{eq:RM supercurvature undotted}\\
& R(M)_{\mu\nu}{}^{\dC\dD} &&= R(\omega)_{\mu\nu}{}^{\dC\dD} - 4 f_{[\mu\,C}{}^{(\dC}e_{\nu]}{}^{C\,\dD)} - i\big(\Psi_{[\mu\,i}{}^{\dC}S_{\nu]}{}^{i\dD} + \Psi_{[\mu\,i}{}^{\dD}S_{\nu]}{}^{i\dC}\big) \nn
 & &&\quad -\frac{i}{8}\big( \Psi_{[\mu\,iA}\Psi_{\nu]}{}^{iA} + \Psi_{[\mu\,i}{}^{\dA}\Psi_{\nu]}{}^{i}{}_{\dA} \big)T^{\dC\dD} -\frac{3i}{\sq}\Psi_{[\mu\,iB}e_{\nu]}{}^{B\dB}\big(\delta_{\dB}{}^{\dC}\Xi^{i\dD} + \delta_{\dB}{}^{\dD}\Xi^{i\dC}\big) \nn
 & &&\quad + i\sq\big( \Psi_{[\mu\,iB}e_{\nu]}{}^{B\dB} R(Q)^{\dC\dD}{}^{,i}{}_{\dB} - \Psi_{[\mu\,i\dB}e_{\nu]}{}^{B\dB}R(Q)^{\dC\dD}{}^{,i}{}_{B} \big) ~. \label{eq:RM supercurvature dotted}
}
Note that $R(M)_{\mu\nu}{}^{C\dC,D\dD}$ depends algebraically on the special conformal connnection $f_{\mu}{}^{a}.$

The expressions for $R(K)_{\mu\nu}{}^{a}$, $R(D)_{\mu\nu}$ and $\bm{R}(\bm{S})_{\mu\nu}{}^{i}$ are not required in this paper, but can be found in the standard references \cite{deWit:2017cle,Freedman:2012zz,Lauria:2020rhc,deWit:1983xhu,deWit:1983xe,deWit:1984rvr,deWit:1979dzm,deWit:1980lyi,deWit:1984wbb}. Note that these curvatures do not enter the supersymmetry transformation laws of the elementary fields of the Weyl multiplet, nor do they enter the supercurvature constraints.

\eject
\subsection{Twisted Theory}\label{app:misc-expr-twisted}
\noindent \ul{\textbf{Post truncation + twist ($\bm{Ψ_{μ}{}^{AB} = 0}$ and $\bm{T_{AB} = 0}$) but prior to simplifications from $\bm{δA_{μ}^{{\rm(}\mathsf{R}{\rm)}}} = 0$}}.\\\\
Eqns. \eqref{eq:Delta1}--\eqref{eq:Delta4} reduce to
\eqas{
  &Δ_{AB}{}^{CD} &&= \frac{1}{2}\big(δ_{A}{}^{D}\Xi^{C}{}_{B} + δ_{B}{}^{D}\Xi^{C}{}_{A}\big) ~,\\
  &Δ_{\dA\dB}{}^{CD} &&= 0 ~,\\
  &Δ_{AB}{}^{C}{}_{\dD} &&= \CJ_{AB}{}^{C}{}_{\dD} ~,\\
  &Δ_{\dA\dB}{}^{C}{}_{\dD} &&= \CJ_{\dA\dB}{}^{C}{}_{\dD} + \frac{1}{2}\big(\veps_{\dA\dD}\Xi^{C}{}_{\dB} + \veps_{\dB\dD}\Xi^{C}{}_{\dA}\big) ~.
}
Eqns. \eqref{eq:DeltaCont1} and \eqref{eq:DeltaCont2} yield, respectively,
\eqas{
 &Δ_{AB}{}^{CA} &&= \frac{3}{2}\Xi^{C}{}_{B} ~,\\
 &Δ_{\dA\dB}{}^{C\dA} &&= \CJ_{\dA\dB}{}^{C\dA} + \frac{3}{2}\Xi^{C}{}_{\dB} ~.
}
Therefore \eqref{eq:conformal gravitino 2-comp left} and \eqref{eq:conformal gravitino 2-comp right} respectively read
\eqa{
\label{eq:SgravTrunc-1}&S_{μ}{}^{AB} &&= \sq\left[\CJ^{B}{}_{C}{}^{A}{}_{\dC} - \frac{1}{3}\delta_{C}{}^{B}\left(\CJ_{\dA\dC}{}^{A\dA} + \frac{3}{2}\Xi^{A}{}_{\dC}\right)\right]e_{μ}{}^{C\dC} ~,\\
\label{eq:SgravTrunc-2}&S_{μ}{}^{A\dA} &&= -\frac{1}{2}\Xi^{A}{}_{C}δ_{\dC}{}^{\dA}e_{μ}{}^{C\dC} ~.
}

\noindent \ul{\textbf{Post truncation + twist ($\bm{Ψ_{μ}{}^{AB} = 0}$ and $\bm{T_{AB} = 0}$) and after including simplifications from $\bm{δA_{μ}^{{\rm(}\mathsf{R}{\rm)}}} = 0$}}.\\\\
Using \eqref{eq:chi vec} and \eqref{eq:chi nonvec}, \eqref{eq:SgravTrunc-1} and \eqref{eq:SgravTrunc-2} reduce to
\eqa{
\label{eq:SgravTrunc-3}&S_{μ}{}^{AB} &&= \sq\left[\CJ^{B}{}_{C}{}^{A}{}_{\dC} - \frac{1}{4}\delta_{C}{}^{B}\left(\CJ_{\dA\dC}{}^{A\dA} + \CJ^{AE}{}_{E\dC}\right)\right]e_{μ}{}^{C\dC} ~,\\
\label{eq:SgravTrunc-4}&S_{μ}{}^{A\dA} &&= 0 ~,\quad \text{or equivalently,} \quad S_{μν} = 0 ~.
}
It is important to note that \eqref{eq:SgravTrunc-3} is not symmetric in $A, B$. The trace of \eqref{eq:SgravTrunc-3} yields
\eqa{
\label{eq:SgravTrunc-5}&S_{μ} &&:= S_{μ,A}{}^{A} = S_{μ}{}^{AB}\veps_{AB} = \frac{3}{\sq}\Xi_{μ} ~.
}
Moreover, from \eqref{eq:SgravTrunc-3}, 
\eqa{
   &S_{\mu,(BC)} &&= \frac{1}{\sq}\left[\CJ_{AB,C\dA} + \CJ_{AC,B\dA}\right]e_{\mu}{}^{A\dA} \nn
   & &&\quad + \frac{1}{4\sq}\left[\veps_{AB}\big(\CJ_{\dA}{}^{\dE}{}_{C\dE} - \CJ^{E}{}_{C,E\dA}\big) + \veps_{AC}\big(\CJ_{\dA}{}^{\dE}{}_{B\dE} - \CJ^{E}{}_{B,E\dA}\big)\right]e_{\mu}{}^{A\dA} ~,\label{eq:SgravTrunc-6} 
}
and the self-dual 2-form S-gravitino is given by
\eqa{
   &S_{\mu,[\rho\sigma]}^{+} &&= \frac{1}{2}S_{\mu,(BC)} e_{\rho}{}^{B}{}_{\dB}e_{\sigma}{}^{C\dB} ~. \label{eq:SgravTrunc-7} 
}
Also, from \eqref{eq:SgravTrunc-3},
\eqa{
   S_{C\dC,AB} + S_{B\dC,AC} &= 2\sq\,\CJ_{BC,A\dC} ~. \label{eq:Ssum for twisted RQ}
}
Using \eqref{eq:SgravTrunc-3}--\eqref{eq:SgravTrunc-4} and \eqref{eq:Ssum for twisted RQ}, \eqref{eq:RQ undotted selfdual}--\eqref{eq:RQ dotted antiselfdual} yield
\eqa{
 & R(Q)_{AB,CD} &&= 0 ~, \label{eq:RQ twisted ABCD}\\
 & R(Q)_{\dA\dB,CD} &&= 0 ~, \label{eq:RQ twisted AdotBdotCD}\\
 & R(Q)_{AB,C\dC} &&= 0 ~, \label{eq:RQ twisted ABCCdot}\\
 & R(Q)_{\dA\dB,C\dC} &&= 2\,\CJ_{\dA\dB,C\dC} - \tfrac{1}{2}\left[\veps_{\dA\dC}\big(\CJ^{E}{}_{C,E\dB} + \CJ_{\dB\dE,C}{}^{\dE}\big) + \veps_{\dB\dC}\big(\CJ^{E}{}_{C,E\dA} + \CJ_{\dA\dE,C}{}^{\dE}\big) \right] ~. \label{eq:RQ twisted AdotBdotCCdot} 
}
In particular, exactly one component of the $Q$-supercurvature components survives in the twisted theory. It is useful to recast \eqref{eq:RQ twisted AdotBdotCCdot} using the decomposition \eqref{eq:2comp} of the gravitino into irreducible components. The result is
\eqa{
&R(Q)_{\dA\dB,C\dC} &&= 2\n_{A(\dA}\wh{\Psi}^{A}{}_{C\,\dB)\dC} - \n_{C(\dA}\Psi_{\dB)\dC} + \tfrac{1}{2}\eps_{\dC(\dA}\n_{C\,\dB)}\Psi - \n_{C\dE}\veps_{\dC(\dA}\Psi_{\dB)}{}^{\dE} ~.\label{eq:RQ twisted 2-component irred} 
}
It is also useful to recast \eqref{eq:RQ twisted AdotBdotCCdot} into 4-component notation. The result can be written in three equivalent ways:
\eqas{
  & {R(Q)_{\mu\nu}^{-}}{}^{\rho} &&= 2\, {\CJ_{\mu\nu}^{-}}^{\rho} - \frac{1}{2}\left[\delta_{\mu}^{\rho}\big(\CJ_{\nu\kappa,\sigma}^{+} - \CJ_{\nu\kappa,\sigma}^{-}\big)g^{\kappa\sigma} - \delta_{\nu}^{\rho}\big(\CJ_{\mu\kappa,\sigma}^{+} - \CJ_{\mu\kappa,\sigma}^{-}\big)g^{\kappa\sigma} \right]^{-} \\
  & &&= 2\,\big(\n_{[\mu}\Psi_{\nu]}{}^{\rho}\big)^{-} - \frac{1}{2}\left[\delta_{\mu}^{\rho}\left(\n^{\sigma}\Psi_{[\nu\sigma]}^{+} -  \n^{\sigma}\Psi_{[\nu\sigma]}^{-}\right) - \delta_{\nu}^{\rho}\left(\n^{\sigma}\Psi_{[\mu\sigma]}^{+} -  \n^{\sigma}\Psi_{[\mu\sigma]}^{-}\right) \right]^{-} ~\\
  & &&= 2\big(\n_{[\mu}\wh{\Psi}_{\nu]}{}^{\rho}\big)^{-} - \tfrac{1}{2}\big(\delta_{[\mu}^{\rho}\n_{\nu]}^{\,}\Psi_{\sigma}{}^{\sigma}\big)^{-} +2\big(\n_{[\mu}^{\,}\Psi^{-}_{\nu]}{}^{\rho}\big)^{-}  + 2\big(\delta_{[\mu}^{\rho}\n^{\sigma}\Psi^{-}_{\nu]\sigma}\big)^{-} ~.
  \label{eq:RQ twisted 4-component}
}
In the second equality, we used \eqref{eq:diff contJbig} and in the last equality we used the decomposition \eqref{eq:4comp} of the gravitino. Note that the last equality clarifies that $R(Q)_{\mu\nu}^{-}{}^{\rho}$ is independent of the SD part $\Psi_{[\mu\nu]}^{+}$, a conclusion that is not obvious from the preceding line, but is consistent with the absence of any $\Psi_{AB}$ contribution in \eqref{eq:RQ twisted 2-component irred}. Another useful version of the last equation in \eqref{eq:RQ twisted 4-component} is:
\eqa{
&R(Q)_{\mu\nu,\rho}^{-} &&= 2\big(\n_{[\mu}\wh{\Psi}_{\nu]\rho}\big)^{-} - \tfrac{1}{2}\big(g_{\rho[\mu}\n_{\nu]}\Psi_{\sigma}{}^{\sigma}\big)^{-} + 2\big(\n_{[\mu}\Psi_{\nu]\rho}^{-}\big)^{-} + 2\big(g_{\rho[\mu}\n^{\sigma}\Psi^{-}_{\nu]\sigma}\big)^{-} ~. \label{eq:RQ twisted 4-component irred}
}
\noindent \ul{\textbf{Twisted theory with a symmetric gravitino}}.\\\\
For a symmetric gravitino, \eqref{eq:diff contJbig} vanishes. In this case, \eqref{eq:chi vec rewrite} reduces to
\eqas{
  &\Xi_{A\dA} &&= -\frac{1}{3}\n_{E\dE}\wh{\Psi}_{A}{}^{E}{}_{\dA}{}^{\dE} + \frac{1}{4}\n_{A\dA}\Psi ~,\\
  &\Xi_{\mu} &&= -\frac{1}{3}\n_{\rho}\wh{\Psi}_{\mu}{}^{\rho} + \frac{1}{4}\n_{\mu}\Psi_{\rho}{}^{\rho} = -\frac{1}{3}\n_{\rho}\Psi_{\mu}{}^{\rho} + \frac{1}{3}\n_{\mu}\Psi_{\rho}{}^{\rho} ~.
}
Due to \eqref{eq:symmetric-gravitino-simp-Jcont}, \eqref{eq:SgravTrunc-5} reduces to
\eqa{
  &S_{\mu} &&:= S_{\mu,A}{}^{A} = \frac{3}{\sq}\Xi_{\mu} =  \frac{1}{\sq}\big(\n_{\mu}\Psi_{\rho}{}^{\rho} - \n_{\rho}\Psi_{\mu}{}^{\rho}\big) = \sq \n_{[\mu}\Psi_{\rho]}{}^{\rho} ~, \label{eq:sym-grav-0-form-S-susy-connection}
}
whereas \eqref{eq:SgravTrunc-6} reduces to 
\eqa{
  &S_{\mu,(BC)} &&= \frac{1}{\sq}\left[\CJ_{AB,C\dA} + \CJ_{AC,B\dA}\right]e_{\mu}{}^{A\dA} ~. \label{eq:SgravTrunc-6-Simp}
}
Therefore \eqref{eq:SgravTrunc-7} can be written as
\eqas{
  &S_{\mu,[\rho\sigma]}^{+} &&= \sq \CJ_{\rho\sigma,\mu}^{+} + 2\sq\left(-\frac{1}{4}g_{\mu[\rho}\n^{\alpha}\Psi_{\sigma]\alpha} + \frac{1}{4}g_{\mu[\rho}\n_{\sigma]}\Psi_{\alpha}{}^{\alpha}\right)^{+} ~,\\  
 &  &&= \frac{1}{\sq}\big(\n_{\rho}\Psi_{\sigma\mu} - \n_{\sigma}\Psi_{\rho\mu}\big)^{+} + 2\sq\left(-\frac{1}{4}g_{\mu[\rho}\n^{\alpha}\Psi_{\sigma]\alpha} + \frac{1}{4}g_{\mu[\rho}\n_{\sigma]}\Psi_{\alpha}{}^{\alpha}\right)^{+} ~. \label{eq:sym-grav-2-form-S-susy-connection}
}
Finally, \eqref{eq:RQ twisted 4-component} reduces simply to
\eqa{
  & {R(Q)_{\mu\nu}^{-}}{}^{\rho} &&= 2\, {\CJ_{\mu\nu}^{-}}^{\rho} = \big(\n_{\mu}\Psi_{\nu}{}^{\rho} - \n_{\nu}\Psi_{\mu}{}^{\rho}\big)^{+} ~. \label{eq:sym-grav-RQ-twisted-4-component}
}
\noindent \underline{\textbf{Proof of \eqref{eq:contraction-identity-S-susy-connections}}}. 
From \eqref{eq:SgravTrunc-6-Simp},
\eqa{
& S_{A\dA,B\dB,C\dC} &&= \frac{1}{2\sq}\veps_{\dB\dC}\big( \CJ_{AB,C\dA} + \CJ_{AC,B\dA} \big) ~, 
}
so that
\eqa{
& g^{\mu\rho}S_{\mu,[\nu\rho]}e^{\nu}{}_{B\dB} &&= \veps^{AC}\veps^{\dA\dC}S_{A\dA,B\dB,C\dC} \nonumber\\
& &&= -\frac{1}{2\sq}\CJ_{B}{}^{A}{}_{\dA\dB} \nonumber\\
& &&\stackrel{\eqref{eq:contJbig1-sym}}{=\joinrel=\joinrel=} \frac{1}{4\sq}\n_{E\dE}\wh{\Psi}_{B}{}^{E,\dE}{}_{\dB} - \frac{3}{16\sq}\n_{B\dB}\Psi ~,
}
and from \eqref{eq:sym-grav-0-form-S-susy-connection},
\eqa{ 
& S_{B\dB} &&\stackrel{\eqref{eq:spinorial-decomp-sym-gravitino}}{=\joinrel=\joinrel=}  -\frac{3}{16\sq}\n_{B\dB}\Psi + \frac{1}{4\sq}\n_{E\dE}\wh{\Psi}_{B}{}^{E,\dE}{}_{\dB} ~.
}
Thus we have shown that $S_{A\dA,B\dB}{}^{A\dA} = -\frac{1}{4}S_{B\dB}$, and hence proved \eqref{eq:contraction-identity-S-susy-connections}.
\\\\
\noindent \ul{\textbf{Superconformal d'Alembertians}}.\\\\
Here we describe how to write the superconformal d'Alembertians $\bm{\Box}_{C}\phi$ and $\bm{\Box}_{C}\lambda$. For this purpose, we must use the supergravity transformations \eqref{eq:delta-twisted-conformal-sugra} and \eqref{eq:delta-twisted-vec lambda}--\eqref{eq:delta-twisted-vec aux-field} to write down all intermediate expressions \underline{before} imposing that the gravitino is symmetric through \eqref{eq:sym-grav-T}. Let us begin with $\phi$. From \eqref{eq:scd phi}, the supercovariant + gauge covariant derivative on $\phi$ is
 \eqa{
   &\scd_{\mu} \phi &&= D_{\mu}\phi + \frac{1}{2}\Psi_{\mu}{}^{\rho}\psi_{\rho} = \partial_{\mu}\phi + [A_\mu, \phi] + \frac{1}{2}\Psi_{\mu}{}^{\rho}\psi_{\rho} ~. 
 }
Therefore, the susy variation of $\scd_{\mu}\phi$ is
\eqa{
  &\delta\big(\scd_{\mu} \phi\big) &&= -\textcolor{red}{\big(\n_{\mu}\eps^{\rho}\big)\psi_{\rho}} - \eps^{\rho}\big(D_{\mu}\psi_{\rho}\big) + \eps[\psi_\mu, \phi] + \eps^{\rho}[\chi_{\rho\mu}, \phi] - \eps_{\mu}[\eta, \phi] - \textcolor{blue}{\eps^{\rho}\Psi_{\mu\rho}[\lambda, \phi]} \nn
  & &&\quad -\frac{1}{2}\eps\,\Psi^{(\rho\nu)}\Psi_{\mu\nu}\psi_{\rho} - \frac{1}{4}\eps\,\Psi_{\mu}{}^{\sigma}\Psi^{\rho}{}_{\sigma}\psi_{\rho} - \frac{1}{2}\eps\,T_{\mu}{}^{\rho-}\psi_{\rho} + \textcolor{red}{\big(\n_{\mu}\eps^{\rho}\big)\psi_{\rho}} - \frac{1}{2}\bn_{\mu}{}^{\rho}\psi_{\rho} + \frac{1}{2}\bn_{0}\psi_{\mu} \nn
  & &&\quad + \frac{1}{2}\eps\,\Psi_{\mu}{}^{\rho}D_{\rho}\phi - \frac{1}{2}\eps^{\sigma}\Psi_{\mu}{}^{\rho}\big(\wh{F}_{\sigma\rho}^{-} + \lambda T_{\sigma\rho}^{-} + D_{\sigma\rho}\big) + \frac{1}{2}\textcolor{blue}{\eps_{\rho}\Psi_{\mu}{}^{\rho}[\lambda,\phi]} \nn
  & && = - \eps^{\rho}\big(D_{\mu}\psi_{\rho}\big) + \eps[\psi_\mu, \phi] + \eps^{\rho}[\chi_{\rho\mu}, \phi] - \eps_{\mu}[\eta, \phi] - \frac{1}{2}\eps^{\rho}\Psi_{\mu\rho}[\lambda, \phi] \nn
  & &&\quad -\frac{1}{2}\eps\,\Psi^{(\rho\nu)}\Psi_{\mu\nu}\psi_{\rho} - \frac{1}{4}\eps\,\Psi_{\mu}{}^{\sigma}\Psi^{\rho}{}_{\sigma}\psi_{\rho} - \frac{1}{2}\eps\,T_{\mu}{}^{\rho-}\psi_{\rho} - \frac{1}{2}\bn_{\mu}{}^{\rho}\psi_{\rho} + \frac{1}{2}\bn_{0}\psi_{\mu} \nn
  & &&\quad + \frac{1}{2}\eps\,\Psi_{\mu}{}^{\rho}D_{\rho}\phi - \frac{1}{2}\eps^{\sigma}\Psi_{\mu}{}^{\rho}\big(\wh{F}_{\sigma\rho}^{-} + \lambda T_{\sigma\rho}^{-} + D_{\sigma\rho}\big) ~.
}
This implies that
\eqa{
  &\scd_{\sigma}\scd_{\mu}\phi &&= D_{\sigma}\scd_{\mu}\phi + \frac{1}{2}\Psi_{\sigma}{}^{\rho}D_{\mu}\psi_{\rho} - \frac{1}{2}\Psi_{\sigma}{}^{\rho}[\chi_{\rho\mu}, \phi] + \frac{1}{2}\Psi_{\sigma\mu}[\eta, \phi] + \frac{1}{4}\Psi_{\sigma}{}^{\rho}\Psi_{\mu\rho}[\lambda,\phi] \nn
  & &&\quad + \frac{1}{\sq}S_{\sigma,\mu}{}^{\rho}\psi_{\rho} - \frac{1}{4\sq}S_{\sigma}\psi_{\mu} + \frac{1}{4}\Psi_{\sigma}{}^{\beta}\Psi_{\mu}{}^{\rho}\big(\wh{F}_{\beta\rho}^{-} + \lambda T_{\beta\rho}^{-} + D_{\beta\rho}\big) ~.
}
We can \emph{now} require that the gravitino be symmetric (and use, for instance, \eqref{eq:contraction-identity-S-susy-connections}). This yields,
\eqa{
  &\bm{\Box}_{C}\phi &&= \scd_{\mu}\scd^{\mu}\phi \nn
  & &&= D_{\mu}D^{\mu}\phi + \frac{1}{2}\big(\n_{\mu}\Psi^{\mu\nu}\big)\psi_{\nu} + \Psi^{\mu\nu}\big(D_{\mu}\psi_{\nu}) + \frac{1}{2}\Psi_{\rho}{}^{\rho}[\eta,\phi] + \frac{1}{4}\Psi^{\rho\mu}\Psi_{\rho}{}^{\nu}\big(\wh{F}_{\mu\nu}^{-} + \lambda T_{\mu\nu}^{-} + D_{\mu\nu}\big) ~.
}
This can be simplified further, of course, using \eqref{eq:sym-grav-T} and \eqref{eq:tsugra-supercov-ym-curvature}. Using \eqref{eq:Fhat minus lambda-Tminus}, $\wh{F}_{\mu\nu}^{-} + \lambda\,T_{\mu\nu}^{-} + D_{\mu\nu} = F_{\mu\nu}^{-} + D_{\mu\nu} + \big(\Psi_{[\mu}{}^{\rho}\chi_{\nu]\rho}\big)^{-} + 2\,\lambda\big(\n_{[\mu}\Phi_{\nu]}\big)^{-}$, and so, we finally have
\beqas{
 &\bm{\Box}_{C}\phi &&=D_{\mu}D^{\mu}\phi + \frac{1}{2}\big(\n_{\mu}\Psi^{\mu\nu}\big)\psi_{\nu} + \Psi^{\mu\nu}\big(D_{\mu}\psi_{\nu}) + \frac{1}{2}\Psi_{\rho}{}^{\rho}[\eta,\phi] + \frac{1}{4}\Psi_{\mu}{}^{\rho}\Psi_{\nu\rho}\big(F^{\mu\nu-} + D^{\mu\nu}\big) \\
& &&\quad + \frac{1}{4}\Psi_{\mu}{}^{\rho}\Psi_{\nu\rho}\big(\Psi^{\sigma[\mu}\chi^{\nu]}{}_{\sigma}\big)^{-} + \frac{1}{2}\Psi_{\mu}{}^{\rho}\Psi_{\nu\rho}\big(\n^{[\mu}\Phi^{\nu]})^{-}\lambda ~.
}
Next, consider $\lambda$. Recall from \eqref{eq:scd lambda} that 
\eqa{
  \scd_{\mu}\lambda &= D_{\mu}\lambda = \partial_{\mu}\lambda + [A_{\mu}, \lambda] ~. 
}
Therefore, the susy variation of $\scd_{\mu}\lambda$ is
\eqa{
   &\delta\big(\scd_{\mu}\lambda\big) &&= \eps D_{\mu}\eta + \eps[\psi_{\mu}, \lambda] + \eps^{\rho}[\chi_{\rho\mu}, \lambda] - \eps_{\mu}[\eta, \lambda] ~,
}
and hence,
\eqa{
   &\scd_{\sigma}\scd_{\mu}\lambda &&= D_{\sigma}\big(D_{\mu}\lambda\big) - \frac{1}{2}\Psi_{\sigma}{}^{\rho}[\chi_{\rho\mu}, \lambda] + \frac{1}{2}\Psi_{\sigma\mu}[\eta, \lambda] ~.
}
Again, since the gravitino is symmetric, we have
\beqa{
  & \bm{\Box}_{C}\lambda &&= \scd_{\mu}\scd^{\mu}\lambda = D_{\mu}D^{\mu}\lambda + \frac{1}{2}\Psi_{\rho}{}^{\rho}[\eta, \lambda] ~.
}

\section{Detailed Computations}\label{app:DetailedComputations}
\subsection{Closure Of The Cartan Model Algebra}\label{app:supporting-EqCoh-closure}
\subsubsection{$\widetilde{\sfd}^2\omega^+$ vs. $\widetilde{\sfd}^2 \chi$}\label{app:d2Omega-d2chi}
One of the subtleties of the localization multiplet is that the fields $\chi$ and $H$ are self-dual. This has some interesting implications on the closure of the algebra. Recall that the closure of the algebra is dictated by what the equivariant differential squares to. Consider a generic two-form field $\omega$. Its self-dual projection is
\eqa{
&\omega_{\mu\nu}^+ &&= \frac{1}{2}\omega_{\mu\nu} + \frac{1}{4}\sqrt{g}\epsilon_{\mu\nu\alpha\beta}g^{\alpha\alpha'}g^{\beta\beta'}\omega_{\alpha'\beta'} ~.
}
If we assume that $\widetilde{\sfd}\omega=0$, then we have
\eqa{
 &\widetilde{\sfd}\omega_{\mu\nu}^+ &&= \frac{1}{2}\sqrt{g}\epsilon_{\mu\nu\alpha\beta}\left(\frac{1}{4}\Psi^{\sigma}{}_{\sigma}g^{\alpha\alpha'}g^{\beta\beta'} - \Psi^{\alpha\alpha'}g^{\beta\beta'}\right)\omega_{\alpha'\beta'} ~.
}
Acting again with $\widetilde{\sfd}$ gives
\eqa{
& \widetilde{\sfd}^2\omega_{\mu\nu}^+ &&=\frac{1}{4}\sqrt{g}\epsilon_{\mu\nu\alpha\beta}\Psi^{\rho}{}_{\rho}\left(\frac{1}{4}\Psi^{\sigma}{}_{\sigma}g^{\alpha\alpha'}g^{\beta\beta'} - \Psi^{\alpha\alpha'}g^{\beta\beta'}\right)\omega_{\alpha'\beta'} \nn
& &&\quad +\frac{1}{2}\sqrt{g}\epsilon_{\mu\nu\alpha\beta}\left(\frac{1}{2}(\n^\sigma\Phi_{\sigma})g^{\alpha\alpha'}g^{\beta\beta'} - (\n^{\alpha}\Phi^{\alpha'}+\n^{\alpha'}\Phi^{\alpha})g^{\beta\beta'}\right)\omega_{\alpha'\beta'} \nn
& &&\quad +\frac{1}{2}\sqrt{g}\epsilon_{\mu\nu\alpha\beta}\left(\frac{1}{2}\Psi^\sigma{}_{\sigma}\Psi^{\alpha\alpha'}g^{\beta\beta'}+\Psi^{\alpha\alpha'}\Psi^{\beta\beta'}\right)\omega_{\alpha'\beta'} \nn
& &&=\frac{1}{2}\sqrt{g}\epsilon_{\mu\nu\alpha\beta}\left(\frac{1}{2}(\n_{\sigma}\Phi^{\sigma})g^{\alpha\alpha'}g^{\beta\beta'} - (\n^{\alpha}\Phi^{\alpha'}+\n^{\alpha'}\Phi^{\alpha})g^{\beta\beta'}\right)\omega_{\alpha'\beta'} ~.
}
This is the expected result: the right hand side involves the Lie derivative of the Hodge star operator along the vector field $\Phi$.

On the other hand, consider a field $\chi$ that is entirely self-dual. We have
\eqa{
&\widetilde{\sfd}\chi_{\mu\nu} &&= \frac{1}{2}\sqrt{g}\epsilon_{\mu\nu\alpha\beta}\left(\frac{1}{4}\Psi^{\sigma}{}_{\sigma}g^{\alpha\alpha'}g^{\beta\beta'} - \Psi^{\alpha\alpha'}g^{\beta\beta'}\right)\chi_{\alpha'\beta'} ~.
}
Unlike the case of general $\omega$, the field $\chi_{\alpha'\beta'}$ on the right hand side is self-dual and thus not $\widetilde{\sfd}$-closed. Therefore, when we compute $\widetilde{\sfd}^2\chi$, we obtain
\eqa{
& \widetilde{\sfd}^2\chi_{\mu\nu} &&=\frac{1}{2}\sqrt{g}\epsilon_{\mu\nu\alpha\beta}\left(\frac{1}{2}(\n_{\sigma}\Phi^{\sigma})g^{\alpha\alpha'}g^{\beta\beta'} - (\n^{\alpha}\Phi^{\alpha'}+\n^{\alpha'}\Phi^{\alpha})g^{\beta\beta'}\right)\chi_{\alpha'\beta'}\nn
& &&\quad - \frac{1}{2}\sqrt{g}\epsilon_{\mu\nu\alpha\beta}\left(\frac{1}{4}\Psi^{\sigma}{}_{\sigma}g^{\alpha\alpha'}g^{\beta\beta'} - \Psi^{\alpha\alpha'}g^{\beta\beta'}\right)\widetilde{\sfd}\chi_{\alpha'\beta'} ~.
}
Since the first line is the desired result, one would like the second line to vanish. Instead, we find that the second line evaluates to
\eqa{
& (\widetilde{\sfd}^2\chi_{\mu\nu})_{\Psi\Psi} &&: =- \frac{1}{2}\sqrt{g}\epsilon_{\mu\nu\alpha\beta}\left(\frac{1}{4}\Psi^{\sigma}{}_{\sigma}g^{\alpha\alpha'}g^{\beta\beta'} - \Psi^{\alpha\alpha'}g^{\beta\beta'}\right)\widetilde{\sfd}\chi_{\alpha'\beta'} 
=\frac{1}{2}\Psi^{\rho\sigma}\Psi_{\rho[\mu}\chi_{\nu]\sigma} ~.
}
Hence, we find, 
\eqa{
& \widetilde{\sfd}^2 \chi_{\mu\nu} &&= \frac{1}{2}\sqrt{g}\epsilon_{\mu\nu\alpha\beta}\left(\frac{1}{2}(\n_\sigma \Phi^\sigma)g^{\alpha\alpha'}g^{\beta\beta'} - (\n^\alpha \Phi^{\alpha'} + \n^{\alpha'}\Phi^{\alpha})g^{\beta\beta'}\right)\chi_{\alpha'\beta'} +\frac{1}{2}\Psi^{\rho\sigma}\Psi_{\rho[\mu}\chi_{\nu]\sigma} ~,
}
This extra term will be compensated thanks to the $\bm{\Delta}_{H}$ differential in our full model. 
The computation of $\widetilde{\sfd}^2$ is actually measuring the curvature of the projected connection -- see Appendix \ref{app:variation-of-self-dual-fields}. The closure of the algebra of the full semidirect product model (which includes purely self-dual adjoint-valued 2-form fields $\chi$ and $H$ or $\chi$ and $D$) relies on a happy cancellation of the $\Psi\Psi$ terms -- this ensures that $\IQ$ squares correctly.

\subsubsection{$\{\widetilde{\sfd}, \mathsf{K}\}H=-\{\widetilde{\sfd}, \mathbf{\Delta}_{H}\}H$}\label{app:DKH}
We begin by recording presentations of our transformations that do not contain self-dual or anti-self-dual projections. We have 
\eqa{
& \widetilde{\sfd}\chi_{\mu\nu} &&= -(\Psi^{\sigma}{}_{[\mu} \chi_{\nu]\sigma})^- \stackrel{ \eqref{eq:identB}}{=\joinrel=\joinrel=} -\Psi^{\sigma}{}_{[\mu}\chi_{\nu]\sigma} - \tfrac{1}{4}\Psi^\sigma{}_{\sigma}\chi_{\mu\nu} ~,\\
& \widetilde{\sfd}H_{\mu\nu}    &&= -(\Psi^{\sigma}{}_{[\mu} H_{\nu]\sigma})^- \stackrel{ \eqref{eq:identB}}{=\joinrel=\joinrel=} -\Psi^{\sigma}{}_{[\mu}H_{\nu]\sigma} - \tfrac{1}{4}\Psi^\sigma{}_{\sigma}H_{\mu\nu} ~,
}
and
\eqa{
&\mathsf{K}H_{\mu\nu} &&= \Phi^\sigma D_\sigma \chi_{\mu\nu} +\left((\n_\mu \Phi^\sigma)\chi_{\sigma\nu} + (\n_\nu\Phi^\sigma)\chi_{\mu\sigma}\right)^+ \nn
& &&\stackrel{ \eqref{eq:identB general}}{=\joinrel=\joinrel=}\Phi^\sigma D_\sigma \chi_{\mu\nu} +\tfrac{1}{2}(\n_\sigma \Phi^\sigma) \chi_{\mu\nu} -\tfrac{1}{2}(\n_\mu \Phi^\sigma - \n^\sigma \Phi_\mu)\chi_{\nu\sigma} + \tfrac{1}{2}(\n_\nu \Phi^\sigma - \n^\sigma \Phi_{\nu})\chi_{\mu\sigma} ~. \label{eq:KH}
}
We compute
\eqa{
&\{\widetilde{\sfd}, \bm{\Delta}_{H}\}H_{\mu\nu}&&=\widetilde{\sfd}\left(-\tfrac{1}{2}\Psi^{\rho\sigma}\Psi_{\rho[\mu}\chi_{\nu]\sigma}\right) +\bm{\Delta}_{H}\left(-\Psi^{\sigma}_{[\mu}H_{\nu]\sigma} - \tfrac{1}{4}\Psi^\sigma{}_{\sigma}H_{\mu\nu}\right) \nn
& &&=-\tfrac{1}{2}(\n^\rho \Phi^\sigma +\n^\sigma \Phi^\rho)\Psi_{\rho[\mu} \chi_{\nu]\sigma} + \tfrac{1}{2}\Psi^{\rho\sigma}(\n_{\rho}\Phi_{[\mu} + \n_{[\mu} \Phi_{\rho})\chi_{\nu]\sigma} \nn
& &&\quad - \tfrac{1}{4}\Psi^{\rho\sigma}\Psi_{\rho\mu}(-\Psi^{\alpha}{}_{[\nu}\chi_{\sigma]\alpha} - \tfrac{1}{4}\Psi^\alpha{}_{\alpha}\chi_{\nu\sigma})+ \tfrac{1}{4}\Psi^{\rho\sigma}\Psi_{\rho\nu}\left(-\Psi^{\alpha}{}_{[\mu}\chi_{\sigma]\alpha} -\tfrac{1}{4}\Psi^{\alpha}{}_{\alpha}\chi_{\mu\sigma}\right) \nn
& &&\quad - \tfrac{1}{4}\Psi^\sigma{}_{\mu}\Psi^{\rho\alpha}\Psi_{\rho[\nu}\chi_{\sigma]\alpha} +\tfrac{1}{4}\Psi^\sigma{}_{\nu}\Psi^{\rho\alpha}\Psi_{\rho[\mu}\chi_{\sigma]\alpha}-\tfrac{1}{8}\Psi^\alpha{}_{\alpha}\Psi^{\rho\sigma}\Psi_{\rho[\mu}\chi_{\nu]\sigma} \nn
& &&=-\tfrac{1}{2}(\n^\rho \Phi^\sigma +\n^\sigma \Phi^\rho)\Psi_{\rho[\mu} \chi_{\nu]\sigma} + \tfrac{1}{2}\Psi^{\rho\sigma}(\n_{\rho}\Phi_{[\mu} + \n_{[\mu} \Phi_{\rho})\chi_{\nu]\sigma} ~. \label{eq:d DeltaH H}
}
Next, we compute $\{\widetilde{\sfd}, \mathsf{K}\}H_{\mu\nu}$ in parts.
\eqa{
&(\mathsf{K}\widetilde{\sfd}) H_{\mu\nu} &&=\mathsf{K}\left(-\Psi^{\sigma}_{[\mu}H_{\nu]\sigma} - \tfrac{1}{4}\Psi^\sigma{}_{\sigma}H_{\mu\nu}\right) \nn
& &&=\Psi^{\sigma}{}_{[\mu}\left(\Phi^\rho D_{\rho} \chi_{\nu]\sigma} +\tfrac{1}{2}(\n_\rho \Phi^\rho) \chi_{\nu]\sigma} -\tfrac{1}{2}(\n_{\nu]}\Phi^\rho - \n^\rho \Phi_{\nu]})\chi_{\sigma\rho} + \tfrac{1}{2}(\n_\sigma \Phi^\rho - \n^{\rho}\Phi_{\sigma})\chi_{\nu]\rho}\right) \nn
& &&\quad +\tfrac{1}{4}\Psi^\sigma{}_{\sigma}\left(\Phi^\rho D_\rho \chi_{\mu\nu} + \tfrac{1}{2}(\n_\rho \Phi^\rho)\chi_{\mu\nu} - \tfrac{1}{2}(\n_\mu \Phi^\rho - \n^\rho \Phi_\mu)\chi_{\nu\rho} +\tfrac{1}{2}(\n_\nu \Phi^\rho - \n^\rho \Phi_{\nu})\chi_{\mu\sigma}\right) ~. \label{eq:K dH}
}
For the other direction, we compute the $\widetilde{\sfd}$ action on each term of $\mathsf{K}H_{\mu\nu}$ from \eqref{eq:KH}. Additionally, we need the variation of the Christoffel connection, which reads
\eqa{
& \widetilde{\sfd}\Gamma^\rho{}_{\mu\nu} &&=g^{\rho\sigma}\left(\n_{(\mu}\Psi_{\nu)\sigma} - \tfrac{1}{2}\n_{\sigma}\Psi_{\mu\nu}\right) ~.
}
This leads to 
\eqa{
&\widetilde{\sfd}(\n_\mu \Phi^\sigma) &&=g^{\sigma\alpha}(\n_{(\mu}\Psi_{\rho)\alpha}-\tfrac{1}{2}\n_\alpha \Psi_{\mu\rho})\Phi^\rho ~,\\
&\widetilde{\sfd}(\n_{\sigma}\Phi_{\mu}) &&=\Psi_{\mu\rho}\n_\sigma \Phi^\rho +(\n_{(\sigma}\Psi_{\rho)\mu}-\tfrac{1}{2}\n_\mu \Psi_{\sigma\rho})\Phi^\rho ~,\\
&\widetilde{\sfd}(\n_\sigma \Phi^\sigma) &&= \tfrac{1}{2}(\n_{\rho}\Psi^\sigma{}_{\sigma})\Phi^\rho ~.
}
Term by term, we find
\eqa{
&\widetilde{\sfd}(\Phi^\sigma D_\sigma \chi_{\mu\nu}) &&=\Phi^\sigma D_\sigma\left(-\Psi^{\rho}{}_{[\mu}\chi_{\nu]\rho} -\tfrac{1}{4}\Psi^\rho{}_{\rho}\chi_{\mu\nu}\right) -\Phi^\sigma(\widetilde{\sfd} \Gamma^\rho{}_{\sigma\mu})\chi_{\rho\nu} - \Phi^\sigma(\widetilde{\sfd}\Gamma^{\rho}{}_{\sigma\nu})\chi_{\mu\rho} \nn
& &&=-\Phi^\sigma \Psi^\rho{}_{[\mu} D_\sigma \chi_{\nu]\rho} -\tfrac{1}{4}\Phi^\sigma\Psi^\rho{}_{\rho}D_\sigma \chi_{\mu\nu} -\Phi^\sigma (\n_{\sigma} \Psi^\rho{}_{[\mu})\chi_{\nu]\rho} -\tfrac{1}{4}\Phi^\sigma (\n_\sigma \Psi^\rho{}_{\rho})\chi_{\mu\nu} \nn
& &&\quad-\Phi^\rho(g^{\rho\alpha}(\n_{(\sigma}\Psi_{\nu)\alpha} - \tfrac{1}{2}\n_{\alpha}\Psi_{\sigma\mu}))\chi_{\rho\nu} - \Phi^\sigma\left(g^{\rho\alpha}\left(\n_{(\sigma}\Psi_{\nu)\alpha} - \tfrac{1}{2}\n_{\alpha}\Psi_{\sigma\nu}\right)\right)\chi_{\mu\rho} \nn
& &&=-\Psi^\sigma{}_{[\mu}(\Phi^\rho D_\sigma \chi_{\nu]\sigma}) -\tfrac{1}{4}\Psi^{\sigma}{}_{\sigma}(\Phi^\rho D_\rho \chi_{\mu\nu}) - \tfrac{1}{4}(\n_\rho\Psi^\sigma{}_{\sigma})\Phi^\rho \chi_{\mu\nu}\nonumber\\
& &&\quad +(\n_{[\mu}\Psi_{\sigma\rho})\Phi^\sigma \chi_{\nu]}{}^\rho - (\n_\rho \Psi_{\sigma[\mu})\Phi^\sigma \chi_{\nu]}{}^\rho ~.\\
& \widetilde{\sfd}\left(\tfrac{1}{2}(\n_\sigma\Phi^\sigma)\chi_{\mu\nu}\right) &&=\tfrac{1}{4}(\n_\rho\Psi^\sigma{}_{\sigma})\Phi^\rho\chi_{\mu\nu} + \tfrac{1}{2}(\n_\sigma \Phi^\sigma)\left(-\Psi^\rho_{[\mu}\chi_{\nu]\rho} - \tfrac{1}{4}\Psi^\rho{}_{\rho}\chi_{\mu\nu}\right) \nn
& &&=\tfrac{1}{4}(\n_\rho \Psi^\sigma{}_{\sigma})\Phi^\rho\chi_{\mu\nu} - \tfrac{1}{2}\Psi^\sigma{}_{[\mu}(\n_\rho\Phi^\rho)\chi_{\nu]\sigma} - \tfrac{1}{4}\Psi^\sigma{}_{\sigma}(\n_\rho \Phi^\rho)\chi_{\mu\nu} ~.
}
\eqa{
&\widetilde{\sfd}\left(-\tfrac{1}{2}(\n_\mu \Phi^\sigma)\chi_{\nu\sigma}\right) &&=-\tfrac{1}{2}(g^{\sigma\alpha}(\n_{(\nu}\Psi_{\rho)\alpha}-\tfrac{1}{2}\n_{\alpha}\Psi_{\mu\rho})\Phi^\rho\chi_{\nu\sigma}) - \tfrac{1}{2}(\n_{\mu}\Phi^\sigma)\left(-\Psi^{\alpha}{}_{[\nu}\chi_{\sigma]\alpha} - \tfrac{1}{4}\Psi^\alpha{}_{\alpha}\chi_{\nu\sigma}\right) \nn
& &&=-\tfrac{1}{4}(\n_{\mu}\Psi_{\sigma\rho})\Phi^\sigma \chi_\nu{}^\rho -\tfrac{1}{4}(\n_\sigma \Psi_{\mu\rho})\Phi^\sigma \chi_\nu{}^\rho +\tfrac{1}{4}(\n_\rho \Psi_{\mu\sigma})\Phi^\sigma \chi_\nu{}^\rho \nn
& &&\quad +\tfrac{1}{2}\Psi^\sigma{}_{[\nu}(\n_\mu \Phi^\rho)\chi_{\rho]\sigma} +\tfrac{1}{8}\Psi^\sigma{}_{\sigma}(\n_\mu \Phi^\rho)\chi_{\nu\rho} ~.\\
& \widetilde{\sfd}\left(\tfrac{1}{2}(\n_\nu \Phi^\sigma)\chi_{\mu\sigma}\right) &&=\tfrac{1}{4}(\n_{\nu}\Psi_{\sigma\rho})\Phi^\sigma \chi_\mu{}^\rho +\tfrac{1}{4}(\n_\sigma \Psi_{\nu\rho})\Phi^\sigma \chi_\mu{}^\rho -\tfrac{1}{4}(\n_\rho \Psi_{\nu\sigma})\Phi^\sigma \chi_\mu{}^\rho \nn
& &&\quad -\tfrac{1}{2}\Psi^\sigma{}_{[\mu}(\n_\nu \Phi^\rho)\chi_{\rho]\sigma} -\tfrac{1}{8}\Psi^\sigma{}_{\sigma}(\n_\nu \Phi^\rho)\chi_{\mu\rho} ~.\\
%
& \widetilde{\sfd}\left(\tfrac{1}{2}(\n_\sigma \Phi_\mu)\chi_\nu{}^\sigma\right) &&=\tfrac{1}{2}\left(\Psi_{\mu\rho}\n_\sigma \Phi^\rho + (\n_{(\sigma}\Psi_{\rho)\mu} -\tfrac{1}{2}\n_\mu \Psi_{\sigma\rho})\Phi^\rho\right)\chi_\nu{}^\sigma -\tfrac{1}{2}\Psi^{\sigma\rho}(\n_\sigma \Phi_\mu)\chi_{\nu\rho} \nn
& &&\quad +\tfrac{1}{2}(\n_\sigma \Phi_\mu)g^{\sigma\rho}\left(-\Psi^\alpha{}_{[\nu}\chi_{\rho]\alpha}-\tfrac{1}{4}\Psi^\alpha{}_{\alpha}\chi_{\nu\rho}\right) \nn 
& &&=\tfrac{1}{2}\Psi^\sigma{}_{\mu} (\n^\rho \Phi_\sigma)\chi_{\nu\sigma} -\tfrac{1}{4}\Psi^{\sigma\rho}(\n_\sigma \Phi_\mu)\chi_{\nu\rho} -\tfrac{1}{4}\Psi^\sigma{}_{\nu}(\n_\rho \Phi_\mu)\chi^\rho{}_{\sigma} - \tfrac{1}{8}\Psi^\sigma{}_{\sigma}(\n^\rho \Phi_\mu)\chi_{\nu\rho} \nn
& &&\quad +\tfrac{1}{4}(\n_\rho \Psi_{\sigma\mu})\Phi^\sigma \chi_\nu{}^\rho +\tfrac{1}{4}(\n_\sigma \Psi_{\rho\mu})\Phi^\sigma \chi_\nu{}^\rho -\tfrac{1}{4}(\n_\mu \Psi_{\sigma\rho})\Phi^\sigma \chi_\nu{}^\rho ~.\\
%
& \widetilde{\sfd}\left(-\tfrac{1}{2}(\n_\sigma \Phi_\nu)\chi_\mu{}^\sigma\right) &&=-\tfrac{1}{2}\Psi^\sigma{}_{\nu} (\n^\rho \Phi_\sigma)\chi_{\mu\sigma} +\tfrac{1}{4}\Psi^{\sigma\rho}(\n_\sigma \Phi_\nu)\chi_{\mu\rho} +\tfrac{1}{4}\Psi^\sigma{}_{\mu}(\n_\rho \Phi_\nu)\chi^\rho{}_{\sigma} + \tfrac{1}{8}\Psi^\sigma{}_{\sigma}(\n^\rho \Phi_\nu)\chi_{\mu\rho} \nn
& &&\quad -\tfrac{1}{4}(\n_\rho \Psi_{\sigma\nu})\Phi^\sigma \chi_\mu{}^\rho -\tfrac{1}{4}(\n_\sigma \Psi_{\rho\nu})\Phi^\sigma \chi_\mu{}^\rho +\tfrac{1}{4}(\n_\nu \Psi_{\sigma\rho})\Phi^\sigma\chi_\mu{}^\rho ~.
}
Putting everything together, we find
\eqa{
& \widetilde{\sfd}\left(-(\n_{[\mu}\Phi^\sigma)\chi_{\nu]\sigma} + (\n_\sigma \Phi_{[\mu})\chi_{\nu]}{}^\sigma\right) &&=-(\n_{[\mu}\Psi_{\sigma\rho})\Phi^\sigma \chi_{\nu]}{}^\rho + (\n_\rho \Psi_{\sigma[\mu})\Phi^\sigma \chi_{\nu]}{}^\rho \nonumber\\
& &&\quad-\tfrac{1}{4}\Psi^{\sigma}{}_{\sigma}(-\tfrac{1}{2}(\n_\mu \Phi^\rho -\n^\rho \Phi_{\mu})\chi_{\nu\rho} +\tfrac{1}{2}(\n_{\nu}\Phi^\rho - \n^\rho \Phi_\nu)\chi_{\mu\sigma})\nonumber\\
& &&\quad +\tfrac{1}{2}\Psi^\sigma{}_{[\mu}(\n_{\nu]}\Phi^\rho)\chi_{\sigma\rho} - \tfrac{1}{2}\Psi^{\sigma\rho}(\n_{[\mu}\Phi_\rho)\chi_{\nu]\sigma} +\Psi^{\sigma}{}_{[\mu}(\n^\rho \Phi_\sigma)\chi_{\nu]\rho}\nonumber\\
& &&\quad -\tfrac{1}{2}\Psi^\sigma{}_{[\mu}(\n^\rho \Phi_{\nu]})\chi_{\sigma\rho} - \tfrac{1}{2}\Psi^{\sigma\rho}(\n_\sigma\Phi_{[\mu})\chi_{\nu]\rho} ~.
}
Therefore,
\eqa{
&(\widetilde{\sfd}\mathsf{K})H_{\mu\nu} &&=-\Psi^{\sigma}{}_{[\mu}(\Phi^\rho D_\rho \chi_{\nu]\sigma}) -\tfrac{1}{4}\Psi^\sigma{}_{\sigma}(\Phi^\rho D_\rho \chi_{\mu\nu}) - \tfrac{1}{2}\Psi^{\sigma}{}_{[\mu}(\n_\rho \Phi^\rho)\chi_{\nu]\sigma}\nonumber\\
& &&\quad -\tfrac{1}{4}\Psi^\sigma{}_\sigma ((\n_\rho \Phi^\rho)\chi_{\mu\nu} - \tfrac{1}{2}(\n_\mu \Phi^\rho - \n^\rho \Phi_\mu)\chi_{\nu\rho} + \tfrac{1}{2}(\n_\nu \Phi^\rho - \n^\rho \Phi_\nu)\chi_{\mu\sigma})\nonumber\\
& &&\quad + \Psi^\sigma{}_{[\mu}(\tfrac{1}{2}(\n_{\nu]}\Phi^\rho)\chi_{\sigma\rho} + (\n^\rho \Phi_{\sigma})\chi_{\nu]\rho}-\tfrac{1}{2}(\n^\rho \Phi_{\nu]})\chi_{\sigma\rho})\nonumber\\
& &&\quad - \tfrac{1}{2}\Psi^{\sigma\rho}((\n_{[\mu}\Phi_\rho) + (\n_\rho \Phi_{[\mu}))\chi_{\nu]\sigma} ~. \label{eq:d KH}
}
Combining \eqref{eq:K dH} and \eqref{eq:d KH}, we find
\eqa{
&\{\widetilde{\sfd}, \mathsf{K}\} &&= -\tfrac{1}{2}\Psi^{\sigma\rho}(\n_{[\mu}\Phi_{\rho} + \n_\rho \Phi_{[\mu})\chi_{\nu]\sigma} + \tfrac{1}{2}(\n^\rho \Phi_\sigma + \n_\sigma \Phi^\rho)\Psi^{\sigma}{}_{[\mu}\chi_{\nu]\rho} ~.
}
Comparing this with \eqref{eq:d DeltaH H}, we find that indeed, $\{\widetilde{\sfd}, \mathsf{K}\}H=-\{\widetilde{\sfd}, \bm{\Delta}_{H}\}H$. $\hfill \blacksquare$

\subsection{Consistency Of The Topological Twist}\label{sec:ConsistencyTwist}
The topological twist is defined by three constraints, namely \eqref{eq:toptwist sugra}, \eqref{eq:Dscr} and \eqref{eq:chi vec}, which we reiterate for the reader's convenience,
\eqa{
   \omega_{\mu}{}^{AB} &= V_{\mu}{}^{AB} ~, \quad \mathscr{D} &&= \frac{1}{6}\mathscr{R} ~, \quad  \Xi_{μ} &&= \CJ_{μρ}^{+}{}^{ρ} + \frac{1}{3}\CJ_{μρ}^{-}{}^{ρ} ~.
}
In this section, we will prove that these constraints are robust under a supersymmetry variation. We use a combination of two- and four- component methods for the computations in this section. Readers interested in following the details may find it useful to review Appendix \ref{app:conventions}. In this section we use the transformation laws \eqref{eq:delta-twisted-conformal-sugra}.

\subsubsection{The $\omega_{+} = V$ Constraint}
\paragraph{Computation of $\delta \omega$.} The supersymmetry variation of the vielbein is
\eqa{
  & \delta e_{\mu}{}^{A\dA} &&= \frac{1}{2}\eps\,\Psi_{\mu}{}^{A\dA} ~. \label{eq:susy variation twisted vielbein}
}
Since a chiral half of the gravitino is set to zero as part of the twist, the four-dimensional torsion vanishes identically, and hence the vielbein postulate reads\footnote{The sign of the second term in the vielbein postulate differs from the conventions of \protect{\cite{Freedman:2012zz}}.}
\eqa{
    &de^{a} - \omega^{a}{}_{b} \wedge e^{b} &&= 0 ~,
}
where $e^{a}$ denotes the vielbein one-form, and $\omega^{a}{}_{b}$ the spin connection one-form. Therefore under a variation of the vielbein,
\eqa{
	&d\delta e^{a} - \omega^{a}{}_{b} \wedge \delta e^{b} - \delta \omega^{a}{}_{b} \wedge e^{b} &&= 0 ~.\label{eq:cartan delta}
}
In local coordinates, the covariant curl of the variation of the vielbein is
\eqa{
    &\n_{[\mu}\delta e_{\nu]}{}^{a} &&= \partial_{[\mu}\delta e_{\nu]}{}^{a} - {\omega_{[\mu}{}^{a}}_{b}\delta e_{\nu]}{}^{b} ~,
}
so \eqref{eq:cartan delta} (the local form of which is $\partial_{[\mu}\delta e_{\nu]}{}^{a} - {\omega_{[\mu}{}^{a}}_{b}\delta e_{\nu]}{}^{b} - \delta {\omega_{[\mu}{}^{a}}_{b} e_{\nu]}{}^{b} = 0$) can be written as
\eqa{
   &\n_{[\mu}\delta e_{\nu]}{}^{a} - \big(\delta {\omega_{[\mu}{}^{a}}_{b}\big) e_{\nu]}{}^{b} &&= 0 ~. \label{eq:covariant curl of vielbein}
}
Writing \eqref{eq:covariant curl of vielbein} three times with cyclically permuted pairs of indices $(\mu, \nu)$, $(\nu, \rho)$ and $(\rho, \mu)$, adding the first and third equation so obtained and subtracting from this the second, we arrive at the following expression for the variation of the spin connection:
\eqa{
	&\delta \omega_{\mu,\ha\hb} &&=
	\left(\n_{[\mu}\delta e_{\nu]}{}^{D\dD}e_{\rho,D\dD} - \n_{[\nu}\delta e_{\rho]}{}^{D\dD} e_{\mu,D\dD} + \n_{[\rho}\delta e_{\mu]}{}^{D\dD}e_{\nu,D\dD}\right)e^{\rho}{}_{[\ha} e^{\nu}{}_{\hb]} \nonumber\\
	& &&= \frac{1}{2}\eps\left(\n_{[\mu}\Psi_{\nu]}{}^{D\dD}e_{\rho,D\dD} - \n_{[\nu} \Psi_{\rho]}{}^{D\dD} e_{\mu,D\dD} + \n_{[\rho}\Psi_{\mu]}{}^{D\dD}e_{\nu,D\dD}\right)e^{\rho}{}_{[\ha} e^{\nu}{}_{\hb]} ~.
}
In the second equality, we used \eqref{eq:susy variation twisted vielbein}. In terms of the curl of the gravitino, which we denote by $\CJ_{[\mu\nu],\rho} = \n_{[\mu}\Psi_{\nu]\rho}$ or its two-component version $\CJ_{[A\dA,B\dB],C\dC} = \n_{[A\dA}\Psi_{B\dB],C\dC}$, the preceding equation can be written in two-component form as
\eqa{
	&\delta\omega_{\mu,E\dE,F\dF} &&= -\frac{1}{2}\eps\left(\CJ_{A\dA,E\dE,F\dF} -\CJ_{A\dA,F\dF,E\dE} + \CJ_{F\dF,E\dE,A\dA} \right)e_{\mu}{}^{A\dA} ~.
}
The self- and anti-self- dual parts of this with respect to $E\dE, F\dF$ are, respectively
\eqa{
 & \delta\omega_{\mu,EF} &&= \frac{1}{4}\eps \left[\CJ_{AE,F\dA} + \CJ_{AF,E\dA} - \veps_{AE}\CJ_{\dA\dE,F}{}^{\dE} - \veps_{AF}\CJ_{\dA\dE,E}{}^{\dE} + 2\,\CJ_{EF,A\dA}\right]e_{\mu}{}^{A\dA} ~, \label{eq:twisted delta omega SD} \\
& \delta \omega_{\mu,\dE\dF} &&= \frac{1}{4}\eps \left[\CJ_{\dA\dE,A\dF} + \CJ_{\dA\dF,A\dE} - \veps_{\dA\dE}\CJ_{AE}{}^{E}{}_{\dF} - \veps_{\dA\dF}\CJ_{AE}{}^{E}{}_{\dE} + 2\,\CJ_{\dE\dF,A\dA}\right]e_{\mu}{}^{A\dA} \label{eq:twisted delta omega ASD} ~.
}

\paragraph{Computation of $\delta V$.} After twisting and using \eqref{eq:chi vec} and \eqref{eq:SgravTrunc-3}, the variation of the $\mathfrak{su(2)}_{\mathsf{R}}$ symmetry connection \eqref{eq:csugra-susy-rsym-su2-2comp} reads 
\eqa{
 &\delta V_{\mu,EF} &&= \frac{1}{4}\eps \left[2(\CJ_{EA,F\dA} + \CJ_{FA,E\dA}) + \veps_{AE}(\CJ^{\dF}{}_{\dA,F\dF} - {\CJ_{DF}{}^{D}}_{\dA}) + \veps_{AF}(\CJ^{\dF}{}_{\dA,E\dF} - {\CJ_{DE}{}^{D}}_{A})\right]e_{\mu}{}^{A\dA} ~. \label{eq:twisted delta V final}
}

\paragraph{Computation of $\delta \omega - \delta V$.} 
We begin by stating a standard identity (the ``exchange identity'') involving two objects with spinor indices:
\eqa{
	&\sfX_A \sfY_{B}{}^{B} &&= \sfX_{B}\sfY_{A}{}^{B} + \sfX^{B}\sfY_{BA}~.\label{eq:standard identity}
}
To prove this identity, observe that the right hand side can be written as $\sfX^B(\sfY_{BA} - \sfY_{AB})$, and that $\sfY_{AB}-\sfY_{BA} = \veps_{AB}\sfY_{C}{}^{C}$. This completes the proof. Note that this identity does not require $\sfY_{AB}$ to be symmetric or antisymmetric.

Consider the quantity $\veps_{AB}{\CJ_{DC,}}^{D}{}_{\dA}$. We can apply the identity \eqref{eq:standard identity} to $\sfX_{A} \rightarrow \veps_{AB}$ and $\sfY_{D}{}^{D} \rightarrow \CJ_{DC,}{}^{D}{}_{\dA}$, i.e., to the indices $B$ and $D$ while treating the indices $A$, $C$, and $\dA$ inert. Doing so yields the identity
\eqa{
	&\veps_{AB}{\CJ_{DC,}}^{D}{}_{\dA} &&= \veps_{AD}\CJ_{BC,}{}^{D}{}_{\dA} + \delta_{A}{}^{D}\CJ_{DC,B\dA} = -\CJ_{BC,A\dA} + \CJ_{AC,B\dA}~.\label{eq:identity 1}
}
We rewrite this identity for $B \leftrightarrow C$:
\eqa{
	&\veps_{AC}{\CJ_{DB,}}^{D}{}_{\dA} &&= -\CJ_{CB,A\dA} + \CJ_{AB,C\dA} ~.\label{eq:identity 2}
}
Adding \eqref{eq:identity 1} and \eqref{eq:identity 2} and using the fact that $\CJ_{BC,A\dA} = \CJ_{CB,A\dA}$, we get
\eqa{
&  2 \, \CJ_{BC,A\dA}  &&= \CJ_{AB,C\dA} + \CJ_{AC,B\dA} - \veps_{AB}J_{DC,}{}^{D}{}_{\dA} - \veps_{AC}J_{DB,}{}^{D}{}_{\dA} ~. \label{eq:identity we need}
}
Let us evaluate $\delta \omega_{\mu,EF} - \delta V_{\mu,EF}$ using \eqref{eq:twisted delta omega SD} and \eqref{eq:twisted delta V final}. This vanishes by a direct application of \eqref{eq:identity we need}:
\eqa{
	&\delta \omega_{\mu,EF} - \delta V_{\mu,EF} &&= \frac{1}{4}\eps\left[2\CJ_{EF,A\dA}-\CJ_{EA,F\dA} - \CJ_{FA,E\dA} + \veps_{AE}\CJ_{DF,}{}^{D}{}_{\dA} + \veps_{AF}\CJ_{DE,}{}^{D}{}_{\dA}\right]e_{\mu}{}^{A\dA} \nonumber\\
	& && \stackrel{ \eqref{eq:identity we need}}{=\joinrel=\joinrel=}0~.
}
Therefore (relabeling and raising free indices),
\eqa{
	&\delta \omega_{\mu}{}^{AB} &&= \delta V_{\mu}{}^{AB} ~.	
}

\subsubsection{The $\mathscr{D} = \frac{1}{6}\mathscr{R}$ Constraint}

\paragraph{Computation of $\delta \mathscr{R}$.} From the twist condition $\omega_{\mu}{}^{EF} = V_{\mu}{}^{EF}$, it follows that
\eqa{
	&R(V)_{\mu\nu}{}^{EF} &&= R(\omega)_{\mu\nu}{}^{EF} \label{eq:twistcurv} ~,
}
and so, using \eqref{eq:contracted R},
\eqas{
	&R(V)_{EF}{}^{EF} &&= \veps^{\dE\dF}R(V)_{\mu\nu}{}^{EF}e^{\mu}{}_{E\dE}e^{\nu}{}_{F\dF} \\
	& && =  \veps^{\dE\dF}R(\omega)_{\mu\nu}{}^{EF}e^{\mu}{}_{E\dE}e^{\nu}{}_{F\dF} \\
	& &&= R(\omega)_{EF}{}^{EF} \\
	& &&=  R(\omega)_{\dE\dF}{}^{\dE\dF} \\
	& && = \veps^{EF}R(\omega)_{\mu\nu}{}^{\dE\dF}e^{\mu}{}_{E\dE}e^{\nu}{}_{F\dF} ~.
}
The key point is that $\delta R(\omega)_{EF}{}^{EF}$ equals $\delta R(\omega)_{\dE\dF}{}^{\dE\dF}$. Let us compute $\delta R(V)_{EF}{}^{EF}$ in the two ways suggested by this equivalence. We use the Palatini identity $\delta R(\omega)_{\mu\nu}{}^{\bullet\bullet} = 2\,\n_{[\mu}\delta\omega_{\nu]}{}^{\bullet\bullet}$, where the bullets can be a pair of dotted or undotted indices.\footnote{The identity as written holds only when the lower index on $\delta\omega$ is a \emph{curved} spacetime index.} Let us begin with $\delta R(\omega)_{EF}{}^{EF}$. 
\eqa{
	&\delta R(\omega)_{EF}{}^{EF} &&= \delta\left(\veps^{\dE\dF}R(\omega)_{\mu\nu}{}^{EF}e^{\mu}{}_{E\dE}e^{\nu}{}_{F\dF}\right)\nonumber\\
	& &&= \veps^{\dE\dF}e^{\mu}{}_{E\dE}e^{\nu}{}_{F\dF}\delta R(\omega)_{\mu\nu}{}^{EF} + \veps^{\dE\dF}\big(-\tfrac{\eps}{2}\Psi_{E\dE}{}^{G\dG}e^{\mu}{}_{G\dG}\big)e^{\nu}{}_{F\dF}R(\omega)_{\mu\nu}{}^{EF}\nonumber\\
	 & && \quad + \veps^{\dE\dF}e^{\mu}{}_{E\dE}\big(-\tfrac{\eps}{2}\Psi_{F\dF}{}^{G\dG}e^{\nu}{}_{G\dG}\big)R(\omega)_{\mu\nu}{}^{EF}\nonumber\\
	& &&= \veps^{\dE\dF}e^{\mu}{}_{E\dE}e^{\nu}{}_{F\dF}\delta R(\omega)_{\mu\nu}{}^{EF} -\frac{\eps}{2}\veps^{\dE\dF}\left[\Psi_{E\dE}{}^{G\dG}R(\omega)_{G\dG,F\dF}{}^{EF} + \Psi_{F\dF}{}^{G\dG}R(\omega)_{E\dE,G\dG}{}^{EF}\right]\label{eq:rw 1} ~.
}
Using \eqref{eq:twisted delta omega SD},
\eqa{
	&e^{\mu}{}_{E\dE}e^{\nu}{}_{F\dF}\delta R(\omega)_{\mu\nu}{}^{EF} &&= e^{\mu}{}_{E\dE}e^{\nu}{}_{F\dF}\left(\n_{\mu}\delta\omega_{\nu}{}^{EF} - \n_{\nu}\delta\omega_{\mu}{}^{EF}\right)\nonumber\\
	& &&= \frac{\eps}{4} \left[\n_{E\dE}\left(\CJ^{EF}{}_{F\dF} - 3\,\CJ_{\dF\dG}{}^{E\dG}\right) - \n_{F\dF}\left(\CJ^{EF}{}_{E\dE} - 3 \,\CJ_{\dE\dG}{}^{F\dG}\right)\right] ~.
}
Therefore, the first term in \eqref{eq:rw 1} evaluates to
\eqa{
	&\veps^{\dE\dF}e^{\mu}{}_{E\dE}e^{\nu}{}_{F\dF}\delta R(\omega)_{\mu\nu}{}^{EF} &&= \frac{\eps}{2} \left(\n_{E\dE}{\CJ^{EF}}_{F}{}^{\dE} - 3\,\n_{E\dE}\CJ^{\dE}{}_{\dF}{}^{E\dF}\right)\label{eq:first term in rw 1} ~.
}
After decomposing the curvatures $R(\omega)_{A\dA,B\dB}{}^{CD}$ with respect to self- and anti-self- dual components (with respect to the lower indices $A\dA, B\dB$), the second term in \eqref{eq:rw 1} is proportional to
\eqa{
	&\veps^{\dE\dF}\left[\Psi_{E\dE}{}^{G\dG}R(\omega)_{G\dG,F\dF}{}^{EF} + \Psi_{F\dF}{}^{G\dG}R(\omega)_{E\dE,G\dG}{}^{EF}\right] &&= \left(\Psi_{E\dE}{}^{G\dE}{R(\omega)_{GF}}^{EF} - \Psi_{E\dE}{}^{G\dG}{R(\omega)_{\dG}{}^{\dE,E}}_{G}\right) \nn
	& &&= \left(\frac{\mathscr{R}}{2}\Psi_{E\dE}{}^{E\dE}- \Psi_{E\dE}{}^{G\dG}{R(\omega)_{\dG}{}^{\dE,E}}_{G}\right)	\label{eq:psiterm1} ~,
}
where we used \eqref{eq:contracted R} in the second equality. Next, consider $\delta R(\omega)_{\dE\dF}{}^{\dE\dF}$.
\eqa{
	&\delta R(\omega)_{\dE\dF}{}^{\dE\dF} &&= \delta\left(\veps^{EF}R(\omega)_{\mu\nu}{}^{\dE\dF}e^{\mu}{}_{E\dE}e^{\nu}{}_{F\dF}\right)\nonumber\\
	& &&= \veps^{EF}e^{\mu}{}_{E\dE}e^{\nu}{}_{F\dF}\delta R(\omega)_{\mu\nu}{}^{\dE\dF} + \veps^{EF}\big(-\tfrac{\eps}{2}\Psi_{E\dE}{}^{G\dG}e^{\mu}{}_{G\dG}\big)e^{\nu}{}_{F\dF}R(\omega)_{\mu\nu}{}^{\dE\dF} \nonumber\\
	& &&\quad + \veps^{EF}e^{\mu}{}_{E\dE}\big(-\tfrac{\eps}{2}\Psi_{F\dF}{}^{G\dG}e^{\nu}{}_{G\dG}\big)R(\omega)_{\mu\nu}{}^{\dE\dF}\nonumber\\
	& &&= \veps^{EF}e^{\mu}{}_{E\dE}e^{\nu}{}_{F\dF}\delta R(\omega)_{\mu\nu}{}^{\dE\dF} -\frac{\eps}{2}\veps^{EF}\left[\Psi_{E\dE}{}^{G\dG}R(\omega)_{G\dG,F\dF}{}^{\dE\dF} + \Psi_{F\dF}{}^{G\dG}R(\omega)_{E\dE,G\dG}{}^{\dE\dF}\right]\label{eq:rw 2} ~.
}
Using \eqref{eq:twisted delta omega ASD},
\eqa{
	& e^{\mu}{}_{E\dE}e^{\nu}{}_{F\dF}\delta R(\omega)_{\mu\nu}{}^{\dE\dF} &&= e^{\mu}{}_{E\dE}e^{\nu}{}_{F\dF}\left(\n_{\mu}\delta\omega_{\nu}{}^{\dE\dF} - \n_{\nu}\delta\omega_{\mu}{}^{\dE\dF}\right)\nonumber\\
	& &&= \frac{\eps}{4}\left[\n_{E\dE}\left(\CJ^{\dE\dF}{}_{F\dF} - 3\,\CJ_{FG}{}^{G\dE}\right) - \n_{F\dF}\left(\CJ^{\dE\dF}{}_{E\dE} - 3\,\CJ_{EG}{}^{G\dF}\right)\right] ~.
}
Therefore, the first term in \eqref{eq:rw 2} evaluates to
\eqa{
	&\veps^{EF}e^{\mu}{}_{E\dE}e^{\nu}{}_{F\dF}\delta R(\omega)_{\mu\nu}{}^{\dE\dF} &&= \frac{\eps}{2} \left(\n_{E\dE}{\CJ^{\dE\dF,E}}{}_{\dF} - 3\,\n_{E\dE}\CJ^{E}{}_{F}{}^{F\dE}\right)\label{eq:first term in rw 2} ~.
}
The second term in \eqref{eq:rw 2}, upon a decomposition of curvatures into self- and anti-self- dual parts, is proportional to
\eqa{
	&\veps^{EF}\left[\Psi_{E\dE}{}^{G\dG}R(\omega)_{G\dG,F\dF}{}^{\dE\dF} + \Psi_{F\dF}{}^{G\dG}R(\omega)_{E\dE,G\dG}{}^{\dE\dF}\right] &&= \left(\Psi_{E\dE}{}^{E\dG}R(\omega)_{\dG\dF}{}^{\dE\dF} - \Psi_{E\dE}{}^{G\dG}{R(\omega)_{G}{}^{E,\dE}}_{\dG}\right)\nn
	& &&= \left(\frac{\mathscr{R}}{2}\Psi_{E\dE}{}^{E\dE} - \Psi_{E\dE}{}^{G\dG}{R(\omega)_{G}{}^{E,\dE}}_{\dG}\right)	\label{eq:psiterm2} ~,
}
where we used \eqref{eq:contracted R} in the second equality.

Therefore, using \eqref{eq:first term in rw 1} and \eqref{eq:psiterm1} in \eqref{eq:rw 1}, we get
\eqa{
	&\delta R(\omega)_{EF}{}^{EF} &&=  \frac{\eps}{2}\left(\n_{E\dE}{\CJ^{EF}}_{F}{}^{\dE} - 3\,\n_{E\dE}\CJ^{\dE}{}_{\dF}{}^{E\dF}\right) - \frac{\eps}{2}\left(\frac{\mathscr{R}}{2}\Psi_{E\dE}{}^{E\dE}- \Psi_{E\dE}{}^{G\dG}{R(\omega)_{\dG}{}^{\dE,E}}_{G}\right)	\label{eq:Rw method 1} ~,
}

Similarly, using \eqref{eq:first term in rw 2} and \eqref{eq:psiterm2} in \eqref{eq:rw 2}, we get
\eqa{
      &\delta R(\omega)_{\dE\dF}{}^{\dE\dF} &&= \frac{\eps}{2}\left(\n_{E\dE}{\CJ^{\dE\dF,E}}{}_{\dF} - 3\,\n_{E\dE}\CJ^{E}{}_{F}{}^{F\dE}\right)- \frac{\eps}{2} \left(\frac{\mathscr{R}}{2}\Psi_{E\dE}{}^{E\dE} - \Psi_{E\dE}{}^{G\dG}{R(\omega)_{G}{}^{E,\dE}}_{\dG}\right) \label{eq:Rw method 2} ~.
}
Since the background is torsion-free, we have ${R(\omega)_{\dG}{}^{\dE,E}}_{G} = {R(\omega)_{G}{}^{E,\dE}}_{\dG}$. Therefore, equating \eqref{eq:Rw method 1} and \eqref{eq:Rw method 2} yields the identity,
\eqa{
	&\n_{E\dE}\CJ^{\dE\dF,E}{}_{\dF} &&= \n_{E\dE}{\CJ^{EF}}_{F}{}^{\dE}	\label{eq:questionable identity} ~.
}
Let us pause to prove \eqref{eq:questionable identity}. We refer the reader to the useful decompositions \eqref{eq:contJbig1}--\eqref{eq:diff contJbig}. In particular, from \eqref{eq:diff contJbig}, we find that the difference of the two sides of \eqref{eq:questionable identity} is equal to a sum of four terms, each of which separately vanishes:
\eqa{
  &\n_{E\dE}\big( \CJ^{EF}{}_{F}{}^{\dE} - \CJ^{\dE\dF,E}{}_{\dF}\big) &&= -2\n^{\mu}\big(\n^{\nu}\Psi_{[\nu\mu]}^{+} - \n^{\nu}\Psi_{[\nu\mu]}^{-}\big) \nn
  & &&= [\n^{\mu}, \n^{\nu}]\Psi_{[\mu\nu]}^{+} - [\n^{\mu}, \n^{\nu}]\Psi_{[\mu\nu]}^{-} \nn
  & &&= g^{\mu\rho}g^{\nu\sigma}[\n_{\mu}, \n_{\nu}]\Psi_{[\rho\sigma]}^{+} - g^{\mu\rho}g^{\nu\sigma}[\n_{\mu}, \n_{\nu}]\Psi_{[\rho\sigma]}^{-} \nn
  & &&= g^{\mu\rho}g^{\nu\sigma}R_{\mu\nu\rho}{}^{\kappa}\Psi_{[\kappa\sigma]}^{+} + g^{\mu\rho}g^{\nu\sigma}R_{\mu\nu\sigma}{}^{\kappa}\Psi_{[\rho\kappa]}^{+} \nn
  & &&\quad  - g^{\mu\rho}g^{\nu\sigma}R_{\mu\nu\rho}{}^{\kappa}\Psi_{[\kappa\sigma]}^{-} - g^{\mu\rho}g^{\nu\sigma}R_{\mu\nu\sigma}{}^{\kappa}\Psi_{[\rho\kappa]}^{-}\nn
  & &&= 0 ~.
}
(Each term in the sum is the contraction of the Ricci tensor with an antisymmetric tensor, and thus vanishes identically: $R_{(\mu\nu)}\Psi^{\pm[\mu\nu]} = 0$.) Therefore, using \eqref{eq:questionable identity} in \eqref{eq:Rw method 1}, we get
\eqa{
  &\frac{1}{6}\delta R(\omega)_{EF}{}^{EF} &&= \frac{1}{3}\eps\,\n_{A\dA}{\mathcal{J}^{AE}}_{E}{}^{\dA} - \frac{\mathscr{R}}{24}\eps\,\Psi_{A\dA}{}^{A\dA} + \frac{1}{12} \eps\,R(\omega)^{\dA\dB,AB}\Psi_{A\dA,B\dB} ~. \label{eq:final delta RV}
}

\paragraph{Computation of $\delta \mathscr{D}$.} From \eqref{eq:csugra-susy-auxiliary-scalar-2comp}, the twisted susy transformation law for $\mathscr{D}$ is
\eqa{
   &\delta\mathscr{D} &&= \frac{1}{2}\eps\, \scd_{\mu}\Xi^{\mu} =  \frac{1}{2}\eps\, \scd_{A\dA}\Xi^{A\dA} ~. \label{eq:susy twisted Dscr}
}
To expand the supercovariant divergence of $\Xi^{A\dA}$, we need the form of the supercovariant derivative. From \eqref{eq:csugra-susy-C-right-2comp}, the susy transformation law of $\Xi^{A\dA}$ is
\eqa{
   &\delta \Xi^{A\dA} &&= \frac{\mathscr{R}}{6}\eps^{A\dA} + \frac{1}{3}\eps\,\scd^{A}{}_{\dB}T^{\dA\dB} - \frac{1}{3}R(\omega)^{\dA\dB,AB}\eps_{B\dB} ~, \label{eq:delta CAAdot}
}
where we used $\mathscr{D} = \frac{1}{6}\mathscr{R}$ in the first term. Therefore, the supercovariant derivative on $\Xi^{A\dA}$ is
\eqa{
   &\scd_{E\dE}\Xi^{A\dA} &&= \n_{E\dE}\Xi^{A\dA} - \frac{1}{12}\mathscr{R}\Psi_{E\dE}{}^{A\dA} + \frac{1}{6}R(\omega)^{\dA\dB,AB}\Psi_{E\dE,B\dB} ~,
}
and thus the supercovariant divergence is
\eqa{
   &\scd_{A\dA}\Xi^{A\dA} &&= \n_{A\dA}\Xi^{A\dA} - \frac{1}{12}\mathscr{R}\Psi_{A\dA}{}^{A\dA} + \frac{1}{6}R(V)^{\dA\dB,AB}\Psi_{A\dA,B\dB} ~. \label{eq:supercovariant-divergence-chi-vec}
}
Substituting the form of $\Xi^{A\dA}$ from \eqref{eq:chi vec}, namely $\Xi^{A\dA} = \frac{1}{2}\CJ^{AE}{}_{E}{}^{\dA} + \frac{1}{6}\CJ^{\dA\dE,A}{}_{\dE}$, and using \eqref{eq:supercovariant-divergence-chi-vec} in \eqref{eq:susy twisted Dscr}, we get
\eqa{
&\delta \mathscr{D} &&= \frac{1}{2}\eps\,\n_{A\dA}\left(\frac{1}{2}\CJ^{AE}{}_{E}{}^{\dA} + \frac{1}{6}\CJ^{\dA\dE,A}{}_{\dE}\right) - \frac{\mathscr{R}}{24}\eps\,\Psi_{A\dA}{}^{A\dA} + \frac{1}{12}\eps\,R(\omega)^{\dA\dB,AB}\Psi_{A\dA,B\dB} ~,
}
where we used \eqref{eq:twistcurv} in the last term. However, the first term simplifies upon using \eqref{eq:questionable identity}, and thus one is left with
\eqa{
&\delta \mathscr{D} &&= \frac{1}{3}\eps\,\n_{A\dA}\CJ^{AE}{}_{E}{}^{\dA} - \frac{\mathscr{R}}{24}\eps\,\Psi_{A\dA}{}^{A\dA} + \frac{1}{12}\eps\,R(\omega)^{\dA\dB,AB}\Psi_{A\dA,B\dB} ~, \label{eq:final delta Dscr}
}
which is identical to \eqref{eq:final delta RV}. Therefore,
\eqa{
  & \delta \mathscr{D} &&= \frac{1}{6}\delta \mathscr{R} ~.
}

\subsubsection{The $\Xi_{\mu}$ Constraint}
From \eqref{eq:chi vec}, $\Xi_{μ} = \CJ_{μρ}^{+}{}^{ρ} + \frac{1}{3}\CJ_{μρ}^{-}{}^{ρ}$ or $\Xi_{A\dA} = \frac{1}{2}{\CJ^{E}}_{A,E\dA} + \frac{1}{6}\CJ_{\dA}{}^{\dE}{}_{A\dE}$. We intend to show that this is compatible with the transformation law \eqref{eq:delta CAAdot}. Under vector supersymmetry, the spin-connection terms inside $\n_{\mu}\Psi_{\nu}{}^{C\dC}$ are invariant\footnote{Note that $\n_{\mu}\Psi_{\nu}{}^{C\dC} = \partial_{\mu}\Psi_{\nu}{}^{C\dC} - \Gamma_{\mu\nu}{}^{\rho}\Psi_{\rho}{}^{C\dC} + \frac{1}{2}\omega_{\mu}{}^{C}{}_{D}\Psi_{\nu}{}^{D\dC} + \frac{1}{2}\omega_{\mu}{}^{\dC}{}_{\dD}\Psi_{\nu}{}^{C\dD}$.} and hence
\eqa{
  &\delta_{{\rm vec}}(\,\vec{\eps}\,)\CJ_{\mu\nu}{}^{C\dC} = 2\n_{[\mu}\n_{\nu]}\eps^{C\dC} &&= [\n_{\mu}, \n_{\nu}]\eps^{C\dC} \nn
  & &&= \frac{1}{2}{R(\omega)_{\mu\nu}{}^{C}}_{D}\eps^{D\dC} + \frac{1}{2}{R(\omega)_{\mu\nu}{}^{\dC}}_{\dD}\eps^{C\dD} ~,
}
where we have used \eqref{eq:omega curv sd} and \eqref{eq:omega curv asd}. Since the vielbein does not transform under vector susy, we can directly extract the self- and anti-self- dual parts, and write
\eqa{
 &\delta_{{\rm vec}}(\,\vec{\eps}\,)\CJ_{AB}{}^{C\dC} &&= \frac{1}{2}{R(\omega)_{AB}{}^{C}}_{D}\eps^{D\dC} + \frac{1}{2}{R(\omega)_{AB}{}^{\dC}}_{\dD}\eps^{C\dD} ~,\\
  &\delta_{{\rm vec}}(\,\vec{\eps}\,)\CJ_{\dA\dB}{}^{C\dC} &&= \frac{1}{2}{R(\omega)_{\dA\dB}{}^{C}}_{D}\eps^{D\dC} + \frac{1}{2}{R(\omega)_{\dA\dB}{}^{\dC}}_{\dD}\eps^{C\dD} ~.
}
Using these variations in $\Xi^{A\dA} = \frac{1}{2}\CJ^{EA}{}_{E}{}^{\dA} + \frac{1}{6}\CJ^{\dA\dE,A}{}_{\dE}$, we find
\eqa{
  &\delta_{{\rm vec}}(\,\vec{\eps}\,)\Xi^{A\dA} &&= -\frac{1}{4}{{R(\omega)^{A}}_{C}{}^{C}}_{D}\eps^{D\dA} - \frac{1}{4}{{R(\omega)^{A}}_{C}{}^{\dA}}_{\dC}\eps^{C\dC} - \frac{1}{12}{{R(\omega)^{\dA}}_{\dC}{}^{A}}_{C}\eps^{C\dC} \nn 
  & &&\quad - \frac{1}{12}R(\omega)^{\dA}{}_{\dC}{}^{\dC}{}_{\dD}\eps^{A\dD} ~. \label{eq:delta vector chi vec}
}
Using \eqref{eq:contracted R}, and the exchange property $R(V)^{\dA\dB,AB} = R(\omega)^{\dA\dB,AB} = R(\omega)^{AB,\dA\dB}$, \eqref{eq:delta vector chi vec} reduces to
\eqa{
&\delta_{{\rm vec}}(\,\vec{\eps}\,)\Xi^{A\dA} &&= \frac{\mathscr{R}}{6}\eps^{A\dA} - \frac{1}{3}R(\omega)^{\dA\dB,AB}\eps_{B\dB} ~,
}
which agrees with the vector transformation in \eqref{eq:delta CAAdot}.

\subsection{Closure Of The Twisted Supergravity Algebra}\label{app:supporting-twisted-sugra-closure}
In this section we will use the transformation laws \eqref{eq:CartanSugra-1}--\eqref{eq:CartanSugra-7} to carry out certain computations that are useful for the verification of the closure of the twisted supergravity algebra. We use the supergravity parametrization of matter fields, as used in section \ref{sec:SuperconformalGrav-twistedSYM}. The gravitino is assumed to be symmetric.

We will use the symbols $\sfA$, $\sfB$, $\sfC$, $\sfX$, $\sfY$ with varying index structures as placeholders for different expressions under study, and we hope the use of these symbols locally will cause no confusion with their occurrences in the main text.
\paragraph{\underline{Computation of $\IQ\big(\Psi_{[\mu}{}^{\sigma}\chi_{\nu]\sigma}\big)^{-}$}.} Using \eqref{eq:identA} and \eqref{eq:identB} and defining $\sfX_{\mu\nu} := \Psi_{[\mu}{}^{\sigma}\chi_{\nu]\sigma}$, we have
\eqa{
	&\IQ \big(\Psi_{[\mu}{}^{\sigma}\chi_{\nu]\sigma}\big)^{-} &&= \big[\IQ \big(\Psi_{[\mu}{}^{\sigma}\chi_{\nu]\sigma}\big)\big]^{-} + \tfrac{1}{2}\big(\Psi_{\mu}{}^{\sigma}\sfX_{\nu\sigma}^{+} - \Psi_{\nu}{}^{\sigma}\sfX_{\mu\sigma}^{+}\big)^{-} - \tfrac{1}{2}\big(\Psi_{\mu}{}^{\sigma}\sfX_{\nu\sigma}^{-} - \Psi_{\nu}{}^{\sigma}\sfX_{\mu\sigma}^{-}\big)^{+} \nn
	& &&=  \big[\IQ \big(\Psi_{[\mu}{}^{\sigma}\chi_{\nu]\sigma}\big)\big]^{-} + \tfrac{1}{8}\Psi_{\sigma}{}^{\sigma}\big(\Psi_{\mu}{}^{\sigma}\chi_{\nu\sigma} - \Psi_{\nu}{}^{\sigma}\chi_{\mu\sigma}\big)^{-} \nn
	& &&\quad - \tfrac{1}{2}\big(\Psi_{\mu}{}^{\sigma}\big(\Psi_{[\nu}{}^{\rho}\chi_{\sigma]\rho}\big)^{-} - \Psi_{\nu}{}^{\sigma}\big(\Psi_{[\mu}{}^{\rho}\chi_{\sigma]\rho}\big)^{-}\big)^{+} \nn
	& &&= \big[\IQ \big(\Psi_{[\mu}{}^{\sigma}\chi_{\nu]\sigma}\big)\big]^{-} + \tfrac{1}{4}\Psi_{\sigma}{}^{\sigma}\big(\Psi_{[\mu}{}^{\sigma}\chi_{\nu]\sigma}\big)^{-} \nn
    & &&\quad - \tfrac{1}{2}\big[\Psi_{\mu}{}^{\sigma}\big(\Psi_{[\nu}{}^{\rho}\chi_{\sigma]\rho}\big)^{-} - \Psi_{\nu}{}^{\sigma}\big(\Psi_{[\mu}{}^{\rho}\chi_{\sigma]\rho}\big)^{-}\big]^{+}  ~.\label{eq:H2-33} 
}
Using \eqref{eq:CartanSugra-6},
\eqa{
	&\IQ \big(\Psi_{\mu}{}^{\sigma}\chi_{\nu\sigma}\big) &&= \big(\Psi_{\mu\rho}\Psi^{\rho\sigma} + \n_{\mu}\Phi^{\sigma} + \n^{\sigma}\Phi_{\mu}\big)\chi_{\nu\sigma} - \Psi_{\mu}{}^{\sigma}\IQ \chi_{\nu\sigma} \nn
	& &&= \big(\Psi_{\mu\rho}\Psi^{\rho\sigma} + \n_{\mu}\Phi^{\sigma} + \n^{\sigma}\Phi_{\mu}\big)\chi_{\nu\sigma} +\Psi_{\mu}{}^{\sigma}\big(\Psi_{[\nu}{}^{\rho}\chi_{\sigma]\rho}\big)^{-} + 4\Psi_{\mu}{}^{\sigma}\big(\Phi_{[\nu}D_{\sigma]}\lambda\big)^{+} \nn
	& &&\quad - \Psi_{\mu}{}^{\sigma}\big(F_{\nu\sigma}^{+} - D_{\nu\sigma}^{+}\big) - 2\Psi_{\mu}{}^{\sigma}\big(\n_{[\nu}\Phi_{\sigma]}\big)^{+}\lambda ~.
}
Therefore,
\eqa{
	& \IQ \big(\Psi_{[\mu}{}^{\sigma}\chi_{\nu]\sigma}\big) &&= -\Psi_{[\mu}{}^{\sigma}\big(F_{\nu]\sigma}^{+} - D_{\nu]\sigma}\big)\nn
	& &&\quad  -\Psi^{\rho\sigma}\Psi_{\rho[\mu}\chi_{\nu]\sigma} + \big(\n_{[\mu}\Phi^{\sigma} + \n^{\sigma}\Phi_{[\mu}\big)\chi_{\nu]\sigma} + \tfrac{1}{2}\big[\Psi_{\mu}{}^{\sigma}\big(\Psi_{[\nu}{}^{\rho}\chi_{\sigma]\rho}\big)^{-} - \Psi_{\nu}{}^{\sigma}\big(\Psi_{[\mu}{}^{\rho}\chi_{\sigma]\rho}\big)^{-}  \big] \nn
	& &&\quad + 2 \big[ \Psi_{\mu}{}^{\sigma}\big(\Phi_{[\nu}D_{\sigma]}\lambda\big)^{+} - \Psi_{\nu}{}^{\sigma}\big(\Phi_{[\mu}D_{\sigma]}\lambda\big)^{+}\big] \nn
    & &&\quad -\big[\Psi_{\mu}{}^{\sigma}\big(\n_{[\nu}\Phi_{\sigma]}\big)^{+} - \Psi_{\nu}{}^{\sigma}\big(\n_{[\mu}\Phi_{\sigma]}\big)^{+}\big]\lambda ~.
}
So the first term in \eqref{eq:H2-33} evaluates to
\eqa{
	&\big[\IQ \big(\Psi_{[\mu}{}^{\sigma}\chi_{\nu]\sigma}\big)\big]^{-} &&= -\big(\Psi_{[\mu}{}^{\sigma}F_{\nu]\sigma}^{+}\big)^{-} + \big(\Psi_{[\mu}{}^{\sigma}D_{\nu]\sigma}\big)^{-} \nn
	& &&\quad  -\Psi^{\rho\sigma}\big(\Psi_{\rho[\mu}\chi_{\nu]\sigma}\big)^{-} + \big[\big(\n_{[\mu}\Phi^{\sigma}\big)\chi_{\nu]\sigma}\big]^{-} + \big[\big(\n^{\sigma}\Phi_{[\mu}\big)\chi_{\nu]\sigma}\big]^{-} \nn
	& &&\quad + \dashuline{\tfrac{1}{2}\big[\Psi_{\mu}{}^{\sigma}\big(\Psi_{[\nu}{}^{\rho}\chi_{\sigma]\rho}\big)^{-} - \Psi_{\nu}{}^{\sigma}\big(\Psi_{[\mu}{}^{\rho}\chi_{\sigma]\rho}\big)^{-}  \big]^{-}} \nn
	& &&\quad + 2 \big[ \Psi_{\mu}{}^{\sigma}\big(\Phi_{[\nu}D_{\sigma]}\lambda\big)^{+} - \Psi_{\nu}{}^{\sigma}\big(\Phi_{[\mu}D_{\sigma]}\lambda\big)^{+}\big]^{-} \nn
	& &&\quad -\big[\Psi_{\mu}{}^{\sigma}\big(\n_{[\nu}\Phi_{\sigma]}\big)^{+} - \Psi_{\nu}{}^{\sigma}\big(\n_{[\mu}\Phi_{\sigma]}\big)^{+}\big]^{-}\lambda \nn
	& &&\stackrel{\eqref{eq:identB}}{=\joinrel=\joinrel=} -\big(\Psi_{[\mu}{}^{\sigma}F_{\nu]\sigma}^{+}\big)^{-} + \big(\Psi_{[\mu}{}^{\sigma}D_{\nu]\sigma}\big)^{-} \nn
	& &&\,\,\qquad  -\Psi^{\rho\sigma}\big(\Psi_{\rho[\mu}\chi_{\nu]\sigma}\big)^{-} + \big[\big(\n_{[\mu}\Phi^{\sigma}\big)\chi_{\nu]\sigma}\big]^{-} + \big[\big(\n^{\sigma}\Phi_{[\mu}\big)\chi_{\nu]\sigma}\big]^{-} \nn
	& &&\,\,\qquad -\dashuline{\tfrac{1}{4}\Psi_{\sigma}{}^{\sigma}\big(\Psi_{[\mu}{}^{\rho}\chi_{\nu]\rho}\big)^{-}} + 2 \big[ \Psi_{\mu}{}^{\sigma}\big(\Phi_{[\nu}D_{\sigma]}\lambda\big)^{+} - \Psi_{\nu}{}^{\sigma}\big(\Phi_{[\mu}D_{\sigma]}\lambda\big)^{+}\big]^{-} \nn
	& &&\,\,\qquad -\big[\Psi_{\mu}{}^{\sigma}\big(\n_{[\nu}\Phi_{\sigma]}\big)^{+} - \Psi_{\nu}{}^{\sigma}\big(\n_{[\mu}\Phi_{\sigma]}\big)^{+}\big]^{-}\lambda ~.
}
Using this in \eqref{eq:H2-33}, we get
\eqa{
&\IQ \big(\Psi_{[\mu}{}^{\sigma}\chi_{\nu]\sigma}\big)^{-} &&=  -\big(\Psi_{[\mu}{}^{\sigma}F_{\nu]\sigma}^{+}\big)^{-} + \big(\Psi_{[\mu}{}^{\sigma}D_{\nu]\sigma}\big)^{-}\nonumber\\
	& &&\qquad  -\Psi^{\rho\sigma}\big(\Psi_{\rho[\mu}\chi_{\nu]\sigma}\big)^{-} + \big[\big(\n_{[\mu}\Phi^{\sigma}\big)\chi_{\nu]\sigma}\big]^{-} + \big[\big(\n^{\sigma}\Phi_{[\mu}\big)\chi_{\nu]\sigma}\big]^{-} \nonumber\\
	& &&\qquad -\dashuline{\tfrac{1}{4}\Psi_{\sigma}{}^{\sigma}\big(\Psi_{[\mu}{}^{\rho}\chi_{\nu]\rho}\big)^{-}} + 2 \big[ \Psi_{\mu}{}^{\sigma}\big(\Phi_{[\nu}D_{\sigma]}\lambda\big)^{+} - \Psi_{\nu}{}^{\sigma}\big(\Phi_{[\mu}D_{\sigma]}\lambda\big)^{+}\big]^{-} \nonumber\\
	& &&\qquad -\big[\Psi_{\mu}{}^{\sigma}\big(\n_{[\nu}\Phi_{\sigma]}\big)^{+} - \Psi_{\nu}{}^{\sigma}\big(\n_{[\mu}\Phi_{\sigma]}\big)^{+}\big]^{-}\lambda \nonumber\\
	& &&\qquad + \dashuline{\tfrac{1}{4}\Psi_{\sigma}{}^{\sigma}\big(\Psi_{[\mu}{}^{\sigma}\chi_{\nu]\sigma}\big)^{-}}  - \tfrac{1}{2}\big[\Psi_{\mu}{}^{\sigma}\big(\Psi_{[\nu}{}^{\rho}\chi_{\sigma]\rho}\big)^{-} - \Psi_{\nu}{}^{\sigma}\big(\Psi_{[\mu}{}^{\rho}\chi_{\sigma]\rho}\big)^{-}\big]^{+}  \nonumber \\
%
%
	& &&= -\big(\Psi_{[\mu}{}^{\sigma}F_{\nu]\sigma}^{+}\big)^{-} + \big(\Psi_{[\mu}{}^{\sigma}D_{\nu]\sigma}\big)^{-} - \tfrac{1}{2}\big[\Psi_{\mu}{}^{\sigma}\big(\Psi_{[\nu}{}^{\rho}\chi_{\sigma]\rho}\big)^{-} - \Psi_{\nu}{}^{\sigma}\big(\Psi_{[\mu}{}^{\rho}\chi_{\sigma]\rho}\big)^{-}\big]^{+} \nn
	& &&\quad + \big[\big(\n_{[\mu}\Phi^{\sigma}\big)\chi_{\nu]\sigma}\big]^{-} + \big[\big(\n^{\sigma}\Phi_{[\mu}\big)\chi_{\nu]\sigma}\big]^{-} + 2 \big[ \Psi_{\mu}{}^{\sigma}\big(\Phi_{[\nu}D_{\sigma]}\lambda\big)^{+} - \Psi_{\nu}{}^{\sigma}\big(\Phi_{[\mu}D_{\sigma]}\lambda\big)^{+}\big]^{-} \nn
	& &&\quad -\big[\Psi_{\mu}{}^{\sigma}\big(\n_{[\nu}\Phi_{\sigma]}\big)^{+} - \Psi_{\nu}{}^{\sigma}\big(\n_{[\mu}\Phi_{\sigma]}\big)^{+}\big]^{-}\lambda ~. \label{eq:H2-37} 
}

\paragraph{\underline{Computation of $\IQ \big(\mathbf{\nabla}_{[\mu}\Phi_{\nu]}\big)^{\pm}$}.}
The affine connection and its variation, being symmetric, do not contribute to the variation of the curl:
\eqa{
&\IQ \big(\n_{[\mu}\Phi_{\nu]}\big) &&= \big(\n_{[\mu}\Phi^{\rho}\big)\Psi_{\nu]\rho} + \Phi^{\rho}\n_{[\mu}\Psi_{\nu]\rho}  ~.
}
Therefore, from \eqref{eq:identA refined},
\eqa{
&\IQ  \big(\n_{[\mu}\Phi_{\nu]}\big)^{+} &&= \big(\IQ \n_{[\mu}\Phi_{\nu]}\big)^{+} - \tfrac{1}{2}\big(\Psi_{\mu}{}^{\rho}\big(\n_{[\nu}\Phi_{\rho]}\big)^{+} - \Psi_{\nu}{}^{\rho}\big(\n_{[\mu}\Phi_{\rho]}\big)^{+}\big)^{-}\nonumber\\
& &&\quad  + \tfrac{1}{2}\big(\Psi_{\mu}{}^{\rho}\big(\n_{[\nu}\Phi_{\rho]}\big)^{-} - \Psi_{\nu}{}^{\rho}\big(\n_{[\mu}\Phi_{\rho]}\big)^{-}\big)^{+} \nonumber\\
& &&= \big[\big(\n_{[\mu}\Phi^{\rho}\big)\Psi_{\nu]\rho}\big]^{+} + \Phi^{\rho}\big[\n_{[\mu}\Psi_{\nu]\rho}\big]^{+} - \tfrac{1}{2}\big[\Psi_{\mu}{}^{\rho}\big(\n_{[\nu}\Phi_{\rho]}\big)^{+} - \Psi_{\nu}{}^{\rho}\big(\n_{[\mu}\Phi_{\rho]}\big)^{+}\big]^{-}\nonumber\\
& &&\quad  + \tfrac{1}{2}\big[\Psi_{\mu}{}^{\rho}\big(\n_{[\nu}\Phi_{\rho]}\big)^{-} - \Psi_{\nu}{}^{\rho}\big(\n_{[\mu}\Phi_{\rho]}\big)^{-}\big]^{+} ~, \label{eq:dt curl phi SD}
}
and
\eqa{
&\IQ  \big(\n_{[\mu}\Phi_{\nu]}\big)^{-} &&= \big(\IQ \n_{[\mu}\Phi_{\nu]}\big)^{-} + \tfrac{1}{2}\big(\Psi_{\mu}{}^{\rho}\big(\n_{[\nu}\Phi_{\rho]}\big)^{+} - \Psi_{\nu}{}^{\rho}\big(\n_{[\mu}\Phi_{\rho]}\big)^{+}\big)^{-}\nonumber\\
& &&\quad  - \tfrac{1}{2}\big(\Psi_{\mu}{}^{\rho}\big(\n_{[\nu}\Phi_{\rho]}\big)^{-} - \Psi_{\nu}{}^{\rho}\big(\n_{[\mu}\Phi_{\rho]}\big)^{-}\big)^{+} \nonumber\\
& &&= \big[\big(\n_{[\mu}\Phi^{\rho}\big)\Psi_{\nu]\rho}\big]^{-} + \Phi^{\rho}\big[\n_{[\mu}\Psi_{\nu]\rho}\big]^{-} + \tfrac{1}{2}\big[\Psi_{\mu}{}^{\rho}\big(\n_{[\nu}\Phi_{\rho]}\big)^{+} - \Psi_{\nu}{}^{\rho}\big(\n_{[\mu}\Phi_{\rho]}\big)^{+}\big]^{-}\nonumber\\
& &&\quad  - \tfrac{1}{2}\big[\Psi_{\mu}{}^{\rho}\big(\n_{[\nu}\Phi_{\rho]}\big)^{-} - \Psi_{\nu}{}^{\rho}\big(\n_{[\mu}\Phi_{\rho]}\big)^{-}\big]^{+} ~. \label{eq:dt curl phi ASD}
}

\paragraph{\underline{Computation of $\IQ \big(\Phi_{[\mu}D_{\nu]}\lambda\big)^{\pm}$.}}
Using \eqref{eq:CartanSugra-3},
\eqa{
	&\IQ \big(\Phi_{\mu}D_{\nu}\lambda\big) &&= \Phi^{\rho}\Psi_{\rho\mu}D_{\nu}\lambda + \Phi_{\mu}D_{\nu}\eta + \Phi_{\mu}[\IQ A_{\nu},\lambda] \nonumber\\
	& &&= \Phi^{\rho}\Psi_{\rho\mu}D_{\nu}\lambda + \Phi_{\mu}D_{\nu}\eta + \Phi_{\mu}[\psi_{\nu},\lambda] +  \Phi_{\mu}\Phi^{\rho}[\chi_{\rho\nu},\lambda] - \Phi_{\mu}\Phi_{\nu}[\eta,\lambda] ~.
}
Therefore,
\eqa{
  &\IQ \big(\Phi_{[\mu}D_{\nu]}\lambda\big) &&= \Phi^{\rho}\Psi_{\rho[\mu}D_{\nu]}\lambda + \Phi_{[\mu}D_{\nu]}\eta  + \Phi_{[\mu}[\psi_{\nu]}, \lambda] - \Phi^{\rho}\Phi_{[\mu}[\chi_{\nu]\rho},\lambda] ~.
}
Using \eqref{eq:identA refined},
\eqa{
	&\IQ \big(\Phi_{[\mu}D_{\nu]}\lambda\big)^{+} &&= +\big(\Phi_{[\mu}[\psi_{\nu]},\lambda]\big)^{+} + \big(\Phi_{[\mu}D_{\nu]}\eta\big)^{+} \nonumber\\
	& &&\quad +\tfrac{1}{2}\big(\Psi_{\mu}{}^{\rho}\big(\Phi_{[\nu}D_{\rho]}\lambda\big)^{-} - \Psi_{\nu}{}^{\rho}\big(\Phi_{[\mu}D_{\rho]}\lambda\big)^{-}\big)^{+}  -\tfrac{1}{2}\big(\Psi_{\mu}{}^{\rho}\big(\Phi_{[\nu}D_{\rho]}\lambda\big)^{+} - \Psi_{\nu}{}^{\rho}\big(\Phi_{[\mu}D_{\rho]}\lambda\big)^{+}\big)^{-} \nonumber\\
	& &&\quad + \Phi^{\rho}\big(\Psi_{\rho[\mu}D_{\nu]}\lambda\big)^{+} - \Phi^{\rho}\big(\Phi_{[\mu}[\chi_{\nu]\rho},\lambda]\big)^{+} ~, \label{eq:dt Phi del lambda SD}
}
and similarly,
\eqa{
	&\IQ \big(\Phi_{[\mu}D_{\nu]}\lambda\big)^{-} &&= +\big(\Phi_{[\mu}[\psi_{\nu]},\lambda]\big)^{-} + \big(\Phi_{[\mu}D_{\nu]}\eta\big)^{-} \nonumber\\
	& &&\quad -\tfrac{1}{2}\big(\Psi_{\mu}{}^{\rho}\big(\Phi_{[\nu}D_{\rho]}\lambda\big)^{-} - \Psi_{\nu}{}^{\rho}\big(\Phi_{[\mu}D_{\rho}\lambda\big)^{-}\big)^{+} +\tfrac{1}{2}\big(\Psi_{\mu}{}^{\rho}\big(\Phi_{[\nu}D_{\rho]}\lambda\big)^{+} - \Psi_{\nu}{}^{\rho}\big(\Phi_{[\mu}D_{\rho]}\lambda\big)^{+}\big)^{-} \nonumber\\
	& &&\quad + \Phi^{\rho}\big(\Psi_{\rho[\mu}D_{\nu]}\lambda\big)^{-} - \Phi^{\rho}\big(\Phi_{[\mu}[\chi_{\nu]\rho},\lambda]\big)^{-} ~. \label{eq:dt Phi del lambda ASD}
}

\paragraph{\underline{Computation of $\IQ F_{\mu\nu}^{\pm}$.}} The $\IQ$-variations of the self- and anti-self-dual components of the SYM field strength, computed using \eqref{eq:CartanSugra-3} are:
\eqa{
	&\IQ F_{\mu\nu}^{+} &&= 2\big(D_{[\mu}\psi_{\nu]}\big)^{+} - \big(\Psi_{[\mu}{}^{\sigma}F_{\nu]\sigma}^{+}\big)^{-} + \big(\Psi_{[\mu}{}^{\sigma}F_{\nu]\sigma}^{-}\big)^{+} \nonumber\\
	& &&\quad -2\big[\big(\n_{[\mu}\Phi^{\rho}\big)\chi_{\nu]\rho}\big]^{+} -2 \Phi^{\rho}\big(D_{[\mu}\chi_{\nu]\rho}\big)^{+} - 2\big(\n_{[\mu}\Phi_{\nu]}\big)^{+}\eta \nonumber\\
	& &&\quad + 2\big(\Phi_{[\mu}D_{\nu]}\eta\big)^{+} -2 \big[\big(\n_{[\mu}\Phi^{\rho}\big)\Psi_{\nu]\rho}\big]^{+}\lambda -2\Phi^{\rho}\big(\n_{[\mu}\Psi_{\nu]\rho}\big)^{+}\lambda + 2\Phi^{\rho}\big[\Psi_{\rho[\mu}D_{\nu]}\lambda\big]^{+} ~. \label{eq:tsugra-sym-grav-YM-Fmunu-SD} \\
	&\IQ F_{\mu\nu}^{-} &&= 2\big(D_{[\mu}\psi_{\nu]}\big)^{-} + \big(\Psi_{[\mu}{}^{\sigma}F_{\nu]\sigma}^{+}\big)^{-} - \big(\Psi_{[\mu}{}^{\sigma}F_{\nu]\lambda}^{-}\big)^{+} \nonumber\\
	& &&\quad -2\big[\big(\n_{[\mu}\Phi^{\rho}\big)\chi_{\nu]\rho}\big]^{-} -2 \Phi^{\rho}\big(D_{[\mu}\chi_{\nu]\rho}\big)^{-} - 2\big(\n_{[\mu}\Phi_{\nu]}\big)^{-}\eta \nonumber\\
	& &&\quad + 2\big(\Phi_{[\mu}D_{\nu]}\eta\big)^{-} -2 \big[\big(\n_{[\mu}\Phi^{\rho}\big)\Psi_{\nu]\rho}\big]^{-}\lambda -2\Phi^{\rho}\big(\n_{[\mu}\Psi_{\nu]\rho}\big)^{-}\lambda + 2\Phi^{\rho}\big[\Psi_{\rho[\mu}D_{\nu]}\lambda\big]^{-} ~.  \label{eq:tsugra-sym-grav-YM-Fmunu-ASD}
}

\paragraph{\underline{Computation of $\IQ \big(F_{\mu\nu}^{\pm} \mp D_{\mu\nu}\big)$.}}
Using \eqref{eq:tsugra-sym-grav-YM-Fmunu-SD}, \eqref{eq:tsugra-sym-grav-YM-Fmunu-ASD}, and \eqref{eq:CartanSugra-7},
\eqa{
	&\IQ \big(F_{\mu\nu}^{+} - D_{\mu\nu}\big) && = +  [\phi,\chi_{\mu\nu}] - 4\big(\Phi_{[\mu}[\lambda,\psi_{\nu]}]\big)^{+}  -\big(\Psi_{[\mu}{}^{\rho}F_{\nu]\rho}^{+}\big)^{-} + \big(\Psi_{[\mu}{}^{\rho}D_{\nu]\rho}\big)^{-} -2\big[\big(\n_{[\mu}\Phi^{\rho}\big)\chi_{\nu]\rho}\big]^{+} \nonumber\\
    & &&\quad - 2\Phi^{\rho}\big(D_{[\mu}\chi_{\nu]\rho}\big)^{+} - 2\big(\Phi_{[\mu}D^{\rho}\chi_{\nu]\rho}\big)^{+} - 2\big(\n_{[\mu}\Phi_{\nu]}\big)^{+}\eta  + 4\big(\Phi_{[\mu}D_{\nu]}\eta\big)^{+} \nonumber\\
	& &&\quad + \tfrac{1}{2}\big[\Psi_{\mu}{}^{\rho}\big(\Psi_{[\rho}{}^{\sigma}\chi_{\nu]\sigma}\big)^{-} - \Psi_{\nu}{}^{\rho}\big(\Psi_{[\rho}{}^{\sigma}\chi_{\mu]\sigma}\big)^{-}\big]^{+} \nonumber\\
	& &&\quad -2\big[\big(\n_{[\mu}\Phi^{\rho}\big)\Psi_{\nu]\rho}\big]^{+}\lambda + \big[\Psi_{\mu}{}^{\rho}\big(\n_{[\rho}\Phi_{\nu]}\big)^{-}\lambda - \Psi_{\nu}{}^{\rho}\big(\n_{[\rho}\Phi_{\mu]}\big)^{-}\lambda\big]^{+} - 2\Phi^{\rho}\big(\n_{[\mu}\Psi_{\nu]\rho}\big)^{+}\lambda \nonumber\\
	& &&\quad + 2\Phi^{\rho}\big[\Psi_{\rho[\mu}D_{\nu]}\lambda\big]^{+}  - 2\big(\Phi_{[\mu}\Psi_{\nu]}{}^{\rho}\big)^{+}D_{\rho}\lambda + \Psi_{\rho}{}^{\rho}\big(\Phi_{[\mu}D_{\nu]}\lambda\big)^{+} \label{eq:dt Fplus minus D} ~.\\
	&\IQ \big(F_{\mu\nu}^{-} + D_{\mu\nu}\big) &&= 2\big(D_{[\mu}\psi_{\nu]}\big) + \big(\Psi_{[\mu}{}^{\sigma}F_{\nu]\sigma}^{+}\big)^{-} - \big(\Psi_{[\mu}{}^{\rho}D_{\nu]\rho}\big)^{-}  -2\big[\big(\n_{[\mu}\Phi^{\rho}\big)\chi_{\nu]\rho}\big]^{-} -2 \Phi^{\rho}\big(D_{[\mu}\chi_{\nu]\rho}\big)^{-} \nn
    & &&\quad + 2\big(\Phi_{[\mu}D^{\rho}\chi_{\nu]\rho}\big)^{+}  - 2\big(\n_{[\mu}\Phi_{\nu]}\big)^{-}\eta + 2\big(\Phi_{[\mu}D_{\nu]}\eta\big)^{-} - 2\big(\Phi_{[\mu}D_{\nu]}\eta\big)^{+} \nn
	& &&\quad -2 \big[\big(\n_{[\mu}\Phi^{\rho}\big)\Psi_{\nu]\rho}\big]^{-}\lambda -\big[\Psi_{\mu}{}^{\rho}\big(\n_{[\rho}\Phi_{\nu]}\big)^{-}\lambda - \Psi_{\nu}{}^{\rho}\big(\n_{[\rho}\Phi_{\mu]}\big)^{-}\lambda\big]^{+} -2\Phi^{\rho}\big(\n_{[\mu}\Psi_{\nu]\rho}\big)^{-}\lambda \nn
	& &&\quad + 2\Phi^{\rho}\big[\Psi_{\rho[\mu}D_{\nu]}\lambda\big]^{-}  + 2\big(\Phi_{[\mu}\Psi_{\nu]}{}^{\rho}\big)^{+}D_{\rho}\lambda - \Psi_{\rho}{}^{\rho}\big(\Phi_{[\mu}D_{\nu]}\lambda\big)^{+} \nn
	& &&\quad  - \tfrac{1}{2}\big[\Psi_{\mu}{}^{\rho}\big(\Psi_{[\rho}{}^{\sigma}\chi_{\nu]\sigma}\big)^{-} - \Psi_{\nu}{}^{\rho}\big(\Psi_{[\rho}{}^{\sigma}\chi_{\mu]\sigma}\big)^{-}\big]^{+} \nn
	& &&\quad -  [\phi,\chi_{\mu\nu}] + 4\big(\Phi_{[\mu}[\lambda,\psi_{\nu]}]\big)^{+} ~. \label{eq:dt Fminus plus D}
}

\subsubsection{\label{sec:proof of vanishing of del chi terms in dt sq psimu}Cancellation Of $D\chi$ Terms In $\IQ^2\psi_{\mu}$}
In this section, we prove the identity,
\eqa{
  &-2\Phi^{\mu}\Phi^{\rho}\big(D_{[\mu}\chi_{\nu]\rho}\big)^{-}  +2\Phi^{\mu}\big(\Phi_{[\mu}D^{\rho}\chi_{\nu]\rho}\big)^{+} &&= 0 ~,
}
which is crucial for demonstrating the cancellation of $D\chi$ terms in $\IQ^2\psi_{\mu}$.
\\\\
(Caveat: Note that $+2\Phi^{\mu}\big(\Phi_{[\mu}D^{\rho}\chi_{\nu]\rho}\big)^{+} = +\Phi^{\mu}\Phi_{[\mu}D^{\rho}\chi_{\nu]\rho}$ because of the identity \eqref{eq:contracted SD equals ASD}.)

Let $\sfA_{\mu\nu} = \Phi^{\rho}D_{[\mu}\chi_{\nu]\rho}$. In two-component notation, 
\eqa{
& \sfA_{A\dA,B\dB} &&= \tfrac{1}{4}\big(-\Phi^{C}{}_{\dB}D_{A\dA}\chi_{BC} + \Phi^{C}{}_{\dA}D_{B\dB}\chi_{AC}\big) ~,\\
& \sfA_{\dA\dB}    &&= \varepsilon^{AB}\sfA_{A\dA,B\dB} =  \tfrac{1}{4}\big(\Phi_{C\dB}D_{A\dA}\chi^{AC} + \Phi_{C\dA}D_{A\dB}\chi^{AC}\big) ~.
}
Therefore, 
\eqa{
 &-2\Phi^{\mu}\Phi^{\rho}\big(D_{[\mu}\chi_{\nu]\rho}\big)^{-} &&= -2\Phi^{A\dA}\big(\tfrac{1}{2}\sfA_{\dA\dB}\varepsilon_{AB}\big)e_{\nu}{}^{B\dB} = -\Phi_{B}{}^{\dA}\sfA_{\dA\dB}e_{\nu}{}^{B\dB} \nonumber\\
  & &&= -\tfrac{1}{4}\big(\Phi_{B}{}^{\dA}\Phi_{C\dB}D_{E\dA}\chi^{EC} + \Phi_{B}{}^{\dA}\Phi_{C\dA}D_{E\dB}\chi^{EC}\big)e_{\nu}{}^{B\dB} ~. \label{eq:term 1}
}
Using the identities $\Phi_{B}{}^{\dA}\Phi_{C\dA} = \Phi_{C}{}^{\dA}\Phi_{B\dA} - \varepsilon_{BC}\Phi^2$ and $\Phi_{C}{}^{\dA}\Phi_{B\dA} = \frac{1}{2}\varepsilon_{BC}\Phi^2$, \eqref{eq:term 1} reads
\beqa{
&-2\Phi^{\mu}\Phi^{\rho}\big(D_{[\mu}\chi_{\nu]\rho}\big)^{-} &&= -\tfrac{1}{4}\big(\Phi_{B}{}^{\dA}\Phi_{C\dB} D_{E\dA}\chi^{EC} + \tfrac{1}{2}\Phi^2 D_{E\dB}\chi^{E}{}_{B} \big)e_{\nu}{}^{B\dB} ~. \label{eq:term 1 final}
}
Next, define $\sfB_{\mu\nu} = \Phi_{[\mu}\n^{\rho}\chi_{\nu]\rho}$. In two-component notation,
\eqa{
 &  \sfB_{A\dA,B\dB} &&= \tfrac{1}{4}\big(-\Phi_{A\dA}D^{C}{}_{\dB}\chi_{BC} + \Phi_{B\dB}D^{C}{}_{\dA}\chi_{AC}\big)e_{\nu}{}^{B\dB} ~,\\
 &  \sfB_{AB}        &&= \varepsilon^{\dA\dB}\sfB_{A\dA,B\dB} = \tfrac{1}{4}\big(-\Phi_{A\dA}D^{C\dA}\chi_{BC} - \Phi_{B\dA}D^{C\dA}\chi_{AC}\big) ~. 
}
Therefore,
\eqa{
& +2\Phi^{\mu}\big(\Phi_{[\mu}\n^{\rho}\chi_{\nu]\rho}\big)^{+} &&= +2\Phi^{A\dA}\big(\tfrac{1}{2}\sfB_{AB}\varepsilon_{\dA\dB}\big)e_{\nu}{}^{B\dB} = \Phi^{A}{}_{\dB}\sfB_{AB}e_{\nu}{}^{B\dB} \nonumber\\
& &&= -\tfrac{1}{4}\big(-\Phi^{A}{}_{\dB}\Phi_{A\dA}D^{C\dA}\chi_{BC} + \Phi^{A}{}_{\dB}\Phi_{B\dA}D^{C\dA}\chi_{AC}\big)e_{\nu}{}^{B\dB} ~. \label{eq:term 2}
}
Using the identities $\Phi^{A}{}_{\dB}\Phi_{A\dA} = -\Phi_{A\dA}\Phi^{A}{}_{\dB} + \varepsilon_{\dA\dB}\Phi^2$ and $\Phi_{A\dA}\Phi^{A}{}_{\dB} = \frac{1}{2}\varepsilon_{\dA\dB}\Phi^2$, \eqref{eq:term 2} is
\eqa{
&+2\Phi^{\mu}\big(\Phi_{[\mu}D^{\rho}\chi_{\nu]\rho}\big)^{+} &&= -\tfrac{1}{4}\big(\Phi^{A}{}_{\dB}\Phi_{B\dA}D^{C\dA}\chi_{AC} + \tfrac{1}{2}\Phi^2 D^{C}{}_{\dB}\chi_{BC}\big)e_{\nu}{}^{B\dB} \nonumber\\
& &&= -\tfrac{1}{4}\big(-\Phi_{B}{}^{\dA}\Phi_{C\dB}D_{E\dA}\chi^{CE} - \tfrac{1}{2}\Phi^2 D_{E\dB}\chi^{E}{}_{B}\big)e_{\nu}{}^{B\dB} ~.\nonumber
}
That is,
\beqa{
	+2\Phi^{\mu}\big(\Phi_{[\mu}D^{\rho}\chi_{\nu]\rho}\big)^{+} &= -\tfrac{1}{4}\big(-\Phi_{B}{}^{\dA}\Phi_{C\dB}D_{E\dA}\chi^{CE} - \tfrac{1}{2}\Phi^2 D_{E\dB}\chi^{E}{}_{B}\big)e_{\nu}{}^{B\dB} ~. \label{eq:term 2 final}
 }
From \eqref{eq:term 1 final} and \eqref{eq:term 2 final}, it follows that
\beqa{
 -2\Phi^{\mu}\Phi^{\rho}\big(D_{[\mu}\chi_{\nu]\rho}\big)^{-} +2\Phi^{\mu}\big(\Phi_{[\mu}D^{\rho}\chi_{\nu]\rho}\big)^{+} &= 0 ~. 
 }

\subsubsection{\label{sec:proof of vanishing of del lambda terms in dt sq psimu}Cancellation Of $D\lambda$ Terms In $\IQ^2\psi_{\mu}$}
The terms involving a derivative of $\lambda$ in $\IQ^2\psi_{\mu}$ are
\eqa{
& \sfX_{\mu} &&= 2\Phi^{\sigma}\Phi^{\rho}\big[\Psi_{\rho[\sigma}D_{\mu]}\lambda\big]^{-} + 2\Phi^{\sigma}\big(\Phi_{[\sigma}\Psi_{\mu]}{}^{\rho}\big)^{+}D_{\rho}\lambda - \Phi^{\sigma}\Psi_{\rho}{}^{\rho}\big(\Phi_{[\sigma}D_{\mu]}\lambda\big)^{+} \nonumber\\
& &&\quad + 2\Phi^{\sigma}\big[\Psi_{\sigma}{}^{\rho}\big(\Phi_{[\mu}D_{\rho]}\lambda\big)^{+} - \Psi_{\mu}{}^{\rho}\big(\Phi_{[\sigma}D_{\rho]}\lambda\big)^{+}\big]^{-} ~.
}
In this section, we will prove that $\sfX_{\mu} = 0$.  First, note that
\eqa{
	&2\Phi^{\sigma}\big[\Psi_{\sigma}{}^{\rho}\big(\Phi_{[\mu}D_{\rho]}\lambda\big)^{+} - \Psi_{\mu}{}^{\rho}\big(\Phi_{[\sigma}D_{\rho]}\lambda\big)^{+}\big]^{-}\nonumber\\
	&= 2\Phi^{\sigma}\big[\Psi_{\sigma}{}^{\rho}\big(\Phi_{[\mu}D_{\rho]}\lambda\big)^{+} - \Psi_{\mu}{}^{\rho}\big(\Phi_{[\sigma}D_{\rho]}\lambda\big)^{+}\big] - 2\Phi^{\sigma}\big[\Psi_{\sigma}{}^{\rho}\big(\Phi_{[\mu}D_{\rho]}\lambda\big)^{+} - \Psi_{\mu}{}^{\rho}\big(\Phi_{[\sigma}D_{\rho]}\lambda\big)^{+}\big]^{+}\nonumber\\
	&= 2\Phi^{\sigma}\big[\Psi_{\sigma}{}^{\rho}\big(\Phi_{[\mu}D_{\rho]}\lambda\big)^{+} - \Psi_{\mu}{}^{\rho}\big(\Phi_{[\sigma}D_{\rho]}\lambda\big)^{+}\big] + \Phi^{\sigma}\Psi_{\rho}{}^{\rho}\big(\Phi_{[\sigma}D_{\mu]}\lambda\big)^{+} ~.
}
Therefore,
\eqa{
	&\sfX_{\mu} &&= \underbrace{2\Phi^{\sigma}\Phi^{\rho}\big[\Psi_{\rho[\sigma}D_{\mu]}\lambda\big]^{-}}_{\sfX_{\mu}^{(1)}} + \underbrace{2\Phi^{\sigma}\big(\Phi_{[\sigma}\Psi_{\mu]}{}^{\rho}\big)^{+}D_{\rho}\lambda}_{\sfX_{\mu}^{(2)}}  + \underbrace{2\Phi^{\sigma}\big[\Psi_{\sigma}{}^{\rho}\big(\Phi_{[\mu}D_{\rho]}\lambda\big)^{+} - \Psi_{\mu}{}^{\rho}\big(\Phi_{[\sigma}D_{\rho]}\lambda\big)^{+}\big]}_{\sfX_{\mu}^{(3)}} \nonumber\\
 & &&\equiv \sfX_{\mu}^{(1)} +  \sfX_{\mu}^{(2)} +  \sfX_{\mu}^{(3)} ~.
}
We would like to show that $\sfX_{\mu} = 0$. Let us begin with $\sfX_{\mu}^{(1)} \equiv 2\Phi^{\sigma}\Phi^{\rho}\big[\Psi_{\rho[\sigma}D_{\mu]}\lambda\big]^{-}$. Define
\eqa{
	& \sfA_{\sigma\mu} &&:= \Phi^{\rho}\Psi_{\rho[\sigma}D_{\mu]}\lambda ~.
}
In two-component notation,
\eqa{
  &\sfA_{A\dA,B\dB} &&= \tfrac{1}{2}	\Phi^{C\dC}\big(\Psi_{C\dC,A\dA}D_{B\dB}\lambda - \Psi_{C\dC,B\dB}D_{A\dA}\lambda\big) ~,\\
  &\sfA_{\dA\dB}  &&= \tfrac{1}{2}\Phi^{C\dC}\big(\Psi_{C\dC,A\dA}D^{A}{}_{\dB}\lambda + \Psi_{C\dC,A\dB}D^{A}{}_{\dA}\lambda\big) ~.
}
Therefore,
\eqa{
&\sfX_{\mu}^{(1)} &&= 2\Phi^{A\dA}\big(\tfrac{1}{2}\sfA_{\dA\dB}\varepsilon_{AB}\big)e_{\mu}{}^{B\dB} = \Phi_{B}{}^{\dA}\sfA_{\dA\dB}e_{\mu}{}^{B\dB}  \nonumber\\
& &&= \tfrac{1}{2}\big(\Phi_{B}{}^{\dA}\Phi^{C\dC}\Psi_{C\dC,A\dA}D^{A}{}_{\dB}\lambda + \Phi_{B}{}^{\dA}\Phi^{C\dC}\Psi_{C\dC,A\dB}D^{A}{}_{\dA}\lambda\big)e_{\mu}{}^{B\dB} ~. \label{eq:X nu 1 intermediate}
}
The second term of \eqref{eq:X nu 1 intermediate} can be written as
\eqa{
	&\Phi_{B}{}^{\dA}\Phi^{C\dC}\Psi_{C\dC,A\dB}D^{A}{}_{\dA}\lambda &&= \Phi_{B}{}^{\dA}\Phi^{C\dC}\Psi_{C\dC,A\dA}D^{A}{}_{\dB}\lambda - \Phi_{B\dB}\Phi^{C\dC}\Psi_{C\dC,E\dE}D^{E\dE}\lambda ~.
}
Therefore, \eqref{eq:X nu 1 intermediate} becomes
\eqa{
	&\sfX_{\mu}^{(1)} &&= \tfrac{1}{2}\big(2 \Phi_{B}{}^{\dA}\Phi^{C\dC}\Psi_{C\dC,A\dA}D^{A}{}_{\dB}\lambda - \Phi_{B\dB}\Phi^{C\dC}\Psi_{C\dC,E\dE}D^{E\dE}\lambda\big)e_{\mu}{}^{B\dB} \nonumber\\
	& &&= \tfrac{1}{2}\big(2 \Phi_{B\dA}\Phi_{C\dC}\Psi^{A\dA,C\dC}D_{A\dB}\lambda - \Phi_{A\dA}\Phi_{B\dB}\Psi^{A\dA,C\dC}D_{C\dC}\lambda\big)e_{\mu}{}^{B\dB} ~. \label{eq:X nu 1 final}
}
Next, consider $\sfX_{\mu}^{(2)} \equiv 2\Phi^{\sigma}\big(\Phi_{[\sigma}\Psi_{\mu]}{}^{\rho}\big)^{+}D_{\rho}\lambda$. Define 
\eqa{
	&\sfB_{\sigma\mu} &&:= \Phi_{[\sigma}\Psi_{\mu]}{}^{\rho}D_\rho\lambda ~.
}
In two-component notation,
\eqa{
	&\sfB_{A\dA,B\dB} &&= \tfrac{1}{2}\big(\Phi_{A\dA}\Psi_{B\dB}{}^{C\dC} - \Phi_{B\dB}\Psi_{A\dA}{}^{C\dC}\big)D_{C\dC}\lambda ~,\\
	&\sfB_{AB}        &&= \tfrac{1}{2}\big(\Phi_{A\dA}\Psi_{B}{}^{\dA,C\dC} + \Phi_{B\dA}\Psi_{A}{}^{\dA,C\dC}\big)D_{C\dC}\lambda ~.
}
Therefore,
\eqa{
	&\sfX_{\mu}^{(2)} &&= 2\Phi^{A\dA}\big(\tfrac{1}{2}\sfB_{AB}\varepsilon_{\dA\dB}\big)e_{\mu}{}^{B\dB} = \Phi^{A}{}_{\dB}\sfB_{AB}e_{\mu}{}^{B\dB} \nonumber\\
	& &&= \tfrac{1}{2}\big(\Phi^{A}{}_{\dB}\Phi_{A\dA}\Psi_{B}{}^{\dA,C\dC}D_{C\dC}\lambda + \Phi^{A}{}_{\dB}\Phi_{B\dA}\Psi_{A}{}^{\dA,C\dC}D_{C\dC}\lambda\big)e_{\mu}{}^{B\dB} ~. \label{eq:X nu 2 intermediate}
}
Using the identities $\Phi^{A}{}_{\dB}\Phi_{A\dA} = \frac{1}{2}\varepsilon_{\dA\dB}\Phi^2$ and $\Phi^{A}{}_{\dB}\Phi_{B\dA} = \Phi^{A}{}_{\dA}\Phi_{B\dB} + \frac{1}{2}\delta_{B}{}^{A}\varepsilon_{\dA\dB}\Phi^2$, \eqref{eq:X nu 2 intermediate} becomes
\eqa{
	&\sfX_{\mu}^{(2)} &&= \tfrac{1}{2}\big(\tfrac{1}{2}\Phi^2\Psi_{B\dB}{}^{C\dC}D_{C\dC}\lambda + \tfrac{1}{2}\Phi^2\Psi_{B\dB}{}^{C\dC}D_{C\dC}\lambda - \Phi_{A\dA}\Phi_{B\dB}\Psi^{A\dA,C\dC}D_{C\dC}\lambda\big)e_{\mu}{}^{B\dB} \nonumber\\
	& &&= \tfrac{1}{2}\big(\Phi^2\Psi_{B\dB}{}^{C\dC}D_{C\dC}\lambda - \Phi_{A\dA}\Phi_{B\dB}\Psi^{A\dA,C\dC}D_{C\dC}\lambda\big)e_{\mu}{}^{B\dB} ~. \label{eq:X nu 2 final}
}
(In fact, $2\Phi^{\sigma}\big(\Phi_{[\sigma}\Psi_{\mu]}{}^{\rho}\big)^{+}D_{\rho}\lambda = \Phi^{\sigma}\big(\Phi_{[\sigma}\Psi_{\mu]}{}^{\rho}\big)D_{\rho}\lambda$ due to \eqref{eq:contracted SD equals ASD}. This is just the four-component version of  \eqref{eq:X nu 2 final}.) So far, from \eqref{eq:X nu 1 final} and \eqref{eq:X nu 2 final}, 
\eqa{
	&\sfX_{\mu}^{(1)} + \sfX_{\mu}^{(2)} &&= \big(\Phi_{B\dA}\Phi_{C\dC}\Psi^{A\dA,C\dC}D_{A\dB}\lambda + \tfrac{1}{2}\Phi^2\Psi_{B\dB}{}^{C\dC}D_{C\dC}\lambda - \Phi_{A\dA}\Phi_{B\dB}\Psi^{A\dA,C\dC}D_{C\dC}\lambda\big)e_{\mu}{}^{B\dB} \nonumber\\
	& &&= \big(\Phi_{A\dA}\Phi_{B\dB}\Psi^{A\dA,C\dC}D_{C\dC}\lambda + \Phi_{B\dE}\Phi_{C\dC}\Psi^{A}{}_{\dB}{}^{C\dC}D_{A}{}^{\dE}\lambda + \tfrac{1}{2}\Phi^2\Psi_{B\dB}{}^{C\dC}D_{C\dC}\lambda \nonumber\\
	& &&\qquad\qquad - \Phi_{A\dA}\Phi_{B\dB}\Psi^{A\dA,C\dC}D_{C\dC}\lambda\big)e_{\mu}{}^{B\dB} \nonumber\\
	& &&= \big(\Phi^{A\dA}\Phi_{B\dE}\Psi^{C}{}_{\dB,A\dA}D_{C}{}^{\dE}\lambda + \tfrac{1}{2}\Phi^2\Psi_{B\dB}{}^{C\dC}D_{C\dC}\lambda\big)e_{\mu}{}^{B\dB} ~. \label{eq:sum of X nu 1 and X nu 2}
}
Finally, consider $\sfX_{\mu}^{(3)} \equiv 2\Phi^{\sigma}\big[\Psi_{\sigma}{}^{\rho}\big(\Phi_{[\mu}D_{\rho]}\lambda\big)^{+} - \Psi_{\mu}{}^{\rho}\big(\Phi_{[\sigma}D_{\rho]}\lambda\big)^{+}\big]$. Define 
\eqa{
	&\sfC_{\sigma\mu} &&= \Phi_{[\sigma}D_{\mu]}\lambda ~.
}
Note that $\sfX_{\mu}^{(3)} = 4\Phi^{\sigma}\big(\Psi_{[\sigma}{}^{\rho}\sfC_{\mu]\rho}^{+}\big)$.
In two-component notation,
\eqa{
&	\sfC_{A\dA,B\dB} &&= \tfrac{1}{2}\big(\Phi_{A\dA}\n_{B\dB}\lambda - \Phi_{B\dB}\n_{A\dA}\lambda\big) ~,\\
&	\sfC_{AB} &&= \tfrac{1}{2}\big(\Phi_{A\dA}\n_{B}{}^{\dA}\lambda + \n_{B\dA}\n_{A}{}^{\dA}\lambda\big) ~.
}
Therefore,
\eqa{
	&\sfX_{\mu}^{(3)} &&= 2\Phi^{A\dA}\big(\Psi_{A\dA}{}^{C\dC}\tfrac{1}{2}\sfC_{BC}\varepsilon_{\dB\dC} - \Psi_{B\dB}{}^{C\dC}\tfrac{1}{2}\sfC_{AC}\varepsilon_{\dA\dC}\big)e_{\mu}{}^{B\dB} \nonumber\\
	& &&= \big(-\Phi^{A\dA}\Psi_{A\dA}{}^{C}{}_{\dB}\sfC_{BC} + \Phi^{A\dA}\Psi_{B\dB}{}^{C}{}_{\dA}\sfC_{AC}\big)e_{\mu}{}^{B\dB} \nonumber\\
	& &&= \tfrac{1}{2}\big[-\Phi^{A\dA}\Phi_{B\dE}\Psi_{A\dA}{}^{C}{}_{\dB}D_{C}{}^{\dE}\lambda -\Phi^{A\dA}\Phi_{C\dE}\Psi_{A\dA}{}^{C}{}_{\dB}D_{B}{}^{\dE}\lambda \nonumber\\
	& &&\qquad\,\, + \Phi^{A\dA}\Phi_{A\dE}\Psi_{B\dB}{}^{C}{}_{\dA}D_{C}{}^{\dE}\lambda + \Phi^{A\dA}\Phi_{C\dE}\Psi_{B\dB}{}^{C}{}_{\dA}D_{A}{}^{\dE}\lambda\big]e_{\mu}{}^{B\dB} \nonumber\\
	& &&= \tfrac{1}{2}\big[-\Phi^{A\dA}\Phi_{B\dE}\Psi_{A\dA}{}^{C}{}_{\dB}D_{C}{}^{\dE}\lambda -\Phi^{A\dA}\Phi_{C\dE}\Psi_{A\dA}{}^{C}{}_{\dB}D_{B}{}^{\dE}\lambda \nonumber\\
	& &&\qquad\,\, - \tfrac{1}{2}\Phi^2\Psi_{B\dB}{}^{C\dC}D_{C\dC}\lambda + \Phi^{A\dA}\Phi_{C\dE}\Psi_{B\dB}{}^{C}{}_{\dA}D_{A}{}^{\dE}\lambda\big]e_{\mu}{}^{B\dB} \nonumber\\
	& &&= \tfrac{1}{2}\big[-2\Phi^{A\dA}\Phi_{B\dE}\Psi_{A\dA}{}^{C}{}_{\dB}D_{C}{}^{\dE}\lambda - \Phi^{A\dA}\Phi_{E\dE}\Psi_{A\dA,B\dB}D^{E\dE}\lambda \nonumber \\
	& &&\qquad\,\, - \tfrac{1}{2}\Phi^2\Psi_{B\dB}{}^{C\dC}D_{C\dC}\lambda + \Phi^{A\dA}\Phi_{E\dE}\Psi_{B\dB}{}^{E}{}_{\dA}D_{A}{}^{\dE}\lambda\big]e_{\mu}{}^{B\dB} \nonumber\\
	& &&= \tfrac{1}{2}\big[-2\Phi^{A\dA}\Phi_{B\dE}\Psi_{A\dA}{}^{C}{}_{\dB}D_{C}{}^{\dE}\lambda - \Phi^{A\dA}\Phi_{E\dE}\Psi_{A\dA,B\dB}D^{E\dE}\lambda \nonumber \\
	& &&\qquad\,\, - \tfrac{1}{2}\Phi^2\Psi_{B\dB}{}^{C\dC}D_{C\dC}\lambda - \tfrac{1}{2}\Phi^2\Psi_{B\dB}{}^{C\dC}D_{C\dC}\lambda + \Phi^{A\dA}\Phi_{E\dE}\Psi_{A\dA,B\dB}D^{E\dE}\lambda\big]e_{\mu}{}^{B\dB} \nonumber\\
	& &&= \big(-\Phi^{A\dA}\Phi_{B\dE}\Psi_{A\dA}{}^{C}{}_{\dB}D_{C}{}^{\dE}\lambda - \tfrac{1}{2}\Phi^2 \Psi_{B\dB}{}^{C\dC}D_{C\dC}\lambda\big)e_{\mu}{}^{B\dB} \\
	& &&= -\sfX_{\mu}^{(1)} - \sfX_{\mu}^{(2)} ~ \quad \text{(from \eqref{eq:sum of X nu 1 and X nu 2})} ~.
}
And hence,
\begin{empheq}[box=\fbox]{align}
	\sfX_{\mu} &\equiv \sfX_{\mu}^{(1)} + \sfX_{\mu}^{(2)} + \sfX_{\mu}^{(3)} \equiv 0 ~.
\end{empheq}
\hfill $\blacksquare$

\subsubsection{\label{sec:blue terms in dtsquared chi}$\Phi D\chi$ Terms In $\IQ^2\chi_{\mu\nu}$} 
In this section, we show that the \textcolor{blue}{\text{blue}} terms in \eqref{eq:dtsquared chi intermediate} combine to give $\Phi^{\rho}D_{\rho}\chi_{\mu\nu}$, the gauge-covariantized transport term in the Lie derivative of $\chi_{\mu\nu}$. 

Define $\sfA_{\mu\nu} := -2\Phi^{\rho}\big(D_{[\mu}\chi_{\nu]\rho}\big)$. In two-component notation,
\eqa{
	& \sfA_{A\dA,B\dB} &&= \tfrac{1}{2}\big(\Phi^{C}{}_{\dB}\n_{A\dA}\chi_{BC} - \Phi^{C}{}_{\dA}\n_{B\dB}\chi_{BC}\big) ~,\\
	& \sfA_{AB} &&= \varepsilon^{\dA\dB}\sfA_{A\dA,B\dB} = \tfrac{1}{2}\Phi_{C\dC}\big(\n_{A}{}^{\dC}\chi_{B}{}^{C} + \n_{B}{}^{\dC}\chi_{A}{}^{C}\big) ~.
}
Define $\sfB_{\mu\nu} := -2\Phi_{[\mu}D^{\rho}\chi_{\nu]\rho}$. In two-component notation,
\eqa{
 & \sfB_{A\dA,B\dB} &&= \tfrac{1}{2}\big(\Phi_{A\dA}D^{C}{}_{\dB}\chi_{BC} - \Phi_{B\dB}D^{C}{}_{\dA}\chi_{AC}\big) ~,\\
 & \sfB_{AB} &&= \varepsilon^{\dA\dB}\sfB_{A\dA,B\dB} = \tfrac{1}{2}\big(\Phi_{A\dA}D^{C\dA}\chi_{BC} + \phi_{B\dA}D^{C\dA}\chi_{AC}\big) ~.
}
Therefore,
\eqa{
	&\sfA_{AB} + \sfB_{AB} &&= \tfrac{1}{2}\big(\Phi_{C\dC}D_{A}{}^{\dC}\chi_{B}{}^{C} - \Phi_{A\dC}D_{C}{}^{\dC}\chi_{B}{}^{C}\big) + \tfrac{1}{2}\big(\Phi_{C\dC}D_{B}{}^{\dC}\chi_{A}{}^{C} - \Phi_{B\dC}D_{C}{}^{\dC}\chi_{A}{}^{C}\big) \nonumber\\
	& &&= \tfrac{1}{2}\varepsilon_{CA}\Phi_{E\dE}D^{E\dE}\chi_{B}{}^{C} + \tfrac{1}{2}\varepsilon_{CB}\Phi_{E\dE}D^{E\dE}\chi_{A}{}^{C} = \Phi^{E\dE}D_{E\dE}\chi_{AB} ~.
}
But this is precisely the two-component version of $\Phi^{\rho}D_{\rho}\chi_{\mu\nu}$, the transport term in the gauge-covariantized Lie derivative. \hfill $\blacksquare$

\subsubsection{\label{sec:red terms in dtsquared chi}$\big(\mathbf{\nabla}\Phi\big)\chi$ Terms In $\IQ ^2\chi_{\mu\nu}$} 
In this section, we show that the \textcolor{red}{\text{red}} terms in \eqref{eq:dtsquared chi intermediate} combine to give the non-transport terms in the Lie derivative of $\chi_{\mu\nu}$. We can write these terms as
\eqa{
	&-\big[\big(\n_{[\mu}\Phi^{\rho}\big)\chi_{\nu]\rho}\big]^{-} - \big[\big(\n^{\rho}\Phi_{[\mu}\big)\chi_{\nu]\rho}\big]^{-} -2\big[\big(\n_{[\mu}\Phi^{\rho}\big)\chi_{\nu]\rho}\big] + 2\big[\big(\n_{[\mu}\Phi^{\rho}\big)\chi_{\nu]\rho}\big]^{-} \nonumber\\
	&= \underbrace{\big[\big(\n_{[\mu}\Phi^{\rho} - \n^{\rho}\Phi_{[\mu}\big)\chi_{\nu]\rho}\big]^{-}}_{=\,0 \text{ (due to \eqref{eq:identB antisym general})}} -2\big[\big(\n_{[\mu}\Phi^{\rho}\big)\chi_{\nu]\rho}\big] \nonumber\\
	&= \big(\n_{\mu}\Phi^{\rho}\big)\chi_{\rho\nu} - \big(\n_{\nu}\Phi^{\rho}\big)\chi_{\rho\mu} ~.
}
This is precisely the non-transport part of the Lie derivative. \hfill $\blacksquare$

\subsubsection{\label{sec:green terms in dtsquared chi}Cancellation of $\Psi \Phi D\lambda $ Terms In $\IQ ^2\chi_{\mu\nu}$} 
In this section, we show that the \textcolor{OliveGreen}{\text{green}} terms in \eqref{eq:dtsquared chi intermediate} add up to zero. That is, we show that
\eqa{
	&\quad - 2 \big[\Psi_{\mu}{}^{\rho}\big(\Phi_{[\nu}D_{\rho]}\lambda\big)^{-} - \Psi_{\nu}{}^{\rho}\big(\Phi_{[\mu}D_{\rho]}\lambda\big)^{-}\big]^{+} \nonumber\\
	&\quad - 2\Phi^{\rho}\big[\Psi_{\rho[\mu}D_{\nu]}\lambda\big]^{+} - 2\big(\Phi_{[\mu}\Psi_{\nu]}{}^{\rho}\big)^{+}D_{\rho}\lambda + \Psi_{\rho}{}^{\rho}\big(\Phi_{[\mu}D_{\nu]}\lambda\big)^{+}  = 0 ~. \label{eq:eks plus wye identity}
}
Let us denote the first line by $\sfX_{\mu\nu}^{+}$ and the second line by $\sfY_{\mu\nu}^{+}$. Specifically, we define
\eqa{
 & \sfX_{\mu\nu}^{+} &&:=  - 2 \big[\Psi_{\mu}{}^{\rho}\big(\Phi_{[\nu}D_{\rho]}\lambda\big)^{-} - \Psi_{\nu}{}^{\rho}\big(\Phi_{[\mu}D_{\rho]}\lambda\big)^{-}\big]^{+} ~,\\
 & \sfY_{\mu\nu}^{+} &&:= - 2\Phi^{\rho}\big[\Psi_{\rho[\mu}D_{\nu]}\lambda\big]^{+} - 2\big(\Phi_{[\mu}\Psi_{\nu]}{}^{\rho}\big)^{+}D_{\rho}\lambda + \Psi_{\rho}{}^{\rho}\big(\Phi_{[\mu}D_{\nu]}\lambda\big)^{+}  ~.
}
In two-component notation, the self dual $\sfX_{\mu\nu}^{+}$ reads
\eqa{
	&\sfX_{AB} &&= \tfrac{1}{2}\big(\Psi_{A\dA,B}{}^{\dC} + \Psi_{B\dA,A}{}^{\dC}\big)\big(\Phi_{E}{}^{\dA}D^{E}{}_{\dC}\lambda + \Phi_{E\dC}D^{E\dA}\lambda\big) ~.
}
By decomposing the symmetric gravitino as $\Psi_{A\dA,B\dB} = \wh{\Psi}_{(AB),(\dA\dB)} + \tfrac{1}{4}\varepsilon_{AB}\varepsilon_{\dA\dB}\Psi$ (see \eqref{eq:spinorial-decomp-sym-gravitino}), this can be written as
\eqa{
	&\sfX_{AB} &&= -2\Phi_{E\dC}\big(D^{E}{}_{\dA}\lambda\big)\wh{\Psi}_{AB}{}^{\dA\dC} ~. \label{eq:xab final}
}
As for $\sfY_{AB}$, it is a bit more complicated, and reads
\eqa{
	&\sfY_{AB} &&= -\Phi^{C\dC}\big[\Psi_{C\dC,A\dA}D_{B}{}^{\dA}\lambda + \Psi_{C\dC,B\dA}D_{A}{}^{\dA}\lambda\big] -\big[\Phi_{A\dA}\Psi_{B}{}^{\dA,C\dC}D_{C\dC}\lambda + \Phi_{B\dA}\Psi_{A}{}^{\dA,C\dC}D_{C\dC}\lambda\big] \nonumber\\
	& &&\quad + \tfrac{1}{2}\Psi\big[\Phi_{A\dA}D_{B}{}^{\dA}\lambda + \Phi_{B\dA}D_{A}{}^{\dA}\lambda\big] ~.
}
After using the substitution \eqref{eq:spinorial-decomp-sym-gravitino}, one finds that
\eqa{
	&\sfY_{AB} &&= \big(\Phi_{C\dC}\wh{\Psi}_{A}{}^{C,\dA\dC}D_{B\dA}\lambda - \Phi_{B\dA}\wh{\Psi}_{A}{}^{C,\dA\dC}D_{C\dC}\lambda\big)\nonumber\\
	& &&\quad + \big(\Phi_{C\dC}\wh{\Psi}_{B}{}^{C,\dA\dC}D_{A\dA}\lambda - \Phi_{A\dA}\wh{\Psi}_{B}{}^{C,\dA\dC}D_{C\dC}\lambda\big) \nonumber\\
	& &&= \varepsilon_{CB}\Phi_{E\dC}\big(D^{E}{}_{\dA}\lambda\big)\wh{\Psi}_{A}{}^{C,\dA\dC} + \varepsilon_{CA}\Phi_{E\dC}\big(D^{E}{}_{\dA}\lambda\big)\wh{\Psi}_{B}{}^{C,\dA\dC} \nonumber\\
	& &&= \Phi_{E\dC}\big(D^{E}{}_{\dA}\lambda\big)\wh{\Psi}_{AB}{}^{\dA\dC} + \Phi_{E\dC}\big(D^{E}{}_{\dA}\lambda\big)\wh{\Psi}_{BA}{}^{\dA\dC} \nonumber\\
	& &&= +2\Phi_{E\dC}\big(D^{E}{}_{\dA}\lambda\big)\wh{\Psi}_{AB}{}^{\dA\dC} ~. \label{eq:yab final}
}
Clearly, from \eqref{eq:xab final} and \eqref{eq:yab final}, it follows that $\sfX_{AB} + \sfY_{AB} = 0$, and hence
\eqa{
	&\sfX_{\mu\nu}^{+} + \sfY_{\mu\nu}^{+} &&= 0 ~,
}
which completes the proof. \hfill $\blacksquare$

\eject
\subsection{\label{app:ActionComparisonDetails}Computations For The Action Comparison}
Here we collect some details for the action comparison of section \ref{sec:CompareActions}. The superscripts `orig.' and `shifted' distinguish between the expressions in sugra and Cartan field parametrizations.
\\\\
\noindent \underline{\textbf{Shifting $\left.e\sfC_{-}\right|_{\text{deg}\,=\,0}^{\text{orig.}}$.}}
The underlined terms shift.
\eqa{
  &\left.e\sfC_{-}\right|_{\text{deg}\,=\,0}^{\text{orig}} &&= \partial_{\mu}\big(-2\,e\,\CF_{I}D^{\mu}\lambda^{I}\big) + 2\,e\,\CF_{IJ}\big(D_{\mu}\lambda^I)\dashuline{\big(D^{\mu}\phi^{J}\big)} + 2\,e\,\CF_{I}\left[\lambda,\left[\phi, \lambda\right]\right]^{I} \nn
& &&\quad  + \frac{e}{2}\CF_{IJ}[\phi,\chi_{\mu\nu}]^{I}\chi^{\mu\nu}{}^{J} + 2\,e\,\CF_{I}[\eta,\eta]^{I} + 2\,e\,\CF_{IJ}\dashuline{\psi_{\mu}{}^{I}[\psi^{\mu},\lambda]^{J}}\nn
  & &&\quad - \frac{e}{2}\CF_{IJ}\dashuline{\left( F^{-}_{\mu\nu}{}^{I}F_{-}^{\mu\nu}{}^{J} - D_{\mu\nu}{}^{I}D_{\mu\nu}{}^{J} - 4\big(D_{\rho}\chi_{\sigma}{}^{\rho}{}^{I}\big)\psi^{\sigma,J} + 4\big(D_{\rho}\eta^{I}\big)\psi^{\rho,J}\right)} \nn
  & &&\quad + e\,\CF_{IJK}\dashuline{\big(F^{\mu\nu}_{-}{}^{I} - D^{\mu\nu,I}\big)\psi_{\mu}{}^{J}\psi_{\nu}{}^{K}} + \frac{1}{12}\CF_{IJKL}\veps^{\mu\nu\rho\sigma}\dashuline{\psi_{\mu}{}^{I}\psi_{\nu}{}^{J}\psi_{\rho}{}^{K}\psi_{\sigma}{}^{L}} ~.
  }
Under the shift, $D^{\mu}\phi \mapsto D^{\mu}\Phi - \Phi^{\mu}[\lambda, \phi]$. Therefore,
\eqa{
&2\,e\,\CF_{IJ}\big(D_{\mu}\lambda^{I}\big)\big(D^{\mu}\phi^{J}\big) &&\mapsto 2\,e\,\CF_{IJ}\big(D_{\mu}\lambda^{I}\big)\big(D^{\mu}\phi^{J}\big)-2\,e\,\Phi^{\mu}\CF_{IJ}\big(D_{\mu}\lambda^{I}\big)[\lambda,\phi]^{J} ~.
}
\noindent Next, 
\eqa{
& \CF_{IJ}\psi_{\mu}{}^{I}[\psi^{\mu},\lambda]^{J} &&\mapsto \CF_{IJ}\big(\psi_{\mu}{}^{I} - \Phi^{\rho}\chi_{\rho\mu}{}^{I}\big)[\psi^{\mu} - \Phi^{\sigma}\chi_{\sigma}{}^{\mu}, \lambda]^{J} \nonumber\\
& &&\quad = \CF_{IJ}\psi_{\mu}{}^{I}[\psi^{\mu},\lambda]^{J} - \Phi^{\rho}\CF_{IJ}\psi_{\mu}{}^{I}[\chi_{\rho}{}^{\mu},\lambda]^{J} - \Phi^{\rho}\CF_{IJ}\chi_{\rho\mu}{}^{I}[\psi^{\mu},\lambda]^{J} \nonumber\\
& &&\quad \quad + \CF_{IJ}\Phi^{\rho}\Phi^{\sigma}\chi_{\rho\mu}{}^{I}[\chi_{\sigma}{}^{\mu}, \lambda]^{J} ~.
}
Next,
\eqa{
&F_{\mu\nu}^{-} &&\mapsto F_{\mu\nu}^{-} - 2\big(\n_{[\mu}\Phi_{\nu]}\big)^{-}\lambda + 2\big(\Phi_{[\mu}D_{\nu]}\lambda\big)^{-} ~.
}
So,
\eqa{
& \CF_{IJ}F_{\mu\nu}^{-}{}^{I}F_{-}^{\mu\nu}{}^{J} &&\mapsto \CF_{IJ}F_{\mu\nu}^{-}{}^{I}F_{-}^{\mu\nu}{}^{J} + 4\big[(\n_{[\mu}\Phi_{\nu]})^{-}\big]^2\CF_{IJ}\lambda^{I}\lambda^{J} \nonumber\\
& &&\quad + 4\,\CF_{IJ}\big(\Phi^{[\mu}D^{\nu]}\lambda^{I}\big)^{-}\big(\Phi_{[\mu}D_{\nu]}\lambda^{J}\big)^{-} \nonumber\\
& &&\quad  - 4\big(\n_{[\mu}\Phi_{\nu]}\big)^{-}\CF_{IJ}F_{-}^{\mu\nu}{}^{I}\lambda^{J} + 4\,\CF_{IJ}F_{\mu\nu}^{-}{}^{I}\big(\Phi^{[\mu}D^{\nu]}\lambda^{J}\big)^{-} \nonumber\\
& &&\quad - 8\big(\n_{[\mu}\Phi_{\nu]}\big)^{-}\CF_{IJ}\lambda^{I}\big(\Phi^{[\mu}D^{\nu]}\lambda^{J}\big)^{-} ~, \\
& \CF_{IJ}D_{\mu\nu}{}^{I}D^{\mu\nu}{}^{J} &&\mapsto \CF_{IJ}D_{\mu\nu}{}^{I}D^{\mu\nu}{}^{J} + 4\,\CF_{IJ}\big(\Phi_{[\mu}D_{\nu]}\lambda^{I}\big)^{+}\big(\Phi^{[\mu}D^{\nu]}\lambda^{J}\big)^{+} \nonumber\\
& &&\quad - 4\,\CF_{IJ}D_{\mu\nu}{}^{I}\big(\Phi^{[\mu}D^{\nu]}\lambda^{J}\big)^{+} ~.
}
Therefore,
\eqa{
& \CF_{IJ}\big(F_{\mu\nu}^{-}{}^{I}F_{-}^{\mu\nu}{}^{J} - D_{\mu\nu}{}^{I}D^{\mu\nu}{}^{J}\big) && \mapsto \CF_{IJ}\big(F_{\mu\nu}^{-}{}^{I}F^{\mu\nu}_{-}{}^{J} - D_{\mu\nu}{}^{I}D^{\mu\nu}{}^{J}\big) -4\big(\n_{[\mu}\Phi_{\nu]}\big)^{-}\CF_{IJ}F_{-}^{\mu\nu}{}^{I}\lambda^{J} \nonumber\\
& &&\quad + 4\,\CF_{IJ}\big(F_{\mu\nu}^{-}{}^{I} + D_{\mu\nu}{}^{I}\big)\big(\Phi^{[\mu}D^{\nu]}\lambda^{J}\big)  + 4\big[(\n_{[\mu}\Phi_{\nu]})^{-}\big]^2 \CF_{IJ}\lambda^{I}\lambda^{J} \nonumber\\
& &&\quad + 4\,\CF_{IJ}\underbrace{\big[ \big(\Phi^{[\mu}D^{\nu]}\lambda^{I}\big)^{-}\big(\Phi_{[\mu}D_{\nu]}\lambda^{J}\big)^{-} - \big(\Phi^{[\mu}D^{\nu]}\lambda^{I}\big)^{+}\big(\Phi_{[\mu}D_{\nu]}\lambda^{J}\big)^{+}\big]}_{=\,0} \nonumber\\
& &&\quad - 8\big(\n_{[\mu}\Phi_{\nu]}\big)^{-}\CF_{IJ}\lambda^{I}\big(\Phi^{[\mu}D^{\nu]}\lambda^{J}\big)^{-} ~.
}
The vanishing of the underbraced term follows from the use of the following identity
\eqa{
& \CF_{IJ}\big(\sfA_{\mu\nu}^{+}{}^{I}\sfA^{\mu\nu}_{+}{}^{J} - \sfA_{\mu\nu}^{-}{}^{I}\sfA^{\mu\nu}_{-}{}^{J}\big) &&= \frac{1}{2\sqrt{g}}\veps^{\mu\nu\rho\sigma}\CF_{IJ}\sfA_{\mu\nu}{}^{I}\sfA_{\rho\sigma}{}^{J} ~,
}	
with $\sfA_{\mu\nu}{}^{I} = \Phi_{[\mu}D_{\nu]}\lambda^{I}$.
Next,
\eqa{
& \CF_{IJ}\big(D_{\rho}\chi_{\sigma}{}^{\rho}{}^{I}\big)\psi^{\sigma}{}^{J} &&\mapsto \CF_{IJ}\big(D_{\rho}\chi_{\sigma}{}^{\rho}{}^{I}\big)\psi^{\sigma}{}^{J} - \Phi^{\kappa}\CF_{IJ}\big(D_{\rho}\chi_{\sigma}{}^{\rho}{}^{I}\big)\chi_{\kappa}{}^{\sigma}{}^{J} - \Phi^{\rho}\CF_{IJ}[\lambda, \chi_{\sigma\rho}]^{I}\psi^{\sigma}{}^{J} \nonumber\\
& &&\quad + \Phi^{\rho}\Phi^{\kappa}\CF_{IJ}[\lambda, \chi_{\sigma\rho}]^{I}\chi_{\kappa}{}^{\sigma}{}^{J} ~,\\
& \CF_{IJ}\big(D_{\rho}\eta^{I}\big)\psi^{\rho}{}^{J} &&\mapsto \CF_{IJ}\big(D_{\rho}\eta^{I}\big)\psi^{\rho}{}^{J} - \Phi^{\sigma}\CF_{IJ}\big(D_{\rho}\eta^{I}\big)\chi_{\sigma}{}^{\rho}{}^{J} - \Phi^{\sigma}\CF_{IJ}[\lambda, \eta]^{I}\psi_{\sigma}{}^{J} ~.
}
Therefore,
\eqa{
& \CF_{IJ}\big(D_{\rho}\chi_{\sigma}{}^{\rho}{}^{I}\big)\psi^{\sigma}{}^{J} - \CF_{IJ}\big(D_{\rho}\eta^{I}\big)\psi^{\rho}{}^{J}  &&\mapsto \CF_{IJ}\big(D_{\rho}\eta^{I}\big)\psi^{\rho}{}^{J} - \CF_{IJ}\big(D_{\rho}\eta^{I}\big)\psi^{\rho}{}^{J}  \nonumber\\
& &&\quad  - \Phi^{\kappa}\CF_{IJ}\big(D_{\rho}\chi_{\sigma}{}^{\rho}{}^{I}\big)\chi_{\kappa}{}^{\sigma}{}^{J} - \Phi^{\rho}\CF_{IJ}[\lambda, \chi_{\sigma\rho}]^{I}\psi^{\sigma}{}^{J} \nonumber\\
& &&\quad + \Phi^{\sigma}\CF_{IJ}\big(D_{\rho}\eta^{I}\big)\chi_{\sigma}{}^{\rho}{}^{J} + \Phi^{\sigma}\CF_{IJ}[\lambda, \eta]^{I}\psi_{\sigma}{}^{J} \nonumber\\
& &&\quad + \Phi^{\rho}\Phi^{\kappa}\CF_{IJ}[\lambda, \chi_{\sigma\rho}]^{I}\chi_{\kappa}{}^{\sigma}{}^{J} ~.
}
Next,
\eqa{
& F_{\mu\nu}^{-} - D_{\mu\nu} &&\mapsto F_{\mu\nu}^{-} - D_{\mu\nu} - 2\big(\n_{[\mu}\Phi_{\nu]}\big)^{-}\lambda + 2\big(\Phi_{[\mu}D_{\nu]}\lambda\big) ~.
}
Therefore,
\eqa{
& \CF_{IJK}\big(F_{-}^{\mu\nu}{}^{I} - D^{\mu\nu}{}^{I}\big)\psi_{\mu}{}^{J}\psi_{\nu}{}^{K} &&\mapsto \CF_{IJK}\big(F_{-}^{\mu\nu}{}^{I} - D^{\mu\nu}{}^{I}\big)\psi_{\mu}{}^{J}\psi_{\nu}{}^{K} \nonumber\\
& &&\quad -2\,\Phi^{\alpha}\CF_{IJK}\big(F_{-}^{\mu\nu}{}^{I} - D^{\mu\nu}{}^{I}\big)\chi_{\alpha\mu}{}^{J}\psi_{\nu}{}^{K} \nonumber\\
& &&\quad -2\big(\n^{[\mu}\Phi^{\nu]}\big)^{-}\CF_{IJK}\lambda^{I}\psi_{\mu}{}^{J}\psi_{\nu}{}^{K}\nonumber\\
& &&\quad + 2\,\Phi^{\mu}\CF_{IJK}\big(D^{\nu}\lambda^{I}\big)\psi_{\mu}{}^{J}\psi_{\nu}{}^{K} \nonumber\\
& &&\quad + \Phi^{\alpha}\Phi^{\beta}\CF_{IJK}\big(F_{-}^{\mu\nu}{}^{I} - D^{\mu\nu}{}^{I}\big)\chi_{\alpha\mu}{}^{I}\chi_{\beta\nu}{}^{K} \nonumber\\
& &&\quad + 4\big(\n^{[\mu}\Phi^{\nu]}\big)^{-}\Phi^{\alpha}\CF_{IJK}\lambda^{I}\chi_{\alpha\mu}{}^{J}\psi_{\nu}{}^{K} \nonumber\\
& &&\quad - 4\,\Phi^{\mu}\Phi^{\alpha}\CF_{IJK}\big(D^{\nu}\lambda^{I}\big)\chi_{\alpha[\mu}{}^{J}\psi_{\nu]}{}^{K} \nonumber\\
& &&\quad - 2\big(\n^{[\mu}\Phi^{\nu]}\big)^{-}\Phi^{\alpha}\Phi^{\beta}\CF_{IJK}\lambda^{I}\chi_{\alpha\mu}{}^{J}\chi_{\beta\nu}{}^{K} \nonumber\\
& &&\quad + 2\,\underbrace{\Phi^{\mu}\Phi^{\alpha}\Phi^{\beta}\CF_{IJK}\big(D^{\nu}\lambda^{I}\big)\chi_{\alpha\mu}{}^{I}\chi_{\beta\nu}{}^{K}}_{=\,0} ~.
}
The underbraced term vanishes because it is the contraction of a tensor symmetric in $\mu, \alpha$ with one antisymmetric in these indices. 

Finally,
\eqa{
& \veps^{\mu\nu\rho\sigma}\CF_{IJKL}\psi_{\mu}{}^{I}\psi_{\nu}{}^{J}\psi_{\rho}{}^{K}\psi_{\sigma}{}^{L} &&\mapsto \veps^{\mu\nu\rho\sigma}\CF_{IJKL}\psi_{\mu}{}^{I}\psi_{\nu}{}^{J}\psi_{\rho}{}^{K}\psi_{\sigma}{}^{L} \nonumber\\
& &&\quad -4\,\Phi^{\alpha}\veps^{\mu\nu\rho\sigma}\CF_{IJKL}\psi_{\mu}{}^{I}\psi_{\nu}{}^{J}\psi_{\rho}{}^{K}\chi_{\alpha\sigma}{}^{L} \nonumber\\
& &&\quad + 6\,\Phi^{\alpha}\Phi^{\beta}\veps^{\mu\nu\rho\sigma}\CF_{IJKL}\psi_{\mu}{}^{I}\psi_{\nu}{}^{J}\chi_{\alpha\rho}{}^{K}\chi_{\beta\sigma}{}^{L} \nonumber\\
& &&\quad -4\,\Phi^{\alpha}\Phi^{\beta}\Phi^{\gamma}\veps^{\mu\nu\rho\sigma}\CF_{IJKL}\psi_{\mu}{}^{I}\chi_{\alpha\nu}{}^{J}\chi_{\beta\rho}{}^{K}\chi_{\gamma\sigma}{}^{L} \nonumber\\
& &&\quad + \underbrace{\Phi^{\alpha}\Phi^{\beta}\Phi^{\gamma}\Phi^{\delta}\veps^{\mu\nu\rho\sigma}\CF_{IJKL}\chi_{\alpha\mu}{}^{I}\chi_{\beta\nu}{}^{J}\chi_{\gamma\rho}{}^{K}\chi_{\delta\sigma}{}^{L}}_{=\,0} ~.
}
The vanishing of the underbraced term follows from the simple fact that on a 4-manifold $\IX$,
\eqa{
& \CF_{IJKL}\chi^{I} \wedge \iota_{\Phi}\chi^{J} \wedge \iota_{\Phi}\chi^{K} \wedge \iota_{\Phi}\chi^{L} &&= 0 ~,
}
since the LHS is a five-form. Thus,
\eqa{
& \CF_{IJKL}\iota_{\Phi}\chi^{I} \wedge \iota_{\Phi}\chi^{J} \wedge \iota_{\Phi}\chi^{K} \wedge \iota_{\Phi}\chi^{L} &&= 0 ~,
}
which in components, reads
\eqa{
\Phi^{\alpha}\Phi^{\beta}\Phi^{\gamma}\Phi^{\delta}\veps^{\mu\nu\rho\sigma}\CF_{IJKL}\chi_{\alpha\mu}{}^{I}\chi_{\beta\nu}{}^{J}\chi_{\gamma\rho}{}^{K}\chi_{\delta\sigma}{}^{L} &&= 0 ~.
}

\noindent \underline{\textbf{Shifting $\left.e\sfC_{-}\right|_{\text{deg}\,=\,1}^{\text{orig.}}$.}} The underlined terms shift.
 \eqa{
  &\left.e\sfC_{-}\right|_{\text{deg}\,=\,1}^{\text{orig}} &&= -e\,\CF_{I}\Psi_{\sigma}{}^{\sigma}[\eta,\lambda]^{I} - e\,\CF_{IJ}\dashuline{F_{\mu\nu}^{-}{}^{I}}\big(\Psi^{\rho[\mu}\chi^{\nu]}{}_{\rho}{}^{J}\big)^{-} + 2\,e\,\CF_{IJ}\Psi_{\sigma}{}^{\rho}\big(D_{\rho}\lambda^{I}\big)\dashuline{\psi^{\sigma,J}}\nn
  & &&\quad - e\,\CF_{IJ}\Psi_{\rho}{}^{\rho}\big(D_{\sigma}\lambda^{I}\big)\dashuline{\psi^{\sigma,J}} + e\,\CF_{IJK}\big(\Psi^{\rho[\mu}\chi^{\nu]}{}_{\rho}{}^{I}\big)^{-}\dashuline{\psi_{\mu}{}^{J}\psi_{\nu}{}^{K}} ~.
  }
Now,
\eqa{
&-e\,\CF_{IJ}F_{\mu\nu}^{-}{}^{I}\big(\Psi^{\rho[\mu}\chi^{\nu]}{}_{\rho}{}^{J}\big)^{-} &&\mapsto -e\,\CF_{IJ}F_{\mu\nu}^{-}{}^{I}\big(\Psi^{\rho[\mu}\chi^{\nu]}{}_{\rho}{}^{J}\big)^{-}  \nonumber\\
& &&\quad + 2\,e\,\CF_{IJ}\big(\n_{[\mu}\Phi_{\nu]}\big)^{-}\lambda^{I}\big(\Psi^{\rho[\mu}\chi^{\nu]}{}_{\rho}{}^{J}\big)^{-} \nonumber\\
& &&\quad - 2\,e\,\CF_{IJ}\big(\Phi_{[\mu}D_{\nu]}\lambda^{I}\big)^{-}\big(\Psi^{\rho[\mu}\chi^{\nu]}{}_{\rho}{}^{J}\big)^{-} ~.
}
\eqa{
& 2\,e\,\CF_{IJ}\Psi_{\sigma}{}^{\rho}\big(D_{\rho}\lambda^{I}\big)\psi^{\sigma,J} &&\mapsto 2\,e\,\CF_{IJ}\Psi_{\sigma}{}^{\rho}\big(D_{\rho}\lambda^{I}\big)\psi^{\sigma,J} - 2\,e\,\CF_{IJ}\Psi^{\mu\nu}\Phi^{\sigma}\big(D_{\mu}\lambda^{I}\big)\chi_{\sigma\nu}{}^{J} ~.
}
\eqa{
& -e\,\CF_{IJ}\Psi_{\rho}{}^{\rho}\big(D_{\sigma}\lambda^{I}\big)\psi^{\sigma,J} &&\mapsto -e\,\CF_{IJ}\Psi_{\rho}{}^{\rho}\big(D_{\sigma}\lambda^{I}\big)\psi^{\sigma,J} + e\,\CF_{IJ}\Psi_{\rho}{}^{\rho}\Phi^{\sigma}\big(D_{\mu}\lambda^{I}\big)\chi_{\sigma}{}^{\mu}{}^{J} ~.
}
\eqa{
& + e\,\CF_{IJK}\big(\Psi^{\rho[\mu}\chi^{\nu]}{}_{\rho}{}^{I}\big)^{-}\psi_{\mu}{}^{J}\psi_{\nu}{}^{K} &&\mapsto + e\,\CF_{IJK}\big(\Psi^{\rho[\mu}\chi^{\nu]}{}_{\rho}{}^{I}\big)^{-}\psi_{\mu}{}^{J}\psi_{\nu}{}^{K} \nonumber\\
& &&\quad -2\,e\,\Phi^{\alpha}\CF_{IJK}\big(\Psi^{\rho[\mu}\chi^{\nu]}{}_{\rho}{}^{I}\big)^{-}\chi_{\alpha\mu}{}^{J}\psi_{\nu}{}^{K} \nonumber\\
& &&\quad + e\,\Phi^{\alpha}\Phi^{\beta}\CF_{IJK}\big(\Psi^{\rho[\mu}\chi^{\nu]}{}_{\rho}{}^{I}\big)^{-}\chi_{\alpha\mu}{}^{J}\chi_{\beta\nu}{}^{K} ~.
}

\noindent \textbf{\underline{Shifting $\left.e\sfC_{-}\right|_{\text{deg}\,=\,2}^{\text{orig.}}$.}} The underlined terms shift.
 \eqa{
  &\left.e\sfC_{-}\right|_{\text{deg}\,=\,2}^{\text{orig}} &&= -2\,e\big(\n^{[\mu}\Phi^{\nu]}\big)^{-}\,\CF_{IJ}\dashuline{F_{\mu\nu}^{-}{}^{I}}\lambda^{J} - \frac{e}{2}\CF_{IJ}\big(\Psi_{[\mu}{}^{\rho}\chi_{\nu]\rho}{}^{I}\big)^{-}\big(\Psi^{\sigma[\mu}\chi^{\nu]}{}_{\sigma}{}^{J}\big)^{-} \nn
  & &&\quad + 2\,e\big(\n^{[\mu}\Phi^{\nu]}\big)^{-}\CF_{IJK}\lambda^{I}\dashuline{\psi_{\mu}{}^{J}\psi_{\nu}{}^{K}} ~. \label{eq:eCminus orig deg 2 repeated}
  }
  Now,
  \eqa{
  & -2\,e\big(\n^{[\mu}\Phi^{\nu]}\big)^{-}\,\CF_{IJ}F_{\mu\nu}^{-}{}^{I}\lambda^{J} &&\mapsto -2\,e\big(\n^{[\mu}\Phi^{\nu]}\big)^{-}\,\CF_{IJ}F_{\mu\nu}^{-}{}^{I}\lambda^{J} \nonumber\\
  & &&\quad + 4\,e\,\big[(\n^{[\mu}\Phi^{\nu]})^{-}\big]^2 \CF_{IJ}\lambda^{I}\lambda^{J} \nonumber\\
  & &&\quad -4\,e\,\big(\n^{[\mu}\Phi^{\nu]}\big)^{-}\CF_{IJ}\lambda^{I}\Phi_{[\mu}D_{\nu]}\lambda^{J} ~.
  }
\eqa{
  & 2\,e\big(\n^{[\mu}\Phi^{\nu]}\big)^{-}\CF_{IJK}\lambda^{I}\psi_{\mu}{}^{J}\psi_{\nu}{}^{K} &&\mapsto 2\,e\big(\n^{[\mu}\Phi^{\nu]}\big)^{-}\CF_{IJK}\lambda^{I}\psi_{\mu}{}^{J}\psi_{\nu}{}^{K} \nonumber\\
  & &&\quad - 4\,e\,\big(\n^{[\mu}\Phi^{\nu]}\big)^{-}\Phi^{\alpha}\CF_{IJK}\lambda^{I}\chi_{\alpha\mu}{}^{J}\psi_{\nu}{}^{K} \nonumber\\
  & &&\quad + 2\,e\,\big(\n^{[\mu}\Phi^{\nu]}\big)^{-}\Phi^{\alpha}\Phi^{\beta}\CF_{IJK}\lambda^{I}\chi_{\alpha\mu}{}^{J}\chi_{\beta\nu}{}^{K} ~.
}

\noindent \textbf{\underline{Summary: Shifted $e\sfC_-$.}} At degree $=0$,
  \beqas{
    &\left.e\sfC_{-}\right|_{\text{deg}\,=\,0}^{\text{shifted}} &&= \partial_{\mu}\big(-2\,e\,\CF_{I}D^{\mu}\lambda^{I}\big) + 2\,e\,\CF_{IJ}\big(D_{\mu}\lambda^I)\big(D^{\mu}\phi^{J}\big) + 2\,e\,\CF_{I}\left[\lambda,\left[\phi, \lambda\right]\right]^{I} \nn
& &&\quad  + \frac{e}{2}\CF_{IJ}[\phi,\chi_{\mu\nu}]^{I}\chi^{\mu\nu}{}^{J} + 2\,e\,\CF_{I}[\eta,\eta]^{I} + 2\,e\,\CF_{IJ}\psi_{\mu}{}^{I}[\psi^{\mu},\lambda]^{J}\nn
  & &&\quad - \frac{e}{2}\CF_{IJ}\left( F^{-}_{\mu\nu}{}^{I}F^{\mu\nu,-J} - D_{\mu\nu}{}^{I}D_{\mu\nu}{}^{J} - 4\big(D_{\rho}\chi_{\sigma}{}^{\rho,I}\big)\psi^{\sigma,J} + 4\big(D_{\rho}\eta^{I}\big)\psi^{\rho,J}\right) \nn
  & &&\quad + e\,\CF_{IJK}\big(F^{\mu\nu,-I} - D^{\mu\nu,I}\big)\psi_{\mu}{}^{J}\psi_{\nu}{}^{K} + \frac{1}{12}\CF_{IJKL}\veps^{\mu\nu\rho\sigma}\psi_{\mu}{}^{I}\psi_{\nu}{}^{J}\psi_{\rho}{}^{K}\psi_{\sigma}{}^{L} ~.
}
At degree $=1$,
  \beqas{
    &\left.e\sfC_{-}\right|_{\text{deg}\,=\,1}^{\text{shifted}} &&= -e\,\CF_{I}\Psi_{\sigma}{}^{\sigma}[\eta,\lambda]^{I} - e\,\CF_{IJ}F_{\mu\nu}^{-}{}^{I}\big(\Psi^{\rho[\mu}\chi^{\nu]}{}_{\rho}{}^{J}\big)^{-} + 2\,e\,\CF_{IJ}\Psi_{\sigma}{}^{\rho}\big(D_{\rho}\lambda^{I}\big)\psi^{\sigma,J}\nn
  & &&\quad - e\,\CF_{IJ}\Psi_{\rho}{}^{\rho}\big(D_{\sigma}\lambda^{I}\big)\psi^{\sigma,J} + e\,\CF_{IJK}\big(\Psi^{\rho[\mu}\chi^{\nu]}{}_{\rho}{}^{I}\big)^{-}\psi_{\mu}{}^{J}\psi_{\nu}{}^{K} ~.
}
At degree $=2$, the new terms we found above are
\eqa{
& \text{new deg. 2} &&= -2\,e\,\Phi^{\mu}\CF_{IJ}\big(D_{\mu}\lambda^{I}\big)[\lambda,\phi]^{J} -2\,e\,\Phi^{\rho}\CF_{IJ}\psi_{\mu}{}^{I}[\chi_{\rho}{}^{\mu},\lambda]^{J} - 2\,e\,\Phi^{\rho}\CF_{IJ}\chi_{\rho\mu}{}^{I}[\psi^{\mu}, \lambda]^{J} \nonumber\\
& &&\quad +\highlightYellow{2\,e\big(\n_{[\mu}\Phi_{\nu]}\big)^{-}\CF_{IJ}F_{-}^{\mu\nu}{}^{I}\lambda^{J}} - 2\,e\,\CF_{IJ}\big(F_{\mu\nu}^{-}{}^{I} + D_{\mu\nu}{}^{I}\big)\big(\Phi^{[\mu}D^{\nu]}\lambda^{J}\big) \nonumber\\
& &&\quad -2\,e\,\Phi^{\kappa}\CF_{IJ}\big(D_{\rho}\chi_{\sigma}{}^{\rho}{}^{I}\big)\chi_{\kappa}{}^{\sigma}{}^{J} - 2\,e\,\Phi^{\rho}\CF_{IJ}[\lambda, \chi_{\sigma\rho}]^{I}\psi^{\sigma}{}^{J} \nonumber\\
& &&\quad + 2\,e\,\Phi^{\sigma}\CF_{IJ}\big(D_{\rho}\eta^{I}\big)\chi_{\sigma}{}^{\rho}{}^{J} + 2\,e\,\Phi^{\sigma}\CF_{IJ}[\lambda, \eta]^{I}\psi_{\sigma}{}^{J} \nonumber\\
& &&\quad -2\,e\,\Phi^{\alpha}\CF_{IJK}\big(F_{-}^{\mu\nu}{}^{I} - D^{\mu\nu}{}^{I}\big)\chi_{\alpha\mu}{}^{J}\psi_{\nu}{}^{K}  - \highlightYellow{2\,e\,\big(\n^{[\mu}\Phi^{\nu]}\big)^{-}\CF_{IJK}\lambda^{I}\psi_{\mu}{}^{J}\psi_{\nu}{}^{K}} \nonumber\\
& &&\quad + 2\,e\,\Phi^{\mu}\CF_{IJK}\big(D^{\nu}\lambda^{I}\big)\psi_{\mu}{}^{J}\psi_{\nu}{}^{K} -\frac{1}{3}\Phi^{\alpha}\veps^{\mu\nu\rho\sigma}\CF_{IJKL}\psi_{\mu}{}^{I}\psi_{\nu}{}^{J}\psi_{\rho}{}^{K}\chi_{\alpha\sigma}{}^{L} ~,
}
whereas we recall
 \eqa{
  &\left.e\sfC_{-}\right|_{\text{deg}\,=\,2}^{\text{orig}} &&= -\highlightYellow{2\,e\big(\n^{[\mu}\Phi^{\nu]}\big)^{-}\,\CF_{IJ}F_{\mu\nu}^{-}{}^{I}\lambda^{J}} - \frac{e}{2}\CF_{IJ}\big(\Psi_{[\mu}{}^{\rho}\chi_{\nu]\rho}{}^{I}\big)^{-}\big(\Psi^{\sigma[\mu}\chi^{\nu]}{}_{\sigma}{}^{J}\big)^{-} \nn
  & &&\quad + \highlightYellow{2\,e\big(\n^{[\mu}\Phi^{\nu]}\big)^{-}\CF_{IJK}\lambda^{I}\psi_{\mu}{}^{J}\psi_{\nu}{}^{K}} ~. \label{eq:eCminus orig deg 2 repeated again}
  }
  Therefore,
  \beqas{
    &\left.e\sfC_{-}\right|_{\text{deg}\,=\,2}^{\text{shifted}} &&= -2\,e\,\Phi^{\mu}\CF_{IJ}\big(D_{\mu}\lambda^{I}\big)[\lambda,\phi]^{J} - 2\,e\,\Phi^{\rho}\CF_{IJ}\chi_{\rho\mu}{}^{I}[\psi^{\mu}, \lambda]^{J} \nonumber\\
    & &&\quad  - 2\,e\,\CF_{IJ}\big(F_{\mu\nu}^{-}{}^{I} + D_{\mu\nu}{}^{I}\big)\big(\Phi^{[\mu}D^{\nu]}\lambda^{J}\big) -2\,e\,\Phi^{\kappa}\CF_{IJ}\big(D_{\rho}\chi_{\sigma}{}^{\rho}{}^{I}\big)\chi_{\kappa}{}^{\sigma}{}^{J} \nonumber\\
    & &&\quad + 2\,e\,\Phi^{\sigma}\CF_{IJ}\big(D_{\rho}\eta^{I}\big)\chi_{\sigma}{}^{\rho}{}^{J} + 2\,e\,\Phi^{\sigma}\CF_{IJ}[\lambda, \eta]^{I}\psi_{\sigma}{}^{J} \nonumber\\
    & &&\quad  -2\,e\,\Phi^{\alpha}\CF_{IJK}\big(F_{-}^{\mu\nu}{}^{I} - D^{\mu\nu}{}^{I}\big)\chi_{\alpha\mu}{}^{J}\psi_{\nu}{}^{K} + 2\,e\,\Phi^{\mu}\CF_{IJK}\big(D^{\nu}\lambda^{I}\big)\psi_{\mu}{}^{J}\psi_{\nu}{}^{K} \nonumber\\
    & &&\quad -\frac{1}{3}\Phi^{\alpha}\veps^{\mu\nu\rho\sigma}\CF_{IJKL}\psi_{\mu}{}^{I}\psi_{\nu}{}^{J}\psi_{\rho}{}^{K}\chi_{\alpha\sigma}{}^{L} - \frac{e}{2}\CF_{IJ}\big(\Psi_{[\mu}{}^{\rho}\chi_{\nu]\rho}{}^{I}\big)^{-}\big(\Psi^{\sigma[\mu}\chi^{\nu]}{}_{\sigma}{}^{J}\big)^{-} ~.
  }
At degree $=3$, the new terms we found above are
\eqa{
   \text{new deg. 3} & &&= +\highlightYellow{2\,e\,\CF_{IJ}\big(\n_{[\mu}\Phi_{\nu]}\big)^{-}\lambda^{I}\big(\Psi^{\rho[\mu}\chi^{\nu]}{}_{\rho}{}^{J}\big)^{-}} - 2\,e\,\CF_{IJ}\big(\Phi_{[\mu}D_{\nu]}\lambda^{I}\big)^{-}\big(\Psi^{\rho[\mu}\chi^{\nu]}{}_{\rho}{}^{J}\big)^{-} \nonumber\\
& &&\quad - 2\,e\,\CF_{IJ}\Psi^{\mu\nu}\Phi^{\sigma}\big(D_{\mu}\lambda^{I}\big)\chi_{\sigma\nu}{}^{J} + e\,\CF_{IJ}\Psi_{\rho}{}^{\rho}\Phi^{\sigma}\big(D_{\mu}\lambda^{I}\big)\chi_{\sigma}{}^{\mu}{}^{J} \nonumber\\
& &&\quad -2\,e\,\Phi^{\alpha}\CF_{IJK}\big(\Psi^{\rho[\mu}\chi^{\nu]}{}_{\rho}{}^{I}\big)^{-}\chi_{\alpha\mu}{}^{J}\psi_{\nu}{}^{K} ~,
  }
  whereas we recall
  \eqa{
  &\left.e\sfC_{-}\right|_{\text{deg}\,=\,3}^{\text{orig}} &&= -\highlightYellow{2\,e\big(\n^{[\mu}\Phi^{\nu]}\big)^{-}\CF_{IJ}\lambda^{I}\big(\Psi_{[\mu}{}^{\rho}\chi_{\nu]\rho}{}^{J}\big)^{-}} ~.\label{eq:eCminus orig deg 3 repeated again}
}
Therefore,
  \beqas{
    &\left.e\sfC_{-}\right|_{\text{deg}\,=\,3}^{\text{shifted}} &&= - 2\,e\,\CF_{IJ}\big(\Phi_{[\mu}D_{\nu]}\lambda^{I}\big)^{-}\big(\Psi^{\rho[\mu}\chi^{\nu]}{}_{\rho}{}^{J}\big)^{-} - 2\,e\,\CF_{IJ}\Psi^{\mu\nu}\Phi^{\sigma}\big(D_{\mu}\lambda^{I}\big)\chi_{\sigma\nu}{}^{J}\nonumber\\
    & &&\quad + e\,\CF_{IJ}\Psi_{\rho}{}^{\rho}\Phi^{\sigma}\big(D_{\mu}\lambda^{I}\big)\chi_{\sigma}{}^{\mu}{}^{J} -2\,e\,\Phi^{\alpha}\CF_{IJK}\big(\Psi^{\rho[\mu}\chi^{\nu]}{}_{\rho}{}^{I}\big)^{-}\chi_{\alpha\mu}{}^{J}\psi_{\nu}{}^{K} ~.
}
At degree $=4$, the new terms we found are
\eqa{
&\text{new deg. 4} &&= - \highlightYellow{2\,e\big[(\n_{[\mu}\Phi_{\nu]})^{-}\big]^2 \CF_{IJ}\lambda^{I}\lambda^{J}} \nonumber\\
& &&\quad + \highlightOrange{4\,e\big(\n_{[\mu}\Phi_{\nu]}\big)^{-}\CF_{IJ}\lambda^{I}\big(\Phi^{[\mu}D^{\nu]}\lambda^{J}\big)^{-}} \nonumber\\
& &&\quad +e\,\Phi^{\alpha}\Phi^{\beta}\CF_{IJK}\big(F_{-}^{\mu\nu}{}^{I} - D^{\mu\nu}{}^{I}\big)\chi_{\alpha\mu}{}^{I}\chi_{\beta\nu}{}^{K} \nonumber\\
& &&\quad + \highlightOliveGreen{4\,e\,\big(\n^{[\mu}\Phi^{\nu]}\big)^{-}\Phi^{\alpha}\CF_{IJK}\lambda^{I}\chi_{\alpha\mu}{}^{J}\psi_{\nu}{}^{K}} \nonumber\\
& &&\quad -2\,e\,\CF_{IJK}\Phi^{\sigma}\psi_{\sigma}{}^{I}\big(\Phi_{[\mu}D_{\nu]}\lambda^{J}\big)\chi^{\mu\nu}{}^{K} \nonumber\\
& &&\quad +\frac{1}{2}\Phi^{\alpha}\Phi^{\beta}\veps^{\mu\nu\rho\sigma}\CF_{IJKL}\psi_{\mu}{}^{I}\psi_{\nu}{}^{J}\chi_{\alpha\rho}{}^{K}\chi_{\beta\sigma}{}^{L} \nonumber\\
& &&\quad +\highlightYellow{4\,e\,\big[(\n^{[\mu}\Phi^{\nu]})^{-}\big]^2 \CF_{IJ}\lambda^{I}\lambda^{J}} - \highlightOrange{4\,e\,\big(\n^{[\mu}\Phi^{\nu]}\big)^{-}\CF_{IJ}\lambda^{I}\Phi_{[\mu}D_{\nu]}\lambda^{J}} \nonumber\\
& &&\quad - \highlightOliveGreen{4\,e\,\big(\n^{[\mu}\Phi^{\nu]}\big)^{-}\Phi^{\alpha}\CF_{IJK}\lambda^{I}\chi_{\alpha\mu}{}^{J}\psi_{\nu}{}^{K}}  ~,
}
whereas we recall,
\eqa{
  &\left.e\sfC_{-}\right|_{\text{deg}\,=\,4}^{\text{orig}} &&= -\highlightYellow{2\,e\big[\big(\n_{[\mu}\Phi_{\nu]}\big)^{-}\big]^{2}\CF_{IJ}\lambda^{I}\lambda^{J}} ~.\label{eq:eCminus orig deg 4 repeated again}
}
Therefore,
  \beqa{
    &\left.e\sfC_{-}\right|_{\text{deg}\,=\,4}^{\text{shifted}} &&= +e\,\Phi^{\alpha}\Phi^{\beta}\CF_{IJK}\big(F_{-}^{\mu\nu}{}^{I} - D^{\mu\nu}{}^{I}\big)\chi_{\alpha\mu}{}^{I}\chi_{\beta\nu}{}^{K} \nonumber\\
& &&\quad -2\,e\,\CF_{IJK}\Phi^{\sigma}\psi_{\sigma}{}^{I}\big(\Phi_{[\mu}D_{\nu]}\lambda^{J}\big)\chi^{\mu\nu}{}^{K} \nonumber\\
& &&\quad +\frac{1}{2}\Phi^{\alpha}\Phi^{\beta}\veps^{\mu\nu\rho\sigma}\CF_{IJKL}\psi_{\mu}{}^{I}\psi_{\nu}{}^{J}\chi_{\alpha\rho}{}^{K}\chi_{\beta\sigma}{}^{L} ~.
}
There were no degree $=5$ or degree $=6$ terms in $\left.e\sfC_{-}\right|^{\text{orig.}}$ so the only ones in the shifted $e\sfC_{-}$ are the ones spawned by the shifts. In particular,
\beqa{
    &\left.e\sfC_{-}\right|_{\text{deg}\,=\,5}^{\text{shifted}} &&= + e\,\Phi^{\alpha}\Phi^{\beta}\CF_{IJK}\big(\Psi^{\rho[\mu}\chi^{\nu]}{}_{\rho}{}^{I}\big)^{-}\chi_{\alpha\mu}{}^{J}\chi_{\beta\nu}{}^{K} ~,\\
    &\left.e\sfC_{-}\right|_{\text{deg}\,=\,6}^{\text{shifted}} &&= -\frac{1}{3}\Phi^{\alpha}\Phi^{\beta}\Phi^{\gamma}\veps^{\mu\nu\rho\sigma}\CF_{IJKL}\psi_{\mu}{}^{I}\chi_{\alpha\nu}{}^{J}\chi_{\beta\rho}{}^{K}\chi_{\gamma\sigma}{}^{L} ~.
}
\noindent \textbf{\underline{Final Comparisons.}} To begin, note that
\beqa{
&\left.e\sfC_{-}\right|_{\text{deg}\,=\,0}^{\text{shifted}} - \left(\IQ \mathsf{V} + \sfC\right)_{\text{deg.}\,=\,0}^{\text{unbarred}} &&= \partial_{\mu}\left(-2\,e\,\CF_{I}D^{\mu}\lambda^{I} + 2\,e\,\CF_{IJ}\psi_{\nu}{}^{I}\chi^{\mu\nu}{}^{J}\right) ~, \label{eq:deg-0-diff-final}\\
&\left.e\sfC_{-}\right|_{\text{deg}\,=\,1}^{\text{shifted}} - \left(\IQ \mathsf{V} + \sfC\right)_{\text{deg.}\,=\,1}^{\text{unbarred}} &&= 0 ~. \label{eq:deg-1-diff-final}
}
Next, we examine the difference at degree $=2$.
It is useful to express $\left.e\sfC_{-}\right|_{\text{deg}\,=\,2}^{\text{shifted}}$ as 
\eqa{
 &\left.e\sfC_{-}\right|_{\text{deg}\,=\,2}^{\text{shifted}} &&= \sqrt{g}\Phi^{\sigma}Z^{\text{sugra-unbarred}}_{\sigma} - \sqrt{g}\frac{1}{2}\CF_{IJ}\big(\Psi_{[\mu}{}^{\rho}\chi_{\nu]\rho}{}^{I}\big)^{-}\big(\Psi^{\sigma[\mu}\chi^{\nu]}{}_{\sigma}{}^{J}\big)^{-} ~,
}
where
\eqa{
& Z^{\text{sugra-unbarred}}_{\sigma} &&= -2\,\CF_{IJ}\big(D_{\sigma}\lambda^{I}\big)[\lambda,\phi]^{J} - 2\,\CF_{IJ}\chi_{\sigma}{}^{\mu}{}^{I}[\psi_{\mu}, \lambda]^{J} \nonumber\\
& &&\quad - 2\,\CF_{IJ}\big(F_{\sigma\nu}^{-}{}^{I} + D_{\sigma\nu}{}^{J}\big)D^{\nu}\lambda^{J} - 2\,\CF_{IJ}\big(D_{\rho}\chi^{\kappa\rho}{}^{I}\big)\chi_{\sigma\kappa}{}^{J} \nonumber\\
& &&\quad + 2\,\CF_{IJ}\big(D_{\rho}\eta^{I}\big)\chi_{\sigma}{}^{\rho}{}^{J} + 2\,\CF_{IJ}[\lambda, \eta]^{I}\psi_{\sigma}{}^{J} \nonumber\\
& &&\quad - 2\,\CF_{IJK}\big(F_{-}^{\mu\nu}{}^{I} - D^{\mu\nu}{}^{I}\big)\chi_{\sigma\mu}{}^{J}\psi_{\nu}{}^{K} + 2\,\CF_{IJK}\big(D^{\nu}\lambda^{I}\big)\psi_{\sigma}{}^{J}\psi_{\nu}{}^{K} \nonumber\\
& &&\quad + \frac{e^{-1}}{3}\veps^{\mu\nu\rho\alpha}\CF_{IJKL}\psi_{\mu}{}^{I}\psi_{\nu}{}^{J}\psi_{\rho}{}^{K}\chi_{\alpha\sigma}{}^{L} ~. \label{eq:z-sugra-unbarred}
}
On the other hand,
\eqa{
& \big(\IQ\mathsf{V} + \sfC\big)_{\text{deg.}\,=2}^{\text{unbarred}} &&= \sqrt{g}\,\Phi^{\sigma}Z_{\sigma}^{\text{Cartan-unbarred}} + \sqrt{g}\,\Psi^{\mu\sigma}\Psi^{\nu}{}_{\sigma}\Upsilon_{\mu\nu}^{\text{Cartan-unbarred}} + \partial_{\mu}\left[ - \sqrt{g}\,\CF_{IJ}\Phi^{\sigma}\chi_{\nu\sigma}{}^{I}\chi^{\mu\nu}{}^{J}\right] ~,
}
where
\eqa{
 &Z_{\sigma}^{\text{Cartan-unbarred}} &&= \frac{1}{2}\CF_{IJK}\psi_{\sigma}{}^{I}(F^{+}_{\mu\nu}{}^{J} + D_{\mu\nu}{}^{J}\big)\chi^{\mu\nu}{}^{K} + \frac{1}{2}\CF_{IJ} \big(D_{\sigma}\chi_{\mu\nu}{}^{I}\big)\chi^{\mu\nu}{}^{J} \nn
 & &&\quad + \CF_{IJK}(D_{\mu}\phi^{I})\chi_{\nu\sigma}{}^{J}\chi^{\mu\nu}{}^{K} + \CF_{IJ}\big(D_{\mu}\chi_{\nu\sigma}{}^{I}\big)\chi^{\mu\nu}{}^{J} \nn
 & &&\quad + \CF_{IJ}\chi_{\nu\sigma}{}^{I}D_{\mu}\chi^{\mu\nu}{}^{J} + 2\,\CF_{IJK}\psi_{\sigma}{}^{I}\psi_{\rho}{}^{J}D^{\rho}\lambda^{K}  \nn
 & &&\quad - 2\,\CF_{IJ}F_{\sigma\rho}{}^{I}D^{\rho}\lambda^{J} - \CF_{IJKL}\psi_{\sigma}{}^{I}\psi_{\mu}{}^{J}\psi_{\nu}{}^{K}\chi^{\mu\nu}{}^{L}\nn
 & &&\quad + 2\,\CF_{IJK}F_{\sigma\mu}{}^{I}\psi_{\nu}{}^{J}\chi^{\mu\nu}{}^{K} - 2\,\CF_{IJ}\psi_{\sigma}{}^{I}[\lambda, \eta]^{J} + 2\,\CF_{I}[\lambda, D_{\sigma}\lambda]^{I} ~, \label{eq:z-cartan-unbarred}\\
 &\Upsilon_{\mu\nu}^{\text{Cartan-unbarred}} &&= \frac{1}{4}\CF_{IJ}\chi_{\mu\rho}{}^{I}\chi_{\nu}{}^{\rho}{}^{J} ~.
}
The relevant difference term is
\eqa{
& \left.e\sfC_{-}\right|_{\text{deg}\,=\,2}^{\text{shifted}} - \big(\IQ\mathsf{V} + \sfC\big)_{\text{deg.}\,=2}^{\text{unbarred}} &&= \partial_{\mu}\left[\sqrt{g}\,\CF_{IJ}\Phi^{\sigma}\chi_{\nu\sigma}{}^{I}\chi^{\mu\nu}{}^{J}\right] + \sqrt{g}\Phi^{\sigma}\big(Z_{\sigma}^{\text{sugra-unbarred}} - Z_{\sigma}^{\text{Cartan-unbarred}}\big) \nonumber\\
& &&\quad - \sqrt{g}\frac{1}{2}\CF_{IJ}\big(\Psi_{[\mu}{}^{\rho}\chi_{\nu]\rho}{}^{I}\big)^{-}\big(\Psi^{\sigma[\mu}\chi^{\nu]}{}_{\sigma}{}^{J}\big)^{-} \nonumber\\
& &&\quad - \sqrt{g}\,\Psi^{\mu\sigma}\Psi^{\nu}{}_{\sigma}\Upsilon_{\mu\nu}^{\text{Cartan-unbarred}} \\
& &&= \partial_{\mu}\left[\sqrt{g}\,\CF_{IJ}\Phi^{\sigma}\chi_{\nu\sigma}{}^{I}\chi^{\mu\nu}{}^{J}\right] + \sqrt{g}\Phi^{\sigma}\big(Z_{\sigma}^{\text{sugra-unbarred}} - Z_{\sigma}^{\text{Cartan-unbarred}}\big) \nonumber\\
& &&\quad - \sqrt{g}\frac{1}{2}\CF_{IJ}\big(\Psi_{[\mu}{}^{\rho}\chi_{\nu]\rho}{}^{I}\big)^{-}\big(\Psi^{\sigma[\mu}\chi^{\nu]}{}_{\sigma}{}^{J}\big)^{-} \nonumber\\
& &&\quad - \sqrt{g}\,\Psi^{\mu\sigma}\Psi^{\nu}{}_{\sigma}\frac{1}{4}\CF_{IJ}\chi_{\mu\rho}{}^{I}\chi_{\nu}{}^{\rho}{}^{J} ~.
}
Consider
\eqa{
&\sqrt{g}\Phi^{\sigma}\big(Z_{\sigma}^{\text{sugra-unbarred}} - Z_{\sigma}^{\text{Cartan-unbarred}}\big) ~.
}
We begin with the $\Psi \chi \chi$ terms. The difference is
\eqa{
& - \sqrt{g}\frac{1}{2}\CF_{IJ}\big(\Psi_{[\mu}{}^{\rho}\chi_{\nu]\rho}{}^{I}\big)^{-}\big(\Psi^{\sigma[\mu}\chi^{\nu]}{}_{\sigma}{}^{J}\big)^{-}  - \sqrt{g}\,\Psi^{\mu\sigma}\Psi^{\nu}{}_{\sigma}\frac{1}{4}\CF_{IJ}\chi_{\mu\rho}{}^{I}\chi_{\nu}{}^{\rho}{}^{J} ~.\label{eq:diff-deg-2-term0}
}
First of all, from \eqref{eq:identB}, we have $\big(\Psi_{[\mu}{}^{\rho}\chi_{\nu]\rho}\big)^{+} = -\frac{1}{4}\Psi_{\sigma}{}^{\sigma}\chi_{\mu\nu}$, and so $\CF_{IJ}\big(\Psi_{[\mu}{}^{\rho}\chi_{\nu]\rho}{}^{I}\big)^{+}\big(\Psi^{\sigma[\mu}\chi^{\nu]}{}_{\sigma}{}^{J}\big)^{+} = 0$ and therefore, by straightforward expansion,
\eqa{
& -\frac{1}{2}\CF_{IJ}\big(\Psi_{[\mu}{}^{\rho}\chi_{\nu]\rho}{}^{I}\big)^{-}\big(\Psi^{\sigma[\mu}\chi^{\nu]}{}_{\sigma}{}^{J}\big)^{-} &&= -\frac{1}{2}\CF_{IJ}\big(\Psi_{[\mu}{}^{\rho}\chi_{\nu]\rho}{}^{I}\big)\big(\Psi^{\sigma[\mu}\chi^{\nu]}{}_{\sigma}{}^{J}\big) \nonumber\\
& &&= \frac{1}{8}\CF_{IJ}\big(\Psi_{\mu}{}^{\rho}\Psi^{\sigma\mu}\chi_{\nu\rho}{}^{I}\chi^{\nu}{}_{\sigma}{}^{J} - \Psi_{\mu}{}^{\rho}\Psi^{\sigma\nu}\chi_{\nu\rho}{}^{I}\chi^{\mu}{}_{\sigma}{}^{J} \nonumber\\
& &&\qquad\qquad  - \Psi_{\nu}{}^{\rho}\Psi^{\sigma\mu}\chi_{\mu\rho}{}^{I}\chi^{\nu}{}_{\sigma}{}^{J} + \Psi_{\nu}{}^{\rho}\Psi^{\sigma\nu}\chi_{\mu\rho}{}^{I}\chi^{\mu}{}_{\sigma}{}^{J}\big) \nonumber\\
& &&= -\frac{1}{4}\CF_{IJ}\big(\Psi^{\rho\sigma}\Psi_{\mu\rho}\chi_{\nu\sigma}{}^{I}\chi^{\mu\nu}{}^{J} + \Psi_{\mu}{}^{\rho}\Psi^{\sigma\nu}\chi_{\nu\rho}{}^{I}\chi^{\mu}{}_{\sigma}{}^{J}\big) ~.
}
In the last equality, we have performed some index relabelings. Next, we use the self-duality of $\chi_{\nu\rho}{}^{I}$ in the second term to write it as
\eqa{
& \CF_{IJ}\Psi_{\mu}{}^{\rho}\Psi^{\sigma\nu}\chi_{\nu\rho}{}^{I}\chi^{\mu}{}_{\sigma}{}^{J} &&= \frac{e}{2}\CF_{IJ}\Psi^{\mu}{}^{\rho}\Psi^{\sigma\nu}\veps_{\nu\rho\alpha\beta}\chi^{\alpha\beta}{}^{I}\chi_{\mu\sigma}{}^{J} = 0 ~,
}
which vanishes because the $\veps$ symbol antisymmetrizes over $\nu, \rho$ resulting in the gravitino bilinear being replaced by $\frac{1}{2}\big(\Psi^{\mu\rho}\Psi^{\sigma\nu} - \Psi^{\mu\nu}\Psi^{\sigma\rho}\big) = \frac{1}{2}\big(\Psi^{\mu\rho}\Psi^{\sigma\nu} + \Psi^{\sigma\rho}\Psi^{\mu\nu}\big)$, which being manifestly symmetric in $\mu, \sigma$, is killed by contraction with $\chi_{\mu\sigma}{}^{J}$. Therefore,
\eqa{
& -\frac{1}{2}\CF_{IJ}\big(\Psi_{[\mu}{}^{\rho}\chi_{\nu]\rho}{}^{I}\big)^{-}\big(\Psi^{\sigma[\mu}\chi^{\nu]}{}_{\sigma}{}^{J}\big)^{-} &&= -\frac{1}{4}\CF_{IJ}\Psi^{\rho\sigma}\Psi_{\mu\rho}\chi_{\nu\sigma}{}^{I}\chi^{\mu\nu}{}^{J} \nonumber\\
& &&= +\frac{1}{4}\CF_{IJ}\Psi^{\mu\sigma}\Psi^{\nu}{}_{\sigma}\chi_{\mu\rho}{}^{I}\chi_{\nu}{}^{\rho}{}^{J} ~.
}
Therefore 
\beqa{
& \eqref{eq:diff-deg-2-term0} &&= 0 ~.\label{eq:diff-deg-2-term0-final}
}
Next, consider the $\psi\psi\psi\chi$ terms. From \eqref{eq:z-sugra-unbarred} and \eqref{eq:z-cartan-unbarred}, the difference is
\eqa{
&\frac{e^{-1}}{3}\Phi^{\sigma}\veps^{\mu\nu\rho\alpha}\CF_{IJKL}\psi_{\mu}{}^{I}\psi_{\nu}{}^{J}\psi_{\rho}{}^{K}\chi_{\alpha\sigma}{}^{L} + \Phi^{\sigma}\CF_{IJKL}\psi_{\sigma}{}^{I}\psi_{\mu}{}^{J}\psi_{\nu}{}^{K}\chi^{\mu\nu}{}^{L} ~.\label{eq:diff-deg-2-term1}
}
Note that
\eqa{
&\CF_{IJKL}\veps^{\mu\nu\rho\sigma}\psi_{\mu}{}^{I}\psi_{\nu}{}^{J}\psi_{\rho}{}^{K}\Phi^{\sigma}\chi_{\alpha\sigma}{}^{L} d^{4}x &&= -\CF_{IJKL}\psi^{I}\wedge\psi^{J}\wedge\psi^{K}\wedge \iota_{\Phi}\chi^{L} ~,\\
&\CF_{IJKL}\Phi^{\sigma}\psi_{\sigma}{}^{I}\psi_{\mu}{}^{J}\psi_{\nu}{}^{K}\chi^{\mu\nu}{}^{L}d^{4}x &&= e^{-1}\CF_{IJKL}\big(\iota_{\Phi}\psi^{I}\big)\psi^{I}\wedge \psi^{J} \wedge \chi^{K} ~,
}
and since $\psi \wedge \psi \wedge \psi \wedge \chi = 0$ on a four-manifold, we have the identity
\eqa{
&\CF_{IJKL}\big(\iota_{\Phi}\psi^{I}\big)\psi^{J}\wedge\psi^{K}\wedge\chi^{L} &&= \frac{1}{3}\CF_{IJKL}\psi^I \wedge\psi^J \wedge \psi^K\wedge\iota_{\Phi}\chi^{L} ~.
}
Therefore, \eqref{eq:diff-deg-2-term1} vanishes:
\beqa{
&\bigg(\frac{e^{-1}}{3}\Phi^{\sigma}\veps^{\mu\nu\rho\alpha}\CF_{IJKL}\psi_{\mu}{}^{I}\psi_{\nu}{}^{J}\psi_{\rho}{}^{K}\chi_{\alpha\sigma}{}^{L} + \Phi^{\sigma}\CF_{IJKL}\psi_{\sigma}{}^{I}\psi_{\mu}{}^{J}\psi_{\nu}{}^{K}\chi^{\mu\nu}{}^{L}\bigg)d^{4}x \nonumber\\
&= -\frac{e^{-1}}{3}\CF_{IJKL}\psi^{I}\wedge\psi^{J}\wedge\psi^{K}\wedge \iota_{\Phi}\chi^{L} + \frac{e^{-1}}{3}\CF_{IJKL}\psi^I \wedge\psi^J \wedge \psi^K\wedge\iota_{\Phi}\chi^{L}  \nonumber\\
&= 0 ~. \label{eq:diff-deg-2-term1-final}
}
Next, consider the $D\psi\chi$ terms (where $D$ is the auxiliary field). These are
\beqa{
& 2\Phi^{\sigma}\CF_{IJK}D^{\mu\nu}{}^{I}\chi_{\sigma\mu}{}^{J}\psi_{\nu}{}^{K} - \frac{1}{2}\Phi^{\sigma}\CF_{IJK}\psi_{\sigma}{}^{I}D_{\mu\nu}{}^{J}\chi^{\mu\nu}{}^{K} \nonumber\\
&= \frac{1}{2}\Phi^{\sigma}\CF_{IJK}D^{\mu\nu}{}^{I}\left(4\chi_{\sigma\mu}{}^{I}\psi_{\nu}{}^{K} - \psi_{\sigma}{}^{I}\chi_{\mu\nu}{}^{K} \right) ~. \label{eq:diff-deg-2-term2-final}
}
Next, consider the $F\psi\chi$ terms. These are
\beqa{
& -2\,\CF_{IJK}\Phi^{\sigma}F_{-}^{\mu\nu}{}^{I}\chi_{\sigma\mu}{}^{I}\psi_{\nu}{}^{K} - \frac{1}{2}\Phi^{\sigma}\CF_{IJK}\psi_{\sigma}{}^{I}F_{\mu\nu}^{+}{}^{J}\chi^{\mu\nu}{}^{K} - 2\,\Phi^{\sigma}\CF_{IJK}F_{\sigma\mu}{}^{I}\psi_{\nu}{}^{J}\chi^{\mu\nu}{}^{K} \nonumber\\
&=\frac{1}{2}\CF_{IJK}\Phi^{\sigma}\big(-4 F_{\mu\nu}^{-}{}^{I}\psi^{\mu}{}^{J}\chi_{\sigma}{}^{\nu}{}^{K} - F_{\mu\nu}^{+}{}^{I}\psi_{\sigma}{}^{J}\chi^{\mu\nu}{}^{K} - 4 F_{\sigma\mu}{}^{I}\psi_{\nu}{}^{J}\chi^{\mu\nu}{}^{K}\big) \nonumber\\
&= \frac{1}{2}\CF_{IJK}\Phi^{\sigma}\big(F_{\mu\nu}^{+}{}^{I}\psi_{\sigma}{}^{J}\chi^{\mu\nu}{}^{K} - 4 F_{\mu\nu}^{+}{}^{I}\psi^{\mu}{}^{J}\chi_{\sigma}{}^{\nu}{}^{K}\big) ~. \label{eq:diff-deg-2-term3-final}
}
In the last equality, we have used the following fact:
Since $\CF_{IJK}F^{I} \wedge \psi^{J} \wedge \chi^{K} = 0$ on a four-manifold, we have the identity, 
\eqa{
&\CF_{IJK}\big(\iota_{\Phi}F^{I}\big)\wedge \psi^{J} \wedge \chi^{K} + \CF_{IJK}\big(\iota_{\Phi}\psi^{I}\big)F^{J} \wedge \chi^{K} - \CF_{IJK}F^{I}\wedge \psi^{J} \wedge \big(\iota_{\Phi}\chi^{K}\big) &&= 0 ~,
}	
which in component form reads
\eqa{
&\Phi^{\sigma}\CF_{IJK}\left(F_{\sigma\mu}{}^{I}\psi_{\nu}{}^{J}\chi^{\mu\nu}{}^{K} + \frac{1}{2}F_{\mu\nu}^{+}{}^{I}\psi_{\sigma}{}^{J}\chi^{\mu\nu}{}^{K} - F_{\mu\nu}^{+}{}^{I}\psi^{\mu}{}^{J}\chi_{\sigma}{}^{\nu}{}^{K} + F_{\mu\nu}^{-}{}^{I}\psi^{\mu}{}^{J}\chi_{\sigma}{}^{\nu}{}^{K}\right) &&= 0 ~. \label{eq:Fpsichi identity 1}
}
\noindent Next, let us consider the $\chi D\chi$ terms (where $D$ is the covariant derivative). These are
\eqa{
&-2\,\Phi^{\sigma}\CF_{IJ}\big(D_{\rho}\chi^{\kappa\rho}{}^{I}\big)\chi_{\sigma\kappa}{}^{J} - \frac{1}{2}\CF_{IJ}\Phi^{\sigma}\big(D_{\sigma}\chi_{\mu\nu}{}^{I}\big)\chi^{\mu\nu}{}^{J}\nonumber\\
& - \CF_{IJ}\Phi^{\sigma}\big(D_{\mu}\chi_{\nu\sigma}{}^{I}\big)\chi^{\mu\nu}{}^{J} - \CF_{IJ}\Phi^{\sigma}\chi_{\nu\sigma}{}^{I}D_{\mu}\chi^{\mu\nu}{}^{J} - \CF_{IJK}\Phi^{\sigma}\big(D_{\mu}\phi^{I}\big)\chi_{\nu\sigma}{}^{J}\chi^{\mu\nu}{}^{K} \nonumber\\
& = + 2\,\Phi^{\sigma}\CF_{IJ}\chi_{\nu\sigma}{}^{I}\big(D_{\mu}\chi^{\mu\nu}{}^{I}\big) - \frac{1}{2}\CF_{IJ}\Phi^{\sigma}\big(D_{\sigma}\chi_{\mu\nu}{}^{I}\big)\chi^{\mu\nu}{}^{J}\nonumber\\
& - \CF_{IJ}\Phi^{\sigma}\big(D_{\mu}\chi_{\nu\sigma}{}^{I}\big)\chi^{\mu\nu}{}^{J} - \CF_{IJ}\Phi^{\sigma}\chi_{\nu\sigma}{}^{I}D_{\mu}\chi^{\mu\nu}{}^{J} - \CF_{IJK}\Phi^{\sigma}\big(D_{\mu}\phi^{I}\big)\chi_{\nu\sigma}{}^{J}\chi^{\mu\nu}{}^{K} \nonumber\\
&= -\frac{1}{2}\CF_{IJ}\Phi^{\sigma}\big(D_{\sigma}\chi_{\mu\nu}{}^{I}\big)\chi^{\mu\nu}{}^{J} - \CF_{IJ}\Phi^{\sigma}\big(D_{\mu}\chi_{\nu\sigma}{}^{I}\big)\chi^{\mu\nu}{}^{J} + \CF_{IJ}\Phi^{\sigma}\chi_{\nu\sigma}{}^{I}D_{\mu}\chi^{\mu\nu}{}^{J} \nonumber\\
& \quad + \CF_{IJK}\Phi^{\sigma}\big(D_{\mu}\phi^{I}\big)\chi_{\sigma\nu}{}^{I}\chi^{\mu\nu}{}^{K} ~, \label{eq:diff-deg-2-term-4}
}
Since $\CF_{IJ}D\chi^{I}\wedge \chi^{J} = 0$ on a four-manifold, we have the identity
\eqa{
& \CF_{IJ}\big(\iota_{\Phi}D\chi^{I}\big)\wedge \chi^{J} - \CF_{IJ} D\chi^{I}\wedge \big(\iota_{\Phi}\chi^{J}\big) &&= 0 ~. \label{eq:ident Dchi chi}
}	
Note that $D_{[\sigma}\chi_{\mu\nu]}{}^{I} = \frac{1}{3}\big(D_{\sigma}\chi_{\mu\nu}{}^{I} + D_{\mu}\chi_{\nu\sigma}{}^{I} + D_{\nu}\chi_{\sigma\mu}{}^{I}\big)$, $D\chi^{I} = \frac{1}{3!}\big(3\cdot D_{[\mu}\chi_{\nu\rho]}{}^{I}\big)dx^{\mu} \wedge dx^{\nu} \wedge dx^{\rho}$, and $\iota_{\Phi}D\chi^{I} = \frac{3}{2}\Phi^{\sigma}D_{[\sigma}\chi_{\mu\nu]}{}^{I}dx^{\mu}\wedge dx^{\nu}$, so
\eqa{
& \CF_{IJ}\big(\iota_{\Phi}D\chi^{I}\big) \wedge \chi^{J} &&= \frac{3}{2}\sqrt{g}\Phi^{\sigma}\big(D_{[\sigma}\chi_{\mu\nu]}{}^{I}\big)\chi^{\mu\nu}{}^{J} d^{4}x ~,\\
& \CF_{IJ} D\chi^{I} \wedge \big(\iota_{\Phi}\chi^{J}\big) &&= \sqrt{g}\,\CF_{IJ}\Phi^{\sigma}\big(D_{\mu}\chi^{\mu\nu}{}^{I}\big)\chi_{\sigma\nu}{}^{J} d^{4}x ~,
} 
so the component form of \eqref{eq:ident Dchi chi} reads
\eqa{
&  \frac{1}{2}\Phi^{\sigma}\big(D_{\sigma}\chi_{\mu\nu}{}^{I} + D_{\mu}\chi_{\nu\sigma}{}^{I} + D_{\nu}\chi_{\sigma\mu}{}^{I}\big)\chi^{\mu\nu}{}^{J} + \Phi^{\sigma}\big(D_{\mu}\chi^{\mu\nu}{}^{I}\big)\chi_{\nu\sigma}{}^{J} &&= 0 ~,
}
which reduces to
\eqa{
& \frac{1}{2}\Phi^{\sigma}\big(D_{\sigma}\chi_{\mu\nu}{}^{I}\big)\chi^{\mu\nu}{}^{J} + \Phi^{\sigma}\big(D_{\mu}\chi_{\nu\sigma}{}^{I}\big)\chi^{\mu\nu}{}^{J} - \Phi^{\sigma}\chi_{\nu\sigma}{}^{I}\big(D_{\mu}\chi^{\mu\nu}{}^{I}\big) &&= 0 ~.
}
So the first line of \eqref{eq:diff-deg-2-term-4} actually vanishes. Therefore, \eqref{eq:diff-deg-2-term-4} reduces to a single term:
\beqa{
 &\eqref{eq:diff-deg-2-term-4} &&= \CF_{IJK}\Phi^{\sigma}\big(D_{\mu}\phi^{I}\big)\chi_{\sigma\nu}{}^{I}\chi^{\mu\nu}{}^{K} ~. \label{eq:diff-deg-2-term-4-final}
 }
 Next, consider the $F D \lambda$ terms (where $D$ is the covariant derivative). These are
 \beqa{
  & -2\Phi^{\sigma}\CF_{IJ}F_{\sigma\nu}^{-}{}^{I}D^{\nu}\lambda^{J} + 2 \CF_{IJ} F_{\sigma}{}^{I}D^{\rho}\lambda^{J} &&= 2\Phi^{\sigma}\CF_{IJ}F_{\sigma\nu}^{+}D^{\nu}\lambda^{I} ~. \label{eq:diff-deg-2-term-5-final}
 }
 Last, but not the least, consider the $D_{\sigma}\lambda\cdot[\lambda,\phi]$ and $[\lambda, D_{\sigma}\lambda]$ terms. Their difference is
 \eqa{
 -2\,e\,\Phi^{\sigma}\CF_{IJ}\big(D_{\sigma}\lambda^{I}\big)[\lambda,\phi]^{J} - 2\,e\,\Phi^{\sigma} \CF_{I}[\lambda, D_{\sigma}\lambda]^{I} ~.\label{eq:marvel-of-homogeneity}
 }
 We claim that \eqref{eq:marvel-of-homogeneity} vanishes! First, let us recall an important fact. 
 The prepotential $\CF$ is gauge-invariant:
 \eqa{
  & \CF_{I}[\Theta, \phi]^{I} &&\equiv \CF_{I} \Theta^{J}\phi^{K}f^{I}{}_{JK} = 0 ~,
 }
 (where $f^{I}{}_{JK}$ are the structure constants of $\mathsf{Lie}(G)$), and of homogeneity degree $2$:
 \eqa{
  & 2\CF &&= \CF_{I}\phi^{I} ~, \quad \CF_{I} = \CF_{IJ}\phi^{J} ~, \quad \CF_{IJK}\phi^{K} = 0 ~.
 }
 Combining these properties, we find
 \beqa{
  & \CF_{IJ}f^{I}{}_{LK} + \CF_{IL}f^{I}{}_{JK} &&= 0 ~.\label{eq:marvelous-identity}
 }
The first term in \eqref{eq:marvel-of-homogeneity} can be written as
\eqa{
&-2\,e\,\Phi^{\sigma}\CF_{IJ}\big(D_{\sigma}\lambda^{I}\big)[\lambda,\phi]^{J} &&= -2\,e\,\Phi^{\sigma}\CF_{IJ}\big(D_{\sigma}\lambda^{I}\big)f^{J}{}_{KL}\lambda^{K}\phi^{L} \nonumber\\
& &&= -2\,e\,\Phi^{\sigma}\big(\CF_{IJ}f^{I}{}_{KL}\big)\big(D_{\sigma}\lambda^{J}\big)\lambda^{K}\phi^{L} &&\quad \text{(symmetry of $\CF_{IJ}$)} \nonumber\\
& &&= +2\,e\,\Phi^{\sigma}\big(\CF_{IJ}f^{I}{}_{LK}\big)\big(D_{\sigma}\lambda^{J}\big)\lambda^{K}\phi^{L} &&\quad \text{($f^{I}{}_{JK} = -f^{I}{}_{KJ}$)}\nonumber\\
& &&= -2\,e\,\Phi^{\sigma}\big(\CF_{IL}f^{I}{}_{JK}\big)\big(D_{\sigma}\lambda^{J}\big)\lambda^{K}\phi^{L}&&\quad \text{(using \eqref{eq:marvelous-identity})} \nonumber\\
& &&= -2\,e\,\Phi^{\sigma}\CF_{IL}[D_{\sigma}\lambda, \lambda]^{I}\phi^{L} \nonumber\\
& &&= +2\,e\,\Phi^{\sigma}\CF_{IJ}\phi^{I}[\lambda, D_{\sigma}\lambda]^{J} ~,
}
which is precisely the second term of \eqref{eq:marvel-of-homogeneity} with the opposite sign. Consequently,
\eqa{
& \eqref{eq:marvel-of-homogeneity} &&= 0 ~.
}
Therefore, the complete degree $=2$ difference is
 \beqa{
 & \left.e\sfC_{-}\right|_{\text{deg}\,=\,2}^{\text{shifted}} - \big(\IQ\mathsf{V} + \sfC\big)_{\text{deg.}\,=2}^{\text{unbarred}} \nonumber\\
 &= \partial_{\mu}\left[\sqrt{g}\,\CF_{IJ}\Phi^{\sigma}\chi_{\nu\sigma}{}^{I}\chi^{\mu\nu}{}^{J}\right] \nonumber\\
&\quad + \sqrt{g}\Phi^{\sigma}\bigg(- 2\,\CF_{IJ}\chi_{\sigma}{}^{\mu}{}^{I}[\psi_{\mu}, \lambda]^{J} + 2\,\CF_{IJ}(F_{\sigma\nu}^{+} - D_{\sigma\nu})D^{\nu}\lambda^{I} - 2\,\CF_{IJ}\chi_{\sigma\mu}{}^{I}D^{\mu}\eta^{J} \nonumber \\
&\qquad\qquad\qquad + \frac{1}{2}\CF_{IJK} D^{\mu\nu}{}^{I}\left(4\chi_{\sigma\mu}{}^{I}\psi_{\nu}{}^{K} - \psi_{\sigma}{}^{I}\chi_{\mu\nu}{}^{K} \right) \nonumber\\
&\qquad\qquad\qquad + \frac{1}{2}\CF_{IJK}F_{\mu\nu}^{+}{}^{I}\big(\psi_{\sigma}{}^{J}\chi^{\mu\nu}{}^{K} - 4 \psi^{\mu}{}^{J}\chi_{\sigma}{}^{\nu}{}^{K}\big) \nonumber\\
&\qquad\qquad\qquad + \CF_{IJK}\big(D_{\mu}\phi^{I}\big)\chi_{\sigma\nu}{}^{I}\chi^{\mu\nu}{}^{K}\bigg) ~. \label{eq:deg-2-diff-final}
}
At degrees $\geq 3$, there is no contribution from $\IQ\mathsf{V}+\sfC$, and consequently,
\beqa{
 & \left.e\sfC_{-}\right|_{\text{deg}\,=\,3}^{\text{shifted}} - \big(\IQ\mathsf{V} + \sfC\big)_{\text{deg.}\,=3}^{\text{unbarred}} \nonumber\\
 & = \left.e\sfC_{-}\right|_{\text{deg}\,=\,3}^{\text{shifted}}\nonumber\\
 & =  - 2\,e\,\CF_{IJ}\big(\Phi_{[\mu}D_{\nu]}\lambda^{I}\big)^{-}\big(\Psi^{\rho[\mu}\chi^{\nu]}{}_{\rho}{}^{J}\big)^{-} - 2\,e\,\CF_{IJ}\Psi^{\mu\nu}\Phi^{\sigma}\big(D_{\mu}\lambda^{I}\big)\chi_{\sigma\nu}{}^{J}\nonumber\\
& \quad + e\,\CF_{IJ}\Psi_{\rho}{}^{\rho}\Phi^{\sigma}\big(D_{\mu}\lambda^{I}\big)\chi_{\sigma}{}^{\mu}{}^{J} -2\,e\,\Phi^{\alpha}\CF_{IJK}\big(\Psi^{\rho[\mu}\chi^{\nu]}{}_{\rho}{}^{I}\big)^{-}\chi_{\alpha\mu}{}^{J}\psi_{\nu}{}^{K} ~. \label{eq:deg-3-diff-final}
}
\beqa{
 & \left.e\sfC_{-}\right|_{\text{deg}\,=\,4}^{\text{shifted}} - \big(\IQ\mathsf{V} + \sfC\big)_{\text{deg.}\,=4}^{\text{unbarred}} \nonumber\\
 & = \left.e\sfC_{-}\right|_{\text{deg}\,=\,4}^{\text{shifted}}\nonumber\\
 & = +e\,\Phi^{\alpha}\Phi^{\beta}\CF_{IJK}\big(F_{-}^{\mu\nu}{}^{I} - D^{\mu\nu}{}^{I}\big)\chi_{\alpha\mu}{}^{I}\chi_{\beta\nu}{}^{K}  -2\,e\,\CF_{IJK}\Phi^{\sigma}\psi_{\sigma}{}^{I}\big(\Phi_{[\mu}D_{\nu]}\lambda^{J}\big)\chi^{\mu\nu}{}^{K} \nonumber\\
& \quad +\frac{1}{2}\Phi^{\alpha}\Phi^{\beta}\veps^{\mu\nu\rho\sigma}\CF_{IJKL}\psi_{\mu}{}^{I}\psi_{\nu}{}^{J}\chi_{\alpha\rho}{}^{K}\chi_{\beta\sigma}{}^{L} ~. \label{eq:deg-4-diff-final}
 }
 \beqa{
 & \left.e\sfC_{-}\right|_{\text{deg}\,=\,5}^{\text{shifted}} - \big(\IQ\mathsf{V} + \sfC\big)_{\text{deg.}\,=5}^{\text{unbarred}} \nonumber\\
& = \left.e\sfC_{-}\right|_{\text{deg}\,=\,5}^{\text{shifted}}\nonumber\\
& = + e\,\Phi^{\alpha}\Phi^{\beta}\CF_{IJK}\big(\Psi^{\rho[\mu}\chi^{\nu]}{}_{\rho}{}^{I}\big)^{-}\chi_{\alpha\mu}{}^{J}\chi_{\beta\nu}{}^{K} ~. \label{eq:deg-5-diff-final}
 }
  \beqa{
 & \left.e\sfC_{-}\right|_{\text{deg}\,=\,6}^{\text{shifted}} - \big(\IQ\mathsf{V} + \sfC\big)_{\text{deg.}\,=6}^{\text{unbarred}} \nonumber\\
& = \left.e\sfC_{-}\right|_{\text{deg}\,=\,6}^{\text{shifted}}\nonumber\\
& = -\frac{1}{3}\Phi^{\alpha}\Phi^{\beta}\Phi^{\gamma}\veps^{\mu\nu\rho\sigma}\CF_{IJKL}\psi_{\mu}{}^{I}\chi_{\alpha\nu}{}^{J}\chi_{\beta\rho}{}^{K}\chi_{\gamma\sigma}{}^{L} ~. \label{eq:deg-6-diff-final}
 }
 It remains to be seen whether these differences add up to a $\IQ$-exact term, up to a total derivative.
 \\\\
 We will work our way down in gravity degree, starting from the degree $=6$ term. We begin by recording a useful identity. 
 Since $\CF_{IJKL}\chi^{I} \wedge (\iota_{\Phi}\chi^{J}) \wedge (\iota_{\Phi}\chi^{K}) \wedge \psi^{L} = 0$ on a four-manifold, we have the identity
 \eqa{
& \CF_{IJKL}\big(\iota_{\Phi}\chi^{I}\big)\wedge\big(\iota_{\Phi}\chi^{J}\big)\wedge\big(\iota_{\Phi}\chi^{K}\big)\wedge \psi^{L} + \CF_{IJKL}\chi^{I}\wedge\big(\iota_{\Phi}\chi^{J}\big)\wedge\big(\iota_{\Phi}{\chi}^{K}\big)\wedge\big(\iota_{\Phi}\psi^{L}\big) &&= 0 ~,
 }	
 which, in components, reads
 \eqa{
& \CF_{IJKL}\veps^{\mu\nu\rho\sigma}\Phi^{\alpha}\Phi^{\beta}\Phi^{\gamma}\left(\psi_{\mu}{}^{I}\chi_{\alpha\nu}{}^{J}\chi_{\beta\rho}{}^{K}\chi_{\gamma\sigma}{}^{L} - \frac{1}{2}\psi_{\gamma}{}^{I}\chi_{\mu\nu}{}^{J}\chi_{\alpha\rho}{}^{K}\chi_{\beta\sigma}{}^{L}\right) &&= 0 ~. \label{eq:ident-for-deg-6}
 }
 Consider 
 \eqa{
 & \IQ\left(\frac{1}{6}\CF_{IJKL}\veps^{\mu\nu\rho\sigma}\Phi^{\alpha}\Phi^{\beta}\chi_{\mu\nu}{}^{I}\chi_{\alpha\rho}{}^{J}\chi_{\beta\sigma}{}^{K}\right) \nonumber\\
 & = -\frac{1}{6}\CF_{IJKL}\veps^{\mu\nu\rho\sigma}\Phi^{\alpha}\Phi^{\beta}\Phi^{\gamma}\psi_{\gamma}{}^{I}\chi_{\mu\nu}{}^{J}\chi_{\alpha\rho}{}^{K}\chi_{\beta\sigma}{}^{L} \nonumber\\
 &\quad + \frac{1}{6}\CF_{IJK}\veps^{\mu\nu\rho\sigma}\Phi^{\alpha}\Phi^{\beta}\left(F_{\mu\nu}^{+}{}^{I} - D_{\mu\nu}{}^{I} - \big(\Psi_{[\mu}{}^{\kappa}\chi_{\nu]\kappa}{}^{I}\big)^{-}\right)\chi_{\alpha\rho}{}^{J}\chi_{\beta\sigma}{}^{K} \nonumber\\
 &\quad - \frac{1}{3}\CF_{IJK}\veps^{\mu\nu\rho\sigma}\Phi^{\alpha}\Phi^{\beta}\chi_{\mu\nu}{}^{I}\left(F_{\alpha\rho}{}^{I} - D_{\alpha\rho}{}^{J} - \big(\Psi_{[\alpha}{}^{\kappa}\chi_{\rho]\kappa}{}^{J}\big)^{-}\right)\chi_{\beta\sigma}{}^{K} \nonumber\\
  & = -\frac{1}{3}\CF_{IJKL}\veps^{\mu\nu\rho\sigma}\Phi^{\alpha}\Phi^{\beta}\Phi^{\gamma}\psi_{\mu}{}^{I}\chi_{\alpha\nu}{}^{J}\chi_{\beta\rho}{}^{K}\chi_{\gamma\sigma}{}^{L} \nonumber\\
 &\quad + \frac{1}{6}\CF_{IJK}\veps^{\mu\nu\rho\sigma}\Phi^{\alpha}\Phi^{\beta}\left(F_{\mu\nu}^{+}{}^{I} - D_{\mu\nu}{}^{I} - \big(\Psi_{[\mu}{}^{\kappa}\chi_{\nu]\kappa}{}^{I}\big)^{-}\right)\chi_{\alpha\rho}{}^{J}\chi_{\beta\sigma}{}^{K} \nonumber\\
 &\quad - \frac{1}{3}\CF_{IJK}\veps^{\mu\nu\rho\sigma}\Phi^{\alpha}\Phi^{\beta}\chi_{\mu\nu}{}^{I}\left(F_{\alpha\rho}^{+}{}^{I} - D_{\alpha\rho}{}^{J} - \big(\Psi_{[\alpha}{}^{\kappa}\chi_{\rho]\kappa}{}^{J}\big)^{-}\right)\chi_{\beta\sigma}{}^{K} ~, \label{eq:Q-deg-4}
 }
 where we used \eqref{eq:ident-for-deg-6} in the first term, which is identical to \eqref{eq:deg-6-diff-final}. Consider the degree $=4$ terms in \eqref{eq:Q-deg-4}. These are:
 \eqa{
 & \frac{1}{6}\CF_{IJK}\veps^{\mu\nu\rho\sigma}\Phi^{\alpha}\Phi^{\beta}\left(F_{\mu\nu}^{+}{}^{I}  - D_{\mu\nu}{}^{I}\right)\chi_{\alpha\rho}{}^{J}\chi_{\beta\sigma}{}^{K} - \frac{1}{3}\CF_{IJK}\veps^{\mu\nu\rho\sigma}\Phi^{\alpha}\Phi^{\beta}\chi_{\mu\nu}{}^{I}\left(F_{\alpha\rho}^{+}{}^{I} - D_{\alpha\rho}{}^{J}\right)\chi_{\beta\sigma}{}^{K} \nonumber\\
 & = \frac{e}{3}\CF_{IJK}\Phi^{\alpha}\Phi^{\beta}\big(F_{+}^{\rho\sigma}{}^{I} - D^{\rho\sigma}{}^{I}\big)\chi_{\alpha\rho}{}^{J}\chi_{\beta\sigma}{}^{K} - \frac{2e}{3}\CF_{IJK}\Phi^{\alpha}\Phi^{\beta}\left(F_{\alpha\rho}^{+}{}^{I} - D_{\alpha\rho}{}^{I}\right)\chi^{\rho\sigma}{}^{J}\chi_{\beta\sigma}{}^{K} ~.\label{eq:deg-4-terms-in-Q-deg-4}
 }
 Since $\CF_{IJK}\big(F_{+}{}^{I} - D^{I}\big)\wedge \big(\iota_{\Phi}\chi^{J}\big) \wedge \chi^{K} = 0$ on a four-manifold, we have the identity
 \eqa{
& \CF_{IJK}\iota_{\Phi}\big(F_{+}^{I} - D^{I}\big) \wedge \big(\iota_{\Phi}\chi^{J}\big)\wedge \chi^{K} - \CF_{IJK}\big(F_{+}^{I} - D^{I}\big)\wedge\big(\iota_{\Phi}{\chi}^{J}\big)\wedge \big(\iota_{\Phi}\chi^{K}\big) &&= 0 ~,
 }	
 which, in components, reads
 \eqa{
& -e\CF_{IJK}\Phi^{\alpha}\Phi^{\beta}\left(F_{\alpha\rho}^{+}{}^{I} - D_{\alpha\rho}{}^{I}\right)\chi^{\rho\sigma}{}^{J}\chi_{\beta\sigma}{}^{K} - e\CF_{IJK}\Phi^{\alpha}\Phi^{\beta}\big(F_{+}^{\rho\sigma}{}^{I} - D^{\rho\sigma}{}^{I}\big)\chi_{\alpha\rho}{}^{J}\chi_{\beta\sigma}{}^{K} &&= 0 ~.\label{eq:ident2-for-deg-4}
 }
Using \eqref{eq:ident2-for-deg-4}, \eqref{eq:deg-4-terms-in-Q-deg-4} reduces to
\eqa{
& \eqref{eq:deg-4-terms-in-Q-deg-4} &&= e\CF_{IJK}\Phi^{\alpha}\Phi^{\beta}\big(F_{+}^{\rho\sigma}{}^{I} - D^{\rho\sigma}{}^{I}\big)\chi_{\alpha\rho}{}^{J}\chi_{\beta\sigma}{}^{K} \nonumber\\
& &&= e \CF_{IJK}\Phi^{\alpha}\Phi^{\beta}\big(F_{+}^{\mu\nu}{}^{I} - D^{\mu\nu}{}^{I}\big)\chi_{\alpha\mu}{}^{J}\chi_{\beta\nu}{}^{K} ~. \label{eq:deg-4 Fchichi term in Q-deg-4}
}
Modulo the SD projection, this is similar to a term in \eqref{eq:deg-4-diff-final}. 

Consider the degree $=5$ terms in \eqref{eq:Q-deg-4}. These are:
\eqa{
& -\frac{1}{6}\CF_{IJK}\veps^{\mu\nu\rho\sigma}\Phi^{\alpha}\Phi^{\beta}\big(\Psi_{[\mu}{}^{\kappa}\chi_{\nu]\kappa}{}^{I}\big)^{-}\chi_{\alpha\rho}{}^{J}\chi_{\beta\sigma}{}^{K} + \frac{1}{3}\CF_{IJK}\veps^{\mu\nu\rho\sigma}\Phi^{\alpha}\Phi^{\beta}\chi_{\mu\nu}{}^{I}\big(\Psi_{[\alpha}{}^{\kappa}\chi_{\rho]\kappa}{}^{J}\big)^{-}\chi_{\beta\sigma}{}^{K} \nonumber\\
&= \frac{e}{3}\CF_{IJK}\Phi^{\alpha}\Phi^{\beta}\big(\Psi^{\kappa[\rho}\chi^{\sigma]}{}_{\kappa}{}^{I}\big)^{-}\chi_{\alpha\rho}{}^{J}\chi_{\beta\sigma}{}^{K} + \frac{2e}{3}\CF_{IJK}\Phi^{\alpha}\Phi^{\beta}\chi^{\rho\sigma}{}^{I}\big(\Psi_{[\alpha}{}^{\kappa}\chi_{\rho]\kappa}{}^{J}\big)^{-}\chi_{\beta\sigma}{}^{K} ~.\label{eq:deg-5-terms-in-Q-deg-4}
}
Let $\sfA_{\mu\nu}{}^{I} := \big(\Psi_{[\mu}{}^{\sigma}\chi_{\nu]\sigma}{}^{I}\big)^{-}$, so $\sfA^I$ is an anti-self-dual 2-form. Since $\CF_{IJK}\sfA^I \wedge \big(\iota_{\Phi}\chi^{J}\big)\wedge \chi^{K} = 0$ on a four-manifold, we have the identity
\eqa{
& \CF_{IJK}\big(\iota_{\Phi}\sfA^I\big)\wedge\big(\iota_{\Phi}\chi^{J}\big)\wedge \chi^{K} - \CF_{IJK}\sfA^I \wedge \big(\iota_{\Phi}\chi^{J}\big)\wedge\big(\iota_{\Phi}\chi^{K}\big) &&= 0 ~,
}
which, in components, reads
\eqa{
&-e\CF_{IJK}\Phi^{\alpha}\Phi^{\beta}\chi^{\rho\sigma}{}^{I}\big(\Psi_{[\alpha}{}^{\kappa}\chi_{\rho]\kappa}{}^{J}\big)^{-}\chi_{\beta\sigma}{}^{K} + e\CF_{IJK}\Phi^{\alpha}\Phi^{\beta}\big(\Psi^{\kappa[\rho}\chi^{\sigma]}{}_{\kappa}{}^{I}\big)^{-}\chi_{\alpha\rho}{}^{J}\chi_{\beta\sigma}{}^{K} &&= 0 ~.\label{eq:ident-for-deg-5}
}
Notice that this identity is similar to \eqref{eq:ident-for-deg-4}, the only difference being a sign, which is due to the anti-self-duality of the 2-form $\sfA$ in this case, as opposed to the self-duality of $(F_{+} - D)$.\\\\ 
Using \eqref{eq:ident-for-deg-5}, \eqref{eq:deg-5-terms-in-Q-deg-4} reduces to
\eqa{
& \eqref{eq:deg-5-terms-in-Q-deg-4} &&= e\CF_{IJK}\Phi^{\alpha}\Phi^{\beta}\big(\Psi^{\kappa[\rho}\chi^{\sigma]}{}_{\kappa}{}^{I}\big)^{-}\chi_{\alpha\rho}{}^{J}\chi_{\beta\sigma}{}^{K} = e\,\Phi^{\alpha}\Phi^{\beta}\CF_{IJK}\big(\Psi^{\rho[\mu}\chi^{\nu]}{}_{\rho}{}^{I}\big)^{-}\chi_{\alpha\mu}{}^{J}\chi_{\beta\nu}{}^{K} ~,
}
which is identical to \eqref{eq:deg-5-diff-final}. To summarize, \eqref{eq:Q-deg-4} can be written as
\beqa{
& \IQ\left(\frac{1}{6}\CF_{IJKL}\veps^{\mu\nu\rho\sigma}\Phi^{\alpha}\Phi^{\beta}\chi_{\mu\nu}{}^{I}\chi_{\alpha\rho}{}^{J}\chi_{\beta\sigma}{}^{K}\right) \nonumber\\
&= -\frac{1}{3}\CF_{IJKL}\veps^{\mu\nu\rho\sigma}\Phi^{\alpha}\Phi^{\beta}\Phi^{\gamma}\psi_{\mu}{}^{I}\chi_{\alpha\nu}{}^{J}\chi_{\beta\rho}{}^{K}\chi_{\gamma\sigma}{}^{L} + e\,\Phi^{\alpha}\Phi^{\beta}\CF_{IJK}\big(\Psi^{\rho[\mu}\chi^{\nu]}{}_{\rho}{}^{I}\big)^{-}\chi_{\alpha\mu}{}^{J}\chi_{\beta\nu}{}^{K}\nonumber\\
&\quad + e\,\Phi^{\alpha}\Phi^{\beta} \CF_{IJK}\big(F_{+}^{\mu\nu}{}^{I} - D^{\mu\nu}{}^{I}\big)\chi_{\alpha\mu}{}^{J}\chi_{\beta\nu}{}^{K} ~. \label{eq:Q-deg-4-final}
}
So we have been able to reproduce exactly the degree $=5$ and degree $=6$ differences, \eqref{eq:deg-5-diff-final} and \eqref{eq:deg-6-diff-final} respectively, an encouraging sign! Next, consider
\eqa{
& \IQ\left(\frac{1}{2}\CF_{IJK}\veps^{\mu\nu\rho\sigma}\Phi^{\alpha}\chi_{\alpha\mu}{}^{I}\chi_{\nu\rho}{}^{J}\psi_{\sigma}{}^{K}\right) \nonumber\\
& = -\frac{1}{2}\CF_{IJKL}\veps^{\mu\nu\rho\sigma}\Phi^{\alpha}\Phi^{\beta}\chi_{\alpha\mu}{}^{I}\chi_{\nu\rho}{}^{J}\psi_{\beta}{}^{K}\psi_{\sigma}{}^{L} \nonumber\\
&\quad + \frac{1}{2}\CF_{IJK}\veps^{\mu\nu\rho\sigma}\Phi^{\alpha}\big(F_{\alpha\mu}^{+}{}^{I} - D_{\alpha\mu}{}^{I} - \big(\Psi_{[\alpha}{}^{\kappa}\chi_{\mu]\kappa}{}^{I}\big)^{-}\big)\chi_{\nu\rho}{}^{J}\psi_{\sigma}{}^{K} \nonumber\\
&\quad - \frac{1}{2}\CF_{IJK}\veps^{\mu\nu\rho\sigma}\Phi^{\alpha}\chi_{\alpha\mu}{}^{I}\big(F_{\nu\rho}^{+}{}^{J} - D_{\nu\rho}{}^{J} - \big(\Psi_{[\nu}{}^{\kappa}\chi_{\rho]\kappa}{}^{I}\big)^{-}\big)\psi_{\sigma}{}^{K} \nonumber\\
&\quad + \frac{1}{2}\CF_{IJK}\veps^{\mu\nu\rho\sigma}\Phi^{\alpha}\chi_{\alpha\mu}{}^{I}\chi_{\nu\rho}{}^{J}\big(-D_{\sigma}\phi^{K} + \Phi^{\beta}F_{\beta\sigma}{}^{K}\big) ~.\label{eq:Q-deg-2}
}
Consider the degree $=4$ term in \eqref{eq:Q-deg-2}. This is
\eqa{
& -\frac{1}{2}\CF_{IJKL}\veps^{\mu\nu\rho\sigma}\Phi^{\alpha}\Phi^{\beta}\chi_{\alpha\mu}{}^{I}\chi_{\nu\rho}{}^{J}\psi_{\beta}{}^{K}\psi_{\sigma}{}^{L} ~. \label{eq:deg-4-term-in-Q-deg-2}
}
Since $\CF_{IJKL}\big(\iota_{\Phi}\chi^{I}\big)\wedge \chi^{J} \wedge \psi^{K} \wedge \psi^{L} = 0$ on a four-manifold, we have the identity
\eqa{
&-\CF_{IJKL}\big(\iota_{\Phi}\chi^{I}\big)\wedge\big(\iota_{\Phi}\chi^{J}\big)\wedge \psi^{K} \wedge \psi^{L} - 2\,\CF_{IJKL}\big(\iota_{\Phi}\chi^{I}\big)\wedge\chi^{J}\wedge\big(\iota_{\Phi}\psi^{K}\big)\psi^{L} &&= 0 ~,
}	
which, in components, reads
\eqa{
&-\CF_{IJKL}\veps^{\mu\nu\rho\sigma}\Phi^{\alpha}\Phi^{\beta}\chi_{\alpha\mu}{}^{I}\chi_{\beta\nu}{}^{J}\psi_{\rho}{}^{K}\psi_{\sigma}{}^{L} - \CF_{IJKL}\veps^{\mu\nu\rho\sigma}\Phi^{\alpha}\Phi^{\beta}\chi_{\alpha\mu}{}^{I}\chi_{\nu\rho}{}^{J}\psi_{\beta}{}^{K}\psi_{\sigma}{}^{L} &&= 0 ~. \label{eq:ident-for-deg-4}
}
Using \eqref{eq:ident-for-deg-4}, \eqref{eq:deg-4-term-in-Q-deg-2} reduces to
\eqa{
& \eqref{eq:deg-4-term-in-Q-deg-2} &&= +\frac{1}{2}\Phi^{\alpha}\Phi^{\beta}\veps^{\mu\nu\rho\sigma}\CF_{IJKL}\Phi^{\alpha}\Phi^{\beta}\chi_{\alpha\mu}{}^{I}\chi_{\beta\nu}{}^{J}\psi_{\rho}{}^{K}\psi_{\sigma}{}^{L} \nonumber\\
& &&= +\frac{1}{2}\Phi^{\alpha}\Phi^{\beta}\veps^{\mu\nu\rho\sigma}\CF_{IJKL}\psi_{\mu}{}^{I}\psi_{\nu}{}^{J}\chi_{\alpha\rho}{}^{K}\chi_{\beta\sigma}{}^{L} ~,
}
which is precisely the last degree $=4$ term in \eqref{eq:deg-4-diff-final}, another encouraging sign! 

Consider the $D_{\sigma}\phi$ term in \eqref{eq:Q-deg-2}. This is 
\eqa{
& -\frac{1}{2}\CF_{IJK}\veps^{\mu\nu\rho\sigma}\Phi^{\alpha}\chi_{\alpha\mu}{}^{I}\chi_{\nu\rho}{}^{J}D_{\sigma}\phi^{K} &&= -e\,\CF_{IJK}\Phi^{\alpha}\chi_{\alpha\mu}{}^{I}\chi^{\mu\sigma}{}^{J}\big(D_{\sigma}\phi^{K}\big) \nonumber\\
& &&= e\,\CF_{IJK}\Phi^{\sigma}\big(D_{\mu}\phi^{I}\big)\chi_{\sigma\nu}{}^{J}\chi^{\mu\nu}{}^{K} ~,
}
which is identical to the last degree $=2$ term in \eqref{eq:deg-4-diff-final}.

Next, consider the $F\psi\chi$ terms in \eqref{eq:Q-deg-2}. These are
\eqa{
& + \frac{1}{2}\CF_{IJK}\veps^{\mu\nu\rho\sigma}\Phi^{\alpha}F_{\alpha\mu}^{+}{}^{I}\chi_{\nu\rho}{}^{J}\psi_{\sigma}{}^{K} - \frac{1}{2}\CF_{IJK}\veps^{\mu\nu\rho\sigma}\Phi^{\alpha}\chi_{\alpha\mu}{}^{I}F_{\nu\rho}^{+}{}^{J}\psi_{\sigma}{}^{K} ~. \label{eq:Fpsichi in Q-deg-2}
}
Since $\CF_{IJK}F_{+}^{I}\wedge \psi^{J} \wedge \chi^{K} = 0$ on a four-manifold, we have the identity
\eqa{
&\CF_{IJK}\big(\iota_{\Phi}F_{+}{}^{I}\big)\wedge \psi^{J} \wedge \chi^{K} + \CF_{IJK}F_{+}{}^{I}\wedge\big(\iota_{\Phi}\psi^{J}\big)\wedge \chi^{K} - \CF_{IJK}F_{+}{}^{I}\wedge \psi^{J} \wedge \big(\iota_{\Phi}\chi^{K}\big) &&= 0 ~,
}	
which, in components, reads
\eqa{
e\,\Phi^{\sigma}\CF_{IJK}F_{\sigma\mu}^{+}{}^{I}\psi_{\nu}{}^{J}\chi^{\mu\nu}{}^{J} + \frac{1}{2}e\,\CF_{IJK}\Phi^{\sigma}F_{\mu\nu}^{+}{}^{I}\psi_{\sigma}{}^{J}\chi^{\mu\nu}{}^{J} - e\,\Phi^{\sigma}\CF_{IJK}F_{\mu\nu}^{+}{}^{I}\psi^{\mu}{}^{J}\chi_{\sigma}{}^{\nu}{}^{K} &&= 0 ~. \label{eq:ident-for-deg-2}
}
This identity should be compared with \eqref{eq:Fpsichi identity 1}.\footnote{Note that a similar identity clearly holds with $F_{\mu\nu}^{+}$ replaced by $D_{\mu\nu}$.}\\\\
Using \eqref{eq:ident-for-deg-2}, \eqref{eq:Fpsichi in Q-deg-2} reduces to
\eqa{
& \eqref{eq:Fpsichi in Q-deg-2} &&= \frac{1}{2}e\,\Phi^{\sigma}\CF_{IJK}F_{\mu\nu}^{+}{}^{I}\big(\psi_{\sigma}{}^{J}\chi^{\mu\nu}{}^{K} - 4 \psi^{\mu}{}^{J}\chi_{\sigma}{}^{\nu}{}^{K}\big) ~,
}
which is identical to the penultimate degree $=2$ term in \eqref{eq:deg-2-diff-final}.

Next, consider the $D\psi\chi$ terms in \eqref{eq:Q-deg-2}, where $D$ is the auxiliary field. These are
\eqa{
& -\frac{1}{2}\CF_{IJK}\veps^{\mu\nu\rho\sigma}\Phi^{\alpha}D_{\alpha\mu}{}^{I}\chi_{\nu\rho}{}^{J}\psi_{\sigma}{}^{K} + \frac{1}{2}\CF_{IJK}\veps^{\mu\nu\rho\sigma}\Phi^{\alpha}\chi_{\alpha\mu}{}^{I}D_{\nu\rho}{}^{J}\psi_{\sigma}{}^{K} \nonumber\\
&= -e\,\CF_{IJK}\Phi^{\alpha}D_{\alpha\mu}{}^{I}\chi^{\mu\sigma}{}^{J}\psi_{\sigma}{}^{K} + e\,\CF_{IJK}\Phi^{\alpha}\chi_{\alpha\mu}{}^{I}D^{\mu\sigma}{}^{J}\psi_{\sigma}{}^{K} \nonumber\\
&= 2\,e\,\CF_{IJK}\Phi^{\sigma}\chi_{\sigma\mu}{}^{I}D^{\mu\nu}{}^{J}\psi_{\nu}{}^{K} - \frac{1}{2}e\,\CF_{IJK}\Phi^{\sigma}D^{\mu\nu}{}^{I}\psi_{\sigma}{}^{J}\chi_{\mu\nu}{}^{K} \nonumber\\
&= \frac{1}{2}e\,\CF_{IJK}D^{\mu\nu}{}^{I}\big(4\chi_{\sigma\mu}{}^{J}\psi_{\nu}{}^{K} - \psi_{\sigma}{}^{J}\chi_{\mu\nu}{}^{K}\big) ~,
}
where, in the penultimate step, we used \eqref{eq:ident-for-deg-2} (for $F_{\mu\nu}^{+} \to D_{\mu\nu}$). The result is identical to the third to last degree $=2$ term in \eqref{eq:deg-2-diff-final}. By now, we have reproduced the last three lines of the degree $=2$ terms in \eqref{eq:deg-2-diff-final} through $\IQ$-exact contributions.

Next, consider the degree $=3$ terms in \eqref{eq:Q-deg-2}. These are
\eqa{
& -\frac{1}{2}\CF_{IJK}\veps^{\mu\nu\rho\sigma}\Phi^{\alpha}\big(\Psi_{[\alpha}{}^{\kappa}\chi_{\mu]\kappa}{}^{I}\big)^{-}\chi_{\nu\rho}{}^{J}\psi_{\sigma}{}^{K} + \frac{1}{2}\CF_{IJK}\veps^{\mu\nu\rho\sigma}\Phi^{\alpha}\chi_{\alpha\mu}{}^{I}\big(\Psi_{[\nu}{}^{\kappa}\chi_{\rho]\kappa}{}^{J}\big)^{-}\psi_{\sigma}{}^{K} \nonumber\\
&= -e\,\CF_{IJK}\Phi^{\alpha}\big(\Psi_{[\alpha}{}^{\kappa}\chi_{\mu]\kappa}{}^{I}\big)^{-}\chi^{\mu\sigma}{}^{J}\psi_{\sigma}{}^{K} - \CF_{IJK}\Phi^{\alpha}\chi_{\alpha\mu}{}^{I}\big(\Psi^{\kappa[\mu}\chi^{\sigma]}{}_{\kappa}{}^{J}\big)^{-}\psi_{\sigma}{}^{K} ~.\label{eq:deg-3 in Q-deg-2}
}
For a generic 2-form $\sfA$ on a four-manifold, $\CF_{IJK}\sfA^{I}\wedge\chi^{J}\wedge \psi^{K} = 0$, and hence
\eqa{
&\CF_{IJK}\big(\iota_{\Phi}\sfA^I\big)\wedge \chi^{J} \wedge \psi^{K} + \CF_{IJK}\sfA^{I} \wedge \big(\iota_{\Phi}\chi^{J}\big)\wedge \psi^{K} + \CF_{IJK}\sfA^{I} \wedge \chi^{J}\wedge \big(\iota_{\Phi}\psi^{K}\big) &&= 0 ~.
}	
With $\sfA_{\mu\nu} := \big(\Psi_{[\mu}{}^{\sigma}\chi_{\nu]\sigma}\big)^{-}$, the third term vanishes as it involves a contraction of a self-dual and an anti-self-dual tensor. In this case, this identity reduces to
\eqa{
& e\,\CF_{IJK}\Phi^{\alpha}\big(\Psi_{[\alpha}{}^{\kappa}\chi_{\mu]\kappa}{}^{I}\big)^{-} &&= e\,\CF_{IJK}\Phi^{\alpha}\chi_{\alpha\mu}{}^{J}\big(\Psi^{\kappa[\mu}\chi^{\nu]}{}{\kappa}{}^{J}\big)^{-}\psi_{\sigma}{}^{K} ~. \label{eq:ident for deg-3 in Q-deg-2}
}
Using \eqref{eq:ident for deg-3 in Q-deg-2}, \eqref{eq:deg-3 in Q-deg-2} reduces to
\eqa{
& \eqref{eq:deg-3 in Q-deg-2} &&= - 2\,e\,\CF_{IJK}\Phi^{\alpha}\chi_{\alpha\mu}{}^{I}\big(\Psi^{\kappa[\mu}\chi^{\sigma]}{}_{\kappa}{}^{J}\big)^{-}\psi_{\sigma}{}^{K} ~\nonumber\\
& &&= -2\,e\,\Phi^{\alpha}\CF_{IJK}\big(\Psi^{\rho[\mu}\chi^{\nu]}{}_{\rho}{}^{I}\big)^{-}\chi_{\alpha\mu}{}^{J}\psi_{\nu}{}^{K} ~,
}
which is identical to the last degree $=3$ term in \eqref{eq:deg-3-diff-final}.

The last degree $=4$ term ($\Phi\Phi F$) in \eqref{eq:Q-deg-2} is
\eqa{
& \frac{1}{2}\CF_{IJK}\veps^{\mu\nu\rho\sigma}\Phi^{\alpha}\Phi^{\beta}\chi_{\alpha\mu}{}^{I}\chi_{\nu\rho}{}^{J}F_{\beta\sigma}{}^{K} \nonumber\\
&= e\,\CF_{IJK}\Phi^{\alpha}\Phi^{\beta}\chi_{\alpha\mu}{}^{I}\chi^{\mu\sigma}{}^{J}F_{\beta\sigma}{}^{K} \nonumber\\
&= e\,\CF_{IJK}\Phi^{\alpha}\Phi^{\beta}F_{\alpha\rho}{}^{I}\chi_{\beta\mu}{}^{J}\chi^{\mu\rho}{}^{K} \nonumber\\
&= e\,\CF_{IJK}\Phi^{\alpha}\Phi^{\beta}F_{\alpha\rho}{}^{I}\chi^{\rho\sigma}{}^{J}\chi_{\beta\sigma}{}^{K} \nonumber\\
&= e\,\CF_{IJK}\Phi^{\alpha}\Phi^{\beta}F_{\alpha\rho}^{+}{}^{I}\chi^{\rho\sigma}{}^{J}\chi_{\beta\sigma}{}^{K} + e\,\CF_{IJK}\Phi^{\alpha}\Phi^{\beta}F_{\alpha\rho}^{-}{}^{I}\chi^{\rho\sigma}{}^{J}\chi_{\beta\sigma}{}^{K} \nonumber\\
&= -e\,\CF_{IJK}\Phi^{\alpha}\Phi^{\beta}F_{+}^{\mu\nu}{}^{I}\chi_{\alpha\mu}{}^{J}\chi_{\beta\nu}{}^{K} + e\,\CF_{IJK}\Phi^{\alpha}\Phi^{\beta}F_{-}^{\mu\nu}{}^{I}\chi_{\alpha\mu}{}^{J}\chi_{\beta\nu}{}^{K} ~,
}
where, in the final step, we have used a version of \eqref{eq:ident2-for-deg-4} on the first term, and a version of \eqref{eq:ident-for-deg-5} on the second term. Note that this, when added to the contribution \eqref{eq:deg-4 Fchichi term in Q-deg-4} from \eqref{eq:Q-deg-4}, actually cancels the $F_{+}$-dependence of this term, and produces the desired $(F_{-} - D)$ dependence seen in \eqref{eq:deg-4-diff-final}. To summarize, \eqref{eq:Q-deg-2} is
\beqa{
& \IQ\left(\frac{1}{2}\CF_{IJK}\veps^{\mu\nu\rho\sigma}\Phi^{\alpha}\chi_{\alpha\mu}{}^{I}\chi_{\nu\rho}{}^{J}\psi_{\sigma}{}^{K}\right) \nonumber\\
&= \frac{1}{2}\Phi^{\alpha}\Phi^{\beta}\veps^{\mu\nu\rho\sigma}\CF_{IJKL}\psi_{\mu}{}^{I}\psi_{\nu}{}^{J}\chi_{\alpha\rho}{}^{K}\chi_{\beta\sigma}{}^{L}  -e\,\CF_{IJK}\Phi^{\alpha}\Phi^{\beta}F_{+}^{\mu\nu}{}^{I}\chi_{\alpha\mu}{}^{J}\chi_{\beta\nu}{}^{K} \nonumber\\
&\quad + e\,\CF_{IJK}\Phi^{\alpha}\Phi^{\beta}F_{-}^{\mu\nu}{}^{I}\chi_{\alpha\mu}{}^{J}\chi_{\beta\nu}{}^{K} -2\,e\,\Phi^{\alpha}\CF_{IJK}\big(\Psi^{\rho[\mu}\chi^{\nu]}{}_{\rho}{}^{I}\big)^{-}\chi_{\alpha\mu}{}^{J}\psi_{\nu}{}^{K}\nonumber\\
&\quad + \frac{1}{2}e\,\Phi^{\sigma}\CF_{IJK}F_{\mu\nu}^{+}{}^{I}\big(\psi_{\sigma}{}^{J}\chi^{\mu\nu}{}^{K} - 4 \psi^{\mu}{}^{J}\chi_{\sigma}{}^{\nu}{}^{K}\big) \nonumber\\
&\quad + \frac{1}{2}e\,\Phi^{\sigma}\CF_{IJK}D^{\mu\nu}{}^{I}\big(4\chi_{\sigma\mu}{}^{J}\psi_{\nu}{}^{K} - \psi_{\sigma}{}^{J}\chi_{\mu\nu}{}^{K}\big) \nonumber\\
&\quad + e\,\CF_{IJK}\Phi^{\sigma}\big(D_{\mu}\phi^{I}\big)\chi_{\sigma\nu}{}^{J}\chi^{\mu\nu}{}^{K} ~. \label{eq:Q-deg-2-final}
}
Next, consider
\eqa{
&\IQ\left(\CF_{IJ}\Phi_{\mu}\veps^{\mu\nu\rho\sigma}\chi_{\rho\sigma}{}^{I}D_{\nu}\lambda{}^{J}\right) &&= -\CF_{IJK}\Phi^{\alpha}\Phi_{\mu}\veps^{\mu\nu\rho\sigma}\psi_{\alpha}{}^{I}\chi_{\rho\sigma}{}^{I}D_{\nu}\lambda^{J} + \CF_{IJ}\Phi^{\alpha}\Psi_{\alpha\mu}\veps^{\mu\nu\rho\sigma}\chi_{\rho\sigma}{}^{I}D_{\nu}\lambda^{J} \nonumber\\
& &&\quad + \CF_{IJ}\Phi_{\mu}\veps^{\mu\nu\rho\sigma}\big( F_{\rho\sigma}^{+}{}^{I} - D_{\rho\sigma}{}^{I} - \big( \Psi_{[\rho}{}^{\kappa}\chi_{\sigma]\kappa}{}^{I}\big)^{-} \big)D_{\nu}\lambda^{J} \nonumber\\
& &&\quad - \CF_{IJ}\Phi_{\mu}\veps^{\mu\nu\rho\sigma}\chi_{\rho\sigma}{}^{I}\big(D_{\nu}\eta^{J} + [\psi_{\nu}, \lambda]^{J} \big)\nonumber\\
& &&= -2\,e\,\CF_{IJK}\Phi^{\alpha}\Phi_{\mu}\psi_{\alpha}{}^{I}\chi^{\mu\nu}{}^{I}D_{\nu}\lambda^{J} + 2\,e\,\CF_{IJ}\Phi^{\alpha}\Psi_{\alpha\mu}\chi^{\mu\nu}{}^{I}D_{\nu}\lambda^{J} \nonumber\\
& &&\quad + 2\,e\,\CF_{IJ}\Phi_{\mu}\big(F_{+}^{\mu\nu}{}^{I} - D^{\mu\nu}{}^{I} + \big(\Psi^{\kappa[\mu}\chi^{\nu]}{}_{\kappa}{}^{I}\big)^{-} \big)D_{\nu}\lambda^{J} \nonumber\\
& &&\quad - 2\,e\,\CF_{IJ}\Phi_{\mu}\chi^{\mu\nu}{}^{I}\big(D_{\nu}\eta^{J} + [\psi_{\nu}, \lambda]^{J}\big) ~. \label{eq:Q-deg-2-II}
}
The degree $=4$ term in \eqref{eq:Q-deg-2-II} is
\eqa{
& -2\,e\,\CF_{IJK}\Phi^{\alpha}\Phi_{\mu}\psi_{\alpha}{}^{I}\chi^{\mu\nu}{}^{I}D_{\nu}\lambda^{J} &&= -2\,e\,\CF_{IJK}\Phi^{\sigma}\psi_{\sigma}{}^{I}\big(\Phi_{[\mu}D_{\nu]}\lambda^{I}\big)\chi^{\mu\nu}{}^{J} ~,
}
which is identical to the second degree $=4$ term in \eqref{eq:deg-4-diff-final}.
\\\\
Next, consider the degree $=3$ terms in \eqref{eq:Q-deg-2-II}. These are
\eqa{
& + 2\,e\,\CF_{IJ}\Phi^{\alpha}\Psi_{\alpha\mu}\chi^{\mu\nu}{}^{I}D_{\nu}\lambda^{J} + 2\,e\,\CF_{IJ}\Phi_{\mu}\big(\Psi^{\sigma[\mu}\chi^{\nu]}{}_{\sigma}{}^{I}\big)^{-}D_{\nu}\lambda^{J} ~. \label{eq:deg-3 in Q-deg-2-II}
}
We claim that the sum of these two terms is precisely equal to the sum of the first three degree $=3$ terms in \eqref{eq:deg-3-diff-final}. First, we describe an obvious rewriting of a degree $=3$ term that is common to both sets of expressions. 
Note that $\CF_{IJ}\sfA_{\mu\nu}^{-}{}^{I}\sfB^{\mu\nu}_{-}{}^{J} = \CF_{IJ}\sfA_{\mu\nu}{}^{I}\sfB^{\mu\nu}{}^{J} - \CF_{IJ}\sfA_{\mu\nu}^{+}{}^{I}\sfB^{\mu\nu}_{+}{}^{J}$ for two 2-forms $\sfA$ and $\sfB$. Applying this with $\sfA_{\mu\nu} := \Phi_{[\mu}D_{\nu]}\lambda$ and $\sfB_{\mu\nu} := \Psi_{[\mu}{}^{\rho}\chi_{\nu]\rho}$, we have
\eqa{
& -2\,e\,\CF_{IJ}\big(\Phi_{[\mu}D_{\nu]}\lambda^{I}\big)^{-}\big(\Psi^{\rho[\mu}\chi^{\nu]}{}_{\rho}{}^{J}\big)^{-} \nonumber\\
&= -2\,e\,\CF_{IJ}\Phi_{\mu}\big(\Psi^{\rho[\mu}\chi^{\nu]}{}_{\rho}{}^{I}\big)D_{\nu}\lambda^{I} + 2\,e\,\CF_{IJ}\big(\Phi_{[\mu}D_{\nu]}\lambda^{I}\big)^{+}\big(-\tfrac{1}{4}\Psi_{\rho}{}^{\rho}\chi^{\mu\nu}{}^{J}\big) \nonumber\\
&= -e\,\CF_{IJ}\Phi_{\mu}\Psi^{\rho\mu}\chi^{\nu}{}_{\rho}{}^{I}D_{\nu}\lambda^{J} + e\,\CF_{IJ}\Phi_{\mu}\Psi^{\rho\nu}\chi^{\mu}{}_{\rho}{}^{I}D_{\nu}\lambda^{J} - \frac{e}{2}\CF_{IJ}\Phi_{\mu}\big(D_{\nu}\lambda^{I}\big)\Psi_{\rho}{}^{\rho}\chi^{\mu\nu}{}^{J} \nonumber\\
&= e\,\CF_{IJ}\Phi^{\alpha}\Psi_{\alpha\mu}\chi^{\mu\nu}{}^{I}D_{\nu}\lambda^{J} + e\,\CF_{IJ}\Psi^{\mu\nu}\Phi^{\sigma}\big(D_{\mu}\lambda^{I}\big)\chi_{\sigma\nu}{}^{J} - \frac{e}{2}\CF_{IJ}\Psi_{\rho}{}^{\rho}\Phi^{\sigma}\big(D_{\mu}\lambda^{I}\big)\chi_{\sigma}{}^{\mu}{}^{J} ~. \label{eq:ident-for-deg-3 in Q-deg-2-II}
}
Using \eqref{eq:ident-for-deg-3 in Q-deg-2-II}, the sum of the first three degree $=3$ terms in \eqref{eq:deg-3-diff-final} evaluates to
\eqa{
& e\,\CF_{IJ}\Phi^{\alpha}\Psi_{\alpha\mu}\chi^{\mu\nu}{}^{I}D_{\nu}\lambda^{J} - e\,\CF_{IJ}\Psi^{\mu\nu}\Phi^{\sigma}\big(D_{\mu}\lambda^{I}\big)\chi_{\sigma\nu}{}^{J} + \frac{e}{2}\CF_{IJ}\Psi_{\rho}{}^{\rho}\Phi^{\sigma}\big(D_{\mu}\lambda^{I}\big)\chi_{\sigma}{}^{\mu}{}^{J} ~. \label{eq:first 3 deg 3}
}
And using \eqref{eq:ident-for-deg-3 in Q-deg-2-II}, \eqref{eq:deg-3 in Q-deg-2-II} reduces to
\eqa{
& \eqref{eq:deg-3 in Q-deg-2-II} &&= -e\,\CF_{IJ}\Phi^{\alpha}\Psi_{\alpha\mu}\chi^{\mu\nu}{}^{I}D_{\nu}\lambda^{J} - e\,\CF_{IJ}\Psi^{\mu\nu}\Phi^{\sigma}\big(D_{\mu}\lambda^{I}\big)\chi_{\sigma\nu}{}^{J} + \frac{e}{2}\CF_{IJ}\Psi_{\rho}{}^{\rho}\Phi^{\sigma}\big(D_{\mu}\lambda^{I}\big)\chi_{\sigma}{}^{\mu}{}^{J} \nonumber\\
& && \quad + 2\,e\,\CF_{IJ}\Phi^{\alpha}\Psi_{\alpha\mu}\chi^{\mu\nu}{}^{I}D_{\nu}\lambda^{J} \nonumber\\
& &&= e\,\CF_{IJ}\Phi^{\alpha}\Psi_{\alpha\mu}\chi^{\mu\nu}{}^{I}D_{\nu}\lambda^{J} - e\,\CF_{IJ}\Psi^{\mu\nu}\Phi^{\sigma}\big(D_{\mu}\lambda^{I}\big)\chi_{\sigma\nu}{}^{J} + \frac{e}{2}\CF_{IJ}\Psi_{\rho}{}^{\rho}\Phi^{\sigma}\big(D_{\mu}\lambda^{I}\big)\chi_{\sigma}{}^{\mu}{}^{J} ~,
}
which is identical to \eqref{eq:first 3 deg 3}. So the degree $=3$ terms in \eqref{eq:Q-deg-2-II} are as they should be.

Finally consider the degree $=2$ terms in \eqref{eq:Q-deg-2-II}. These are 
\eqa{
& 2\,e\,\CF_{IJ}\Phi_{\mu}\big(F_{+}^{\mu\nu}{}^{I} - D^{\mu\nu}{}^{I}\big)D_{\nu}\lambda^{J} - 2\,e\,\CF_{IJ}\Phi_{\mu}\chi^{\mu\nu}{}^{I}D_{\nu}\eta^{J} - 2\,e\,\CF_{IJ}\Phi_{\mu}\chi^{\mu\nu}{}^{I}[\psi_{\nu},\lambda]^{J} \nonumber\\
&= e\,\Phi^{\sigma}\left(2\,\CF_{IJ}\big(F^{+}_{\sigma\nu}{}^{I} - D_{\sigma\nu}{}^{I}\big)D^{\nu}\lambda^{J} - 2\,\CF_{IJ}\chi_{\sigma\mu}{}^{I}D^{\mu}\eta^{J} - 2\,\CF_{IJ}\chi_{\sigma}{}^{\mu}{}^{I}[\psi_{\mu},\lambda]^{J} \right) ~,
}
reproducing the second, fourth and fifth degree $=2$ terms in \eqref{eq:deg-2-diff-final}. Therefore, \eqref{eq:Q-deg-2-II} is
\beqa{
& \IQ\left(\CF_{IJ}\Phi_{\mu}\veps^{\mu\nu\rho\sigma}\chi_{\rho\sigma}{}^{I}D_{\nu}\lambda{}^{J}\right) \nonumber\\
&= -2\,e\,\CF_{IJK}\Phi^{\sigma}\psi_{\sigma}{}^{I}\big(\Phi_{[\mu}D_{\nu]}\lambda^{I}\big)\chi^{\mu\nu}{}^{J} \nonumber\\
& \quad - 2\,e\,\CF_{IJ}\big(\Phi_{[\mu}D_{\nu]}\lambda^{I}\big)^{-}\big(\Psi^{\rho[\mu}\chi^{\nu]}{}_{\rho}{}^{J}\big)^{-} - 2\,e\,\CF_{IJ}\Psi^{\mu\nu}\Phi^{\sigma}\big(D_{\mu}\lambda^{I}\big)\chi_{\sigma\nu}{}^{J}\nonumber\\
& \quad + e\,\CF_{IJ}\Psi_{\rho}{}^{\rho}\Phi^{\sigma}\big(D_{\mu}\lambda^{I}\big)\chi_{\sigma}{}^{\mu}{}^{J} \nonumber\\
&\quad + e\,\Phi^{\sigma}\left(2\,\CF_{IJ}\big(F^{+}_{\sigma\nu}{}^{I} - D_{\sigma\nu}{}^{I}\big)D^{\nu}\lambda^{J} - 2\,\CF_{IJ}\chi_{\sigma\mu}{}^{I}D^{\mu}\eta^{J} - 2\,\CF_{IJ}\chi_{\sigma}{}^{\mu}{}^{I}[\psi_{\mu},\lambda]^{J} \right) ~, \label{eq:Q-deg-2-II-final}
}
where we used \eqref{eq:first 3 deg 3} to write the degree $=3$ terms in the same form as they appear in \eqref{eq:deg-3-diff-final}.

To summarize the findings of \eqref{eq:deg-0-diff-final}, \eqref{eq:deg-1-diff-final} \eqref{eq:deg-2-diff-final}, \eqref{eq:deg-3-diff-final}, \eqref{eq:deg-4-diff-final}, \eqref{eq:deg-5-diff-final}, \eqref{eq:deg-6-diff-final}, and \eqref{eq:Q-deg-4-final}, \eqref{eq:Q-deg-2-final}, \eqref{eq:Q-deg-2-II-final}, we have
\beqas{
&\left.e\sfC_{-}\right|^{\text{shifted}} - \big(\IQ\mathsf{V} + \sfC\big)^{\text{unbarred}} &&= \partial_{\mu}\mathscr{P}^{\mu} + \IQ\Delta_{\mathsf{diff}} ~,
}
where
\beqas{\label{eq:DiffActionAppendix}
& \widetilde{\mathscr{P}}^{\mu} &&:= -2\,e\,\CF_{I}D^{\mu}\lambda^{I} + 2\,e\,\CF_{IJ}\psi_{\nu}{}^{I}\chi^{\mu\nu}{}^{J} + e\,\CF_{IJ}\Phi^{\sigma}\chi_{\nu\sigma}{}^{I}\chi^{\mu\nu}{}^{J} ~,\\
& \widetilde{\mathsf{\Delta}}_{\mathsf{diff}} &&:= 2\,e\,\CF_{IJ}\Phi_{\mu}\chi^{\mu\nu}{}^{I}D_{\nu}\lambda{}^{J} + e\,\CF_{IJK}\Phi^{\alpha}\chi_{\alpha\mu}{}^{I}\chi^{\mu\sigma}{}^{J}\psi_{\sigma}{}^{K} + \tfrac{1}{3}e\,\CF_{IJKL}\Phi^{\alpha}\Phi^{\beta}\chi^{\rho\sigma}{}^{I}\chi_{\alpha\rho}{}^{J}\chi_{\beta\sigma}{}^{K} ~.
}
In Section \ref{sec:CompareActions}, we define $\mathscr{P}^{\mu} := e^{-1}\widetilde{\mathscr{P}}^{\mu}$ and $\Delta_{\mathsf{diff}} := \int_{\IX}d^{4}x\,\widetilde{\Delta}_{\mathsf{diff}}$.

\cleardoublepage
\phantomsection
\addcontentsline{toc}{section}{References}
\bibliographystyle{plainurl}  
\bibliography{FamilyDonaldsonRefs}    

\end{document}